\documentclass[
12pt, 
english, 
singlespacing, 
headsepline, 
openany]{MastersDoctoralThesis} 
\usepackage{newclude}
\usepackage[square,super, comma, numbers, sort&compress]{natbib}
\usepackage[utf8]{inputenc} 
\usepackage[T1]{fontenc} 
\usepackage{babel}
\usepackage{slantsc}

\usepackage{mathpazo} 
\usepackage{xcolor}
\colorlet{linkequation}{blue}
\usepackage[colorlinks]{hyperref}

\usepackage{etoolbox}
\usepackage[nottoc,notlot,notlof]{tocbibind}
\usepackage[retainorgcmds]{IEEEtrantools}
\usepackage{bm}
\usepackage{enumitem}
\usepackage{amssymb,amsmath,amsfonts,amsthm}
\usepackage{xspace}
\usepackage[sectionbib]{chapterbib}

\usepackage{minitoc}
\usepackage[Lenny]{fncychap}
\ChNameVar{\Large\fontfamily{put}\selectfont\color{black}}
\ChTitleVar{\Huge\fontfamily{put}\selectfont\color{black}}
\usepackage{float,graphicx}
\usepackage{afterpage}
\usepackage{titling,lipsum}

\usepackage[autostyle=true]{csquotes} 
\usepackage{setspace}
\usepackage[margin=0.5cm]{caption}
\usepackage{hyperref, bookmark}
\usepackage{csquotes}
\usepackage{array}
\usepackage{bigdelim}
\usepackage{changepage}
\usepackage{pifont}
\newcolumntype{P}[1]{>{\centering\arraybackslash}p{#1}}
\newcolumntype{M}[1]{>{\centering\arraybackslash}m{#1}}
\usepackage{makecell} 

\AtBeginDocument{
\hypersetup{pdftitle=\ttitle} 
\hypersetup{pdfauthor=\authorname} 
\hypersetup{pdfkeywords=\keywordnames} 
}

\thesistitle{Equilibrium and non-equilibrium properties of active matter systems} 
\supervisor{Prof. Raja Paul} 
\degree{Doctor of Philosophy} 
\author{Mintu Karmakar} 
\subject{Physics} 
\begin{document}

\thispagestyle{empty}
\begin{titlingpage}
\begin{center}
%
\definecolor{brightmaroon}{rgb}{0.59, 0.0, 0.09}

\color{brightmaroon}{\Large\textbf\textsc{\ttitle}\par}\vspace{3.0cm} 
\color{black}\HRule \\ 
{\large \textbf\textsc{A Thesis submitted for the degree of\\ Doctor of Philosophy (Science)}}\\
\vspace{-0.3cm}\color{black}\HRule \\ 
\vspace{2cm}%
{\textit{by}\large}\\
\vspace{0.5cm}%
\color{brightmaroon}{\Large \textbf\textsc{Mintu Karmakar}}\\
\vspace{2.0cm}%
\begin{center}
	\includegraphics[scale=0.15]{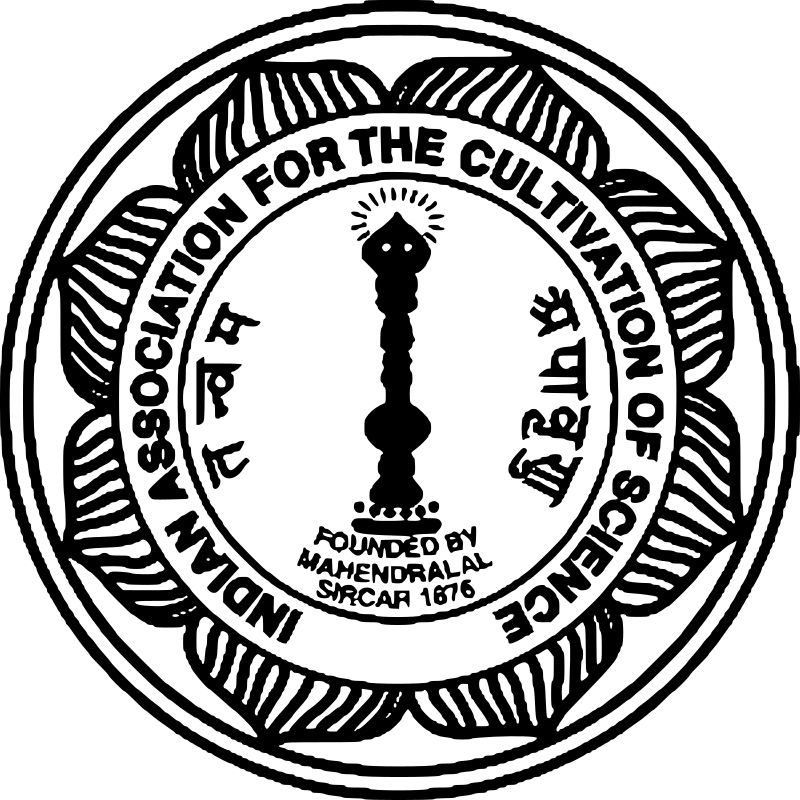}
\end{center}
\vspace{1.0cm}%
School of Mathematical \& Computational Sciences\\
{\Large\textbf\textsc{Indian Association for the Cultivation of Science}}\\
\textbf\textsc{(Deemed to be University Under Section 3 of UGC Act, 1956)}\\
{\textbf\textsc{2024}\large}
\end{center}
\end{titlingpage}

\thispagestyle{empty}
\pdfbookmark[1]{Certificate}{Certificate}
\bigskip
\par\noindent\rule{\textwidth}{1.5pt}
\vspace{-0.5 cm}
\begin{center}
	{\Large \textit{\textbf{Indian Association for the Cultivation of Science}}}\\
\end{center}
\vspace{-0.5cm}
\begin{flushleft}
		\textit{\textbf{Professor Raja Paul}}\\
		School of Mathematical \& Computational Sciences\\
		Jadavpur, Kolkata-700032, India \\
		E-mail: ssprp@iacs.res.in\\
		Phone: +91 33 2473 4971 (Ext 1310)
\end{flushleft}
\vspace*{-3.3cm}
\begin{flushright}	
\includegraphics[width=3.0cm]{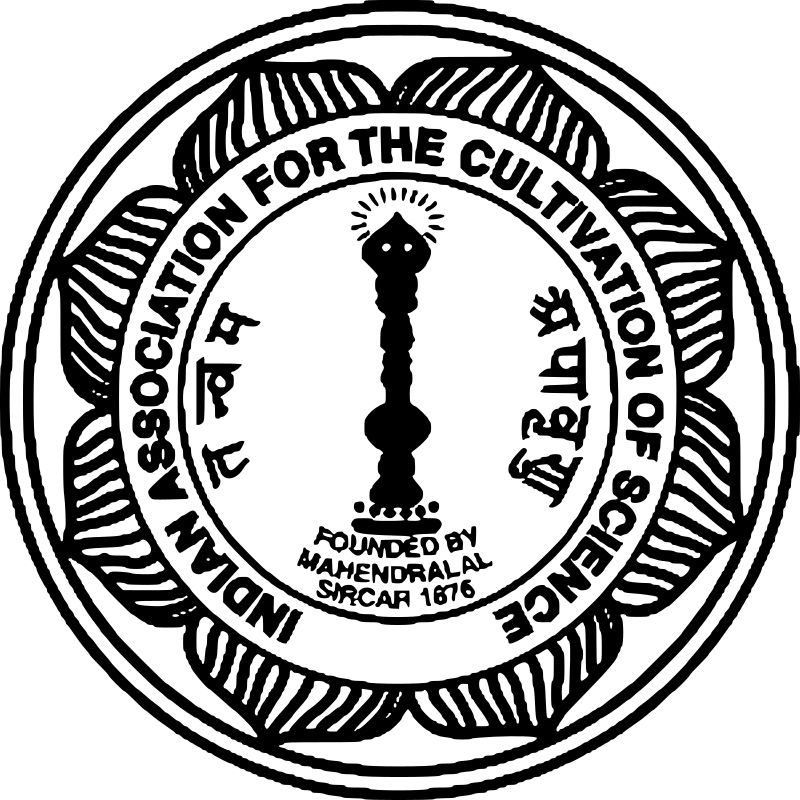}\\
\end{flushright}

\vspace{-1.0 cm}
\par\noindent\rule{\textwidth}{1.5pt}
\begin{flushright}
	Date: \textbf{22/02/2024}               
\end{flushright}
\bigskip
\begin{center}
	{\Large \underline{\textit{Certificate from the Supervisor}}}
\end{center}
I hereby certify that the research work described in the thesis entitled, \textbf{\enquote{Equilibrium and non-equilibrium properties of active matter systems}} has been carried out by
Mr. \textbf{Mintu Karmakar} (Ph.D. Registration No: \textbf{2020 03 05 01 02 053}, dated  \textbf{18/12/2020}) at the School of Mathematical \& Computational Sciences, Indian Association for the Cultivation of Science, Kolkata, India, under my supervision and that neither any part of the thesis nor the whole of the thesis has been submitted to any University or Institution for obtaining any degree /diploma /academic award.
\vspace{2.5cm}
\begin{flushleft}
	\textit{2A \& 2B Raja S. C. Mullick Road}\\
	\textit{Jadavpur, Kolkata-700032}\\
	\textit{India, 2024}
\end{flushleft}

\vspace*{-4.8cm}

\begin{flushright}
	\begin{tabular}{m{5.5cm}}
		\includegraphics[width=5cm]{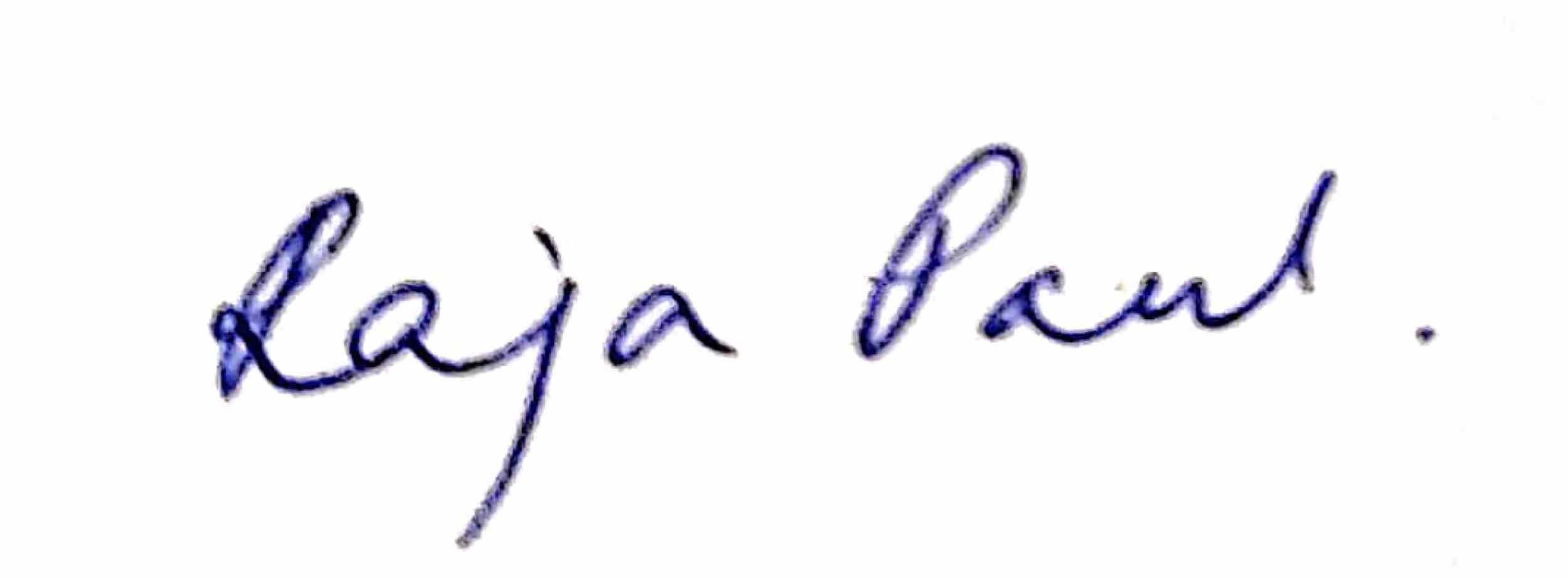}
		\\ \hline
		\centering{\large Professor Raja Paul}
		\\
		(Supervisor)
	\end{tabular}
\end{flushright}


\thispagestyle{empty}
\pdfbookmark[1]{Declaration}{Declaration}
\begin{center}
	{\large\slshape\textbf{DECLARATION}}
\end{center}
\vspace{0.5 cm}
I declare that the thesis entitled \textbf{\enquote{Equilibrium and non-equilibrium properties of active matter systems}}, submitted for the
PhD degree, is the outcome of research work carried out by me at the School of Mathematical \& Computational Sciences, Indian Association for the Cultivation of Science, Kolkata, India, under the supervision of \textbf{Prof. Raja Paul} and that neither any part of the thesis nor the whole
of the thesis has been submitted to any University or Institution for obtaining any degree /diploma /academic award. I declare that during writing this thesis, I have adhered to the principles of academic integrity and have not misrepresented or fabricated or falsified any idea/data/fact/source in my submission and the thesis does not contain any plagiarized contents. In keeping with the general practice in reporting scientific observations, due
acknowledgements and citations have been made and copyright permission has been obtained (wherever applicable), whenever the work described based on the findings of other investigators. I shall be solely responsible for any dispute or plagiarism issue arising out of this doctoral thesis.

\vspace{0.1 cm}
\begin{flushright}
	Signature:\hspace{0.2cm}\underline{\includegraphics[width=5cm]{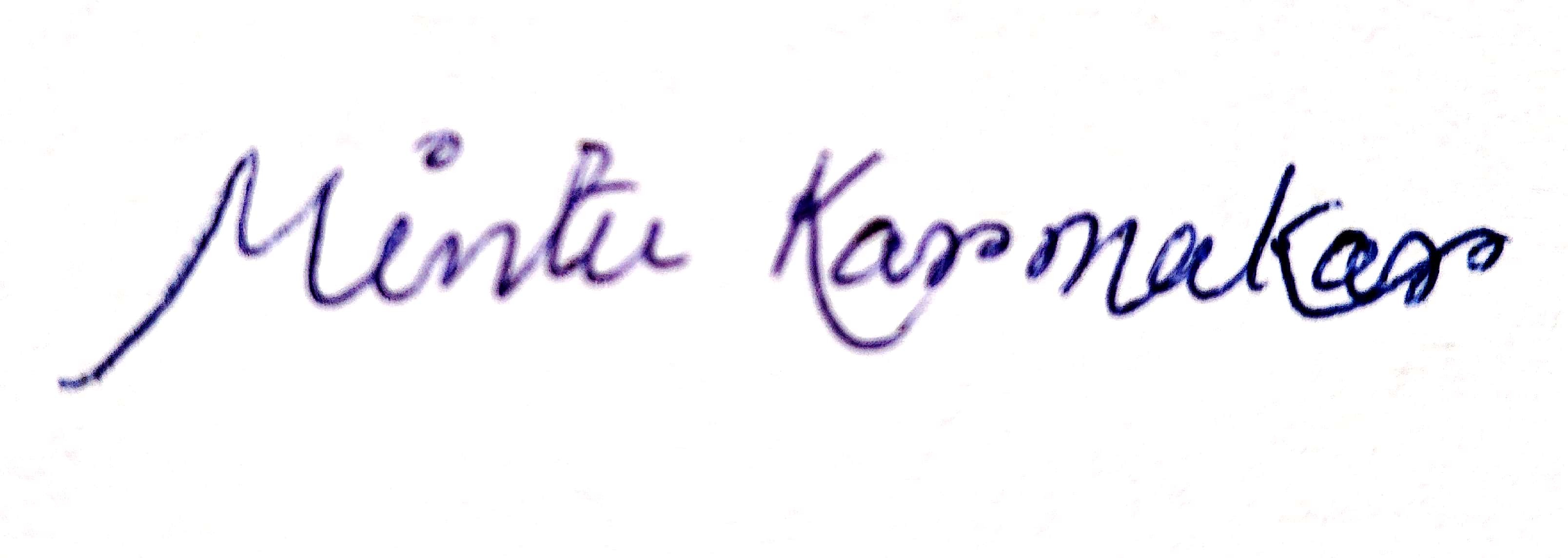}}\\
	\vspace{0.4 cm}
	Name of the Student: {\slshape \textbf{MINTU KARMAKAR}}\\
	\vspace{0.4 cm}
	Registration Number: {\slshape \textbf{2020 03 05 01 02 053}}\\
        Dated: {\slshape \textbf{18/12/2020}}
\end{flushright}

\vspace{-3.5cm}
\begin{flushleft}
Date \& Place: \textbf{22/02/2024} 
        \\ \vspace{0.4 cm}
	\textit{2A \& 2B Raja S. C. Mullick Road}\\
	\textit{Jadavpur, Kolkata-700032}\\
	\textit{India, 2024}
\end{flushleft}


\thispagestyle{empty}

\pdfbookmark[1]{Dedication}{Dedication}
\vspace*{5cm}

{\Huge{\begin{center}
			{\huge\color[rgb]{0.59, 0.0, 0.09}{\textit{\textbf{Dedicated To}}}}
			\\ \color[rgb]{0.59, 0.0, 0.09}{\textit{\textbf{My Ever Loving Parents}}}
			\\ \color[rgb]{0.59, 0.0, 0.09}{\textit{\textbf{And My Teachers}}}
\end{center}}}

\dominitoc
\frontmatter 
\pagestyle{headings} 

\addcontentsline{toc}{chapter}{Acknowledgement}\mtcaddchapter
\chapter*{Acknowledgments\markboth{Acknowledgments}{Acknowledgments}}
{\it Embarking on the journey of a Ph.D. is an odyssey marked by challenges, growth, and shared experiences. As I stand at the culmination of this academic endeavor, I am deeply grateful to the countless individuals and entities who have been instrumental in shaping this trajectory.

Foremost, I express my heartfelt gratitude to my esteemed Ph.D. guide, Prof. Raja Paul. His role in my academic development cannot be overstated. Beyond being a guide, he has been a source of inspiration, challenging me to think critically and pushing me to reach new heights. The countless hours spent in discussions, the insightful feedback provided, and the unwavering belief in my capabilities have been instrumental in shaping the trajectory of my research.

I express my deep gratitude to my collaborators, namely Prof. Heiko Rieger(Saarland University, Saarbrücken, Germany) Dr. Swarnajit Chatterjee, Dr. Matthieu Mangeat, Dr. Suman Mondal, and Prof. Subham Majumdar(Indian Association for the Cultivation of Science (IACS), Kolkata), for enhancing my research with their shared expertise and collaborative efforts. I particularly want to acknowledge the invaluable insights provided by Dr. Swarnajit Chatterjee, which have greatly enriched my work and expanded the horizons of my academic pursuits.

My academic journey commenced in the classrooms of my school, where committed educators laid the groundwork for my intellectual curiosity. Biswanath Basak, in particular, played a pivotal role by consistently encouraging me to explore diverse avenues such as quiz competitions, debates, and science fairs beyond the prescribed syllabus. Moreover, he introduced physics in an exciting manner, sparking my love for the subject. During my college years, I was fortunate to encounter teachers like Prabir Haldar, who continued to nurture my curiosity and deepen my understanding of physics. 

Gratitude extends to my university professors—Prof. Chanchal Chaudhuri, Dr. Ankur Sensharma, Dr. Subrata Sarkar, Dr. Atul Bandyopadhyay, and Dr. Soma Mitra (Banerjee). Their guidance and support have played a pivotal role in shaping my development as a learner and researcher. In particular, I am indebted to Dr. Ankur Sensharma for his enduring guidance in learning quantum mechanics, and his invaluable instruction in computational physics during my master's project. Each lesson and piece of advice provided by these mentors has been a fundamental building block in constructing my academic foundation. I am especially grateful to Prof. Chanchal Chaudhuri, whose few hours of class on mathematical physics and nuclear physics worked like magic in solidifying my understanding of these subjects.

I want to express my heartfelt thanks to my friends, Dr. Sourav Paul, Rejuan Islam, Bibhas Roy, and Nimay Roy, who have been steadfast companions throughout this academic journey. Your unwavering support, shared laughter, and moments of relief have made the challenges more manageable and the triumphs more joyous. Additionally, I extend my gratitude to others such as Joydev, Laxman, Chandan, and many more who have contributed to the camaraderie that has enriched this experience.

To my friends at IACS, it is challenging to single out names as there are countless individuals, and a few pages would not be enough to cover everyone. The shared experiences at Khelar Math, Pukur Par, OAT, along with late-night discussions and mutual support, have forged a community that is both intellectually stimulating and emotionally comforting. The numerous trips from sea to mountain with a large group hold a special place in my heart. This five-year journey feels like just a few days, and our collective experience stands as a testament to the power of collaboration and friendship in academia. I consider myself fortunate to have made friends like you during this incredible journey.

I express my gratitude to my lab mates, including Dr. Apurba Sarkar, Pinaki, Aditya, Suravi, Prasanta, and Dr. Subhendu Som, along with former lab mates Dr. Sabyasachi Sutradhar, Dr. Saikat Basu, Dr. Swarnajit Chatterjee, and Dr. Saptarshi Chatterjee. Their collaborative spirit and shared passion for research have fostered a dynamic and enriching environment in our lab. The collective efforts in our laboratory have played a crucial role in the success of our individual pursuits. A special acknowledgment goes to Dr. Apurba Sarkar, who has consistently been supportive, not only in discussing work-related challenges but also in extending assistance in personal matters. I would also like to express my appreciation to the members of Prof. Ankan Paul's group, including Debu da, Muniya di, Boyli di, Anuj, Ishita, Amit, Rahul, and Rounak. As we share the same lab space, these years have been filled with memorable times and a great atmosphere. I especially cherish the countless nights spent with Debu da and Anuj, indulging in music and creating a vibrant atmosphere in the lab.

I extend my deepest gratitude to my family, the silent pillars of support throughout my academic journey. Your understanding of the demands of this pursuit and unwavering encouragement have been profoundly appreciated. The sacrifices you've made and your steadfast belief in my abilities have served as a constant source of inspiration.

I extend my thanks to all taxpayers in India, whose contributions fund the infrastructure, resources, and opportunities that make academic pursuits like mine possible. Your investment in education has a far-reaching impact on the growth of knowledge and the development of future researchers.

I want to convey my gratitude to the members of the research assessment committee—Prof. Subham Majumdar, Prof. Soumitra Sengupta, and Prof. Abhik Basu—for generously dedicating their time, sharing their expertise, and providing constructive feedback. Your careful scrutiny and valuable guidance have played a crucial role in upholding the academic rigor of my work, contributing significantly to the shaping of the final outcome.

Finally, {\bf Council of Scientific and Industrial Research (CSIR), India} is gratefully acknowledged
for research fellowships during the JRF and SRF stages. I would also like to express my sincere thanks to {\bf Indian Association for the Cultivation of Science (IACS), Kolkata} for the financial, infrastructural, and administrative support. 

In conclusion, this acknowledgment is a collective expression of gratitude to all who have played a role, big or small, in this academic journey. Each one of you has left an indelible mark, and I am sincerely thankful for the shared moments and collaborative efforts.}
\vspace{1.5cm}
\begin{flushleft}
	\textit{2A \& 2B Raja S. C. Mullick Road}\\
	\textit{Jadavpur, Kolkata-700032}\\
	\textit{India, 2024}
\end{flushleft}

\vspace{-3cm} 

\begin{flushright}
	Signature:\hspace{0.2cm}\underline{\includegraphics[width=5cm]{Declaration/St_signature.jpg}}\\
\large{\bf Mintu Karmakar}\\
\end{flushright}

\addcontentsline{toc}{chapter}{Publications}\mtcaddchapter
\pdfbookmark[1]{Publications}{publications}
\chapter*{List of Publications}
\chaptermark{Publications}
{\it
\textbf{Parts of this Thesis have been Published in the Following Publications:}

\begin{enumerate}
\item [1.] \enquote{Jamming and flocking in the restricted active Potts model,} \\
{\textbf{\underline{Mintu Karmakar}}}, Swarnajit Chaterjee,  Matthieu Mangeat, Heiko Rieger, \& Raja Paul, \textit{\textbf{Physical Review E}} {\bf 108}, 014604 (2023). 

[\href{https://doi.org/10.1103/PhysRevE.108.014604}{https://doi.org/10.1103/PhysRevE.108.014604}]

\item [2.] \enquote{Consequence of anisotropy on flocking: the discretized Vicsek model,} \\
{\textbf{\underline{Mintu Karmakar}}}, Swarnajit Chaterjee, Raja Paul, \& Heiko Rieger,\\ \textit{\textbf{New J. Phys.}} \textbf{26}, 043023(2024). 

[\href{ https://doi.org/10.1088/1367-2630/ad3ea7}{ https://doi.org/10.1088/1367-2630/ad3ea7}]

\item [3.] \enquote{Ordering kinetics in the active Ising model,} \\
Sayam Bandyopadhyay, Swarnajit Chaterjee, Aditya Kumar Dutta,\\  {\textbf{\underline{Mintu Karmakar}}}, Heiko Rieger, and Raja Paul,\\ \textit{\textbf{Physical Review E}} {\bf 109}, 064143 (2024).

[\href{https://doi.org/10.1103/PhysRevE.109.064143}{https://doi.org/10.1103/PhysRevE.109.064143}]

\item [4.] \enquote{Polar flocks in disordered media,} \\
{\textbf{\underline{Mintu Karmakar}}}, Swarnajit Chaterjee,  Matthieu Mangeat, Heiko Rieger, \& Raja Paul, \textit{Manuscript under preparation}.
\end{enumerate}

\clearpage
\textbf{Other publications:}
\begin{enumerate}
\item [5.] \enquote{Self organized criticality of magnetic avalanches in disordered ferrimagnetic material,}\\
Suman Mandal, {\textbf{\underline{Mintu Karmakar}}}, Prabir Dutta, Saurav Giri, Subham Majumdar, \& Raja Paul, \textit{\textbf{Physical Review E}} {\bf 107}, 034106 (2023).

[\href{https://doi.org/10.1103/PhysRevE.107.034106}{https://doi.org/10.1103/PhysRevE.107.034106}].

\item [6.] \enquote{SEIRD model to study the asymptomatic growth during COVID-19 pandemic in India,} \\
Saptarshi Chatterjee, Apurba Sarkar, {\textbf{\underline{Mintu Karmakar}}}, Swarnajit Chatterjee, and Raja Paul, \textit{\textbf{Indian Journal of Physics}} (2020). 

[\href{https://doi.org/10.1007/s12648-020-01928-8}{https://doi.org/10.1007/s12648-020-01928-8}].

\item [7.] \enquote{Studying the progress of COVID-19 outbreak in India using SIRD model,} \\
Saptarshi Chatterjee, Apurba Sarkar, Swarnajit Chatterjee, {\textbf{\underline{Mintu Karmakar}}}, and Raja Paul, \textit{\textbf{Indian Journal of Physics}} (2020). 

[\href{https://doi.org/10.1007/s12648-020-01766-8}{https://doi.org/10.1007/s12648-020-01766-8}].

\item [8.] \enquote{Quarantine as a delay, not a definitive solution,} \\
{\textbf{\underline{Mintu Karmakar}}}, \textit{arXiv:2310.11121} (2023).

[\href{https://doi.org/10.48550/arXiv.2310.11121}{https://doi.org/10.48550/arXiv.2310.11121}].

\item [9.] \enquote{Metastability of polar flocks in a few active matter systems,} \\
Swarnajit Chaterjee, {\textbf{\underline{Mintu Karmakar}}}, Matthieu Mangeat, Raja Paul, and Heiko Rieger, \textit{Manuscript under preparation}.

\item [10.] \enquote{Jamming and flocking in the rAIM Hydrodynamic description,} \\
Matthieu Mangeat, {\textbf{\underline{Mintu Karmakar}}}, Swarnajit Chaterjee, Raja Paul, \& Heiko Rieger, \textit{Manuscript under preparation}.
\end{enumerate}}
{
	\hypersetup{linkcolor=[rgb]{0.59, 0.0, 0.09}}
	\pdfbookmark[section]{\contentsname}{toc}
	\tableofcontents

}
\mainmatter 
\pagestyle{thesis} 

\chapter{General Introduction}\label{chap:Chap1}
~~~~~Active matter systems~\cite{toner2005hydrodynamics,juelicher2007active,joanny2009active,ramaswamy2010mechanics} encompass natural and artificially created systems that contain many active particles. These particles actively consume energy to propel themselves~\cite{schweitzer2003brownian} or exert mechanical forces, leading to complex behaviors and a diverse range of collective motions~\cite{vicsek2012collective}. Collective motion refers to the spontaneous emergence of organized movement in large clusters of self-propelled particles (SPPs), typically much larger than the size of an individual particle. Energy consumption takes place at the individual particle level, where interactions occur directly between particles or through disturbances propagated in the surrounding medium. This intricate interplay gives rise to complex motions within the particle assembly, exhibiting various forms of collective behavior~\cite{vicsek2012collective}. 
\begin{figure}[!htb]
\centering
\includegraphics[width=0.8\columnwidth]{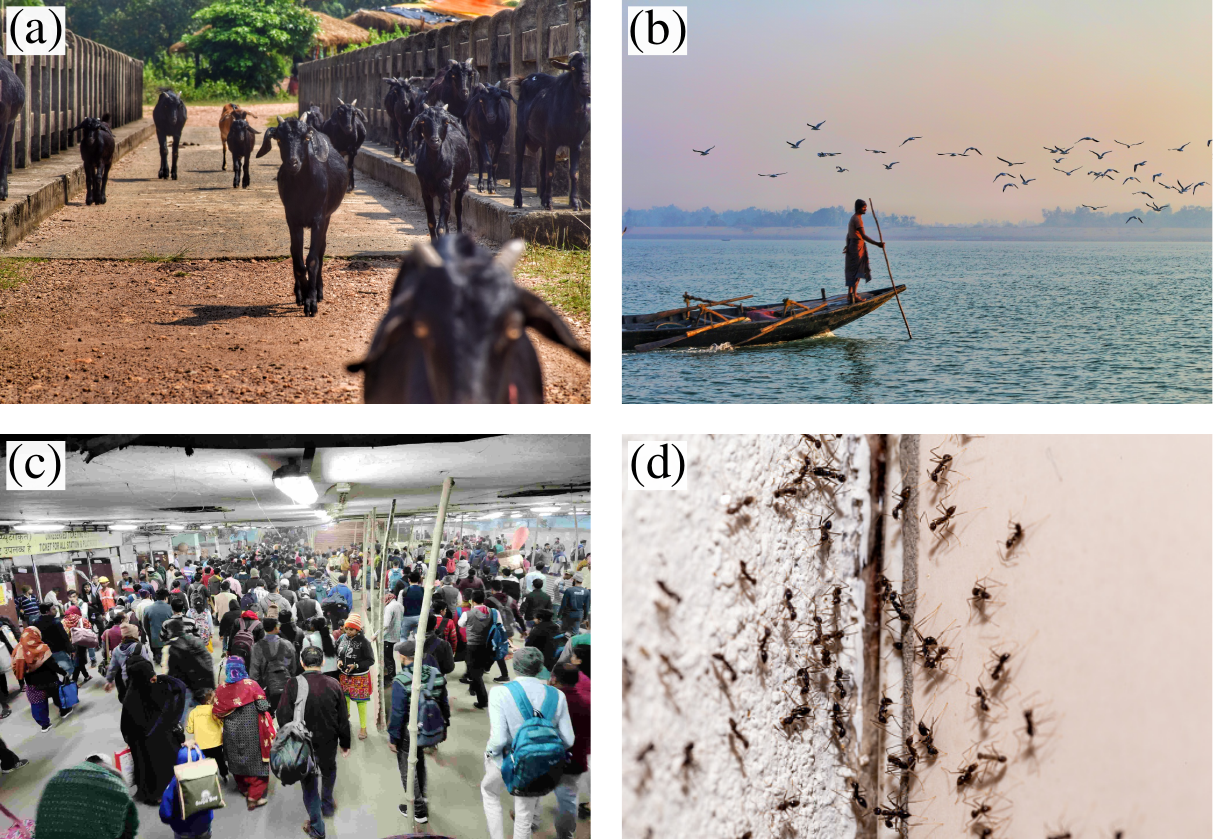}\\
\caption{{\bf{Examples of different natural active matter systems:}} (a) mammalian herds, (b) flock of birds, (c) human crowds moving underneath a tunnel, and (d) colony of ants.}
\label{1_active}
\end{figure}
Active matter systems, characterized by self-driven particles, have gained significant attention due to their fascinating collective behavior and potential applications in physics, biology, ecology, social sciences, and neurosciences~\cite{strogatz2004sync,de2015introduction,krishnan2010rheology,bottinelli2016emergent,helbing1995social,garcimartin2015flow,marchetti2013hydrodynamics,ballerini2008interaction,becco2006experimental,calovi2014swarming,peruani2012collective,schaller2010polar,sumino2012large,sanchez2012spontaneous}. To achieve this spontaneous collective motion called flocks, self-propelled agents extract energy from the systems, leading to thermally non-equilibrium states~\cite{ramaswamy2001physics, ramaswamy2000nonequilibrium}. Flocking transitions, as depicted in Figure~\ref{1_active}, are abundant in nature. Examples include mammalian herds~\cite{garcimartin2015flow}, bird flocks~\cite{ballerini2008interaction}, human crowds~\cite{bottinelli2016emergent,helbing1995social}, colony of ants, fish schools~\cite{becco2006experimental,calovi2014swarming}, and even unicellular organisms such as amoebae and bacteria~\cite{peruani2012collective}. Moreover, individual cells and self-organizing biopolymers like microtubules and actins within the cell's cytoskeleton exhibit remarkable active reorganization~\cite{de2015introduction,menon2010active,marchetti2013hydrodynamics,peruani2012collective,schaller2010polar,sumino2012large,sanchez2012spontaneous,surrey2001physical}. The physics of flocking phenomena can also be observed in nonliving substances like vertically agitated rods on a horizontal surface, self-propelled liquid droplets, liquid crystal hydrodynamics, and rolling colloids~\cite{deseigne2010collective,bricard2013emergence,thutupalli2011swarming}.

Living matter, which is inherently out of equilibrium, is an excellent example of an active matter system. The non-equilibrium nature of biological systems arises from the hydrolysis of GTP or ATP into GDP or ADP, releasing energy in the process~\cite{menon2010active}. Consequently, an intrinsic resemblance exists between internally driven biological systems and self-driven active matter systems. As a result, active matter has become an important technique or tool for figuring out the mechanical underpinnings of living matter systems~\cite{menon2010active}. Thus, active matter physics has appropriately been referred to as \enquote{The physics of life}~\cite{popkin2016physics}. Over the past three decades extensive research has been dedicated to elucidating the underlying physics of various biological phenomena and mechanisms~\cite{prost2015active,needleman2017active,xi2019material,trepat2018mesoscale,dell2018growing,perez2019active}. Despite the substantial differences in aggregation scales among different active matter systems, the observed similarities in patterns suggest the existence of a general principle governing flocking behavior.

Due to active living systems inherent complexity and heterogeneity, obtaining a comprehensive theoretical description of their overall structure and framework remains challenging. Thus, universal principles such as conservation laws and symmetries restrict the dynamical behaviors of cells and organisms. Therefore, quantifying an active system's dynamical organization and motion is the first step toward understanding these concepts.

\section{Motivation for the study of active matter systems}
~~~~~Biological inspiration plays a significant role in motivating the study of active matter. Of particular interest are two essential properties exhibited by these systems: flocking transition~\cite{ramaswamy2010mechanics,marchetti2013hydrodynamics} and motility-induced phase separation (MIPS)~\cite{cates2015motility}. These have become a focal point for researchers across disciplines, including physics, biology, and applied mathematics. The flocking transition has been extensively studied in the literature and how it is influenced by various parameters such as density, noise, and self-propulsion velocity~\cite{ramaswamy2010mechanics,marchetti2013hydrodynamics}. On the other hand, MIPS refers to the segregation of active particles into distinct dense regions based on their differing motility levels. Understanding the underlying mechanisms driving these phenomena is not only crucial for unraveling the fundamental principles governing the behavior of these systems but also holds promise for applications in fields such as swarm robotics~\cite{brambilla2013swarm,dorigo2014swarm}. While significant progress has been made in understanding flocking transition and MIPS in active matter systems, several challenges remain. For instance, the precise mechanisms that trigger the interplay between flocking and MIPS. Similarly, the dependence of flocking behavior on other factors, such as volume exclusion of SPPs, external fields, complex environments, discretized orientations of SPPs, etc., requires further investigation. Also, the origin of those large polar flocks in active matter systems must be understood clearly.

In this Ph.D. thesis, we primarily aim to develop theoretical frameworks to investigate the behavior of active matter systems under steady-state conditions. We will explore the influence of external factors such as field, inhomogeneous medium, and spatial anisotropy~\cite{gompper20202020,wensink2012meso}, which may provide new insights into the behavior of active matter systems. We will identify some of the gaps and limitations of the present literature and highlight the areas where our research will significantly contribute. The following sections will comprehensively review the existing literature on active matter systems. This review will cover key concepts, theories, and methodologies employed in previous studies.

\section{Overview of self-propelled particles (SPPs) and their flocking behavior}
\subsection{Vicsek Model (VM)} \label{1_VM}
Researchers have extensively studied the flocking transition in active matter systems, developing numerous theoretical frameworks. Among these frameworks, the most popular one proposed by Vicsek~\textit{et al.} in 1995 has gained significant attention~\cite{vicsek1995novel}. This model describes collective motion in SPPs that align their velocities with neighboring particles. Researchers have used the VM to study various phenomena, including flocking, schooling, and swarming. A polar point particle system with $N$ particles is proposed within a periodic square box. The model is characterized by the positions $\mathbf{r}_i(t)$ and self-propulsion directions $\theta_i(t)$ of the particles at discrete time $t$, where $i = 1, \ldots, N$. The particles exhibit a collective behavior by aligning their directions with those of neighboring particles within a circle of radius $R$. The particles take account of directional errors during this alignment process, represented by $\Delta\theta_i(t)$. It is assumed that $\Delta\theta_i(t)$ follows a uniform white noise distribution with zero mean, i.e., $\langle\Delta\theta_i(t)\rangle = 0$, and $\langle\Delta\theta_i(t)\Delta\theta_j(t_0)\rangle = \eta\delta_{ij}\delta(t-t_0)$. The update rule for the angle $\theta_i$ of the $i$th particle at time $t+\Delta t$ is defined as follows:

\begin{figure}[!htbp]
\centering
\includegraphics[width=0.8\columnwidth]{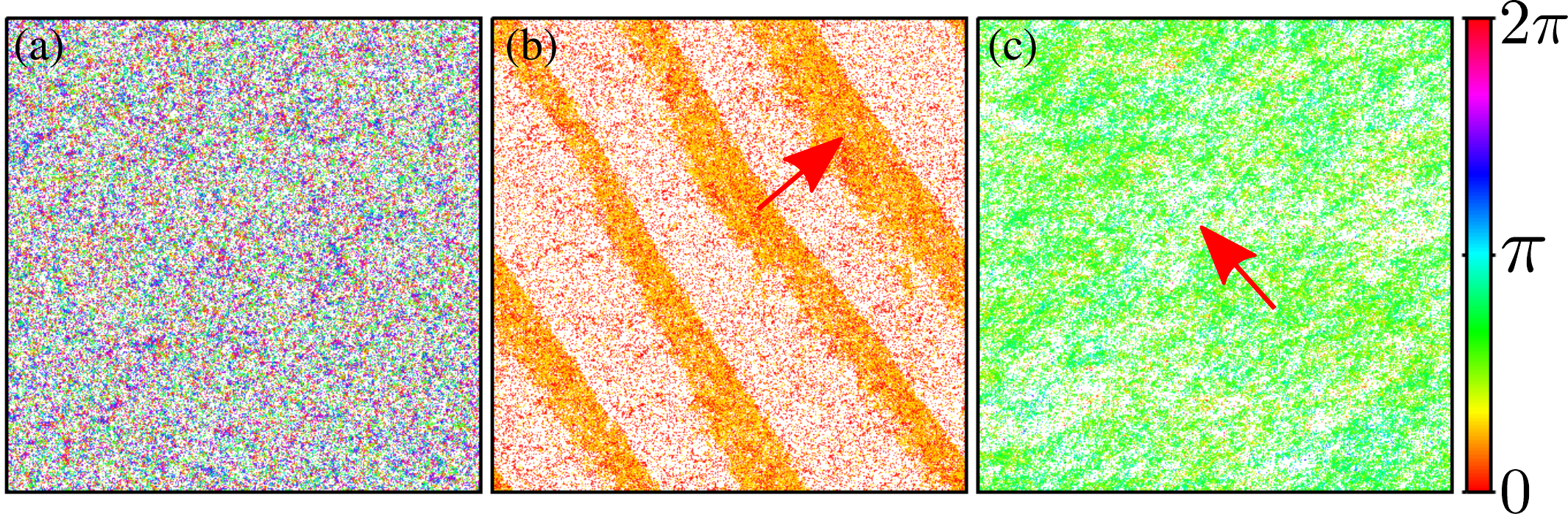}\\
\caption{Steady state snapshots of the VM simulation. (a) Disordered gas phase at high noise $\eta=0.5$. (b) Phase-separated coexistence phase at intermediate noise $\eta=0.4$. (c) Polar ordered liquid phase at low noise $\eta=0.2$. The color bar represents the angle of particle orientations. Parameters: $L=1024$, $v_0=0.5$, and $\rho_0=2.$}\label{1_vicsek_snap}
\end{figure}

The following equations govern the dynamic evolution of the polar point particle system;

\begin{equation}
\theta_i(t+\Delta t) = \text{arctan}\left[\frac{\langle\sin{\theta(t)}\rangle_i^R}{\langle\cos{\theta(t)}\rangle_i^R}\right]+\eta \xi_i(t)
\end{equation}

where angular bracket $\langle*\rangle_i^R$ denotes the average of an observable over all $n_i^R$ neighboring particles within the radius $R=1$, expressed as:

\begin{equation*}
\langle\sin{\theta(t)}\rangle_i^R=\frac{1}{n_i^R}\sum_{j=1}^{n_i^R}\sin{\theta_j(t)};\quad\langle\cos{\theta(t)}\rangle_i^R=\frac{1}{n_i^R}\sum_{j=1}^{n_i^R}\cos{\theta_j(t)}    
\end{equation*}

The term $\eta \xi_i(t)$ introduces random fluctuations to the angle ($\Delta\theta_i(t)$), where $\xi_i(t)$ is a random number chosen uniformly from the range $[-\pi,\pi]$. The parameter $\eta\in[0,1]$ acts as a temperature-like quantity in ferromagnetic interactions. The particle's position update rule is given by:

\begin{equation}
\mathbf{r}_i(t+\Delta t) = \mathbf{r}_i(t) + v_0\Delta t\begin{pmatrix}\cos(\theta_i(t))\\ \sin(\theta_i(t))\end{pmatrix}
\end{equation}

Here, $\mathbf{r}_i(t)$ represents the position of the $i$th particle at time $t$, and $v_0$ is the constant speed.

The VM is often referred to as a \enquote{dynamical XY model}~\cite{toner1995long,toner1998flocks,toner2012reanalysis}, as the particles can align in any direction within a two-dimensional plane, reminiscent of the ferromagnetic alignment of XY spins. This model is a fundamental component in the field of active matter physics as it exhibits a phase transition to collective motion based on particle density or noise, similar to the role played by the Ising model in paramagnetic to ferromagnetic phase transitions in equilibrium systems~\cite{ginelli2016physics}. Through simulations of the VM in the density-noise parameter space, several significant behavioral observations have been made:

\begin{figure}[!htbp]
\centering
\includegraphics[width=0.6\columnwidth]{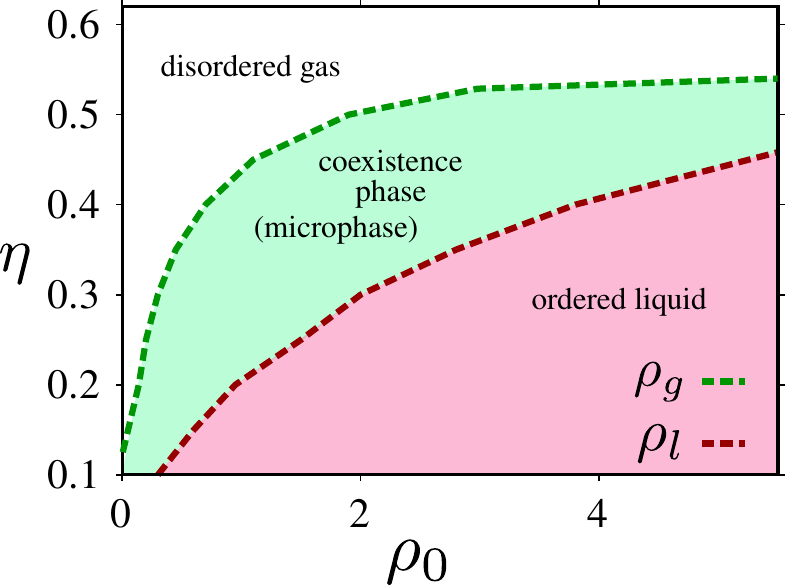}\\
\caption{Phase diagram of the VM in the density($\rho_0$)-noise($\eta$) parameter space. The phase transition from a disordered gas phase to a polar ordered liquid phase via intermediate coexistence regime as the noise decreases or the number density increases.}\label{1_vicsek}
\end{figure}

For fixed intermediate densities, at high noise, particles exhibit random and independent motion, indicating a disordered phase (see Fig.~\ref{1_vicsek_snap} (a)). At intermediate noise, particles form several bands and flock together in a disordered gaseous background referred to as the coexistence phase (see Fig.~\ref{1_vicsek_snap} (b)). At low noise, all particles move collectively in the same spontaneously-selected direction, exhibiting orderly motion (see Fig.~\ref{1_vicsek_snap} (c)).

Figure~\ref{1_vicsek}~\cite{solon2015phase,fodor2018statistical} depicts the phase diagram of the VM in the density-noise parameter space, represented by $\rho_0$ and $\eta$. Like continuous-spin magnets, the VM displays a distinct phase transition from a disordered phase to a coherent flock as the noise parameter decreases or the number density increases. The Vicsek family of models exhibits this well-defined phase transition. The transition from high noise/low density in the gas phase to the low noise/high density in the polar-ordered Toner-Tu phase, exhibiting long-range order through the collective movement of all particles, is characterized as a first-order transition~\cite{gregoire2004onset}, previously thought second order. The coexistence phase of the VM displays smectic arrays of multiple liquid bands of collectively moving particles, known as micro-phase separation~\cite{solon2015phase, solon2015pattern}. However, a recent study discovered an interesting polar-ordered cross-sea phase between the polar liquid and the coexistence regimes~\cite{kursten2020dry}. The polar-ordered cross-sea phase can only be seen when a minimum number of bands cross each other and form a wave-like structure.

The simplicity of the dynamical rule in the VM allows for numerous generalizations and variations, which have found applications in diverse domains. Subsequent numerical investigations have played a significant role in unraveling the steady state properties of the VM~\cite{gregoire2004onset,chate2008collective,solon2013revisiting}.

\subsection{Toner and Tu's Model}\label{1_TT}
In 1995, Toner and Tu proposed a continuum effective theory for the flocking model originally introduced by Vicsek \textit{et al}.~\cite{toner1995long}. The VM describes a system consisting of SPPs with a fixed speed and interactions that align their directions with some noise. It exhibits a non-equilibrium phase transition from a disordered state at low density or high noise to an ordered, coherently moving state at high density or low noise strength. This transition is particularly remarkable because it occurs in a two-dimensional system, defying the Mermin-Wagner theorem, which states that no ferromagnetic state should exist in a system with dimensions $d \leqslant 2$.

The Mermin-Wagner theorem dictated that fluctuations play a dominant role in lower dimensions and hinder the emergence of long-range order (LRO). However, this theorem assumes that the system is in a state of thermal equilibrium. In the case of non-equilibrium phenomena such as the flocking phase transition, the emergence of LRO is not surprising. The self-propulsion of particles is responsible for the collective coherent motion observed in the flocking model, often referred to as the Toner-Tu model~\cite{toner1995long}. Notably, when the self-propulsion velocity $v$ approaches zero, the system behaves akin to the XY model in two dimensions ($d=2$).

In order to gain a comprehensive understanding of the physics underlying the emergence of LRO, Toner and Tu developed a continuum model by considering the symmetries of the VM and treating the flock as a continuous hydrodynamic system~\cite{toner1998flocks}. The equations can be written as follows,
\begin{equation}
    \frac{\partial \mathbf{u}}{\partial t} + (\mathbf{u} \cdot \nabla) \mathbf{u} = (\alpha -\beta |\mathbf{u}|^2)\mathbf{u}-\nabla P(\rho) + D\nabla (\nabla \cdot \mathbf{u}) + \mathbf{f}
    \label{toner_eq1}
\end{equation}
\begin{equation}
    \frac{\partial \rho}{\partial t} + \nabla \cdot (\rho \mathbf{u}) = 0
    \label{toner_eq2}
\end{equation}
Phenomenologically, they derived stochastic differential equations to describe the velocity and density fields Eq.~\ref{toner_eq1} and Eq.~\ref{toner_eq2}, drawing analogies from magnetic systems and liquid crystals. This continuum model aims to capture the dynamics at large length scales, focusing on the coarse-grained properties in the asymptotic regime.

\subsection{Active Ising Model (AIM)} \label{1_AIM}
Developed by A. P. Solon and J. Tailleur, the AIM offers valuable insights into the dynamics of flocking behavior and its connection to the physics observed in Vicsek-like models~\cite{solon2013revisiting,solon2015flocking}. Unlike the VM, the AIM is a lattice-based model, which simplifies both numerical and analytical investigations. This model employs self-propelled Ising spins, displaying phenomenological similarities to conventional flocking models.

The active Ising model (AIM) incorporates the essential elements necessary for a flocking transition: biased diffusion and local aligning interactions. This is a purely local interaction: particles align with other particles only at the same site, and the model is equivalent to $L^2$ independent fully connected Ising models in the absence of particle hopping. A particle with spin $s$ on-site \textit{i} flips its spin at a rate of; 
\begin{equation}\label{eq:2}
    W_{\rm flip}(s \to -s) = \gamma \exp{\Big{[}-s\beta\Big{(}\frac{n_i^{s} - n_i^{-s} - 1}{\rho _ i}\Big{)}\Big{]}}
\end{equation}

where $\gamma$ is the rate of particle flipping and $\beta$, denoted as \enquote{inverse temperature}, $\beta=T^{-1}$, controls the flip noise strength. Each lattice site $i$ can accommodate an arbitrary number of particles $n_i^{s}$ with spin $s =\pm 1$. In contrast to the continuous rotational symmetry of the VM, the AIM introduces the discrete symmetry of the Ising model, where each particle assumes a spin state of either $+1$ or $-1$. In the AIM, particles diffuse on a two-dimensional plane ($d=2$), but their motion is biased in one of two possible directions. Specifically, particles with a spin state of $+1$ are biased to hop in the right direction, while particles with a spin state of $-1$ are biased to hop in the left direction. The hopping rates along the upward and downward directions remain unchanged. Consequently, the AIM can be considered a dynamical Ising model with discrete symmetry among the particle's movement. AIM is designed so that self-propulsion can be tuned; here, it is through the parameter $\epsilon$. The limit of vanishing self-propulsion $\epsilon \to 0$ is well defined as spins continue to diffuse on the lattice. This dynamics allow us to interpolate continuously between totally self-propelled $(\epsilon = 1)$, self-propelled $[\epsilon \in (0,1)]$, weakly self-propelled $[\epsilon \sim 1/L]$, and purely diffusive $(\epsilon = 0)$~\cite{solon2013revisiting,solon2015flocking}. 

\begin{figure}[!htbp]
\centering
\includegraphics[width=0.8\columnwidth]{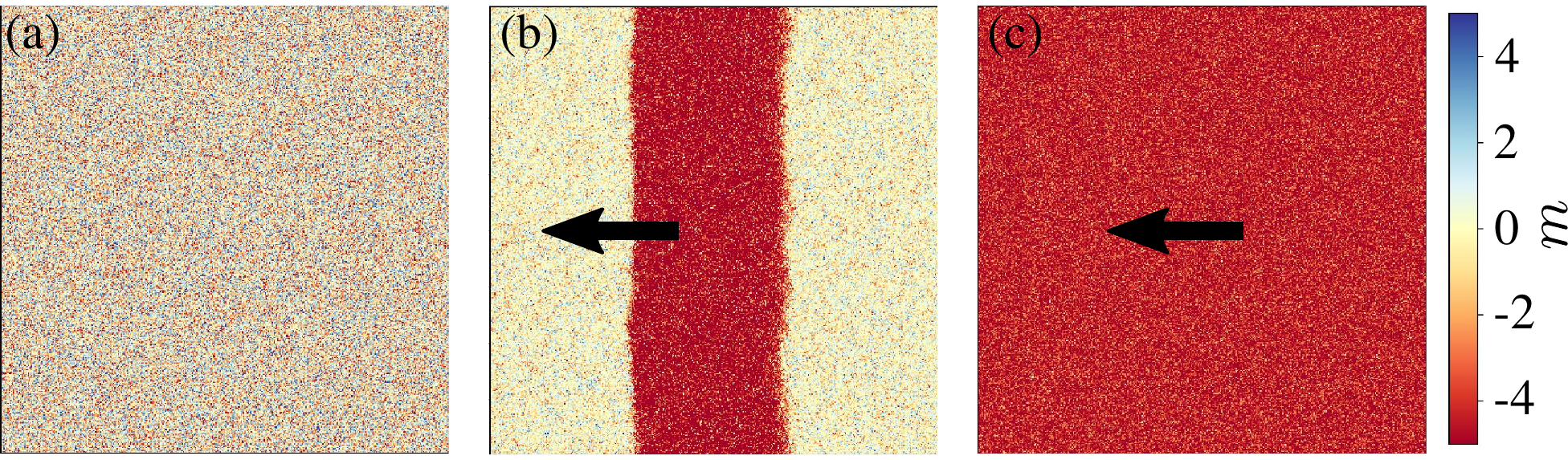}\\
\caption{Steady state snapshots of the AIM simulation. (a) Disordered gas phase at high temperature $T=1$. (b) Phase-separated coexistence phase at intermediate temperature $T=0.7$. (c) Polar ordered liquid phase at low temperature $T=0.5$. The color bar represents local magnetization. Parameters: $L=400$, $\epsilon=1$, and $\rho_0=4.$}\label{1_aim_snap}
\end{figure}
\begin{figure}[!htbp]
\centering
\includegraphics[width=0.8\columnwidth]{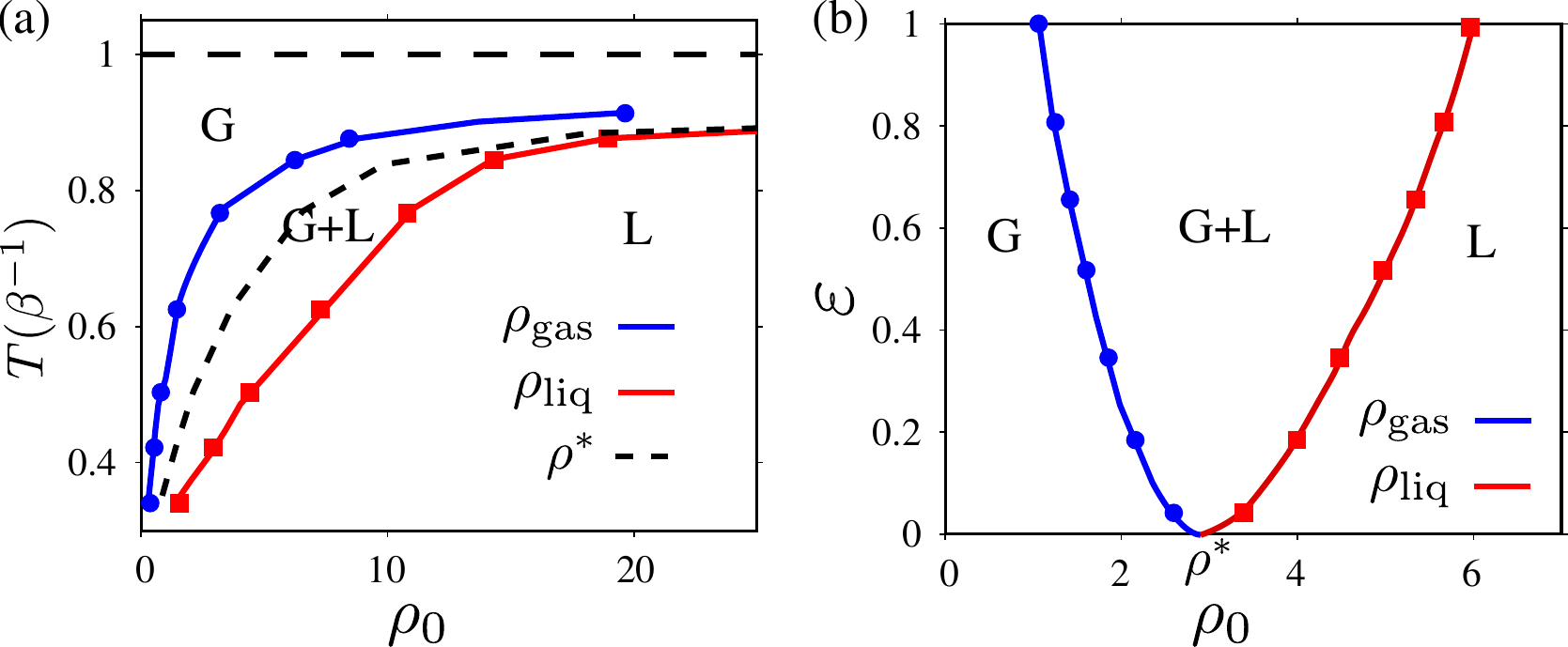}\\
\caption{(a) $T-\rho_0$ phase diagram at a constant $\epsilon=0.9$. (b) $\epsilon-\rho_0$ phase diagram at a constant $T=0.52$. The blue ($\rho_g$) and red ($\rho_l$) lines depict the region of existence of phase-separated profiles. $T$, $\epsilon$, and $\rho_0$ represent the noise, self-propulsion velocity, and average density, respectively. \textbf{G} represents the gaseous portion of the phase space, \textbf{G+L} represents the coexistence phase, and \textbf{L} corresponds to the ordered liquid phase. The black dashed line indicates the critical points at $\epsilon=0$.}
\label{1_aim}
\end{figure}

Three different phases are observed in AIM, shown in Fig.~\ref{1_aim_snap}. Figure~\ref{1_aim} showcases the phase diagrams of the active Ising model (AIM), providing the following insights:

\textbf{(a)} At high thermal noise and low densities, the system is in a disordered gaseous phase.

\textbf{(b)} A polar liquid phase is observed in the system at high densities and low temperatures. 

\textbf{(c)} For intermediate densities $(\rho_0 \in [\rho_g,\rho_l])$, a coexistence phase of liquid and gas is observed. 

\textbf{(d)} At $\epsilon=0$, a critical point $\rho^*$ exists, representing a direct liquid-gas phase transition without the coexistence phase $G+L$.

The active Ising model (AIM) has been extensively investigated using numerical simulations, microscopic derivations, and developing a hydrodynamic theory. One of the notable features of this model is that it exhibits universal properties of the conventional Ising model in two dimensions when the self-propulsion parameter tends to zero. Moreover, the AIM captures the essential physics of the VM while offering additional insights into the flocking transition in a simpler framework.

\subsection{Active Potts Model (APM)}\label{I_apm}
Recently, Chatterjee and Mangeat \textit{et al.} investigated the dynamics of flocking by introducing an extended version of the AIM known as the active Potts model (APM)~\cite{chatterjee2020flocking,mangeat2020flocking}. The APM incorporates an internal spin structure with $q$ spin states, where the AIM corresponds to $q = 2$. In their investigation, APM was studied on $2d$ lattices with a coordination number $q$, such as a square lattice for $q = 4$ or a triangular lattice for $q = 6$~\cite{chatterjee2020flocking,mangeat2020flocking}.
\begin{figure}[!htbp]
\centering
\includegraphics[width=0.9\columnwidth]{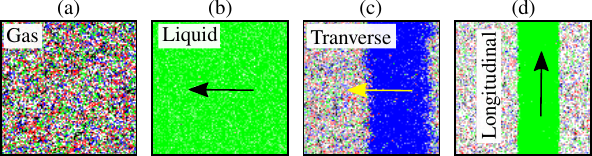}
\caption{Steady state snapshots of the APM simulation. (a) Disordered gas phase at high temperature $T=2.2$. $\epsilon=0.9$. (b) Polar ordered liquid phase at low temperature $T=1$. $\epsilon=0.9$. (c-d) Phase-separated coexistence phase at intermediate temperature $T=1.4$ for self-propulsion velocity $\epsilon=0.9$ and $2.7$, respectively. For increasing values of the self-propulsion velocity, the band formation in the steady state has a transverse to longitudinal orientation. The color code for the liquid phase is red ($q = 1$), green ($q = 2$), blue ($q = 3$), and black ($q = 4$). Parameters: $L=200$, and $\rho_0=4.$}\label{1_apm_snap}
\end{figure}
This model's essential components for flocking include local ferromagnetic alignment between neighboring particles, motivated by the standard $q$-state Potts model. Self-propulsion is achieved by biased hopping to neighboring lattice sites; repulsive interactions are not considered. The spin state of the $k$-th particle on lattice site $i$ is represented by $\sigma^k_i$, taking integer values within the range $[1, q]$. The count of particles in state $\sigma$ at site $i$ is denoted as $n^\sigma_i$. The local density at site $i$ is defined as $\rho_i = \sum_{\sigma=1}^q n^\sigma_i$, representing the total number of particles on that site. The Hamiltonian of a $q$-state APM is given by $H^{\rm APM} = \sum_i H_i^{\rm APM}$, where it is decomposed into the sum of local Hamiltonians $H_i$~\cite{chatterjee2020flocking, mangeat2020flocking}:
\begin{equation}
H_i^{\rm APM}=-\frac{J}{2\rho_i}\sum_{k=1}^{\rho_i}\sum_{l\ne k}(q\delta_{\sigma_i^k,\sigma_i^l}-1),
\end{equation}
In this system, the coupling between neighboring sites is denoted by $J=1$. They consider the case where $q=4$. A particle situated at site $i$ in state $\sigma$ can either change to another state $\sigma^\prime$ or move to any neighboring site, following the specific constraints imposed by different restriction protocols. The local magnetization associated with state $\sigma$ at site $i$ is defined as $m_i^{\sigma}$:
\begin{equation}
m_i^{\sigma}=\sum_{j=1}^{\rho_i}\frac{q\delta_{\sigma,\sigma_i^j}-1}{q-1} \, .
\end{equation} 
The authors determined the steady state of the APM, characterized by collective motion observed in significant regions of the phase diagram, by developing a coarse-grained theory of hydrodynamics and then analyzing it with microscopic Monte Carlo simulations. The study yielded several significant findings, which are summarized as follows:

\begin{figure}[!htbp]
\centering
\includegraphics[width=0.8\columnwidth]{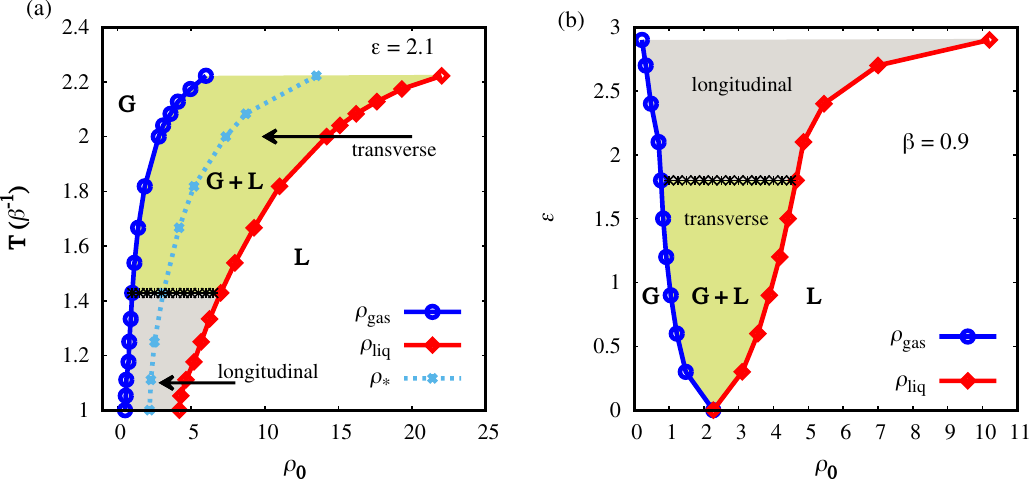}\\
\caption{(a) $T-\rho_0$ phase diagram at a constant $\epsilon=2.1$. (b) $\epsilon-\rho_0$ phase diagram at a constant $T=1.1$. The blue ($\rho_g$) and red ($\rho_l$) lines depict the region of existence of phase-separated profiles. $T$, $\epsilon$, and $\rho_0$ respectively represent the noise, self-propulsion velocity, and average density. {\bf{G}} signifies the gaseous portion of the phase space, {\bf{G+L}} represents the co-existence phase, and {\bf{L}} is the ordered liquid phase. The green dashed line indicates the critical points at $\epsilon=0$. Source:~\cite{chatterjee2020flocking}.}
\label{1_apm}
\end{figure}

\textbf{(a)} The flocking transition observed in the active Potts model (APM) exhibits characteristics similar to a liquid-gas phase transition via an intermediate coexistence phase shown in Fig.~\ref{1_apm_snap}, akin to what is observed in the active Ising model (AIM) with temperature $T=\beta^{-1}$, average particle density $\rho_0$, and self-propulsion velocity $\epsilon$ as governing parameters. 

\textbf{(b)} In the coexistence phase of the APM, the liquid domains undergo a unique structural arrangement, forming stripe-like structures. These stripes can be oriented either transversely (referred to as bands) or longitudinally (referred to as lanes), and their motion can be either perpendicular or parallel to the internal spin state that controls the stripes of the liquid phase. Reorientation occurs as a result of this behavior, depending on the self-propulsion velocity $\epsilon$ (see Fig.~\ref{1_apm_snap}(c-d)). The distinctive characteristic of the APM lies in its reorientation transition, a phenomenon notably absent in both the VM and the AIM.

\textbf{(c)} The critical point at $\epsilon=0$ in the APM is characterized as a first-order phase transition, distinct from the standard $q$-state Potts model. The transition from a high-density ordered phase to a low-density disordered phase is indicated by this critical point for APM with $q\geq4$. The phase diagrams of APM in Fig.~\ref{1_apm}(a-b) provide the above insights in detail.

\subsection{Active Clock Model (ACM)}\label{I_acm}
It is crucial to emphasize that the VM~\cite{vicsek1995novel} proposes a continuous spectrum of possible particle motion directions. However, exploring the sensitivity of the active system to anisotropy, specifically in the presence of fixed preferred directions in space, remains relatively uncharted. A natural discretization of the VM (in $2d$) is the $2d$ $q$-state active clock model (ACM). In ACM, particles possess a spin-state or clock angle $\theta=\frac{2\pi\sigma}{q}$ with $\sigma \in [0,q-1]$, allowing them to either transition to a different spin-state $\theta^\prime$ or probabilistically move to a neighboring lattice site. This behavior is analogous to the AIM~\cite{solon2015flocking} or APM~\cite{chatterjee2020flocking}. The flip probabilities within the ACM are determined through a ferromagnetic Hamiltonian $H^{\rm ACM} = \sum_i H_i^{\rm ACM}$, where it is expressed as the summation of local Hamiltonians $H_i^{\rm ACM}$~\cite{chatterjee2022polar}:
\begin{equation}\label{H_acm}
\begin{split}
 H_i^{\rm ACM} &=  - \frac{J}{2\rho_i} \sum_{k=1}^{\rho_i} \sum_{l\ne k} \cos(\theta_i^k-\theta_i^l) \\
&= - \frac{J}{2\rho_i} \sum_{k=1}^{\rho_i} \sum_{l\ne k} \cos \frac{2\pi}{q} (\sigma_i^k-\sigma_i^l) \, ,  
\end{split}
\end{equation}
Including the prefactor $1/2\rho_i$ renders the Hamiltonian intensive, preventing the double counting of interactions. Additionally, $J$ represents the coupling constant governing interactions between particles. Notably, when $q=2$, the Hamiltonian aligns with that defined for the AIM~\cite{solon2015flocking}. The investigation presented in Ref.~\cite{chatterjee2022polar} yielded important results, summarized as follows:

\textbf{(a)} For all values of the number of states $q$, the flocking transition in the $q$-state ACM is described as a liquid-gas phase transition, akin to the VM~\cite{vicsek1995novel, solon2015phase}, AIM~\cite{solon2015flocking}, and APM~\cite{chatterjee2020flocking, mangeat2020flocking}.

\textbf{(b)} Macrophase separation is seen in the coexistence phase for small directions or low $q$-values, resembling the AIM~\cite{solon2015flocking} and APM~\cite{chatterjee2020flocking, mangeat2020flocking}.

\textbf{(c)} Giant fluctuations observed for large $q$-values prevent bulk phase separation, breaking large liquid domains into narrow periodic traveling bands. Consequently, impeding further coarsening, resulting in microphase separation, reminiscent of the VM~\cite{solon2015phase}.

However, an interesting observation is proposed by a recent study in Ref.~\cite{solon2022susceptibility} with different dynamical rules. In the liquid and coexistence regions, the transition from VM to AIM physics is argued to occur only beyond a characteristic length scale, diverging for vanishingly small anisotropy and large $q$.

\subsection{Multi-species flocking model} 
Multi-species flocking refers to the collective motion and coordinated behavior exhibited by groups of individuals from different species. In this phenomenon, diverse species come together to form cohesive and organized patterns of movement, often resembling the synchronized motion observed in single-species flocks or swarms. Multi-species flocking can be observed in various biological contexts, including birds, fish, and microorganisms. The study of multi-species flocking offers insights into the complex dynamics of interspecies interactions, communication, and cooperation, shedding light on the fascinating ways different organisms collaborate for mutual benefit in nature. A recent experimental investigation examined the collective dynamics of mixed swarming bacterial populations~\cite{peled2021heterogeneous}. These populations consisted of cells of the same species but with different phenotypes, specifically varying in aspect ratios (length). In contrast to a homogeneous system, the mixed population did not exhibit macroscopic phase separation. Instead, long cells acted as nucleation sites, facilitating the formation of aggregates around which short, rapidly moving cells could gather. This phenomenon resulted in an observed enhancement in swarming speeds. In a recent study~\cite{TSVM2023}, researchers explored the dynamics of two unfriendly species moving in tandem. The findings reveal the existence of two separate dynamical states within the coexistence region. The Parallel Flocking (PF) state was identified by bands of both species moving in the same direction. In contrast, the Antiparallel Flocking (APF) state exhibited bands of species A and B moving in opposite directions. Stochastic transitions between PF and APF states were detected in the low-density segment of the coexistence region.

\section{Brief introduction on motility-induced phase separation (MIPS)}\label{1_MIPS}
MIPS, on the other hand, gained prominence with the discovery of systems where active particles segregate into dense, collectively moving clusters and dilute, diffusive regions~\cite{cates2015motility}. These phenomena have been observed in a variety of experimental setups~\cite{geyer2019freezing}, ranging from self-propelled colloidal particles to biological systems like bacterial suspensions. Theoretical and computational studies have shed light on the role of self-propulsion, steric interactions, and hydrodynamic flows in driving MIPS~\cite{cates2015motility,fily2012athermal}. Self-propelled entities with repulsive interactions, such as active Brownian particles (ABPs)~\cite{romanczuk2012active}, exhibit an alternative cluster state at larger density and high Péclet numbers. This state, distinguished from flocking through alignment, is called MIPS~\cite{cates2015motility}. Consequently, the interplay between alignment and repulsive interactions could lead to complex phase diagrams, as was demonstrated for the VM with repulsive particle interactions~\cite{gregoire2003moving} or ABPs with alignment interactions~\cite{martin2018collective,sese2018velocity}.

\subsection{Traffic Jam and Gliders}
\begin{figure}[!htb]
\centering
\includegraphics[width=\columnwidth]{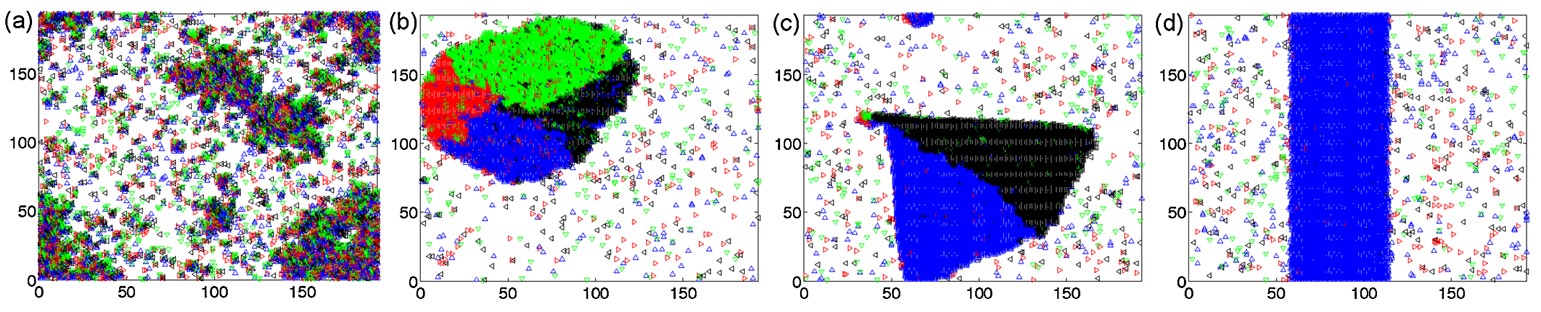}\\
\caption{The following are some examples of self-organized spatial patterns: (a) disordered aggregates, (b) traffic jams, (c) gliders, and (d) bands. The small triangles, color-coded as follows: down (green), up (blue), left (black), and right (red), indicate the direction of the particles. Source:~\cite{peruani2011traffic}}
\label{1_traffic}
\end{figure}

Peruani \textit{et al.}~\cite{peruani2011traffic} investigated a swarming model on a two-dimensional lattice where SPPs demonstrate ferromagnetic alignment. The model incorporates volume exclusion, allowing particles to hop only to neighboring empty nodes. Their study reveals diverse self-organized spatial patterns arising from these volume exclusion effects. At very low alignment sensitivity, the system initially forms a disordered phase (see Fig.~\ref{1_traffic}(a)). Upon increasing the alignment sensitivity, particles aggregate into high-density locally ordered regions and transform into traffic jams, Fig.~\ref{1_traffic}(b). For enhanced alignment, the traffic jams develop into gliders, triangular regions of high particle density that migrate in a specific direction shown in Fig.~\ref{1_traffic}(c). The system achieves maximum order when elongated high-density regions, known as bands, traverse the entire system (see Fig.~\ref{1_traffic}(d)).
For example, static traffic jams are likely to be a common feature in systems where stagnation can occur. Jamming SPPs has revealed unexpected self-organized structures in two dimensions, such as dynamic traffic jams to gliders. The introduction of an alignment mechanism promotes local orientational order, and when particles are correctly oriented, density waves of stagnant particles can emerge. The findings presented in this study represent an initial advancement in comprehending the potential phenomena that arise from such coupling between particle alignment and density waves.

\subsection{Active Lattice Gas (ALG)}
Recently, Tailleur \textit{et al}.~\cite{kourbane2018exact} investigated a microscopic lattice gas displaying MIPS. In their model, $N$ particles evolve on a discrete ring containing $\alpha L$ sites, where $\alpha$ is the rescaled parameter. The particles come in two types, and each site can be occupied by at most one particle. $\sigma_i$ is the occupation numbers at site $i$ are used to represent configurations; values are taken from $\{-1, 0, 1\}$. Self-propulsion is introduced in the simulation by applying a weak drift to the right experienced by $``+"$ particles, while the $``-"$ particles undergo a weak drift to the left, in addition to a symmetric diffusive motion. Additionally, particles have the ability to switch, changing signs at a fixed rate.
\begin{figure}[!htb]
\centering
\includegraphics[width=0.6\columnwidth]{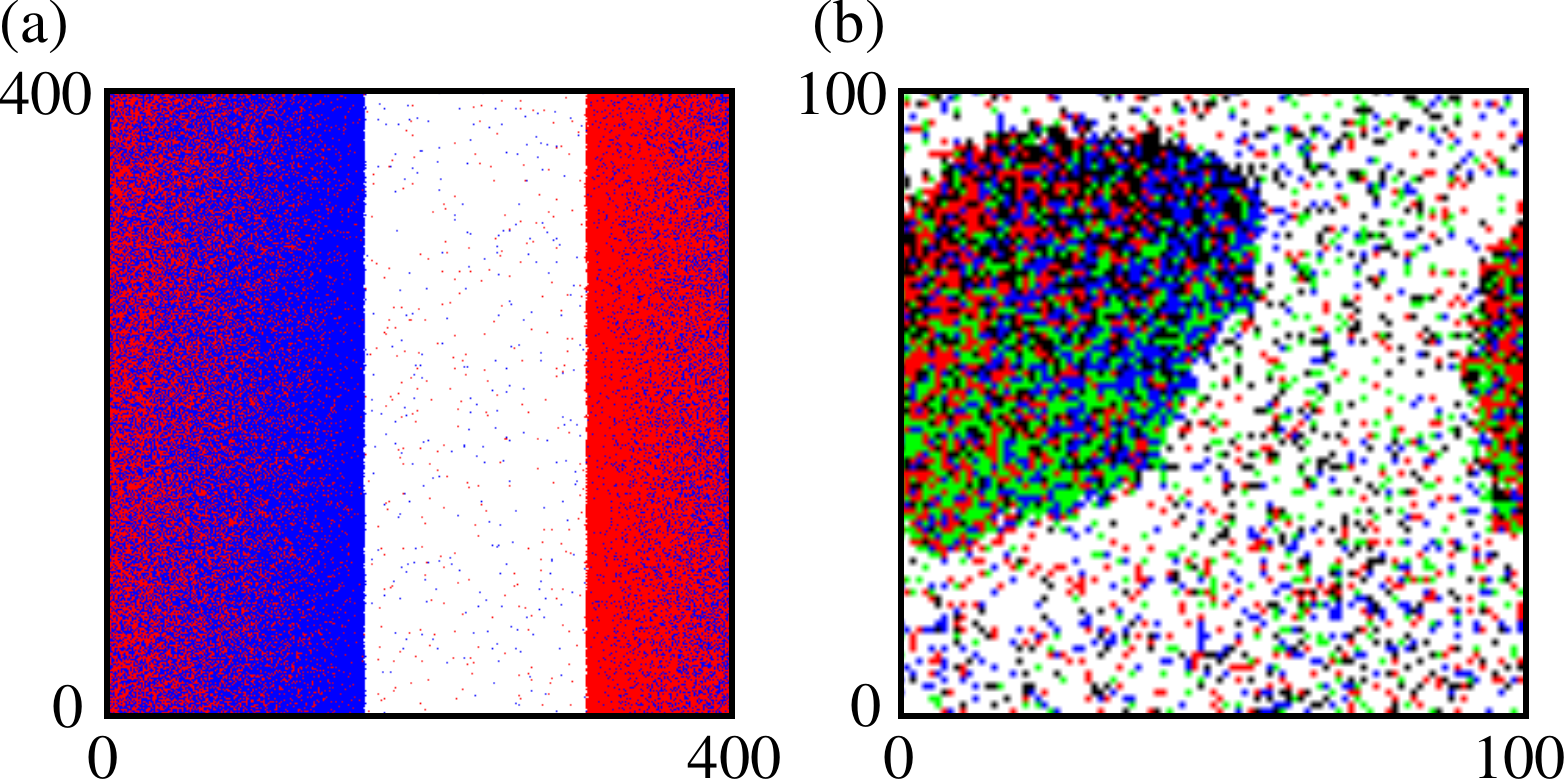}\\
\caption{Snapshots of microscopic simulations in two-dimension showing MIPS for lattices of $\alpha L\times \alpha L$ sites, with $L=100$. (a) Particles are biased only along $\hat{x}$ ($\alpha=4$, $\rho_0=0.65$) or (b) can point along the four lattice bonds ($\alpha=1$, $\rho_0=0.65$) determines the symmetry of the dense phase. A particle's direction can be inferred from its color of sites: red indicates right, green indicates up, blue indicates left, and black indicates down. The simulation parameters are $\lambda=40$, $\gamma=10$, and $D=1$.}
\label{1_alg}
\end{figure}
The total number of particles is expressed as $N \equiv \rho_0 \alpha L$, where $\rho_0 \in [0, 1]$ represents the mean density. The system exhibits homogeneity for small values of $\rho_0$ or $\lambda$, while instability in the homogeneous phases occurs at higher densities and drift. The results of $2d$ numerical simulations show that the dense and diluted phases typical of MIPS coexist in Fig.~\ref{1_alg}. Different symmetries for the coexistence phases are observed depending on whether the $2d$ case is constructed with a purely left-right bias or if biases along each of the four directions are considered.

\section{Ordering kinetics in flocking models}
While substantial advancements have been achieved to understand the steady state properties of various active systems~\cite{toner2005hydrodynamics,giomi2013defect,cates2015motility,solon2015phase,criticalABP2018,capillary2020adam,chatterjee2020flocking,chatterjee2020flocking,kursten2020dry,fruchart2021NR,solon2022susceptibility,chatterjee2022polar,codina2022small, TSVM2023,karmakar2023jamming}, there remains ample exploration potential within the domain of ordering kinetics in active systems transitioning to a non-equilibrium steady state (NESS). The comprehension of the inherent non-equilibrium dynamics steering an active system toward its steady state holds both fundamental and practical significance. In contrast to active systems, the study of ordering kinetics in non-equilibrium passive systems has been conducted for numerous decades~\cite{alan_bray,bray1993theory,sanjay_puri,aging2012,puri2014rfim,kumar2017ordering,rbcm}. Domain growth in passive systems with non-conserved scalar order parameters follows the Lifshitz-Cahn-Allen (LCA) growth law: $R(t) \sim t^{1/2}$ (\textit{Model A} of order-parameter kinetics) whereas passive systems with conserved order parameter follow a  Lifshitz-Slyozov-Wagner (LSW) growth law: $R(t) \sim t^{1/3}$ (\textit{Model B} of order-parameter kinetics). Several active systems have been explored using tools that quantify the kinetics of passive systems. These include Active Model B~\cite{wittkowski2014scalar,pattanayak2021AMB,pattanayak2021domain}, active nematics~\cite{mishra2014aspects}, SPPs in disordered medium~\cite{das2018ordering}, Model B with nonreciprocal activity~\cite{saha2020scalar}, Kuramoto oscillators~\cite{rouzaire2022dynamics}, active Brownian particles~\cite{dittrich2023growth} and motility-induced phase separated (MIPS) clusters~\cite{caporusso2023dynamics}. Moreover, an interesting observation of multiple coarsening length scales was made in the prototypical VM~\cite{Coarsening2020VM}, where velocities are found to align over a faster-growing length scale compared to density. Another intriguing result of an active system with a non-conserved vector order parameter following the growth law of the non-conserved scalar order parameter field~\cite{Dikshit_2023}. 

The two-point equal-time $(t)$ correlation function of the local scalar order parameter($\psi$) is used to study the morphology of the system during phase ordering kinetics. The notion utilizes the spatial fluctuations in the density and magnetization fields to estimate~\cite{Dikshit_2023}:
\begin{equation}\label{correlation_density}
    C_{\psi\psi}(r,t) =  \frac{1}{L^2} \sum_{i=1}^{L^2} \langle \Delta \psi_{i,t} \Delta \psi_{i+r,t} \rangle 
\end{equation}  

Where $\langle \cdots \rangle$ denotes averaging over independent initial realizations, $\Delta \psi_{i,t} = \psi_i - \psi_0$ is the local fluctuations in order parameter from the mean. The above definition of $C_{\psi\psi}$ characterizes the morphology of the spatial structures among the evolving structures separated by a distance $r$. Following a temperature quench from a random initial configuration into the ordered state, domains appear and grow with time. Similar morphology of the evolving domains with average domain size $R(t)$ would correspond to a dynamical scaling relation \cite{sanjay_puri,alan_bray,bray1993theory}: 
\begin{equation}\label{correlation}
    C(r,t) =  f \left (\frac{r}{R(t)} \right)
\end{equation}
where $f(x)$ is a time-independent scaling function. $R(t)$, estimated from the decay of $C(r,t)$ generally show a power-law growth \cite{sanjay_puri,alan_bray,bray1993theory}:
\begin{equation}\label{lengthscale}
    R(t) \sim t^\theta \, ,
\end{equation}
with $\theta$ as the growth exponent. Usually, scattering experiments are used to study the morphology of an ordering system. These experiments measure the structure factor $S(k,t)$, which is defined by the correlation function $C(r,t)$'s Fourier transform:
\begin{equation} \label{sf}
    S(k,t)=\int_{-\infty}^{\infty} C(r,t) e^{ikr} dr \, ,
\end{equation}
Furthermore, has a dynamical scaling form in $d$ dimensions:
\begin{equation}\label{sfscaling}
    S(k,t) =  R(t)^{d} g\left[ kR(t) \right] \, .
\end{equation}
The structure factor scaling functions short-distance (large-$k$) behavior for scalar order parameters, such as the density field, is given by Porod's law (for domains with smooth boundaries or scattering off sharp domain interfaces), which corresponds to $g(k) \sim k^{-(d+1)}$.   

\section{Methodologies}
~~~~~This thesis deals with computer simulations of different microscopic active systems. The emergence of macroscopic or large-scale properties in a system is anticipated to arise from the interactions among its microscopic constituents. Statistical mechanics acts as a link between the microscopic and macroscopic worlds. To refine the description, it is essential to formalize the preceding discussions, leading to the development of the theory of thermodynamic ensembles. If we denote the microscopic state of a system as $\mathcal{S}$ in the phase space (determined by both the positions and momenta of particles), the probability of the system being in a state $\mathcal{S}$ at time $t$ is represented by $P(\mathcal{S}, t)$. The temporal evolution of $P(\mathcal{S}, t)$ is determined by the master equation,
\begin {eqnarray}\label{1_MC_master}
\frac{\partial P(\mathcal{S}, t)}{\partial t}=- \sum_{\mathcal{S} \ne \mathcal{S^\prime}} [P(\mathcal{S}, t)W_{\mathcal{S}\rightarrow \mathcal{S^\prime}}-P(\mathcal{S^\prime}, t)W_{\mathcal{S^\prime}\rightarrow \mathcal{S}}],
\end {eqnarray}
The transition rate, denoted as $W_{\mathcal{S}\rightarrow \mathcal{S^\prime}}$, represents the rate at which a system transitions from one state $\mathcal{S}$ to another state $\mathcal{S^\prime}$. In a state of equilibrium, the probability distribution no longer varies with time, and thus, $P(\mathcal{S}, t)$ becomes a steady-state probability denoted as $P(\mathcal{S})$. Additionally, at steady state, the left-hand side of Eq.~\ref{1_MC_master} is zero, meaning that the partial derivative of $P(\mathcal{S}, t)$ with respect to time, $\frac{\partial P(\mathcal{S}, t)}{\partial t}$, equals zero. Furthermore, under conditions of equilibrium, the probability of the system transitioning from a specific state $\mathcal{S}$ to another state $\mathcal{S^\prime}$ is equal to the probability of transitioning from $\mathcal{S^\prime}$ to $\mathcal{S}$. This principle, also known as detailed balance, is expressed as,
\begin {eqnarray}\label{1_detail_balnc}
 P(\mathcal{S})W_{\mathcal{S}\rightarrow \mathcal{S^\prime}}=P(\mathcal{S^\prime})W_{\mathcal{S^\prime}\rightarrow \mathcal{S}}
\end {eqnarray}
Further, from Eq.~\ref{1_detail_balnc} we have,
\begin{gather}
  \frac{P(\mathcal{S})}{P(\mathcal{S^\prime})}=\frac{W_{\mathcal{S^\prime}\rightarrow \mathcal{S}}}{W_{\mathcal{S}\rightarrow \mathcal{S^\prime}}}=\frac{\exp[-\beta E(\mathcal{S})]}{\exp[-\beta E(\mathcal{S^\prime})]}  
\end{gather}
where $\exp[-\beta E(\mathcal{S})]$ represents the Boltzmann-Gibbs factor associated with the probability distribution of an equilibrium state $\mathcal{S}$. Here, $E(\mathcal{S})$ denotes the energy function of the same state, and $\beta$ is the inverse temperature. Unlike equilibrium states, where a known prescription governs the probability distribution, non-equilibrium states lack such a priori information. Solving Eq.~\ref{1_MC_master} is required to determine the steady state, a challenging task for systems driven out of equilibrium. Consequently, non-equilibrium systems fall outside the established framework of statistical mechanics built upon the familiar Boltzmann-Gibbs approach. Furthermore, non-equilibrium systems exhibit non-zero currents, such as macroscopic energy or particle currents, and probability currents within the state space. These currents arise as a consequence of an external energy supply. The presence of non-vanishing currents in the state space leads to a violation of detailed balance. In contrast to the insights gained from equilibrium statistical mechanics, many of the intuitions and principles derived from such equilibrium systems do not apply to understanding out-of-equilibrium systems. The challenge of characterizing driven many-particle systems has gained significant attention in recent decades, as highlighted in works by Ramaswamy \textit{et al}. ~\cite{ramaswamy2001physics,ramaswamy2000nonequilibrium}. In this thesis, our focus is exclusively on systems with a steady, non-equilibrium state characterized by stochastic dynamics. Below, we discuss the simulation techniques in a general context, e.g., the main algorithms and computational details that are used to simulate our systems of interest. 

\subsection{Metropolis Algorithm} 
~~~~~In the Metropolis algorithm, (1953)~\cite{metropolis1953equation,newman1999monte,landau2021guide}, configurations are produced from a preceding state through non-deterministic time evolution, employing a transition probability linked to the energy difference between the initial and final states. Any transition rate that satisfies the detailed balance is acceptable. For the Metropolis algorithm, these are as follows:
\begin {eqnarray}\label{1Metropolis_}
~~~ W_{\mathcal{S}\rightarrow \mathcal{S^\prime}} &=& \exp(-\Delta E/k_BT) ~~~~\Delta E >0\\
                 &=& 1~~~~~~\Delta E <0,
\end {eqnarray}
where $\Delta E = E_\mathcal{S^\prime} -E_\mathcal{S}$, and $k_B$ is the Boltzmann constant. The argument $\Delta E$ in the exponential is justified, considering 
that the relative probability between two states depends upon it. Note that we have set the time required for a 
trial move to unity. The implementation of this algorithm in computer simulations involves the following steps.\\
\textbf{(1)} A particle $i$ is randomly chosen.\\
\textbf{(2)} Particle $i$ is given a random displacement or exchanged with one of its 
neighbors or its identity is changed (equivalent to flipping a spin) depending upon the type 
of dynamics one is interested in.\\
\textbf{(3)} Difference in energy $\Delta E$ between the trial state and old state is calculated.\\
\textbf{(4)} A random number, denoted as $r$, is generated within the range of $0 < r < 1$.\\
\textbf{(5)} If the generated random number $r$ is less than $\exp(-\Delta E/k_BT)$, then the new state is retained.\\
\textbf{(6)} Go to the next site and go to \textbf{(3)}.\\
One Monte Carlo time step (MCS) means $N$ such attempts. The variable $N$ represents the total number of spins or particles present in the system.

In practice, Monte Carlo simulations can be adapted and extended to solve many problems beyond simple integrals, optimization, and partial differential equations. The implementation of Monte Carlo simulations are discussed in chapter {\bf Chapter~\ref{chap:Chap2}}, {\bf Chapter~\ref{chap:Chap3}}, {\bf Chapter~\ref{chap:Chap4}}, and {\bf Chapter~\ref{chap:Chap5}} respectively.

\subsection{Computational details}
The simulation tasks are carried out using the C programming language. C codes are developed and simulated on the Linux operating system (in Intel(R) Xeon(R) CPU E5-2630 v4 having clock speed 2.20 GHz, RAM 62 GB) with 40 CPU cores. GNUPLOT and Grace are employed to visualize the simulation results. All the figures are edited by Inkscape software.

\section{Outline of the thesis}
~~~~~~This Ph.D. thesis work is primarily devoted to studying active matter systems, specifically focusing on the collective behavior or flocking dynamics of SPPs~\cite{ramaswamy2010mechanics,marchetti2013hydrodynamics} and the MIPS~\cite{cates2015motility}. Additionally, it also provides the ordering kinetics of one of the minimal discrete flocking models to understand how large polar flocks originated. Overall, the thesis contributes to the broader understanding of active matter physics and sheds light on SPPs intriguing and diverse behavior in different scenarios and complex environments. The remainder of this thesis is divided into four main chapters, followed by a general summary and outlook.

In {\bf Chapter~\ref{chap:Chap2}}, we present a comprehensive Monte Carlo study that showed a discretized flocking model that can control particle mobility and produce a variety of self-organized patterns. The results showed that particle speed, temperature, and density all impact the local flux of particles, with increasing speed and density leading to increased particle flow. However, increased temperature rapidly changes the particle state and reduces effective flow. Significant volume exclusion leads to suppressed flow and promotes a jammed state. Results showed that volume exclusion or increasing particle density reduces mobility due to cluster formation and local jamming, while increased thermal fluctuations increase particle transport, but the system becomes effectively jammed at high activity. The jammed state is a robust feature that resembles real-world systems, such as animal groups, pedestrian traffic, and robot swarms, which cause high-density clogs that quickly arrest into a frozen state. This work is published in \textbf{Physical Review E} entitled as \enquote{Jamming and flocking in the restricted active Potts model},{\textbf{\underline{Mintu Karmakar}}}, Swarnajit Chaterjee,  Matthieu Mangeat, Heiko Rieger, \& Raja Paul, \textit{\textbf{Physical Review E}} {\bf 108}, 014604 (2023).

{\bf Chapter~\ref{chap:Chap3}} of the thesis explores the behavior of SPPs in disordered media employing the $q$-state active Potts model, viz., the Random Field Active Potts Model (RFAPM) and Random Diffusion Active Potts Model (RDAPM). Interestingly, a unique feature of the APM is the treadmilling of a longitudinal band opposite to a small unidirectionally applied field. The observed treadmilling behavior of the ordered phase is likely driven by unbiased and biased diffusion of the active particles. Increasing the constant unidirectional local field transforms the coexistence band into a fully liquid state, where particles align with the field direction despite thermal fluctuations. A flocking to MIPS transition can also be observed in the case of a bidirectional field. In contrast, randomly distributed field orientations cause a polar liquid to transition to a disordered gas phase as the field strength increases. The decrease in interaction strength between neighboring sites weakens particle hopping, causing a loss of spatial correlation and transitioning into a disordered phase. We found that the value of $P_{rd}^*$ indicates the critical probability at which the transition from the ordered to the disordered phase occurs. A summary of this research work will be communicated soon as \enquote{Polar flocks in disordered media},
{\textbf{\underline{Mintu Karmakar}}}, Swarnajit Chaterjee,  Matthieu Mangeat, Heiko Rieger, \& Raja Paul, \textit{Manuscript under preparation}.

{\bf Chapter~\ref{chap:Chap4}} provides an extensive numerical study of the discretized Vicsek model (DVM) on an off-lattice two-dimensional domain. The outcome elucidates a transition of the collective motion as the system switches its symmetry from discrete to continuous. The DVM consists of particles able to execute motion on a plane in $q$ discrete angular directions like the active clock model (ACM) and follows dynamical rules of particle alignment and movement inspired by the prototypical VM. We find a novel cluster phase, macrophase for small $q$ and noise. Soon, we observe the formation of microphase and cross-sea in the phase coexistence region as $q$ increases, and the system approaches the VM. These results are quantified by estimating the number fluctuations, correlations, the pinning properties of the orientation vector, and the nature of the ordering of the liquid phase with the characterization of the coexistence region. Although the giant number fluctuations for large $q$ corroborate with the nature of phase separation in the coexistence region, the large length-scale behavior of the direction of the global ordering and correlations do not correspond to it. We also investigate the stability of the ordered liquid phase and find it metastable to the nucleation of droplets of different polarization. Results from this study is under review at \textbf{New Journal of Physics} entitled \enquote{Consequence of anisotropy on flocking: the discretized Vicsek model},
{\textbf{\underline{Mintu Karmakar}}}, Swarnajit Chaterjee, Raja Paul, \& Heiko Rieger,\\ \textit{\textbf{New J. Phys.}} \textbf{26}, 043023 (2024). 

In {\bf Chapter~\ref{chap:Chap5}}, we undertake a detailed numerical study of the ordering kinetics in the two-dimensional active Ising model (AIM), a discrete flocking model with a non-conserved scalar order parameter. For a quench into the liquid-gas coexistence region and in the ordered liquid region, the characteristic length scale of both the density and magnetization domains follow the Lifshitz-Cahn-Allen (LCA) growth law: $R(t) \sim t^{1/2}$, consistent with the growth law of passive systems with scalar order parameter and non-conserved dynamics. The system morphology is analyzed with the two-point correlation function, its Fourier transform, and the structure factor conforms with the well-known Porod’s law, a manifestation of the coarsening of compact domains with smooth boundaries. We also find the domain growth exponent unaffected by the active particle's noise and self-propulsion velocity. However, transverse diffusion is found to play a key role in the growth kinetics of the AIM. We extract the same growth exponent by solving the hydrodynamic equations of the AIM. This work is under review at \textbf{Physical Review E} entitled \enquote{Ordering kinetics in the active Ising model},
Sayam Bandyopadhyay, Swarnajit Chaterjee, Aditya Kumar Dutta,  {\textbf{\underline{Mintu Karmakar}}}, Heiko Rieger, and Raja Paul, \textit{\textbf{Physical Review E}} {\bf 109}, 064143 (2024).

In {\bf Chapter~\ref{chap:Chap6}}, we conclude the thesis with an overall summary of the investigations presented in the previous chapters. Additionally, we discuss the outlook for the research work presented in this thesis.

\chapter{Transition from flocking to jamming (MIPS) in the restricted Active Potts Model (rAPM)}\label{chap:Chap2}
We study the active Potts model with either site occupancy restriction or on-site repulsion to explore jamming and kinetic arrest in a flocking model. The incorporation of such volume exclusion features leads to a surprisingly rich variety of self-organized spatial patterns. While bands and lanes of moving particles commonly occur without or under weak volume exclusion, strong volume exclusion along with low temperature, high activity, and large particle density facilitates jams due to motility-induced phase separation. Through several phase diagrams, we identify the phase boundaries separating the jammed and free-flowing phases and study the transition between these phases which provide us with both qualitative and quantitative predictions of how jamming might be delayed or dissolved. We further formulate and analyze a hydrodynamic theory for the restricted APM which predicts various features of the microscopic model.


\section{Introduction}
Flocking is a collective phenomenon of active matter \cite{marchetti2013hydrodynamics,shaebani2020computational}, occurs in ensembles of self-propelled particles and denotes the emergence of ordered motion of large clusters, called flocks. It plays a significant role in a wide range of systems across disciplines including physics, biology, ecology, social sciences, and neurosciences, and is abundantly observed in nature: from human crowds \cite{bottinelli2016emergent,helbing1995social}, mammalian herds \cite{garcimartin2015flow}, bird flocks \cite{ballerini2008interaction}, fish schools \cite{becco2006experimental,calovi2014swarming} to unicellular organisms such as amoebae, bacteria \cite{steager2008dynamics,peruani2012collective}, collective cell migration in dense tissues \cite{giavazzi2018flocking}, and sub-cellular structures including cytoskeletal filaments and molecular motors \cite{schaller2010polar,sumino2012large,sanchez2012spontaneous}.

The paradigmatic model to describe the flocking transition is the Vicsek model (VM)~\cite{vicsek1995novel,toner1995long,toner1998flocks,toner2012reanalysis,solon2015phase} in which  individual particles tend to align with the average direction of the motion of their  neighbors. At low noise and high density, the particles cluster and move collectively in a common direction, which is the landmark of flocking. The transition from the gas phase at high noise/low density 
to the polar ordered Toner-Tu phase at low noise/high density, displaying long-range order by coherent motion of all particles, is first order~\cite{gregoire2004onset}. But, in contrast to conventional first-order phase transition scenarios, the coexistence phase of the VM shows either multiple bands of collectively moving particles denoted as micro-phase separation \cite{solon2015phase,solon2015pattern} or a polar ordered cross sea phase \cite{kursten2020dry}. Since the early 2000s, many variations of the original VM have been studied (see Ref.~\cite{shaebani2020computational} for an overview). One of the significant works in the context of this chapter work is done by Peruani et al. in Ref.~\cite{peruani2011traffic}, which shows new emerging features like traffic jams, gliders, and kinetically arrested states. Recently discretized versions of VM have been analyzed: the active Ising model (AIM)~\cite{solon2013revisiting,solon2015flocking,ishibashi2022solitary}, the active Potts model (APM)~\cite{chatterjee2020flocking,mangeat2020flocking,ishibashi2022solitary} and the active clock model (ACM) \cite{chatterjee2022polar,solon2022susceptibility}, in which particles move on a lattice in two (AIM) or more (APM, ACM) discrete directions. Common to those is that the micro-phase separation in the coexistence region is replaced by macro-phase separation - a single collectively moving band~\cite{solon2013revisiting,solon2015flocking,chatterjee2020flocking,mangeat2020flocking,chatterjee2022polar}.

Self-propelled particles with repulsive interactions, like for instance active Brownian particles (ABPs)~\cite{romanczuk2012active}, show at high density and high P\'eclet number another cluster state, different from flocking via alignment, denoted as motility-induced phase separation (MIPS) \cite{cates2015motility}.
Consequently, the interplay between alignment and repulsive interactions could lead to complex phase diagrams, as was demonstrated for the VM with repulsive particle interactions~\cite{gregoire2003moving} or ABPs with alignment interactions~\cite{martin2018collective,sese2018velocity}.

In this chapter, we address the question, what happens to the phase diagram of the APM when on-site interactions between the particles are present, either in the form of hard or soft core repulsion or in the form of a maximal occupancy larger than one of each site, the restricted APM (rAPM). In analogy, the VM with particle-particle repulsion one expects the phase diagram for the rAPM to be enriched by at least one MIPS state, but also other kinetically arrested states can occur as active lattice gas models with repulsive interactions ~\cite{peruani2011traffic}. These are sometimes denoted as ``jammed states'', not to be confused with jamming occurring in active glasses~\cite{berthier2019glassy}.

The chapter is organized as follows. In Sec.~\ref{s2}, we define the restricted active Potts model and provide details of the simulation protocols. Sec.~\ref{s3} presents our results for three different versions of the on-site
repulsion: (i) `maximum particle per site' (MPS) is restricted to one akin to the active lattice gas (ALG)~\cite{kourbane2018exact}, (ii) hard-core restriction or ${\rm MPS} > 1$, and (iii) soft-core repulsion. 
In Sec.~\ref{secHydro}, we present the hydrodynamic description of our model. Finally, we conclude this chapter with a summary and discussion of the results in Sec.~\ref{s4}.

\section{Modeling and simulation details}
\label{s2}
\begin{figure}[!t]
\centering
\includegraphics[width=\columnwidth]{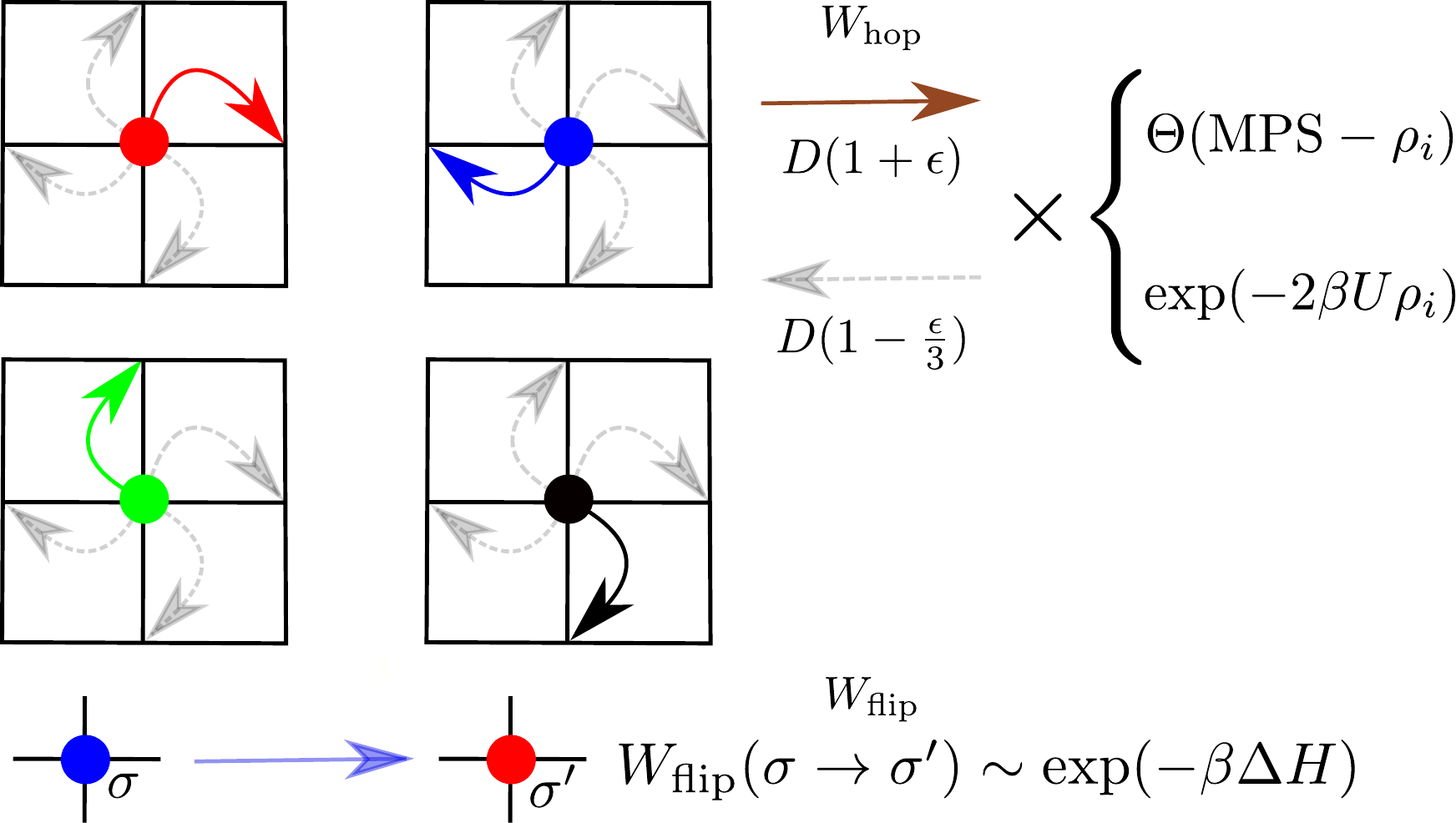}
\caption{(Color online) (above) Sketch of the $q=4$ rAPM showing biased hopping (solid bent arrow) to a neighboring lattice site with rate $D(1+\epsilon)$ and unbiased hopping (faint bent arrow) along the remaining directions with rate $D(1-\epsilon/3)$. hard-core restriction is represented by a Heaviside step function $\Theta({\rm MPS}-\rho_i)$ 
and soft-core repulsion is represented by the probability $\exp(-2\beta U \rho_i)$. Red, green, blue, and black circles represent particles of state $\sigma = \{1, 2, 3, 4\}$, respectively. (below) Illustration of on-site flipping of a particle from $\sigma=3$ to $\sigma=1$ with the flipping rate $W_{\rm flip}$.}
\label{fig1}
\end{figure}

We consider an ensemble of $N$ particles defined on a two-dimensional square lattice of size $L^2$ with periodic boundary conditions applied on both sides, where $L$ is the linear lattice dimension. The average particle density $\rho_0$ is then defined as $\rho_0=N/L^2$. The model is built upon the APM~\cite{chatterjee2020flocking,mangeat2020flocking} in which the dynamics is governed either by the on-site flipping of the internal spin state or by nearest-neighbor hopping. Besides, we now propose restrictions on particle hopping. We suggest three types of mutually exclusive restrictions: a particle is allowed to hop to its neighbor if (a) that neighbor is empty or (b) the population of the neighboring site is less than the maximum occupation per site (hard-core restriction), or (c) the hopping is allowed with a probability (soft-core repulsion). A schematic diagram of this arrangement is shown in Fig.~\ref{fig1}. 

The spin state of the $k$-th particle on lattice site $i$ is denoted by $\sigma^k_i$, with an integer value in $[1,q]$, while the number of particles in state $\sigma$ on-site $i$ is $n^\sigma_i$. The local density on-site $i$ is then defined by $\rho_i=\sum_{\sigma=1}^q n^\sigma_i$, counting the total number of particles on the site. The Hamiltonian of a $q$-state APM is defined as $H_{\rm APM}=\sum_i H_i$ decomposed as the sum of local Hamiltonian $H_i$ \cite{chatterjee2020flocking,mangeat2020flocking}:
\begin{equation}
\label{Hapm}
H_i=-\frac{J}{2\rho_i}\sum_{k=1}^{\rho_i}\sum_{l\ne k}(q\delta_{\sigma_i^k,\sigma_i^l}-1),
\end{equation}
where $J=1$ is the coupling between the neighboring sites. We take $q=4$. A particle at site $i$ with state $\sigma$ either flips to another state $\sigma^\prime$ or hops to any of the neighboring sites (as permitted by the different restriction protocols). The local magnetization corresponding to state $\sigma$ at site $i$ is defined as $m_i^{\sigma}$:
\begin{equation}
\label{Hapm_mag}
m_i^{\sigma}=\sum_{j=1}^{\rho_i}\frac{q\delta_{\sigma,\sigma_i^j}-1}{q-1} \, .
\end{equation}

\subsection{On-site alignment or flipping dynamics}
A particle at site $i$ with state $\sigma$ can flip and align with another state $\sigma^\prime$; therefore, flipping is a purely on-site phenomenon. From Eq.~\eqref{Hapm}, one can calculate the local energy difference before and after the flipping. From Refs.~\cite{chatterjee2020flocking,mangeat2020flocking}, the expression of the energy difference reads:
\begin{equation}
\label{delH}
\Delta H_i=H_{i}^{\rm new}-H_i^{\rm old}=\frac{qJ}{\rho_i}(n_i^{\sigma}-n_i^{\sigma^\prime}-1) \, .
\end{equation}
The flipping is then accepted with the rate:
\begin{align}
\label{flipeq}
W_{\rm flip}(\sigma\to\sigma^\prime)&=\gamma \exp(-\beta \Delta H_i) \nonumber \\
&=\gamma \exp\left[-\frac{q\beta J}{\rho_i}(n_i^\sigma-n_i^{\sigma^\prime}-1)\right] ,
\end{align}
where $\beta=T^{-1}$ is the inverse temperature. It should be noted that for MPS = 1, only one particle is allowed per site and consequently, on-site alignment interaction is absent for this limit of the model. From Eq.~\eqref{flipeq}, $n_i^{\sigma}=1$ and $n_i^{\sigma^\prime}=0$ leads to $\Delta H_i=0$ and hence we have $W_{\rm flip}(\sigma\to\sigma^\prime)= \gamma$ as the flipping rate of particles for MPS = 1. For hard-core restriction and soft-core repulsion, we take $\gamma=1$. 

\subsection{Biased diffusion or hopping dynamics}
The biased diffusion mechanism is similar to the process described in Refs. \cite{chatterjee2020flocking,mangeat2020flocking}. A particle with state $\sigma$ hops to a direction $p$ with rate:
\begin{equation}
\label{whop}
W_{\rm hop}(\sigma,p)=D\left(1+\epsilon\frac{q\delta_{\sigma,p}-1}{q-1}\right) \, ,
\end{equation}
where $\epsilon$ ($0 \leqslant \epsilon \leqslant q-1$) is the self-propulsion parameter. At $\epsilon=q-1$, particles move purely ballistically, resulting in complete self-propulsion, while $\epsilon=0$ corresponds to the absence of self-propulsion. However, with $\epsilon = 0$, particles are not passive and can still diffuse on the lattice (see Eq.~\ref{whop}) but without any bias. This differs from the VM where the zero-velocity limit corresponds to immobile particles with dynamics reminiscent of the XY model. Let us also mention that for $\epsilon>0$, the system is out-of-equilibrium but the model is not at equilibrium even when $\epsilon=0$. Following Ref. \cite{solon2015flocking}, it can be shown using Kolmogorov’s criterion~\cite{kolmogorov1936markov} that the system does not satisfy detailed balance with respect to any distribution i.e. the products
of the forward transition rates differs from that of the reverse order. The system would be effectively an equilibrium system only when the particles would cease to move (being jammed), for instance when the density is high and hopping restrictions are strong.

Under the purely repulsive hard-core exclusion, biased diffusion is subjected to the maximum number of particles allowed per site set by the parameter MPS and Eq.~\eqref{whop} gets modified in the following way:
\begin{equation}
\label{HC_eq}
W^{\rm HC}_{\rm hop}(\sigma,p)=W_{\rm hop}(\sigma,p) \Theta({\rm MPS}-\rho_i) \, ,
\end{equation}
where $\Theta({\rm MPS}-\rho_i)$ is a Heaviside step function and is defined as:
\[
\Theta({\rm MPS}-\rho_i) =
\begin{cases}
1, & \text{for} \quad \rho_i < \text{MPS} \\
0, & \text{otherwise} \, .
\end{cases}
\]
$\rho_i$ is the particle number at a neighboring site $i$ to which a hopping is attempted. MPS = 1 is a special case under the hard-core exclusion category where a move to the neighbor is only possible if that site is empty. Therefore, unlike hard-core and soft-core repulsions, on-site interactions between the particles are absent for MPS = 1. This can also be thought of as an asymmetric simple exclusion process (ASEP) in which particles perform biased random walks under the hard-core repulsion that two particles cannot occupy the same site at a given time (MPS $>1$ is the non-lattice-gas variety of the ASEP). Therefore, MPS = 1 constructs the simple-exclusion ``active lattice gas''~\cite{kourbane2018exact} version of the APM.

A soft-core repulsion would allow a particle to hop to a neighboring site $i$ from a randomly chosen site depending on the change in the local field. The local field is defined by: 
\begin{equation}
\label{BH}
V(\rho_i)=U\rho_i(\rho_i -1) \, ,
\end{equation}
where $U$ is an interaction coefficient that can be attractive ($U<0$) or repulsive ($U>0$). After hopping, the local field with $\rho_i+1$ particles at site $i$ becomes
$V(\rho_i+1)=U\rho_i(\rho_i +1)$.
Particle hopping to site $i$ is then accepted with probability:
\begin{align}
P &= \min[1,\exp(-\beta \Delta V)] \\ \nonumber 
&=\min[1,\exp(-2\beta U \rho_i)] \, ,
\end{align}
where $\Delta V=V(\rho_i+1)-V(\rho_i)=2U\rho_i$. Then, for soft-core repulsion, the modified form of Eq.~\eqref{whop} can be written as:
\begin{equation}
\label{SC_eq}
W^{\rm SC}_{\rm hop}(\sigma,p)= W_{\rm hop}(\sigma,p) \exp(-2\beta U \rho_i) \, .
\end{equation}
$U$ symbolizes the restriction strength that regulates particle accumulation on a lattice site. Note that, $U \leqslant 0$ denotes $\exp(-\beta\Delta V) \geqslant 1$ which physically signifies an attractive field where particles can freely crowd into a site similar to the APM \cite{chatterjee2020flocking}. In this chapter, we only consider repulsive interactions, $U>0$, acting in the limit $U\to\infty$ 
like volume exclusion. 

\subsection{Simulation Details}
\label{simul}
Simulation evolves in the unit of Monte Carlo steps (MCS) $\Delta t$ resulting from a microscopic time $\Delta t/N$, $N$ being the total number of particles. During $\Delta t/N,$ a randomly chosen particle either updates its spin state with probability $p_{\rm flip}=W_{\rm flip}\Delta t$ or hops to one of the neighboring sites with probability $p_{\rm hop}=W_{\rm hop}\Delta t$. For $q$-state APM, an expression for $\Delta t$ can be obtained by minimizing the probability of nothing happens $p_{\rm wait}=1-(p_{\rm hop}+p_{\rm flip})$ \cite{chatterjee2020flocking}, 
\begin{equation}\label{delt}
\Delta t=[qD+\exp(q\beta J)]^{-1} \, .
\end{equation}
This hybrid Monte Carlo dynamics was used previously in the simulations of the AIM~\cite{solon2013revisiting,solon2015flocking}, APM~\cite{chatterjee2020flocking,mangeat2020flocking}, and ACM~\cite{chatterjee2022polar}. Instead of computing $\Delta t$ from the minimum of $p_{\rm wait}$ for systems that have small transition probabilities and therefore large $p_{\rm wait}$ (one has to generate random numbers until the chosen transition is accepted), one can also apply a Gillespie-like algorithm where one computes the time at which the next event will take place in the system.

\section{Numerical Results}
\label{s3}
In this section, we present the numerical simulation results of the $q = 4$ rAPM with MPS = 1, hard-core restriction (${\rm MPS}>1$) and soft-core repulsion. The models are simulated on a square lattice of linear size $L = 100$ with periodic boundary conditions, where individual particle states $\sigma = \{1, 2, 3, 4\}$ correspond to the movement directions right, up, left and down, respectively. Simulations are performed for various control parameters: $D = 1$ is kept constant throughout the simulations, $\beta=1/T$ regulates the noise in the system, $\rho_0=N/L^2$ defines the average particle density, and self-propulsion parameter $\epsilon$ dictates the effective velocity of the particles. Starting from a homogeneous initial condition, the Monte Carlo algorithm (Sec.~\ref{simul}) evolves the system under various control parameters until the stationary distribution is reached. Following this, measurements are carried out and thermally averaged data are recorded.

\subsection{MPS = 1 (ALG version of the rAPM)}
\label{ALG}
In this segment, we present the results for $q=4$ state APM with MPS = 1. Following Ref.~\cite{kourbane2018exact} and Eq.~\eqref{flipeq}, $\gamma$ represents the flipping parameter with flipping probability $\gamma\Delta t$. We then define the P\'eclet number Pe as:
\begin{equation} \label{peclet}
{\rm Pe} = \frac{v}{\sqrt{D\gamma}} \, ,
\end{equation} 
where $v=4 D \epsilon/3$ is the self-propulsion velocity in the hydrodynamic limit of the 4-state APM \cite{chatterjee2020flocking} and we get ${\rm Pe}=(4\epsilon/3)\sqrt{D/\gamma}$. As Pe is proportional to $\epsilon$, for small Pe, diffusion dominates, and the effect of self-propulsion becomes negligible. Conversely, the effect of activity gets more and more pronounced as Pe increases.

\begin{figure}[t]
\centering
\includegraphics[width=\columnwidth]{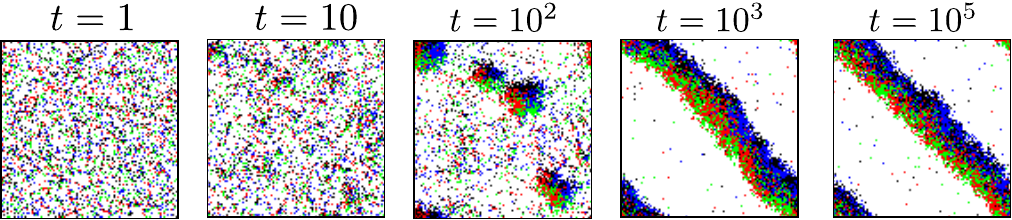}
\caption{(Color online) Time evolution snapshots of the rAPM with MPS = 1 displaying MIPS. The right boundary of the diagonal high-density band is due to particles of state $\sigma=3$ (blue) and $\sigma=4$ (black) and the left boundary is due to particles of state $\sigma=1$ (red) and $\sigma=2$ (green). Parameters: Pe = 50 and $\rho_0=0.3$.}
\label{fig2}
\end{figure}

In Fig.~\ref{fig2}, snapshots demonstrate MIPS via the time evolution of the
rAPM starting from a random initial configuration. Initially, the self-propelled particles (SPPs) nucleate stable clusters (where domains of all the four states can be visible) and coarsen and coalesce at later times ($t=10^5$) to phase-separate into a diagonal solid phase that stabilizes in a steady state and a gas phase, a consequence of MIPS. A careful examination of the diagonal domain ($t=10^5$) reveals that the right (upper) and left (lower) domain boundaries are formed by multiple opposite spin states ($e.g.$ for a diagonal band spanning from the bottom-right corner to the top-left corner, the right domain boundary is always formed by particles with $\sigma=3$ and 4 and the left domain boundary is formed by particles of $\sigma=1$ and 2). A two-state variant of this model having $\sigma=1$ and 3 would result in a vertically jammed band~\cite{kourbane2018exact} and a combination of $\sigma=2$ and 4 would result in a horizontally jammed band. Therefore, the high-density diagonal band arises when the steady-state culminates into orthogonally directed clusters intercepted by oppositely directed clusters.

\begin{figure}[!t]
\centering
\includegraphics[width=\columnwidth]{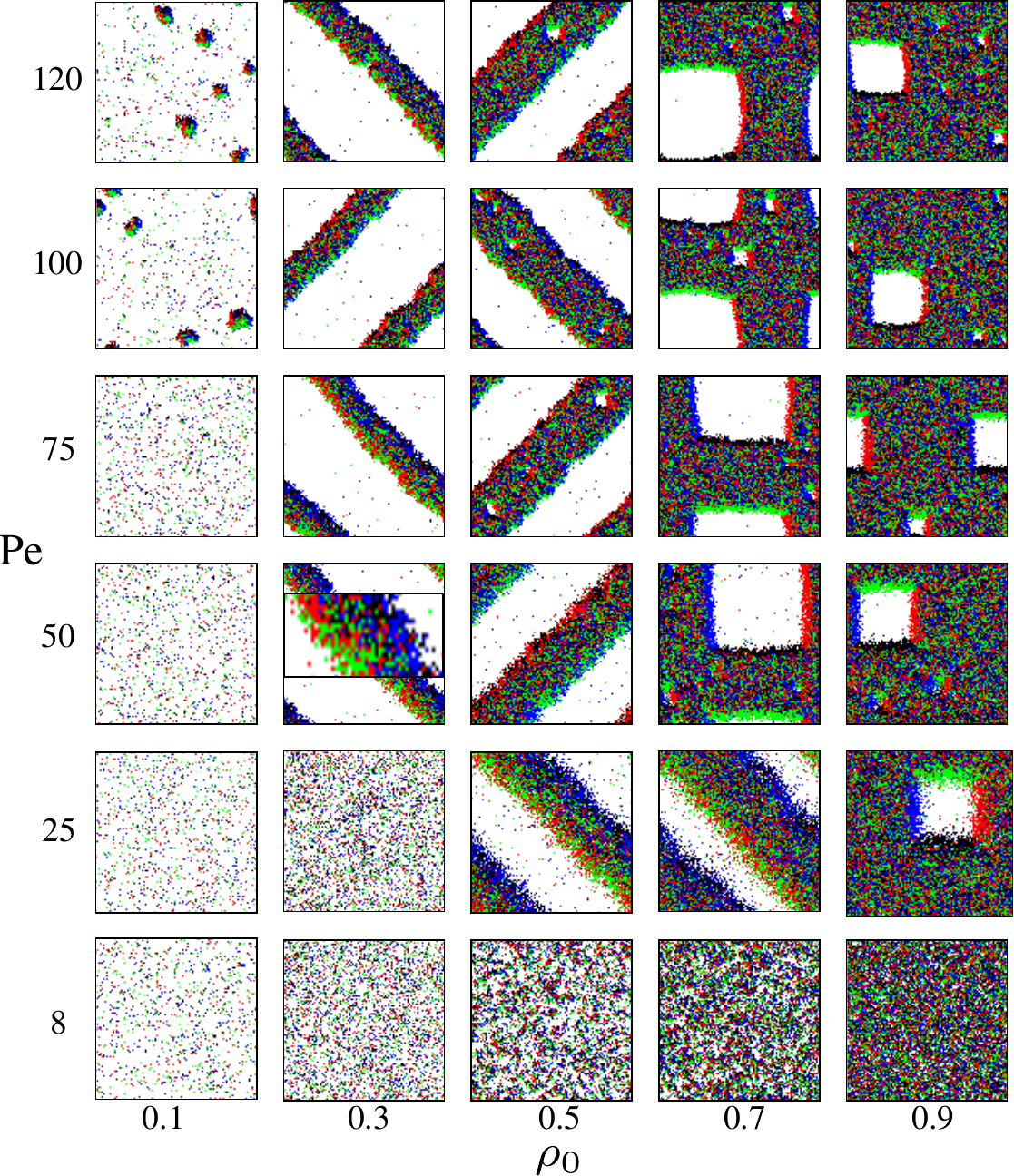}
\caption{(Color online) ${\rm Pe}-\rho_0$ phase diagram (for $\gamma=10$ and varying $\epsilon$) for MPS=1 illustrated by snapshots at time $t=10^5$. As density increases, the system undergoes a transition from the disordered gas phase to a MIPS state~\cite{cates2015motility} for ${\rm Pe}>8$. At the intermediate densities, we observe solid diagonal bands where a small section of the band (Pe = 50, $\rho_0=0.3$) is magnified to provide a detailed structure of the domain.}
\label{fig3}
\end{figure}

The steady-state behavior of the rAPM with MPS = 1 is illustrated in Fig.~\ref{fig3} by representative late-stage snapshots as a function of propulsion strength Pe and average particle density $\rho_0$. As a function of density, the system undergoes a transition from a disordered gaseous phase to the MIPS state at both intermediate and large Pe. At intermediate densities, we observe diagonal high-density bands on a disordered background which we discussed in detail in the context of Fig.~\ref{fig2}. As density is increased, the area of the high-density solid phase also increases by shrinking the area of the gaseous phase. The boundaries of such square gaseous regions are always formed by particles of oppositely moving states ($\sigma=1$: right vertical boundary, $\sigma=3$: left vertical boundary, $\sigma=2$: up horizontal boundary, and $\sigma=4$: down horizontal boundary) whose preferred hopping directions are unavailable due to the restriction. Nevertheless, density fluctuation happens from the domain boundaries to the square void space (therefore, the center of mass of the jammed clusters shifts position with time, though slowly). The internal structure of the high-density phases in both these MIPS states are similar i.e. orientationally disordered (due to the lack of the alignment interactions) and can be described as amorphous solid. As mentioned before, the formation of these jammed amorphous solid phases is a consequence of MIPS. MIPS refers to the spontaneous phase separation of a system of SPPs with purely repulsive interactions (and without any attractive interaction) into coexisting dense and dilute phases. The physics of MIPS can be understood as slowing down of SPPs due to enhanced crowding when the local density of SPPs increases in some part of the system due to fluctuation, as shown in Fig.~\ref{fig3}.

Similar to our model, the model used by Peruani et al. in Ref.~\cite{peruani2011traffic} is a two-dimensional lattice model with periodic boundary conditions where particles can have four possible orientations similar to the 4-state APM and the orientation determines the moving direction of the particle. However, Peruani's model only resembles the MPS = 1 limit of our model; the soft-core and hard-core cases that we discussed in our study were not considered by Peruani and co-authors. Even between the MPS = 1 limit of our model and Peruani's model, there are the following distinct differences in the flipping and hopping dynamics:

(a) In our model, flipping is a purely on-site phenomenon, whereas, in Peruani's model, the flipping or reorientation transition rate depends on the nearest lattice neighbors of the particle. In the MPS = 1 limit of our model, on-site interactions between the particles are totally absent, unlike hard-core and soft-core scenarios, but in Peruani's model, ferromagnetic alignment interaction with the neighbors is possible.

(b) Hopping dynamics in our model depends on the self-propulsion parameter $\epsilon$ $(0 \leqslant \epsilon \leqslant q-1)$ and for $\epsilon<q-1$, we always have a non-zero hopping rate along the non-preferred directions. But, in Peruani's model~\cite{peruani2011traffic}, a particle can not hop to a direction which is not preferred by its orientation and hopping to a neighboring node is only possible if that node is empty. Now, in our model with MPS = 1, we allow hopping to a neighboring site only if that site is empty, but even with MPS = 1, a particle can hop to a non-preferred direction (if the corresponding hopping probability allows it and the site is empty).

The difference in the shape of the jamming state between the present study and Peruani et al.'s study are due to these different underlying dynamics. For MPS = 1, we have observed only MIPS jammed clusters (see Fig.~\ref{fig3}); no traffic jams, gliders, and bands are observed, similar to Peruani's results due to the absence of ferromagnetic alignment interaction with the neighbors. 


\begin{figure}[t]
\centering
\includegraphics[width=\columnwidth]{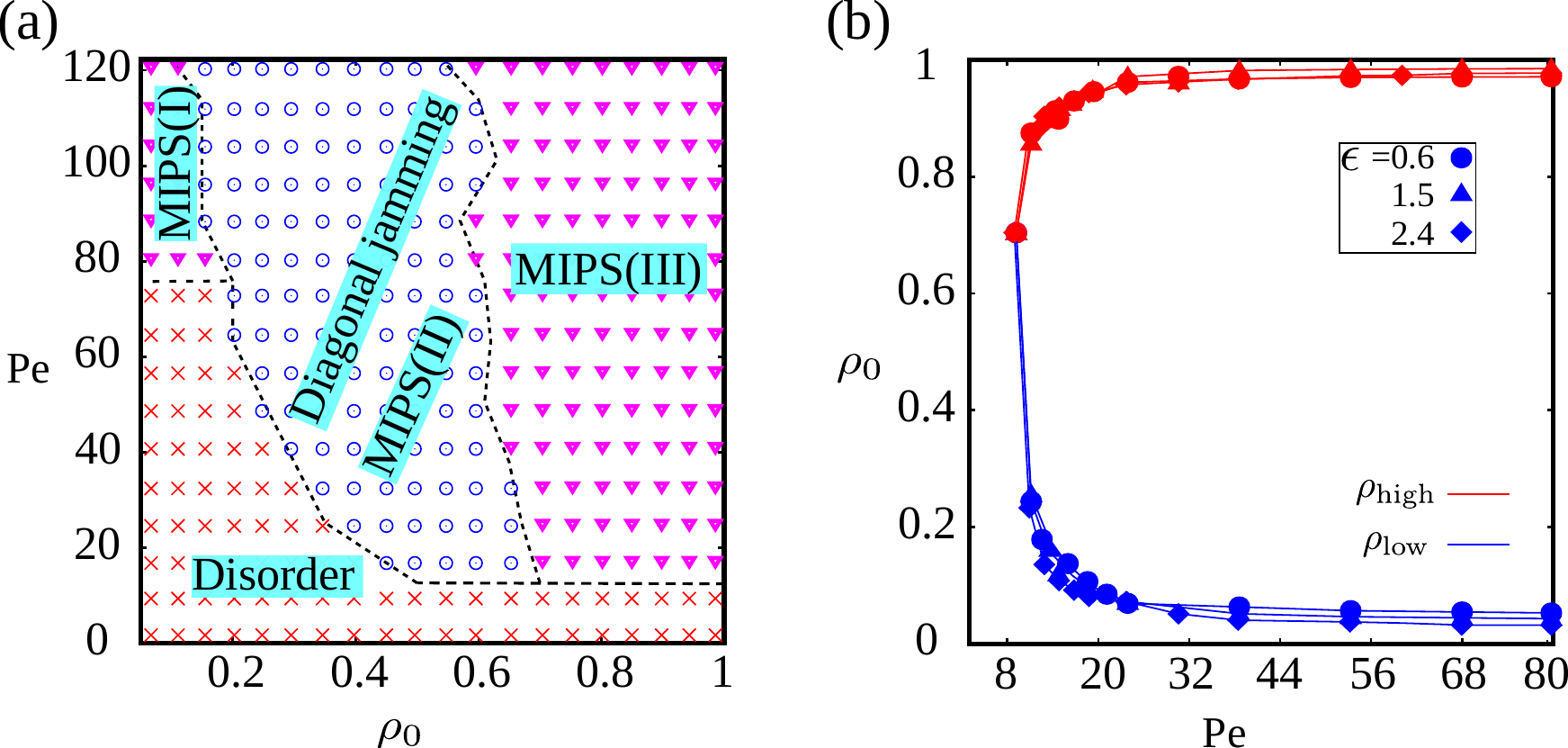}
\caption{(Color online) (a) ${\rm Pe}-\rho_0$ phase diagram for MPS=1 for the snapshots in Fig.~\ref{fig3}. (b) MIPS phase diagram (for $\epsilon=0.6$, 1.5, and 2.4) showing the binodals $\rho_{\rm low}$ and $\rho_{\rm high}$ are independent of $\epsilon$. The critical Pe is estimated as ${\rm Pe}_c \simeq 8$ above which the phase separation occurs.}
\label{fig4}
\end{figure}

Phase diagrams of the rAPM with MPS = 1 are shown in Fig.~\ref{fig4}. Fig.~\ref{fig4}(a) shows the quantitative analog of the diagram shown in Fig.~\ref{fig3} where three different shades signify the three MIPS states having similar internal domain morphology but different shapes. We categorized these three MIPS states as MIPS (I) (small cluster state at large Pe and small $\rho_0$), MIPS (II) (diagonal high-density state at intermediate $\rho_0$), and MIPS (III) (high-density state with a square gaseous domain at large $\rho_0$). At large Pe, as the average density increases, the system transitions from the MIPS (I) state to the MIPS (III) state via the MIPS (II) state of jammed diagonal stripes. The system remains in the gaseous phase at small Pe because of low activity (diffusion dominates self-propulsion) and then transitions to the MIPS (III) state at large densities.

We further compute the binodals $\rho_{\rm low}$ and $\rho_{\rm high}$ for several $\epsilon$ and plot the resulting phase diagram in Fig.~\ref{fig4}(b). The binodals are the coexisting densities and physically signify the average densities of the gas and ordered phases at a given Pe and are estimated by calculating the average densities in different square boxes inside the high and low-density regions. From the diagram we observe that the binodals are independent of $\epsilon$ up to fluctuations and the critical Pe is estimated as ${\rm Pe}_c \simeq 8$ above which phase separation proceeds via spinodal decomposition. The shape of the phase diagram and the qualitative nature of the coexistence lines are similar to the diagram obtained for the active lattice gas~\cite{kourbane2018exact} where the critical Pe was estimated as ${\rm Pe}_c \simeq 4$.

\subsection{Hard-core restriction (${\rm MPS} > 1$)}
\label{HC}

This section presents our findings of the rAPM with hard-core restriction where we restrict the number of allowed particles on a site (${\rm MPS} > 1$). A lower MPS signifies higher restriction on particle movement.

\begin{figure*}[t]
\centering
\includegraphics[width=\textwidth]{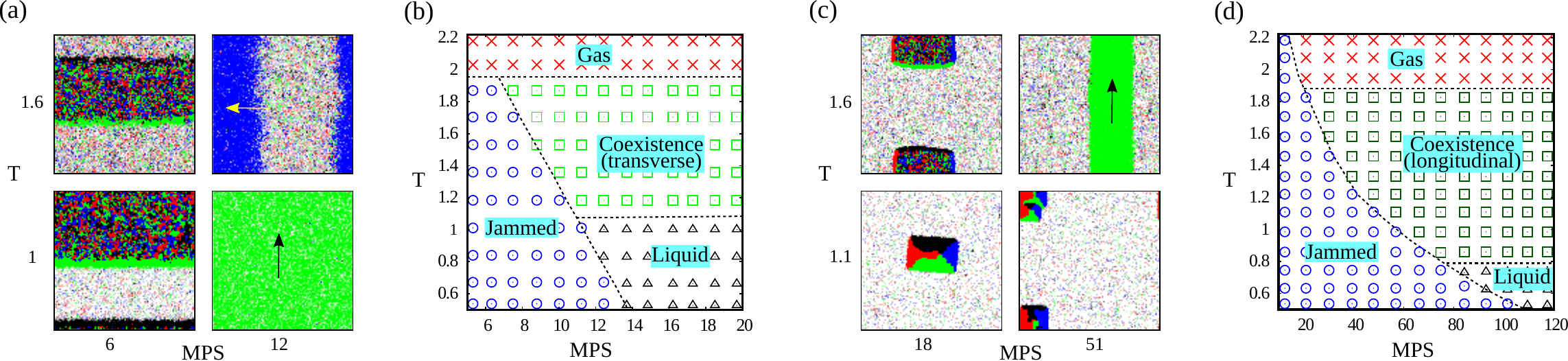}
\caption{(Color online) (a \& c) Steady-state snapshots of the 4-state rAPM with MPS$>$1 in the $T-{\rm MPS}$ plane. (a) For $\epsilon=0.9$, an increase in ${\rm MPS}$ induces a transition from the jammed MIPS state to the liquid phase at lower $T$ and to the coexistence region with a transversely propagating liquid band at a higher $T$. (c) Snapshots for $\epsilon=2.7$ where a jam-to-lane transition occurs at large ${\rm MPS}$ and higher $T$. Here the coexistence band is longitudinal. Legend: red ($\sigma=1$): right; green ($\sigma=2$): up; blue ($\sigma=3$): left; black ($\sigma=4$): down; white: empty. Arrows indicate the direction of motion. (b \& d) $T-{\rm MPS}$ phase diagrams showing the four phases of rAPM where phase transition happens from the jammed state to the free-flowing state as either MPS or $T$ is increased. Parameter: $\rho_0=4$.}
\label{fig5}
\end{figure*}

Fig.~\ref{fig5}(a) and Fig.~\ref{fig5}(c) respectively show steady-state snapshots as a function of temperature and MPS for small ($\epsilon=0.9$) and large ($\epsilon=2.7$) propulsion velocities. In Fig.~\ref{fig5}(a), the system exhibits phase-separated orientationally disordered jammed domains with well-defined ordered domain boundaries for strong repulsion whereas it shows features of unrestricted APM as relaxation on particle movement increases. By jamming we denote a transition from a free-flowing state to a high-density kinetically arrested solid configuration~\cite{liu2010jamming} due to MIPS. Although domain boundaries are ordered, the preferred directions of motion of the particles on the boundaries are inaccessible due to the hard cutoff and as a consequence, the jammed bands are almost immobile. With more freedom of movement, the system manifests a liquid phase at low temperature and a liquid-gas coexistence region (with a transversely moving liquid band on a gaseous background) at a relatively higher temperature typical of unrestricted flocking models~\cite{solon2013revisiting,solon2015flocking,chatterjee2020flocking,chatterjee2022polar}. In this chapter, by liquid, we always mean a liquid phase with respect to orientational order. A further increase in temperature leads to a disordered gaseous phase (not shown in the snapshots). \\

Fig.~\ref{fig5}(c) is analogous to Fig.~\ref{fig5}(a) but for higher particle velocity. Fig.~\ref{fig5}(c) at low temperature demonstrates jammed clusters with four orientationally ordered sub-domains (as $q=4$) in a gridlocked position (where the four quadrants orient against each other in such a way that the cluster is jammed), although the ordering does not extend over a long distance even at long times and the global polarization of such a cluster is zero. Such a configuration is almost immobile although there are fluctuations in the boundaries. With their preferred hopping directions fully crowded, particles from the boundaries of such a cluster can hop to sites in the gaseous region but the probability of such a jump is much lower when the particle velocity is high [see Eq.~\eqref{whop}]. At high temperatures, however, the morphology of the jammed cluster changes to a spatially and structurally disordered phase with well-defined domain boundaries which appears less congested compared to the low-temperature clusters. Both these less and strongly congested jammed clusters are examples of MIPS where the difference in the internal domain structure arises due to the effect of the temperature. MIPS with a phase-separated cluster of ordered domains having finite characteristic length scales  (similar to the snapshot corresponding to $T=1.1$ and ${\rm MPS}=18$) has also been observed for ABPs and active Janus colloids~\cite{digregorio2018full,van2019interrupted,caporusso2020motility}. For ${\rm MPS}>1$ (and also for soft-core repulsion, see Sec.~\ref{SC}), we have local alignment interaction between particles and although MIPS is traditionally defined as a phase separation of repulsively interacting particles without any alignment interaction, a phase separation where particle velocity reduces with increasing local density can also be attributed to MIPS~\cite{geyer2019freezing}. It has also been shown for ABPs with alignment interaction that alignment promotes MIPS {\cite{sese2018velocity,sese2021phase}}. 

In Fig.~\ref{fig5}(c), as the restriction is relaxed, enhanced flipping at high temperature dissolves the congestion observed at low temperature and the system makes a transition from the jammed phase to the coexistence phase exhibiting lane. During this process, the system also exhibits a band-to-lane reorientation transition as a function of the particle self-propulsion velocity, a novel feature of the flocking phenomenon in the unrestricted APM \cite{chatterjee2020flocking,mangeat2020flocking}, where we observe a transverse band [blue band in Fig.~\ref{fig5}(a)] at small particle velocity whereas a longitudinally moving lane [green band in Fig.~\ref{fig5}(c)] at large velocity.\\

Fig.~\ref{fig5}(b) and Fig.~\ref{fig5}(d) shows the $T-{\rm MPS}$ phase diagrams for $\epsilon=0.9$ and $\epsilon=2.7$, respectively. In Fig.~\ref{fig5}(b), the combination of temperature and MPS determines four phases of the rAPM. At large MPS, which facilitates particle hopping to neighboring sites, the rAPM behaves like the unrestricted APM \cite{chatterjee2020flocking,mangeat2020flocking}, and we observe the three phases of the unrestricted APM in the phase diagram: a gaseous phase at high temperatures, an ordered liquid phase at low temperatures, and a coexistence region at intermediate temperatures, where the motion of the ordered liquid band is transverse. At low MPS, the hard-core repulsion prevents collective motion, resulting in a jammed MIPS state for a large range of temperatures. As temperature increases, the jammed region shrinks and at high enough temperatures, the system always remains gaseous. These jammed clusters occur either in the coexistence region or in the liquid region and emerge due to MIPS.

Fig.~\ref{fig5}(d) shows $T-{\rm MPS}$ phase diagram for large particle velocity. The coexistence region is now characterized by lanes \cite{chatterjee2020flocking}, and for small MPS, the jammed phase occurs even at very high temperatures. As portrayed in both the phase diagrams, fluctuations play a crucial role in the transition of the jammed phase to the coexistence or liquid phase by enhancing the probability of flipping. For $\epsilon=0.9$, the hopping rate to non-preferred directions is substantial compared to $\epsilon=2.7$. Thus, moderate self-propulsion and thermal fluctuations help break the jammed configuration more efficiently. Another difference is the MPS range. Particles hop quickly at large $\epsilon$. At higher MPS values, tending toward unrestricted APM, the liquid band of the coexistence region becomes narrower and more populated with increasing $\epsilon$. Unrestricted APM's $\epsilon-\rho_0$ phase diagram shows this \cite{chatterjee2020flocking}. As the liquid binodal value increases with $\epsilon$ (at a fixed $T$), so does the cutoff MPS.

In this chapter, our main goal is to investigate the effect of various repulsive interactions on particle hopping and the consequent flocking dynamics. To do so, we will now compare the temperature-density and velocity-density phase diagrams of the unrestricted APM \cite{chatterjee2020flocking,mangeat2020flocking} with the phase diagrams obtained with the current model.\\
\begin{figure*}[t]
\centering
\includegraphics[width=\textwidth]{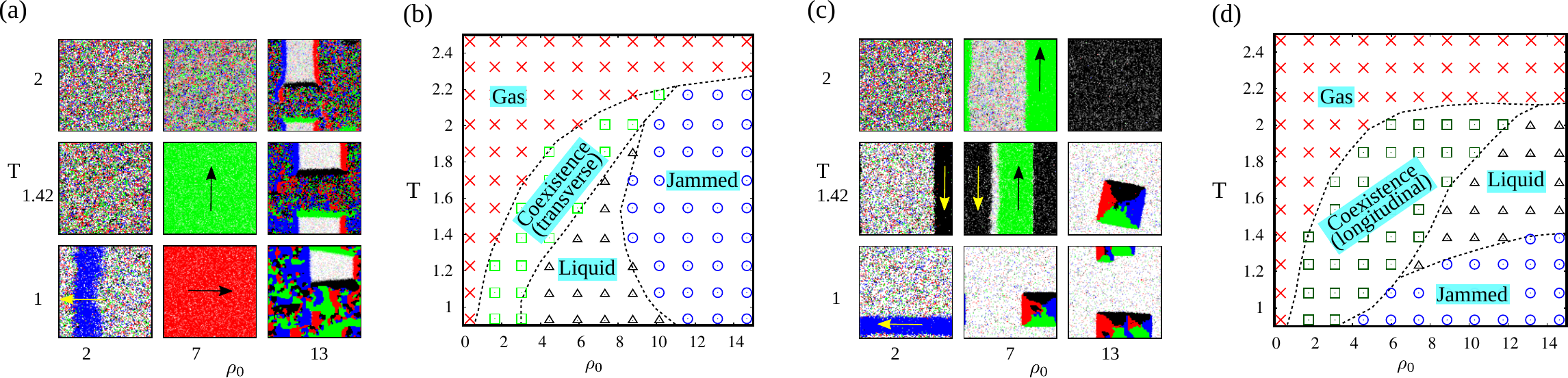}
\caption{(Color online) (a \& c) Steady-state snapshots for MPS$>$1 in the $T-\rho_0$ plane for (a) MPS $=15$ and $\epsilon=0.9$ and (c) MPS $=60$ and $\epsilon=2.4$. (b \& d) $T-\rho_0$ phase diagrams of the rAPM with hard-core restriction for (b) $\epsilon=0.9$ and (d) $\epsilon=2.4$. In (b), the liquid phase appears for intermediate densities but makes a transition to the jammed MIPS state at higher densities whereas in (d), at large activity, jamming appears early and the liquid phase is only observed when the temperature is moderately high.}
\label{fig6} 
\end{figure*}

Fig.~\ref{fig6}(a) and Fig.~\ref{fig6}(c) show steady-state snapshots 
as a function of average particle density and temperature with hard-core repulsion for $\epsilon=0.9$ and $\epsilon=2.4$, respectively. In Fig.~\ref{fig6}(a), at smaller densities, as the temperature is increased, the system shows a transition from the coexistence region and polar ordered phase to the disordered gas phase. No jamming occurs as the allowed MPS merely restricts particle hopping at these densities. At a larger density, however, jamming happens for all values of temperature as a larger density requires higher MPS to avoid jamming. The jamming observed for large densities shows a MIPS state of ordered domains at low temperatures and a MIPS state of disordered internal domains with well-defined ordered domain boundaries at high temperatures. 

Fig.~\ref{fig6}(c) shows the steady-state snapshots with faster moving particles compared to Fig.~\ref{fig6}(a) which was for slow particles. Notice the difference in the domain morphology of the MIPS state in Fig.~\ref{fig6}(a) and Fig.~\ref{fig6}(c) for low temperatures and high density. For $\epsilon=0.9$, hopping probability along the non-biased directions is substantial compared to $\epsilon=2.4$ and therefore the high-density jammed area is larger and less congested for $\epsilon=0.9$. At large particle velocity ($\epsilon=2.4$), the transition to more crowded jamming is enhanced due to large MPS and high velocity where more particles can now gather at a site and also the activity along the non-preferred directions decreases significantly. As a consequence, the jammed cluster of four orientationally ordered sub-domains in a gridlocked position now occupies less area but the congestion is extremely strong.

The corresponding $T-\rho_0$ phase diagrams are shown in Fig.~\ref{fig6}(b) and Fig.~\ref{fig6}(d). Fig.~\ref{fig6}(b) shows that at low temperature and activity, the system mimics the unrestricted APM behavior with a transition from a gaseous to an ordered liquid phase via the liquid-gas coexistence region as density is increased. At higher densities, a phase transition to the jammed phase happens from the coexistence region and the ordered liquid phase due to the hopping restriction through MPS.

With high particle velocity, however, low temperature facilitates jamming even at intermediate densities [Fig.~\ref{fig6}(d)]. Increasing the temperature helps the system to break the congestion and free-flowing phases such as lane and ordered liquid phases emerge. A large $\epsilon$ allows particles to self-propel more along the preferred direction, which is unaltered at low temperatures, and particles of different states meeting at a point stay stuck for a long time, causing a jammed MIPS state at low temperature and high density. A temperature increase partially dissolves this situation by enhancing flipping as switching the state changes the preferred direction of propulsion. Both phase diagrams show a high-temperature gaseous phase unaffected by the control parameters beyond a critical temperature $T_c \sim 2.3$.
\begin{figure}
\centering
\includegraphics[width=\columnwidth]{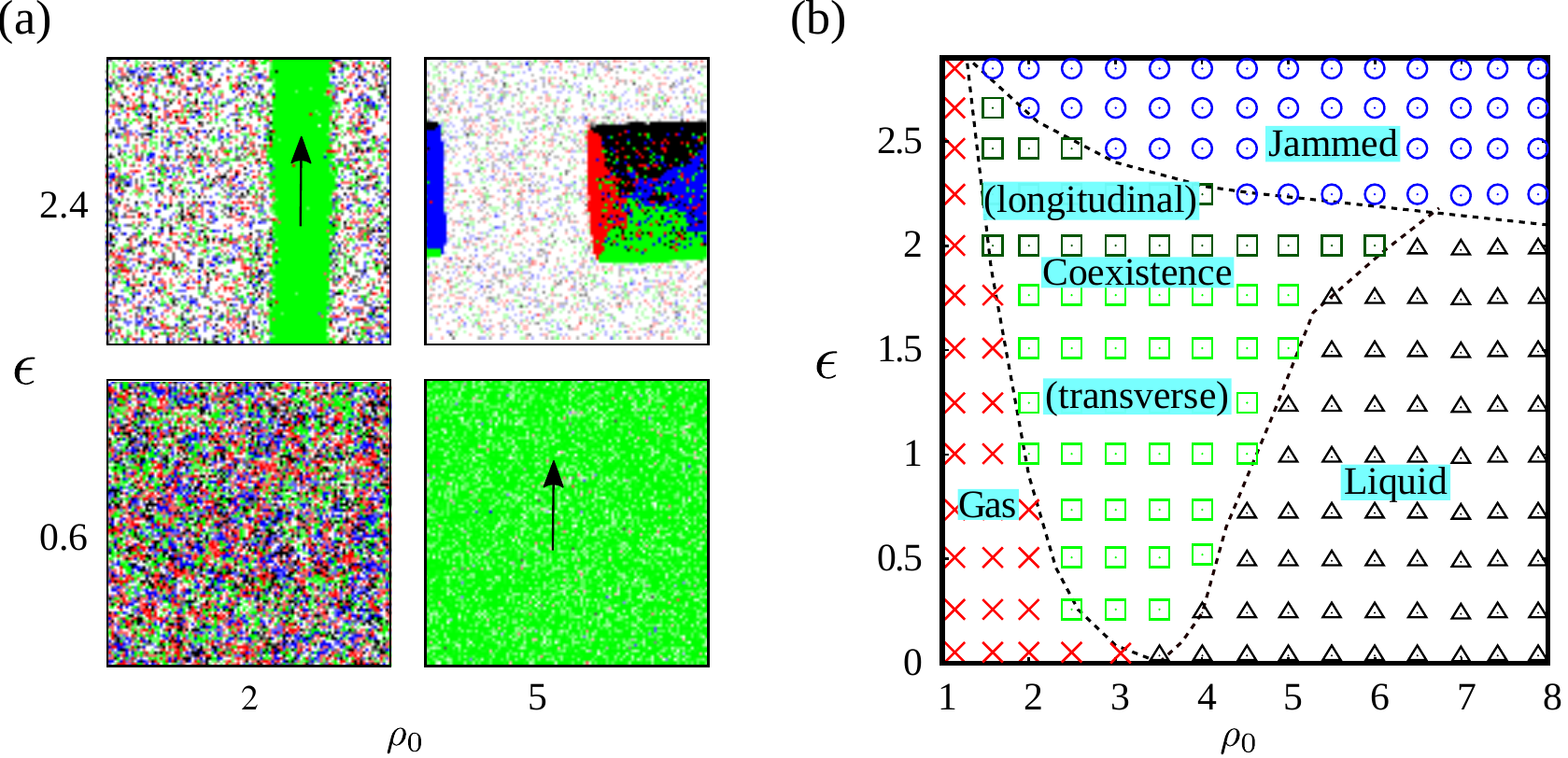}
\caption{(Color online) (a) Steady-state snapshots and (b) $\epsilon-\rho_0$ phase diagram of the rAPM with MPS=20 for $T=1.42$ ($\beta=0.7$). (a) An increase in $\rho_0$ and $\epsilon$ drives the system towards a jammed state. (b) The phase diagram resembles the unrestricted APM~\cite{chatterjee2020flocking,mangeat2020flocking} for $\epsilon<2$ but breaks down at larger $\rho_0$ and $\epsilon$. The reorientation transition (transverse to longitudinal band motion) happens at $\epsilon \sim 2.1$ and a liquid-gas phase transition takes place at $\rho_0 \sim 3.5$ for $\epsilon=0$ \cite{chatterjee2020flocking,mangeat2020flocking}.}
\label{fig7} 
\end{figure}

Next, we will discuss the $\epsilon-\rho_0$ diagram of the rAPM by changing the strength of the self-propulsion $\epsilon$ while keeping the temperature fixed. The resultant steady-state snapshots and the corresponding phase diagram are shown in Fig.~\ref{fig7}(a) and Fig.~\ref{fig7}(b), respectively. The snapshots show that the system exhibits a jam of orientationally ordered sub-domains due to MIPS at high density and motility. High particle motility promotes particle accumulation at a lattice site and since these particles have a very small probability to hop toward the non-biased directions, a higher MPS is required to avoid jamming. The snapshots also exhibit the three typical phases of the unrestricted APM at small density and velocity.

The $\epsilon-\rho_0$ phase diagram in Fig.~\ref{fig7}(b) shows four regions similar to the $T-\rho_0$ phase diagrams shown in Fig.~\ref{fig6}(b) and Fig.~\ref{fig6}(d). Excluding the high-velocity limit, the phase diagram resembles the unrestricted APM diagram~\cite{chatterjee2020flocking,mangeat2020flocking} where the binodals $\rho_{\rm gas}$ and $\rho_{\rm liq}$ delimit the coexistence region from the gas and liquid phases. The reorientation transition, which is a novel feature of the APM and where the system exhibits a transverse band motion at small $\epsilon$ and a longitudinal lane motion at large $\epsilon$, is also observed. The conventional phase diagram however breaks down at large $\epsilon$ where a transition to the jammed MIPS state from the coexistence region and the liquid phase is realized. At $\epsilon=0$, similar to the unrestricted APM \cite{chatterjee2020flocking,mangeat2020flocking}, the system exhibits a direct liquid-gas phase transition around $\rho_0 \sim 3.5$. 
\begin{figure}
\centering
\includegraphics[width=\columnwidth]{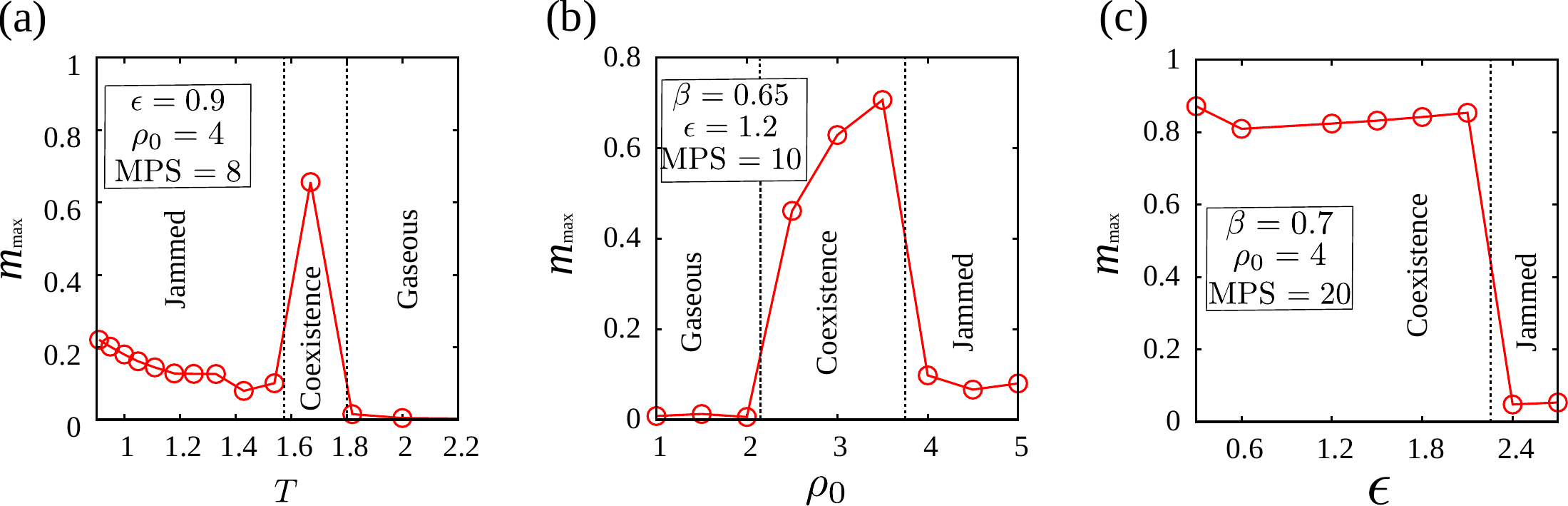}
\caption{(Color online) Normalized maximal magnetization 
$m_{\rm max}$ for the rAPM with MPS$>$1 as a function of $T$, $\rho_0$, and $\epsilon$ in (a), (b), and (c), respectively. $m_{\rm max}$ of the coexistence region is larger than the jammed phase. For the disordered gas phase, $m_{\rm max}=0$ and $m_{\rm max} \simeq 1$ for the liquid phase.}
\label{fig8} 
\end{figure}
\begin{figure*}[t]
\centering
\includegraphics[width=\textwidth]{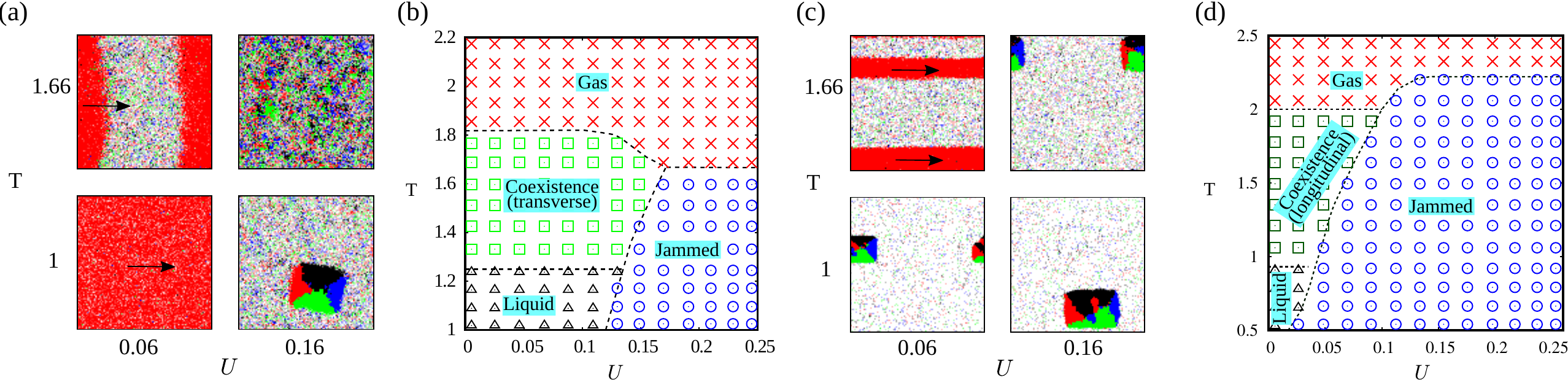}
\caption{(Color online) (a \& c) Steady-state snapshots for the rAPM with soft-core repulsion in the $T-U$ plane for (a) $\epsilon=0.9$ and (c) $\epsilon=2.7$ with fixed $\rho_0=4$. (b \& d) $T-U$ phase diagrams for the parameters of (a) and (c), respectively. Jamming transition happens at low $T$ and large $U$ with the fact that high propulsion facilitates jamming.}
\label{fig9} 
\end{figure*}

The systematic variation of magnetization is an indicator of symmetry breaking. Fig.~\ref{fig8} shows the order parameter against different control parameters where we consider the maximal magnetization $m_{\rm max}$ among the four estimates (see Eq.~\ref{Hapm_mag}). It can be seen from the plots that magnetization changes abruptly across the phase boundaries denoted by dashed vertical lines. A fully ordered liquid state is characterized by $m_{\rm max}\simeq 1$ and $m_{\rm max}= 0$ signifies a disordered gaseous phase. Fig.~\ref{fig8}(a) shows $m_{\rm max}$ versus temperature where the low-temperature region is characterized by a non-zero but small $m_{\rm max}$ signifying a MIPS state (locally ordered sub-domains marginally rise the $m_{\rm max}$ to a non-zero value). As the temperature is increased, phase transition from the MIPS state to the coexistence region happens with a sharp jump in the magnetization indicating a first-order phase transition. Further increase of the temperature shows another first-order phase transition from the coexistence to the disordered gas phase. $m_{\rm max}$ against density in Fig.~\ref{fig8}(b) also presents two first-order phase transitions as density is increased, gas to coexistence and coexistence to MIPS, respectively. The MIPS state occurs at a relatively high $\epsilon$ in Fig.~\ref{fig8}(c) because of the large MPS value. Fig.~\ref{fig8}(c) also validates that the transition to the MIPS state is a first-order transition as demonstrated by the discontinuous jump of the order parameter at the transition point. 

As discussed so far, jamming in our model is a kinetically arrested state due to MIPS and MIPS signifies the coexistence of an active low-density gas with a high-density jammed cluster which is reminiscent of equilibrium liquid-gas de-mixing and thus can be seen as a first-order transition. As shown in Fig.~\ref{fig8}, we also find the transition from the ordered/flocking phase to the jammed phase as a discontinuous first-order phase transition. At this point, the system undergoes a sudden jamming transition, leading to the formation of large-scale clusters that are kinetically trapped in a glassy state. It has already been shown in the context of active Brownian particles that MIPS can be described as an equilibrium-like phase transition (not just a dynamically trapped state) and MIPS verifies the characteristic properties of first-order liquid–gas phase transitions ~\cite{levis2017active}. That MIPS-like transitions between polar liquid and amorphous jammed solid is a first-order transition has also been shown in the context of motile colloids both experimentally and theoretically~\cite{geyer2019freezing}. This observation is also consistent outside the field of active matter where experimental data of traffic flow on highways finds the phase transition from free flow to traffic jam as a first-order phase transition \cite{kerner1997experimental}. 

One can further distinguish between a jammed state and a free-flowing state by the mean-square displacement (MSD) of the high-density clusters or by the number fluctuations~\cite {henkes2011active}. 
A strongly suppressed number fluctuation and specific time dependence of the MSD characterize a jammed state. The MSD shows either an oscillatory behavior~\cite{henkes2011active} because of the jammed clusters oscillate around their mean position, or saturate at large times because clusters do not move at all (see the Auxiliary material~\ref{appendix_msd}).

\subsection{Soft-core restriction}
\label{SC}
In this section, we present the results of the rAPM with soft-core repulsion parameterized by the value of $U>0$ in Eq.~\eqref{SC_eq}. Fig.~\ref{fig9}(a) and Fig.~\ref{fig9}(c) depict the late-stage coarsening of the 4-state rAPM in the $T-U$ plane for (a) $\epsilon=0.9$ and (c) $\epsilon=2.7$. For small propulsion, the system exhibits jammed clusters 
composed of four ordered sub-domains at low temperatures and large restrictions similar to the hard-core repulsion. An increase in the temperature (and low $U$) helps to dissolve the jam and we observe the three phases of the unrestricted APM \cite{chatterjee2020flocking,mangeat2020flocking}. Fig.~\ref{fig9}(c) is analogous to Fig.~\ref{fig9}(a) but for a larger velocity. Due to the high propulsion and therefore suppressed particle motion along the non-preferred directions, the liquid phase (observed with $\epsilon=0.9$) becomes a jammed MIPS state. A rise in the thermal fluctuation helps to break the clog for small $U$ but due to high motility, jam persists even when the temperature is high for strong repulsion. As explained before, this jammed phase is a kinetically arrested jammed phase due to MIPS where we observe a reduction of the active particle current as density becomes sufficiently high. The structural transformation of such a kinetically jammed phase with temperature has been demonstrated in the Auxiliary material~\ref{appendix_jam_structure}.


\begin{figure*}[t]
\centering
\includegraphics[width=\textwidth]{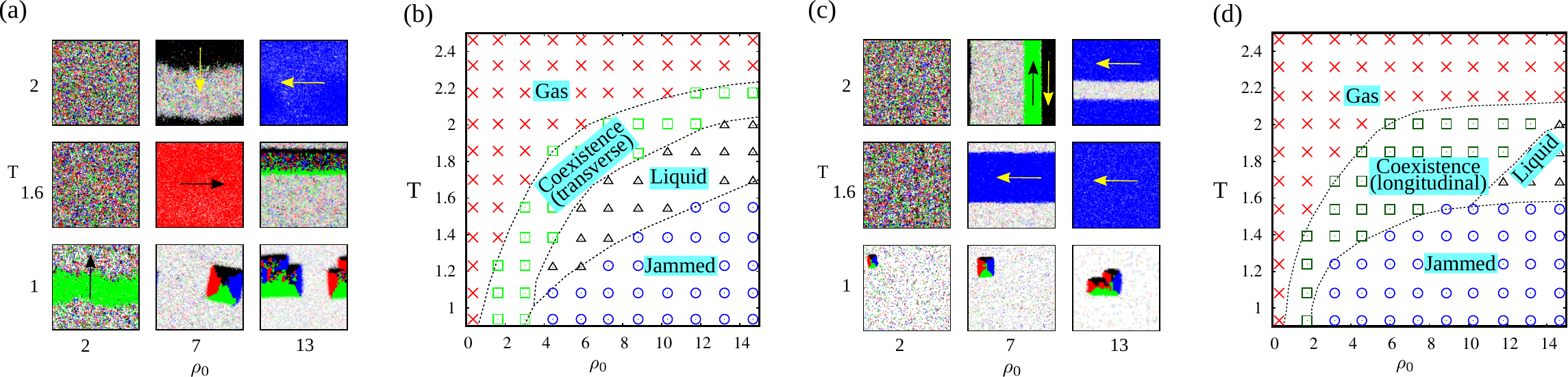}
\caption{(Color online) (a \& c) Snapshots of the rAPM with soft-core repulsion
illustrating the different self-organized patterns as a function of $T$ and $\rho_0$ for (a) small $\epsilon=0.9$ ($U=0.07$) and (c) large $\epsilon=2.7$ ($U=0.02$). The snapshots signify that low temperature, high density and large activity facilitate jamming. (b \& d) $T-\rho_0$ phase diagrams of the rAPM with soft-core repulsion for parameters of (a) and (c), respectively. Apart from the usual gas and liquid binodals observed in the unrestricted APM, we observe a jammed phase adjoining both coexistence and liquid phases at large densities and low temperatures.}
\label{fig10}
\end{figure*}

We find, e.g., square/rectangular shaped kinetically jammed clusters with our model for both hard-core restriction and soft-core repulsion [see Fig.~\ref{fig5}(c) (MPS = 18, $T=1.1$) and Fig.~\ref{fig9}(a) ($U=0.16$, $T=1$)]. The square-shaped structure is due to the 4-state version of our model where particles can only hop to the four nearest neighbors of a square lattice and it is through hopping, the neighbors are connected. Therefore, jammed clusters
have a square/rectangular geometry and contain four locally ordered sub-domains for $q = 4$ and a hexagonal cluster with six locally ordered sub-domains for $q = 6$ (simulated on a triangular lattice, data are not shown here). We have also investigated the AIM ($q = 2$) with soft-core repulsion and observed a vertically jammed MIPS state similar to active lattice gas \cite{kourbane2018exact}. The effect of soft-core repulsion has also been investigated on off-lattice flocking models such as ACM and VM which manifest MIPS (see Auxiliary material~\ref{racm_vm}). 


The corresponding $T-U$ phase diagrams are shown in Fig.~\ref{fig9}(b) and Fig.~\ref{fig9}(d). Fig.~\ref{fig9}(b) shows a liquid-to-gas transition via the coexistence region for small values of $U$ as the temperature is raised. 
An increase in hopping restriction via $U$, however, changes this scenario and we observe a transition to the jammed phase at low and intermediate temperatures. At very high temperatures, the system remains in the gaseous phase. In Fig.~\ref{fig9}(d), due to the high velocity of the particles, jamming dominates the phase diagram and at low temperatures, the system exhibits jamming even at small $U$.

Now we will focus on the $T-\rho_0$ and $\epsilon-\rho_0$ phase diagrams of the rAPM with soft-core repulsion and will compare them with the similar diagrams obtained for hard-core rAPM and unrestricted APM.

The soft-core $T-\rho_0$ phase diagrams for low and high particle velocities along with the steady-state snapshots are shown in Fig.~\ref{fig10}. The snapshots in Fig.~\ref{fig10}(a) and Fig.~\ref{fig10}(c) tell a story of jamming transition that is consistent with the findings of hard-core rAPM where we observe jamming for low temperatures and high densities. Notice the difference in the internal structure of the jammed clusters in Fig.~\ref{fig10}(a) and Fig.~\ref{fig10}(c) where for a small activity, we observe a MIPS state of orientationally disordered active particles having ordered domain boundaries whereas the MIPS state exhibits a gridlock of ordered sub-domains at large activity. The system behaves as the unrestricted APM at high temperatures.
\begin{figure}[t]
\centering
\includegraphics[width=\columnwidth]{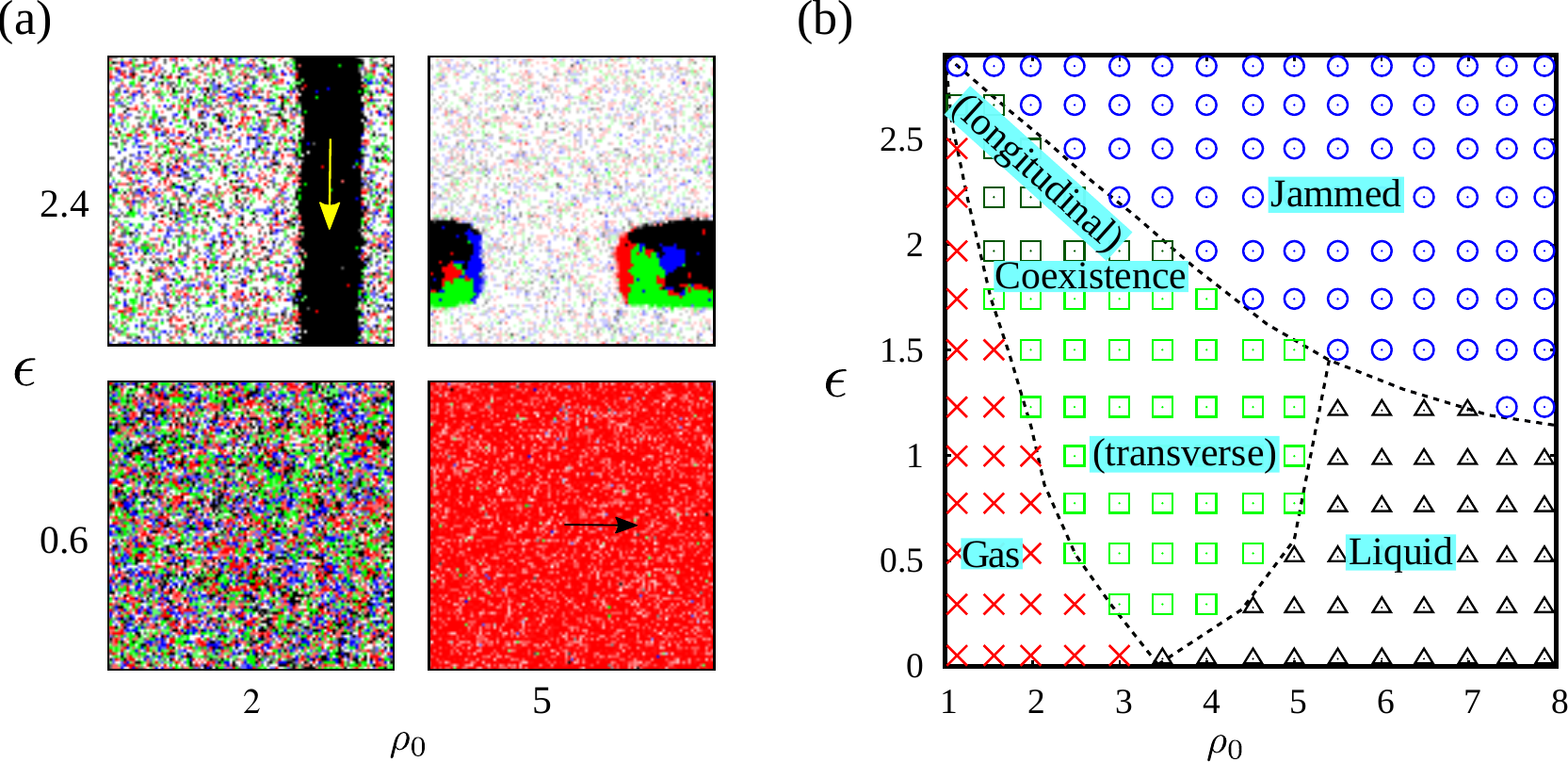}
\caption{(Color online) (a) Steady-state snapshots of the rAPM with 
soft-core repulsion in the $\epsilon-\rho_0$ plane for $T=1.42$ ($\beta=0.7$) and $U=0.07$. (b) The soft-core $\epsilon-\rho_0$ phase diagram qualitatively resembles the hard-core phase diagram of Fig.~\ref{fig7}(b) showing that the jamming threshold decreases with density and activity.}
\label{fig11} 
\end{figure}
\begin{figure}[t]
\centering
\includegraphics[width=\columnwidth]{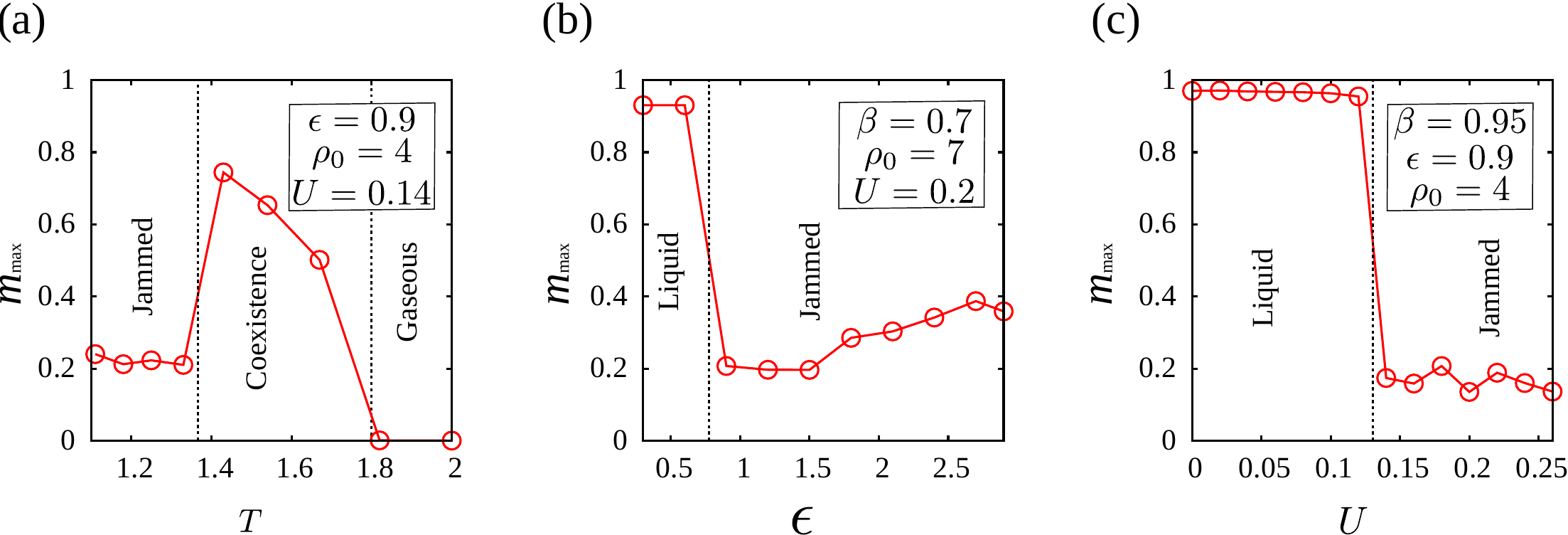}
\caption{(Color online) (a--c) 
Normalized maximal magnetization $m_{\rm max}$ of the rAPM with soft-core repulsion as a function of $T$, $\epsilon$, and $U$. A small nonzero $m_{\rm max}$ is indicative of a jammed phase and a comparatively high $m_{\rm max}<1$ characterizes a coexistence region.}
\label{fig12} 
\end{figure}
These observations are presented in the form of phase diagrams in Fig.~\ref{fig10}(b) and Fig.~\ref{fig10}(d) where we notice that the system replicates unrestricted APM behavior \cite{chatterjee2020flocking,mangeat2020flocking} for intermediate and large temperatures. Temperature reduction, on the other hand, leads to an increase in the jammed region with increasing densities. 
For sufficiently high temperatures, the system is always in a gaseous phase that is unaffected by the control parameters and we find the critical temperature $T_c \sim 2.3$ is also independent of the nature of restriction [see Fig.~\ref{fig6}(b) and Fig.~\ref{fig6}(d)]. A comparison of the phase diagram in Fig.~\ref{fig10}(d) and the $T-\rho_0$ phase diagram of the unrestricted APM \cite{chatterjee2020flocking} reveals that restriction has shifted the unrestricted phase diagram to the low-density region (similar for the hard-core diagrams too) where the jamming phase emerges as a fourth phase at low temperatures, keeping the $T_c$ same.

In Fig.~\ref{fig11}, we show the phase snapshots and phase diagrams in the $\epsilon-\rho_0$ plane. The qualitative nature of the snapshots shown in Fig.~\ref{fig11}(a) is similar to our findings for the hard-core restriction shown in Fig.~\ref{fig7}(a) where at large velocity, jams emerge when density is increased. At small velocities, the system sequentially shows the gas, coexistence and liquid phases as density is increased.  

The corresponding phase diagram is shown in Fig.~\ref{fig11}(b). The system behaves nearly diffusively for very small $\epsilon$; therefore, jamming does not exist at this limit. However, with high activity, due to the restriction on propulsion, a jammed phase occurs even at small and intermediate densities where the cutoff $\epsilon$ value to jamming decreases with $\rho_0$. This physically signifies that for small densities, high activity is needed to create a jam whereas, for large densities, jamming can happen at a lesser speed. The conventional $\epsilon-\rho_0$ phase diagram of the unrestricted APM \cite{chatterjee2020flocking} breaks down at the large velocity-density limit in Fig.~\ref{fig11}(b) whereas in the $\epsilon \leqslant 1$ limit, the two diagrams are analogous. At the zero activity limit ($\epsilon=0$), the system behaves like the unrestricted APM at small repulsion where we see a direct liquid-gas transition around $\rho_0 \sim 3.5$ whereas, at strong repulsion, orientationally disordered self-segregated domains are observed instead of the ordered liquid phase (see Auxiliary material~\ref{appendix_diffusive} for details).


Fig.~\ref{fig12} is analogous to Fig.~\ref{fig8} but for soft-core repulsion. Similar to Fig.~\ref{fig8}(a), $m_{\rm max}$ versus temperature in Fig.~\ref{fig12}(a) shows two first-order transitions (jammed to coexistence and coexistence to gas) as the temperature is increased. $m_{\rm max}$ versus $\epsilon$ in Fig.~\ref{fig12}(b) shows a jamming transition from an ordered liquid phase. At small velocities, due to slow propulsion and large density, the system exhibits a fully ordered liquid phase but at large velocities, particles quickly accumulate locally, and due to large restriction ($U$), particle movements are heavily restricted and therefore we observe a jammed state. Fig.~\ref{fig12}(c) shows the variation of $m_{\rm max}$ with $U$ where small $U$ facilitates particle hopping which together with slow particles and a relatively large density gives rise to an ordered liquid phase. As restriction is enhanced, the system makes an unsurprised transition to the jammed phase. Similar to Fig.~\ref{fig8}, the jamming transition with the soft-core repulsion is also first-order signifying that the specific origin of the restriction imposed on the particle movement does not alter the nature of the phase transition.


\section{Hydrodynamic theory}
\label{secHydro}

In this section, we formulate the hydrodynamic continuum theory for the microscopic rAPM. From the microscopic hopping and flipping rates, we derive the equation for the probability density function $\rho_\sigma({\bf x};t) \sim \langle n_i^\sigma(t) \rangle$ for a particle to be at the position ${\bf x}$, corresponding to the site $i$, and in the state $\sigma$ at the time $t$. We only keep the first-order terms in the $|{\bf m}_i| \ll \rho_i$ expansion of the flipping rate~\eqref{flipeq}. To represent the different hopping restrictions, we introduce a function $f(\rho)$ where $\rho$ is the total density at the arrival position. The form of this function is $f(\rho) = 1 - \zeta \rho$ for $\zeta = 1/{\rm MPS}$, and $f(\rho) = \exp(-s\rho)$ for the soft-core rAPM where $s=2\beta U$. In Appendix~\ref{hydro}, we derive the hydrodynamic equations:
\begin{equation}
\partial_t \rho_\sigma = - \partial_\parallel J_{\sigma \parallel} - \partial_\perp J_{\sigma \perp} + \sum_{\sigma' \ne \sigma} K_{\sigma \sigma'} (\rho_\sigma - \rho_{\sigma'}) \label{eqhydro},
\end{equation}
where the current is
\begin{gather}
J_{\sigma \parallel} = -D_\parallel [f(\rho)\partial_\parallel \rho_\sigma -f'(\rho) \rho_\sigma \partial_\parallel \rho] + vf(\rho) \rho_\sigma,\\
J_{\sigma \perp} = -D_\perp [f(\rho)\partial_\perp \rho_\sigma -f'(\rho) \rho_\sigma \partial_\perp \rho],
\end{gather}
with $D_\parallel = D(1+\epsilon/3)$, $D_\perp=D(1-\epsilon/3)$, and $v=4D\epsilon/3$, and the flipping interaction term $K_{\sigma \sigma'} = - \gamma$ for ${\rm MPS}=1$ and
\begin{equation}
K_{\sigma \sigma'} = \frac{4\beta J}{\rho}(\rho_\sigma + \rho_{\sigma'}) - 1 - \frac{r}{\rho} - \alpha \frac{(\rho_\sigma - \rho_{\sigma'})^2}{\rho^2}, \label{Iflip}
\end{equation}
with $\alpha = 8(\beta J)^2(1-2\beta J/3)$, for ${\rm MPS}>1$ and soft-core rAPM. The different expressions of $f(\rho)$ and $K_{\sigma \sigma'}$ lead to three different hydrodynamic equations for the three different studied models, discussed in the next subsections. Without any restriction for the particle density, i.e. setting $f(\rho) =1$ in Eq.~(\ref{eqhydro}), the unrestricted APM hydrodynamic equations derived in Ref.~\cite{chatterjee2020flocking} are recovered. Without any loss of generality, we set $D=1$, $r=1$ and $J=1$ defining the scales of time, density, and temperature.

We solve Eq.~(\ref{eqhydro}) numerically using FreeFEM++~\cite{hecht2012new}, a software package based on the finite element method~\cite{zienkiewicz1977finite}. The equations are integrated over discrete time $t_n = n \Delta t$, at which the density is denoted as $\rho_\sigma^{(n)}({\bf x})$. The initial density $\rho^{(0)}({\bf x})$ is taken as a high-density bubble or stripe on a low-density background. The final time is denoted as $t_{\rm max}$. The weak formulation of Eq.~(\ref{eqhydro}) is the integral equation:
\begin{gather}
\int_\Omega d{\bf x} \  \sum_\sigma  \left[ w_\sigma \rho_\sigma^{(n+1)} - \Delta t \left(\partial_\parallel w_\sigma \partial_\parallel J_{\sigma \parallel} + \partial_\perp w_\sigma \partial_\perp J_{\sigma \perp} \right) \right] \nonumber\\
- \Delta t \sum_{\sigma' > \sigma } (w_\sigma - w_{\sigma'}) K_{\sigma \sigma'} \left[\rho_\sigma^{(n+1)}-\rho_{\sigma'}^{(n+1)} \right]\nonumber\\
 = \int_\Omega d{\bf x} \ \sum_\sigma w_\sigma \rho_\sigma^{(n)},
\end{gather}
where $\rho_\sigma^{(n)}({\bf x})$ is the known particle density at time $t_n$, $\rho_\sigma^{(n+1)}({\bf x})$ is the unknown particle density at time $t_{n+1}$, and $w_\sigma({\bf x})$ is a test function. $K_{\sigma \sigma'}$ and the restriction terms in $J_{\sigma \parallel}$ and $J_{\sigma \perp}$ are calculated at time $t_n$ to have a linear equation of $\rho_\sigma^{(n+1)}({\bf x})$. This integral equation is solved over a triangular mesh-grid with ${\cal N}$ vertices on the boundaries. The densities are calculated on the nodes of the mesh-grid and interpolated linearly over the space with Lagrange polynomials. The precision of the numerical solution is increased for a narrow grid (${\cal N}\gg 1$) and small time increments ($\Delta t \ll 1$), and the computational time has a complexity proportional to ${\cal N}^2/\Delta t$. It takes about $48$ hours for ${\cal N}=75$ and $t_{\rm max}/\Delta t = 50000$ time steps, on a 4 GHz processor. The FreeFEM++ codes used to compute the numerical solutions are available in Ref.~\cite{karmakar2023zenodo}.

\subsection{MPS = 1 (ALG version of the rAPM)}

\begin{figure}[t]
\centering
\includegraphics[width=\columnwidth]{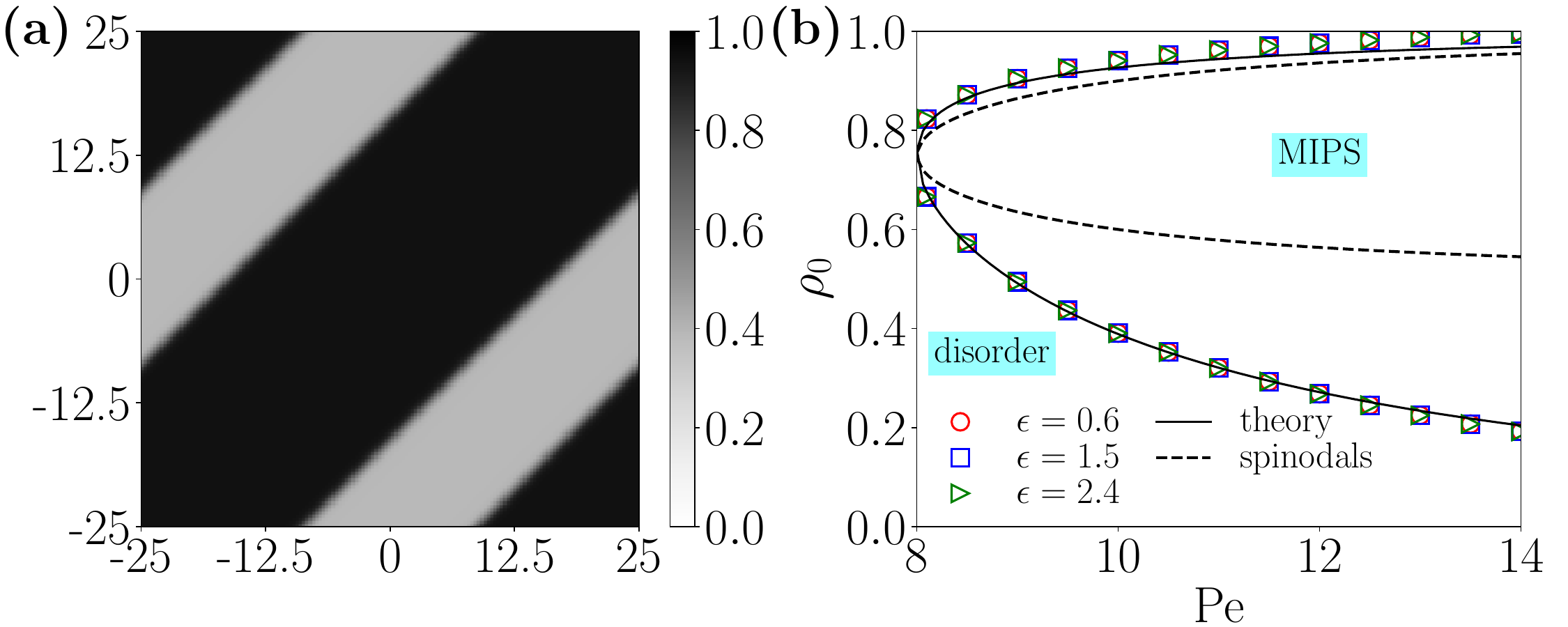}
\caption{(Color online) (a)~MIPS density profile for ${\rm MPS}=1$, $\rho_0=0.75$ and ${\rm Pe}=10$, obtained numerically with FreeFEM++ for ${\cal N}=100$, $\Delta t = 0.01$ and $t_{\rm max}=1000$. The densities of the two disordered phases are $\rho_{\rm low}\simeq 0.391$ and $\rho_{\rm high}\simeq 0.941$. (b)~Velocity-density phase diagram for ${\rm MPS}=1$, computed with the numerical solutions of the Eq.~\eqref{eqmps1}, in a square domain of linear size $L=50$, for three values of $\epsilon$. The dotted line shows the spinodals given by Eq.~\eqref{spinodal}, and the solid line represents the theoretical value of the binodals calculated in Appendix~\ref{ls_mps1}.}
\label{fig_hydro_MPS1}
\end{figure}

For ${\rm MPS} = 1$, with $f(\rho) = 1 - \rho$ and $K_{\sigma \sigma'} = - \gamma$, Eq.~(\ref{eqhydro}) becomes:
\begin{align}
\partial_t \rho_\sigma &= D_\parallel \partial_\parallel[ (1-\rho) \partial_\parallel \rho_\sigma + \rho_\sigma \partial_\parallel \rho ] \nonumber\\
&+ D_\perp \partial_\perp [ (1-\rho) \partial_\perp \rho_\sigma + \rho_\sigma \partial_\perp \rho ] \nonumber\\
&- v \partial_\parallel \left[ (1 - \rho) \rho_\sigma \right] - \gamma(4\rho_\sigma - \rho) \label{eqmps1}.
\end{align}
The only homogeneous solution is $\rho_\sigma = \rho_0/4$, corresponding to a disordered phase. In Appendix~\ref{ls_mps1}, we perform the linear stability analysis of this solution, leading to the spinodal:
\begin{equation}
\label{spinodal}
\varphi_\pm = \frac{3}{4} \pm \frac{1}{4}\sqrt{1 - \frac{64}{{\rm Pe}^2}},
\end{equation}
where ${\rm Pe} = v/\sqrt{D\gamma}$ is the P\'eclet number. Note that the P\'eclet number must be larger than ${\rm Pe}_c = 8$, to observe MIPS. The MIPS state is a high-density diagonal band on a low-density phase, whose densities are denoted by $\rho_{\rm high}$ and $\rho_{\rm low}$, respectively. Fig.~\ref{fig_hydro_MPS1}(a) shows the numerically obtained phase-separated density profile for $\rho_0=0.75$ and ${\rm Pe} = 10$. The right boundary of the diagonal high-density band is mainly populated by state $\sigma=2$ (top) and $\sigma=3$ (left) and the left boundary by state $\sigma=1$ (right) and $\sigma=4$ (down). This leads to the relation $\rho_1+\rho_3 = \rho_2 + \rho_4$, irrespective of the direction in which the diagonal band is formed.

In Appendix~\ref{ls_mps1}, we also derive two relations linking implicitly $\rho_{\rm low}$ and $\rho_{\rm high}$, and demonstrate that the binodals are independent of $\epsilon$. The demonstration is similar to the one made in Ref.~\cite{kourbane2018exact}, for an active lattice gas with a slightly different hydrodynamic equation. From these two relations and for ${\rm Pe} = 10$, we get $\rho_{\rm low}^{\rm th}\simeq 0.389$ and $\rho_{\rm high}^{\rm th}\simeq 0.927$, comparable to the densities numerically obtained in Fig.~\ref{fig_hydro_MPS1}(a). At the large P\'eclet limit, we derive the asymptotic behavior: $\rho_{\rm low} \simeq -(8/{\rm Pe}^2) \ln (16/3{\rm Pe}^2)$ and $\rho_{\rm high} \simeq 1 - 16/3{\rm Pe}^2$. Fig.~\ref{fig_hydro_MPS1}(b) shows the velocity-density phase diagram computed with the phase-separated density profiles, validating the independence of the binodal densities $\rho_{\rm low}$ and $\rho_{\rm high}$ on $\epsilon$. The spinodal and binodal lines obtained analytically are also represented.

\subsection{Hard-core restriction (${\rm MPS} > 1$)}
\label{hydrohc}

\begin{figure}[t]
\centering
\includegraphics[width=\columnwidth]{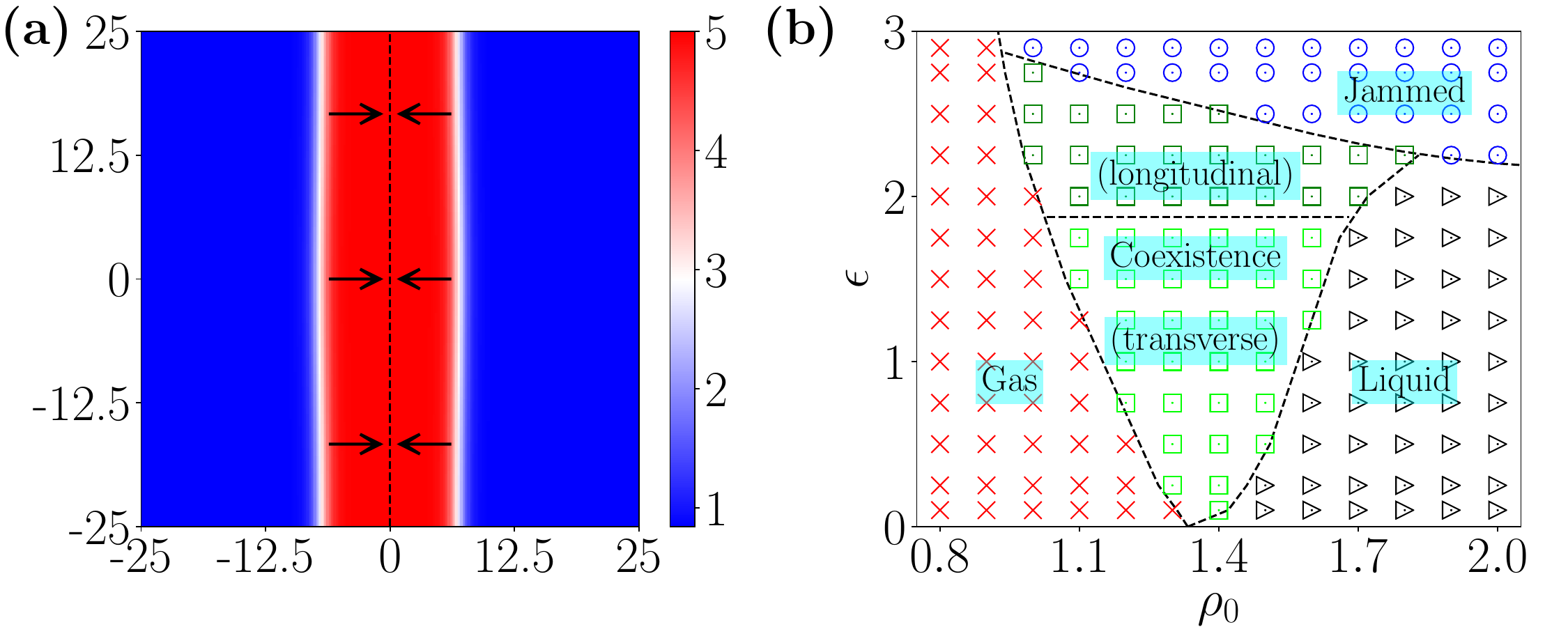}
\caption{(Color online) (a)~Jammed density profile for the hard-core rAPM
with ${\rm MPS}=5$, for $\beta=0.75$, $\rho_0=2$, $\epsilon=2.7$, obtained numerically with FreeFEM++ for ${\cal N}=75$, $\Delta t = 0.02$ and $t_{\rm max}=1000$. (b)~Velocity-density phase diagram for the hard-core rAPM, computed with the numerical solutions of the Eq.~\eqref{eqmps}, for $\beta=0.75$ and ${\rm MPS}=6$ in a square domain of linear size $L=50$.}
\label{fig_hydro_MPS}
\end{figure}

For ${\rm MPS} > 1$, with $f(\rho) = 1 - \zeta\rho$, Eq.~(\ref{eqhydro}) becomes:
\begin{align}
\partial_t \rho_\sigma &= D_\parallel \partial_\parallel \left[ (1-\zeta\rho) \partial_\parallel \rho_\sigma + \zeta \rho_\sigma \partial_\parallel \rho \right] \nonumber \\
&+ D_\perp \partial_\perp \left[(1-\zeta\rho) \partial_\perp \rho_\sigma + \zeta \rho_\sigma \partial_\perp \rho  \right]\nonumber\\
&- v \partial_\parallel \left[ (1-\zeta\rho) \rho_\sigma \right] + \sum_{\sigma'\ne\sigma} K_{\sigma \sigma'} (\rho_\sigma - \rho_{\sigma'}),\label{eqmps}
\end{align}
with $K_{\sigma \sigma'}$ defined in Eq.~(\ref{Iflip}) and $\zeta=1/{\rm MPS}$. The homogeneous solutions, given by the solutions of $K_{\sigma \sigma'} (\rho_\sigma - \rho_{\sigma'}) = 0$, are those of the unrestricted APM derived in Refs.~\cite{chatterjee2020flocking,mangeat2020flocking}. The disordered homogeneous solution $\rho_\sigma = \rho_0 / 4$ corresponds to a gas phase. The ordered homogeneous solution is given by the relation $K_{\sigma \sigma'} = 0$, corresponding to a liquid phase moving in a given direction. For a polar liquid along the state $\sigma = 1$, the densities are $\rho_1 = \rho_0 (1+3M) / 4$ and $\rho_{2,3,4} = \rho_0 (1-M) / 4$ with the magnetization $M$:
\begin{equation}
M = \frac{\beta J}{\alpha} \left[ 1 \pm \sqrt{1 + \frac{\alpha \mu_0}{\beta J}} \right],
\end{equation}
with $\mu_0 = 2 \beta J -1 -r/\rho_0$. This ordered homogeneous solution exists only when $\alpha \mu_0 + (\beta J)^2>0$ i.e. for density larger than
\begin{equation}
\rho_* = \frac{8(1-2\beta J/3) r}{1 + 8(2\beta J - 1)(1-2\beta J/3)}.
\end{equation}
Additionally, the temperature must be below $T_c = (1 - \sqrt{22}/8)^{-1} \simeq 2.417$. For $\epsilon=0$, the transition between the gas phase ($M=0$) and the liquid phase ($M>0$) is discontinuous at density $\rho_*$. For $\epsilon>0$, a first-order liquid-gas phase transition occurs, with a phase separation made by a liquid stripe on a gas background. Two kinds of phase-separated profiles can be observed: a transverse or longitudinal band motion for which the liquid phase is mainly populated by one state, and a jammed state formed by two locked liquid bands mainly populated by oppositely moving states ($\sigma=1,3$ or $\sigma=2,4$). Fig.~\ref{fig_hydro_MPS}(a) shows a numerically obtained jammed density profile, for $\beta=0.75$, $\rho_0=2$, $\epsilon=2.7$ and ${\rm MPS}=5$. For $x<0$, the liquid band is mainly populated by the right state $\sigma=1$, while for $x>0$ it is mainly populated by the left state $\sigma=3$, explaining why the band is jammed. 

Fig.~\ref{fig_hydro_MPS}(b) shows the velocity-density phase diagram determined by the numerical solutions of the Eq.~\eqref{eqmps} for ${\rm MPS}=6$ and $\beta=0.75$. The topology of this phase diagram agrees well with what we obtained for the microscopic model shown in Fig.~\ref{fig7}(b), with jammed state at large velocities and densities, and flocking band motion at low velocities and densities (transverse band motion for $\epsilon<1.9$ and longitudinal band motion for $\epsilon>1.9$). Note that the difference in the relevant density $\rho_0$ region is due to the different  MPS values used (MPS=20 in Fig.~\ref{fig7}(b), MPS=6 here).

In Appendix~\ref{ls_hc}, we perform the linear stability analysis of the disordered and ordered homogeneous solutions of the Eq.~\eqref{eqmps}. These eigenvalues allow the determination of the velocity for which the reorientation transition occurs, denoted $\epsilon_*$, as a function of ${\rm MPS}$ and $T<T_c$. $\epsilon_*$ is a decreasing function of ${\rm MPS}$, meaning that $\epsilon_*$ increases with the restriction. However, the existence of the jammed state cannot be derived from the linear stability analysis of a homogeneous solution.

\subsection{Soft-core restriction}

\begin{figure}[t]
\centering
\includegraphics[width=\columnwidth]{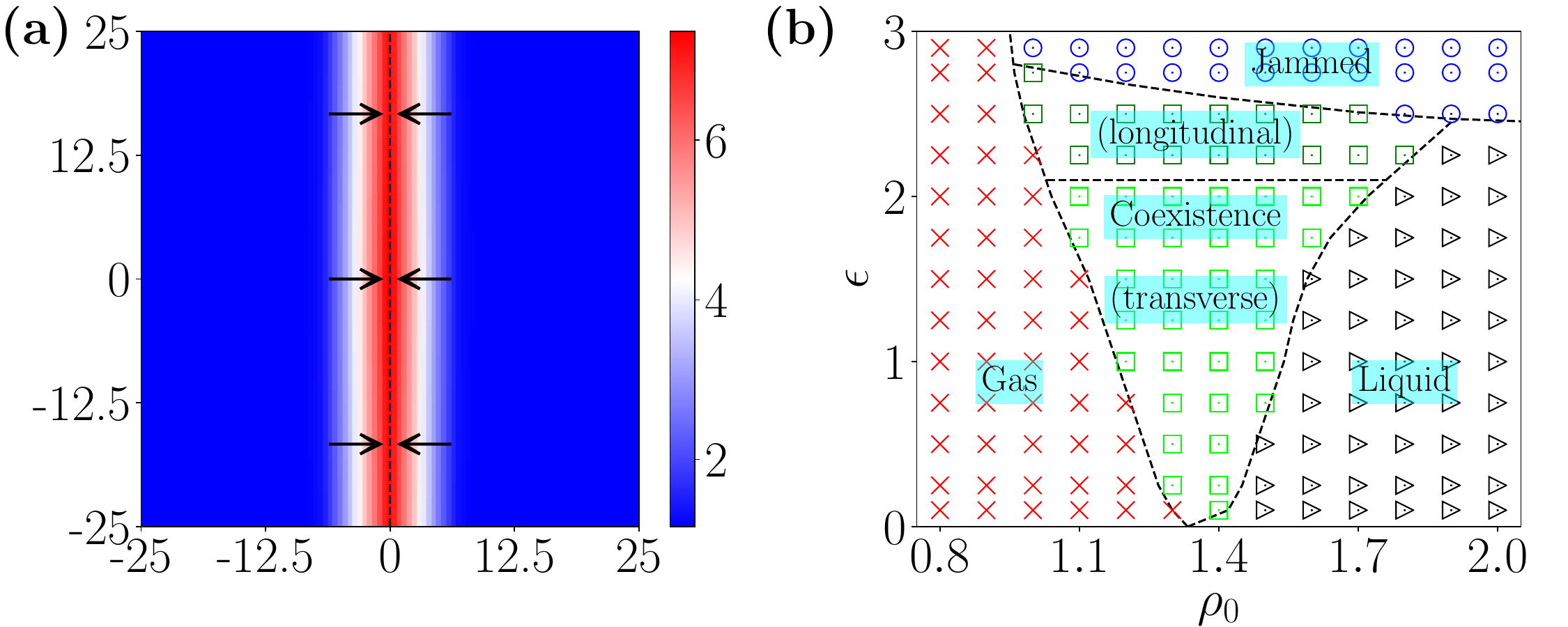}
\caption{(Color online) (a)~Jammed density profile for the soft-core rAPM
with $U=1$, for $\beta=0.75$, $\rho_0=2$, $\epsilon=2.7$, obtained numerically with FreeFEM++ for ${\cal N}=75$, $\Delta t = 0.1$ and $t_{\rm max}=5000$. (b)~Velocity-density phase diagram for the soft-core rAPM, computed with the numerical solutions of the Eq.~\eqref{eqsc}, for $\beta=0.75$ and $U=0.25$ in a square domain of linear size $L=50$.}
\label{fig_hydro_softcore}
\end{figure}

For soft-core rAPM, with $f(\rho) = \exp(-s \rho)$, Eq.~(\ref{eqhydro}) becomes:
\begin{align}
\partial_t \rho_\sigma &= D_\parallel \partial_\parallel \left[ \exp(-s\rho) \left( \partial_\parallel \rho_\sigma + s \rho_\sigma \partial_\parallel \rho \right) \right] \nonumber \\
&+ D_\perp \partial_\perp \left[ \exp(-s\rho) \left( \partial_\perp \rho_\sigma + s \rho_\sigma \partial_\perp \rho \right) \right]\nonumber\\
&- v \partial_\parallel \left[ \exp(-s\rho) \rho_\sigma \right] + \sum_{\sigma'\ne\sigma} K_{\sigma \sigma'} (\rho_\sigma - \rho_{\sigma'}),\label{eqsc}
\end{align}
with $K_{\sigma \sigma'}$ defined in Eq.~(\ref{Iflip}) and $s=2\beta U$. The homogeneous solutions, given by the solutions of $K_{\sigma \sigma'} (\rho_\sigma - \rho_{\sigma'}) = 0$, are those of the unrestricted APM derived in Refs.~\cite{chatterjee2020flocking,mangeat2020flocking}, as well as those explained for the hard-core rAPM in Sec.~\ref{hydrohc}. Similarly, a first-order liquid-gas phase transition occurs, with a phase separation made by a liquid stripe on a gas background. The same two kinds of phase-separated profiles can be observed: a transverse or longitudinal band motion for which the liquid phase is mainly populated by one state, and a jammed state formed by two locked liquid bands mainly populated by oppositely moving states ($\sigma=1,3$ or $\sigma=2,4$). Fig.~\ref{fig_hydro_softcore}(a) shows a numerically obtained jammed density profile, for $U=1$ $\beta=0.75$, $\rho_0=2$, $\epsilon=2.7$. For $x<0$, the liquid band is mainly populated by the right state $\sigma=1$, while for $x>0$ it is mainly populated by the left state $\sigma=3$, explaining why the band is jammed. 

Fig.~\ref{fig_hydro_softcore}(b) shows the velocity-density phase diagram determined by the numerical solutions of the Eq.~\eqref{eqsc}, for $\beta=0.75$ and $U=0.25$. The topology of this phase diagram agrees well with what we obtained for the microscopic model shown in Fig.~\ref{fig11}(b), with a jammed state at large velocities and densities, and flocking band motion at low velocities and densities (transverse band motion for $\epsilon<2.1$ and longitudinal band motion for $\epsilon>2.1$). Note that the difference in the relevant density $\rho_0$ region is due to the different $U$ values used ($U=0.07$ in Fig.~\ref{fig11}(b), $U=0.25$ here).

In Appendix~\ref{ls_sc}, we perform the linear stability analysis of the disordered and ordered homogeneous solutions of the Eq.~\eqref{eqsc}. These eigenvalues allow the determination of the velocity for which the reorientation transition occurs, denoted $\epsilon_*$, as a function of $U$ and $T<T_c$. $\epsilon_*(U)$ increases for small $U$ and decreases to zero at large $U$. Again, the existence of the jammed state cannot be derived from the linear stability analysis of a homogeneous solution.


\section{Summary and Discussion}
\label{s4}
To summarize, we have analyzed a discretized flocking model with volume exclusion and showed that the interplay between alignment and on-site repulsion produces a vast spectrum of self-organized patterns ranging from jammed clusters or bands and argued that they are a manifestation of MIPS, which relies on the reduction of particle velocity with increasing local density~\cite{geyer2019freezing}. Generally, it has been observed that velocity alignment interactions promote MIPS~\cite{sese2018velocity,sese2021phase}. In the rAPM considered here, alignment is even necessary for MIPS to occur, since the jammed clusters disappear in the gas phase for $T \to \infty$, i.e. vanishing alignment. For increasing alignment, i.e. small $T$, either orientationally ordered domains appear in the jammed clusters, arranged in such a way that the cluster configuration is kinetically arrested (up to fluctuations), or, depending on density and self-propulsion strength, the jammed clusters dissolve into an orientationally ordered liquid phase, both manifestations of flocking.

The phase diagrams of the rAPM with ${\rm MPS}=1$ and those for ${\rm MPS}>1$ or soft-core repulsion turn out to be different due to the absence of alignment interactions for ${\rm MPS}=1$. The latter model is equivalent to an active lattice gas with persistent walkers instead of diffusing particles. Consequently, the rAPM with ${\rm MPS}=1$ is always in an
orientationally disordered (gas) phase in which various MIPS or jammed states occur. For ${\rm MPS}>1$ or soft-core repulsion in addition to various jammed states the three
typical flocking phases occur: the orientationally disordered gas, liquid-gas coexistence (flocking phase), and the orientationally ordered liquid.

Part of the phenomenology we describe in our chapter has been seen in an experiment of colloidal-roller system~\cite{geyer2019freezing}, where the system undergoes a phase transition from a gas phase to a solid jam phase via flocking as the packing fraction of the rollers is increased. It would be interesting to study the effects of volume exclusion on the transport and jamming of active particles in disordered landscapes. The quenched disorder is abundant in all-natural systems and is known to reduce the effect of local interactions. The altered composition may have an impact on the universal behavior of unrestricted systems, both in equilibrium and out-of-equilibrium conditions \cite{weinrib1983critical,dotsenko1995critical,kumar2017ordering}. It is also known that active matter systems with random quenched disorder undergo activity-induced jamming~\cite{forgacs2021active,reichhardt2014active}. Preliminary results for the restricted and the unrestricted APM with quenched disorder reveal interesting emergent collective behaviors~\cite{karmakarunpublished}. Another extension of our research would be to look into the influence of soft-core constraints on active systems with continuous symmetry, such as the Vicsek model where initial investigation suggests an arrest of the flocking state with the emergence of MIPS jammed clusters. 

\section{Auxiliary material}
\subsection{Mean-squared displacements (MSD) of particles in the jammed state}\label{appendix_msd}
Here, we explore the appearance of arrested states through measurements of the MSD of individual particles. The MSD of $N$ number of particles in the system at time $t$ (which quantifies how the particles move from their initial positions under various volume exclusion effects) is defined as
\begin{align}
R^2(t)=\frac{1}{N}\sum_{i=1}^N |\mathbf{r}_i(t)-\mathbf{r}_i(0)|^2 \, ,
\end{align}
where $\mathbf{r}_i(t)$ is the instantaneous position of the $i$-th particle at time $t$. For ballistic motion, $R^2 \sim t^2$ while for diffusive motion $R^2 \sim t$. For an arrested or jammed state, however, MSD is proportional to $t^x$ with $x \sim 0$~\cite{merrigan2020arrested}.

\begin{figure}[t]
\centering
\includegraphics[width=\columnwidth]{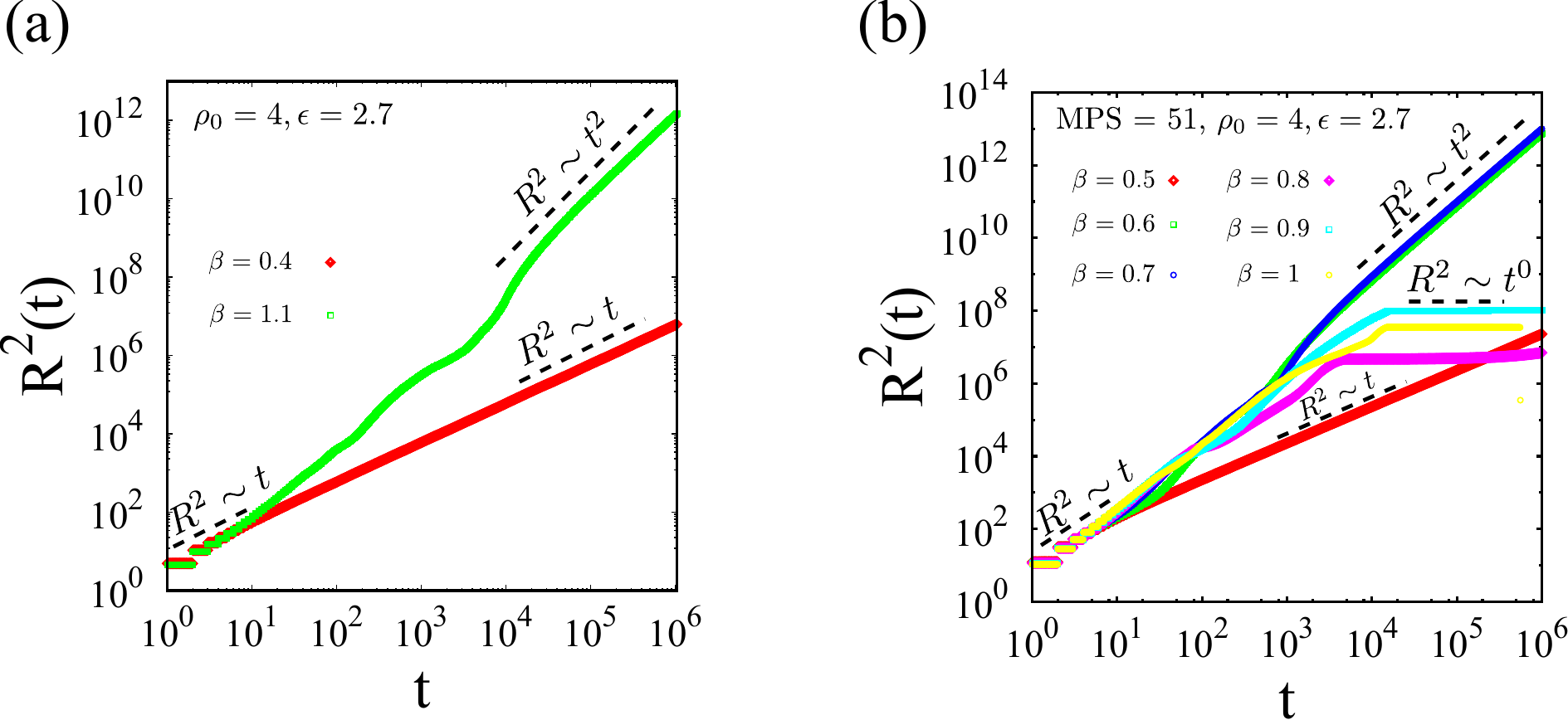}
\caption{(Color online) MSD $R^2(t)$ versus $t$ (on a log-log scale) for several $\beta$. (a) Unrestricted APM and (b) rAPM with hard-core repulsion.}
\label{rapm_appfig1} 
\end{figure}

In Fig.~\ref{rapm_appfig1}(a), we show $R^2$ versus $t$ (on a log-log scale) for the unrestricted APM~\cite{chatterjee2020flocking} as a function of $\beta$ ($T^{-1}$). At small $\beta$, which physically signifies the gaseous phase, the system obeys the diffusive growth $R^2 \sim t$ whereas, for small temperature, where the system exhibits the liquid phase, we observe two distinct regimes in the MSD. The small $t$ limit is characterized by a diffusive regime with $R^2 \sim t$ growth whereas a ballistic growth regime characterized by the power-law $R^2 \sim t^2$ is observed at large $t$. In the liquid phase, advective force (self-propulsion) plays a crucial role as the system exhibits the ballistic growth regime, which is a collision-free regime in which particles travel freely after a majority of the particles switch in the same state.

MSD for the rAPM with hard-core repulsion is shown in Fig.~\ref{rapm_appfig1}(b). At small $\beta$, the MSD shows a diffusive growth as the system is in the gas phase. For large $\beta$, the system exhibits the liquid phase and we observe a crossover from the diffusive growth at small $t$ to a ballistic growth at large $t$. For intermediate $\beta$, the small $t$ diffusive growth regime is followed by a plateau in the MSD at large $t$, which signifies a jammed or arrested state. A similar crossover in MSD can be seen for a random walker confined in a box, where the MSD crosses over from a diffusive growth ($R^2 \sim t$) to a plateau once the walker sticks to the walls~\cite{merrigan2020arrested}.

The MSD plots signifying the jammed phase describe the fact that initially the particles (the system is initially prepared homogeneous) do not feel the effect of the hopping restrictions and move diffusively but as the system coarsens with time, particles feel the restricted environment and finally, the system reaches the steady state jammed phase.

\subsection{Structural characteristics and transformation of a jammed phase}
\label{appendix_jam_structure}
\begin{figure}[t]
\centering
\includegraphics[width=0.9\columnwidth]{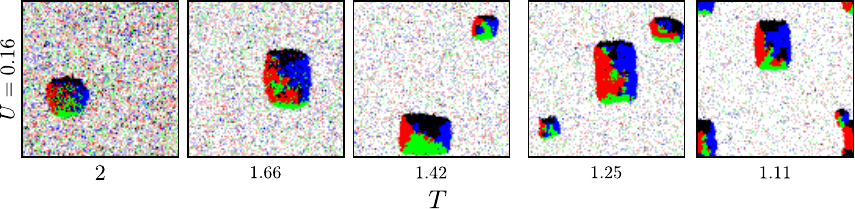}
\caption{(Color online) Morphological characteristics and transformation of jammed clusters in the rAPM with soft-core interactions as a function of temperature. Parameters: $U=0.16$, $\epsilon=2.7$, and $\rho_0=4$.}
\label{rapm_appfig2}
\end{figure}

Fig.~\ref{rapm_appfig2} displays the temperature-dependent structural changes of a jammed cluster. The jammed phase observed in this study is a kinetically arrested phase due to MIPS, where the particles cease to move due to steric repulsion. The active particles in this study only interact repulsively (when the MPS is reached or when dealing with soft-core) and as local density increases, the speed of the motile particles decreases, which results in a phase-separated state with a dilute active gas coexisting with a dense jammed cluster. At low temperatures, this jammed cluster comprises four orientationally ordered sub-domains in a completely gridlocked position but as the temperature is increased, the overall area of the jammed cluster shrinks and the internal structure of the cluster becomes orientationally disordered. This happens because as the on-site ferromagnetic alignment strength between the particles decreases with temperature, the flipping probability increases, which helps the particles change their predominant hopping direction and break from the gridlock.

\subsection{Effect of soft-core repulsion on active models in continuum}\label{racm_vm}
\begin{figure}[t]
\centering
\includegraphics[width=0.9\columnwidth]{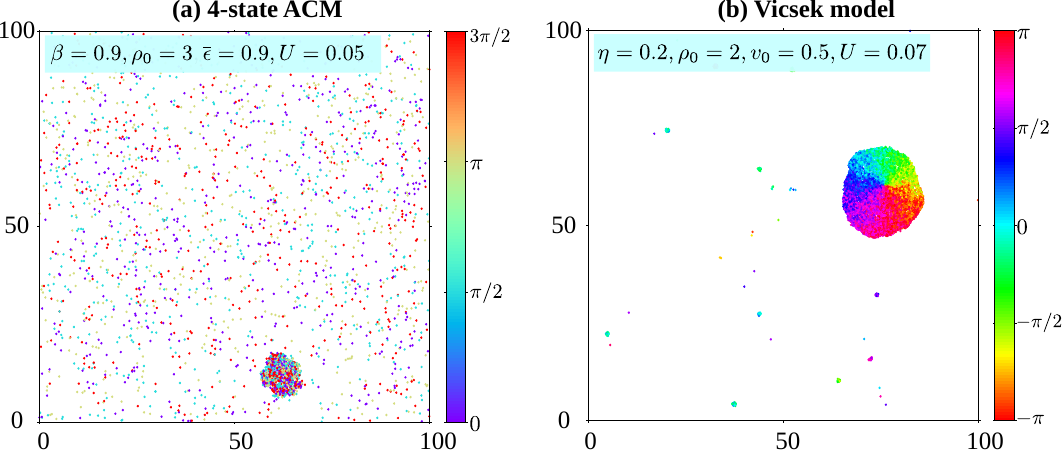}
\caption{(Color online) 
Jammed clusters in the active clock model and Vicsek model with soft-core repulsion: (a) 4-state ACM ~\cite{chatterjee2022polar} ($\beta=0.9$, $\rho_0=3$, $\bar{\epsilon}=0.9$, and $U=0.05$) and (b) Vicsek model \cite{solon2015phase} (noise $\eta=0.2$, $\rho_0=2$, $v_0=0.5$, and $U=0.07$). Colorbar represents particle orientation where for 4-state ACM particles can have only four discrete orientations: 0, $\pi/2$, $\pi$, and 3$\pi/2$ whereas for VM, a particle can have any orientation in $[-\pi,\pi]$.}
\label{rapm_appfig3}
\end{figure}

Our results are not artifacts of the discrete space we consider and to establish that we investigate the effect of soft-core repulsion on active models which are defined on an off-lattice geometry such as the ACM~\cite{chatterjee2022polar} and the VM~\cite{solon2015phase}. We take the 4-state ACM with density-dependent motility where particle hopping is allowed with a probability $\exp(-\beta U n_i)$ ($U$ is the strength of the steric repulsion and $n_i$ is the number of particles in the unit neighborhood of particle $i$ which is trying to propel). For such a setup, a jamming transition reminiscent of the rAPM is observed from a flocking phase to a MIPS-induced jammed phase as the repulsion strength $U$ is increased. The MIPS state of 4-state ACM with a soft-core repulsion is shown in Fig.~\ref{rapm_appfig3}(a) where the particle orientations signify that the internal structure of the cluster is disordered. With an off-lattice geometry, the domain boundaries of the jammed cluster are not as sharp as one observes with the lattice geometry but apart from this apparent difference, the underlying physics of jamming is the same.

Our findings can also be extended to continuum to the active XY model ($q \to \infty$)~\cite{chatterjee2022polar}, or the VM \cite{solon2015phase} with soft-core repulsion. For instance, our study of the VM with soft-core repulsion, where the particle velocity is modified as $v_0 \to v_0\exp(-Un_i)$, demonstrates that such a configuration produces circular/elliptical high-density jammed clusters with particle orientation continuously distributed between $[-\pi,\pi]$ [see Fig.~\ref{rapm_appfig3}(b)]. This jammed phase is a MIPS cluster typical of repulsively interacting active Brownian particles (ABP)~\cite{caprini2020spontaneous} and happens above a threshold value of $U$, below which the system exhibits flocking behavior. A similar kind of density-dependent velocity was introduced in a model of active Brownian particles with alignment interaction \cite{farrell2012pattern} where at large restriction, the system was shown to exhibit an aster-like jammed stationary phase. In both cases, the origin lies in the slowdown of particles due to crowding jamming.

\subsection{Zero activity ($\epsilon=0$) limit of the rAPM}
\label{appendix_diffusive}
\begin{figure}[!t]
\centering
\includegraphics[width=0.9\columnwidth]{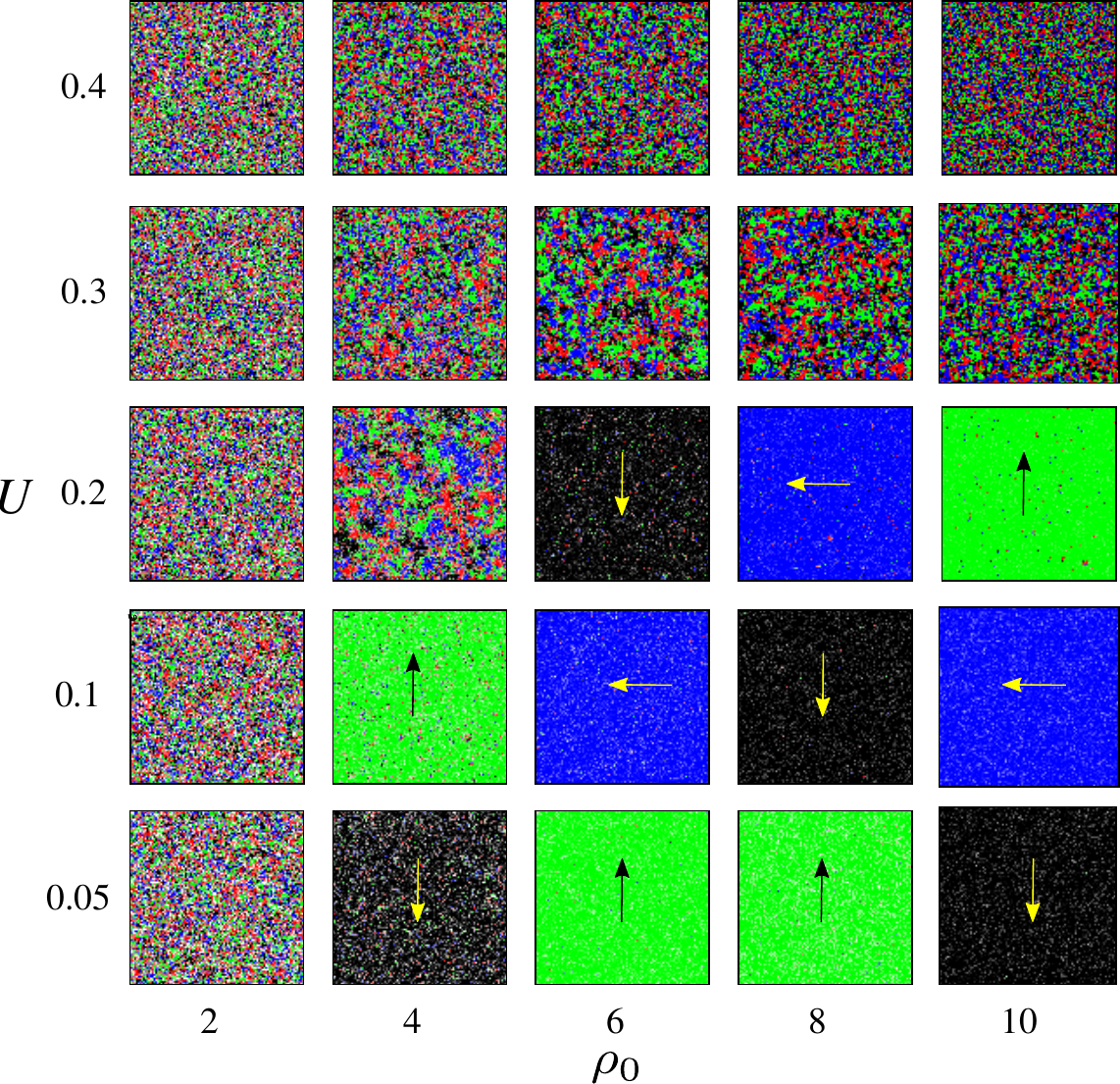}
\caption{(Color online) $U-\rho_0$ phase diagram of the rAPM with soft-core
repulsion with snapshots at $t=10^5$ for $\epsilon=0$ and $\beta=0.7$. For small $U$, similar to the unrestricted APM, we observe a phase transition from a disordered gaseous phase to an ordered liquid phase as density is increased. Due to $q=4$, a steady state liquid phase can be formed by any of the four states, and the following is the color code for the liquid phase: red ($q=1$), green ($q=2$), blue ($q=3$), and black ($q=4$). Arrows indicate the direction of motion. At large $U$, particles can only segregate locally into small clusters forming an orientationally disordered phase at large spatial scales.}
\label{rapm_appfig4}
\end{figure}
In Fig.~\ref{rapm_appfig4}, we show the late-stage representative snapshots of the rAPM with soft-core repulsion in the $U-\rho_0$ plane for the zero velocity ($\epsilon=0$) limit where particle movement is controlled by diffusion (particles hop without any bias). For small repulsion, the system goes through a direct gas-liquid phase transition without a coexistence regime similar to the unrestricted APM \cite{chatterjee2020flocking}. With strong repulsion, however, the system exhibits orientational disorder at high densities, where the system forms a domain state without any long-range order because at large density and strong repulsion, particle hopping is impeded (``jammed") and particles align their states in small clusters. The average size of these domains does not grow in time i.e. the system reaches a stationary state with a specific average domain size which can be quantified following a characteristic length scale analysis.

Please note that at large $U$ and $\rho_0$ limit, the active particles become nearly immobile, and at this limit, our model is equivalent to the equilibrium 4-state Potts model on a random 2d graph which should show a long-range order but our model does not exhibit that. This is because the equilibrium 4-state Potts model is defined with nearest neighbor interaction which helps in domain coarsening whereas in the soft-core rAPM with large $U$ and $\rho_0$, the probability that an edge is present between nodes in the network following the soft-core hopping acceptance probability $\exp(-2\beta U \rho_0)$ is nearly zero. Therefore, at this limit, the system behaves as a disconnected random 2d graph and does not display long-range order.

We have also checked that the order parameter at $\epsilon=0$ for small $U$ values ($U = 0.05$, 0.1, 0.15) shows a discontinuous jump at a critical density ($\rho_* \sim 3.5$ for $\beta = 0.7$) similar to the unrestricted APM.


%

\chapter{Disorder media and its impact on self-propelled particles in discrete flocking model}\label{chap:Chap3}
This study explores the behavior of SPPs in disordered media employing the $q$-state active Potts model, viz., the Random Field Active Potts Model (RFAPM) and Random Diffusion Active Potts Model (RDAPM). Interestingly, a unique feature of the APM is the treadmilling of a longitudinal band opposite to a small unidirectionally applied field driven by non-biased diffusion. However, the self-propulsion velocity for reorientation transition in the coexistence regime remains constant. Increasing a constant unidirectional local field transforms the coexistence band into a fully liquid state, where particles align with the field direction despite larger thermal fluctuations. Conversely, randomly distributed field orientations lead to the destruction of the flocking phase. The decrease in interaction strength between neighboring sites weakens particle hopping, causing a loss of spatial correlation and transitioning into a disordered phase.   

\section{Introduction}
Active matter systems contain self-propel particles (SPPs) that consume energy to move and interact with each other, resulting in complex collective movements~\cite{ramaswamy,vicsek1995,vicsek2012collective,toner1995, GC2004, TT2005,shaebani}. These systems are found in various natural systems, ranging from flocks of birds and schools of fish to human crowds and sub-cellular structures~\cite{strogatz}. The study of active matter is crucial in understanding the mechanics of both biological systems~\cite{prost,needleman2017active,xi2019material,trepat2018mesoscale,dell2018growing,perez2019active} and internally driven active matter systems~\cite{schrodinger1951life,goldenfeld2011life,popkin2016physics}, as they share similarities in their behavior. Flocking, in particular, is of great interest as it occurs in a variety of natural systems and has implications in physics, biology, and neurosciences~\cite{de2015introduction,krishnan2010rheology,bottinelli2016emergent,helbing1995social,garcimartin2015flow,marchetti2013hydrodynamics,ballerini2008interaction,becco2006experimental,calovi2014swarming,peruani2012collective,schaller2010polar,sumino2012large,sanchez2012spontaneous}.

Most of the theoretical studies on active matter systems have focused on SPPs' behavior in homogeneous environments~\cite{vicsek1995,solon2013revisiting,solon2015flocking,chatterjee2020flocking,mangeat2020flocking}. Also, there is experimental work on flocking behavior that does not consider heterogeneous media~\cite{peruani2006nonequilibrium,peruani2012collective,romanczuk2012active,deseigne2010collective,schaller2010polar}. However, active systems are typically heterogeneous as SPPs vary their response to the external environment. For example, at a large scale, mammalian herds (wildebeest migration in Africa~\cite{musiega2004simulating}) migrate long distances traversing rivers, forests, etc~\cite{holdo2011predicted}. In micro-scale, collective cell migration in heterogeneous media~\cite{cai2016modeling,angelini2011glass} in particular, many biological systems that show flocking involve self-propelled particles with heterogeneous environments (e.g., those involving bacteria or enzymes, the environments are disordered~\cite{muddana2010substrate}). However, few investigations have examined the effects of disorder on flocking models~\cite{chepizhko2013optimal,toner2018hydrodynamic,toner2018swarming,das2018polar}. 

Recent studies on Vicsek flocking rules have shown an optimal level of angular noise that maximizes collective motion~\cite{chepizhko2013optimal}. However, disordered environments can destroy this movement and trap or localize flocking active particles~\cite{quint2013swarming,chepizhko2013diffusion}. Despite some experiments and simulations using optical speckle fields or nonsmooth substrates~\cite{volpe2014brownian,paoluzzi2014run}, there is still a lack of understanding of their behavior in the presence of disorder media, which motivates the study of flocking models in heterogeneous environments. A very recent study showed that in the presence of volume exclusions in the active Potts model (APM), SPPs can form MIPS at low thermal noise and high density~\cite{karmakar2023jamming}. APM~\cite{chatterjee2020flocking,mangeat2020flocking} is the generalized version of the active Ising model (AIM)~\cite{solon2013revisiting,solon2015flocking}, in which particles move on a lattice in more discrete directions. Similar to the Vicsek model~\cite{solon2015phase}, macrophase separation replaces microphase separation in the coexistence region, resulting in a single collectively moving band.~\cite{solon2013revisiting,solon2015flocking,chatterjee2020flocking,mangeat2020flocking}. Further study is needed to better our understanding of the behavior of these systems in complex environments. 

The present study compares two models to understand the impact of the disorder in terms of an external field with fixed or random orientations and random interaction strength between neighboring sites. These are the Random Field Active Potts Model (RFAPM) and Random Diffusion Active Potts Model (RDAPM). We ask, what happens when SPPs are forced to encounter an external field environment: (i)  Does a collective behavior emerge when the external field is unidirectional or randomly oriented? (ii) If so, how do the SPPs behave with thermal fluctuations? (iii) How does heterogeneity in diffusion strength between lattice sites affect the order-disorder phase transition? (iv) Is a long-range order (LRO) possible for a large self-propulsion velocity of SPPs? Etc. Here, we address these questions by focusing on the effect of external field and heterogeneity in diffusion strength on SPPs.

This chapter is organized as follows. In Sec.~\ref{s2A} up to Sec.~\ref{s2D}, we have defined the RFAPM and provided details of the simulation protocols. Next, in Sec.~\ref{s3_rfapm}, we present numerical results of extensive Monte Carlo simulations of the microscopic model, where simulation results are discussed. Then, we introduce RDAPM in Sec.~\ref{Mod_RB} and discuss numerical results in Sec.~\ref {Res_RB}. Finally, we conclude this chapter with a summary and discussion of the results in Sec.~\ref{s4_rfapm}.

\section{Random Field APM}\label{Random_Field_APM}
\subsection{\label{s2A}The Model}
\begin{figure*}[!t]
\centering
\includegraphics[width=\textwidth]{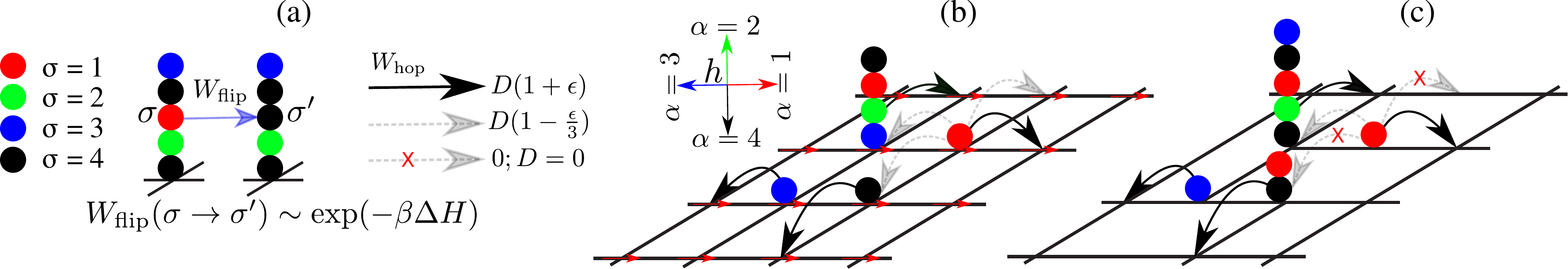}
\caption{\label{fig:Model_fig}(Color online) (a) Sketch of the two possible actions, spin-flipping and hopping, and their associated rates.(b) Model schematic of the RFAPM. The model includes a lattice of spins, where each spin can take on one of the $q$ different states. The spins are subject to an external field with a fixed orientation ($\alpha=1$), which is shown here as straight red arrows. (c) Schematic representation of the RDAPM on a two-dimensional square lattice. The solid black lines indicate the bonds present with probability $1-P_{rd}$. The broken lines indicate the bonds absent with probability $P_{rd}$. The red cross signifies no hopping of particles due to broken lines.}
\end{figure*}

We consider an ensemble of $N$ particles defined on a two-dimensional square lattice of size $L^2$ with periodic boundary conditions applied on both sides, where $L$ is the linear lattice dimension. The average particle density $\rho_0$ is then defined as $\rho_0=N/L^2$. The model is built upon the APM~\cite{chatterjee2020flocking,mangeat2020flocking} in which the dynamics are governed by the on-site flipping of the internal spin state and by nearest-neighbor hopping. Besides, we now applied an external field. Fig.~\ref{fig:Model_fig}(b) shows a schematic diagram of the lattice setup.

The RFAPM is a disordered flocking model, where $\rho_i$ number of particles and quenched (static) random field $h_i$ are placed at each site $i$ of the lattice. Each particle has a spin state $\sigma$ which takes an integer value in $[1,q]$. The Hamiltonian for the RFAPM with $L^2$ sites can be expressed as~\cite{paul2002low}: 
\begin{equation}
\label{Hapm_rfapm}
\centering
    H=-J\sum_{i=1}^{L^2}\frac{1}{2\rho_i}\sum_{j=1}^{\rho_i}\sum_{k\ne j}\left(q\delta_{\sigma_j,\sigma_k}-1\right)-\sum_{i=1}^{L^2} h_i\left(q\delta_{\sigma_j,\alpha_i}-1\right) \, ,
\end{equation}
where $\delta_{\sigma,\sigma^\prime}$ is the Kronecker-delta ($\delta_{\sigma,\sigma^\prime}=1$ for $\sigma=\sigma^\prime$ and $\delta_{\sigma,\sigma^\prime}=0$ otherwise) and $\alpha_i$ denotes the direction of the field at site $i$ along any of the four directions of the square lattice. In this paper, $h_i$ is taken as a constant of sites, i.e. $h_i=h$ and $J=1$ is the strength of the ferromagnetic exchange coupling between the particles. A particle at site $i$ with state $\sigma$ either flips to another state $\sigma^\prime$ or hops to any of the four nearest neighboring sites without any restrictions. The local density at site $i$ is defined by $\rho_i=\sum_{j=1}^{q}n^{\sigma_j}$, whereas, the local magnetization corresponding to state $\sigma$ at site $i$ is denoted by:
\begin{equation}
\label{Hapm_mag_rfapm}
m_i^{\sigma}=\sum_{j=1}^{\rho_i}\frac{\left(q\delta_{\sigma,\sigma_j}-1\right)}{q-1}=\frac{qn_i^{\sigma}-\rho_i}{q-1} \, .
\end{equation}

\subsubsection{On-site alignment of particles}\label{s2B}
From Eq.~\eqref{Hapm_rfapm}, the alignment probability for particles at site $i$ can be defined
by the following local Hamiltonian:
\begin{equation}
\label{H_i}
\centering
    H_i=-\frac{J}{2\rho_i}\sum_{j=1}^{\rho_i}\sum_{k\ne j}\left(q\delta_{\sigma_j,\sigma_k}-1\right)- h\left(q\delta_{\sigma_j,\alpha_i}-1\right) \, .
\end{equation}

If a particle at site $i$ with state $\sigma$ flips its state to $\sigma^\prime$, then the energy difference due to this flip would be:
\begin{equation}
\label{delH1}
    \Delta H_i=H_{i}^{\rm new}-H_i^{\rm old} \nonumber
\end{equation}
\begin{gather}
\label{delH_rfapm}
    \implies\Delta H_i=\frac{qJ}{\rho_i}\left(n^{\sigma}-n^{\sigma^\prime}-1\right)-h\left(q\delta_{\sigma^\prime,\alpha_i}-1\right)\nonumber\\+h\left(q\delta_{\sigma,\alpha_i}-1\right) \nonumber
\end{gather}
\begin{equation}
\label{delH3}
    \implies\Delta H_i=\frac{qJ}{\rho_i}\left(n^{\sigma}-n^{\sigma^\prime}-1\right)+qh\left(\delta_{\sigma,\alpha_i}-\delta_{\sigma^\prime,\alpha_i}\right) \, .
\end{equation}
Therefore, a particle with an initial state $\sigma$ at site $i$ makes a transition to a new state $\sigma^\prime$ with the following rate:
\begin{equation}
\label{flipeq_rfapm}
    W_{\rm flip}(\sigma\to\sigma^\prime)\propto \exp\left(-\beta\Delta H_i\right) \, ,
\end{equation}
where $\beta$ is the inverse temperature ($\beta=T^{-1}$).

\subsubsection{\label{s2C}Biased hopping of particles}
The hopping rate of a particle with state $\sigma$ is not affected by the external field and follows the APM hopping rule~\cite{chatterjee2020flocking,mangeat2020flocking}:
\begin{equation}\label{Whop}
    W_{\rm hop}=D\left[1+\epsilon\frac{(q\delta_{\sigma,p}-1)}{(q-1)}\right] \, .
\end{equation}
Here, $D=1$ controls the diffusion, and $\epsilon \in [0,q-1]$ governs the self-propulsion velocities of the particles. The hopping rate is $W_{\rm hop}=D(1+\epsilon)$ for $\sigma = p$ and $W_{\rm hop}=D[1-\epsilon/(q-1)]$, otherwise. The total hopping rate for a particle is $qD$, which is independent of $\epsilon$. 

\subsubsection{\label{s2D}Simulation Details}
Simulation evolves in the unit of Monte Carlo steps (MCS) $\Delta t$ resulting from a microscopic time $\Delta t/N$, $N$ being the total number of particles. During $\Delta t/N,$ a randomly chosen particle either updates its spin state with probability $p_{\rm flip}=W_{\rm flip}\Delta t$ or hops to one of the neighboring sites with probability $p_{\rm hop}=W_{\rm hop}\Delta t$. An expression for $\Delta t$ can be obtained by minimizing the probability of nothing happens $p_{\rm wait}=1-(p_{\rm hop}+p_{\rm flip})$:
\begin{equation}\label{delt_rfapm}
\Delta t=[qD+\exp\{q\beta(1+h)\}]^{-1} \, .
\end{equation}
For small $h$, Eq.~\eqref{delt_rfapm} can be approximated as: $\Delta t \simeq [qD+\exp\left(q\beta\right)]^{-1}$.

\subsection{Numerical results for homogeneous and unidirectional external field}\label{s3_rfapm}
We first analyze the behavior of randomly distributed (Initial configuration) 4-state Self-Propelled Particles (SPPs) in a homogeneous and unidirectional external field on a two-dimensional square lattice of size $L=100$. Note that the magnitude of the field $h_i=h$ is site independent and the field is directed along $\sigma=1$ state for all lattice sites.

\subsubsection{Collective behaviour \& phase diagrams}\label{sub1}
\begin{figure}[!htbp]
\centering
\includegraphics[width=0.8\columnwidth]{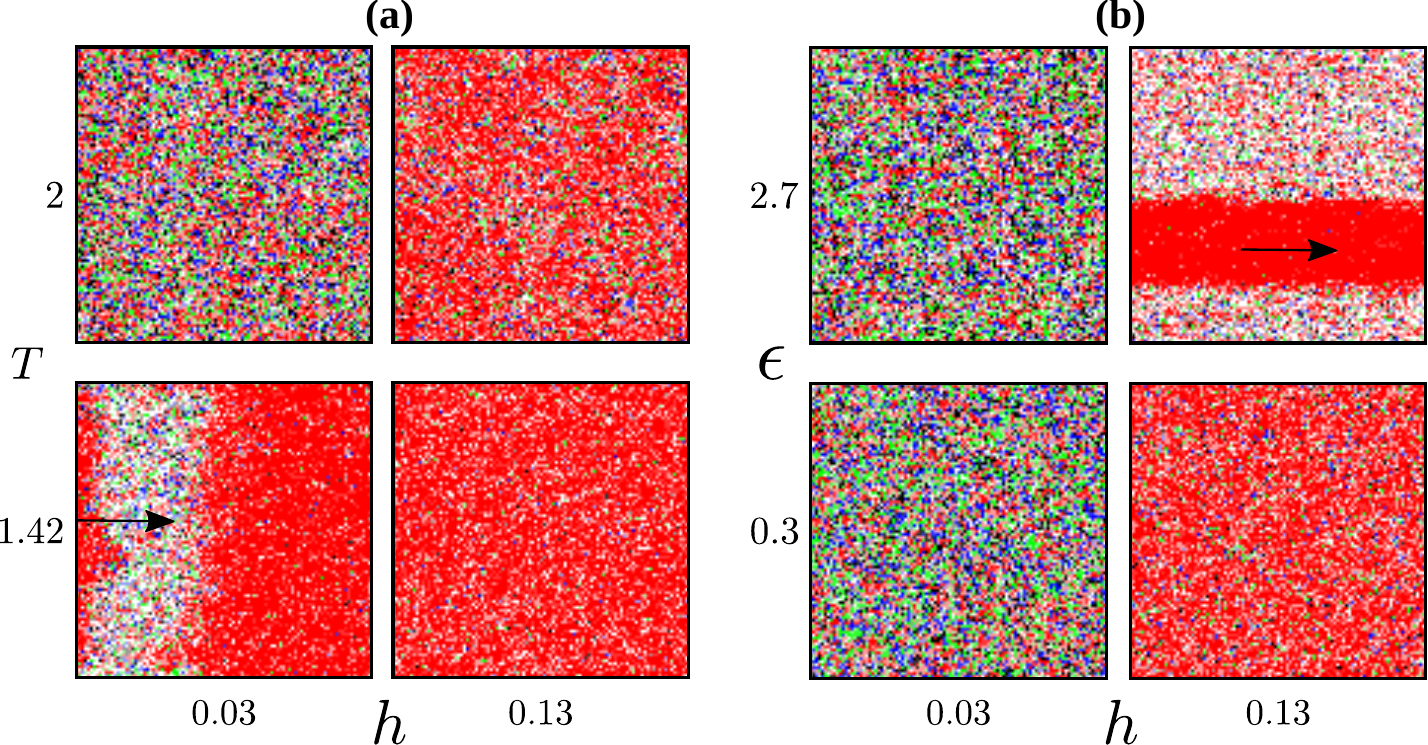}
\caption{\label{H_vs_B-E_Snap}(Color online) Steady-state snapshots of a 4-state RFAPM with a homogeneous and unidirectional field. Color code for state $\sigma=1$ is red, green
($\sigma=2$), blue ($\sigma=3$), and black ($\sigma=4$). (a) Snapshots in the $h-T$ plane at $\epsilon=0.9$, and $\rho_0=3$, while (b) illustrates the effect of increasing $h$ with $\epsilon$, at fixed $T=2$, and $\rho_0=3$. Both (a, and b) demonstrate that the external field promotes an ordered state.}
\end{figure}

Fig.~\ref{H_vs_B-E_Snap} represents steady-state snapshots of a four-state RFAPM. 
In Fig.~\ref{H_vs_B-E_Snap}(a), two different temperatures are considered, and the influence of field strength $h$ on the system is shown at a fixed average density ($\rho_0=3$) and self-propulsion velocity ($\epsilon=0.9$). At large thermal fluctuation (upper panel of Fig.~\ref{H_vs_B-E_Snap}(a)), a transition from a disordered phase to a polar ordered liquid phase is observed as $h$ increases. In contrast, at an intermediate temperature (lower panel of Fig.~\ref{H_vs_B-E_Snap}(a)), the system undergoes a transition from coexistence to liquid as $h$ is increased. According to Eq.~\ref{flipeq_rfapm}, if a particle switches its state($\sigma \to \sigma^\prime$) matches with the field direction $\alpha$, then the flipping rate($W_{\rm flip}$) is increased. Consequently, a larger external field tends to align the particle more in the direction of the field, promoting an ordered phase. At a very high field, most of the particles are aligned along the field direction and a polar liquid phase will form. Similarly, (b) demonstrates how the self-propulsion velocity $\epsilon$ and field strength $h$ affect the system at a constant thermal noise $T=2$. Due to large thermal fluctuations, a disordered gaseous phase is formed for both small and large velocities at low field strength ($h=0.03$). However, at large strength ($h=0.13$) of the field, a transition to a polar liquid phase is observed. When the self-propulsion velocity is large ($\epsilon=2.7$), more particles accumulate in the biased directions of the particle state, and the level of activity in non-preferred directions decreases significantly. Consequently, when $h$ is increased at large self-propulsion velocity, a lane of $\sigma=1$ is formed along the field direction.
\begin{figure}[!htbp]
\centering
\includegraphics[width=0.8\columnwidth]{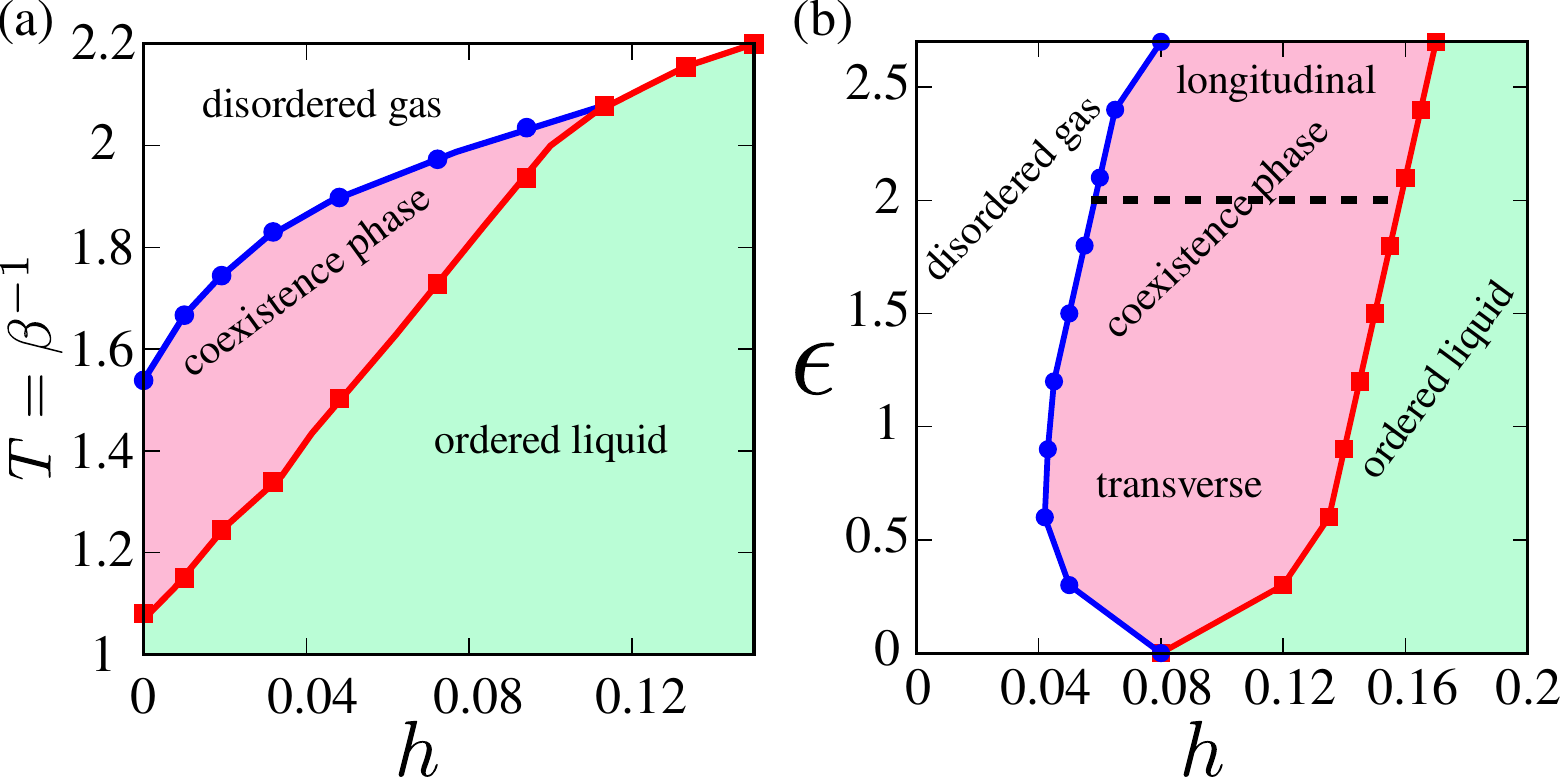}
\caption{\label{H_vs_Beta_Eps}(Color online) Phase diagrams of the RFAPM with homogeneous and unidirectional field. (a) $T-h$ phase diagram at fixed $\epsilon=0.9$. The blue and red dotted lines correspond to the liquid and coexistence phase boundaries, respectively. (b) $\epsilon-h$ phase diagram at fixed thermal noise $T=2$. The reorientation transition limit in the coexistence region is denoted by a dotted line at $\epsilon=2$, unaffected by the change in $h$.}
\end{figure}

The phase diagrams of the RFAPM show the impact of noise, self-propulsion velocity, and field strength on the collective behavior of SPPs (Fig.~\ref{H_vs_Beta_Eps}(a, b)). In the temperature-field strength ($T,h$) phase diagram (Fig.~\ref{H_vs_Beta_Eps}(a)), increasing field strength promotes a disordered gas at relatively high temperature (e.g., $T \sim 1.8$) to transition into a coexistence phase and then to a liquid phase. At very high temperatures (e.g., $T \sim 2.1$), the disordered gas becomes liquid without passing through the coexistence phase at a high field (e.g., $h>0.12$). At lower temperatures (e.g., $T<1.5$) and vanishing fields, the system remains in the coexistence phase and transitions into a fully polar liquid phase with increasing field strength.
\begin{figure}[!htbp]
\centering
\includegraphics[width=0.8\columnwidth]{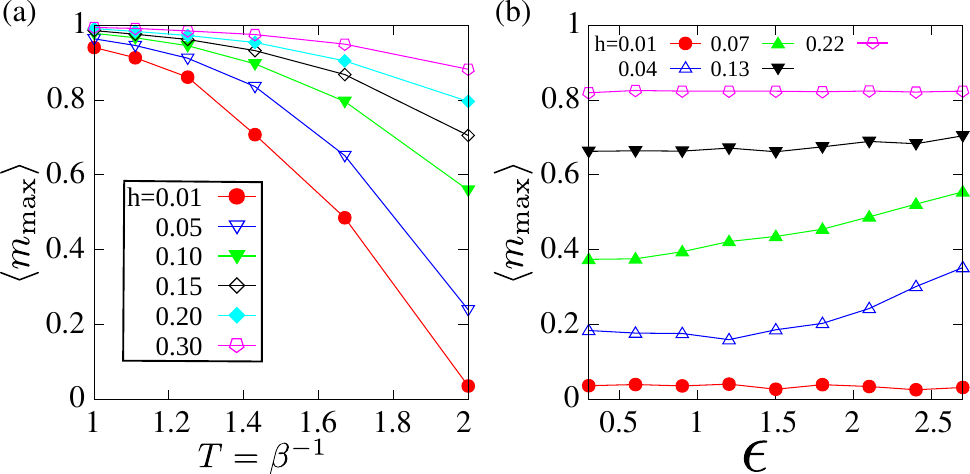}
\caption{\label{OP-RFAPM}(Color online) Order parameter plot. (a) $m_{\rm max}$ vs $T$ for different $h$. Parameters: $\epsilon=0.9$, and $\rho_0=3$. (b) $m_{\rm max}$ vs $\epsilon$ for different $h$. Parameters: $T=2$, and $\rho_0=3$.}
\end{figure}

In the velocity-field strength ($\epsilon,h$) phase diagram (Fig.~\ref{H_vs_Beta_Eps} b), we find a coexistence phase sandwiched between a disordered gas at low field and ordered liquid at high field values. The coexistence regime disappears, and a direct transition from disordered to liquid is observed in the purely diffusive limit (i.e., $\epsilon = 0$) at field $h^*\sim0.08$. However, the threshold values of $\epsilon$ for the reorientation transition become unaltered with field strength shown by the dotted line.

\subsubsection{Order parameter}
The steady-state magnetization characterizes the ordering of the SPPs. Fig.~\ref{OP-RFAPM} shows the maximal magnetization as a function of different control parameters. Where among the four different magnetizations corresponding to four internal states, we calculate the maximal magnetization $m_{\rm max}$ using Eq.~\ref{Hapm_mag_rfapm}, providing a measure of the degree of order in the system. The transition from a disordered gaseous state to a fully ordered liquid state is depicted through a change in magnetization, indicated by a shift from $m_{\rm max} \simeq 0$ to $m_{\rm max} \simeq 1$. Fig.~\ref{OP-RFAPM} (a) shows the progression of $m_{\rm max}$ against temperature ($T$) at different local field strengths ($h$). As thermal fluctuations increase with $T$, the order parameter $m_{\rm max}$ decreases rapidly at low values of $h$ and slowly at high values of $h$. Fig.~\ref{OP-RFAPM} (b) shows the relationship between the order parameter and self-propulsion velocity ($\epsilon$) at different values of $h$ at $T=2$. At low field strengths, the order parameter tends to approach zero, indicative of a disordered gas phase. As the field strength increases, a transition to an ordered phase is marked by a substantial and constant value of $m_{\rm max}$. However, for intermediate field strengths ($h=0.04, 0.07$), the behavior of $m_{\rm max}$ exhibits a relatively modest increase with the rise in $\epsilon$. This phenomenon can be attributed to the fact that at low field strengths, particles are less likely to undergo flipping along the field orientation. Additionally, the influence of unbiased diffusion remains significant for low levels of self-propulsion. Conversely, large self-propulsion facilitates the formation of high-density clusters, primarily aligned along the biased direction, often coinciding with the field direction. As a result, the order parameter experiences a slight increase with larger self-propulsion. Nevertheless, under conditions of high field strength, all particles undergo a transition to align along the field direction, rendering the influence of self-propulsion negligible in this regime.


To distinguish between a disordered phase and a free-flowing phase, one can look at the average mean-square displacement (MSD) of all the particles in the system. A free-flowing phase can be characterized by a ballistic growth behavior at larger times of the MSD (see Auxiliary material~\ref{appendix_msd_rfapm}).

\subsubsection{Diffusion driven treadmilling of the longitudinal band at finite field}
\begin{figure}[!htbp]
\centering
\includegraphics[width=0.8\columnwidth]{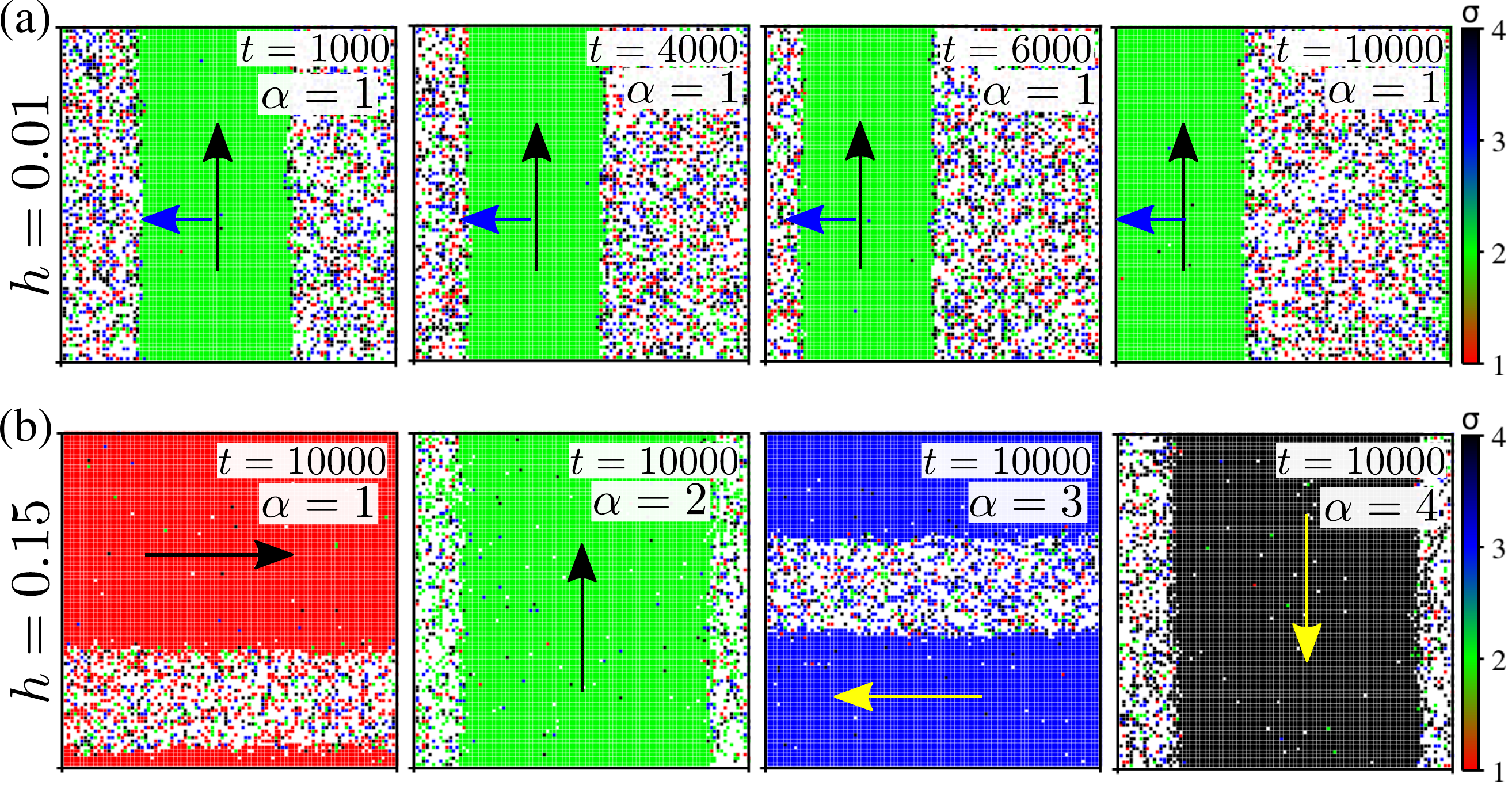}
\caption{\label{tredmilling}(Color online) (a) Time evolution steady state snapshots starting from an initial coexistence longitudinal band of state $\sigma=2$ orient transversely in compared to the field direction ($\alpha=1$) at low field strength $h=0.01$. The band is slowly displaced in the opposite direction of the field. (b) Late time steady state snapshots starting from an initial coexistence longitudinal band of state $\sigma=2$ orient along the field directions ($\alpha=1, 2, 3,$ and 4), and the width of the band is increased at larger field strength $h=0.15$. Parameters: $L=100$, $\rho_0=4$, $\beta=0.7$, and $\epsilon=2.7$}
\end{figure}
\begin{figure}[t]
\centering
\includegraphics[width=0.8\columnwidth]{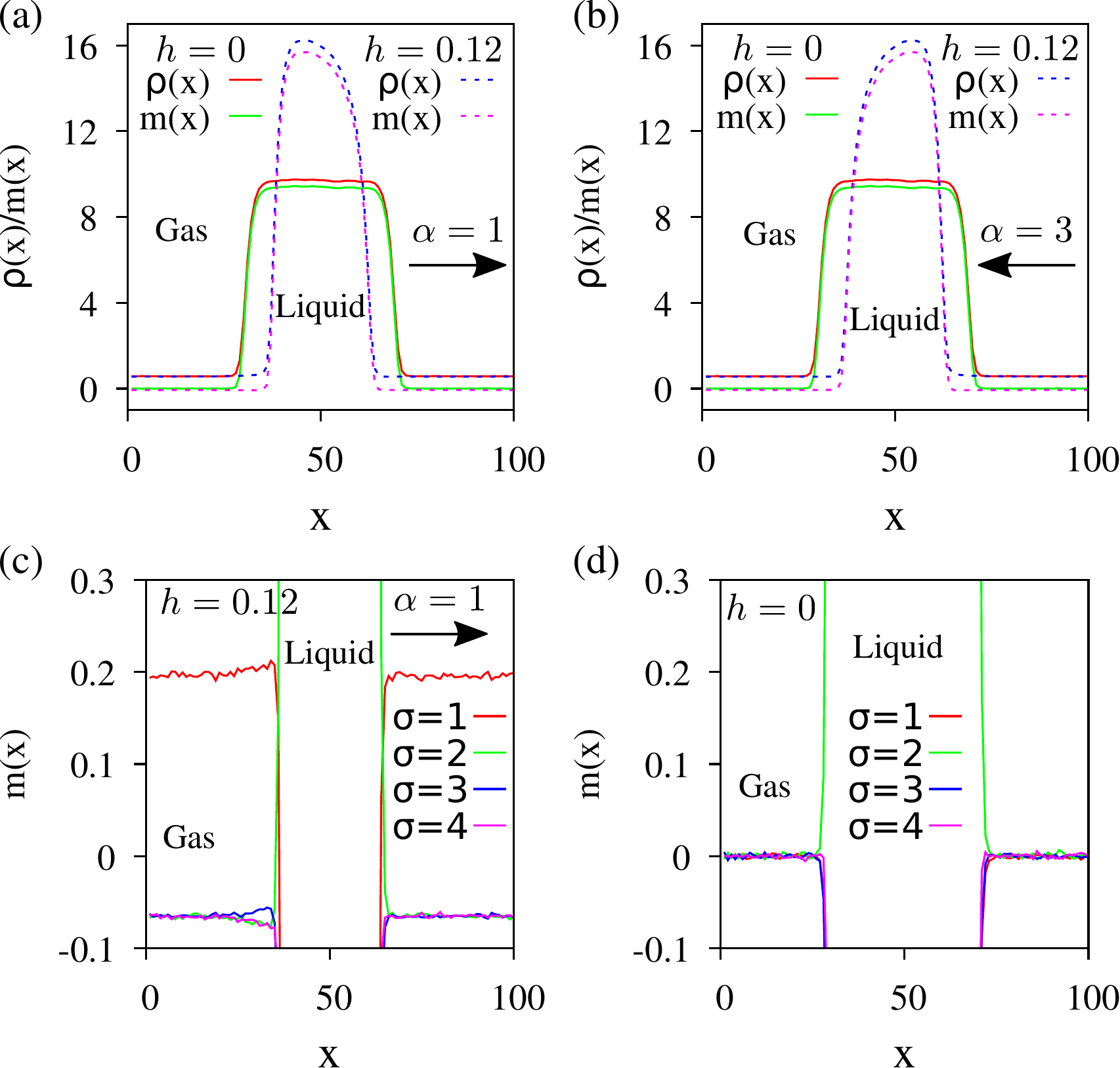}
\caption{\label{den_mag_prof_tredmilling}(Color online) (a-b) Time-averaged density and magnetization profiles at zero field flat liquid profile (solid lines) and finite field $h=0.12$ tilted liquid profile opposite to the field orientation (dashed lines). (c) Time-averaged magnetization profiles of four states exhibit an increase in magnetization in the gaseous profile for particles of $\sigma=1$ along which $h=0.12$ is applied. (d) In the absence of a field, gaseous profiles for each of the $\sigma$ is close to zero. Parameters: $L=100$, $\rho_0=4$, $\beta=0.7$, and $\epsilon=2.7$.}
\end{figure}

A distinctive characteristic emerges when a relatively small transverse field is applied compared to the orientation of the longitudinal band under significant self-propulsion of SPPs. The temporal evolution of steady-state snapshots, as illustrated in Fig.~\ref{tredmilling}(a), reveals a gradual transverse motion of the longitudinal band opposite to the field direction ($\alpha=1$). Notably, an increase in the field strength, depicted in Fig.~\ref{tredmilling}(b), can orient the lane parallel to the field direction. The lane, composed of particles with $\sigma=2$, undergoes a transformative change along the field direction, exhibiting increased thickness attributable to the higher coupling field strength. Importantly, no transverse motion of the lane is observed in this scenario. This phenomenon can be explained from the density and magnetization profiles shown in Fig.~\ref{den_mag_prof_tredmilling}(a-b), both of which exhibit a tilt in the direction opposite to the applied field at a finite strength $h=0.12$. This tilt indicates that when the applied field is insufficient to align the liquid band in the field direction, the band's side opposite to the field becomes denser than the other side. Without the field, we recover the flat liquid portion, APM-like density and magnetization profiles~\cite{chatterjee2020flocking,mangeat2020flocking}. So particles are injected into the side with higher density tilt, and conversely, particles are ejected from the opposite side. The individual magnetization profiles for different states in Fig.~\ref{den_mag_prof_tredmilling}(c) reveal that within the gaseous portion, the magnetization for state $\sigma=1$ surpasses that of the other three states. This discrepancy arises because the applied field aligns predominantly along $\alpha=1$, causing most particles in the gaseous region to transition to state $\sigma=1$. Consequently, these particles are biased towards movement in the $\alpha=1$ direction. On the contrary, the liquid portion of state $\sigma=2$ cannot switch to state $\sigma=1$ due to the dominance of local high-density sites over the coupling field of low strength. Consequently, particles of state $\sigma=1$ accumulate on the left side of the band from the right side within the gaseous portion. As the particles of $\sigma=2$ diffuses at a rate $W_{\rm hop}=D[1-\epsilon/(q-1)]$ from the left portion of the band, the accumulated particles of $\sigma=1$ undergo a state switch, contributing to an increase in bandwidth and density. Simultaneously, in the right portion of the band, diffusion of particles of $\sigma=2$ results in particles being ejected into the gaseous portion at a consistent rate. Most of these particles switch to state $\sigma=1$ due to the influence of the coupling field, subsequently engaging in self-propulsion. This dynamic gives rise to a treadmill-like behavior akin to the movement observed in actin filament dynamics~\cite{wegner1976head} (essential for driving directional movements~\cite{small1995getting}), with the band moving in the direction opposite to the applied field. Notably, in the absence of the field, the magnetization for all states becomes zero in the gaseous portion shown in Fig~\ref{den_mag_prof_tredmilling}(d), precluding the observation of treadmill-like behavior without the presence of the field.

\subsubsection{Bidirectional orientation of external field}
\begin{figure}[!t]
\centering
\includegraphics[width=0.8\columnwidth]{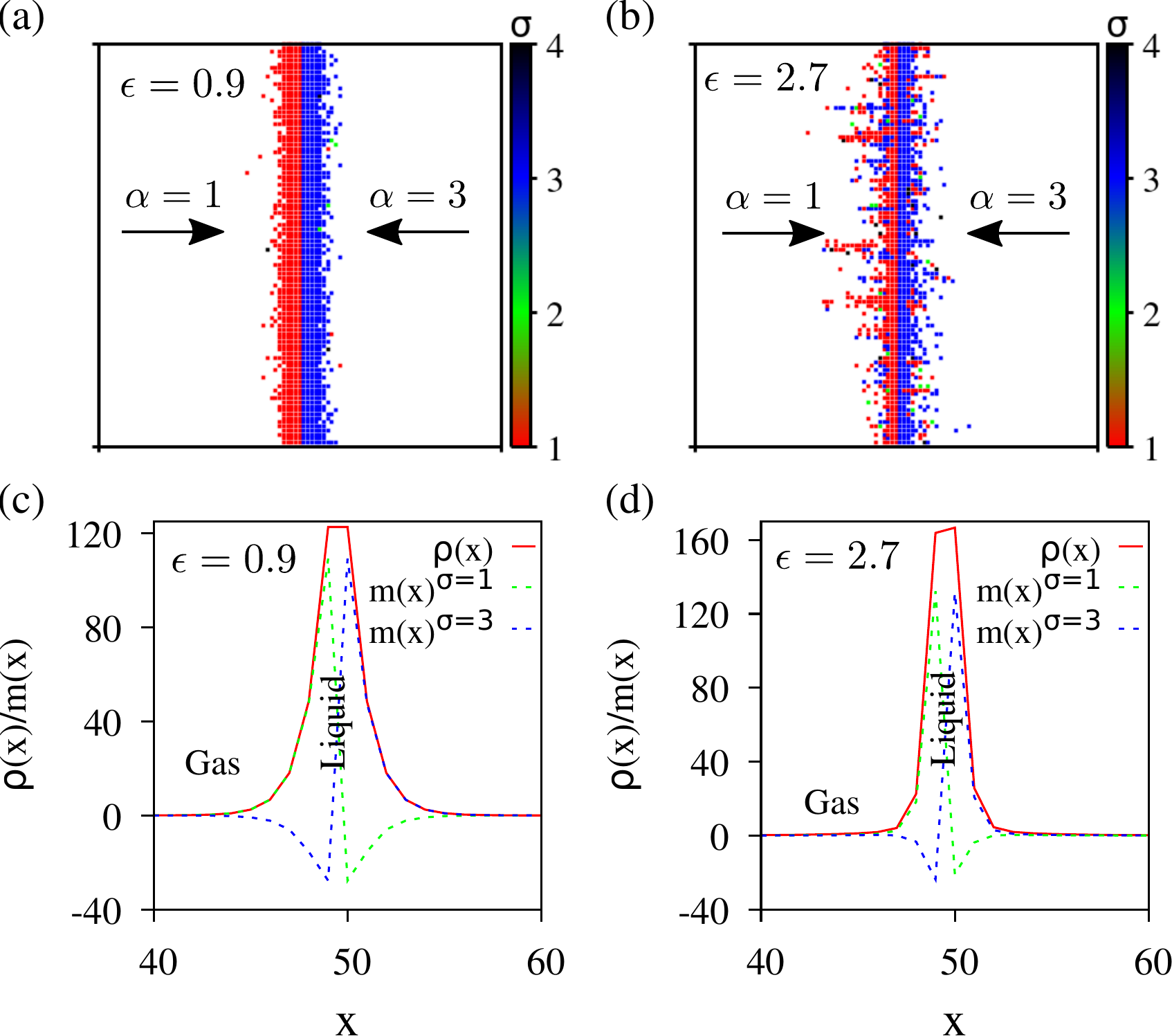}
\caption{\label{den_mag_prof_bidirectional}(Color online) (a-b) Late time steady state snapshot from random homogeneous configuration shows a MIPS of two types of particle ($\sigma=1,2$) as the applied field in the two halves of the square lattice is directed towards each other both for small and large self-propulsion velocity respectively. (c-d) Time-averaged density and magnetization profiles corresponding to (a-b) show a very large density MIPS phase; one half of which is populated by particles of $\sigma=1$ (peak in magnetization profile for $\sigma=1$, green dashed line) and another half by particles of $\sigma=3$ (peak in magnetization profile for $\sigma=3$, blue dashed line). Parameters: $L=100$, $\rho_0=4$, $\beta=1$, and $h=0.43$.}
\end{figure}
A transition from flocking to Motility-Induced Phase Separation (MIPS) is evident when the orientation of the applied field in the two halves of the square lattice is directed toward each other. Specifically, in the left portion of the lattice ($0\to L_x/2$), the field is oriented rightward ($\alpha=1$), while in the right portion of the lattice ($L_x/2 \to L_x$), the field is oriented leftward ($\alpha=3$). Commencing from an initially random and homogeneous configuration, the steady-state snapshots presented in Fig.~\ref{den_mag_prof_bidirectional}(a-b) for small and large self-propulsion velocities, respectively, showcase a distinctive Motility-Induced Phase Separation (MIPS) characterized by the coexistence of states $\sigma=1$ and 3. SPPs originating from the left portion undergo a state transition to $\sigma=1$ due to the influence of the field directed along $\alpha=1$. Conversely, particles from the right portion switch to $\sigma=3$ due to the field acting along $\alpha=3$. As a result, a highly dense Motility-Induced Phase Separation (MIPS) phase forms in the middle region. Importantly, the presence of a large self-propulsion velocity enables particles of state $\sigma=1$ from the left to self-propel and traverse the middle region. Similarly, particles of state $\sigma=3$ from the right exhibit the capability to self-propel across the middle region, as illustrated in Fig.~\ref{den_mag_prof_bidirectional}(b). In the high-density profiles of the respective Motility-Induced Phase Separation (MIPS) phases presented in Fig.~\ref{den_mag_prof_bidirectional}(c-d), the presence of two peaks in the magnetization profiles for states $\sigma=1$ and $\sigma=3$ signifies a jammed structure. As particles traverse the middle portion propelled by self-propulsion, the coupling field acts in the opposite direction, inducing a flip in the particle state aligned with the field direction. This dynamic contributes to the observed jammed structure within the MIPS phases.

\subsection{Numerical results for spatial heterogeneity in orientation of external field}
In this section, the external field is kept constant but oriented randomly along any of the $q=4$ directions at each lattice site denoted by $h_i=h_{ro}$. Steady-state behavior is analyzed under various control parameters.

\begin{figure}[!htbp]
\centering
\includegraphics[width=0.8\columnwidth]{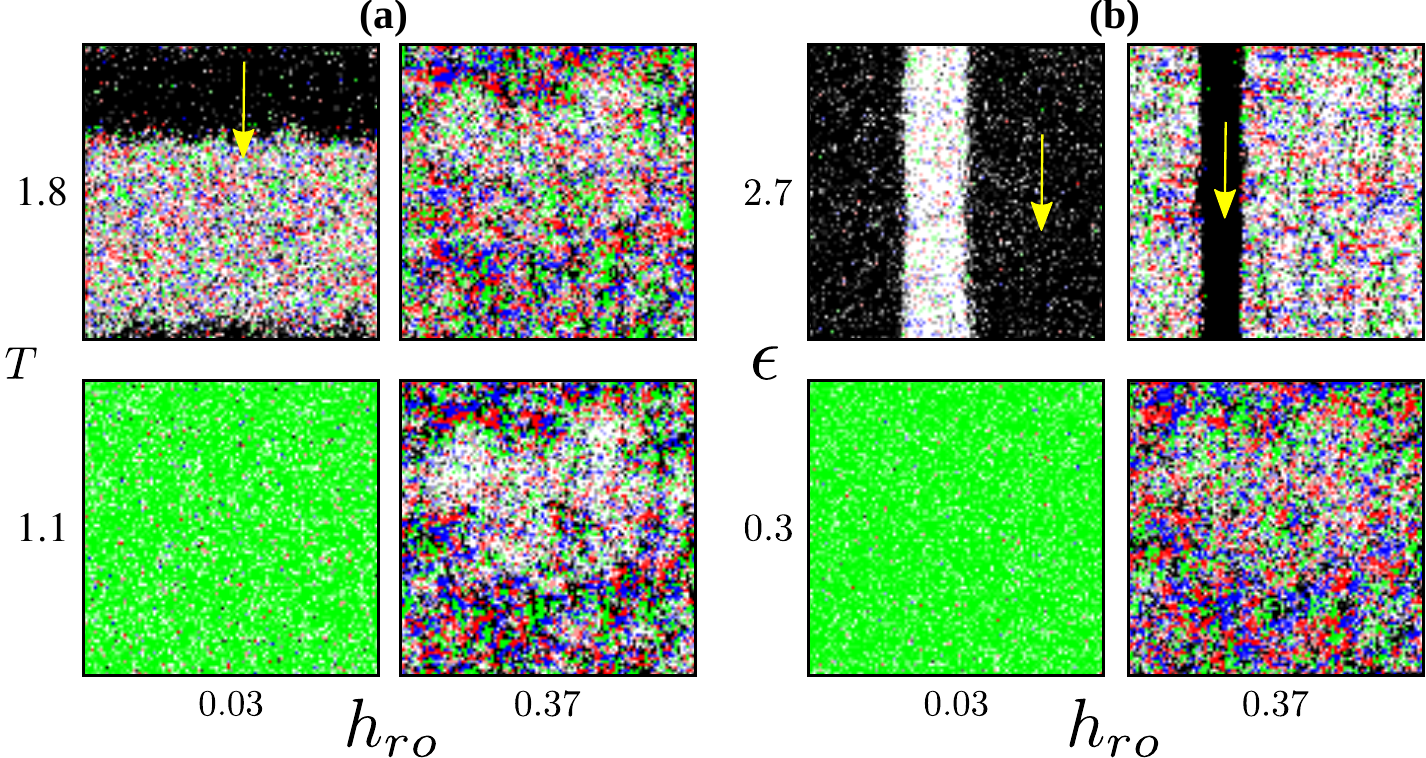}
\caption{\label{H_vs_B-E_Rand_Snap}(Color online) Steady-state snapshots of a four-state RFAPM subjected to a homogeneous but random directional field. (a) Snapshots in the $h_{ro}-T$ plane. Control parameters: $\rho_0=4$, and $\epsilon=0.9$. (b) illustrates how the system responds to an increasing $h_{ro}$ for varying self-propulsion. Control parameters: $T=1$, and $\rho_0=4$. Both (a, and b) demonstrate how the external field destroys the system's order and leads to a disordered phase.}
\end{figure}

Equilibrium snapshots for different temperatures and field strength are represented in Fig.~\ref{H_vs_B-E_Rand_Snap}. In Fig.~\ref{H_vs_B-E_Rand_Snap} (a) at a small field strength ($h_{ro}=0.03$), the system transforms from an ordered liquid phase to a coexistence phase as the temperature is increased. The system becomes disordered at both temperatures when the field strength increases ($h_{ro}=0.37$). This is due to the second term of Eq.~\ref{delH3}, which favors particle alignment along the field direction. However, due to the randomness in the direction of the local field, large domains cannot form. The system is, therefore, more likely to form a disordered gas phase. It is worth noting that a large field switches particle states along the direction of the local field, and changes in the particles' biased direction cause the formation of smaller domains. 
 
Fig.~\ref{H_vs_B-E_Rand_Snap}(b) illustrates how the interplay between the external field and self-propulsion velocity affects the collective state of the particles at a fixed temperature and density mentioned in the caption. At small field ($h_{ro}=0.03$), both at small and large self-propulsion velocities ($\epsilon=0.3, 2.7$) lead to an ordered phase. In contrast, at a large field ($h_{ro}=0.37$), the width of the order phase is decreased, resulting in a disordered phase at small $\epsilon$ and a very thin lane at large $\epsilon$. A stronger field is required to destroy the large domains or longitudinal bands for systems with higher self-propulsion velocity. This behavior can be attributed to the fact that a large self-propulsion velocity helps particles persist along their biased directions, making it more difficult for the external field to disorder the system completely.

\begin{figure}[!htbp]
\centering
\includegraphics[width=0.8\columnwidth]{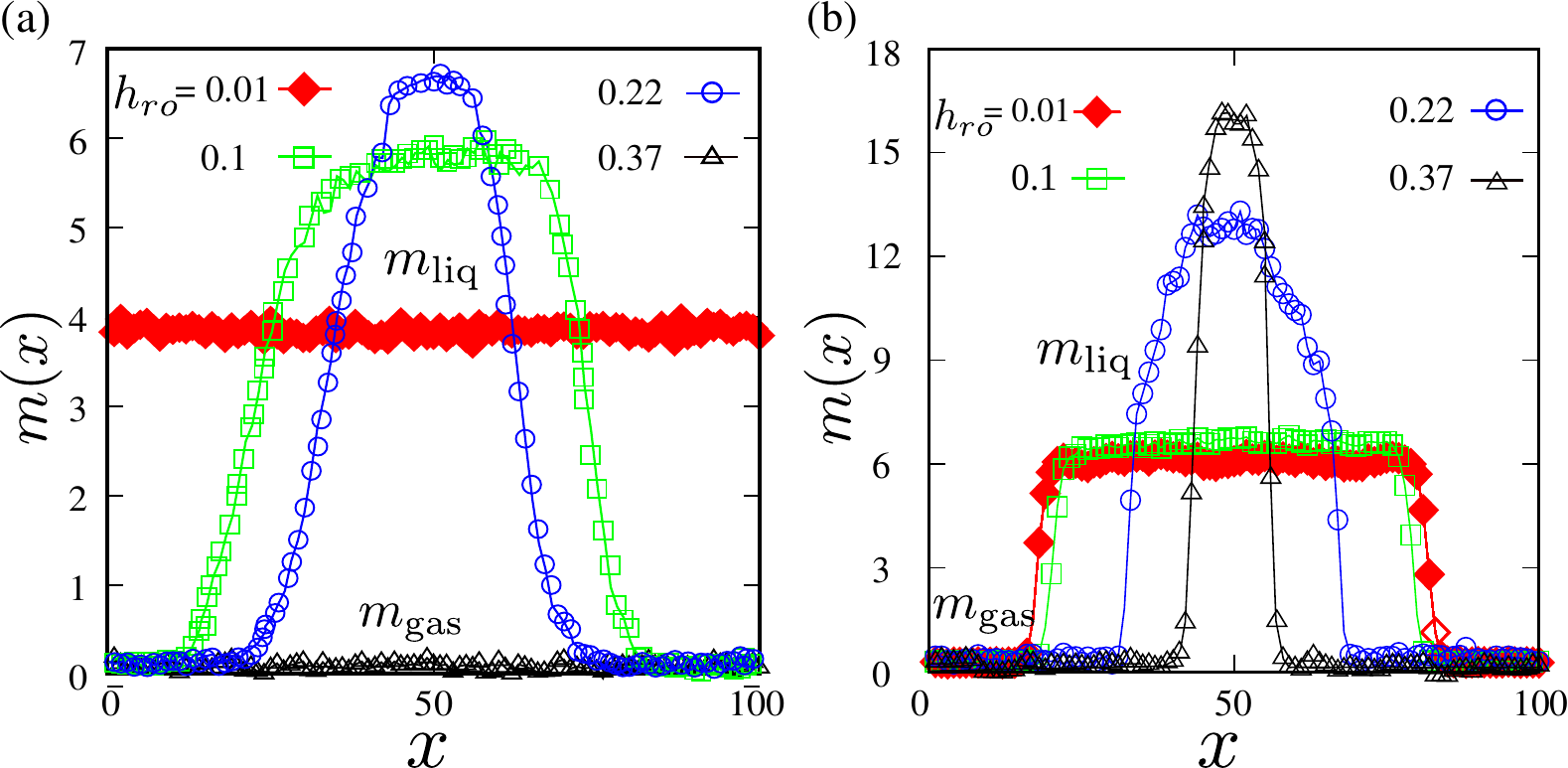}
\caption{\label{H_vs_E_Rand_magprof}(Color online) (a) Magnetization profiles for different $h_{ro}$ at lower self-propulsion velocity ($\epsilon=0.9$). (b) Magnetization profiles for different $h_{ro}$ at larger self-propulsion velocity ($\epsilon=2.7$). Control parameters: $\rho_0=4$, $T=1$.}
\end{figure}

Time-averaged magnetization profiles as a function of field strength 
both at small and large self-propulsion velocity are shown in  Fig.\ref{H_vs_E_Rand_magprof}(a) and (b), respectively.
At low field strength ($h_{ro}=0.01$), average magnetization ($\langle m_i^{\sigma=1}\neq0\rangle$) closely matches the average density, suggesting the presence of a polar liquid phase. Increase in field strength ($h_{ro}=0.1$ and $h_{ro}=0.22$), the width of the profiles is decreased with an increase in the difference between $m_{\rm liq}$ and $m_{\rm gas}$, suggesting a transition from a liquid to coexisting phase to a disordered phase with an average local magnetization of $\langle m_i^{\sigma=1}\sim0\rangle$ at $h_{ro}=0.37$. Fig.\ref{H_vs_E_Rand_magprof}(b) suggests that the width of the polar liquid band decreases with increase in field strength ($h_{ro}$). This also demonstrates the phase transition from the liquid to the co-existence phase to the gas phase for constant thermal noise and increasing field strength.

Next, we show the phase diagram on $T,h_{ro}$ plane at a fixed self-propulsion velocity and phase diagram on $\epsilon,h_{ro}$ plane at a fixed thermal noise in 
\begin{figure}[!htbp]
\centering
\includegraphics[width=0.8\columnwidth]{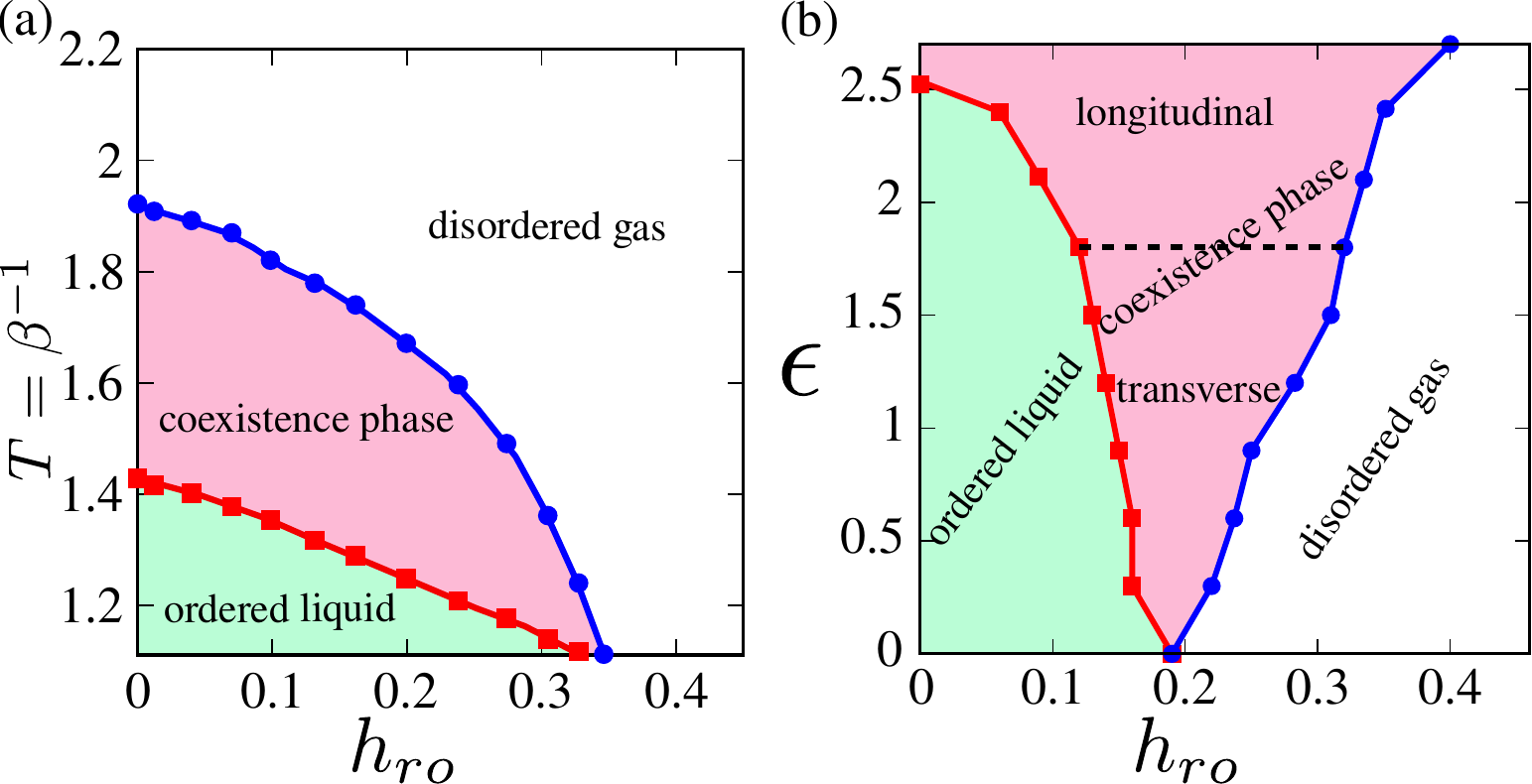}
\caption{\label{H_vs_Beta_Eps-Rand}(Color online) Phase diagrams of the RFAPM for homogeneous \& random directional field at average density $\rho_0=4$. (a) $T-h_{ro}$ phase diagram for fixed self-propulsion velocity ($\epsilon=0.9$). (b) $\epsilon-h_{ro}$ phase diagram for fixed thermal noise ($T=1$). Reorientation transition in the coexistence region is denoted by a dotted line at $\epsilon=1.8$, unaffected by the change in $h_{ro}$.}
\end{figure}
Fig.~\ref{H_vs_Beta_Eps-Rand}(a-b). We noticed a substantial difference in the results between fixed $h$ and randomly oriented $h_{ro}$. At a small field (Fig.~\ref{H_vs_Beta_Eps-Rand}(a)), the disordered gas phase passes through a coexistence phase to an ordered liquid phase as the temperature is lowered. At small thermal noise, the liquid phase transforms into a disordered gas at a field ($h_{ro}^*\simeq 0.34$), passing briefly through a coexistence phase. In Fig.~\ref{H_vs_Beta_Eps-Rand}(b), we see that at a moderate $\epsilon$, an ordered liquid makes a transition into the coexistence phase and then to the disordered gas with increasing $h_{ro}$. In the limit of vanishing $\epsilon$, the coexistence phase disappears, and a direct transition between ordered liquid and disordered gas is observed at $h_{ro}^*=0.2$.

\begin{figure}[!htbp]
\centering
\includegraphics[width=0.8\columnwidth]{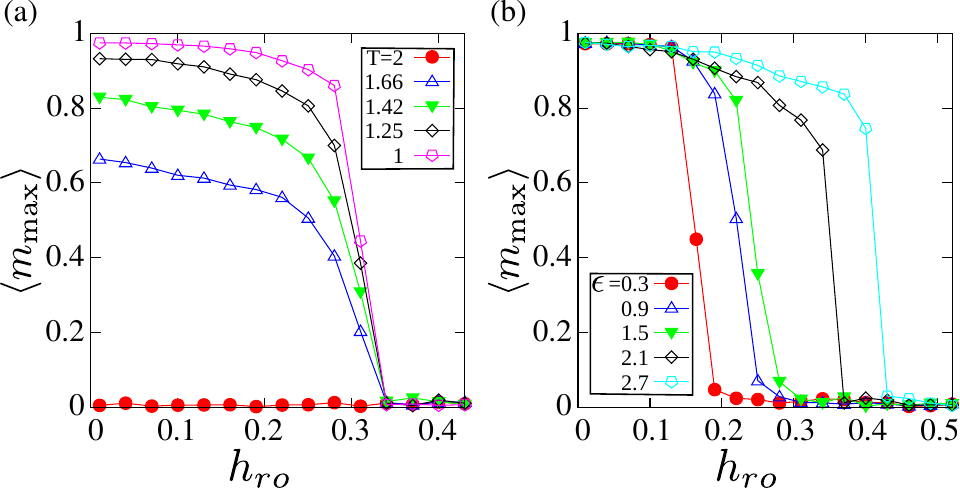}
\caption{\label{OP-RFDAPM}(Color online) (a) $m_{\rm max}$ vs $h_{ro}$ is plotted for different thermal noise. Control parameters: $\rho_0=4$, and $\epsilon=0.9$. (b) $m_{\rm max}$ vs $h_{ro}$ is plotted for different self-propulsion velocity. Control parameters: $\rho_0=4$, and $T=1$.}
\end{figure}

In Fig.~\ref{OP-RFDAPM}(a-b), the time-averaged maximal order parameter ($<m_{\rm max}>$) is plotted as a function of the local field strength ($h_{ro}$), at
different thermal noise and self-propulsion velocity, respectively. In Fig.~\ref{OP-RFDAPM}(a), as the local field strength increases, the order parameter decreases and approaches zero, indicating a first-order transition from an ordered phase to a disordered phase at $h_{ro}^*\sim0.33$. However, a transition from an ordered phase to a disordered phase happens at a larger field for large self-propulsion velocity in Fig.~\ref{OP-RFDAPM}(b).

\section{Random Diffusion APM}
\subsection{\label{Mod_RB}Modelling \& Simulation}
Next, we present the Random Diffusion Active Potts Model (RDAPM) where the corresponding Hamiltonian is the same as described in~\cite{chatterjee2020flocking,mangeat2020flocking}, and particles hop according to the Eq.~\ref{Whop}. However, diffusion constant $D_{ij}$ takes binary values of 0 or 1, indicating the diffusivity between neighboring sites $i$ and $j$. When $D_{ij}$ is equal to 0, then the system is passive (independent on $\epsilon$). To define $D_{ij}$, we use a random diffusion probability denoted as $P_{rd}$, where $D_{ij}=0$ with probability $P_{rd}$ and $D_{ij}=1$ with probability $(1-P_{rd})$. A schematic representation of RDAPM is shown in Fig.~\ref{fig:Model_fig}(c).

In this study, the Monte Carlo simulation begins by randomly setting the diffusivity strength ($D_{ij}$) to 0 with probability $P_{rd}$ for a square lattice with PBC of dimension $L$. Afterward, particles of different states were randomly distributed on the sites to create a high-temperature disordered configuration before the quench. The particles could flip their spin state with probability $W_{\rm flip}\Delta t$ or hop to one of the nearby sites with probability $W_{\rm hop}\Delta t$. The time step $\Delta t$ is defined as:
\begin{equation}
\Delta t=[q+\exp(q\beta J]^{-1},
\end{equation}
where $\beta$ is the inverse temperature and $J$ is the coupling constant.

\subsection{\label{Res_RB}Numerical results}
\begin{figure}[!htbp]
\centering
\includegraphics[width=0.8\columnwidth]{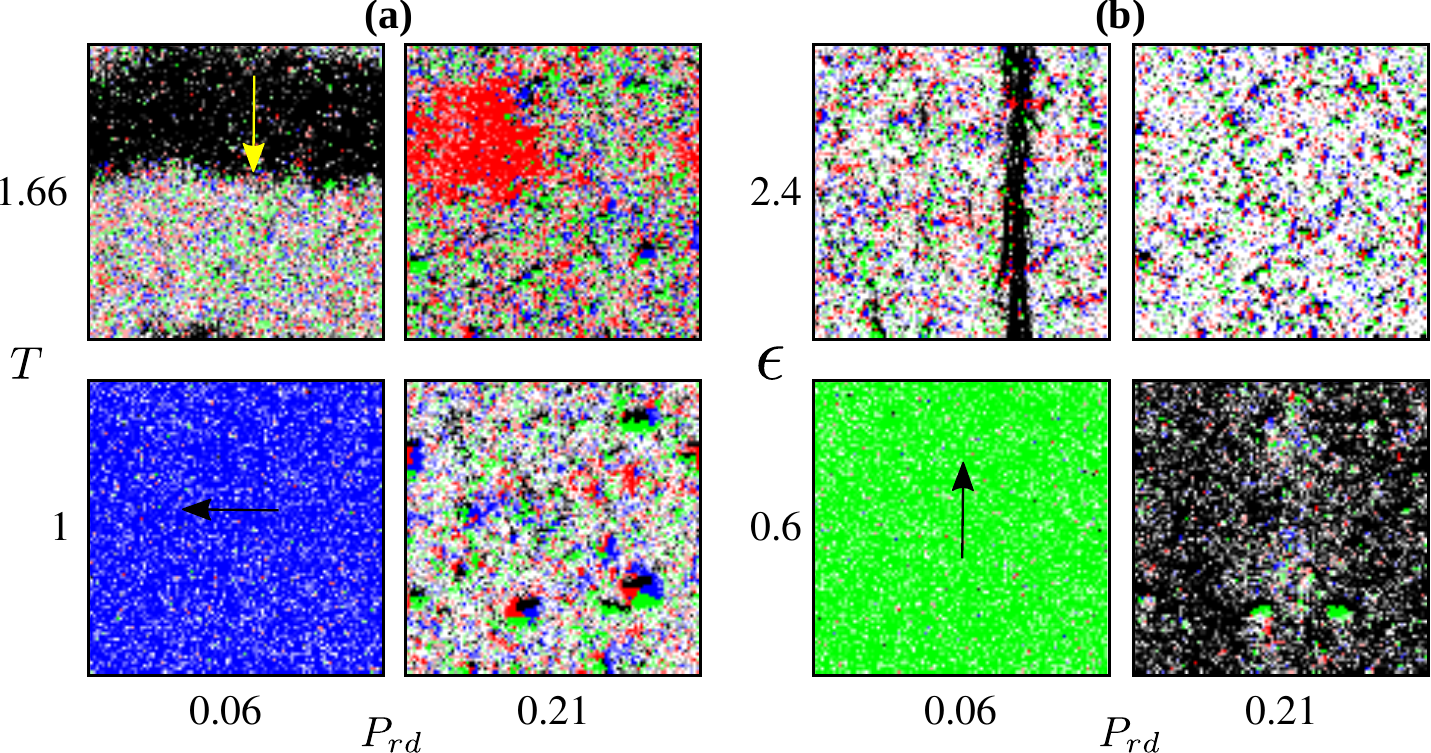}
\caption{\label{RBAPM_snap}(Color online) Steady-state snapshots of a four-state RDAPM. (a) Snapshots in the $P_{rd}-T$ plane. Control parameters: $\rho_0=4$, and $\epsilon=0.9$. (b) illustrates how the system responds to an increasing $P_{rd}$ value for various self-propulsion velocities. Control parameters: $T=1$, and $\rho_0=4$.}
\end{figure}

Steady-state snapshots in Fig.~\ref{RBAPM_snap}(a-b) illustrate the system's behavior with thermal noise and self-propulsion velocity under different random diffusion probabilities respectively. For small random diffusion probability ($P_{rd}=0.06$), an ordered liquid phase transforms into a coexistence phase as the temperature increases. However, at larger random diffusion probability ($P_{rd}=0.21$), the system becomes disordered at both temperatures. Larger random diffusion probability results in most sites having zero $D_{ij}$, hindering particle hopping and collective motion and leading to the presence of the disordered phase. Moreover, larger $P_{rd}$ gives rise to small local clusters at certain isolated sites where particles become trapped.
At small $P_{rd}=0.06$ and $\epsilon=0.6$, Fig.~\ref{RBAPM_snap}(b) exhibits a liquid phase. However, increasing self-propulsion velocity causes particles to move predominantly along their biased directions. As a result, particles tend to trap into those sites and can't hop further. Thus, we get a thinner longitudinal band even with small $P_{rd}$. Conversely, at larger $P_{rd}=0.21$, a mixture of liquid and local small cluster phases is observed for small $\epsilon$. Further increasing $\epsilon$ transforms these phases into a disordered gas phase.

\begin{figure*}[!htbp]
\centering
\includegraphics[width=\columnwidth]{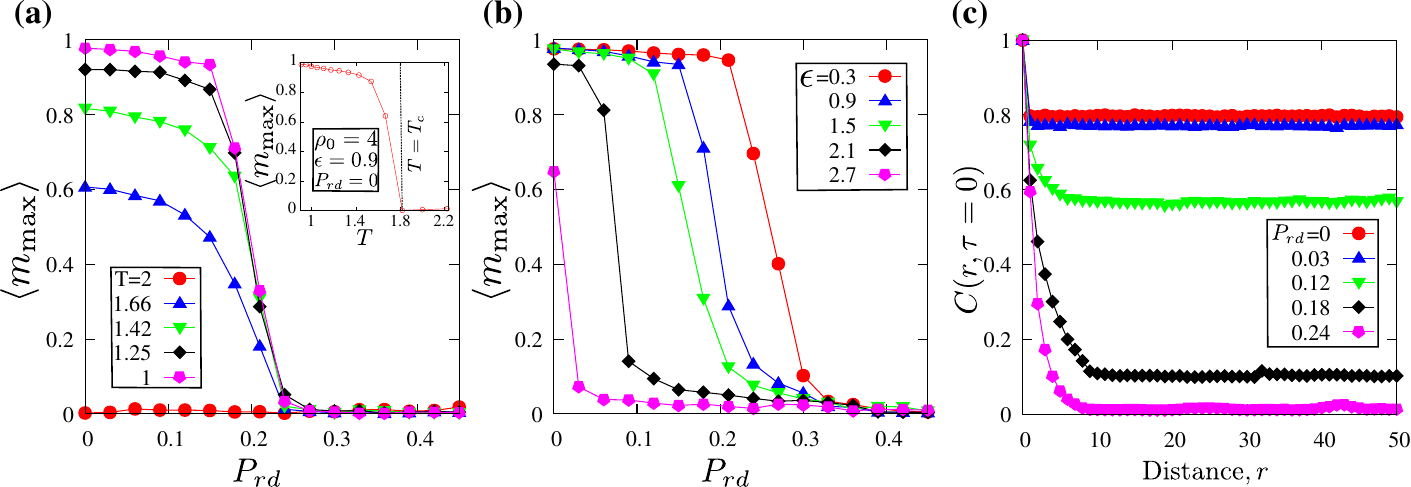}
\caption{\label{OP-Corr-RBAPM}(Color online) (a) Order parameter ($m_{\rm max}$) as a function of $P_{rd}$ is plotted for different temperature. Control parameters: $\rho_0=4$, and $\epsilon=0.9$. Inset shows the variation of the order parameter against temperature at $P_{rd}=0$, exhibiting a sudden jump to 0 at $T_c=1.8$. (b) $m_{\rm max}$ vs $P_{rd}$ for different self-propulsion. Control parameters: $\rho_0=4$, and $T=1$. (c) Equal-time spatial correlation function $C(r)$ is plotted for varying $P_{rd}$. Control parameters: $T=1$, $\epsilon=0.9$, and $\rho_0=4$. As $P_{rd}$ approaches the critical random diffusion probability $P_{rd}^*=0.21$, the system approaches a disordered phase and loses LRO.}
\end{figure*}
In Fig.~\ref{OP-Corr-RBAPM}(a), we show the order parameter, $m_{\rm max}$, as a function of the random diffusion probability, $P_{rd}$, at various thermal noise. As $P_{rd}$ increases, a critical value of $P_{rd}^*=0.21\pm0.02$ triggers a sudden jump, indicating a first-order transition to the disordered phase. The inset figure demonstrates the order parameter variation against temperature at $P_{rd}=0$, revealing a sudden jump to zero at the critical temperature, $T^*_c=1.8$, indicating a first-order phase transition from ordered to disordered phase. Figure.~\ref{OP-Corr-RBAPM}(b) shows behaviour of $m_{\rm max}$ against $P_{rd}$ for different $\epsilon$ values. Lower self-propulsion values cause the transition of $m_{\rm max}$ from 1 to 0 at higher $P_{rd}$. Conversely, higher $\epsilon$ values lead to this transition at smaller $P_{rd}$. This observation suggests that even at high self-propulsion velocities, a small proportion of sites with zero diffusive strength with other neighbor sites greatly impacts collective motion.

\subsubsection{\label{corr_func}Correlation function calculation:}
In Fig.~\ref{OP-Corr-RBAPM}(c), we calculated equal time spatial spin-spin correlation function $C(r)$. The equal-time spatial spin-spin correlation function measures the degree of correlation between the spin orientations of particles at different positions in the system. It is defined as;
\begin{equation}
    C(r) = \frac{1}{N} \sum_{i,j} \langle \delta_{\sigma_i,\sigma_j} \rangle
\end{equation}
where $N$ is the total number of particles in the system, $r$ is the distance between two particles $i$ and $j$, $\delta_{\sigma_i,\sigma_j}$ is the Kronecker delta function which takes a value of 1 if $\sigma_i$ and $\sigma_j$ are the same and 0 otherwise, and $\langle \delta_{\sigma_i,\sigma_j} \rangle$ is the ensemble average of $\delta_{\sigma_i,\sigma_j}$ over many realizations of the system. The decay of $C(r)$ with increasing random diffusion probability, $P_{rd}$, indicates a decrease in the spatial correlation between spins, and as $P_{rd}$ approaches the critical random diffusion probability, $P_{rd}^*=0.21\pm 0.02$, the system approaches a state of randomness, and LRO is lost. This result supports the idea that random diffusion leads to the loss of spatial correlation and the formation of a disordered phase.

\begin{figure}[!htbp]
\centering
\includegraphics[width=0.8\columnwidth]{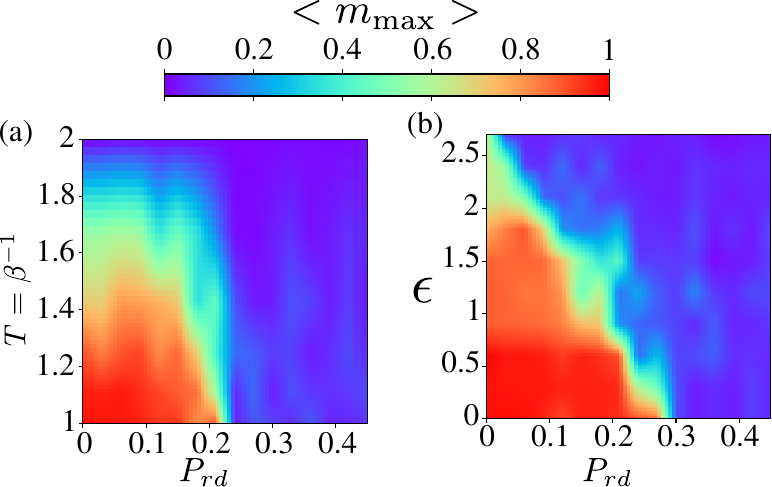}
\caption{\label{Pc_vs_T-E_RBAPM}(Color online) Phase diagrams of the RDAPM. (a) $T$ vs $P_{rd}$ phase diagram. Control parameters: $\rho_0=4$ and $\epsilon=0.9$. (b) $\epsilon$ vs $P_{rd}$ phase diagram. Control parameters: $\rho_0=4$, and $T=1$. The color bar represents the maximal magnetization ranging from 0 to 1, indicating a disorder-to-order phase transition.}
\end{figure}
In Fig.~\ref{Pc_vs_T-E_RBAPM}(a-b), we present the phase diagrams in the $P_{rd}-T$ and $P_{rd}-\epsilon$ planes, respectively. In Fig.~\ref{Pc_vs_T-E_RBAPM}(a), at small values of $P_{rd}$ and $T$, the system exhibits a highly ordered phase characterized by a high maximal magnetization. As we increase the value of $P_{rd}$ at fixed thermal noise, a transition occurs from an ordered to a disordered phase, decreasing the maximal magnetization. In Fig.~\ref{Pc_vs_T-E_RBAPM}(b), the system resides in an ordered state with a high maximal magnetization for small values of self-propulsion velocity. However, as we increase the random diffusion probability, a transition emerges, resulting in a disordered gas phase. At purely diffusive limit ($\epsilon=0$), critical $P^*_{rd}$ is $\sim0.21 \pm 0.02$ obtained.


\section{Summary and Discussion}\label{s4_rfapm}
In this study, we investigated the behavior of self-propelled particles (SPPs) in disordered media using the $q$-state APM~\cite{mangeat2020flocking,chatterjee2020flocking}. We study two models, the Random Field Active Potts Model (RFAPM) and the Random Diffusion Active Potts Model (RDAPM), to understand the impact of the disorder on the behavior of SPPs. Interestingly, a unique feature of the APM is the movement of a longitudinal band opposite to a small unidirectionally applied field. We found that this movement is a treadmilling behavior~\cite{wegner1976head,small1995getting} driven by non-biased diffusion. Despite the movement of the longitudinal band at a small field, the self-propulsion velocity remains constant during the transition in the coexistence regime. The study reveals that the coexistence band expands and transforms into a fully liquid state at higher field strengths with an increase in the constant unidirectional local field. Intriguingly, particles in the liquid state align themselves with the field direction, even in the presence of substantial thermal fluctuations. On the other hand, a flocking to MIPS transition can also be seen in the case of a bidirectional field. Conversely, when local field orientations are randomly distributed, the system transitions from a polar liquid to a disordered gas phase as the field strength increases, even at lower temperatures. This emphasizes the crucial role of field orientation in dictating the system's behavior. The impact of decreasing interaction strength between neighboring sites results in weakened and less probable particle hopping. This reduction in spatial correlation leads to a loss of LRO and culminates in a disordered phase. The critical probability for random diffusion, denoted as $P_{rd}^* \approx 0.21\pm 0.02$, is a pivotal indicator of the transition from the ordered to the disordered phase. Below this threshold, the system maintains a degree of LRO, while beyond it, a complete loss of LRO occurs, marking the transition into a fully disordered state.

Understanding the behavior of SPPs in disordered environments is relevant to biological systems such as bacterial colonies or cell migration in tissues, where the presence of obstacles and heterogeneity can impact the collective behavior of cells~\cite{muddana2010substrate}. In conclusion, this study not only uncovers striking phenomena in flocking models but also provides valuable insights into the role of field orientation, interaction strength, and random diffusion in shaping the collective behavior of particles. The findings contribute significantly to our understanding of SPPs' behavior in complex environments. Future directions for this work could include exploring the impact of different types of self-propulsion, such as persistent random walks or active Brownian motion, which further insights into the collective behavior of SPPs in disordered media.

\section{Auxiliary material}
\subsection{Mean-squared displacements (MSD) of particles}\label{appendix_msd_rfapm}
\begin{figure*}[!htbp]
\centering
\includegraphics[width=\columnwidth]{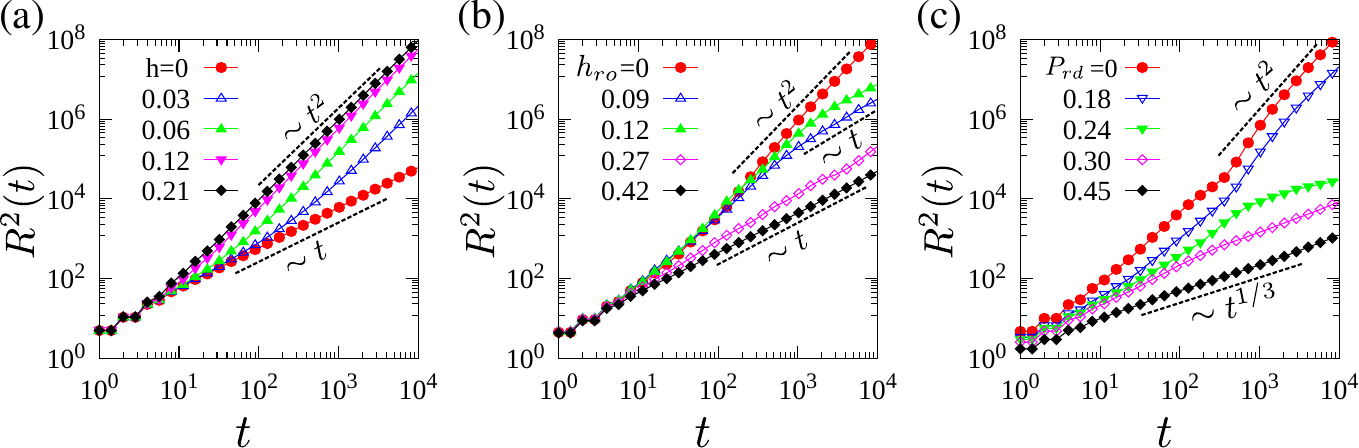}
\caption{(Color online) Average MSD of $N$ particles versus $t$ (on a log-log scale) at fixed control parameters: $\rho_0=3$, $\epsilon=0.9$, and $T=2$ in (a), $\rho_0=4$, $\epsilon=0.9$, and $T=1$ in (b), and (c) respectively.}
\label{MSD-RF-RB} 
\end{figure*}
Here, we explore the appearance of arrested states through measurements of the MSD of individual particles. The MSD of $N$ number of particles in the system at time $t$ (which quantifies how the particles move from their initial positions under various volume exclusion effects) is defined as
\begin{align}
R^2(t)=\frac{1}{N}\sum_{i=1}^N |\mathbf{r}_i(t)-\mathbf{r}_i(0)|^2 \, ,
\end{align}
where $\mathbf{r}_i(t)$ is the instantaneous position of the $i$-th particle at time $t$. For ballistic motion, $R^2 \sim t^2$ while for diffusive motion $R^2 \sim t$. For the suppressed state, however, MSD is proportional to $t^x$ where $x < 1$ (sub-diffusive)~\cite{merrigan2020arrested}.

In Fig.~\ref{MSD-RF-RB}(a), we show $R^2$ versus $t$ (on a log-log scale) for RFAPM as a function of $h$. At small $h$, which physically signifies the gaseous phase at large temperature, the system obeys the diffusive growth $R^2 \sim t$ whereas, for larger $h$, where the system exhibits the liquid phase, we observe two distinct regimes in the MSD. The small $h$ limit is characterized by a diffusive regime with $R^2 \sim t$ growth, whereas a ballistic growth regime characterized by the power-law $R^2 \sim t^2$ is observed at large $h$. In the liquid phase, the external field ($h$) plays a crucial role as the system exhibits the ballistic growth regime, which is a collision-free regime in which particles travel freely after most of the particles switch in the same state. In the case of a randomly oriented field, at small $h_{ro}$ and low temperature, the system exhibits a late-time ballistic growth regime characterized by the power-law $R^2 \sim t^2$ shown in Fig.~\ref{MSD-RF-RB}(b). However, as $h_{ro}$ increases, a crossover to a diffusive regime with $R^2 \sim t$ growth is observed. In the case of RDAPM in Fig.~\ref{MSD-RF-RB}(c), at the small limit of random diffusion probability $P_{rd}$ due to low temperature, a late time ballistic regime with $R^2 \sim t^2$ growth is observed. However, as $P_{rd}$ increases, we observe a suppressed state characterized by $R^2 \sim t^{1/3}$, which signifies very strong repulsion as particles are less likely to hop due to the broken bonds.

\chapter{Consequence of anisotropy on flocking: the Discretized Vicsek Model (DVM)}\label{chap:Chap4}
We numerically study a discretized Vicsek model (DVM) with particles orienting in $q$ possible orientations in two dimensions. The study probes the significance of anisotropic orientation and microscopic interaction on the macroscopic behavior. The DVM is an off-lattice flocking model like the active clock model [ACM; EPL {\bf 138}, 41001 (2022)] but the dynamical rules of particle alignment and movement are inspired by the prototypical Vicsek model (VM). The DVM shows qualitatively similar properties as the ACM for intermediate noise strength where a transition from macrophase to microphase separation of the coexistence region is observed as $q$ is increased. But for small $q$ and noise strength, the liquid phase appearing in the ACM at low temperatures is replaced in the DVM by a configuration of multiple clusters with different polarization which does not exhibit any long-range order. We find that the dynamical rules have a profound influence on the overarching features of the flocking phase. We further identify the metastability of the ordered liquid phase subjected to a perturbation.


\section{Introduction}
\label{introduction}
Active matter systems represent a fascinating class of materials composed of self-propelled entities that convert energy into mechanical motion, giving rise to complex and often out-of-equilibrium behaviors~\cite{ramaswamy2010mechanics,marchetti2013hydrodynamics,gompper20202020,needleman2017active}. The emergent phenomena in active matter systems has gained significant attention in recent years due to their potential applications in various physical, biological, and engineering systems~\cite{marchetti2013hydrodynamics,vernerey2019biological,ghosh2021enzymes}. Active matter exhibits dynamic behaviors like collective motion~\cite{marchetti2013hydrodynamics}, pattern formation~\cite{bar2020self,shaebani2020computational}, and even the ability to exhibit controlled transport~\cite{sanchez2012spontaneous}. These systems encompass a wide range of physical, chemical, and biological entities, from swimming bacteria~\cite{peruani2012collective}, mammalian herds~\cite{garcimartin2015flow}, fish schools~\cite{becco2006experimental,calovi2014swarming}, and sterling flocks to amoeba and bacteria colonies~\cite{steager2008dynamics}, to the cooperative behavior of cytoskeletal filaments and molecular motors in living cells~\cite{schaller2010polar,sumino2012large,sanchez2012spontaneous} or in vitro environments to synthetic colloidal particles equipped with motors~\cite{veigel2011moving,wong2016synthetic}. To understand and unravel the fundamental principles governing active matter systems, new models~\cite{shaebani2020computational} have been developed in the last two decades. 

The Vicsek model (VM), introduced by Vicsek and collaborators in 1995~\cite{vicsek1995novel}, provides a fundamental framework for studying the collective behavior of particles under aligning interactions. In this model, particles adjust their velocities to align with the average velocities of neighboring particles, leading to the emergence of coherent motion and ordered patterns. The VM has played a crucial role in advancing our understanding of flocking dynamics~\cite{toner1995long,toner1998flocks,toner2012reanalysis,solon2015phase}. At low particle density or high noise, particles move in random directions, and no long-range order is observed. The transition from the gas phase at high noise and low density to the polar ordered Toner-Tu phase at low noise and high density, displaying long-range order (LRO) by a coherent motion of all particles, is first order~\cite{chate2008collective}. But, in contrast to conventional first-order phase transition scenarios, the coexistence phase of the VM can manifest as either multiple bands of particles moving collectively, a phenomenon known as microphase separation~\cite{solon2015phase,solon2015pattern}, or a polar-ordered cross sea phase~\cite{kursten2020dry}, primarily driven by giant number fluctuations (GNF)~\cite{solon2015phase}.

Nonetheless, it is important to note that the VM posits a continuous range of possible directions for the motion of particles. However, when considering a scenario in which particles are constrained to discrete, equidistant angular orientations within a two-dimensional plane, such as in the active clock model (ACM)~\cite{solon2022susceptibility,chatterjee2022polar}, the VM-inspired dynamical principles governing particle alignment and movement remain uncharted territory. In a recent study on the ACM~\cite{solon2022susceptibility}, it was revealed that in large systems, any values of discrete orientations result in significant changes in phenomenology when compared to the VM. These changes include the loss of long-range correlations, the pinning of global order, and the transformation of coexistence bands into a single moving domain. Additionally, another study on the ACM~\cite{chatterjee2022polar} with different dynamical rules shows that for a small number of directions, the ACM mirrors the active Potts model (APM)~\cite{chatterjee2020flocking,mangeat2020flocking}, exhibiting macrophase separation and reorientation transition. Conversely, with more directions, the ACM transitions towards the VM, displaying microphase separation and transverse bands without reorientation. Remarkably, the transition in the $q\to\infty$ limit of the ACM~\cite{chatterjee2022polar}, known as the active XY model, shares the same universality class as the VM. Motivated by these findings, in this chapter, we undertake an extensive computational investigation that examines in detail a $q$-state discretized version of the Vicsek model (DVM) where the rules of particle alignment and movement follow the Vicsek protocol.
 
We ask several intriguing questions that persist within the context of the DVM, e.g., (a) How does discretizing the directions of particles in the VM, affect the overall diverse collective dynamics and steady-state phases? (b) What is the impact of $q$ and system size on the coexistence region (micro- or macro-phase separation)? (c) How do the behavior of the density fluctuations, the direction of the system's global order, and the behavior of correlation functions in the liquid phase correspond with the self-organized patterns in the phase-coexistence region? (d) What is the nature of the DVM liquid phase as a function of $q$? To answer these questions, we study the DVM in an off-lattice domain, focusing on the three key factors: the anisotropy and degeneracy parameter $q$, noise level $\eta$, and system size. 

The chapter is organized as follows: after introducing the model in Sec.~\ref{model}, we present our numerical results in Sec.~\ref{result}. Finally, we summarize and discuss the implications of
our findings in Sec.~\ref{discussion}.

\section{Model}
\label{model}
We consider $N$ self-propelled particles within a two-dimensional off-lattice domain of size $L_x \times L_y$ ($L_x > L_y$ for rectangular domain and $L_x=L_y=L$ for square domain) with periodic boundary conditions. Akin to the two-dimensional  VM, each point particle $i$ is endowed with an off-lattice position vector $\bm{r}_i=(x_i,y_i)$ and moves with a constant speed $v_0$ in individual directions given by a unit orientation vector ${\bm \sigma}_i=(\cos \theta_i,\sin \theta_i)$ with an orientation angle $\theta_i \in (0,2\pi)$ where 
\begin{equation}
\label{discrete}
\theta_i = \frac{2\pi n_i}{q} \, ,
\end{equation}
and $n_i = \{0, 1, 2, \cdots, (q-1)\}$ denote discrete orientations of the particles. $q$ denotes the ground state degeneracy where each particle can only have discrete orientations allowed by the $q$ value and therefore, the continuous $U(1)$ symmetry of the VM is replaced by the discrete $Z_q$ symmetry.  

At each discrete time step $\Delta t=1$, a particle $i$ moves a fixed distance $v_0 \Delta t$ and interacts with $\mathcal{N}_i$ neighboring particles within a circle of unit radius around it. The position evolves in the following way:
\begin{equation}
\label{r}
\bm{r}_i^{t+\Delta t}=\bm{r}_i^t+ v_0 \bm{\sigma}_i^{t+\Delta t} \Delta t \, , 
\end{equation} 
while the new orientation is determined by a projection of the updated orientation proposed by the Vicsek rule onto one of the $q$ allowed directions:
\begin{equation}
\theta_i^{t+\Delta t}= \mathbb{P}(\bar{\theta}_i^t + \eta \xi_i^t) \,
,\label{sigma} 
\end{equation} 
where $\mathbb{P}$ is the projection and $\bar{\theta}_i^t$ is the orientation angle of a spin-weighted sum
\begin{equation}
\label{sigma2}
\bar{\bm{\sigma}}_i^t = \sum_{j\in \mathcal{N}_i} \bm{\sigma}_j^t,
\end{equation}
of orientation vectors of neighboring particles. $\xi_i^t\in[-\pi,\pi]$ is a
scalar noise uniformly distributed and uncorrelated for all sites and times. Such noise is often called $white$ since it has a flat Fourier spectrum. $\eta$ is the noise amplitude. 

\begin{figure}[!t]
    \centering
    \includegraphics[width=1\columnwidth]{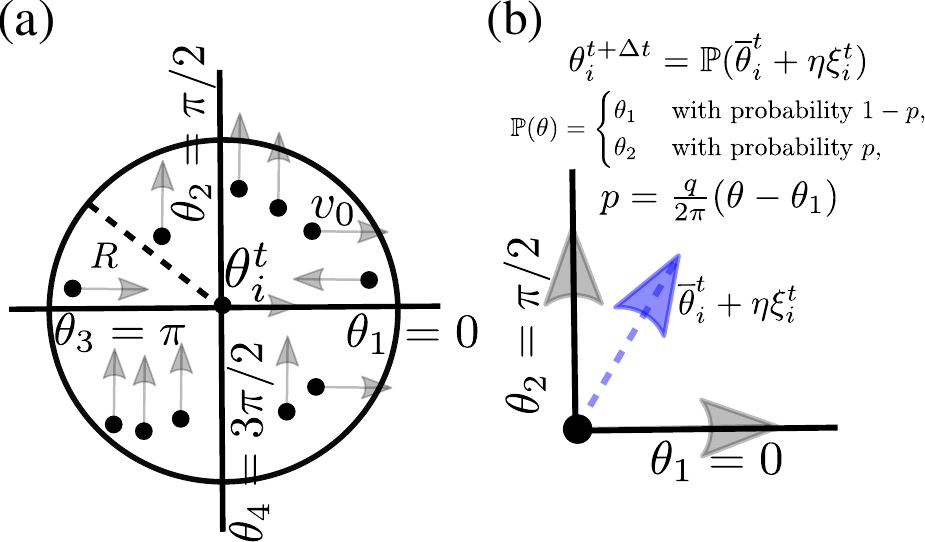}
    \caption{(Color online) Schematic of the DVM for $q=4$. (a) Four allowed orientations for the particle are $0, \pi/2, \pi,$ and $3\pi/2$ with $\theta_i^t=0$. The circular neighborhood (of radius $R$) of particle $i$ contains neighboring particles to calculate $\theta_i^{t+\Delta t}$. (b) The new orientation proposed by the Vicsek rule (blue dotted arrow) will be projected either along $\theta_1=0$ or $\theta_2=\pi/2$ probabilistically.}
    \label{schematic}
\end{figure}

We define the projection onto the allowed directions probabilistically by
\begin{equation}
\mathbb{P}(\theta) =
\begin{cases}
\theta_1 & \text{with probability} \ 1-p \, ,\\
\theta_2 & \text{with probability} \ p \, ,
\end{cases}
\end{equation}

where $\theta_1$ and $\theta_2$ are the two allowed directions which
are closest to $\theta$, such that $\theta_1=2 \pi n/q < \theta$
and $\theta_2=2 \pi (n+1)/q > \theta$ for some $n$. The probability $p\in[0,1]$ is  given by $p=\frac{q}{2\pi}(\theta-\theta_1)$, i.e. minimal (0) for $\theta$ close to $\theta_1$ and maximal (1) for $\theta$ close to $\theta_2$ (c.f. Fig.~\ref{schematic}). Note that then for all particles going into the direction, say, $\theta_1$, the probability to turn stochastically into another direction, say $\theta_2$, is of the order $\sim$ $\frac{q}{2}\eta$, i.e. small for small $q$ and noise $\eta$. 

It should be mentioned here that the dynamical rules governing DVM are different from the dynamical rules of the $q$-state ACM \cite{chatterjee2022polar}. In the DVM, as per Eq.~\eqref{r}, the hopping probability of a particle with orientation angle $\theta$ along another discrete direction $\phi$ is zero whereas the probability of hopping along non-preferred directions is nonzero in the ACM and depends on the self-propulsion velocity of the particle. Therefore, the transverse fluctuations, say in $q=4$ ACM, are stronger than the $q=4$ DVM. In the ACM, transverse fluctuations mainly originate from the nonzero hopping probability of a particle along its non-preferred directions. Here thermal fluctuation, through the inverse temperature $\beta$, also plays an important role as it controls the flipping dynamics. However, in the large $\beta$ limit of the ACM, similar to the small $\eta$ limit of the DVM, the probability that all particles moving in a particular direction will flip their orientation to another direction is also very small. But, unlike DVM, an ACM particle can move along a direction different than its orientation angle. This crucial difference in the hopping dynamics, as we will see, plays an essential role in the steady-state pattern formation of the DVM at low $\eta$ and $q$. 

DVM control parameters are the average particle density $\rho_0 = N/L_xL_y$, noise strength $\eta$, particle velocity $v_0=0.5$ (unless mentioned otherwise), and the measure of anisotropy $q$. According to Eq.~\eqref{discrete} a large $q$ signifies weak anisotropy while a small $q$ signifies strong anisotropy.

We performed numerical simulations of the stochastic process with parallel updates. The initial configuration is prepared homogeneously by assigning random orientations and positions to the particles as defined in  Eq.~\eqref{discrete} and Eq.~\eqref{r}, respectively. After the initialization, we let the system evolve under various control parameters for $t_{\rm eq} \sim 10^5$ to reach the steady state followed by measurements of various quantities until the maximum simulation time $t_{\rm max} \sim 10^6$. 

\section{Numerical Results}\label{result}

\subsection{Collective motion \& phase diagram.} 
\begin{figure}[!t]
    \centering
    \includegraphics[width=\columnwidth]{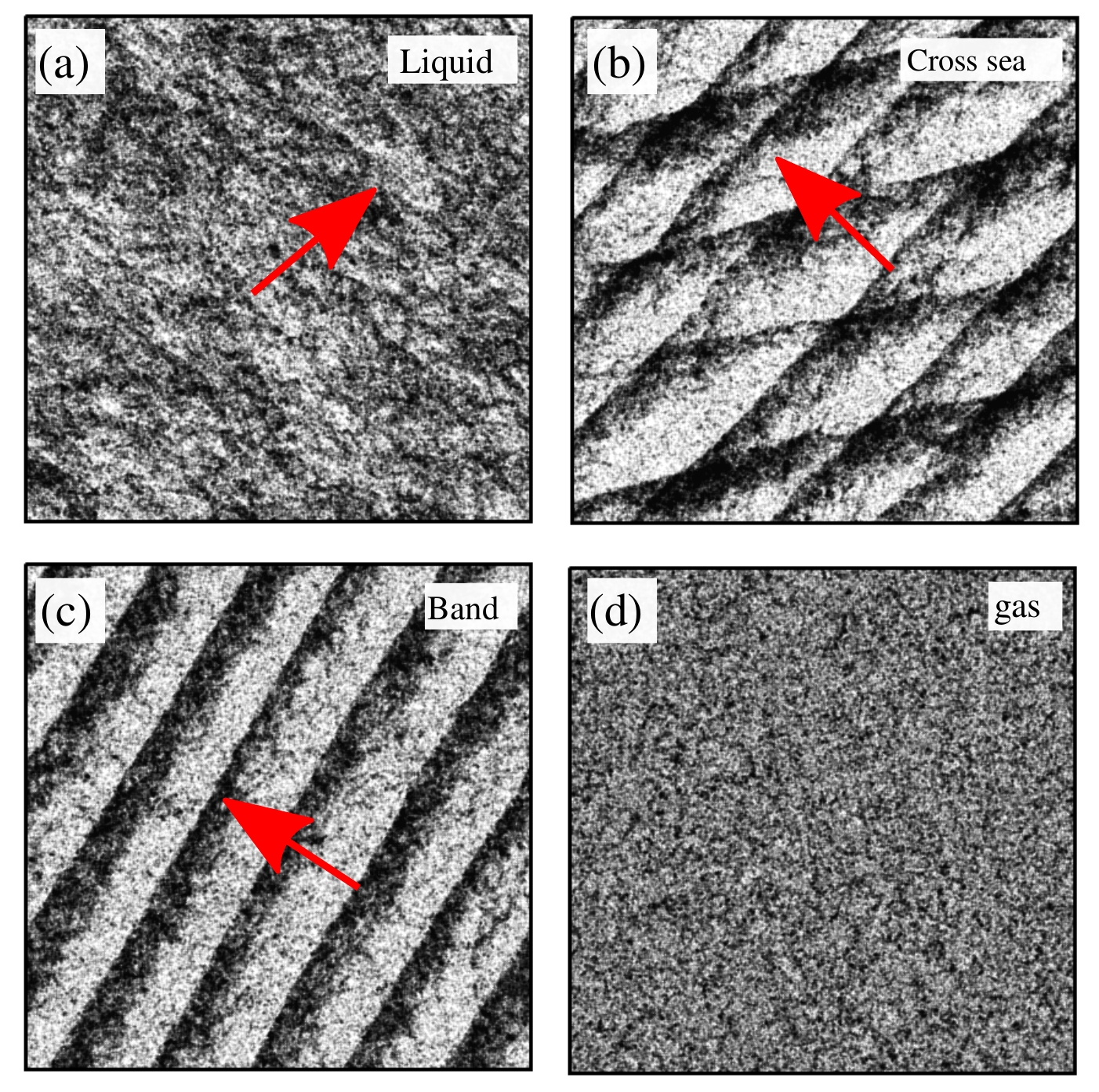}
    \caption{(Color online) Snapshots on a square domain of size $1024^2$ exhibiting the four phases of the DVM for $q=9$. (a) Polar ordered liquid phase ($\eta=0.1$). (b) Cross-sea state ($\eta=0.3$). (c) Band state ($\eta=0.35$). (d) Disordered gas phase ($\eta = 0.5$). Dark color represents high particle density and red arrows indicate the average direction of motion. Parameters: $\rho_0=2$, $v_0=0.5$.}
    \label{TDVM_q9_L1024}
\end{figure}

We present the typical non-equilibrium steady-state configurations of the DVM in Fig.~\ref{TDVM_q9_L1024} for $q=9$ and density $\rho_0=2$. The system exhibits phases of polar ordered liquid (a), liquid-gas coexistence (b--c), and disordered gas (d) as noise strength $\eta$ is increased from 0.1 to 0.5. The phase-coexistence region is characterized by a low-noise cross-sea phase (b) and a high-noise band phase (c). The cross-sea phase has particle density much higher at the crossing points than anywhere else and has recently been reported as the fourth phase of the VM \cite{kursten2020dry}. We notice that such patterns assemble spontaneously without an external drive at certain parameter values. Conversely, the band phase is a collection of high-density bands moving parallelly along a specific direction at a constant speed. Polar flocks [the homogeneous ordered liquid phase shown in Fig.~\ref{TDVM_q9_L1024}(a)] can be observed in a large class of active matter systems and have been considered robust to fluctuations. But recent studies have argued that liquid polar flocks are metastable to the presence of a small obstacle \cite{codina2022small} or to the nucleation of an opposite-phase droplet \cite{benvegnen2023metastability} and triggers counter-propagating dense clusters leading to the reversal of the liquid flow. In light of these observations, in the subsequent part of this chapter, we will investigate the stability of the DVM liquid phase. Fig.~\ref{TDVM_q9_L1024} suggests that the system exists in distinct phases, which we will characterize next.

\begin{figure*}[!t]
    \centering
    \includegraphics[width=\textwidth]{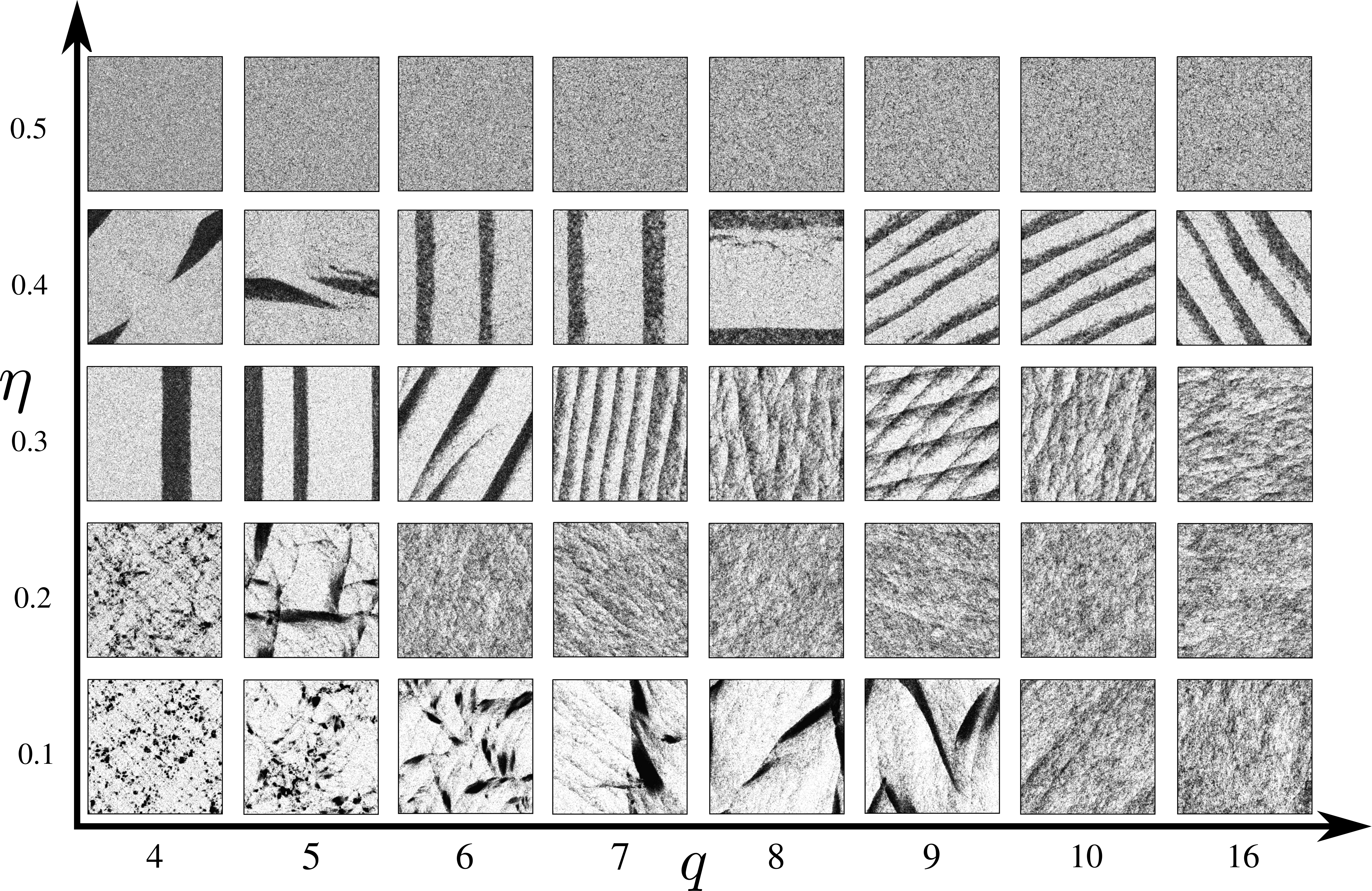}
    \caption{(Color online) $\eta-q$ phase diagram of the DVM illustrated by snapshots on a $1024^2$ domain at time $t = 10^5$. As a function of $\eta$ and $q$, we observe six distinct self-organized patterns: cluster ($\eta=0.1$, $q=4$), macrophase ($\eta=0.3$, $q=4$), microphase ($\eta=0.4$, $q=6 \to 16$), cross-sea ($\eta=0.3$, $q=8 \to 10$), ordered liquid ($\eta=0.2$, $q=6 \to 16$), and disordered gas ($\eta=0.5$, $q=4 \to 16$). Parameters: $\rho_0=2$, $v_0=0.5$.}
    \label{TDVM_L1024}
\end{figure*}

The non-equilibrium steady-state behavior of the DVM is illustrated by representative late-stage snapshots as a function of noise strength $\eta$ and anisotropy parameter $q$ (see Fig.~\ref{TDVM_L1024}). We observe six distinct self-organized patterns in the ($q$, $\eta$) plane which completely describe the DVM. At low noise and $q$, we observe a locally ordered cluster phase. Although each cluster is highly dense and polar, the system as a whole does not possess any net polarization (see Auxiliary material~\ref{appA}). This cluster phase, which has not been observed earlier in any flocking models, appears only for small $q$ and $\eta$ when the system is strongly discretized and the impact of fluctuation is insignificant. 

The appearance of a cluster phase in the $q=4$ DVM for small $(q,\eta)$ and high density instead of a polar ordered phase similar to the 4-state APM \cite{chatterjee2020flocking,mangeat2020flocking} and ACM \cite{solon2022susceptibility,chatterjee2022polar} can be attributed to the absence of transverse fluctuations through hopping. In DVM, similar to the VM, particle movement is solely controlled by the orientation $\theta$ [see Eq.~\eqref{r}]. Therefore, $q=2$ DVM only manifests one-dimensional movement (along the $x$-axis) of high-density clusters of self-propelled particles having orientations $\theta = 0$ and $\theta=\pi$. We observe that these clusters never coalesce due to the lack of transverse fluctuations and thus never form a band or polar liquid phase. Similar observations are made when the constant transverse diffusion is switched off in the two-dimensional active Ising model (AIM) \cite{solon2015flocking}, although the one-dimensional AIM \cite{solon1dAIM} exhibits a flocking state consisting of a single dense ordered aggregate. This signifies that diffusion along the non-preferred hopping directions plays a crucial role in the formation of large liquid domains. For APM and ACM at large $\beta$ and small $q$ and DVM with small $\eta$ and $q$, the probability of transverse flipping is very small as fluctuations are weak. However, the nonzero finite hopping rates along the unbiased directions facilitate the formation of large liquid domains in the APM and ACM whereas an absence of this forms a cluster phase in the DVM. Finally, it is the interplay of $q$ and $\eta$ that determines the fate of the cluster phase. If $q$ is small but fluctuation is large, the rigid cluster phase can relax and form a large ordered domain (see the snapshot for $q=6$ and $\eta=0.2$). On the contrary, if $\eta$ is small but $q$ is large, the weak anisotropy helps the cluster phase merge into a large ordered domain (see the snapshot for $q=10$ and $\eta=0.1$). This also explains why cluster size increases with $q$ for a fixed $\eta$ (see the snapshots for $\eta=0.1$). 

Beyond the cluster phase, for $\eta=0.2$, we observe the emergence of the polar ordered liquid phase ($q \geqslant 6$). For intermediate noise strength ($\eta=0.3-0.4$), a transition from macrophase separation ($q=4$, a single liquid band moving through the gaseous background) to microphase separation (multiple bands moving parallelly or in a cross-sea pattern through the gas phase) in the coexistence region is observed where the number of bands increases with $q$. This is a consequence of having more allowed particle orientations through $q$. The cross-sea phase, where the interactions become more intensive for the characteristics of the band structure, appears between the polar liquid phase and the band state \cite{xue2023machine} for a fixed $q$. This phase is not simply a superposition of two band waves, but an independent self-organized complex pattern with an inherently selected crossing angle. One should note here that the cross-sea phase can not be formed with the single band macrophase separation, a single cross-sea pattern at least requires two bands where they cross each other approximately at an angle $\sim \pi/4$. It can be noticed that for $\eta=0.3$, as we increase $q$ from $q=7$ to $q=16$, we observe that the self-organized patterns change from band structure $(q=7)$ to cross-sea pattern $(q=8$, 9, 10) to a pattern showing the outset of the polar liquid phase ($q=16$). For the same values of the control parameters, the VM $(q \to \infty)$ exhibits a similar phase to the $q=16$ DVM where the phase point on the $(\eta,\rho_0)$ diagram almost lies on the liquid binodal \cite{solon2015phase}. This happens because anisotropy becomes weaker with $q$ and, for $q \geqslant 16$, the characteristic of the system becomes similar to the VM. The VM does exhibit a cross-sea phase but at a different parameter regime \cite{kursten2020dry}. For very large noise, e.g., $\eta=0.5$, we observe a disordered gas phase irrespective of $q$. 

Since the emergence of different phases in the DVM depends on the spatial anisotropy, in Auxiliary material~\ref{appB}, we show that the cluster phase is ubiquitous for small ($q,\eta$) and present a cluster size analysis of the $q=4$ DVM for varying noise. See also Auxiliary material~\ref{appC} and Auxiliary material~\ref{appD} for more discussions on the DVM phases with spatial anisotropy (rectangular domain).

Based on the above observations, we can quantify the DVM phase diagram. Notice that from Fig.~\ref{TDVM_L1024}, we have visually distinguished four different phases so far \cite{kursten2020dry}: the ordered phase, the cross-sea phase, the band phase, and the disordered gas phase. It is, however, challenging to define the phase boundaries by visual inspection. 

\begin{figure}[!t]
    \centering
    \includegraphics[width=0.8\columnwidth]{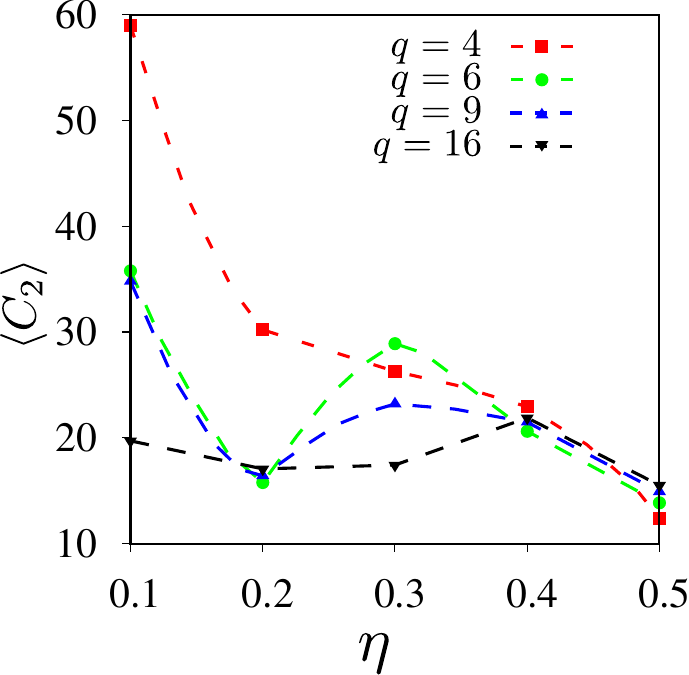}
    \caption{(Color online) The structural order parameter $\langle C_2 \rangle$ versus the noise amplitude $\eta$ for several values of $q$. The dotted curves between two successive points are approximated using spline interpolation. Parameters: $L=1024$, $v_0=0.5$, and $\rho_0=2$.}
    \label{figc2}
\end{figure} 

Earlier, phase-separated density profiles were used to identify the binodals that delimit the gas and liquid phases from the coexistence region~\cite{solon2015phase}. Yet, this technique can not differentiate between the distinct self-organized patterns we observe in the coexistence region (i.e. macrophase separation, microphase separation, and cross-sea) of the DVM. One might also want to distinguish different structures in terms of the global polar order parameter or the global magnetization defined as: 
\begin{align}
\label{GOP}
m = |{\bf m}| &= \frac{1}{N} \left|\sum_i {\bm \sigma}_i\right| \, .
\end{align}
The polar order parameter can be used to study the transition between the disorder gas phase and the band states but can not differentiate between the band and the cross-sea states \cite{kursten2020dry}. The Binder cumulant constructed from this order parameter shows only the transition between the disordered phase and the band state \cite{kursten2020dry}. The polar order parameter $m$ is also insensitive around the cross-sea state. Thus for precise quantification of different phases and their boundaries, we compute the structural order parameter $C_2$ \cite{kursten2020dry,xue2023machine,kursten2020multiple} as follows:
\begin{align}
\label{c2}
C_2 &= N^2 \int G_2({\bf r}_1,{\bf r}_2)d{\bf r}_1 d{\bf r}_2 \Theta(R-\vert{\bf r}_1 \vert) \Theta(R-\vert {\bf r}_2\vert) \nonumber \\
 &= \left(\frac{N}{L^2}\right)^2 \int_{\mathbb{R}^2} \left[g(\vert {\bf r}_1-{\bf r}_2\vert)-1\right] \nonumber \\ 
& \times \Theta(R-\vert{\bf r}_1 \vert) \Theta(R-\vert {\bf r}_2 \vert) d{\bf r}_1 d{\bf r}_2 \, ,
\end{align}
where $G_2({\bf r}_1,{\bf r}_2)=P_2({\bf r}_1,{\bf r}_2)-P_1({\bf r}_1)P_2({\bf r}_2)$ for one- and two-particle probability density functions $P_1$ and $P_2$, $\Theta$ is the Heaviside step function, and $R$ is the distance around an arbitrary fixed point in space. For macroscopically isotropic systems, $G_2$ can be expressed in terms of the pair correlation function $g(r)$ which physically signifies the probability of finding a particle at a distance $r$ relative to that of a given reference particle and provides a statistical description of the local packing and particle density of the system. It has been shown \cite{kursten2020dry,kursten2020multiple} that $C_2$ performs better than $m$ in capturing the structural change and the Binder cumulant of the $C_2$ is also more efficient in distinguishing different features than the Binder cumulant of the order parameter $m$. In practice, to compute $C_2$, we take all pairs of particles, draw circles of radius $R$ around them, and calculate the overlap area of the two circles. The overlap area is given by $A_{\rm overlap} = 2R^2 \cos^{-1} \left(\frac{d}{2R}\right) - \frac{d}{2} \sqrt{4R^2-d^2}$ \cite{kursten2020multiple}, where $d$ is the distance between the centers of the circles. 

\begin{figure}[!htbp]
    \centering
    \includegraphics[width=\columnwidth]{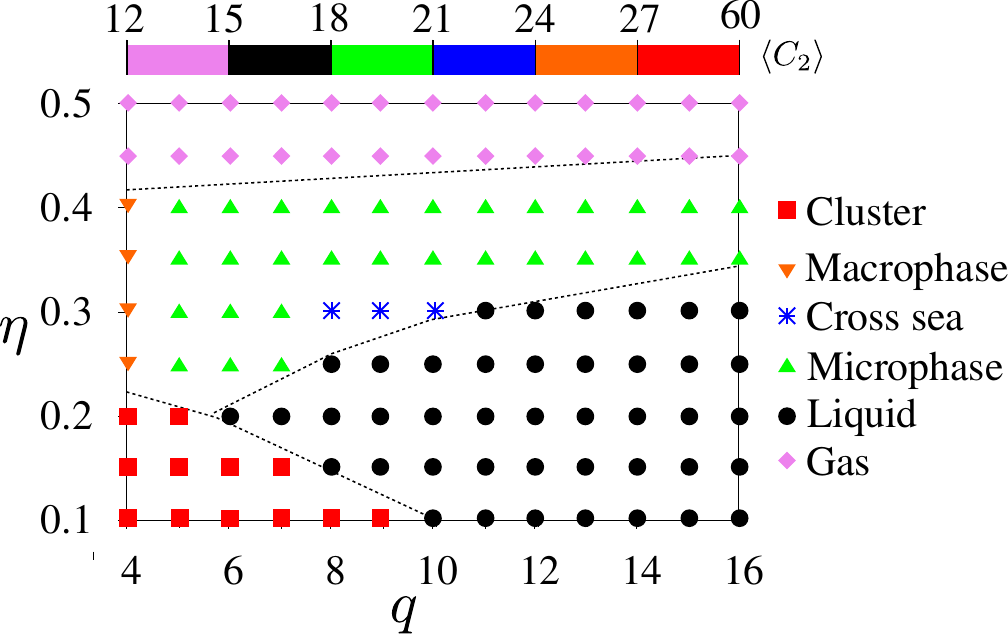}
    \caption{(Color online) $\eta-q$ phase diagram of the DVM by computing $\langle C_2 \rangle$. The colorbar represents the range of $\langle C_2 \rangle$ values for different phases. Parameters: $L=1024$, $v_0=0.5$, and $\rho_0=2$.}
    \label{phasediagram}
\end{figure}

Fig.~\ref{figc2} shows the dependence of the time-averaged structural order parameter $\langle C_2 \rangle$ with respect to $\eta$ for different $q$. The value of $\langle C_2 \rangle$ is the lowest for disorder gas and becomes maximum when particles are clustered. For the band phase, $\langle C_2 \rangle$ values for macrophase separation are larger than the microphase separation and cross-sea phase. Between microphase separation and the cross-sea phase, $\langle C_2 \rangle_{\rm cross} > \langle C_2 \rangle_{\rm micro}$. Although a change in $\langle C_2 \rangle$ is not very significant in these two phases, it is still a better candidate for distinguishing the cross-sea from the microphase separation than the traditional polar order parameter. In Fig.~\ref{phasediagram}, we plot the $\eta-q$ phase diagram by computing $\langle C_2 \rangle$ for the six different phases where in the coexistence region, $\langle C_2 \rangle_{\rm macro}>\langle C_2 \rangle_{\rm cross} > \langle C_2 \rangle_{\rm micro}$. This phase diagram complements Fig.~\ref{TDVM_L1024}, which has been constructed using the density field and depicts the nature of phase separation in the DVM as a function of $q$. 

Now, the nature of the coexistence region for any finite, large $q$ has been a subject of discussion in the context of $q$-state ACM \cite{solon2022susceptibility,chatterjee2022polar}. It was argued in Ref.~\cite{solon2022susceptibility} that spatial anisotropy plays a crucial role in determining the macro/micro-phase separation of the coexistence region in the ACM and one should observe a macrophase separation of the coexistence region for a finite $q$ beyond a characteristic length scale which diverges for large $q$. ACM with a different set of dynamical rules than Ref.~\cite{solon2022susceptibility} was studied in Ref.~\cite{chatterjee2022polar} where the flocking transition in the ACM was argued to be equivalent to the VM at large $q$. The $q$-state DVM is governed by a completely different set of microscopic rules than both the ACM~\cite{solon2022susceptibility,chatterjee2022polar} and we will therefore investigate next the impact of microscopic rules on the DVM steady-state as a function of $q$. We plan to do this through the analysis of number fluctuations, the pinning-unpinning property of the system's global order, and the structure factor manifesting the correlation of polarization.

\subsection{Number fluctuations.}
\begin{figure}[!t]
    \centering
    \includegraphics[width=0.8\columnwidth]{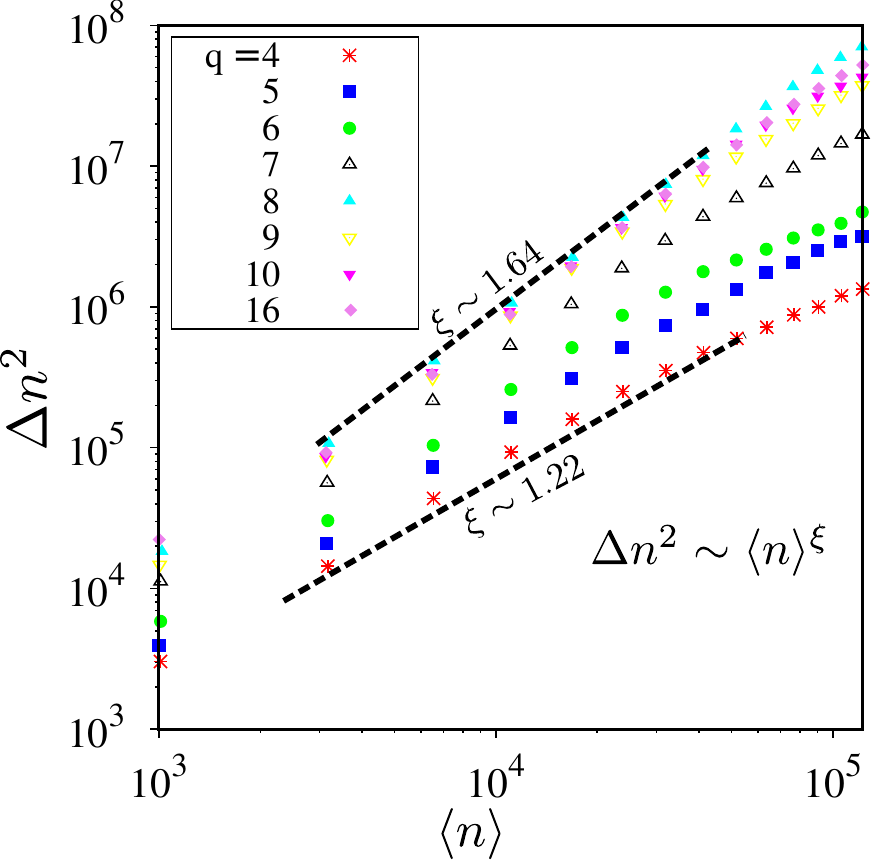}
    \caption{(Color online) Number fluctuations $\Delta n^2=\langle n^2 \rangle-\langle n \rangle^2$ versus average particle number $\langle n \rangle$ for several $q$ values in a $300^2$ domain. Parameters: $\eta=0.3$, $v_0=0.5$, and $\rho_0=6$.}
    \label{NF}
\end{figure}
It is known that fluctuations play an essential role in selecting the phase-separated patterns in flocking models. In the AIM \citep{solon2015flocking}, the number fluctuation $\Delta n^2=\langle n^2 \rangle-\langle n \rangle^2$ in the liquid phase was found to be normal ($\Delta n^2=n$) and the coexistence region shows a macrophase separation. In the VM \cite{solon2015phase}, on the contrary, giant density fluctuations ($\Delta n^2=n^{1.6}$) are observed which break large liquid domains and prevent the bands from coarsening further. This results in a microphase-separated coexistence region. In the $q$-state ACM \cite{solon2022susceptibility,chatterjee2022polar}, which can be thought of as a bridge between the discrete AIM and continuous VM, density fluctuations show a transition from normal to giant fluctuation as $q$ increases and can be explained as a transition from AIM physics to VM physics. 

In Fig.~\ref{NF}, therefore, we show the number fluctuations $\Delta n^2$ versus average particle number $\langle n \rangle$ computed in the ordered liquid phase of the DVM for various $q$. $n(\ell)$ is the number of particles in boxes of different sizes $\ell$ included in a $300^2$ domain (with $\ell \leqslant 150$), with $\langle n \rangle=\rho_0\ell^2$. As shown in Table~\ref{table_exponents}, the number fluctuation behaves like $\langle n \rangle^\xi$ with the fluctuation exponent $\xi$ increasing with $q$, from $\xi \simeq 1.22$ for $q=4$ to $\xi \simeq 1.64$ for large $q$. This transition from normal fluctuations for small $q$ to giant fluctuations for larger $q$ was also observed in the ACM~\cite{solon2022susceptibility,chatterjee2022polar} although $\xi$ for $q=4$ and 5 are moderately larger in the DVM than the ACM. The increase of the density fluctuations with $q$ can be attributed to the fact that for large $q$, particles have more rotational degrees of freedom due to the weak anisotropy and therefore more directional freedom to propel. The existence of giant number fluctuations (GNF) and its connection with the microphase separation in the VM was hypothesized in Ref.~\cite{solon2015phase}. It was argued that GNF ($\xi \simeq 1.6$) breaks bulk liquid domains and produces a smectic-like microphase separation in the coexistence regime whereas the system stabilizes in the bulk phase when the density fluctuations are normal ($\xi \simeq 1$)~\cite{solon2015flocking}. Exploiting the same logic for ACM \cite{chatterjee2022polar},  GNF was argued to be responsible for the microphase separation in the coexistence regime for $q \geqslant 8$, although a direct relation between the existence of GNF in the ordered phase and the microphase separation in the coexistence phase is still ambiguous. Nonetheless, if we compare $\xi$ in Table~\ref{table_exponents} and the snapshots in Fig.~\ref{TDVM_L1024}, we observe a correspondence between the fluctuation exponent and the pattern formation in the coexistence region of DVM. 

One should also consider the finite size effect on the fluctuation exponents as discussed in Ref.~\cite{chatterjee2022polar}. In Fig.~\ref{NF}, the data can be fitted to two different power-law regimes (the extracted exponents depend on the interval along the $x$-axis to which the fits are restricted): (a) $\xi$ tabulated in Table~\ref{table_exponents} in the interval $[10^3,5 \times 10^4]$ and (b) a smaller $\xi$ in the interval $[5 \times 10^4,10^5]$. Around the second interval, the plot shows a ``saturation'' because $\xi$ must decrease with increasing $\langle n \rangle$ due to the finite-size cut-off at $\langle n \rangle=N=\rho_0L^2$, where $\Delta n^2$ vanishes. 

\begin{table}[!htbp]

\begin{center}
\begin{tabular}{ |c|c|c|c|c|c|c|c|c| } 
\hline
$q$ & 4 & 5 & 6 & 7 & 8 & 9 & 10 & 16 \\
\hline
$\xi$ & 1.22 & 1.28 & 1.41 & 1.48 & 1.67 & 1.64 & 1.64 & 1.64 \\
\hline
\end{tabular}
\caption{(Color online) Number fluctuation exponents $\xi$ for several values of $q$, reported from Fig.~\ref{NF}. The typical error on the fluctuation exponents is 0.03. \label{table_exponents}}
\end{center}

\end{table}

\subsection{Pinned property of the order parameter.}
The global order parameter defined in Eq.~\eqref{GOP} quantifies the overall ordering of the particles in the system. In the ACM~\citep{solon2022susceptibility}, at finite size, the direction of the global polar order $\Phi \equiv \rm{arg}\langle {\bf m} \rangle$  exhibits distinct behaviors in the liquid phase depending on the value of $q$: it displays AIM-like properties (pinned along an angle) for small $q$ and VM-like behavior (unpinned over time) for large $q$. 

\begin{figure}[!t]
    \centering
    \includegraphics[width=\columnwidth]{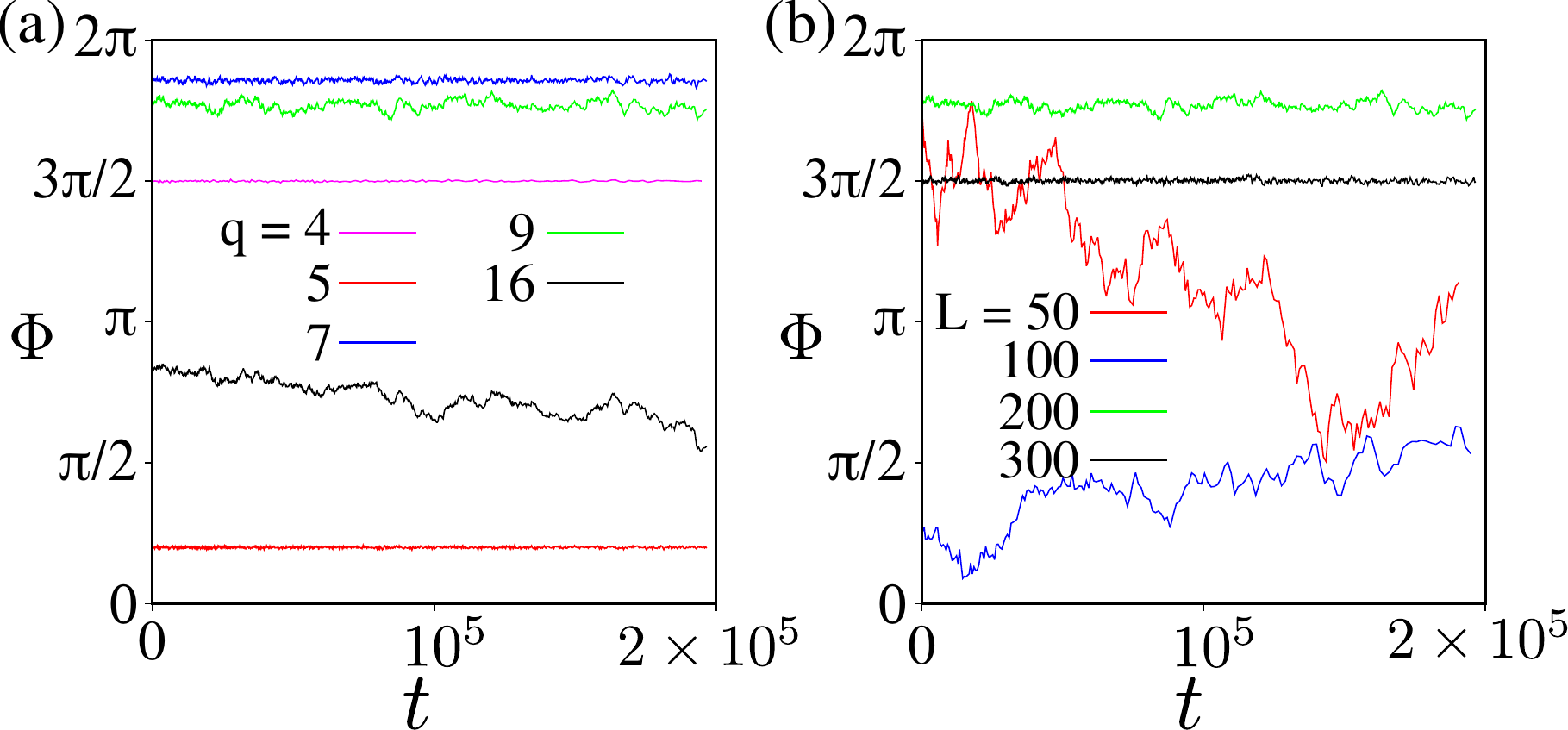}
    \caption{(Color online) (a) Time series of the orientation of global polar order $\Phi$ in the liquid phase showing a transition from unpinned to pinned as $q$ increases ($\eta=0.3$, $\rho_0=6$, $v_0=0.5$, and $L=200$). (b) $\Phi$ shows an unpinning to pinning transition as a function of system size $L$ for $q=9$ ($\eta=0.3$, $v_0=0.5$ and $\rho_0=6$).}
    \label{figGOP}
\end{figure}

In Fig.~\ref{figGOP}(a), we show the time evolution of $\Phi(t)$ in the DVM liquid phase for varying $q$. Akin to the observation made in the ACM \cite{solon2022susceptibility}, $\Phi(t)$ starts wandering slowly with $q$ and becomes an unpinned variable of $t$ for large $q$. In other words, for weak anisotropy, the global ordering does not remain constrained to a specific orientation. While, for small $q$, $\Phi(t)$ remains pinned and tends to exhibit a stable global ordering. Microscopically, this refers to a picture that at large $q$, a proportional number of degrees of freedom allows the particles to choose between adjacent directions facilitated by fluctuation while it is not the obvious choice for particles at small $q$ requiring a significantly larger jump to switch directions. Translating this to the global polar order parameter and comparing Fig.~\ref{figGOP}(a) with Fig.~\ref{NF} we propose that GNF corresponds to the unpinned behavior of $\Phi(t)$. Likewise, the unpinning nature of the direction of the global polar order is a characteristic of microphase separation. 

In addition to $q$, the finite system size also affects the evolution of $\Phi(t)$, which is shown in Fig.~\ref{figGOP}(b) for $q=9$. Similar to the ACM \cite{solon2022susceptibility}, we observe a transition from unpinned behavior to pinned behavior in $\Phi(t)$ as the system size is increased. In larger systems with polar order, a particle interacts with more particles in the neighborhood and correlates over longer distances. Higher connectivity promotes stronger alignment and cooperative motion among the particles. As a result, the direction of the global order becomes more pronounced and persistent in larger systems, leading to the pinned state. Fig.~\ref{figGOP}(b) further signifies that if $\Phi(t)$ is pinned for $L=300$, it must also be pinned for $L=1024$. This is inconsistent with the microphase separation and cross-sea patterns observed in the coexistence region of the $q=9$ DVM (Fig.~\ref{TDVM_L1024}). Evidently, the correlation proposed earlier between the pinned property of the system's ordered liquid phase and system morphology observed in the coexistence region (macro/micro/cross-sea) is not conclusive in the DVM (see Auxiliary material~\ref{appF} for more details).

\subsection{Structure factor.} 
\begin{figure}[!t]
    \centering
    \includegraphics[width=\columnwidth]{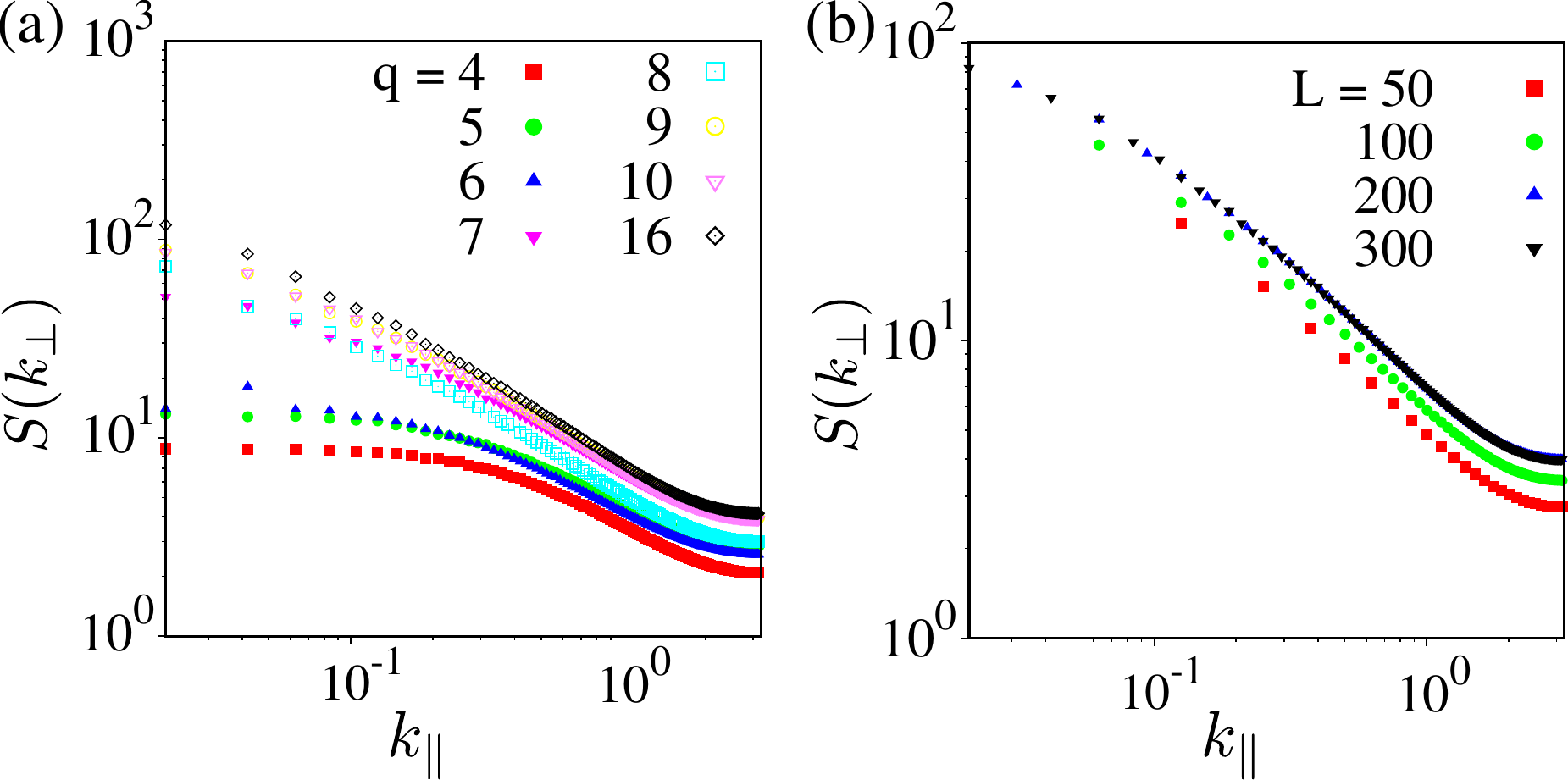}
    \caption{(Color online) Structure factor $S_{\perp}({\bf k})$ vs ${\bf k}=(k_{\parallel},0)$ in the ordered liquid phase for (a) several $q$ values ($L=300$) and (b) varying system size ($q=9$). Parameters: $\eta=0.3$, $\rho_0=6$, and $v_0=0.5$.}
    \label{SF}
\end{figure}

To explore the correlation among the particle polarizations, we consider the transverse magnetization structure factor $S_{\perp}({\bf k})=\langle m_{\perp}({\bf k}) m_{\perp}(\bf{-k}) \rangle$ against wavelength ${\bf k}$ and plot it in Fig.~\ref{SF}. The structure factor has been calculated in the liquid phase on a $300^2$ domain for (a) various $q$ values (the same behavior is observed for the structure factor of the density field) and (b) for a fixed $q=9$ but for various system sizes. The results presented in Fig.~\ref{SF}(a) show that for small $q$, the structure factor $S(k)$ converges to finite values as the wave vector ${\bf k} \to 0$. This convergence indicates an AIM-like behavior or a macrophase separation of the coexistence region \cite{solon2022susceptibility}. However, one can notice that this convergence is achieved only beyond certain length scales and these length scales are functions of $q$. The structure factor captures the correlations between particle orientations and therefore the magnitude of ordering at different length scales. A saturation of $S(k)$ for small $q$ when ${\bf k}$ approaches zero thus signifies a strongly correlated liquid domain whereas for large $q$, due to weak anisotropy or more allowed orientations for ordering, particles inside the liquid domain are not as strongly correlated as for small $q$. Fig.~\ref{SF}(b) shows $S(k)$ for several system sizes and manifests that with larger system sizes $(L \geq 300)$, $S(k)$ tends to converge to a finite value when ${\bf k} \to 0$. Our earlier argument that stronger interactions between particles (with larger $L$) promote robust ordering also applies here. Comparing Fig.~\ref{SF} with Fig.~\ref{figGOP}, we conclude that the pinning behavior (unpinning behavior) of $\Phi(t)$ and the saturation of $S(k)$ (algebraic scaling of $S(k)$) compliment each other and convey the same physics. 

It was argued in Ref.~\citep{solon2022susceptibility} considering the pinning properties of the order parameter and behavior of the structure factor in the liquid phase that for the large $q$ ACM, VM behavior (microphase separation) will be observed only up to large finite sizes. But the asymptotic large length scale behavior will be AIM-like (macrophase separation) where the length scale diverges with $q$ as $\exp(q^2)$. Ref.~\citep{solon2022susceptibility} also showed that in the phase-coexistence region of the ACM, a microphase to macrophase transition happens upon increasing the linear system size along the transverse direction at fixed $q$ which we do not observe (in the DVM, multiple bands do not merge to a single band upon increasing $L_y$ for a fixed $L_x$). We observe a similar behavior of $\Phi(t)$ and $S(k)$ as Ref.~\citep{solon2022susceptibility} but can not draw a conclusive correspondence between the large length-scale liquid phase behavior of $\Phi(t)$ and $S(k)$ to the phase-coexistence behavior of the DVM as shown in Fig.~\ref{TDVM_L1024}.

In Fig.~\ref{TDVM_L1024}, the snapshots on a large square domain (without any spatial anisotropy) show the existence of a microphase separation and cross-sea patterns (for which one needs at least two bands) of the phase-coexistence region for large $q$ values. The number fluctuation plotted in Fig.~\ref{NF} corroborates this observation by exhibiting GNF for those large $q$ values. The large length scale asymptotic behavior (for large $q$) of the direction of global order $\Phi(t)$ (Fig.~\ref{figGOP}) and the structure factor $S(k)$ (Fig.~\ref{SF}) in the ordered liquid phase respectively shows a pinned behavior and saturation for $k_{\parallel}=0$ which as argued in Ref.~\citep{solon2022susceptibility} signifies an AIM phenomenology. However, our numerical investigation of the DVM does not show a cross-over from micro- to macrophase separation for higher $q$-values. Therefore, we argue that the impact of dynamical rules governing flipping and hopping in a flocking model has a significant influence over the system dynamics.

\begin{figure}[!t]
    \centering
    \includegraphics[width=\columnwidth]{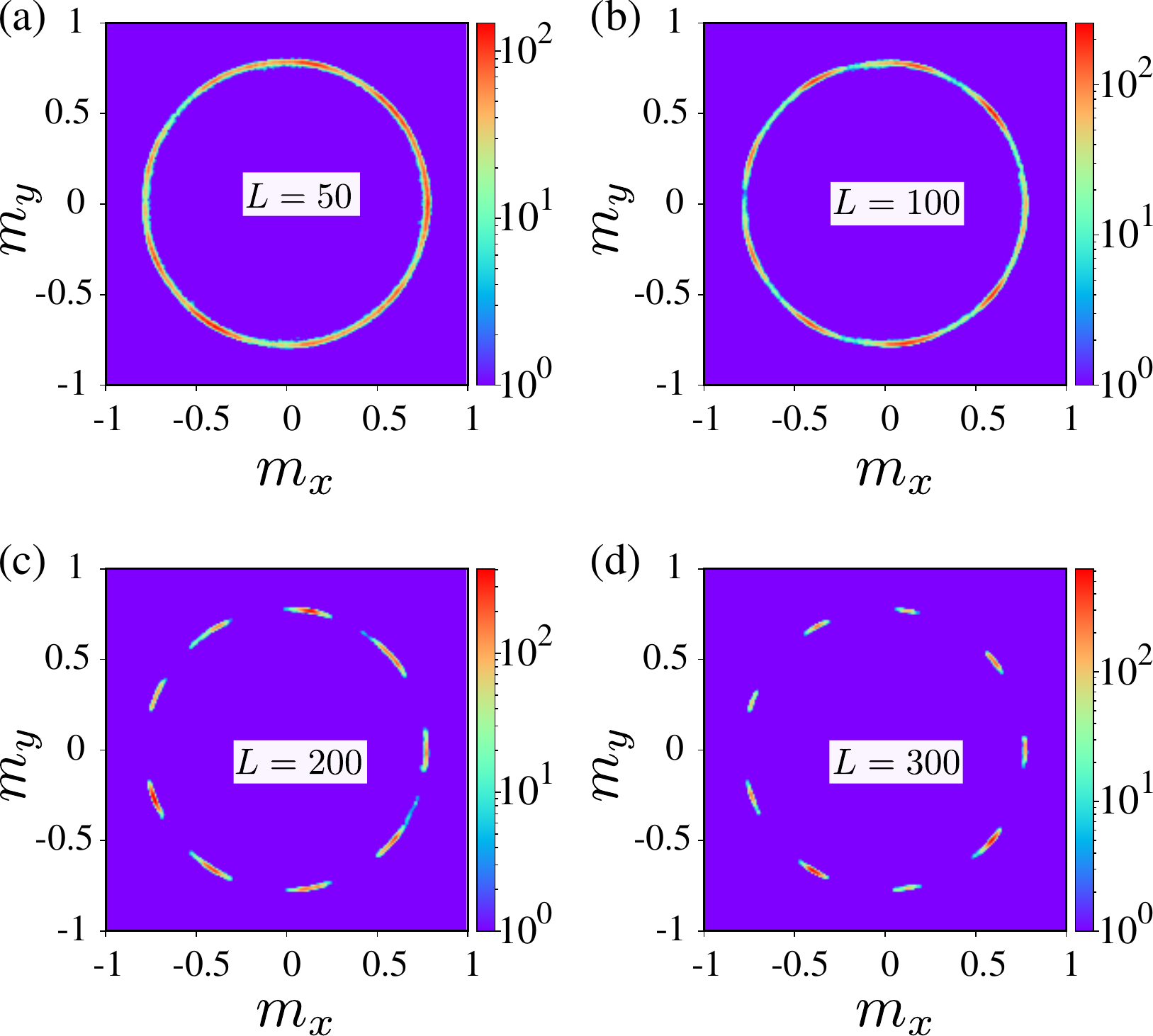}
    \caption{(Color online) Order parameter distributions of the $q=9$ DVM for the liquid phase. Parameters: $\eta=0.3$, $v_0=0.5$ and $\rho_0=6$. Ring-like distributions for smaller system sizes ($L=50$ and $L=100$) in (a) and (b) change to distinct isolated spots for larger system sizes ($L=200$ and $L=300$) in (c) and (d), which correspond to the ordered liquid phase's $q$-fold degeneracy.}
    \label{op_distro}
\end{figure}

\subsection{Order parameter distribution.}
In Vicsek-like models, where particles are active and move with a constant velocity, the ordered state exhibits a true long-range order (LRO) in two dimensions because of spontaneous symmetry breaking due to the out-of-equilibrium nature of the models. In Fig.~\ref{op_distro}, we show the time- and ensemble-averaged distribution of the order parameter $\textbf{m}=(m_x,m_y)$ for increasing system sizes, where $m_x=\frac{1}{N}\sum_{i=1}^N \cos \theta_i$ and $m_y=\frac{1}{N}\sum_{i=1}^N \sin \theta_i$. In Fig.~\ref{op_distro}(a) and Fig.~\ref{op_distro}(b), ring-like distributions (unpinned orientations) characteristic of the quasi long–range ordered (QLRO) phase are observed. But, this is only due to the finite-size effect and similar to the impact of finite-size on $\Phi(t)$ and $S(k)$. We recover the LRO for larger system sizes ($L=200$ and $L=300$) where the distributions display nine distinct isolated spots (pinned orientations) that correspond to the 9-fold degeneracy of the ordered liquid phase, each spot having equal probability. One can expect that the finite size effect will be much weaker for $L=1024$ and the spread of the distribution in the LRO phase around the allowed ordering angles will also be more precise.

The DVM for large $q$ exhibits a QLRO phase when $v_0=0$ (see Auxiliary material~\ref{appE}). We argue that DVM with immobile particles reduces to the two-dimensional $q$-state clock model (with a quenched bond disorder as only particles within a fixed distance interact) which approaches the XY model for large $q$ with vanishing LRO regime~\cite{clockmodel2018}. For $v_0>0$ and a fixed $L$, as flocking directions increase with $q$, we again observe ring-like distributions for large $q$ (see Auxiliary material~\ref{appE}) but beyond a length scale which is proportional to $q$, the order parameter distributions for large $q$ show $q$ isolated spots characteristic of the LRO phase. This is similar to the unpinned to the pinned transition of $\Phi(t)$ and convergence of $S(k)$ to a finite value at ${\bf k} \to 0$ for large $q$ values as the system size is increased. In the VM ($q \to \infty$), even for a large $L$, the order parameter distribution shows a ring-like structure because of the continuous symmetry. 

\subsection{Stability of the ordered liquid phase.}
\begin{figure*}[!htbp]
    \centering
    \includegraphics[width=\textwidth]{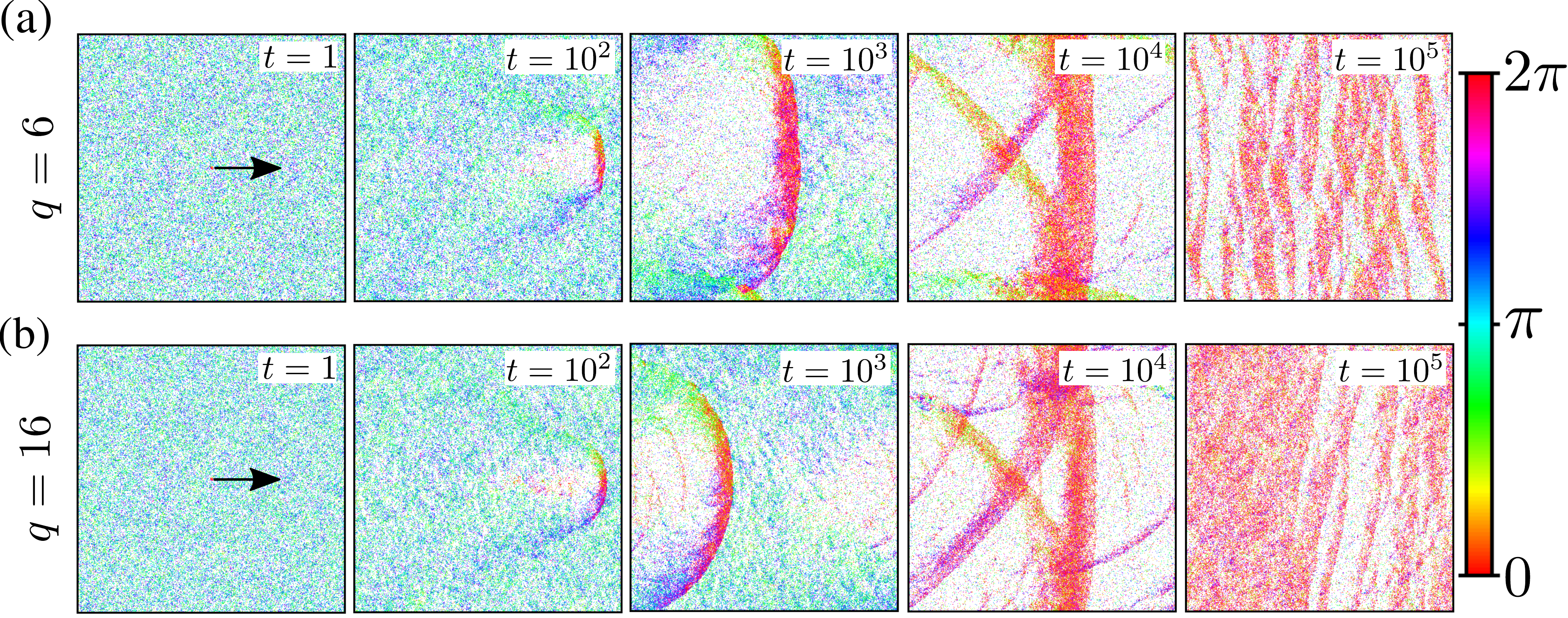}
    \caption{(Color online) Time evolution snapshots of the orientation field showing the reversal of the initial ordered liquid phase in (a) $q=6$ and (b) $q=16$-state DVM following the introduction of a tiny counter-propagating high-density droplet (motion of direction is denoted by black arrows, $t=1$) of radius $r_d=5$ and density $\rho_0^d=10\rho_0$. Colorbar represents the particle orientation field. Parameters: $L=1024$, $\eta=0.4$, $v_0=1$ and $\rho_0=2$.}
    \label{metastable}
\end{figure*}
As discussed in the context of Fig.~\ref{TDVM_q9_L1024}, here we present a brief analysis of the stability of the DVM ordered liquid phase by inserting a small high-density counter-propagating liquid droplet. Initially, the average direction of the particles in the polar liquid phase is aligned along the direction $\Phi_{\rm liq}=\pi$, and the average particle orientation of the liquid droplet is $\Phi_d=0$. The radius of the droplet is $r_d$ and density is $\rho_0^d$ and it is inserted in an ordered phase of density $\rho_0$ ($\rho_0^d\gg\rho_0$). We take several $q$ values and calculate the probability ($P_{\rm rev}$) that the droplet grows against the main order and reverses the ordered phase as a function of $\eta$, $r_d$, and $\rho_0^d$. The simulation protocol follows Ref. \cite{benvegnen2023metastability}, where the initial configuration is prepared by particles with $\theta=\pi$ and we let the system evolve up to a certain time $t$ to reach the steady state. The system retains the average global polarization in the direction $\theta=\pi$. Then, a circular region of radius $r_d$ centered at $(L/2, L/2)$ is selected and an additional $\Delta N = (\rho_0^d-\rho_0)\pi r_d^2$ number of particles are added to make a high-density circular droplet. Finally, the orientation of all the particles within the droplet is changed to $\theta=0$.  

In Fig.~\ref{metastable}, we study the fate of polar flocks in $q=6$ and 16-state DVM by introducing a small high-density counter-propagating droplet against the initial polar ordered liquid phase of the main flow and observe the subsequent time evolution. One should note here that the perturbation through the droplet is very small i.e. the ratio of droplet diameter to the linear length of the simulation box is $\sim 10^{-2}$ ($r_d=5$, $L=1024$). We observe, similar to Ref.~\cite{codina2022small}, that the droplet grows with time leaving behind a dilute region ($t=10^2$) and adds more and more particles as it moves along forming a principal dense, curved band (followed by several other curved bands) that invades the whole system ballistically ($t=10^3$ and $t=10^4$). In the final stage, this principal curved band connects itself over the system boundaries and widens until a steady state liquid phase of a different $\Phi_{\rm liq}$ emerges (at $t=10^5$), signifying the metastability of the DVM liquid phase. The time-evolution is similar for both small and large $q$, which signifies that both discrete and continuous-symmetry flocks are metastable.

The growth pattern of the DVM droplet is similar to the VM~\cite{codina2022small} but distinct from the AIM \cite{benvegnen2023metastability}. In the DVM, after its introduction, the droplet front interacts with the liquid particles outside and creates a curved band of particles having several different orientations (impact of $q$). In this process, the droplet seizes to exist and it is this high-density curved band that destroys the initial flow. In AIM \cite{benvegnen2023metastability}, the droplet grows along the direction transverse to the propagation (due to the constant transverse diffusion) creating a comet-like trail of disordered particles that can not be observed in the DVM.

\begin{figure}[!t]
    \centering
    \includegraphics[width=\columnwidth]{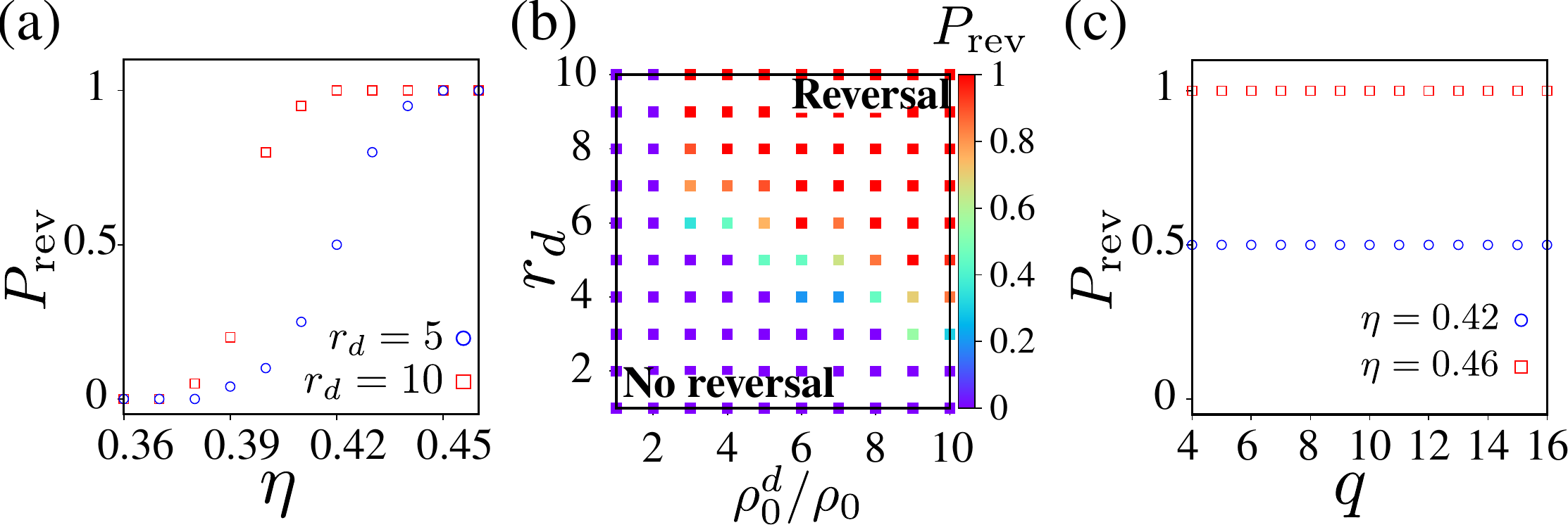}
    \caption{(Color online) (a) Reversal probability $P_{\rm rev}$ versus noise $\eta$ for $q=16$ and two different droplet radius $r_d=5$ and 10. (b) $r_d-\rho_0^d/\rho_0$ phase diagram for $\eta=0.45$ and $q=16$. The colorbar represents reversal probability $P_{\rm rev}$. (c) $P_{\rm rev}$ versus $q$ for various noise, $r_d=5$. Parameters: $L=400$, $\rho_0^d=10\rho_0$ (a \& c), $v_0=1$ and $\rho_0=2$.}
    \label{metastable2}
\end{figure}

Fig.~\ref{metastable2} quantifies $P_{\rm rev}$, the probability of reversing the main flow upon the introduction of a given droplet, for several control parameters. For each set of control parameters, we have taken 20 independent realizations to calculate $P_{\rm rev}$. Akin to Ref.~\cite{codina2022small}, we observe that the noise strength $\eta$ has a strong influence on the reversal dynamics and $P_{\rm rev}$ increases from 0 to 1 as $\eta$ is increased [Fig.~\ref{metastable2}(a)]. This is because for small $\eta$, the ordered phase is very stable, and thus, the counter-propagating dense bands find it difficult to reverse its flow. For large $\eta$, fluctuations are stronger, and therefore, the probability of reversal increases. Fig.~\ref{metastable2}(a) also exhibits that the transitional value of $\eta$ ($P_{\rm rev}=\frac{1}{2}$) above which a droplet triggers a reversal is a decreasing function of $r_d$ although there is a critical $\eta$ ($\eta \sim 0.35$), below which no droplet can trigger a reversal irrespective of its density. The $r_d-\rho_0^d$ phase diagram in Fig.~\ref{metastable2}(b) has been constructed by calculating $P_{\rm rev}$ for several $(r_d,\rho_0^d)$. Unsurprisingly, large $r_d$ combined with large $\rho_0^d$ are shown to facilitate the reversal of the initial liquid phase. The droplet-induced reversal of the liquid flow is found independent of $q$ where $P_{\rm rev}$ is found to behave similarly for each $q$ under certain values of $\eta$ [Fig.~\ref{metastable2}(c)]. 

In addition, a study of the metastability of the DVM liquid phase on rectangular geometry ($L_x=800, L_y=100$), produces a similar outcome. One can also consider droplet movement in other directions than opposite to the global flow, e.g. transversely propagating droplets with $\Phi_d=\pi/2$ for $q=8$ or $\Phi_d=2\pi/3$ for $q=6$ and perform a similar analysis to check whether the liquid phase is susceptible to droplets irrespective of their propagation direction.

\section{Summary and Discussion}
\label{discussion}
Our study motivated by the active clock model~\cite{chatterjee2022polar,solon2022susceptibility}, considers a true $q$-state discrete version of the Vicsek model where $q$ defines the strength of orientation anisotropy. At small $q$, the system is highly anisotropic which however vanishes in the limit $q\rightarrow \infty$ when we recover the Vicsek model. The DVM shows qualitatively similar features as the ACM~\cite{chatterjee2022polar} for intermediate noise strength $\eta$ where a transition from macrophase to microphase separation is observed in the coexistence region as $q$ is increased. But for small $q$ and $\eta$, the liquid phase appearing in the ACM at low temperatures is replaced in the DVM by a cluster phase. The cluster phase consists of multiple clusters with different polarization (see Fig.~\ref{appfig1}) which does not exhibit a long-range order. The clusters grow and merge with increasing $q$ leading to a homogeneous ordered phase at large $q$. For small $q$, a long-range ordered phase can be achieved by increasing the noise strength. For low noise and small $q$, the probability of transverse flipping is very small as fluctuations are weak. In addition to that, transverse fluctuations through hopping are also absent. Consequently, the combined influence of these factors results in clusters failing to grow continuously for low $\eta$ and small $q$, preventing the system from reaching a homogeneous liquid state.

The self-organized patterns in the coexistence region of the discretized VM indicate a transition from AIM-like patterns to VM-like patterns as anisotropy becomes weaker. This observation is corroborated by the giant density fluctuations for large $q$. However, the large length scale behavior of the direction of global order $\Phi(t)$, the structure factor $S(k)$, and the order parameter distribution in the liquid phase do not correspond with the phase-coexistence patterns of the large $q$ DVM. It shows microphase or cross-sea pattern for large length scales without any spatial anisotropy as $q$ increases. 

We also find that the DVM liquid phase is susceptible to perturbation applied through a counter-propagating droplet.  The liquid phase reorients and propagates along the direction of the droplet. The reversal dynamics is significantly impacted by the noise strength $\eta$ ~\cite{codina2022small} but remains independent of $q$. The stability of the high-density flocking ordered phase at low noise is still an open problem and will be addressed in a subsequent study \cite{swarnajit2023metastability}.

As a final remark, we add that the rotational flexibility of the particles and microscopic details of the dynamical rules can significantly impact the macroscopic properties of the ordered phase.
\section{Auxiliary material}
\subsection{Dependency of the cluster phase on the initial condition}\label{appA}
\begin{figure}[!htbp]
    \centering
    \includegraphics[width=0.8\columnwidth]{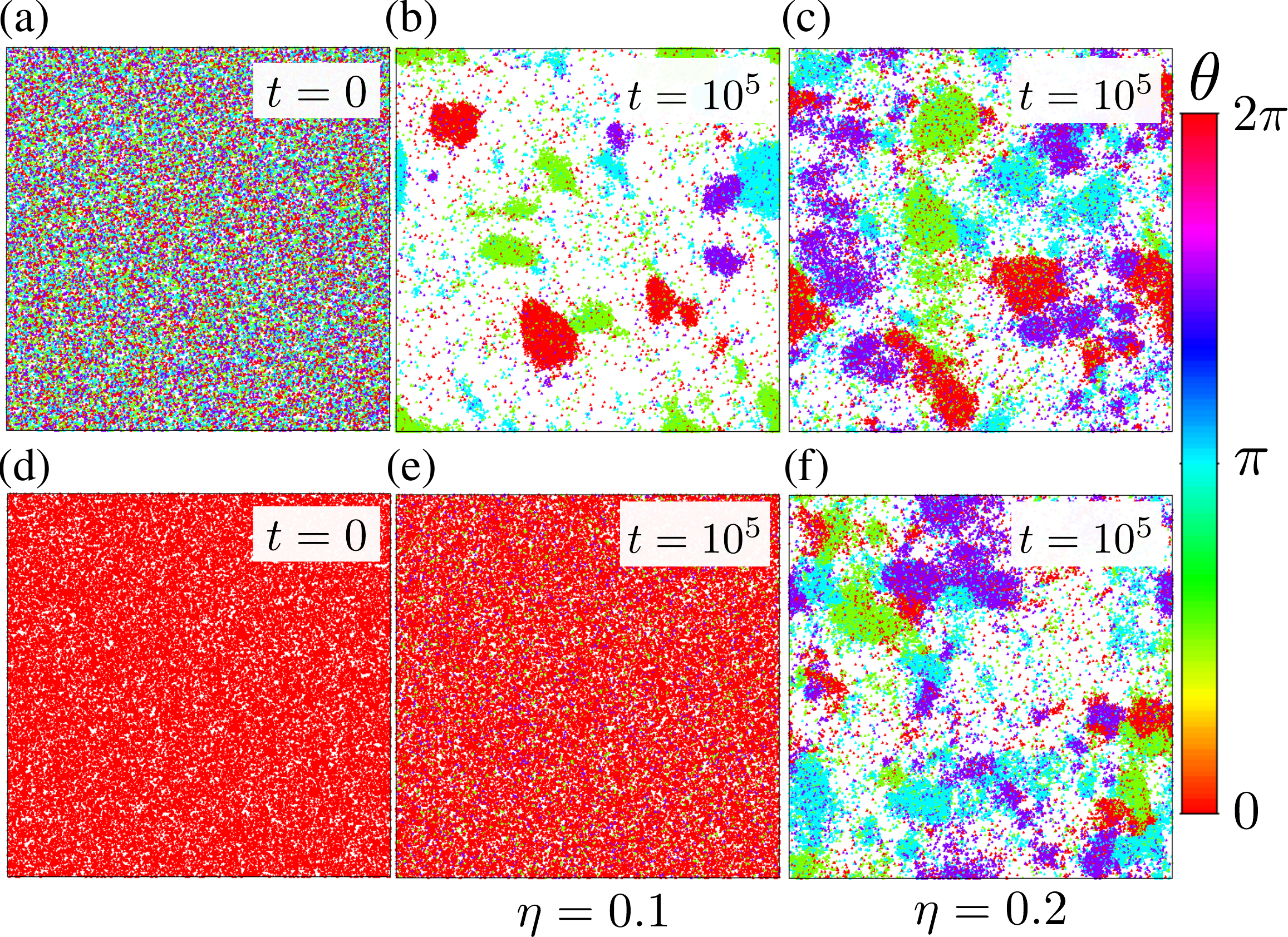}
    \caption{(Color online) Steady-state snapshots (b-c) and (e-f) from two different initial conditions: (a) random disordered and (d) polar ordered. (b, e) $\eta=0.1$ and (c, f) $\eta=0.2$. The colorbar represents particle orientations. Parameters: $q=4$, $L=300$, $v_0=0.5$, and $\rho_0=6$.}
    \label{appfig1}
\end{figure}
Snapshots in Fig.~\ref{appfig1}(b-c) and Fig.~\ref{appfig1}(e-f) illustrate the steady state behavior of the $q = 4$ DVM starting from two different initial conditions: (a) random disordered and (d) polar ordered. In the top panel, the system exhibits cluster phase for $q=4$ for both values of $\eta$ when starting from an unbiased random configuration. The initial coarsening process forms clusters, but they do not merge to create a single large ordered domain due to the absence of transverse fluctuations as discussed in the context of Fig.~\ref{TDVM_L1024}. In the bottom panel, this observation changes for the lowest noise strength when starting from a polar-ordered initial configuration. For $\eta=0.1$, the steady state remains in an ordered liquid phase (similar to the steady state behavior of the 4-state APM or ACM at low temperature) signifying that the fluctuation is weak to alter the broken symmetry phase into a cluster phase. However, with an increase in the noise ($\eta=0.2$), the steady-state cluster phase appears again. This also suggests that for DVM with weak anisotropy and fluctuations, the cluster phase is the non-equilibrium steady-state and the well-known small $\eta$ or large $\beta$ ordered liquid phase can only be achieved by taking a strongly polarized ordered initial condition. It is worth noting that the number of clusters in (c) is higher than in (b) due to more relaxation via the noise. A further increase of the noise will lead to a single large ordered domain as shown in the phase diagram of Fig.~\ref{appfig3}.

\subsection{Cluster size analysis for $q=4$}
\label{appB}
\begin{figure}[!htbp]
    \centering
    \includegraphics[width=0.8\columnwidth]{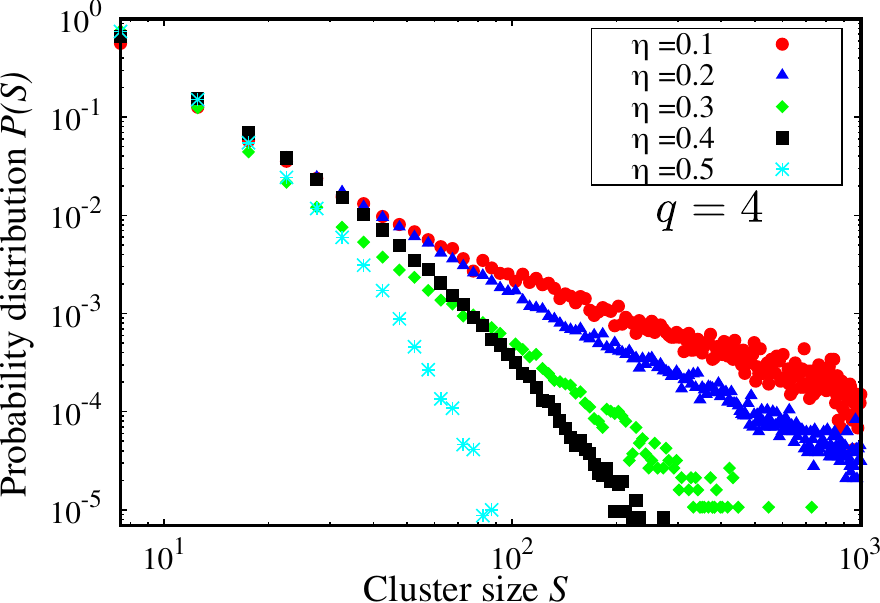}
    \caption{(Color online) Probability distribution $P(S)$ of cluster size $S$ for different noise level. Parameters: $q=4$, $L_x=800, L_y=100, \rho_0=2$, and $v_0=0.5$.}
    \label{appfig4}
\end{figure}
The cluster size analysis for $q=4$ is shown in Fig.~\ref{appfig4} for a rectangular domain of size $800 \times 100$. We use a box-counting method to find a cluster which is described below:

If $L_x$ and $L_y$ are respectively the linear sizes in a rectangular domain, we consider $L_x \times L_y$ as the total number of boxes. For a box $i$, $n_i$ is the number of particles in that box and $c_i$ is the cluster label of the box. If the box does not contain any particles $c_i = 0$ i.e. it is not part of any cluster. Then we use the Depth-first search (DFS) algorithm to find the connected boxes that are not void of particles and label them as a single cluster. For each cluster label $c_j$, we calculate the size of the cluster $S_{c_i}$ as following:
\begin{equation}
    S_{c_i} = \sum_{j=1}^{L_x L_y} \delta_{c_i, c_j} \times n_j \, ,
\end{equation}
where $\delta_{c_i, c_j}$ is the Kronecker delta function that equals 1 if $c_i = c_j$, and 0 otherwise. Then we calculate the cluster size probability distribution denoted as $P(S)$:
\begin{equation}
    P(S) = \frac{\text{Number of clusters with size } S}{\text{Total number of clusters}} \, .
\end{equation}

Fig.~\ref{appfig4} illustrates that for small noise, the probability of larger cluster formation is high. This is because reduced fluctuation in the system facilitates the formation of high-density clusters. However, with noise ($\eta=0.3$), the cluster phase vanishes and the $q=4$ DVM exhibits a macrophase separation, resulting in a decrease in the probability of obtaining large clusters. At large noise ($\eta \geqslant 0.4$), the system becomes disordered, leading to a lesser probability of formation of large clusters. We would also like to mention that for a fixed noise, $P(S)$ versus $S$ for various orientations ($\theta=0, \pi/2, \pi, 3\pi/2$) shows almost identical distributions signifying no preference in orientation in the cluster formation.

\subsection{Impact of spatial anisotropy: steady-states for rectangular domain}
\label{appC}
\begin{figure}[!htbp]
    \centering
    \includegraphics[width=\textwidth]{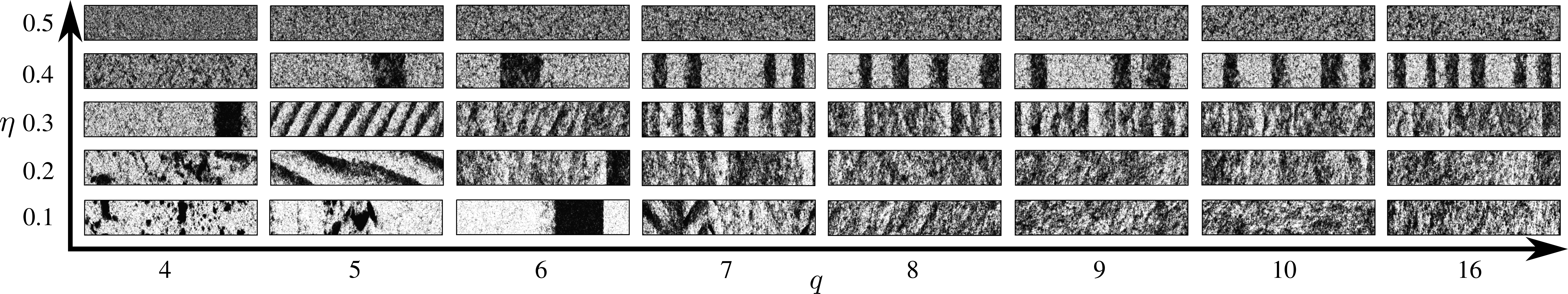}
    \caption{(Color online) $\eta-q$ phase diagram of the DVM illustrated by snapshots on a rectangular domain ($L_x=800$, $L_y=100$). Parameters: $\rho_0=2$, $v_0=0.5$. As a function of $\eta$ and $q$, we observe five distinct self-organized patterns: cluster ($\eta=0.1$, $q=4$), macrophase separation ($\eta=0.3$, $q=4$), microphase separation ($\eta=0.4$, $q=7 \to 16$), ordered liquid ($\eta=0.2$, $q=7 \to 16$), and disordered gas ($\eta=0.5$, $q=4 \to 16$).}
    \label{appfig2}
\end{figure}
In Fig.~\ref{appfig2}, we investigate how spatial anisotropy influences the non-equilibrium steady-state behavior of the DVM if we switch from a square domain to a rectangular domain by analyzing late-stage representative snapshots as a function of noise strength ($\eta$) and discretization parameter ($q$). When noise is low and $q$ is small, we observe the emergence of a locally ordered cluster phase similar to our finding for the square domain. However, when $q$ is small and fluctuations are pronounced, those cluster phases relax and transform into a larger, organized domain (see the snapshot for $q=4$ and $\eta=0.3$) akin to the Fig.~\ref{TDVM_L1024}. Conversely, when $\eta$ is small and $q$ is large, weak anisotropy facilitates the merger of the cluster phase into a larger, well-organized domain (see the snapshot for $q\geqslant8$ and $\eta=0.1$) which is observed at $q=10$ in the absence of spatial anisotropy (see Fig.~\ref{TDVM_L1024}). An increase in the fluctuation for small $q$ might exhibit multiple bands but those are connected by the periodic boundary conditions and should be considered a single band (see the snapshot for $q=5$ and $\eta=0.2, 0.3$). However, no cross-sea phase is observed with the rectangular geometry, which is seen in Ref~\cite{kursten2020dry} probably due to larger particle velocity ($v_0=1$). An increase in anisotropy $q$ also increases the no. of bands in the coexistence region at intermediate noise ($\eta=0.4$) because with $q$, the density fluctuation increases along with the magnetization fluctuation which prompts the breaking of large domains.

\subsection{($\eta-\rho_0$) phase diagram of the DVM}
\label{appD}
\begin{figure}[!t]
    \centering
    \includegraphics[width=\columnwidth]{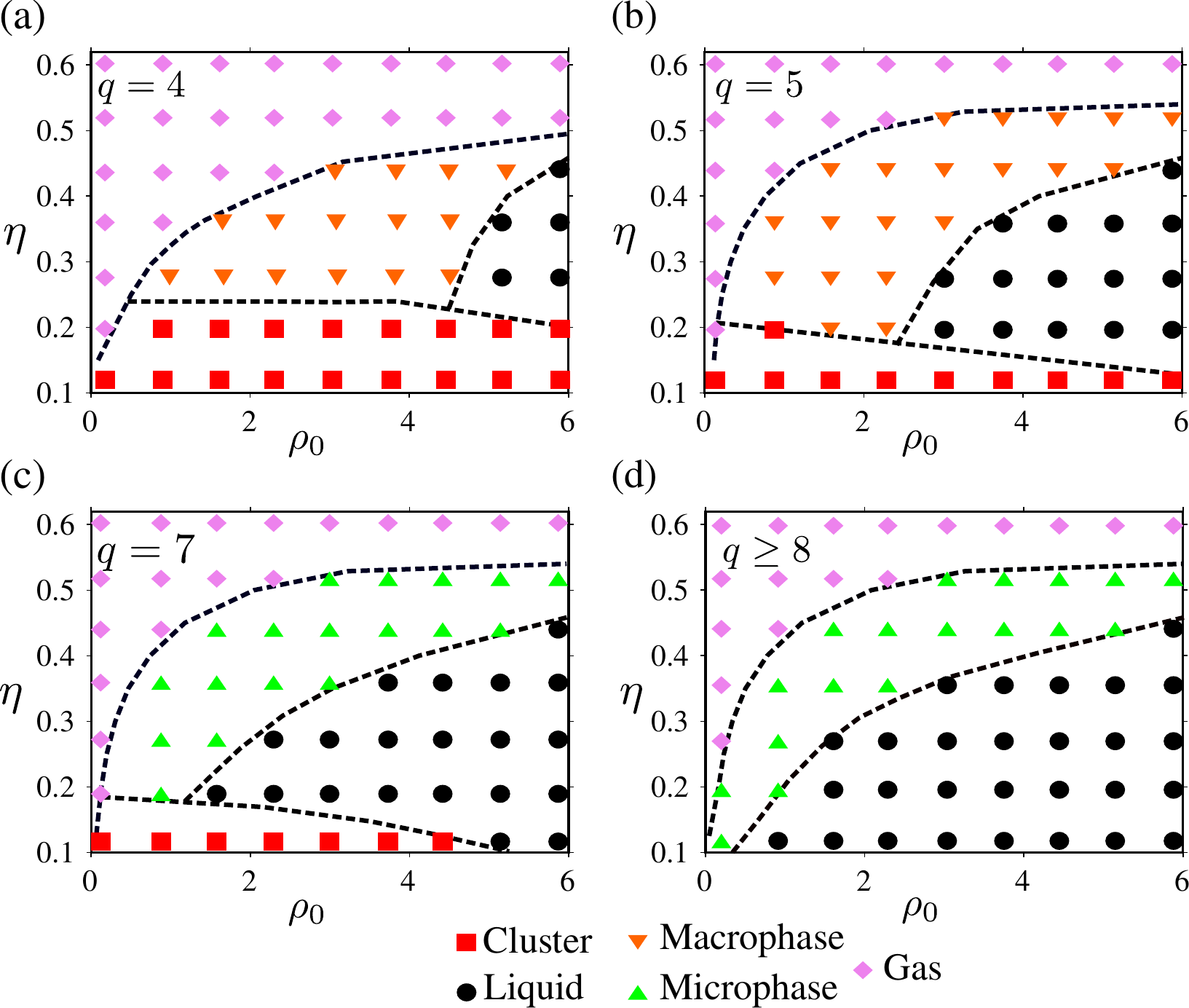}
    \caption{(Color online) Phase diagrams on $\eta-\rho_0$ plane computed on a rectangular domain of $800 \times 100$ for various $q$. A macrophase separation similar to AIM in (a) and (b) transforms to a Vicsek-like phase separation (microphase) in the coexistence region in (c) and (d) at intermediate densities. $v_0=0.5$.}
    \label{appfig3}
\end{figure}
In Fig.~\ref{appfig3}, we plot the noise-density ($\eta-\rho_0$) phase diagrams computed on a rectangular domain of $800 \times 100$ for various $q$. The clustering phase is very prominent for small $q$ values ($q=4$, 5) and exists for high densities. As $q$ is increased, the conventional ordered liquid phase appears at high densities and for $q \geqslant 8$, the cluster phase disappears and we notice the emergence of the typical $\eta-\rho_0$ diagram observed for Vicsek-like systems [Fig.~\ref{appfig3}(d)]. These Vicsek-like phases are characterized by a more global alignment of the particles, leading to coherent motion and the absence of distinct clusters or bands. In this regime, the behavior of the system is predominantly governed by the alignment interactions between the particles rather than the specific value of $q$. It shows clear boundaries between different phases based on varying values of $q$ and $\eta$. At small $q$ and intermediate densities, a macrophase separation similar to AIM is observed. However, as $q$ exceeds a threshold (e.g., $q \geqslant 8$), the system transitions to a Vicsek-like phase separation in the coexistence region. 

\subsection{Direction of global order in the coexistence region}
\label{appF}
\begin{figure}[!t]
    \centering
    \includegraphics[width=\columnwidth]{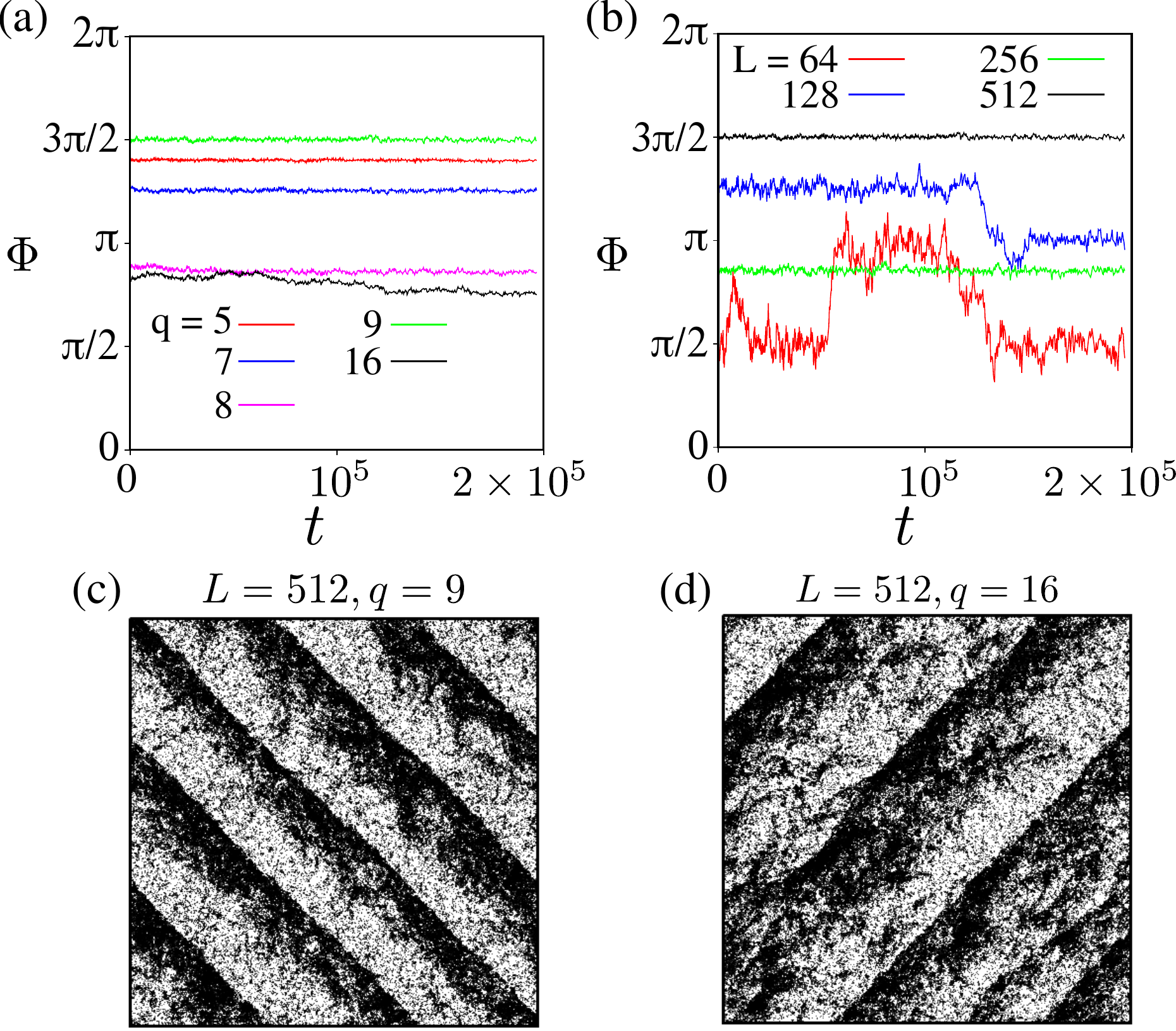}
    \caption{(Color online) Time evolution of the direction of global order $\Phi$ in the coexistence region for (a) $L=512$ and various $q$ and (b) $q=9$ and several values of $L$. (c--d) Snapshots showing multiple bands moving parallelly constituting a microphase-separated coexistence region for $q=9$ and $q=16$, respectively. System size $L=512$. Parameters: $\eta=0.3$, $\rho_0=1.5$, $v_0=0.5$.}
    \label{appfig8}
\end{figure}
Here, we analyze the pinned property of the system in the coexistence region and directly compare the outcome with the steady-state snapshots. In Fig.~\ref{appfig8}, the time series of the orientation of global order $\Phi$ is shown for fixed $L=512$ and various $q$ [Fig.~\ref{appfig8}(a)] and for fixed $q=9$ and several $L$ [Fig.~\ref{appfig8}(b)]. When the system size is fixed, we observe pinning to unpinning transition with $q$ whereas for fixed $q$, the reverse transition happens as the system size increases. Both observations are similar to the observations made regarding the time evolution of $\Phi$ in the DVM polar ordered phase [Fig.~\ref{figGOP}]. The snapshots shown in Fig.~\ref{appfig8}(c--d) without any spatial anisotropy exhibit a coexistence region with multiple bands signifying a microphase-separated region. Comparing the snapshots with the time evolution of $\Phi$ we notice that both the unpinning behavior of the DVM for $q=16$ and the pinning behavior for $q=9$ shows flocking with multiple parallelly moving bands. 

\subsection{Nature of the DVM ordered phase}
\label{appE}
\begin{figure}[!htbp]
    \centering
    \includegraphics[width=\columnwidth]{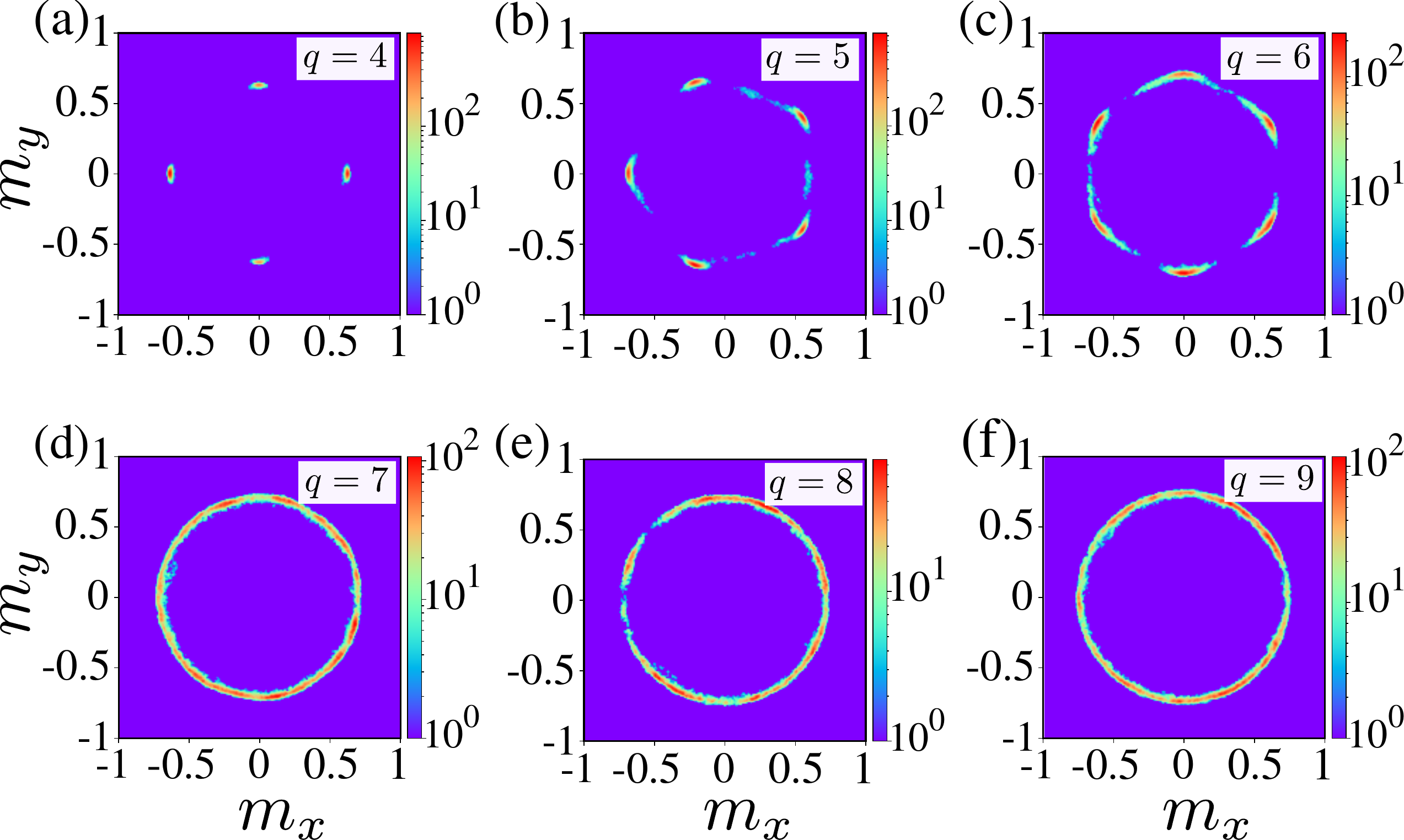}
    \caption{(Color online) Zero-activity limit of the order parameter distributions of the DVM with varying $q$ in the liquid phase. Parameters: $L=100, \eta=0.3$, and $\rho_0=6$. Distinct isolated spots in (a), (b), and (c) indicate LRO while ring-like distributions in (d), (e), and (f) are characteristic of the QLRO phase.}
    \label{appfig5}
\end{figure}

\begin{figure}[!htbp]
    \centering
    \includegraphics[width=\columnwidth]{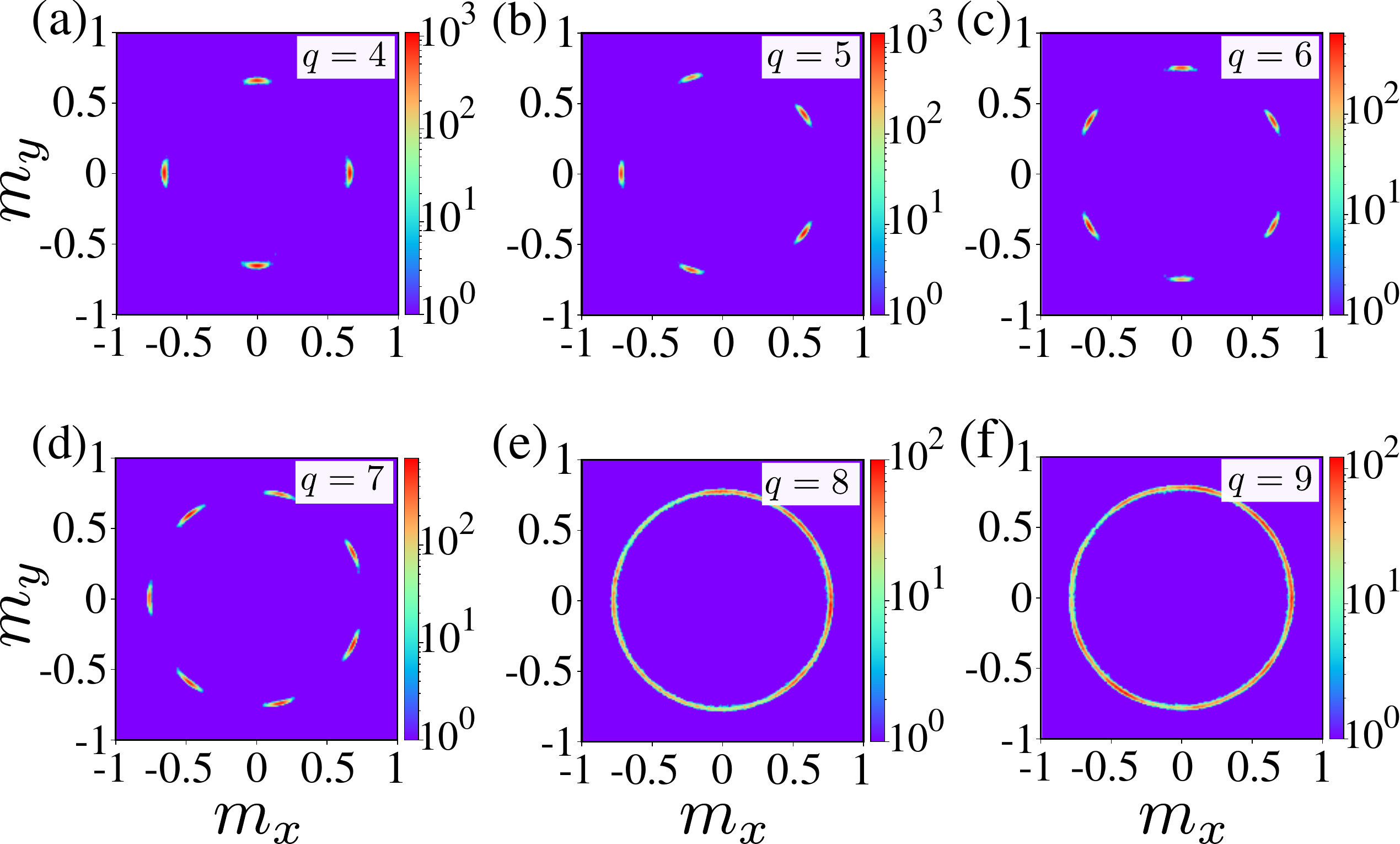}
    \caption{(Color online) Order parameter distributions of the DVM at constant activity $v_0=0.5$ with varying $q$ in the liquid phase. Parameters: $L=100, \eta=0.3$, and $\rho_0=6$.}
    \label{appfig6}
\end{figure}
In Fig.~\ref{appfig5}, we show the order parameter ({\bf m}) distribution on the $(m_x,m_y)$ plane for $v_0=0$. At this limit, particles can no longer move but modify their orientation according to Eq.~\eqref{sigma} and the $q$-state DVM reduces to the two-dimensional $q$-state clock model which shows two distinct phase transitions, one from disordered to QLRO phase at a higher temperature and the other from QLRO to LRO phase at a lower temperature for $q \geqslant 5$ \cite{clockmodel2018}. At large $q$, the LRO phase gradually starts to vanish, and for the two-dimensional XY model ($q \to \infty$), only one phase transition occurs (Kosterlitz-Thouless phase transition) from the disordered phase to the QLRO phase. As shown in Fig.~\ref{appfig5}, at the zero activity limit, the order parameter distribution shows $q$ distinct isolated spots (signifying LRO) for small $q$ but as $q$ increases ($q\geqslant 7$), ring-like distributions characteristic of the QLRO phase appears. These ring-like distributions become more pronounced as the system size is increased for large $q$. In contrast, for $v_0>0$ (see Fig.~\ref{appfig6}), the order parameter distribution exhibits LRO through isolated points of phase ordering as activity facilitates the broken symmetry phase. For $q>7$, a comprehensible LRO phase is observed at a large length scale limit as shown in Fig.~\ref{op_distro}. 
\begin{figure}[!htbp]
    \centering
    \includegraphics[width=\columnwidth]{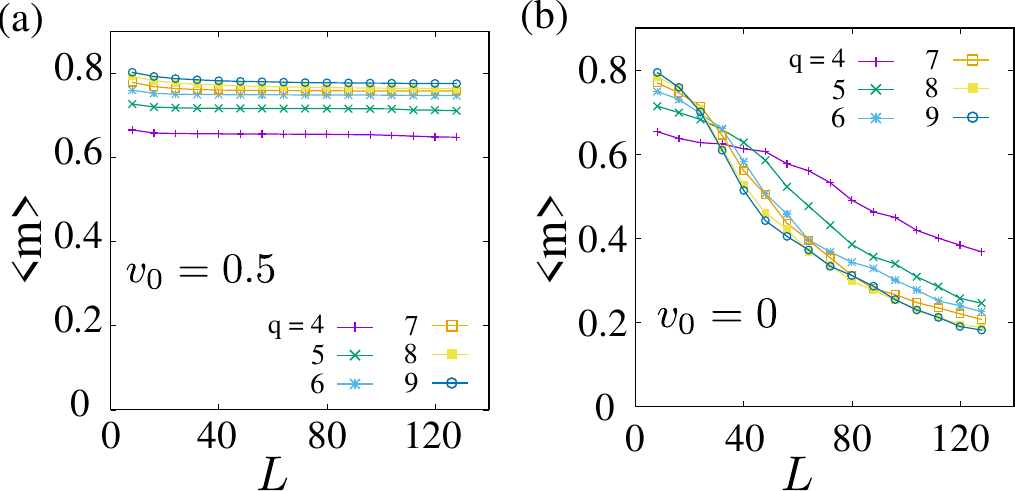}
    \caption{(Color online) Order parameter $\langle m \rangle$ versus system size $L$ for varying $q$ in the ordered liquid phase. (a) $\langle m \rangle$ is independent of $L$ for $v_0 = 0.5$ signifying LRO. (b) $\langle m \rangle$ decays algebraically with $L$ signifying QLRO in the zero-activity limit. Parameters: $\eta=0.3$, $\rho_0=6$.}
    \label{appfig7}
\end{figure}

To understand better the nature of ordering of the DVM liquid phase, we show the order parameter $\langle m \rangle$ against increasing system size $L$ for several $q$ in Fig.~\ref{appfig7}. The data presented are averaged over time and several initial configurations. We note that, $\langle m \rangle$ remains independent of the system size $L$ (actually, $m$ scales with $L$ as $m \sim L^{-\lambda}$, decays much slower than a power law) for all $q$ for $v_0 = 0.5$ [Fig.~\ref{appfig7}(a)] signifying LRO. As a result, the liquid phase of the constant-speed DVM is LRO and the direction of the order parameter exhibits a pinned behavior. For $v_0 = 0$, shown in Fig.~\ref{appfig7}(b), however, $\langle m \rangle$ is expected to algebraically decay to zero for $L\to \infty$ and this effect is more pronounced for larger $q$ because, for large $q$, the $v_0=0$ DVM approaches the two-dimensional XY model where the low-temperature phase is QLRO.

\chapter{Domain growth in a discrete flocking model with nonconserved kinetics}\label{chap:Chap5}
We undertake a numerical study of the ordering kinetics in the two-dimensional ($2d$) active Ising model (AIM), a discrete flocking model with a non-conserved scalar order parameter. We find that for a quench into the liquid-gas coexistence region and in the ordered liquid region, the characteristic length scale of both the density and magnetization domains follows the Lifshitz-Cahn-Allen (LCA) growth law: $R(t) \sim t^{1/2}$, consistent with the growth law of passive systems with scalar order parameter and non-conserved dynamics. The system morphology is analyzed with the two-point correlation function and its Fourier transform, the structure factor, which conforms to the well-known Porod’s law, a manifestation of the coarsening of compact domains with smooth boundaries. We also find the domain growth exponent unaffected by different noise strengths and self-propulsion velocities of the active particles. However, transverse diffusion is found to play the most significant role in the growth kinetics of the AIM. We extract the same growth exponent by solving the hydrodynamic equations of the AIM.

\section{Introduction}\label{introduction_cAIM}
Active matter systems involve the movement of large assemblies of individual active particles that consume energy to self-propel and exhibit collective behavior in a non-equilibrium steady state~\cite{ramaswamy,marchetti2013hydrodynamics,needleman2017active,gompper20202020}. Collective motion is ubiquitous in nature, observed in a wide array of different living systems over a range of scales, from macroscopic fields like fleets of birds~\cite{Ballerini} and schools of fish~\cite{Becco,Calovi} to microscopic scales like hoards of bacteria~\cite{Steager,Peruani}, cytoskeletal filaments and molecular motors~\cite{schaller2010polar,sumino2012large,sanchez2012spontaneous}. It leads to the emergence of ordered motion of large clusters, called flocks, with a typical size larger than an individual~\cite{ramaswamy,collectivemotion,shaebani2020computational,de2015introduction,krishnan2010rheology}. Over the past two decades, new models have emerged to understand the various physical principles governing active matter systems \cite{shaebani2020computational}.

Vicsek and collaborators years ago introduced a minimal model~\cite{Vicsek} of active particles that move with a constant speed and orient via a ferromagnetic interaction with a neighborhood similar to the XY model. In the Vicsek model (VM) activity can stabilize the ordered phase even in two dimensions which is not possible in the $2d$ XY model where the long-range fluctuation destroys the ordered phase following Mermin-Wagner theorem ~\cite{toner1995long,toner1998flocks}. Then, about a decade ago, Solon and Tailleur introduced the active Ising model (AIM)~\cite{AIM_solon_intro,solon2015flocking} where the continuous
rotational symmetry of the VM is replaced by a discrete symmetry. In the AIM, each particle assumes two possible states allowing the particle to propel in a preferred direction which changes upon interaction with other particles at the same lattice site.  The AIM retains the essential part of the VM physics and exhibits flocking behavior with three different phases at steady state: disordered gas at high noise and low densities, polar liquid at low noise and high densities, and a phase-separated liquid-gas coexistence state at intermediate densities. However, a key difference between the VM and the AIM arises in the steady-state behavior of the coexistence region. In this region, AIM shows a {\it macrophase} separation associated with normal density fluctuations whereas the VM is characterized by a {\it microphase} separation with giant density fluctuations. The flocking transition in the AIM is a first-order liquid-gas phase transition similar to the VM, however, for zero activity, despite the dynamics being non-equilibrium, the AIM shows a second-order phase transition belonging to the Ising universality class. 

Although significant progress has been made to understand the steady state properties of various active systems~\cite{toner2005170,giomi2013defect,MIPS,solonVM,criticalABP2018,capillary2020adam,Raja_APM,chatterjee2020flocking,kursten2020dry,fruchart2021NR,solon2022susceptibility,chatterjee2022polar,codina2022small,TSVM2023,karmakar2023jamming}, there is much to explore in the realm of ordering kinetics in active systems that relaxes to a non-equilibrium steady state (NESS). Understanding the intrinsic non-equilibrium dynamics that drive an active system towards its steady state is of fundamental as well as practical relevance. Unlike active systems, ordering kinetics in non-equilibrium passive systems have been studied over several decades~\cite{alan_bray,bray1993theory,sanjay_puri,aging2012,puri2014rfim,kumar2017ordering,rbcm}. Domain growth in passive systems with non-conserved scalar order parameters follows the Lifshitz-Cahn-Allen (LCA) growth law: $R(t) \sim t^{1/2}$ (\textit{Model A} of order-parameter kinetics) whereas passive systems with conserved order parameter follow a  Lifshitz-Slyozov-Wagner (LSW) growth law: $R(t) \sim t^{1/3}$ (\textit{Model B} of order-parameter kinetics), $R$ being average size of domains. Using tools that quantify the kinetics of passive systems, several active systems have been explored. These include Active Model B~\cite{wittkowski2014scalar,pattanayak2021AMB,pattanayak2021domain}, active nematics~\cite{mishra2014aspects}, self-propelled particles in disordered medium~\cite{das2018ordering}, Model B with nonreciprocal activity~\cite{saha2020scalar}, Kuramoto oscillators~\cite{rouzaire2022dynamics}, active Brownian particles~\cite{dittrich2023growth} and motility-induced phase separated (MIPS) clusters~\cite{caporusso2023dynamics}. Moreover, an interesting observation of multiple coarsening length scales was made in the prototypical VM~\cite{Coarsening2020VM} where velocities are found to align over a faster-growing length scale compared to density. Another intriguing result of an active system with a non-conserved vector order parameter following the growth law of the non-conserved scalar order parameter field has also been observed~\cite{Dikshit_2023}. Since AIM is a minimal flocking model with a rich phase behavior, studying the growth kinetics of this model will allow us to interpret the origin of large flocks in terms of microscopic interactions. 

In this chapter, we explore the phase ordering kinetics of the AIM with a non-conserved scalar order parameter. Quenching the AIM inside the spinodal region results in the formation of small positively or negatively magnetized clusters which in the late stage of the coarsening merge to form a single, macroscopic domain of one spin type~\cite{solon2015flocking}. A few questions arise in this context: (a) Does the domain morphology follow the same pattern and growth law for quenches into the coexistence and in the ordered liquid region? (b) Since the order parameter is non-conserved in AIM, how does the growth law relate to the established growth law of similar passive systems? (c) Do the density and magnetization align over the same length scale?~\cite{Coarsening2020VM} (d) What is the impact of noise and particle activity on the domain growth? and (e) What is the role of diffusion in the domain growth dynamics? We address these issues by analyzing the ordering dynamics of the $2d$ AIM on a square lattice via Monte Carlo simulations and by solving the AIM hydrodynamic equations using the finite difference method.

This chapter is organized as follows. In Sec.~\ref{model_cAIM} we discuss the model and then present the details of numerical simulations in Sec.~\ref{simulation}. In Sec.\ref{results}, we present the growth law of the AIM from both numerical simulation and hydrodynamic description. Finally, in Sec.~\ref{conclusion}, we conclude this chapter with a summary and discussion of the results.

\section{Model}\label{model_cAIM}
We consider $N$ particles on a two-dimensional square lattice $L\times L$ with periodic boundary conditions. Thus the average particle density is  $\rho_{0} = N/L^2$. Each lattice site $i$ can accommodate an arbitrary number of particles $n_i^{\sigma}$ with spin $\sigma =\pm 1$. Defining local density $\rho_i=n_i^{+} + n_i^{-}$ and magnetization $m_i = n_i^+ - n_i^-$ we note that $\rho_i$ has no upper bound, while $m_i$ is bounded by $\rho_i$: $-\rho_i \leq m_i \leq \rho_i$.  Each particle with a given spin state $\sigma$ can either flip to $-\sigma$ or jump to a nearest-neighbor site probabilistically.

The flipping rates are derived from a local ferromagnetic Potts Hamiltonian~\cite{Raja_APM,chatterjee2020flocking} defined as:
\begin{equation}\label{hamiltonian}
    H_i = -\frac{J}{2 \rho_i} \sum_{k=1}^ {\rho_i} \sum_{l \ne k} \left(q \delta_{\sigma_i^k, \sigma_i^l}  - 1 \right) \, ,
\end{equation}
where $J$ is the coupling between any two particles at site $i$ and the Kronecker delta $\delta_{\sigma_i^k, \sigma_i^l}$ survives only for $\sigma_i^k=\sigma_i^l$. Eq.~\eqref{hamiltonian} with $q = 2$ is the local Hamiltonian defined for the AIM~\cite{solon2015flocking}. Local interaction implies that a particle can align with the average direction of all other particles at the same site.

The spin-flip transition rates are derived from the Potts Hamiltonian $H_{\rm APM}=\sum_i H_i$ according to the energy difference before and after the spin flip. For $q=2$, a particle with spin $\sigma$ flips its state according to the transition rate ~\citep{Raja_APM,chatterjee2020flocking}:
\begin{equation}\label{flip}
    W_{\rm flip}(\sigma \rightarrow - \sigma) = \gamma \exp\left[-\frac{2\beta J}{\rho_i}\left(\sigma m_i - 1\right)\right] \, ,
\end{equation}
where $\gamma$ is the rate of particle flipping when $\sigma m_i = 1$. In this chapter, we choose $J=1$, and $\gamma=1$ without any loss of generality. The parameter $\beta$, denoted as ``inverse temperature'', $\beta=T^{-1}$, in passive systems, controls the flip noise strength. Although the system under consideration is athermal, we denote the parameter $T$ as ``temperature'' from now on. 

Moreover, each particle performs a biased diffusion on the lattice depending on the spin state $\sigma$. For $q=2$ particles perform a one-dimensional biased hopping along the $\pm x$-direction with the following hopping rate~\cite{Raja_APM,chatterjee2020flocking}:
\begin{equation}\label{hop}
    W_{\rm hop}(\sigma,p) = D\left[1+\epsilon({2\delta_{\sigma,p}-1})\right] \, ,
\end{equation}
where $p$ denotes the direction of bias-hopping ($p=+1$ along the $+x$ direction and $p=-1$ along the $-x$ direction). The presence of other particles does not influence hopping rates and hence is independent of particle density. The hopping rate $D=1$ is constant along the upward and downward directions ($\pm y$-direction). The parameter $\epsilon \in [0, 1]$ $(0 \leq \epsilon \leq q-1)$ controls the asymmetry between the purely diffusive limit $\epsilon = 0$ and the purely ballistic limit $\epsilon = 1$, while $D$ controls the overall hopping rate. On average, a particle drifts with speed $v = 2D\epsilon$ in the direction set by the sign of its spin state (where lattice spacing is 1), while the total hopping rate $4D$, remains constant.

\section{Simulation details}\label{simulation}
A Monte Carlo (MC) simulation of the stochastic process defined above evolves in unit Monte Carlo steps (MCS) $\Delta t$ resulting from a microscopic time $\Delta t/N$. During $\Delta t/N$, a randomly chosen particle with spin $\sigma$ flips with probability $W_{\rm flip}\Delta t$ or hops to one of the neighboring sites with probability $W_{\rm hop}\Delta t$. Consequently, $1 - [4D + W_{\rm flip}]\Delta t$ is the probability that the particle does nothing, and minimizing this we obtain $\Delta t=[4D + \exp(2\beta)]^{-1}$.

To study the morphology of the system during phase ordering kinetics, we use the two-point equal-time $(t)$ correlation function of the local scalar order parameters. The notion utilizes the spatial fluctuations in the density and magnetization fields to estimate~\cite{Dikshit_2023}:
\begin{equation}\label{correlation_density_cAIM}
    C_{\rho\rho}(r,t) =  \frac{1}{L^2} \sum_{i=1}^{L^2} \langle \Delta \rho_{i,t} \Delta \rho_{i+r,t} \rangle 
\end{equation}  
\begin{equation}\label{correlation_mag}
    C_{ mm}(r,t) =  \frac{1}{L^2} \sum_{i=1}^{L^2}  \langle \Delta m_{i,t} \Delta m_{i+r,t} \rangle \, ,
\end{equation}
where $\langle \cdots \rangle$ denotes averaging over independent initial realizations, $\Delta \rho_{i,t} = \rho_i - \rho_0$ and $\Delta m_{i,t}=m_i - m_0$ are the local fluctuations in density and magnetization from the mean, respectively. The above definition of $C_{\rho\rho}$ characterizes the morphology of the spatial structures and $C_{ mm}$ evaluates the correlations in the polar alignment among the evolving structures separated by a distance $r$. Since we observe that $C_{\rho\rho}$ and $C_{mm}$ behave similarly in the AIM (see below), we focus here on $C_{\rho\rho}$, which we denote as $C$ [$C_{\rho\rho}(r,t) \equiv C(r,t)$] from now on. Following a temperature quench from a random initial configuration into the ordered state, domains of both spin types appear and grow with time. Similar morphology of the evolving domains with average domain size $R(t)$ would correspond to a dynamical scaling relation \cite{sanjay_puri,alan_bray,bray1993theory}: 
\begin{equation}\label{correlation_cAIM}
    C(r,t) =  f \left (\frac{r}{R(t)} \right)
\end{equation}
where $f(x)$ is a time-independent scaling function. $R(t)$, estimated from the decay of $C(r,t)$ generally show a power-law growth \cite{sanjay_puri,alan_bray,bray1993theory}:
\begin{equation}\label{lengthscale_cAIM}
    R(t) \sim t^\theta \, ,
\end{equation}
with $\theta$ as the growth exponent. Typically, the morphology of an ordering system is studied by scattering experiments, which measure the structure factor $S(k,t)$, defined by the Fourier transform of the correlation function $C(r,t)$:
\begin{equation} \label{sf_cAIM}
    S(k,t)=\int_{-\infty}^{\infty} C(r,t) e^{ikr} dr \, ,
\end{equation}
and has a dynamical scaling form in $d$ dimensions:
\begin{equation}\label{sfscaling_cAIM}
    S(k,t) =  R(t)^{d} g\left[ kR(t) \right] \, .
\end{equation}
For scalar order parameters like the density field, the short-distance (large-$k$) behavior of the structure factor scaling function is given by Porod's law (for domains with smooth boundaries or scattering off sharp domain interfaces) which corresponds to $g(k) \sim k^{-(d+1)}$. 
Next, we present results for model parameters set by the average particle density $\rho_0$, temperature $T=\beta^{-1}$, self-propulsion velocity $\epsilon$, and diffusion constant $D$. 

\section{Results}\label{results}

\subsection{Phase diagram and domain morphology}
\begin{figure}[!t]
\centering
\includegraphics[width=0.9\columnwidth]{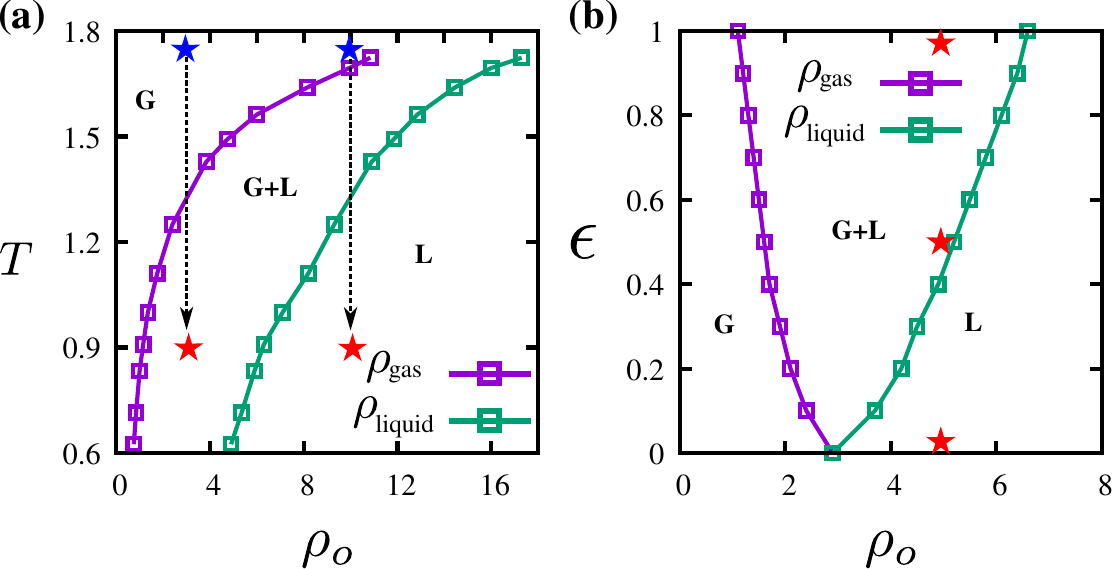}
\caption{(Color online) Phase diagrams of the AIM showing the quench directions at fixed densities. (a) $(T,\rho_0)$ phase diagram for self-propulsion velocity $\epsilon=1$. (b) $(\epsilon,\rho_0)$ phase diagram at fixed temperature $T=0.9$. G and L denote the disordered gas and polar liquid regions whereas G+L denotes the phase-coexistence region. The blue stars indicate the initial high-temperature points in the phase diagram from where the system is quenched (indicated by arrows) into the coexistence region ($\rho_0=3$) and to the liquid region ($\rho_0=10$) [red stars]. $\rho_{\rm gas}$ and $\rho_{\rm liquid}$ are the gas and liquid binodal, respectively.}
\label{fig:phase_diagram}
\end{figure}

The steady-state behavior of the AIM is summarized in the temperature-density ($T - \rho_0$) [Fig.~\ref{fig:phase_diagram}(a)] and velocity-density ($\epsilon - \rho_0$) [Fig.~\ref{fig:phase_diagram}(b)] phase diagrams. The general structure of the phase diagrams consists of a gas phase (G), a liquid phase (L), and a liquid-gas coexistence region (G+L) separated by the gas and liquid binodals. Qualitatively similar results were obtained earlier in Ref.~\cite{solon2015flocking}. However, we obtain a critical temperature $T_c \simeq 2$ (above which no phase separation occurs irrespective of the density, critical density $\rho_c=\infty$) twice as large as in Ref.~\cite{solon2015flocking}. This can be understood if one compares the flipping rate of Ref.~\cite{solon2015flocking}, $W_{\rm flip}(\sigma \rightarrow - \sigma) = \gamma \exp\left(-\sigma \beta \frac{m_i}{\rho_i}\right)$ with Eq.~\eqref{flip} for $J=1$ which denotes that the effective $\beta$ considered in this chapter is roughly twice as large as the $\beta$ of Ref.~\cite{solon2015flocking}. In the $\epsilon - \rho_0$ phase diagram, the two binodals converge at a critical density $(\rho_c \simeq 2.9)$ for vanishing self-propulsion ($\epsilon=0$) which signifies a second-order phase transition belonging to the Ising universality class~\cite{AIM_solon_intro}.

We chose to quench the random initial systems in two different regimes of the temperature-density phase diagram shown by the arrows in Fig.~\ref{fig:phase_diagram}(a). The quench occurs instantaneously from a very high-temperature regime ($T\rightarrow \infty$) shown by the blue stars to the liquid-gas coexistence region or the polar liquid region, denoted by red stars, with the same final temperature $T=0.9$ ($T<T_c$, corresponds to $\beta=1.1$) but different densities.

To quantify the coarsening dynamics, we performed simulations on a square lattice of size $400^2$ with periodic boundary conditions applied on both sides. Following quench at time $t=0$, the system is evolved up to $t=10^5$ using the MC algorithm described in Sec.~\ref{simulation}. All numerical data presented here are averaged over at least 300 independent realizations.

\begin{figure*}[!t]
\centering
    \includegraphics[width=\columnwidth]{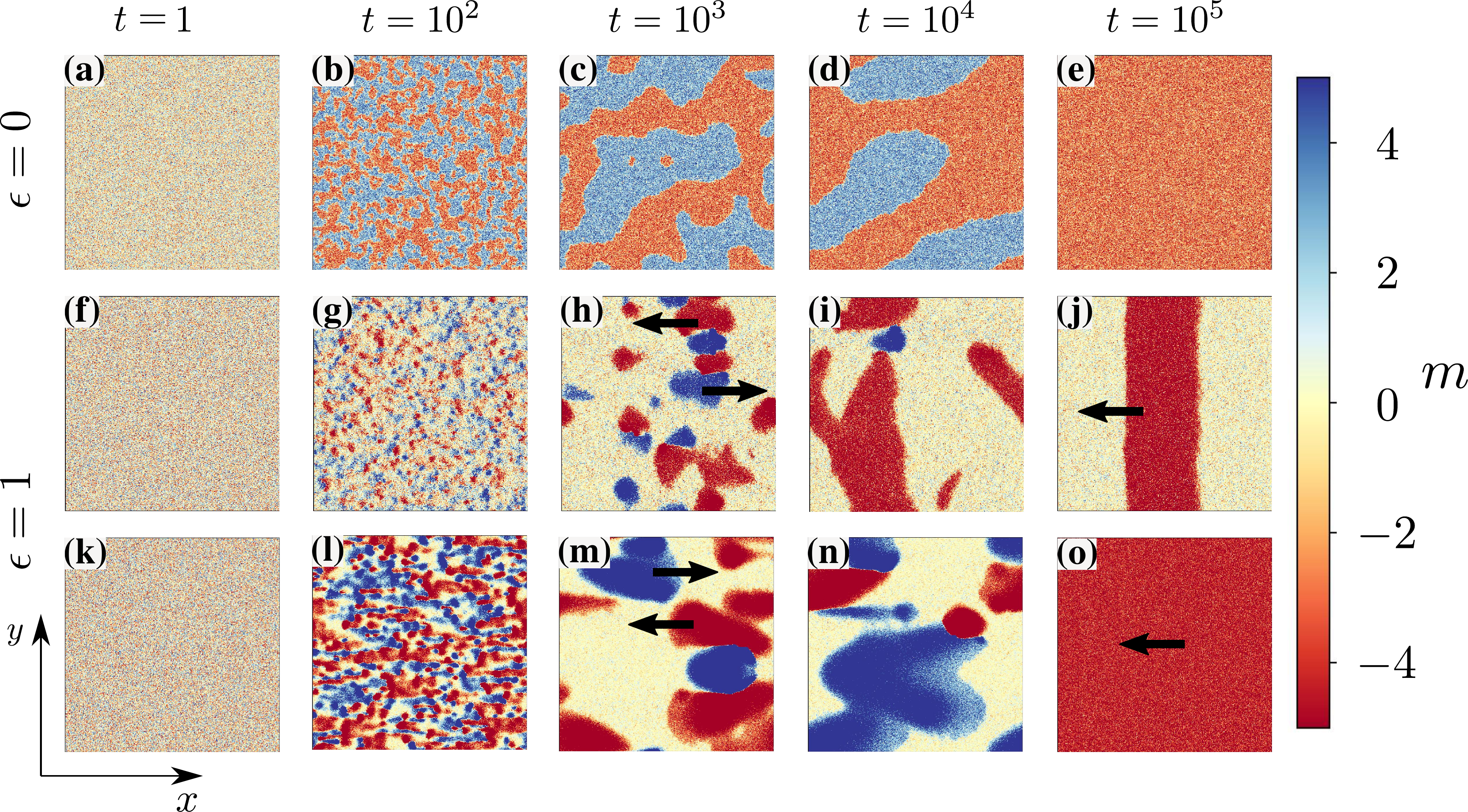}
    \caption{(Color online) Time evolution snapshots of the local magnetization field on a $400^2$ system after a quench from $T=\infty$ to $T=0.9$ for $\epsilon=0$ (top panel) and $\epsilon=1$ (middle and lower panel). The color bar denotes magnetization per site. (a--e) Curvature-driven domain growth of the diffusive Ising model ($\rho_0=5$, $\epsilon=0$). (f--j) Growth dominated by the dynamics of spinodal decomposition after the system is quenched inside the spinodal region ($\rho_0=3$, $\epsilon=1$) and (k--o) domain growth mediated by the merging of high-density clusters of $\sigma=\pm 1$ after the system is quenched deep inside the ordered liquid regime ($\rho_0=10$, $\epsilon=1$). Black arrows denote the direction of the movement of the clusters and bands.}
    \label{fig:dynamics_snap}
\end{figure*}

A typical simulation of coarsening dynamics starts with a homogeneous initial configuration where particles with $\sigma=\pm 1$ are distributed randomly on each lattice site. This can also be described as an equilibrium configuration at infinite temperature because all configurations are equally likely. Then, after a quench well below the critical temperature ($T_c \simeq 2$), the homogeneous initial configuration starts evolving in time, and subsequent dynamics are governed by the formation and growth of $\sigma=1$ and $\sigma=-1$ rich domains. Such a time evolution of the local magnetization field is shown in Fig.~\ref{fig:dynamics_snap} for $\epsilon=0$ and $\epsilon=1$. In the latter scenario, time evolution has been shown for a quench in both the coexistence (middle row) and liquid regions (bottom row). The non-equilibrium steady states (NESS) are characterized by a single band (at lower density) and a polar liquid (at higher density).

The top panel, Fig.~\ref{fig:dynamics_snap}(a--e), depicts the ordering kinetics of a purely diffusive $(\epsilon=0)$ polar liquid at $\rho_0=5$. The domain morphology with increasing time exhibits a close resemblance with the passive Ising model~\cite{2dIsing_growth}. In the Ising model, the driving force for the domain growth is the curvature of the domain wall since the system surface energy can only decrease through a reduction in the total net surface area. In the $\epsilon=0$ limit, particles do not form high-density domains due to the diffusive movement of particles. Therefore, density-wise, the system remains homogeneous as we observe a steady growth of small, high-curvature to large, low-curvature domains. This is unsurprising as the $2d$ AIM for $\epsilon=0$ belongs to the same universality class of the passive $2d$ Ising model~\cite{solon2015flocking}. 

Next, we looked at the evolution of AIM with self-propulsion velocity ($\epsilon=1$)  quenching the system inside the spinodal  [Fig.~\ref{fig:dynamics_snap}(f--j)] and homogeneous ordered regions [Fig.~\ref{fig:dynamics_snap}(k--o)]. The average densities representing these two regions correspond to $\rho_0=3$ and 10, respectively. Inside the spinodal region, the growth dynamics are driven by spinodal decomposition and result in the formation of numerous small clusters of negative and positive spins $(t=10^2)$. The coarsening then stems from the merging of these clusters $(t=10^3$ and $t=10^4)$, until a single, macroscopic domain emerges $(t=10^5)$ in the steady state. A quench deep inside the homogeneous ordered region also results in similar dynamics of cluster formation and coalescence of those clusters into a single large liquid domain (in this chapter, for $\epsilon>0$, the word cluster is used interchangeably with domain). With extreme self-propelled particles ($\epsilon=1$), although the high-density clusters are strongly biased along the horizontal directions, they can also grow along the transverse direction due to the constant transverse diffusion $D$. When these clusters merge into a larger cluster, it always tries to minimize the surface energy by decreasing the surface area. Accordingly, the coarsening process leads to a single band [Fig.~\ref{fig:dynamics_snap}(j)] with domain walls having the lowest curvature (the curvature of a straight line is zero but a liquid band with a perfectly straight domain boundary can only happen when there is no thermal fluctuation). The focus of this study is, therefore, to analyze the coarsening dynamics shown in Fig.~\ref{fig:dynamics_snap} which we will do next.

\subsection{Dynamical scaling}
\begin{figure*}[!t]
\centering
    \includegraphics[width=0.7\columnwidth]{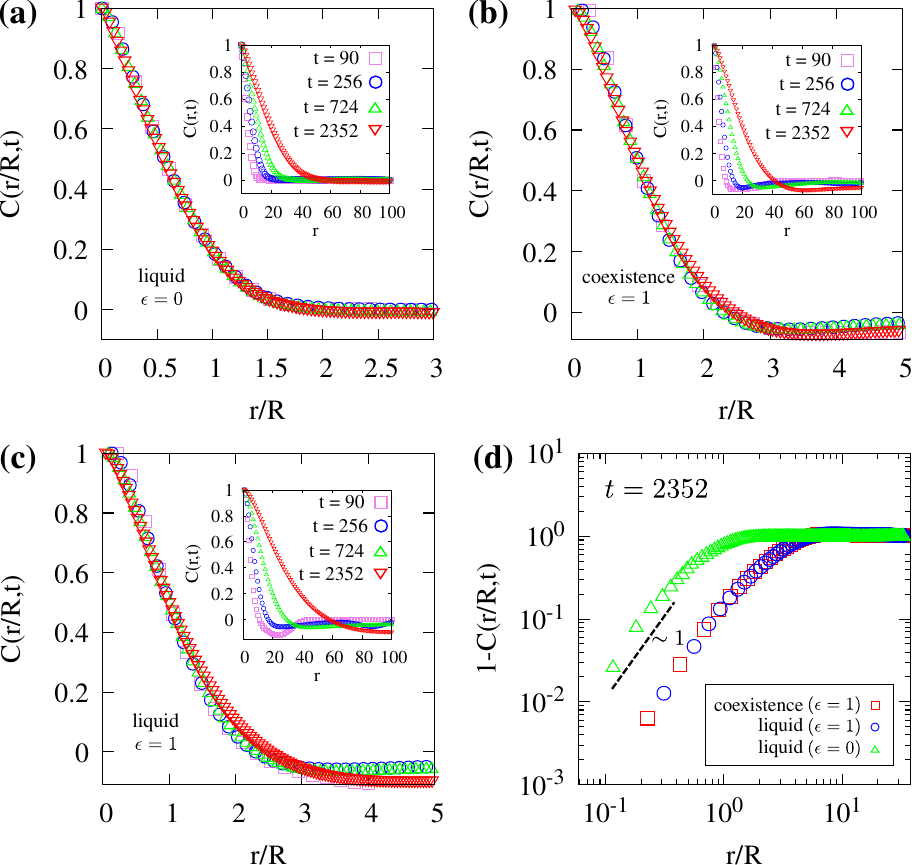}
    \caption{(Color online) (a--c) Scaled correlation functions $C(r/R,t)$ versus $r/R$, for the evolution of the $2d$ AIM shown in Fig.~\ref{fig:dynamics_snap}. Unscaled versions are shown in the inset. ``Liquid'' and ``coexistence'' signify the region where the system is quenched, $\epsilon$ denotes particle speed. (d) Log-log plot of $1-C(r/R)$ versus $r/R$. The cusp exponent is estimated as $\alpha \sim 1$. Parameters correspond to Fig.~\ref{fig:dynamics_snap}.}
    \label{fig:correlation_plot}
\end{figure*}

\begin{figure*}[!t]
    \includegraphics[width=\columnwidth]{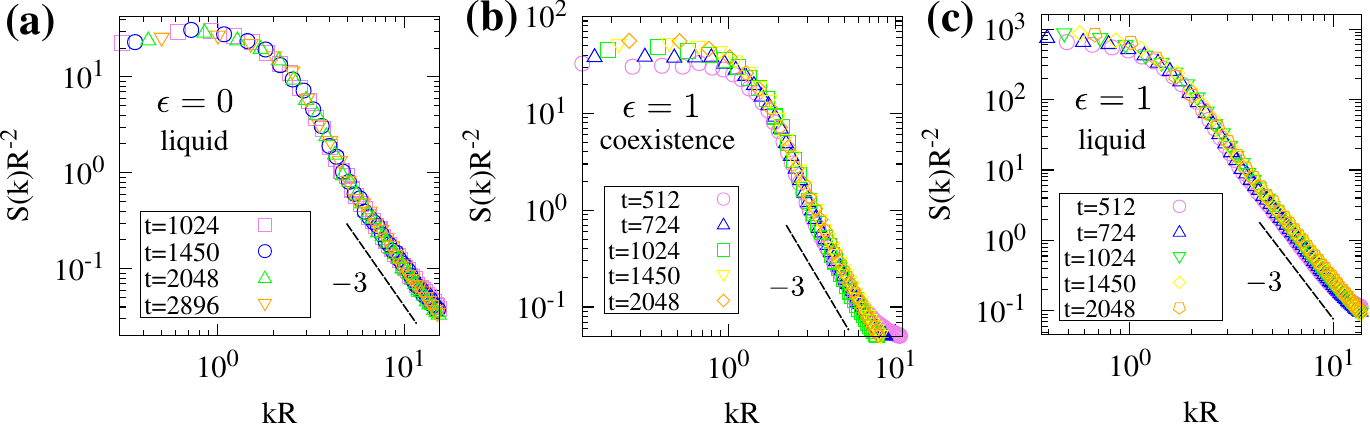}
    \caption{(Color online) (a--c) Scaled structure factors, $S(k,t)R(t)^{-2}$ versus $kR$, for the Fourier transform of the correlation function data sets corresponding to the same values of time.  The line of slope $\simeq -3$ denotes the Porod's law: $S(k)\sim k^{-(d+1)}$ for $d=2$.}
    \label{fig:struc_plot}
\end{figure*}
In the theory of phase-ordering kinetics, the scaling hypothesis states that if the system is characterized by a single length scale $R(t)$ [$R(t)$ is equivalent to the average domain size], the domain morphology is statistically the same at all times, apart from a scale factor. When all domain lengths are measured in units of $R(t)$, the equal-time pair correlation function should exhibit the dynamical-scaling property of Eq.~\eqref{correlation_cAIM}. The equal-time correlation function is a non-equilibrium quantity so as $R(t)$ which can be estimated from the decay of the correlation function. To identify the self-similar behavior of the evolving domains we plot the correlation function in Fig.~\ref{fig:correlation_plot}. The scaled $C(r/R,t)$ and unscaled $C(r,t)$ correlation functions are shown in Fig.~\ref{fig:correlation_plot}(a--c) for  $\epsilon=0$ and $\epsilon=1$. $R(t)$ is determined from the distance over which the correlation function decays to e.g., 0.2 of its maximum value, that is, $C(r,t)=0.2C(0,t)$. 

As the system coarsens, the correlation function decays slowly [insets of Fig.~\ref{fig:correlation_plot}(a--c)], signifying the growth of the characteristic length scale $R(t)$. Upon rescaling the spatial coordinates by this length scale, the correlation function at different times collapses onto a single function $C(r/R,t)$ as shown in Fig.~\ref{fig:correlation_plot}(a--c), thus confirming a universal coarsening behavior with time. Such scaling behavior implies that the structure is time-invariant and consistent with a power-law growth of $R(t)$ with increasing $t$. This scaling hypothesis is satisfied when the growing length is much smaller than the system size to avoid the finite size effect. We further examine the AIM domain morphology by approximating the small distance behavior of the scaled two-point correlation function 
\begin{equation}\label{cusp}
  1-C(r) = \Bar{C}(r) \sim r^\alpha 
\end{equation}
 in Fig.~\ref{fig:correlation_plot}(d) which yields the cusp exponent $\alpha \sim 1$. This signifies the existence of sharp domain interfaces [see Fig.~\ref{fig:dynamics_snap}(h) and Fig.~\ref{fig:dynamics_snap}(m)] and translates into the power-law behavior of the scaled structure factor plotted in Fig.~\ref{fig:struc_plot}~\cite{puri2014rfim}. Fig.~\ref{fig:correlation_plot}(d) also signifies that the domain structure of the AIM for $\epsilon=1$ is statistically self-similar for quenches into the coexistence and liquid region whereas different from the domain morphology of the AIM for $\epsilon=0$, as evident from Fig.~\ref{fig:dynamics_snap}. 

In Fig.~\ref{fig:struc_plot}, we plot the scaled structure factor, $S(k,t)R(t)^{-2}$ versus $kR$, which is the Fourier transform of the correlation function. In Fourier space, Eq.~\eqref{cusp} translates into the following power-law behavior of the structure factor: $S(k)\sim k^{-(d+\alpha)}$ and therefore, the large-$k$ behavior of the structure factor tail generates a slope $\sim -3$ (in log-log plot) for $d=2$ and $\alpha=1$ which denotes ``Porod’s decay'': $S(k)\sim k^{-(d+1)}$, associated with scattering from sharp interfaces~\cite{alan_bray,bray1993theory}. This naturally originates from the long-range ordering in AIM leading to compact high-density clusters with smooth boundaries. Domains with rough morphologies having fractal interfaces do not follow Porod's law and the large-$k$ tail of the scaled structure factor yields a non-integer exponent~\cite{chatterjee2018clock}. A similar violation of the Porod law was also observed in the coarsening of the VM due to the irregular morphology associated with the cluster boundaries \cite{Supravat2012spatstructGNF,Coarsening2020VM}. $C(r,t)$ [and consequently, $S(k,t)$] exhibits two distinct power laws for small and large $r/R$ limits [large and small $kR$ limits] in the VM~\cite{Supravat2012spatstructGNF} which we do not observe in the AIM. Furthermore, density fluctuations might also play a role in determining whether a system will follow or violate Porod’s law. In the VM, giant density fluctuations break large liquid domains and restrict the formation of large compact domains (which eventually manifests in the microphase separation of the coexistence region in the steady-state)~\cite{solonVM} and might be responsible for the non-Porod behavior of the system~\cite{Supravat2012spatstructGNF}. On the other hand, AIM obeys the Porod behavior where density fluctuations are normal in the liquid phase~\cite{solon2015flocking} and the steady-state manifests a bulk phase separation.

\subsection{Growth law}
\begin{figure*}[!t]
    \centering
    \includegraphics[width=\columnwidth]{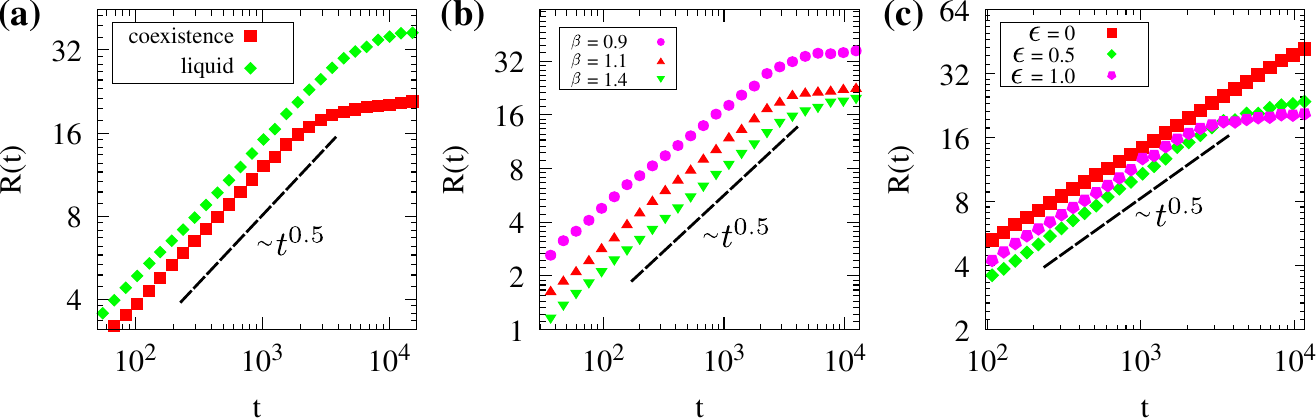}
    \caption{(Color online) (a) $R(t) \sim t^{1/2}$ for a quench into the coexistence and liquid regimes. The late time saturation of $R(t)$ signifies that the system has reached the corresponding NESS. As the density is higher for the liquid quench, the saturation appears late. Parameters: $\rho_0=3$ (coexistence), $\rho=10$ (liquid), $L=400$, $\beta=1.1$ and $\epsilon=1$. (b) $R(t)$ versus $t$ for different thermal noises ($T=\beta^{-1}$, coexistence regime). Parameters: $L=400$, $\rho_0=3$ and $\epsilon=1$. (c) Role of self-propulsion on the characteristic length $R(t)$ as a function of $t$. For a larger $\epsilon$, the system reaches the NESS faster while for $\epsilon=0$, the corresponding dynamics is slow. Parameters: $L=400$, $\beta=1.1$, and $\rho_0=5$. The growth law $R(t)\sim t^{1/2}$ is unaffected by thermal fluctuations and particle activity.}
    \label{fig:growth_plot}
\end{figure*}
In passive systems with non-conserved scalar order parameters, the late-stage domain growth is governed by the {\it diffusive} Lifshitz-Cahn-Allen (LCA) growth law $R(t) \sim t^{1/2}$~\cite{bray1993theory}. For non-conserved systems described by scalar fields such as the Ising model, the growth process is driven by the diffusion of the domain walls (the simplest form of topological defect) caused by the local changes in the order parameter. Diffusion also influences coarsening in non-conserved systems with vector fields, such as the $2d$ XY model, where domain evolution occurs when point topological defects, such as vortices and anti-vortices, diffuse, interact, and annihilate~\cite{yurke1993coarsening}. Both systems exhibit a diffusive growth exponent $\theta=\frac{1}{2}$ (in the XY model, $R(t) \sim (t/\ln t)^{1/2}$, the logarithmic correction is due to the free vortices~\cite{yurke1993coarsening}). Therefore, a 0.5 growth exponent signifies a coarsening process dominated by the diffusion of defects. For example, if the time evolution of a non-conserved scalar order parameter such as the magnetization $m$ follows the diffusive equation $\Dot{m} = D\nabla^2 m$, then the length scale will exhibit a $\sqrt{t}$ dependence.

Therefore, the immediate question one can ask is whether the LCA growth law also governs the late-stage coarsening dynamics of an {\it active} system with non-conserved scalar order parameters such as the AIM and whether the coarsening process is diffusion-dominated. To characterize this, we plot the length scale data with time in Fig.~\ref{fig:growth_plot} for various control parameters and quench regimes. We find that the late-stage growth kinetics of the domain in the coexistence and liquid regimes exhibit a $R(t) \sim t^{1/2}$ growth law [Fig.~\ref{fig:growth_plot}(a)]. We extract the same growth law for different temperatures [Fig.~\ref{fig:growth_plot}(b)], and self-propulsion velocities  [Fig.~\ref{fig:growth_plot}(c)]. Domains identified by the correlation of density and magnetization fields also show similar growth behavior (see Auxiliary material~\ref{appA_cAIM}). An interesting feature of the domain growth kinetics is that the domain size for a quench into the liquid regime is larger than the corresponding domain size in the coexistence regime [see Fig.~\ref{fig:dynamics_snap} at  $t=10^3$ and Fig.\ref{fig:growth_plot}(a)], although the domain morphologies for both these quench regimes are self-similar. The larger domains in the liquid regime are likely to be arising from higher density. Also, notice that a larger $\beta$ in Fig.~\ref{fig:growth_plot}(b) signifies a reduced thermal noise that slows down the local ordering of spins. Thus smaller $\beta$ allows the formation of larger domains. Nevertheless, different thermal fluctuations exhibit the same growth as thermal noise is asymptotically irrelevant for ordering in systems that are free from disorder~\cite{sanjay_puri}.

Fig.~\ref{fig:growth_plot}(c) shows the increasing length scale with time for different $\epsilon$. For $\epsilon=0$, the system is purely diffusive, and domain growth proceeds via the coarsening of connected domains (curvature-driven growth facilitated by diffusing domain wall) [see Fig.~\ref{fig:dynamics_snap}(c--d)]. The domain morphology looks very similar to the evolution of an Ising ferromagnet quenched below the critical temperature. Thus, a $\sim t^{1/2}$ growth law for $\epsilon=0$ similar to the pure Ising model is physically plausible.  But, when $\epsilon>0$, the domains of each spin no longer remain connected and form high-density clusters that self-propel along the horizontal direction. As time progresses, these clusters spread in the transverse direction due to constant diffusion ($D=1$) and merge with other clusters. Therefore, the domain growth for $\epsilon>0$ is again a diffusive phenomenon, and consequently, the growth kinetics exhibits a $R(t) \sim t^{1/2}$ growth law similar to $\epsilon=0$. We have identified this novel mechanism of diffusion-driven domain growth in the AIM after a thorough investigation of the altered diffusion coefficient in Sec.~\ref{diffusion} and Auxiliary material~\ref{appB_cAIM}. 

\begin{figure}[!t]
    \centering
    \includegraphics[width=0.8\columnwidth]{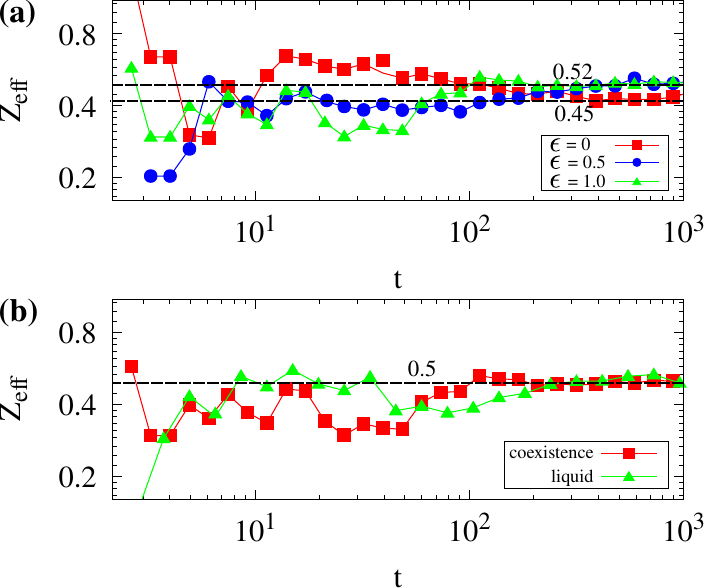}
    \caption{(Color online) Effective growth exponent $Z_{\rm eff}$ versus time $t$ for (a) different self-propulsion ($\epsilon=0, 0.5$, and $1$, $\beta=1.1$ and $\rho_0=3$) and (b) quenching into different NESS ($\rho_0=3$ for coexistence and $\rho_0=10$ for liquid, $\beta=1.1$ and $\epsilon=1$). The dashed lines are a guide to the eyes.}
    \label{fig:zeff_plot}
\end{figure}

It should also be noted that while approaching the NESS via coarsening, the high-density AIM domains of individual spins, upon merging, try to minimize the surface energy by decreasing the surface area similar to the dynamics of the passive non-conserved scalar order parameter. Therefore, although the system is active, the critical mechanisms (diffusion-dominated growth and minimization of surface energy during coarsening) of domain coarsening in AIM are similar to the $2d$ Ising model, and thus, it is not surprising that we extract a $R(t) \sim t^{1/2}$ growth law for both the passive and active models. For a more precise quantification of the asymptotic growth law, we determine the effective growth exponent, defined as:
\begin{equation}\label{zeff}
    Z_{\rm eff} = \frac{d [\ln R(t)]}{d[\ln t]} \, .
\end{equation}
In Fig.~\ref{fig:zeff_plot}(a), we plot $Z_{\rm eff}$ versus $t$ for different $\epsilon$ corresponding to the data in Fig.\ref{fig:growth_plot}(c). All data show extended flat regimes at late times (after $\sim t>10^2$). The effective exponent turns out to be in the regime $0.45<Z_{\rm eff}<0.52$, denoted by the dashed lines. We perform a similar study for the quench into the coexistence and liquid regimes. Fig.~\ref{fig:zeff_plot}(b) shows $Z_{\rm eff}$ vs. $t$ once again confirms $Z_{\rm eff}\sim 0.5$, irrespective of the quench regimes. 

\subsection{Role of transverse diffusion}
\label{diffusion}
\begin{figure*}[!t]
    \centering
    \includegraphics[width=0.7\columnwidth]{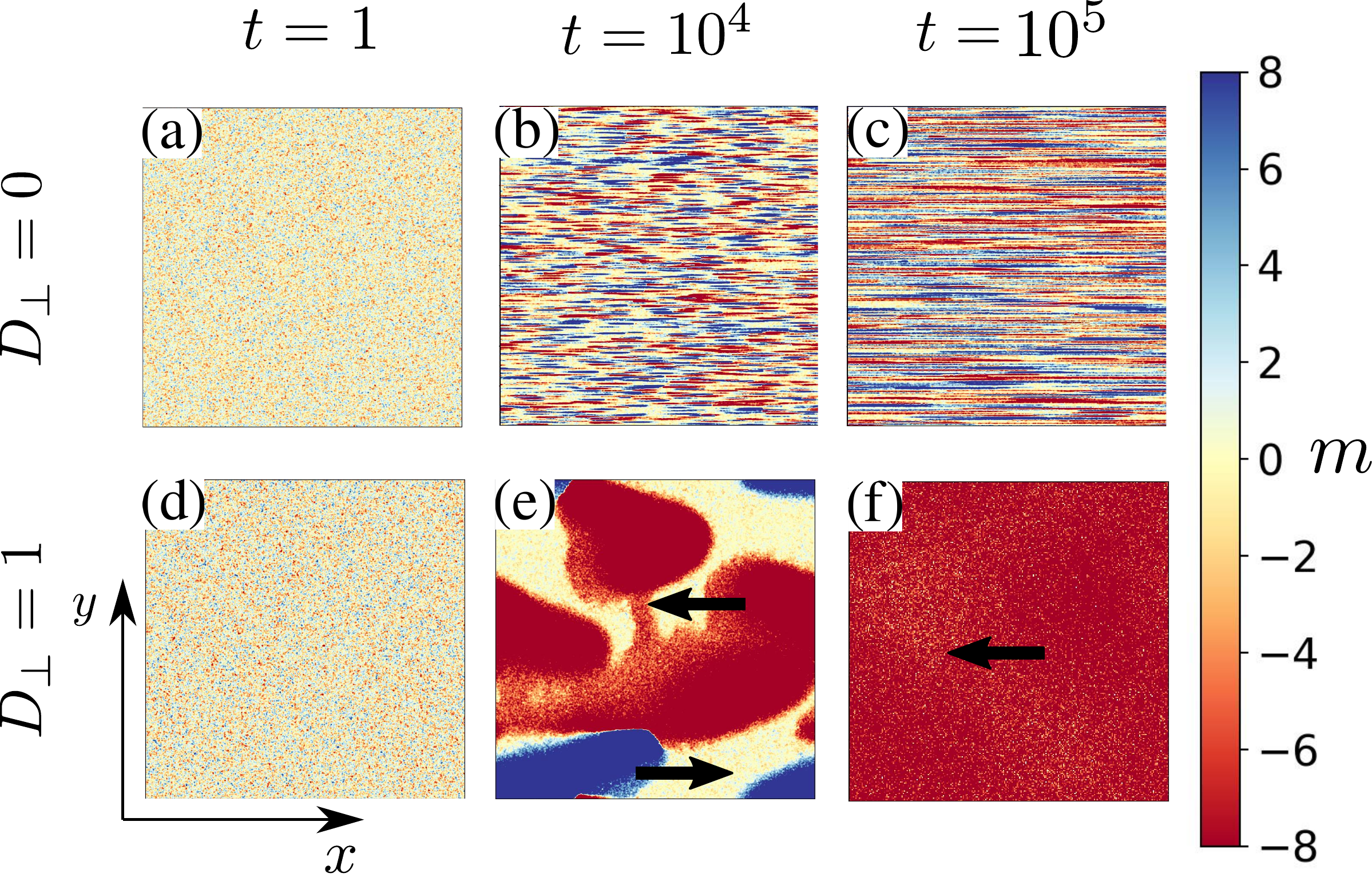}
    \caption{(Color online) Time evolution of the local magnetization field for $D_{\perp}=0$ [top panel, (a--c)] and $D_{\perp}=1$ [bottom panel, (d--f)] after quenching the system from a disordered gaseous phase to a high-density region. Parameters: $L=400, \beta=1.1$, $\rho_0=10$, and $\epsilon=1$.}
    \label{fig:D_zero_snap}
\end{figure*}
In the AIM, particles self-propel only in the horizontal direction with average velocity $2D\epsilon$. The directional hopping of the particles is a function of the spin type ($+\sigma$ or $-\sigma$). To test the hypothesis that the late time domain growth in the AIM is also a diffusion-driven process, we decompose the diffusion into two components, $D_{\perp}$ (along $\pm y$ direction) and $D_{\parallel}$ (along $\pm x$ direction). We vary $D_{\perp}$, keeping $D_{\parallel}=D=1$ henceforth, and show the domain evolution in Fig.\ref{fig:D_zero_snap}. 

The time evolution of the domains for $D_{\perp}=0$ is shown in Fig.~\ref{fig:D_zero_snap}(a--c). Starting from a disordered configuration, the time evolution of the system progresses via the formation of $1d$ rings with domains of alternating polarity along the horizontal direction ($x-$direction). These domains are one lattice unit wide along the transverse direction ($y-$direction) and are independent of the neighboring rings [Fig.~\ref{fig:D_zero_snap}(b)]. As particles can not diffuse along the transverse direction, these domains can only grow horizontally without merging with the neighboring rings. At late times, we observe narrow horizontal stripes of alternating magnetization [Fig.~\ref{fig:D_zero_snap}(c)]. Therefore, AIM with $D_{\perp}=0$ does not manifest the observed domain morphology representative of the growth law $\sim t^{1/2}$. $D_{\perp}=0$ also signifies $L_y$ numbers of one-dimensional periodic rings on which the AIM is defined. Such $1d$ AIM has been found to display flocking of a single dense ordered aggregate at intermediate temperatures (this flock undergoes stochastic reversals of its magnetization with time) but an aster phase consisting of sharp peaks of positive and negative magnetizations in a jammed state at lower temperatures~\cite{benvegnen2022flocking}. The one lattice unit-wide (along $y$) domains in Fig.~\ref{fig:D_zero_snap}(c) manifest the characteristics of the flocking state of $1d$ AIM for the given parameters (see Auxiliary material~\ref{appB_cAIM} for details) but the system does not exhibit flocking as a whole since the $L_y$ number of $1d$ rings do not interact with each other for $D_{\perp}=0$. Altering the transverse diffusion $D_{\perp}$ to a nonzero value, particle diffusion occurs along the transverse $\pm y$ direction. Therefore, the random initial state coarsens to form high-density clusters which coarsen further to give rise to a large flocking domain [Fig.~\ref{fig:D_zero_snap}(d-f)]. 

\begin{figure}[!t]
    \centering
    \includegraphics[width=0.6\columnwidth]{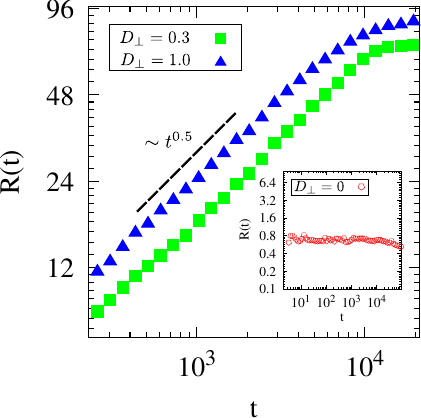}
    \caption{(Color online) $R(t)$ versus $t$ (on a log-log scale) showing a 0.5 growth exponent for different transverse diffusion $D_{\perp}=0.3$ and $D_{\perp}=1$. Inset: $R(t)$ versus $t$ in the absence of any transverse diffusion ($D_{\perp}=0$) showing no growth of domains with time. Parameters: $L=400$, $\beta=1.1$, $\epsilon=1$, and $\rho_0=8$.}
    \label{fig:diff_D}
\end{figure}

To quantify the role of the magnitude of transverse diffusion coefficient ($D_{\perp}$), we plot $R(t)$ versus $t$ in Fig.~\ref{fig:diff_D} for two different values of $D_{\perp}$. The plot shows that the system exhibits the power-law growth of $R(t)\sim t^{1/2}$ for different values of $D_{\perp}$. Interestingly, even a small $D_{\perp}=0.3$ is sufficient to drive the proper domain growth. A larger $D_{\perp}$ only increases the average domain size. As discussed earlier in Fig.~\ref{fig:D_zero_snap}(b--c), a vanishing $D_{\perp}$ can not initiate a domain growth as $R(t)$ remains constant with $t$ (see inset of Fig.~\ref{fig:diff_D}). Therefore, transverse diffusion plays the most crucial role in the AIM growth kinetics.

\subsection{Growth law from the hydrodynamic description of the AIM}
\label{hydro_cAIM}
In this section, we want to investigate whether the continuous description of the AIM also manifests the same time dependency $(\sim t^{1/2})$ of the coarsening length scale as observed in our numerical analysis. We consider {\it refined} mean-field equations similar to Ref.~\cite{solon2015flocking} for the spatiotemporal evolutions of the density $(\rho)$ and magnetization $(m)$ fields:
\begin{gather}
    \Dot{\rho}=D\nabla^2\rho-v\partial_x m \, , \label{rmf1} \\ 
    \Dot{m}=D\nabla^2 m-v\partial_x \rho + 2\left(2\beta-1-\frac{r}{\rho}\right)m-\alpha\frac{m^3}{\rho^2} \, , \label{rmf2}
\end{gather}
where $v=2D\epsilon$, $\alpha=4\beta^2(1-2\beta/3)$, and $r=3\alpha\alpha_m/2$ is a positive function of $\beta$~\cite{solon2015flocking}. In constructing Eq.~\eqref{rmf2}, we slightly modify the flipping rate equation of Eq.~\eqref{flip} to read $W_{\rm flip}=\exp\left(-2\beta\sigma m/\rho\right)$. For the mean-field equations, $r = 0$, which can not capture the AIM physics correctly as the system always exhibits homogeneous profiles (gas and liquid), and the inhomogeneous phase-separated profiles are never observed~\cite{solon2015flocking}. In Eqs.~\eqref{rmf1} and \eqref{rmf2}, local fluctuations are taken into account, which are generally neglected in the mean-field approximations~\cite{solon2015flocking}.
\begin{figure}[!t]
    \centering
    \includegraphics[width=0.8\columnwidth]{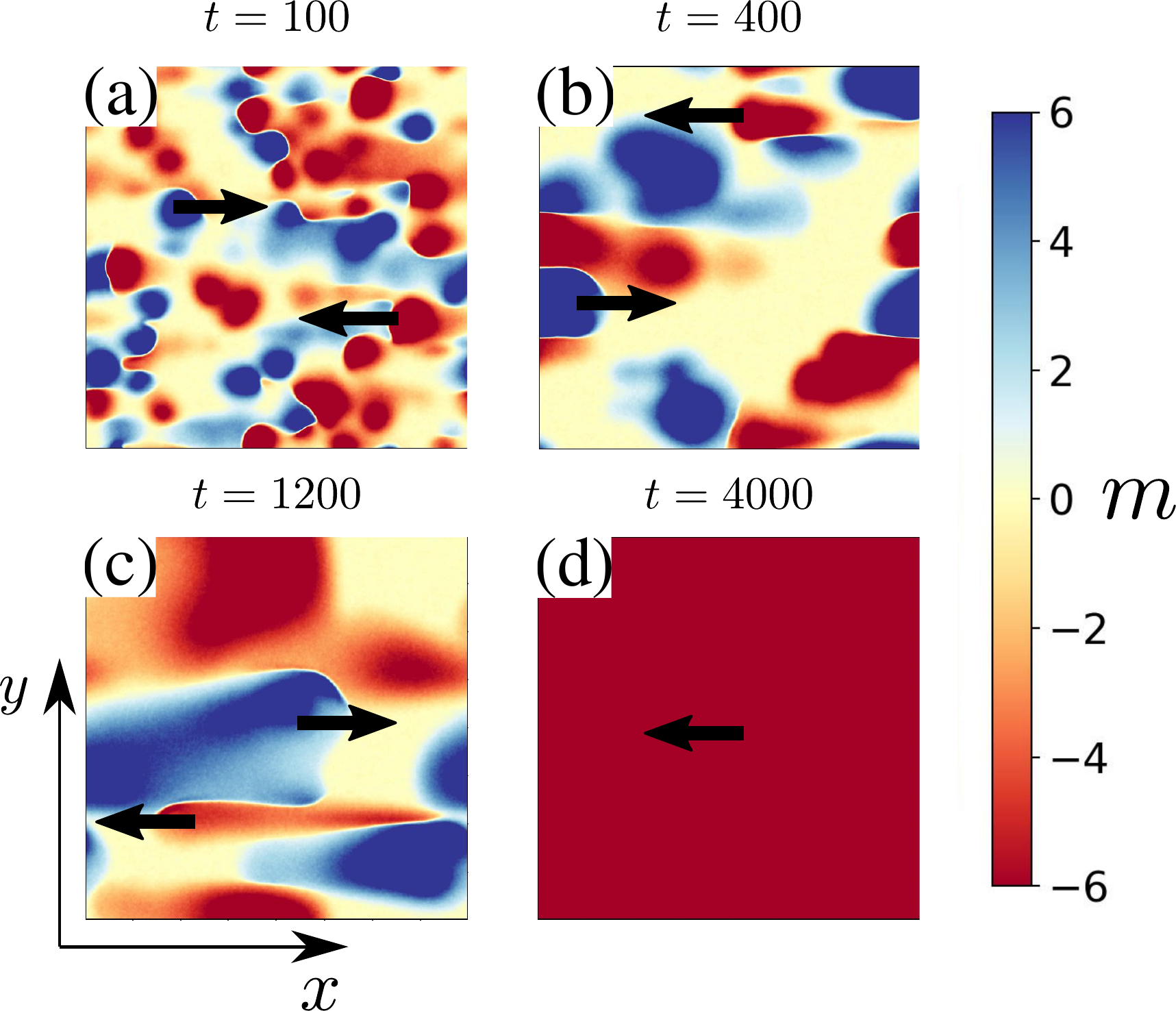}
    \caption{(Color online) Time evolution snapshots of the magnetization field $m$ obtained by solving the AIM hydrodynamic equations Eq.~\eqref{rmf1} and Eq.~\eqref{rmf2} after a quench from the homogeneous gaseous phase to an ordered liquid phase. Parameters: $L=400$, $\beta=0.75$, and $\rho_0=5$.}
    \label{fig:Snap_Hydro}
\end{figure}

Eqs.~\eqref{rmf1} and \eqref{rmf2} immediately show the importance of the transverse diffusion. If we start with a $x$-independent initial condition, for instance, a horizontal thin stripe of width $w$, where: $\rho(y)=\rho_0/w$, $m(y)=m_0=\sqrt{\frac{2\rho_0}{\alpha}\left[\rho_0(2\beta-1)-r\right]}>0$ [from $2\left(2\beta-1-r/\rho_0\right)=\alpha m_0^2/\rho_0^2$] is the stationary solution of Eq.~\eqref{rmf2} for $0<y<w$, and $\rho(y)=\delta \ll \rho_0/w$ and $m_0=0$ otherwise. Then the solution stays $x$-independent at all later times: $\rho(x,y,t)=\Tilde{\rho}(y,t)$ and $m(x,y,t)=\Tilde{m}(y,t)$ where the equations for $\tilde \rho$ and $\tilde m$ are:
\begin{gather}
    \Dot{\Tilde{\rho}}=D\partial_{yy}\Tilde{\rho} \, , \label{rmf5} \\ 
    \Dot{\Tilde{m}}=D\partial_{yy} \Tilde{m} + 2\left(2\beta-1-\frac{r}{\Tilde{\rho}}\right)\Tilde{m}-\alpha\frac{\Tilde{m}^3}{\Tilde{\rho}^2} \, . \label{rmf6}
\end{gather}
Eq.~\eqref{rmf5} shows that the thin stripe expands diffusively in the $y$-direction (transverse direction) yielding a $\sqrt{t}$ dependence of the stripe width. For general initial conditions, we use explicit Euler FTCS (Forward Time Centered Space)~\cite{press2007numerical} differencing scheme to numerically integrate Eqs.~\eqref{rmf1} and \eqref{rmf2}. We solve these two coupled partial differential equations on a square domain of size $L \times L$ with periodic boundary conditions applied in both directions. In our simulation, $L=400$ and the maximum simulation time is $t_{\rm sim}=5 \times 10^6$. To maintain the numerical stability criteria, we set $\Delta x = 1$ as the discretization in space and $\Delta t = 10^{-3}$ as the discretization in time. These discretization parameters satisfy the Courant-Friedrichs-Lewy (CFL) stability condition. In our numerical implementation, we fix $D=r=v=1$ and the initial system is prepared as a high-noise homogeneous gas phase with $\rho=\rho_0$ and $m=0$ by adding a zero-mean scalar Gaussian white noise to Eq.~\eqref{rmf2}~\cite{solonVM}. We then calculate the spatial dependence of the density and magnetization correlation using Eqs.~\eqref{correlation_density_cAIM} and \eqref{correlation_mag} for 25 independent realizations. Finally, $R(t)$ is determined where the ensemble-averaged correlation functions decay to 0.2 of its maximum value. 
\begin{figure}[!t]
    \centering
    \includegraphics[width=0.6\columnwidth]{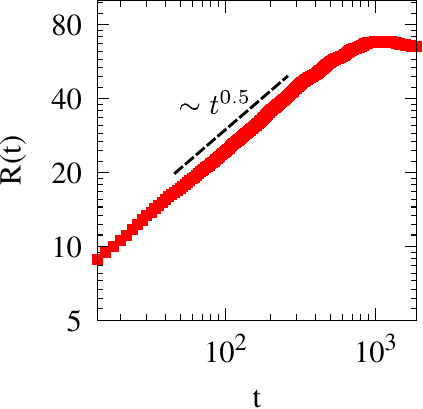}
    \caption{(Color online) $R(t)$ versus $t$ (on a log-log scale) exhibiting a growth exponent $\theta=1/2$ by solving the AIM hydrodynamic equations Eq.~\eqref{rmf1} and Eq.~\eqref{rmf2} using the FTCS scheme. Parameters: $L=400$, $\beta=0.75$, and $\rho_0=5$.}
    \label{fig:R_Hydro}
\end{figure}

In Fig.~\ref{fig:Snap_Hydro}, we plot the time evolution of the magnetization field $m$ by solving Eqs.~\eqref{rmf1} and \eqref{rmf2} via the finite difference FTCS scheme. The formation of self-propelling clusters with smooth interfaces and their growth with time resembles the dynamics shown in Fig.~\ref{fig:dynamics_snap} for the time-evolution of the microscopic model [Eqs.~(\ref{hamiltonian}--\ref{hop})]. We also extract the LCA growth law $R(t) \sim t^{1/2}$ (as shown in Fig.~\ref{fig:R_Hydro} on a logarithmic scale) by solving the hydrodynamic equations after quenching the system from a disordered gaseous phase to an ordered liquid phase. The length scale in Fig.~\ref{fig:R_Hydro} is obtained from the equal-time spatial correlation of the density fields although the length scale obtained from the correlation of the magnetization fields also exhibits the same growth law. Therefore, the self-propulsion terms in the hydrodynamic equations \eqref{rmf1} and \eqref{rmf2} of the AIM do not affect the asymptotic growth law of the non-conserved {\it Model A}.

\section{Summary and Discussion}\label{conclusion}
We conclude this chapter with a summary and discussion of our results. We study the ordering kinetics of the active Ising model (AIM), a flocking model with a non-conserved scalar order parameter, after it is quenched from a disordered high-temperature gaseous phase to the phase-coexistence region and the polar-ordered liquid phase. We observe the formation of connected domains of negative and positive spins similar to the ferromagnetic Ising model in the zero activity diffusive limit of the AIM. But for self-propelled particles, AIM manifests an extensive number of disconnected small clusters of the negative and positive spins which eventually merge to a single, macroscopic liquid domain~\cite{solon2015flocking}. The domain evolution morphology is characterized by the equal-time two-point correlation function and its Fourier transform, the structure factor. The scaling of the correlation function exhibits good data collapse, signifying a self-similar nature of the domain growth. The large-$k$ behavior of the scaled structure factor tail shows the Porod's decay which signifies the smooth spatial structure of the AIM domains. The growth law we extract for the AIM follows the Lifshitz-Cahn-Allen (LCA) growth law~\cite{alan_bray,bray1993theory,sanjay_puri} $R(t)\sim t^{1/2}$ of the non-conserved scalar order parameter and is unaffected by the system control parameters such as temperature, density, and particle velocity. Unlike the VM~\cite{Coarsening2020VM}, in AIM, the density domain aligns over the same length scale as the orientation and we do not observe any activity-induced correction to the growth law of non-conserved scalar order parameter discussed in the context of the active polar fluid~\cite{Dikshit_2023}. We further investigate the role of diffusion on the AIM growth kinetics. We observe that due to the horizontal biased hopping (along $\pm x$-direction) of the AIM clusters, particle diffusion along the vertical $\pm y$-direction is the predominant mechanism through which coarsening happens in the AIM.  This establishes diffusion as the principal growth mechanism rather than activity, leading to a $R(t)\sim t^{1/2}$ growth law for the AIM. We further solve the AIM hydrodynamic equations via a finite difference scheme and the extracted coarsening length scale validates the growth exponent $\theta \simeq 1/2$ observed in the microscopic simulation.

\section{Auxiliary material}
\subsection{Comparison of the growth law for the density and magnetization field}
\label{appA_cAIM}
\begin{figure}
    \centering
    \includegraphics[width=0.6\columnwidth]{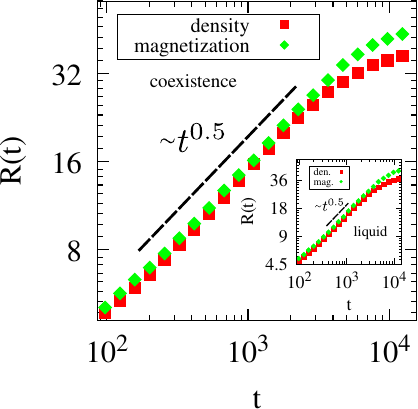}
    \caption{(Color online) $R(t)$ versus $t$ for density and magnetization field yields the same growth exponent 0.5 for a quench in the coexistence regime, $\rho_0=3$. (Inset) A similar growth law is extracted for a quench in the liquid regime, $\rho_0=10$. Parameters: $L=400$, $\beta=1.1$ and $\epsilon=1$.}
    \label{fig:mag_vs_den_compare}
\end{figure}
The ordering kinetics of the AIM discussed in this chapter are studied mainly by extracting the length scale $R(t)$ from the equal-time two-point density correlation function defined in Eq.~\eqref{correlation_density_cAIM}. However, one can also extract $R(t)$ from the magnetization correlation function defined in Eq.~\eqref{correlation_mag}. Now, it was argued in the context of the coarsening dynamics of the VM that, unlike generic coarsening systems which typically exhibit a single dominant length scale, the Vicsek model exhibits distinct coarsening length scales for the density and velocity correlations \cite{Coarsening2020VM}. In VM, despite the density and velocity fields being fully coupled, the velocity length scale grows much faster compared to the density length scale because velocity order extends over longer distances than density clusters due to the irregular fractal morphology of the density clusters. In AIM, however, besides being the density and magnetization fields fully coupled, the clusters are also regularly shaped with smooth boundaries, and therefore, the temporal behavior of the two length scales are found similar as shown in Fig.~\ref{fig:mag_vs_den_compare}.

Fig.~\ref{fig:mag_vs_den_compare} shows, for the coarsening of the AIM, that the two length scales exhibit the same growth law, $R(t) \sim t^{1/2}$ (although the domain size for the density field is marginally larger than the corresponding magnetization field) for a quench into the coexistence region and into the polar ordered liquid regime (see inset of Fig.~\ref{fig:mag_vs_den_compare}). Therefore, we can conclude that the $R(t) \sim t^{1/2}$ growth law is reasonably universal in the AIM as it neither depends on the quenching regime nor the local order parameter (be it density or magnetization). 

\subsection{2\textit{d} AIM with \texorpdfstring{\bm{$D_\perp=0$}}{Lg}}
\label{appB_cAIM}
\begin{figure}
    \centering
    \includegraphics[width=.8\columnwidth]{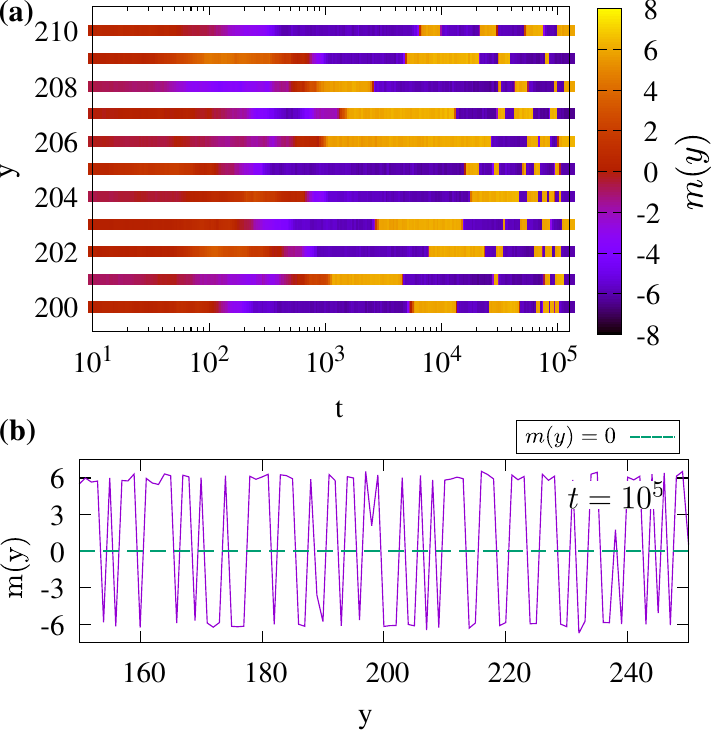}
    \caption{(Color online) (a) $m(y)$ versus $t$ for $L_y\in[200:210]$ starting from a random disordered configuration ($m \sim 0$). The color bar denotes the averaged magnetization along the $x$-axis. (b) Late time averaged magnetization profile as a function of $y$. Parameters: $L=400$, $\rho_0=10$, $\beta=1.1$, and $\epsilon=0.8$.}
    \label{fig:mag_D_zero}
\end{figure}   
The role of transverse diffusion in domain growth of $2d$ AIM is immense. Here we present a more detailed picture of the $D_\perp=0$ scenario. In Fig.~\ref{fig:mag_D_zero}(a) we plot $m(y)= \frac{1}{L_x}\sum^{L_x}_{i=1} m(i,y)$ versus time $t$ for $L_y\in[200:210]$. As the density is very large $(\rho_0=10)$, we see highly magnetized one lattice unit-wide domains that can not merge to create a larger domain as time progresses because we inhibit the diffusive hopping in the transverse direction. These $1d$ domains are single dense ordered aggregates that stochastically reverse their magnetization~\cite{benvegnen2022flocking}. The magnetization profile in Fig.~\ref{fig:mag_D_zero}(b) is averaged over $x$ and shows sharp peaks of positive and negative magnetizations, spread over either one site or a few sites. The density profile of such an arrangement shows a homogeneous profile around the average density which is similar to the density profile of a large liquid domain but the alternating magnetization profile signifies that there is no domain growth at the asymptotic limit in the $2d$ AIM for $D_{\perp}=0$.  

\chapter{Summary and Outlook}\label{chap:Chap6}
~~~~~Active matter is a broad field of study that involves the collective behavior of living systems. Studies in this field make quantitative comparisons with observations possible, even for living systems. There is considerable potential for constructing theoretical approaches to examining the role of activity in the formation of groups known as \enquote{flocks}~\cite{toner2005hydrodynamics,juelicher2007active,joanny2009active,ramaswamy2010mechanics}. From the unique self-driven nature of active particles, a rich set of open problems emerged in this field. In a combination of analytical and numerical work, this thesis provides a quantitative understanding of self-assembly and emergent structures and patterns in active matter systems with discrete rotational symmetry. Using minimal active matter models, we have shown that important aspects of seemingly complex organizational processes can be captured using, for instance, the interplay of activity and noise. Below, we present a summary and discuss future studies likely to emerge from this thesis work.\\

\noindent
~~~~~We begin with a motivation for studying active matter systems in Chapter~\ref{chap:Chap1} and subsequently discuss the physics of flocking transition~\cite{ramaswamy2010mechanics,marchetti2013hydrodynamics}. We extensively review several fundamental theoretical models such as the Vicsek model, Toner-Tu model, active Ising model, active Potts model, and active Clock model and provide the foundation of flocking dynamics~\cite{vicsek1995novel,toner1995long,toner1998flocks,solon2013revisiting,solon2015flocking,chatterjee2020flocking,mangeat2020flocking,chatterjee2022polar,solon2022susceptibility}. We briefly mention how more than one interacting species might alter the flocking patterns observed with a single species~\cite{peled2021heterogeneous,TSVM2023}. Next, we discuss motility-induced phase separation (MIPS)~\cite{cates2015motility} and characteristics of domains exhibited by MIPS~\cite{kourbane2018exact,peruani2011traffic}. We introduce the methodologies such as simulation techniques, and algorithms (the Metropolis algorithm~\cite{metropolis1953equation,newman1999monte,landau2021guide}), which we have extensively used in this thesis. Finally, we conclude Chapter~\ref{chap:Chap1} by discussing the outline of the thesis.\\ 

\noindent 
~~~~~It has been reported earlier that SPPs such as active Brownian particles (ABPs)~\cite{romanczuk2012active} with repulsive interactions, show that at high density and high Péclet number, a cluster state appears which is different from flocking via alignment. The emerging state is due to the motility-induced phase separation (MIPS)~\cite{cates2015motility}. A discretized flocking model with volume exclusion studied in Chapter~\ref{chap:Chap2}, showed that the interplay between alignment and on-site repulsion produces a vast spectrum of self-organized patterns ranging from jammed clusters to band formation. We argued that the formation of such jammed clusters is a manifestation of MIPS, which relies on the reduction of particle velocity with increasing local density~\cite{geyer2019freezing}. Generally, it has been observed that velocity alignment interactions promote MIPS~\cite{sese2018velocity,sese2021phase}. Our study also showed that alignment is even necessary for MIPS to occur since the jammed clusters disappear in the gas phase for $T \to \infty$, i.e., vanishing alignment. For increasing alignment, i.e., small $T$, either orientationally ordered domains appear in the jammed clusters, arranged in such a way that the cluster configuration is kinetically arrested (up to fluctuations), or, depending on density and self-propulsion strength, the jammed clusters dissolve into an orientationally ordered liquid phase, both manifestations of flocking. The phase diagrams with ${\rm MPS}=1$ and those for ${\rm MPS}>1$ or soft-core repulsion turn out to be different due to the absence of alignment interactions for ${\rm MPS}=1$. The model with ${\rm MPS}=1$ is equivalent to an active lattice gas with persistent walkers instead of diffusing particles. Consequently, ${\rm MPS}=1$ is always in an orientationally disordered (gas) phase in which various MIPS or jammed states occur. For ${\rm MPS}>1$ or soft-core repulsion in addition to various jammed states, the three typical flocking phases occur: the orientationally disordered gas, liquid-gas coexistence (flocking phase), and the orientationally ordered liquid.\\  

\noindent
~~~~~We recognize that most of the earlier studies on active matter systems have considered homogeneous environments. However, in reality, active systems are heterogeneous as SPPs vary their response to the external environment. Several recent studies have examined the effects of disorder on flocking models~\cite{chepizhko2013optimal,toner2018hydrodynamic,toner2018swarming,das2018polar}. Chapter~\ref{chap:Chap3} sheds light on the behavior of SPPs in disordered media employing the $q$-state active Potts model, viz., the Random Field Active Potts Model (RFAPM) and Random Diffusion Active Potts Model (RDAPM). Interestingly, a unique feature of the APM is the treadmilling of a longitudinal band opposite to a small unidirectionally applied field driven by non-biased diffusion. Despite the movement of the longitudinal band at a small field, the self-propulsion velocity for the reorientation transition remains constant in the coexistence regime. The study reveals that with the increase of a constant unidirectional local field, the coexistence band expands and transforms into a fully liquid state at higher field strengths. Intriguingly, particles in the liquid state align themselves with the field direction, even in the presence of substantial thermal fluctuations. A flocking to MIPS transition can also be seen in the case of a bidirectional field. However, when local field orientations are randomly distributed, the system transitions from a polar liquid to a disordered gas phase as the field strength increases, even at lower temperatures. The impact of decreasing interaction strength between neighboring sites results in weakened and less probable particle hopping. Consequently, the spin-spin spatial correlation function decreases, resulting in a loss of long-range order and a disordered phase. Below the threshold probability $P^*_{rd}=0.21\pm 0.02$, the system shows LRO, while beyond it, a complete loss of LRO occurs, indicating the transition into a fully disordered state.\\

\noindent
~~~~~Our discussion, so far, has been inspired by the flocking transition and MIPS studied via the discrete flocking model ($4-state$ active Potts model). Nevertheless, a fundamental framework for studying the collective behavior of particles under aligning interactions is the Vicsek Model. However, while considering particles constrained to discrete, equidistant angular orientations within a two-dimensional plane, such as in the active clock model (ACM)~\cite{solon2022susceptibility,chatterjee2022polar}, the VM-inspired dynamical principles governing particle alignment and movement cannot be retrieved in the limit of a larger number of angular orientations. Motivated by the ACM~\cite{solon2022susceptibility,chatterjee2022polar}, our study carried out in Chapter \ref{chap:Chap4} considers a true $q$-state discrete version of the Vicsek model (DVM) where $q$ defines the number of possible orientations or the strength of orientation anisotropy. At small $q$, the system is highly anisotropic, which, however, vanishes in the limit $q\rightarrow \infty$ when we recover the Vicsek model. The DVM shows qualitatively similar features as the ACM~\cite{chatterjee2022polar} for intermediate noise strength $\eta$ where a transition from macrophase to microphase separation is observed in the coexistence region as $q$ is increased. But for small $q$ and low noise, the liquid phase appearing in the ACM at low temperatures is replaced by a cluster phase in the DVM. The cluster phase consists of multiple clusters with different polarization, which does not exhibit a long-range order. The clusters grow and merge with increasing $q$, leading to a homogeneous ordered phase at large $q$. For small $q$, increasing the noise strength can achieve a long-range ordered phase. For low noise and small $q$, the probability of transverse flipping is very small as fluctuations are weak. In addition to that, transverse fluctuations through hopping are also absent. Consequently, the combined influence of these factors results in clusters failing to grow continuously, preventing the system from reaching a homogeneous liquid state. The self-organized patterns in the coexistence region of the discretized VM indicate a transition from AIM-like patterns to VM-like patterns as anisotropy becomes weaker. This observation is corroborated by the giant density fluctuations for large $q$. However, the large length scale behavior of the direction of global order, the structure factor, and the order parameter distribution in the liquid phase do not correspond with the phase-coexistence patterns of the large $q$ DVM. It shows microphase or cross-sea pattern for large length scales without any spatial anisotropy as $q$ increases. We also find that the DVM liquid phase is susceptible to perturbation applied through a counter-propagating droplet.  The liquid phase reorients and propagates along the direction of the droplet. The noise strength $\eta$~\cite{codina2022small} significantly impacts the reversal dynamics but remains independent of $q$. In conclusion, the rotational flexibility of the particles and microscopic details of the dynamical rules can significantly impact the macroscopic properties of the ordered phase. \\

\noindent
~~~~~While the majority of this thesis is devoted to understanding the steady state properties of various active systems~\cite{toner2005hydrodynamics,solonflocking2013,chatterjee2020flocking,TSVM2023,karmakar2023jamming}, an interesting question we might ask now is how an active system relaxes to a non-equilibrium steady state (NESS). In Chapter \ref{chap:Chap5}, we investigated the ordering kinetics of the minimal active system, the Active Ising model (AIM), a flocking model with a non-conserved scalar order parameter. The kinetic study is performed after the AIM is quenched from a disordered high-temperature gaseous phase to the phase-coexistence and the polar-ordered liquid regions. We observe the formation of connected domains of negative and positive spins similar to the ferromagnetic Ising model in the zero activity diffusive limit of the AIM. But for self-propelled particles, AIM manifests an extensive number of disconnected small clusters of the negative and positive spins, which eventually merge to a single, macroscopic liquid domain~\cite{solon2015flocking}. The domain evolution morphology is characterized by the equal-time two-point correlation function and its Fourier transform structure factor. The scaling of the correlation function exhibits good data collapse, signifying a self-similar nature of the domain growth, and the large-$k$ behavior of the scaled structure factor tail shows the Porod's decay, which signifies the smooth spatial structure of the AIM domains. The growth law we extract for the AIM follows the Lifshitz-Cahn-Allen (LCA) growth law~\cite{alan_bray,bray1993theory,sanjay_puri} $R(t)\sim t^{1/2}$ associated with the non-conserved scalar order parameter and is unaffected by the system control parameters such as temperature, density, and particle velocity. Unlike the VM~\cite{Coarsening2020VM}, in AIM, the density domain aligns over the same length scale as the orientation, and we do not observe any activity-induced correction to the growth law of non-conserved scalar order parameter discussed in the context of the active polar fluid~\cite{Dikshit_2023}. We further investigate the role of diffusion on the AIM growth kinetics. We observe that due to the horizontal biased hopping (along $\pm x$-direction) of the AIM clusters, particle diffusion along the vertical $\pm y$-direction is the predominant mechanism through which coarsening happens in the AIM.  This establishes diffusion as the principal growth mechanism rather than activity, leading to a $R(t)\sim t^{1/2}$ growth law for the AIM.  We further solve the hydrodynamic equations of the AIM, which validate the $t^{1/2}$ dependency of the coarsening length scale observed in the microscopic model.

~~~~~In summary, this Ph.D. thesis is primarily devoted to the study of steady-state properties of active matter systems, focusing on the collective behavior such as flocking transition~\cite{ramaswamy2010mechanics,marchetti2013hydrodynamics}and the motility-induced phase separation (MIPS)~\cite{cates2015motility}. Besides, it also provides an insight into the ordering kinetics of a minimal active matter system. The research explores the influence of volume exclusion of active particles, the impact of inhomogeneities present in the surrounding media, and the rotational anisotropy of the SPPs. We found that model parameters such as temperature, self-propulsion velocity, and density of the SPPs affect the overall collective behavior. As a consequence, several interesting features such as jamming, kinetic arrest, motility-induced phase separation, coexisting phases (macro phase and microphase separation), etc., appear. Finally, a discrete flocking model elucidates the evolution kinetics of large polar flocks.

A natural extension of this thesis is to analyze various active matter models in the presence of non-reciprocal interaction between SPPs. When two or more active agents have an asymmetric mutual influence with unequal actions and reactions, this is referred to as non-reciprocal interaction. Such systems are abundant in nature and there is a growing interest in understanding the physics of these active systems~\cite{bowick2022symmetry,shankar2022topological,knevzevic2022collective,dinelli2023non}. Many scientific fields, including active matter~\cite{ivlev2015statistical,soto2014self,fruchart2021non}, ecology~\cite{rieger1989solvable,allesina2012stability,bunin2017ecological}, neuroscience~\cite{rieger1989glauber,sompolinsky1986temporal}, and robotics~\cite{brandenbourger2019non}, have recognised the significance of non-reciprocal interaction. 

Polar flocks are observed in a large class of active matter systems and have been considered robust under external perturbation. However, recent studies have argued that liquid polar flocks are metastable to the presence of small obstacles~\cite{codina2022small} or the nucleation of opposite-phase droplets~\cite{benvegnen2023meta}. One can investigate the robustness of the flocking phase in the Vicsek model under different conditions, such as varying noise levels, system sizes, and interaction ranges. Also, the effects of external perturbations, such as obstacles or localized disturbances, on the stability of the flocking phase can be worthwhile to study. This could involve systematically introducing obstacles or perturbations in different regions of the system and studying their impact on long-range order. The stability of the high-density flocking ordered phase at low noise is still an open problem and will be addressed in a subsequent study~\cite{swarnajit2023metastability}. Also, It would be interesting to compare the model predictions of the DVM with suitable experiments where a finite number of motility directions may control anisotropy in the particle orientation.

Another interesting future perspective of this thesis would be to explore the phase ordering kinetics in flocking models in the presence of disorder, as experimental systems always contain both quenched and mobile impurities. The similarity in the growth law of a passive system and its active counterpart is an interesting result, and therefore, further studies on the coarsening dynamics of active systems, such as the active Potts model or active clock model, where the growth law of the corresponding passive models are well-known~\cite{grest1988domain,chatterjee2018clock,rbcm}, are required to confirm (or contradict) this theoretical observation.

%
\begin{appendix}
\chapter{Hydrodynamic description for rAPM}\label{AppendixA}
In Chapter \ref{chap:Chap2}, we formulate and analyze a hydrodynamic theory for the restricted APM which predicts various features of the microscopic model. In this Appendix, we will briefly discuss the derivation of hydrodynamic equations and linear stability analysis for rAPM. 

\section{Derivation of hydrodynamic equations for rAPM}\label{hydro}
In this section, we will derive the hydrodynamic equations for the $q$-state rAPM. In Ref.~\cite{mangeat2020flocking}, we presented the detail hydrodynamic description of the unrestricted APM and their numerical solutions. In the rAPM, as we have modified only the rule of hopping dynamics of the particles (keeping the flipping rule unchanged), we present here the derivation of the hydrodynamic equations only for the hopping term in details. We introduce a function $f(n_i)$ where $i$ denotes the arrival site to represent the different hopping restrictions. The form of this function is $f(\rho) = 1 - \zeta\rho$ for ${\rm MPS} = 1/\zeta$, and $f(\rho) = \exp(-s \rho)$ for the soft-core rAPM where $s=2\beta U$.

The master equation writes
\begin{gather}
n_i^\sigma(t+dt) = n_i^\sigma(t) \left[ 1 - dt \sum_p W_{\rm hop}(\sigma,p) f(n_{i+p}) - dt \sum_{\sigma \ne \sigma'} W_{\rm flip} (\sigma \to \sigma') \right] \nonumber\\
+ dt \sum_p W_{\rm hop}(\sigma,p) n_{i-p}^\sigma(t) f(n_i) + dt \sum_{\sigma \ne \sigma'} n_i^{\sigma'}(t) W_{\rm flip} (\sigma' \to \sigma).
\end{gather}
Taking the limit $dt \to 0$, we get
\begin{equation}
\partial_t n_i^\sigma = \sum_p W_{\rm hop}(\sigma,p) \left[ n_{i-p}^\sigma f(n_i) - n_i^\sigma f(n_{i+p}) \right] + \sum_{\sigma' \ne \sigma} \left[ n_i^{\sigma'} W_{\rm flip}(\sigma', \sigma)- n_i^{\sigma} W_{\rm flip}(\sigma, \sigma') \right].
\end{equation}
In the following, we decompose the r.h.s of this equation into two terms: the hopping term $I_{\rm hop}$ and the flipping term $I_{\rm flip}$, unchanged under hopping restrictions. We then have $\partial_t n_i^\sigma = I_{\rm hop} + I_{\rm flip}$. Using the definition of $W_{\rm hop}$, given by Eq.~\eqref{whop}, we obtain
\begin{equation}
I_{\rm hop} = D\left(1- \frac{\epsilon}{q-1} \right)\sum_{p} \left[ n_{i-p}^\sigma f(n_i) - n_i^\sigma f(n_{i+p}) \right] + \frac{qD\epsilon}{q-1} \left[ n_{i-\sigma}^\sigma f(n_i) - n_i^\sigma f(n_{i+\sigma}) \right].\label{Ihop1}
\end{equation}

We consider the Taylor expansion of the function $G_{i+p}$ as
\begin{equation}
G_{i+p} = G_i + a \partial_p G_i + \frac{a^2}{2} \partial_p^2 G_i + {\cal O}(a^3),
\end{equation}
which leads to the expression
\begin{equation}
G_{i-p} H_i - G_i H_{i+p} = -a \partial_p[G_i H_i] + \frac{a^2}{2} [H_i \partial_p^2 G_i - G_i \partial_p^2 H_i] + {\cal O}(a^3).
\end{equation}
Knowing that the sum of the derivatives of any function $F$ are
\begin{gather}
\sum_{p=1}^q \partial_p F = \sum_{p=1}^q ({\bf e_p} \cdot {\bf e_x}) \partial_x F + \sum_{p=1}^q ({\bf e_p} \cdot {\bf e_y}) \partial_y F = 0, \\
\sum_{p=1}^q \partial_p^2 F = \sum_{p=1}^q ({\bf e_p} \cdot {\bf e_x})^2 \partial_x^2 F + 2\sum_{p=1}^q ({\bf e_p} \cdot {\bf e_x})({\bf e_p} \cdot {\bf e_y}) \partial_x \partial_y F + \sum_{p=1}^q ({\bf e_p} \cdot {\bf e_y})^2 \partial_y^2 F = \frac{q}{2} \nabla^2 F,
\end{gather}
we finally get
\begin{equation}
\sum_{p=1}^q [G_{i-p} H_i - G_i H_{i+p}] = \frac{qa^2}{4} [H_i \nabla^2 G_i - G_i \nabla^2 H_i] + {\cal O}(a^3).
\end{equation}

Eq.~\eqref{Ihop1} becomes
\begin{gather}
I_{\rm hop} = D\left(1- \frac{\epsilon}{q-1} \right) a^2 \left[ f(n_i) \nabla^2 n_i^\sigma - n_i^\sigma \nabla^2 f(n_i) \right] \nonumber \\ + \frac{qD\epsilon}{q-1} \left\{ - a \partial_\sigma\left[ f(n_i)n_i^\sigma \right] + \frac{a^2}{2} \left[ f(n_i) \partial_\sigma^2 n_i^\sigma - n_i^\sigma \partial_\sigma f(n_i) \right] \right\}.
\end{gather}
We can decompose $a^2 \nabla^2 = \partial_\parallel^2 + \partial_\perp^2$ and $a \partial_\sigma = \partial_\parallel$, and denote $\rho_\sigma = \langle n_i^\sigma \rangle$ as well as $\rho = \langle \rho_i \rangle$, which leads to the expression
\begin{equation}
\langle I_{\rm hop} \rangle = D_\parallel \left[ f(\rho) \partial_\parallel^2 \rho_\sigma - \rho_\sigma \partial_\parallel^2 f(\rho) \right] + D_\perp \left[ f(\rho) \partial_\perp^2 \rho_\sigma - \rho_\sigma \partial_\perp^2 f(\rho) \right] - v \partial_\parallel \left[ f(\rho) \rho_\sigma \right],
\end{equation}
with 
\begin{equation}
D_\parallel = D\left(1+ \frac{\epsilon}{q-1} \right), \quad D_\perp = D\left(1- \frac{\epsilon}{q-1} \right) \quad  {\rm and} \quad  v= \frac{qD\epsilon}{q-1}.
\end{equation}
Note that we can decompose
\begin{equation}
f(\rho) \partial_i^2 \rho_\sigma - \rho_\sigma \partial_i^2 f(\rho) = \partial_i \left[ f(\rho) \partial_i \rho_\sigma - \rho_\sigma \partial_i f(\rho) \right],
\end{equation}
and $\partial_i f(\rho) = f'(\rho)\partial_i \rho$, which yields the drift term of the rAPM equation
\begin{equation}
\langle I_{\rm hop} \rangle = D_\parallel \partial_\parallel \left[ f(\rho) \partial_\parallel \rho_\sigma - f'(\rho)\rho_\sigma \partial_\parallel \rho \right] + D_\perp \partial_\perp \left[ f(\rho) \partial_\perp \rho_\sigma - f'(\rho) \rho_\sigma \partial_\perp \rho \right] - v \partial_\parallel \left[ f(\rho) \rho_\sigma \right].
\end{equation}

The current is then 
\begin{gather}
{J_\sigma}_\parallel = -D_\parallel \left[ f(\rho) \partial_\parallel \rho_\sigma - f'(\rho)\rho_\sigma \partial_\parallel \rho \right] + v f(\rho) \rho_\sigma, \label{Jpara} \\
{J_\sigma}_\perp = -D_\perp \left[ f(\rho) \partial_\perp \rho_\sigma - f'(\rho) \rho_\sigma \partial_\perp \rho \right], \label{Jperp}
\end{gather}
or in the vectorial form: $J_{\sigma i} = f(\rho) J_{\sigma i}^0 - \lambda_i \rho_\sigma \partial_i \rho$,
where ${\bf J}_\sigma^0$ is the current without restriction, and $\lambda_i$ a positive constant since $f(\rho)$ is a strictly decreasing function. The first term corresponds to the current without restriction multiplied by $f(\rho)$ while the second term corresponds to an additional current from high to low densities.

For ${\rm MPS} = 1$, the flipping term $I_{\rm flip}$ is calculated according to $W_{\rm flip}(\sigma \to \sigma') = \gamma$, which gives
\begin{equation}
\langle I_{\rm flip} \rangle = \gamma \sum_{\sigma' \ne \sigma} (\rho_{\sigma'} - \rho_{\sigma}).
\end{equation}
For ${\rm MPS} > 1$ and soft-core rAPM, the flipping term $I_{\rm flip}$ is calculated according to Eq.~\eqref{flipeq}, and equal to the flipping term of the unrestricted APM derived in Ref.~\cite{mangeat2020flocking}:
\begin{equation}
\langle I_{\rm flip} \rangle = \sum_{\sigma' \ne \sigma} \left[ \frac{4\beta J}{\rho}(\rho_\sigma + \rho_{\sigma'}) - 1 - \frac{r}{\rho} - \alpha \frac{(\rho_\sigma - \rho_{\sigma'})^2}{\rho^2} \right] (\rho_{\sigma} - \rho_{\sigma'}),
\end{equation}
with $\alpha = 8(\beta J)^2(1-2\beta J/3)$.

The hydrodynamic equation for the rAPM is then	
\begin{equation}
\partial_t \rho_\sigma = - \partial_\parallel J_{\sigma \parallel} - \partial_\perp J_{\sigma \perp} + \sum_{\sigma' \ne \sigma} K_{\sigma \sigma'} (\rho_\sigma - \rho_{\sigma'}),
\end{equation}
where $J_{\sigma \parallel}$ and $J_{\sigma \perp}$ are given by Eqs.~\eqref{Jpara} and~\eqref{Jperp}, respectively, and the flipping interaction term is $K_{\sigma \sigma'} = - \gamma$ for ${\rm MPS}=1$ and
\begin{equation}
K_{\sigma \sigma'} = \frac{4\beta J}{\rho}(\rho_\sigma + \rho_{\sigma'}) - 1 - \frac{r}{\rho} - \alpha \frac{(\rho_\sigma - \rho_{\sigma'})^2}{\rho^2},
\end{equation}
for ${\rm MPS}>1$ and soft-core rAPM.


\section{Linear stability analysis and binodal calculus for ${\rm MPS}=1$}
\label{ls_mps1}

For ${\rm MPS} = 1$, the hydrodynamic equation is
\begin{gather}
\partial_t \rho_\sigma = D_\parallel \partial_\parallel \left[ (1 - \rho) \partial_\parallel \rho_\sigma + \rho_\sigma \partial_\parallel \rho \right] + D_\perp \partial_\perp \left[ (1 - \rho) \partial_\perp \rho_\sigma + \rho_\sigma \partial_\perp \rho \right]\nonumber\\ - v \partial_\parallel \left[ (1 - \rho) \rho_\sigma \right] - \gamma(4\rho_\sigma - \rho).
\end{gather}

Using dimensionless coordinates $\tau = \gamma t$ and ${\bf X}=\sqrt{\gamma/D} {\bf x}$, the hydrodynamic equation rewrites
\begin{gather}
\partial_\tau \rho_\sigma = D_\parallel \partial_\parallel \left[ (1 - \rho) \partial_\parallel \rho_\sigma + \rho_\sigma \partial_\parallel \rho \right] + D_\perp \partial_\perp \left[ (1 - \rho) \partial_\perp \rho_\sigma + \rho_\sigma \partial_\perp \rho \right]\nonumber\\ - {\rm Pe} \partial_\parallel \left[ (1 - \rho) \rho_\sigma \right] - (4\rho_\sigma - \rho), \label{eqmps1app}
\end{gather}
with $D_\parallel = 1+\epsilon/3$, $D_\perp = 1-\epsilon/3$ and the P\'eclet number ${\rm Pe} = v / \sqrt{D\gamma}$. The only homogeneous solution is $\rho_\sigma = \rho_0/4$ for all states $\sigma$.

{\bf Linear stability analysis.} We consider a linear stability analysis for $\rho_\sigma = \rho_0/4 + \delta \rho_\sigma$, where $\delta \rho_\sigma \ll \rho_0$ is a small perturbation. Keeping only the first order terms in $\delta \rho_\sigma$, the hydrodynamic equation becomes
\begin{gather}
\partial_\tau \delta \rho_\sigma = \left[\left( 1 - \frac{3\rho_0}{4} \right)(D_\parallel \partial_\parallel^2 + D_\perp \partial_\perp^2 ) - {\rm Pe} \left( 1 - \frac{5\rho_0}{4} \right) \partial_\parallel - 3 \right] \delta \rho_\sigma \nonumber \\
+ \left[\frac{\rho_0}{4} (D_\parallel \partial_\parallel^2 + D_\perp \partial_\perp^2 ) + {\rm Pe} \frac{\rho_0}{4} \partial_\parallel + 1 \right] \sum_{\sigma' \ne \sigma} \delta \rho_{\sigma'}.
\end{gather}

Performing a Fourier transform in space, we obtain
\begin{equation}
\partial_\tau \delta \rho_\sigma = A(k_\parallel,k_\perp) \delta \rho_\sigma + B(k_\parallel,k_\perp) \sum_{\sigma' \ne \sigma} \delta \rho_{\sigma'},
\end{equation}
with
\begin{gather}
A(k_\parallel,k_\perp) = \left( 1 - \frac{3\rho_0}{4} \right)(-D_\parallel k_\parallel^2 - D_\perp k_\perp^2 ) + \imath {\rm Pe} \left( 1 - \frac{5\rho_0}{4} \right) k_\parallel - 3 , \\
B(k_\parallel,k_\perp) = \frac{\rho_0}{4} (-D_\parallel k_\parallel^2 - D_\perp k_\perp^2 ) - \imath {\rm Pe} \frac{\rho_0}{4} k_\parallel + 1 .
\end{gather}

The stability of the homogeneous solution is then given by the eigenvalues of the matrix
\begin{equation}
M = \begin{pmatrix}
A(k_x,k_y) & B(k_x,k_y) & B(k_x,k_y) & B(k_x,k_y) \\
B(k_y,-k_x) & A(k_y,-k_x) & B(k_y,-k_x) & B(k_y,-k_x) \\
B(-k_x,-k_y) & B(-k_x,-k_y) & A(-k_x,-k_y) & B(-k_x,-k_y) \\
B(-k_y,k_x) & B(-k_y,k_x) & B(-k_y,k_x) & A(-k_y,k_x)
\end{pmatrix}.
\end{equation}

With the help of Mathematica~\cite{inc2017mathematica}, we get that 3 eigenvalues are always negative and the fourth eigenvalue writes
\begin{equation}
\lambda = \frac{1}{8} \left[ -4 (D_\parallel + D_\perp) + {\rm Pe}^2(1-\rho_0)(2\rho_0-1) \right](k_x^2 + k_y^2) + {\cal O}(k_x^2,k_y^2).
\end{equation}
using $D_\parallel + D_\perp=2$, the homogeneous solution is then stable if and only if $(1-\rho_0)(2\rho_0-1) < 8/{\rm Pe}^2$, leading to the spinodals $\varphi_\pm$:
\begin{equation}
\varphi_\pm = \frac{3}{4} \pm \frac{1}{4}\sqrt{1 - \frac{64}{{\rm Pe}^2}}, \label{spinodalapp}
\end{equation}
and a critical P\'eclet ${\rm Pe}_c = 8$, to observe the MIPS.

\medskip

{\bf Derivation of the binodals.} Now we derive the expression of the binodals, denoted as $\rho_{\rm low}$ and $\rho_{\rm high}$, following the demonstration made in Ref.~\cite{kourbane2018exact}, for an active lattice gas with a slightly different hydrodynamic equation. At steady state, the Eq.~\eqref{eqmps1app} writes
\begin{gather}
0 = D_\parallel \partial_\parallel \left[ (1 - \rho) \partial_\parallel \rho_\sigma + \rho_\sigma \partial_\parallel \rho \right] + D_\perp \partial_\perp \left[ (1 - \rho) \partial_\perp \rho_\sigma + \rho_\sigma \partial_\perp \rho \right]\nonumber\\ - {\rm Pe} \partial_\parallel \left[ (1 - \rho) \rho_\sigma \right] - (4\rho_\sigma - \rho). \label{eqmps1ss}
\end{gather}
We define $\rho_x = \rho_1 + \rho_3$, $\rho_y = \rho_2 + \rho_4$, and the magnetization vector ${\bf m} = (m_x,m_y)$ with $m_x = \rho_1 - \rho_3$ and $m_y = \rho_2 - \rho_4$. From Eq.~\eqref{eqmps1ss}, the equation for $\rho$ writes:
\begin{align}
0 &= D_\parallel \left\{ \partial_x \left[ (1 - \rho) \partial_x \rho_x + \rho_x \partial_x \rho \right] + \partial_y \left[ (1 - \rho) \partial_y \rho_y + \rho_y \partial_y \rho \right] \right\} \nonumber \\
&+ D_\perp \left\{ \partial_y \left[ (1 - \rho) \partial_y \rho_x + \rho_x \partial_y \rho \right] + \partial_x \left[ (1 - \rho) \partial_x \rho_y + \rho_y \partial_x \rho \right] \right\} \nonumber \\
&- {\rm Pe} \left\{ \partial_x \left[ (1 - \rho) m_x \right] + \partial_y \left[ (1 - \rho) m_y \right] \right\} .
\end{align}
From microscopic simulations and numerical solutions of Eq.~\eqref{eqmps1app}, we suppose the relation $\rho_x = \rho_y= \rho/2$. After simplifications and using the relation $D_\parallel + D_\perp = 2$, the equation for $\rho$ becomes
\begin{equation}
0 = \nabla^2 \rho - {\rm Pe} \nabla \cdot \left[ (1-\rho){\bf m} \right] = - \nabla \cdot {\bf J}.
\end{equation}
Since we observe a MIPS state, there is no steady current: ${\bf J} = 0$, then the steady state magnetization is
\begin{equation}
{\bf m}  = \frac{\nabla \rho}{{\rm Pe}(1-\rho)}.
\end{equation}

From Eq.~\eqref{eqmps1ss}, the equation for $m_x$ writes
\begin{equation}
0 = D_\parallel \partial_x \left[ (1 - \rho) \partial_x m_x + m_x \partial_x \rho \right] + D_\perp \partial_y \left[ (1 - \rho) \partial_y m_x + m_x \partial_y \rho \right] - \frac{{\rm Pe}}{2} \partial_x \left[ (1 - \rho) \rho \right] - 4 m_x.
\end{equation}
Using $m_x = \partial_x \rho/{\rm Pe}(1-\rho)$, we get
\begin{equation}
0 =  \partial_x \left[ \frac{D_\parallel}{{\rm Pe}} \partial_{xx} \rho + \frac{2D_\parallel(\partial_x\rho)^2}{{\rm Pe(1-\rho)}} - \frac{{\rm Pe}}{2} (1 - \rho) \rho + \frac{4}{{\rm Pe}} \ln(1-\rho) \right] + \partial_y \left[ \frac{D_\perp}{{\rm Pe}} \partial_{xy} \rho + \frac{2D_\perp(\partial_x\rho) (\partial_y\rho)}{{\rm Pe(1-\rho)}} \right]. \label{eqmx}
\end{equation}

From Eq.~\eqref{eqmps1ss}, the equation for $m_y$ writes
\begin{equation}
0 = D_\parallel \partial_y \left[ (1 - \rho) \partial_y m_y + m_y \partial_y \rho \right] + D_\perp \partial_x \left[ (1 - \rho) \partial_x m_y + m_y \partial_x \rho \right] - \frac{{\rm Pe}}{2} \partial_y \left[ (1 - \rho) \rho \right] - 4 m_y.
\end{equation}
Using $m_y = \partial_y \rho/{\rm Pe}(1-\rho)$, we get
\begin{equation}
0 =  \partial_y \left[ \frac{D_\parallel}{{\rm Pe}} \partial_{yy} \rho + \frac{2D_\parallel(\partial_y\rho)^2}{{\rm Pe(1-\rho)}} - \frac{{\rm Pe}}{2} (1 - \rho) \rho + \frac{4}{{\rm Pe}} \ln(1-\rho) \right] + \partial_x \left[ \frac{D_\perp}{{\rm Pe}} \partial_{xy} \rho + \frac{2D_\perp(\partial_x\rho) (\partial_y\rho)}{{\rm Pe(1-\rho)}} \right]. \label{eqmy}
\end{equation}

Performing $\partial_x$\eqref{eqmx}$-\partial_y$\eqref{eqmy}, we get
\begin{align}
0 &=  (\partial_{xx} - \partial_{yy}) \left\{ \frac{D_\parallel}{2{\rm Pe}} (\partial_{xx}+\partial_{yy}) \rho + \frac{D_\parallel}{{\rm Pe(1-\rho)}} [(\partial_x\rho)^2 + (\partial_y\rho)^2] - \frac{{\rm Pe}}{2} (1 - \rho) \rho + \frac{4}{{\rm Pe}} \ln(1-\rho) \right\} \nonumber \\
&+ (\partial_{xx} + \partial_{yy}) \left\{ \frac{D_\parallel}{2{\rm Pe}} (\partial_{xx}-\partial_{yy}) \rho + \frac{D_\parallel}{{\rm Pe(1-\rho)}} [(\partial_x\rho)^2 - (\partial_y\rho)^2] \right\}.
\end{align}
We take the new coordinates $u=x+y$ and $v=x-y$, for which $\partial_x = (\partial_u + \partial_v)/2$ and $\partial_y = (\partial_u - \partial_v)/2$. We get
\begin{align}
0 &=  \partial_u\partial_v \left\{ \frac{D_\parallel}{8{\rm Pe}} (\partial_{uu}+\partial_{vv}) \rho + \frac{D_\parallel}{4{\rm Pe(1-\rho)}} [(\partial_u\rho)^2 + (\partial_v\rho)^2] - \frac{{\rm Pe}}{2} (1 - \rho) \rho + \frac{4}{{\rm Pe}} \ln(1-\rho) \right\} \nonumber \\
&+ \frac{1}{4}(\partial_{uu} + \partial_{vv}) \left\{ \frac{D_\parallel}{2{\rm Pe}} \partial_{uv} \rho + \frac{D_\parallel}{{\rm Pe(1-\rho)}} [\partial_u\rho\partial_v\rho] \right\}.
\end{align}
The solution is then symmetric under the transformation $u \leftrightarrow v$, as observed from numerical solutions. If we choose a solution such that $\partial_v \rho = 0$, the quantity
\begin{equation}
g(u)= - \frac{D_\parallel}{8{\rm Pe}} \partial_{uu} \rho - \frac{D_\parallel}{4{\rm Pe(1-\rho)}} (\partial_u\rho)^2 + \frac{{\rm Pe}}{2} (1 - \rho) \rho - \frac{4}{{\rm Pe}} \ln(1-\rho),
\end{equation}
is constant. We define the quantities $\kappa(\rho)$, $\Lambda(\rho)$ and $g_0(\rho)$ such that
\begin{equation}
g= - \kappa(\rho) \partial_{uu} \rho + \Lambda(\rho) (\partial_u\rho)^2 + g_0(\rho).\label{defgu}
\end{equation}
Since $g$ is constant, we have a first relation between the two binodals: $g= g_0(\rho_{\rm high}) = g_0(\rho_{\rm low})$. A second relation has to be found to determine the values of $\rho_{\rm high}$ and $\rho_{\rm low}$. We consider the following quantity
\begin{equation}
I = \int_{u_{\rm low}}^{u_{\rm high}} du g(u) \partial_u R(u)
\end{equation}
integrated between the regions of low densities and high densities, where $R(\rho)$ is a monotonic function to determine. Since $R(\rho)$ is monotonic, we can calculate the quantity $I$ by doing a change of variable, and we get $I= g [R(\rho_{\rm high}) - R(\rho_{\rm low})]$. Using the definition of $g(u)$ given by Eq.~\eqref{defgu}, we obtain
\begin{equation}
I = \Phi(R_{\rm high}) - \Phi(R_{\rm low}) + \frac{1}{2}\int_{u_{\rm low}}^{u_{\rm high}} du \left[ 2R'(\rho)\Lambda(\rho) + \kappa(\rho) R''(\rho) + R'(\rho)\kappa'(\rho)\right] (\partial_{uu} \rho)^3,
\end{equation}
with $\Phi'(R) = g_0(R)$. Choosing the function $R(\rho)$ such that $(\kappa R')' = -2R' \Lambda$, the remaining integral vanishes, and $I = \Phi(R_{\rm high}) - \Phi(R_{\rm low})$. We may then define a second relation $h_0(\rho_{\rm high}) = h_0(\rho_{\rm low})$ with $h_0(R) = g_0(R) R - \Phi(R)$. Using the definitions $\kappa(\rho) = D_\parallel/8 {\rm Pe}$ and $\Lambda(\rho) = - D_\parallel/4 {\rm Pe} (1-\rho)$, we can take the monotonic function
\begin{equation}
R(\rho) = \frac{1}{(1-\rho)^3}.
\end{equation}

\begin{figure}[t]
\centering
\includegraphics[width=\columnwidth]{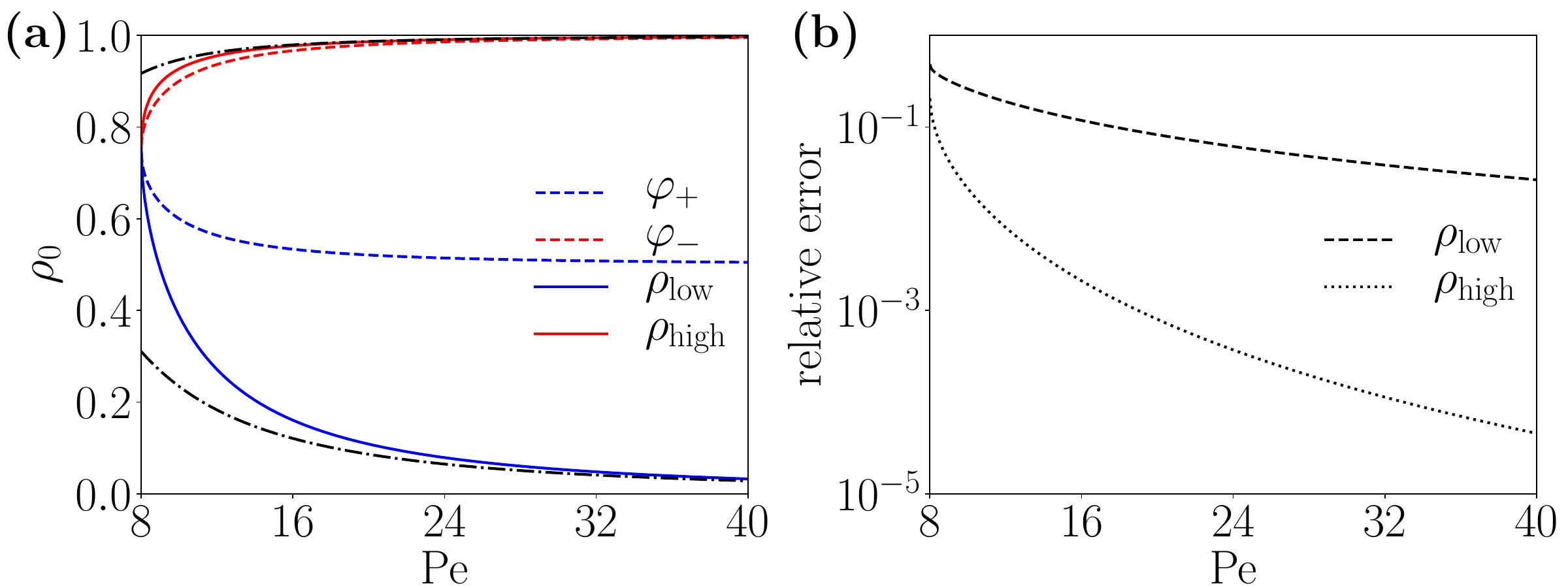}
\caption{(Color online) Spinodals and binodals of the rAPM with MPS=1. {\bf (a)}~The dotted lines show the spinodal lines given by Eq.~\eqref{spinodalapp}, and the solid lines show the binodals calculated from the relations $g_0(\rho_{\rm high}) = g_0(\rho_{\rm low})$ and $h_0(\rho_{\rm high}) = h_0(\rho_{\rm low})$, with Eqs.~\eqref{defg0} and~\eqref{defh0}. The dash-dotted lines display the asymptotic value of the binodals at large P\'eclet, given by Eqs.~\eqref{rhighlim} and~\eqref{rlowlim}. {\bf (b)}~Relative error between the binodal densities and their asymptotic values at large P\'eclet.} 
\label{fig_MPS1_coex}
\end{figure}

After simplifications, we get the relations $g_0(\rho_{\rm high}) = g_0(\rho_{\rm low})$ and $h_0(\rho_{\rm high}) = h_0(\rho_{\rm low})$, with
\begin{gather}
g_0(\rho) = \frac{{\rm Pe}}{2} (1 - \rho) \rho - \frac{4}{{\rm Pe}} \ln(1-\rho),\label{defg0}\\
h_0(\rho) = \frac{{\rm Pe}}{4} \frac{3-4\rho}{(1-\rho)^2} + \frac{4}{3{\rm Pe}} \frac{1}{(1-\rho)^3} \label{defh0}.
\end{gather}

\begin{table}[H]
\begin{center}
\begin{tabular}{ |c|c|c|c|c|c|c|c|c| } 
\hline
Pe & $8$ & $9$ & $10$ & $11$ & $12$ & $13$ & $14$ \\
\hline
$\rho_{\rm low}$ & $0.75$ & $0.491$ & $0.389$ & $0.321$ & $0.271$ & $0.234$ & $0.204$ \\
\hline
$\rho_{\rm high}$ & $0.75$ & $0.896$ & $0.927$ & $0.944$ & $0.955$ & $0.963$ & $0.969$ \\
\hline
\end{tabular}
\caption{Theoretical values of the binodals for MPS=1, for several P\'eclet numbers.\label{tabmps1}}
\end{center}
\end{table}

Table~\ref{tabmps1} shows the values of $\rho_{\rm low}$ and $\rho_{\rm high}$ for small values of ${\rm Pe}$. At the large P\'eclet number, we derive the asymptotic expression of $\rho_{\rm low}$ and $\rho_{\rm high}$ as follows. We know the leading order: $\rho_{\rm low}=0$ and $\rho_{\rm high}=1$. For $\rho_{\rm low}=0$, we get the value $h_0(0) \simeq 3{\rm Pe}/4$. We search the sub-leading order as $\rho_{\rm high}\simeq 1-\xi$, with $\xi \ll 1$. At leading order, we get
\begin{equation}
h_0(1-\xi) \simeq -\frac{{\rm Pe}}{4\xi^2} + \frac{4}{3{\rm Pe} \xi^3} \simeq \frac{3{\rm Pe}}{4},
\end{equation}
telling that the diverging terms in $\xi$ must cancel: $\xi = 16/3{\rm Pe}^2$. Then we have:
\begin{equation}
\rho_{\rm high}\simeq 1-\frac{16}{3{\rm Pe}^2}. \label{rhighlim}
\end{equation}
For $\rho_{\rm low} \ll 1$, we find that $g_0(\rho_{\rm low}) \simeq {\rm Pe} \rho_{\rm low}/2$. Since $g_0(\rho_{\rm high}) \simeq -(4/{\rm Pe}) \ln (16/3{\rm Pe}^2)$, we get
\begin{equation}
\rho_{\rm low}\simeq-\frac{8}{{\rm Pe}^2} \ln \frac{16}{3{\rm Pe}^2}. \label{rlowlim}
\end{equation}

Fig.~\ref{fig_MPS1_coex}(a) shows the spinodals given by Eq.~\eqref{spinodalapp}, and the binodals calculated with Eqs.~\eqref{defg0} and~\eqref{defh0}. The asymptotic value of the binodals at large P\'eclet, given by Eqs.~\eqref{rhighlim} and~\eqref{rlowlim}, are also represented. Fig.~\ref{fig_MPS1_coex}(b) shows the relative error between the binodal densities and their asymptotic values at large P\'eclet, decreasing to zero when ${\rm Pe} \to \infty$.


\section{Linear stability analysis for ${\rm MPS}>1$}
\label{ls_hc}

For soft-core rAPM, the hydrodynamic equation is
\begin{gather}
\partial_t \rho_\sigma = D_\parallel \partial_\parallel \left[ (1-\zeta \rho) \partial_\parallel \rho_\sigma + \zeta \rho_\sigma \partial_\parallel \rho \right] + D_\perp \partial_\perp \left[ (1-\zeta\rho)  \partial_\perp \rho_\sigma + \zeta \rho_\sigma \partial_\perp \rho  \right] - v \partial_\parallel \left[ (1-\zeta\rho) \rho_\sigma \right] \nonumber \\
+ \sum_{\sigma'\ne\sigma} \left[ \frac{4\beta J}{\rho}(\rho_\sigma + \rho_{\sigma'}) - 1 - \frac{r}{\rho} - \alpha \frac{(\rho_\sigma - \rho_{\sigma'})^2}{\rho^2} \right](\rho_\sigma - \rho_{\sigma'}),
\end{gather}
with $\alpha = 8(\beta J)^2(1-2\beta J/3)$ and $\zeta = 1/{\rm MPS}$. The homogeneous solutions are given by:
\begin{equation}
I_{\rm flip}(\sigma,\sigma') = \left[ \frac{4\beta J}{\rho}(\rho_\sigma + \rho_{\sigma'}) - 1 - \frac{r}{\rho} - \alpha \frac{(\rho_\sigma - \rho_{\sigma'})^2}{\rho^2} \right](\rho_\sigma - \rho_{\sigma'}) = 0,
\end{equation}
and are then those of the unrestricted rAPM~\cite{mangeat2020flocking}. The disordered homogeneous solution is $\rho_\sigma = \rho_0 / 4$, and the ordered homogeneous solution (supposed along state $\sigma = 1$) is $\rho_1 = \rho_0 (1+3M) / 4$ and $\rho_{2,3,4} = \rho_0 (1-M) / 4$ with the magnetization $M$ following the equation:
\begin{equation}
\label{eqMag}
2\beta J (1+M) - 1 - \frac{r}{\rho_0} - \alpha M^2 = 0,
\end{equation}
or $M=M_0 \pm M_1 \delta$ with $M_0 = \beta J / \alpha$, $M_1 = \sqrt{r/\alpha \rho_*}$ and $\delta = \sqrt{(\rho_0-\rho_*)/\rho_0}$, where $\rho_*$ defined by
\begin{equation}
\rho_* = \frac{8(1-2\beta J/3) r}{1 + 8(2\beta J - 1)(1-2\beta J/3)},
\end{equation}
is the critical density below which the ordered homogeneous solution does not exist, for a temperature below $T_c = (1 - \sqrt{22}/8)^{-1} \simeq 2.417$.

\medskip

{\bf Linear stability analysis for the disordered homogeneous solution.} We take $\rho_\sigma = \rho_0/4 + \delta \rho_\sigma$ and $\rho = \rho_0 + \delta \rho$, with $\delta \rho = \sum_\sigma \delta \rho_\sigma$. The hopping term writes
\begin{gather}
I_{\rm hop} \simeq \left[ \left( 1 - \frac{3\zeta\rho_0}{4}\right) (D_\parallel \partial_\parallel^2 + D_\perp \partial_\perp^2) - v \left( 1 - \frac{5\zeta\rho_0}{4}\right) \partial_\parallel \right] \delta \rho_\sigma \nonumber \\
+ \frac{\zeta\rho_0}{4} \left[ D_\parallel \partial_\parallel^2 + D_\perp \partial_\perp^2 + v \partial_\parallel \right] \sum_{\sigma' \ne \sigma} \delta \rho_{\sigma'}, \label{IhopMPS}
\end{gather}
and the flipping term writes $I_{\rm flip}(\sigma,\sigma') \simeq \mu_0 (\delta \rho_\sigma - \delta \rho_{\sigma'})$, with $\mu_0 = 2\beta J - 1 - r/\rho_0$. Then, in the Fourier space, the hydrodynamic equation becomes
\begin{equation}
\partial_t \delta\rho_\sigma = \left[A(k_\parallel,k_\perp) + 3 \mu_0 \right] \delta \rho_\sigma + \left[B(k_\parallel,k_\perp)- \mu_0\right] \sum_{\sigma' \ne \sigma} \delta \rho_{\sigma'},
\end{equation}
with
\begin{gather}
A(k_\parallel,k_\perp) = \left( 1 - \frac{3\zeta\rho_0}{4}\right) (-D_\parallel k_\parallel^2 - D_\perp k_\perp^2) + \imath k_\parallel v \left( 1 - \frac{5\zeta\rho_0}{4}\right)  \\
B(k_\parallel,k_\perp) = \frac{\zeta\rho_0}{4} \left[ -D_\parallel k_\parallel^2 - D_\perp k_\perp^2 - \imath k_\parallel v \right] 
\end{gather}

The stability of the homogeneous disordered solution is then given by the eigenvalues of the matrix
\begin{equation}
M_{\rm gas} = \begin{pmatrix}
A(k_x,k_y) + 3 \mu_0 & B(k_x,k_y) - \mu_0 & B(k_x,k_y) - \mu_0 & B(k_x,k_y) - \mu_0\\
B(k_y,-k_x) - \mu_0 & A(k_y,-k_x) + 3 \mu_0 & B(k_y,-k_x) - \mu_0 & B(k_y,-k_x) - \mu_0\\
B(-k_x,-k_y) - \mu_0 & B(-k_x,-k_y) - \mu_0 & A(-k_x,-k_y) + 3 \mu_0 & B(-k_x,-k_y) - \mu_0\\
B(-k_y,k_x) - \mu_0 & B(-k_y,k_x) - \mu_0 & B(-k_y,k_x) - \mu_0 & A(-k_y,k_x) + 3 \mu_0
\end{pmatrix}.
\end{equation}

Supposing $k_x=k$ and $k_y=0$ (since no preferred direction), at leading order in $k \ll 1$, the eigenvalues, calculated with Mathematica~\cite{inc2017mathematica}, are: $\lambda_{\rm gas}^{1,2,3} \simeq 4 \mu_0$ and
\begin{equation}
\lambda_{\rm gas}^4 \simeq \left[-\frac{D_\parallel +D_\perp}{2} + (1- \zeta\rho_0) (1-2\zeta\rho_0) \frac{v^2}{8\mu_0} \right] k^2.
\end{equation}
The disordered homogeneous solution is then stable if $\mu_0<0$ and
\begin{equation}
\label{lambdaGasMPS}
\lambda_{\rm gas} = -D + (1- \zeta \rho_0)(1-2\zeta \rho_0) \frac{v^2}{8\mu_0}<0  .
\end{equation}

\begin{figure}[t]
\centering
\includegraphics[width=\columnwidth]{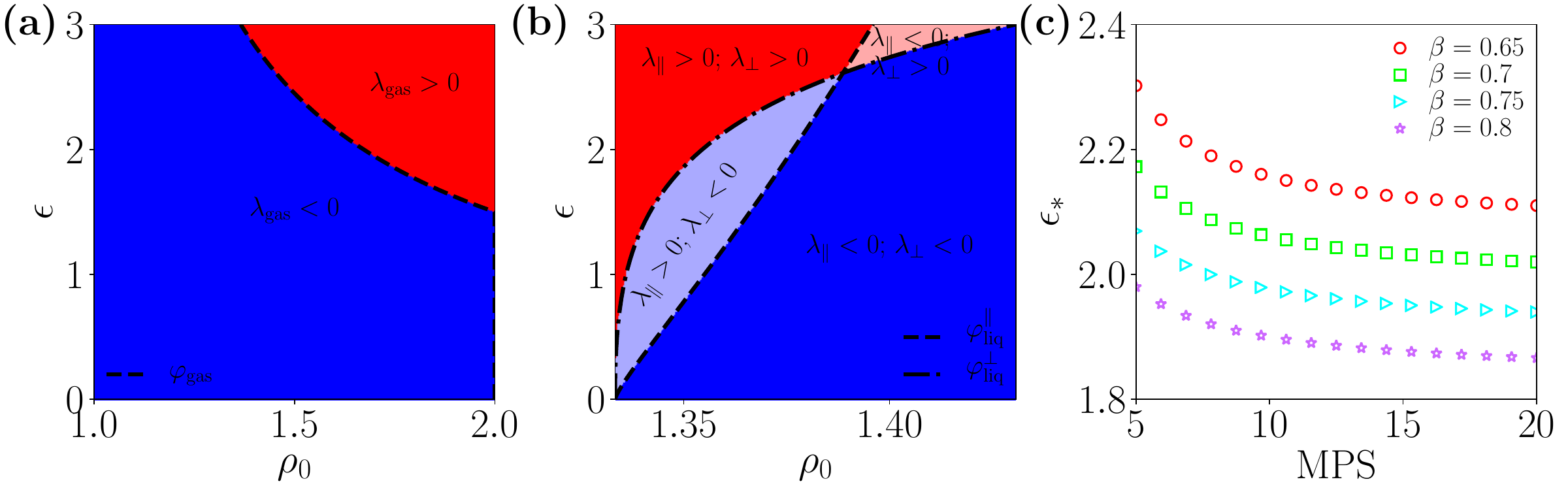}
\caption{(Color online) {\bf (a)}~and {\bf (b)}~Velocity-density stability diagram of the rAPM with hard-core repulsion for the disordered and ordered homogeneous solution for $\beta = 0.75$ for ${\rm MPS}=2$. {\bf (a)} The stability region of the disordered solution is plotted in blue, according to the eigenvalue $\lambda_{\rm gas}$ given by Eq.~\eqref{lambdaGasMPS}. {\bf (b)} The stability region of the ordered solution is plotted in blue, according to the eigenvalues $\lambda_\parallel$ and $\lambda_\perp$ given by Eqs.~\eqref{lambdaParaMPS} and~\eqref{lambdaPerpMPS}, respectively. The ordered solution is only unstable under longitudinal perturbations ($\lambda_\parallel > 0$) in the light blue region, and under transverse perturbations ($\lambda_\perp > 0$) in the light red region. {\bf (c)}~$\epsilon_*$ value for which the reorientation transition occurs in the rAPM with MPS$>$, as a function of MPS for several temperatures $T=\beta^{-1}$.} 
\label{fig_MPS_stab}
\end{figure}

In Fig.~\ref{fig_MPS_stab}(a), we have represented the velocity-density stability diagram for $\beta=0.75$ and ${\rm MPS}=2$ according to the sign of this eigenvalue. Note that the last inequality is always fulfilled when $\zeta=0$, meaning that $\mu_0<0$ only impacts the spinodals: $\varphi_{\rm gas}(\zeta=0)=r/(2\beta J -1)$, independent of $\epsilon$. However, we cannot extract an analytical expression for the spinodal $\varphi_{\rm gas}$ for all ${\rm MPS}$ values.

\medskip

{\bf Linear stability analysis for the ordered homogeneous solution.} We consider the ordered solution along the right state, and we take $\rho_1 = \rho_0 (1+3M)/4 + \delta \rho_1$, $\rho_{2,3,4} = \rho_0 (1+3M)/4 + \delta \rho_{2,3,4}$ and $\rho = \rho_0 + \delta \rho$, with $\delta \rho = \sum_\sigma \delta \rho_\sigma$. The hopping term of the right state $\sigma=1$ writes
\begin{gather}
I_{\rm hop}^{(1)} \simeq \left\{ \left[ 1 - \frac{3\zeta\rho_0}{4}(1-M)\right] (D_\parallel \partial_\parallel^2 + D_\perp \partial_\perp^2) - v \left[ 1 - \frac{\zeta\rho_0}{4}(5+3M)\right] \partial_\parallel \right\} \delta \rho_\sigma \nonumber \\
+ \frac{\zeta \rho_0}{4}(1+3M) \left[ D_\parallel \partial_\parallel^2 + D_\perp \partial_\perp^2 + v \partial_\parallel \right] \sum_{\sigma' \ne \sigma} \delta \rho_{\sigma'},
\end{gather}
and the hopping term of the other states $\sigma\ne 1$ writes
\begin{gather}
I_{\rm hop}^{(2)} \simeq \left\{ \left[ 1 - \frac{\zeta\rho_0}{4}(3+M)\right] (D_\parallel \partial_\parallel^2 + D_\perp \partial_\perp^2) - v \left[ 1 - \frac{\zeta\rho_0}{4}(5-M)\right] \partial_\parallel \right\} \delta \rho_\sigma \nonumber \\
+ \frac{\zeta\rho_0}{4}(1-M) \left[ D_\parallel \partial_\parallel^2 + D_\perp \partial_\perp^2 + v \partial_\parallel \right] \sum_{\sigma' \ne \sigma} \delta \rho_{\sigma'}.
\end{gather}
We may note for $M=0$, we recover the expression of $I_{\rm hop}$, given by Eq.~\eqref{IhopMPS}, calculated for the disordered solution.

The flipping terms implying the right state write
\begin{equation}
I_{\rm flip}(1,\sigma') \simeq M\left\{ (4\beta J - 2\alpha M) \delta \rho_1 + (4\beta J + 2\alpha M) \delta \rho_{\sigma'} - \left[ 2 \beta J (1+M) - \frac{r}{\rho_0} - 2 \alpha M^2\right] \delta \rho \right\}.
\end{equation}
Using Eq.~\eqref{eqMag}, we get
\begin{gather}
I_{\rm flip}(1,\sigma') \simeq M\left[ (4\beta J - 2\alpha M) \delta \rho_1 + (4\beta J + 2\alpha M) \delta \rho_{\sigma'} + (\alpha M^2 -1) \delta \rho \right]\nonumber\\ \equiv \gamma_1 \delta \rho_1 + \gamma_2 \delta \rho_{\sigma'} + \gamma_3 \delta \rho.
\end{gather}
The flipping terms which does not imply the right state write
\begin{equation}
I_{\rm flip}(\sigma,\sigma') \simeq M ( \alpha M - 4 \beta J) (\delta \rho_\sigma - \delta \rho_{\sigma'}) \equiv \gamma_4 (\delta \rho_\sigma - \delta \rho_{\sigma'}).
\end{equation}
Then we have the terms:
\begin{equation}
\sum_{\sigma' \ne \sigma} I_{\rm flip}(\sigma,\sigma') \simeq
\begin{cases}
3(\gamma_1 + \gamma_3) \delta \rho_\sigma + (\gamma_2 + 3 \gamma_3) \sum_{\sigma' \ne \sigma} \rho_{\sigma'}, \qquad &{\rm if}~\sigma=1,\\
- (\gamma_1 + \gamma_3) \delta \rho_1 + (-\gamma_2-\gamma_3+2 \gamma_4) \delta \rho_{\sigma} - (\gamma_3+\gamma_4)\sum_{\sigma' \ne \{1,\sigma\}} \rho_{\sigma'}, \qquad & {\rm if}~\sigma\ne 1.
\end{cases}
\end{equation}

Then, in the Fourier space, the hydrodynamic equation becomes
\begin{small}
\begin{equation}
\partial_t \delta\rho_\sigma =
\begin{cases}
\left[ A_1(k_\parallel,k_\perp) + 3\mu \right] \delta \rho_\sigma + \left[ B_1(k_\parallel,k_\perp) + \nu \right] \sum_{\sigma' \ne \sigma} \delta \rho_{\sigma'}, \qquad &{\rm if}~\sigma=1,\\
\left[ A_2(k_\parallel,k_\perp) + \kappa \right] \delta \rho_{\sigma} + \left[ B_2(k_\parallel,k_\perp) - \mu \right]\delta \rho_1 + \left[ B_2(k_\parallel,k_\perp) - \frac{\kappa + \nu}{2} \right] \sum_{\sigma' \ne \{1,\sigma\}} \delta \rho_{\sigma'}, \qquad & {\rm if}~\sigma\ne 1,
\end{cases}
\end{equation}
\end{small}
with $\mu = \gamma_1 + \gamma_3 = M(4\beta J - 2 \alpha M + \alpha M^2 - 1)$, $\nu = \gamma_2 + 3 \gamma_3 = M(4\beta J + 2 \alpha M + 3\alpha M^2 - 3)$, $\kappa = - \gamma_2 - \gamma_3 + 2 \gamma_4 = M(-12\beta J - \alpha M^2 +1)$, and
\begin{gather}
A_1(k_\parallel,k_\perp) = \left[ 1 - \frac{3\zeta\rho_0}{4}(1-M)\right] (-D_\parallel k_\parallel^2 - D_\perp k_\perp^2) + \imath k_\parallel v \left[ 1 - \frac{\zeta\rho_0}{4}(5+3M)\right], \\
B_1(k_\parallel,k_\perp) = \frac{\zeta\rho_0}{4}(1+3M) \left[ -D_\parallel k_\parallel^2 - D_\perp k_\perp^2 - \imath k_\parallel v \right], \\
A_2(k_\parallel,k_\perp) =  \left[ 1 - \frac{\zeta\rho_0}{4}(3+M)\right] (-D_\parallel k_\parallel^2 - D_\perp k_\perp^2) + \imath k_\parallel v \left[ 1 - \frac{\zeta\rho_0}{4}(5-M)\right], \\
B_2(k_\parallel,k_\perp) = \frac{\zeta\rho_0}{4}(1-M) \left[ -D_\parallel k_\parallel^2 - D_\perp k_\perp^2 - \imath k_\parallel v \right].
\end{gather}

The stability of the homogeneous ordered solution is then given by the eigenvalues of the matrix
\begin{footnotesize}
\begin{equation}
M_{\rm liq} = \begin{pmatrix}
A_1(k_x,k_y) + 3\mu & B_1(k_x,k_y)+\nu & B_1(k_x,k_y)+\nu & B_1(k_x,k_y)+\nu \\
B_2(k_y,-k_x)-\mu & A_2(k_y,-k_x)+\kappa & B_2(k_y,-k_x)- (\kappa + \nu)/2 & B_2(k_y,-k_x)- (\kappa + \nu)/2 \\
B_2(-k_x,-k_y)-\mu & B_2(-k_x,-k_y)- (\kappa + \nu)/2 & A_2(-k_x,-k_y)+\kappa & B_2(-k_x,-k_y)- (\kappa + \nu)/2 \\
B_2(-k_y,k_x)-\mu & B_2(-k_y,k_x)- (\kappa + \nu)/2 & B_2(-k_y,k_x)- (\kappa + \nu)/2 & A_2(-k_y,k_x)+\kappa
\end{pmatrix}.
\end{equation}
\end{footnotesize}

First, we consider a perturbation in the $x$ direction ($k_y=0$). At leading order in $k_x \ll 1$, the real part of the eigenvalues, calculated with Mathematica~\cite{inc2017mathematica}, are: $\lambda_{\rm liq,x}^{1,2} = (3\kappa+\nu)/2$, $\lambda_{\rm liq,x}^3 \simeq (3\mu+\nu)$, and
\begin{small}
\begin{gather}
\lambda_{{\rm liq}, x}^4 \simeq  - \left[ \frac{(D_\parallel + 2D_\perp) \mu - D_\parallel \nu}{3\mu - \nu} + \frac{D_\parallel-D_\perp}{2(3\mu-\nu)}[\mu(1+3M)+\nu(1-M)] \zeta \rho_0 + \frac{(c_1+c_2 \zeta \rho_0)(1-\zeta \rho_0)}{(3\mu-\nu)^3(3\kappa+\nu)} v^2\right] k_x^2,
\end{gather}
\end{small}
with $c_1=4\mu[-3\mu^2+\nu(2\mu+4\kappa+\nu)]$, and
\begin{small}
\begin{equation}
c_2=\kappa\nu^2(1-M) + \mu[3\mu^2(7-3M)-2\kappa\nu(11+3M)-\nu^2(3+5M)+9\kappa\mu(1+3M)+ 2\mu\nu(-7+9M)].\nonumber
\end{equation}
\end{small}

Now we look at a perturbation in the $y$ direction ($k_x=0$). At leading order in $k_y \ll 1$, the real part of the eigenvalues, calculated with Mathematica~\cite{inc2017mathematica}, are: $\lambda_{\rm liq,y}^{1,2} = (3\kappa+\nu)/2$, $\lambda_{\rm liq,y}^3 \simeq (3\mu+\nu)$, and
\begin{small}
\begin{gather}
\lambda_{{\rm liq}, y}^4 \simeq \left[ \frac{-(2 D_\parallel + D_\perp) \mu + D_\perp \nu}{3\mu - \nu} + \frac{D_\parallel-D_\perp}{2(3\mu-\nu)}[\mu(1+3M)+\nu(1-M)] \zeta \rho_0 +  \frac{(d_1+d_2 \zeta \rho_0)(1-\zeta \rho_0)}{(3\mu-\nu)(3\kappa+\nu)}v^2 \right] k_y^2,
\end{gather}
\end{small}
with $d_1=4\mu$, and $d_2=\mu(-7+3M)+\nu(1-M)$.

The ordered homogeneous solution is then stable if $3\kappa+\nu<0$ and $3\mu-\nu<0$ for the two different perturbations. This result was already observed for the unrestricted APM~\cite{chatterjee2020flocking}, and allows the selection of the position magnetization solution: $M=M_0 + M_1 \delta$. However, the stability of the two different perturbations differs from $\lambda_{{\rm liq}, x}^4$ and $\lambda_{{\rm liq}, y}^4$. The perturbation along $x$ is stable only if
\begin{gather}
\lambda_\parallel = \frac{\lambda_{{\rm liq}, x}^4}{k_x^2} = -D + \left[\frac{\mu+\nu}{3\mu-\nu}(1-\zeta \rho_0) - M\zeta \rho_0 \right] \frac{D\epsilon}{3} - \frac{(c_1+c_2 \zeta \rho_0)(1-\zeta \rho_0)}{(3\mu-\nu)^3(3\kappa+\nu)}  \left(\frac{4D\epsilon}{3} \right)^2 \label{lambdaParaMPS}
\end{gather}
is negative and the perturbation along $y$ is stable only if
\begin{equation}
\label{lambdaPerpMPS}
\lambda_\perp = \frac{\lambda_{{\rm liq}, y}^4}{k_y^2} = -D - \left[\frac{\mu+\nu}{3\mu-\nu}(1-\zeta \rho_0) - M\zeta \rho_0 \right] \frac{D\epsilon}{3} +  \frac{(d_1+d_2 \zeta \rho_0)(1-\zeta \rho_0)}{(3\mu-\nu)(3\kappa+\nu)} \left(\frac{4D\epsilon}{3} \right)^2
\end{equation}
is negative. We may note that these eigenvalues are those obtained in Ref.~\cite{chatterjee2020flocking} for $\zeta=0$. In Fig.~\ref{fig_MPS_stab}(b), we have represented the velocity-density stability diagram for $\beta=0.75$ and ${\rm MPS}=2$ according to the sign of these two eigenvalues. Here, we have not derived an analytical expression of $\epsilon_*$, for which the reorientation transition occurs, but we have computed a numerical estimation in Fig.~\ref{fig_MPS_stab}(c). $\epsilon_*$ is an decreasing function of MPS, meaning that $\epsilon_*$ increases with a strong repulsion.


\section{Linear stability analysis for the soft-core rAPM}
\label{ls_sc}

For soft-core rAPM, the hydrodynamic equation is
\begin{small}
\begin{gather}
\partial_t \rho_\sigma = D_\parallel \partial_\parallel \left[ \exp(-s\rho) \left( \partial_\parallel \rho_\sigma + s \rho_\sigma \partial_\parallel \rho \right) \right] + D_\perp \partial_\perp \left[ \exp(-s\rho) \left( \partial_\perp \rho_\sigma + s \rho_\sigma \partial_\perp \rho \right) \right] - v \partial_\parallel \left[ \exp(-s\rho) \rho_\sigma \right] \nonumber \\
+ \sum_{\sigma'\ne\sigma} \left[ \frac{4\beta J}{\rho}(\rho_\sigma + \rho_{\sigma'}) - 1 - \frac{r}{\rho} - \alpha \frac{(\rho_\sigma - \rho_{\sigma'})^2}{\rho^2} \right](\rho_\sigma - \rho_{\sigma'}),
\end{gather}
\end{small}
with $\alpha = 8(\beta J)^2(1-2\beta J/3)$ and $s=2\beta U$. The homogeneous solutions are given by:
\begin{equation}
I_{\rm flip}(\sigma,\sigma') = \left[ \frac{4\beta J}{\rho}(\rho_\sigma + \rho_{\sigma'}) - 1 - \frac{r}{\rho} - \alpha \frac{(\rho_\sigma - \rho_{\sigma'})^2}{\rho^2} \right](\rho_\sigma - \rho_{\sigma'}) = 0,
\end{equation}
and are then those of the unrestricted rAPM~\cite{mangeat2020flocking}. The disordered homogeneous solution is $\rho_\sigma = \rho_0 / 4$, and the ordered homogeneous solution (supposed along state $\sigma = 1$) is $\rho_1 = \rho_0 (1+3M) / 4$ and $\rho_{2,3,4} = \rho_0 (1-M) / 4$ with the magnetization $M$ following the equation:
\begin{equation}
\label{eqMag2}
2\beta J (1+M) - 1 - \frac{r}{\rho_0} - \alpha M^2 = 0,
\end{equation}
or $M=M_0 \pm M_1 \delta$ with $M_0 = \beta J / \alpha$, $M_1 = \sqrt{r/\alpha \rho_*}$ and $\delta = \sqrt{(\rho_0-\rho_*)/\rho_0}$, where $\rho_*$ defined by
\begin{equation}
\rho_* = \frac{8(1-2\beta J/3) r}{1 + 8(2\beta J - 1)(1-2\beta J/3)},
\end{equation}
is the critical density below which the ordered homogeneous solution does not exist, for a temperature below $T_c = (1 - \sqrt{22}/8)^{-1} \simeq 2.417$.

\medskip

{\bf Linear stability analysis for the disordered homogeneous solution.} We take $\rho_\sigma = \rho_0/4 + \delta \rho_\sigma$ and $\rho = \rho_0 + \delta \rho$, with $\delta \rho = \sum_\sigma \delta \rho_\sigma$. The hopping term writes
\begin{gather}
I_{\rm hop} \simeq \exp(-s\rho_0) \left[ \left( 1 + \frac{s\rho_0}{4}\right) (D_\parallel \partial_\parallel^2 + D_\perp \partial_\perp^2) - v \left( 1 - \frac{s\rho_0}{4}\right) \partial_\parallel \right] \delta \rho_\sigma \nonumber \\
+ \exp(-s\rho_0) \frac{s\rho_0}{4} \left[ D_\parallel \partial_\parallel^2 + D_\perp \partial_\perp^2 + v \partial_\parallel \right] \sum_{\sigma' \ne \sigma} \delta \rho_{\sigma'}, \label{Ihop}
\end{gather}
and the flipping term writes $I_{\rm flip}(\sigma,\sigma') \simeq \mu_0 (\delta \rho_\sigma - \delta \rho_{\sigma'})$, with $\mu_0 = 2\beta J - 1 - r/\rho_0$. Then, in the Fourier space, the hydrodynamic equation becomes
\begin{equation}
\partial_t \delta\rho_\sigma = \left[A(k_\parallel,k_\perp) + 3 \mu_0 \right] \delta \rho_\sigma + \left[B(k_\parallel,k_\perp)- \mu_0\right] \sum_{\sigma' \ne \sigma} \delta \rho_{\sigma'},
\end{equation}
with
\begin{gather}
A(k_\parallel,k_\perp) = \exp(-s\rho_0) \left[ \left( 1 + \frac{s\rho_0}{4}\right) (-D_\parallel k_\parallel^2 - D_\perp k_\perp^2) + \imath k_\parallel v \left( 1 - \frac{s\rho_0}{4}\right) \right] \\
B(k_\parallel,k_\perp) = \exp(-s\rho_0) \frac{s\rho_0}{4} \left[ -D_\parallel k_\parallel^2 - D_\perp k_\perp^2 - \imath k_\parallel v \right] 
\end{gather}

The stability of the homogeneous disordered solution is then given by the eigenvalues of the matrix
\begin{equation}
M_{\rm gas} = \begin{pmatrix}
A(k_x,k_y) + 3 \mu_0 & B(k_x,k_y) - \mu_0 & B(k_x,k_y) - \mu_0 & B(k_x,k_y) - \mu_0\\
B(k_y,-k_x) - \mu_0 & A(k_y,-k_x) + 3 \mu_0 & B(k_y,-k_x) - \mu_0 & B(k_y,-k_x) - \mu_0\\
B(-k_x,-k_y) - \mu_0 & B(-k_x,-k_y) - \mu_0 & A(-k_x,-k_y) + 3 \mu_0 & B(-k_x,-k_y) - \mu_0\\
B(-k_y,k_x) - \mu_0 & B(-k_y,k_x) - \mu_0 & B(-k_y,k_x) - \mu_0 & A(-k_y,k_x) + 3 \mu_0
\end{pmatrix}.
\end{equation}

\begin{figure}[t]
\centering
\includegraphics[width=\columnwidth]{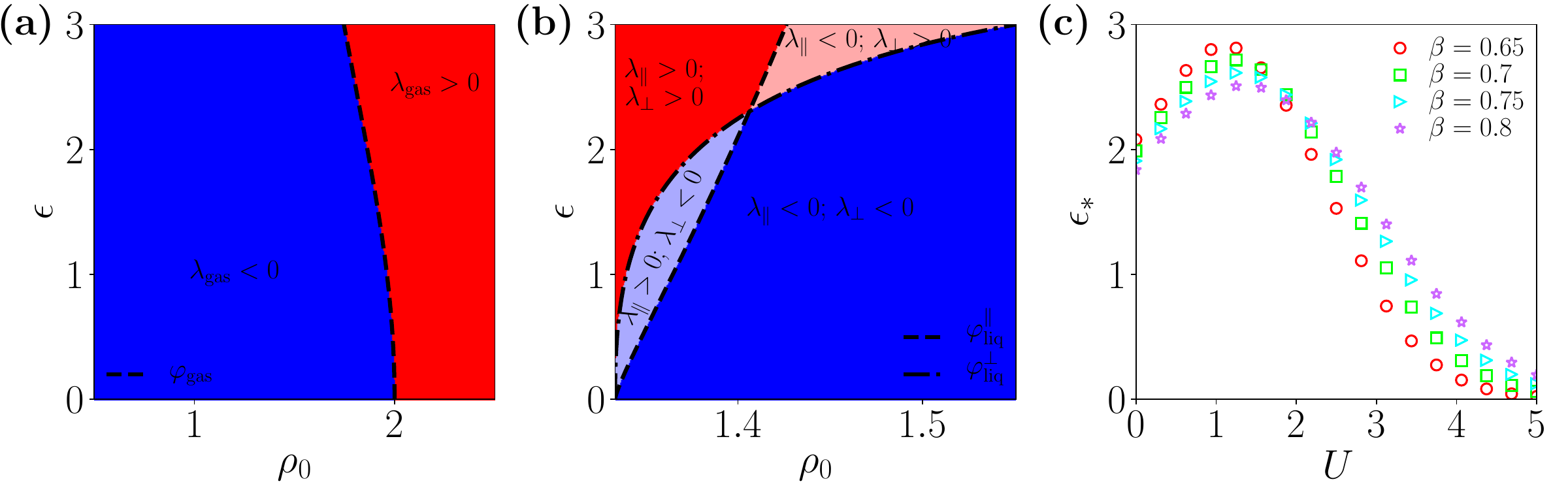}
\caption{(Color online) {\bf (a)}~and {\bf (b)}~Velocity-density stability diagram of the rAPM with soft-core repulsion for the disordered and ordered homogeneous solution for $\beta = 0.75$ and $U=0.5$. {\bf (a)} The stability region of the disordered solution is plotted in blue, according to the eigenvalue $\lambda_{\rm gas}$ given by Eq.~\eqref{lambdaGas}. {\bf (b)} The stability region of the ordered solution is plotted in blue, according to the eigenvalues $\lambda_\parallel$ and $\lambda_\perp$ given by Eqs.~\eqref{lambdaPara} and~\eqref{lambdaPerp}, respectively. The ordered solution is only unstable under longitudinal perturbations ($\lambda_\parallel > 0$) in the light blue region, and under transverse perturbations ($\lambda_\perp > 0$) in the light red region. {\bf (c)}~$\epsilon_*$ value for which the reorientation transition occurs, as a function of $U$ for several temperatures $T=\beta^{-1}$.} 
\label{fig_softcore_stab}
\end{figure}

Supposing $k_x=k$ and $k_y=0$ (since no preferred direction), at leading order in $k \ll 1$, the eigenvalues, calculated with Mathematica~\cite{inc2017mathematica}, are: $\lambda_{\rm gas}^{1,2,3} \simeq 4 \mu_0$ and
\begin{gather}
\lambda_{\rm gas}^4 \simeq \exp(-s\rho_0) \left[-(1+s\rho_0)\frac{D_\parallel +D_\perp}{2} + (1- s\rho_0) \exp(-s\rho_0) \frac{v^2}{8\mu_0} \right] k^2.
\end{gather}
The disordered homogeneous solution is then stable if $\mu_0<0$ and
\begin{equation}
\label{lambdaGas}
\lambda_{\rm gas} = -(1+s\rho_0)D + (1- s\rho_0) \exp(-s\rho_0) \frac{v^2}{8\mu_0}<0.
\end{equation}
In Fig.~\ref{fig_softcore_stab}(a), we have represented the velocity-density stability diagram for $\beta=0.75$ and $U=0.5$ according to the sign of this eigenvalue. Note that the last inequality is always fulfilled when $U=0$, meaning that $\mu_0<0$ only impacts the spinodals: $\varphi_{\rm gas}(U=0)=r/(2\beta J -1)$, independent of $\epsilon$. However, we cannot extract an analytical expression for the spinodal $\varphi_{\rm gas}$ for all $U$ values.

\medskip

{\bf Linear stability analysis for the ordered homogeneous solution.} We consider the ordered solution along the right state, and we take $\rho_1 = \rho_0 (1+3M)/4 + \delta \rho_1$, $\rho_{2,3,4} = \rho_0 (1+3M)/4 + \delta \rho_{2,3,4}$ and $\rho = \rho_0 + \delta \rho$, with $\delta \rho = \sum_\sigma \delta \rho_\sigma$. The hopping term of the right state $\sigma=1$ writes
\begin{gather}
I_{\rm hop}^{(1)} \simeq \exp(-s\rho_0) \left\{ \left[ 1 + \frac{s\rho_0}{4}(1+3M)\right] (D_\parallel \partial_\parallel^2 + D_\perp \partial_\perp^2) - v \left[ 1 - \frac{s\rho_0}{4}(1+3M)\right] \partial_\parallel \right\} \delta \rho_\sigma \nonumber \\
+ \exp(-s\rho_0) \frac{s\rho_0}{4}(1+3M) \left[ D_\parallel \partial_\parallel^2 + D_\perp \partial_\perp^2 + v \partial_\parallel \right] \sum_{\sigma' \ne \sigma} \delta \rho_{\sigma'},
\end{gather}
and the hopping term of the other states $\sigma\ne 1$ writes
\begin{gather}
I_{\rm hop}^{(2)} \simeq \exp(-s\rho_0) \left\{ \left[ 1 + \frac{s\rho_0}{4}(1-M)\right] (D_\parallel \partial_\parallel^2 + D_\perp \partial_\perp^2) - v \left[ 1 - \frac{s\rho_0}{4}(1-M)\right] \partial_\parallel \right\} \delta \rho_\sigma \nonumber \\
+ \exp(-s\rho_0) \frac{s\rho_0}{4}(1-M) \left[ D_\parallel \partial_\parallel^2 + D_\perp \partial_\perp^2 + v \partial_\parallel \right] \sum_{\sigma' \ne \sigma} \delta \rho_{\sigma'}.
\end{gather}
We may note for $M=0$, we recover the expression of $I_{\rm hop}$, given by Eq.~\eqref{Ihop}, calculated for the disordered solution.

The flipping terms implying the right state write
\begin{equation}
I_{\rm flip}(1,\sigma') \simeq M\left\{ (4\beta J - 2\alpha M) \delta \rho_1 + (4\beta J + 2\alpha M) \delta \rho_{\sigma'} - \left[ 2 \beta J (1+M) - \frac{r}{\rho_0} - 2 \alpha M^2\right] \delta \rho \right\}.
\end{equation}
Using Eq.~\eqref{eqMag2}, we get
\begin{gather}
I_{\rm flip}(1,\sigma') \simeq M\left[ (4\beta J - 2\alpha M) \delta \rho_1 + (4\beta J + 2\alpha M) \delta \rho_{\sigma'} + (\alpha M^2 -1) \delta \rho \right]\nonumber\\ \equiv \gamma_1 \delta \rho_1 + \gamma_2 \delta \rho_{\sigma'} + \gamma_3 \delta \rho.
\end{gather}
The flipping term which does not imply the right state write
\begin{equation}
I_{\rm flip}(\sigma,\sigma') \simeq M ( \alpha M - 4 \beta J) (\delta \rho_\sigma - \delta \rho_{\sigma'}) \equiv \gamma_4 (\delta \rho_\sigma - \delta \rho_{\sigma'}).
\end{equation}
Then we have the terms:
\begin{small}
\begin{equation}
I_\sigma = \sum_{\sigma' \ne \sigma} I_{\rm flip}(\sigma,\sigma') \simeq
\begin{cases}
3(\gamma_1 + \gamma_3) \delta \rho_\sigma + (\gamma_2 + 3 \gamma_3) \sum_{\sigma' \ne \sigma} \rho_{\sigma'}, \qquad &{\rm if}~\sigma=1,\\
- (\gamma_1 + \gamma_3) \delta \rho_1 + (-\gamma_2-\gamma_3+2 \gamma_4) \delta \rho_{\sigma} - (\gamma_3+\gamma_4)\sum_{\sigma' \ne \{1,\sigma\}} \rho_{\sigma'}, \qquad & {\rm if}~\sigma\ne 1.
\end{cases}
\end{equation}
\end{small}

Then, in the Fourier space, the hydrodynamic equation becomes
\begin{small}
\begin{equation}
\partial_t \delta\rho_\sigma =
\begin{cases}
\left[ A_1(k_\parallel,k_\perp) + 3\mu \right] \delta \rho_\sigma + \left[ B_1(k_\parallel,k_\perp) + \nu \right] \sum_{\sigma' \ne \sigma} \delta \rho_{\sigma'}, \qquad &{\rm if}~\sigma=1,\\
\left[ A_2(k_\parallel,k_\perp) + \kappa \right] \delta \rho_{\sigma} + \left[ B_2(k_\parallel,k_\perp) - \mu \right]\delta \rho_1 + \left[ B_2(k_\parallel,k_\perp) - \frac{\kappa + \nu}{2} \right] \sum_{\sigma' \ne \{1,\sigma\}} \delta \rho_{\sigma'}, \qquad & {\rm if}~\sigma\ne 1,
\end{cases}
\end{equation}
\end{small}
with $\mu = \gamma_1 + \gamma_3 = M(4\beta J - 2 \alpha M + \alpha M^2 - 1)$, $\nu = \gamma_2 + 3 \gamma_3 = M(4\beta J + 2 \alpha M + 3\alpha M^2 - 3)$, $\kappa = - \gamma_2 - \gamma_3 + 2 \gamma_4 = M(-12\beta J - \alpha M^2 +1)$, and
\begin{gather}
A_1(k_\parallel,k_\perp) = \exp(-s\rho_0) \left\{ \left[ 1 + \frac{s\rho_0}{4}(1+3M)\right] (-D_\parallel k_\parallel^2 - D_\perp k_\perp^2) + \imath k_\parallel v \left[ 1 - \frac{s\rho_0}{4}(1+3M)\right] \right\}, \\
B_1(k_\parallel,k_\perp) = \exp(-s\rho_0) \frac{s\rho_0}{4}(1+3M) \left[ -D_\parallel k_\parallel^2 - D_\perp k_\perp^2 - \imath k_\parallel v \right], \\
A_2(k_\parallel,k_\perp) = \exp(-s\rho_0) \left\{ \left[ 1 + \frac{s\rho_0}{4}(1-M)\right] (-D_\parallel k_\parallel^2 - D_\perp k_\perp^2) + \imath k_\parallel v \left[ 1 - \frac{s\rho_0}{4}(1-M)\right] \right\}, \\
B_2(k_\parallel,k_\perp) = \exp(-s\rho_0) \frac{s\rho_0}{4}(1-M) \left[ -D_\parallel k_\parallel^2 - D_\perp k_\perp^2 - \imath k_\parallel v \right].
\end{gather}

The stability of the homogeneous ordered solution is then given by the eigenvalues of the matrix

\begin{footnotesize}
\begin{equation}
M_{\rm liq} = \begin{pmatrix}
A_1(k_x,k_y) + 3\mu & B_1(k_x,k_y)+\nu & B_1(k_x,k_y)+\nu & B_1(k_x,k_y)+\nu \\
B_2(k_y,-k_x)-\mu & A_2(k_y,-k_x)+\kappa & B_2(k_y,-k_x)- (\kappa + \nu)/2 & B_2(k_y,-k_x)- (\kappa + \nu)/2 \\
B_2(-k_x,-k_y)-\mu & B_2(-k_x,-k_y)- (\kappa + \nu)/2 & A_2(-k_x,-k_y)+\kappa & B_2(-k_x,-k_y)- (\kappa + \nu)/2 \\
B_2(-k_y,k_x)-\mu & B_2(-k_y,k_x)- (\kappa + \nu)/2 & B_2(-k_y,k_x)- (\kappa + \nu)/2 & A_2(-k_y,k_x)+\kappa
\end{pmatrix}.
\end{equation}
\end{footnotesize}

In the following, we will denote $\overline{D_\parallel} = D_\parallel \exp(-s\rho_0)$, $\overline{D_\perp} = D_\perp \exp(-s\rho_0)$ and $\overline{v} = v \exp(-s\rho_0)$. First, we consider a perturbation in the $x$ direction ($k_y=0$). At leading order in $k_x \ll 1$, the real part of the eigenvalues, calculated with Mathematica~\cite{inc2017mathematica}, are: $\lambda_{\rm liq,x}^{1,2} \simeq (3\kappa+\nu)/2$, $\lambda_{\rm liq,x}^3 \simeq (3\mu+\nu)$, and
\begin{gather}
\lambda_{{\rm liq}, x}^4 \simeq - \left[ \frac{(\overline{D_\parallel} + 2\overline{D_\perp}) \mu - \overline{D_\parallel} \nu}{3\mu - \nu} + \frac{\overline{D_\parallel}(1+M)+\overline{D_\perp}(1-M)}{2} s\rho_0 +\frac{c_1+c_2 s\rho_0}{(3\mu-\nu)^3(3\kappa+\nu)} \overline{v}^2\right] k_x^2,
\end{gather}
with $c_1 = 4\mu[-3\mu^2+\nu(2\mu+4\kappa+\nu)]$, and $c_2 = (3\mu-\nu) \{3\mu[\kappa(1+3M)+\mu(1-M)] - \nu[\kappa(1-M)+\mu(1-5M)] \}$.

Now we look at a perturbation in the $y$ direction ($k_x=0$). At leading order in $k_y \ll 1$, the real part of the eigenvalues, calculated with Mathematica~\cite{inc2017mathematica}, are: $\lambda_{\rm liq,y}^{1,2} \simeq (3\kappa+\nu)/2$, $\lambda_{\rm liq,y}^3 \simeq (3\mu+\nu)$, and
\begin{equation}
\lambda_{{\rm liq}, y}^4 \simeq \left[ \frac{-(2\overline{D_\parallel} + \overline{D_\perp}) \mu + \overline{D_\perp} \nu}{3\mu - \nu} - \frac{\overline{D_\parallel}(1-M)+\overline{D_\perp}(1+M)}{2} s\rho_0 + \frac{d_1+d_2 s\rho_0}{(3\mu-\nu)(3\kappa+\nu)}  \overline{v}^2 \right] k_y^2,
\end{equation}
with $d_1 = 4\mu$, and $d_2=-(1-M)(3\mu-\nu)$.

The ordered homogeneous solution is then stable if $3\kappa+\nu<0$ and $3\mu-\nu<0$ for the two different perturbations. This result was already observed for the unrestricted APM~\cite{chatterjee2020flocking}, and allows the selection of the position magnetization solution: $M=M_0 + M_1 \delta$. However, the stability of the two different perturbations differs from $\lambda_{{\rm liq}, x}^4$ and $\lambda_{{\rm liq}, y}^4$. The perturbation along $x$ is stable only if
\begin{gather}
\lambda_\parallel = \frac{\lambda_{{\rm liq}, x}^4}{k_x^2 \exp(-s\rho_0)} = -D(1+s\rho_0) + \left[\frac{\mu+\nu}{3\mu-\nu} - M s\rho_0 \right] \frac{D\epsilon}{3}\nonumber\\ -  \frac{(c_1 + c_2 s\rho_0)\exp(-s\rho_0)}{(3\mu-\nu)^3(3\kappa+\nu)} \left(\frac{4D\epsilon}{3} \right)^2\label{lambdaPara}
\end{gather}
is negative and the perturbation along $y$ is stable only if
\begin{gather}
\label{lambdaPerp}
\lambda_\perp = \frac{\lambda_{{\rm liq}, y}^4}{k_y^2 \exp(-s\rho_0)} = -D(1+s\rho_0) - \left[\frac{\mu+\nu}{3\mu-\nu} - M s\rho_0 \right] \frac{D\epsilon}{3}\nonumber\\ +  \frac{(d_1+d_2 s\rho_0)\exp(-s\rho_0)}{(3\mu-\nu)(3\kappa+\nu)} \left(\frac{4D\epsilon}{3} \right)^2 
\end{gather}
is negative. We may note that these eigenvalues are those obtained in Ref.~\cite{chatterjee2020flocking} for $s=0$. In Fig.~\ref{fig_softcore_stab}(b), we have represented the velocity-density stability diagram for $\beta=0.75$ and $U=0.5$ according to the sign of these two eigenvalues. Here, we have not derived an analytical expression of $\epsilon_*$, for which the reorientation transition occurs, but we have computed a numerical estimation in Fig.~\ref{fig_softcore_stab}(c). $\epsilon_*(U)$ increases for small $U$ and decreases to zero at large $U$.

\end{appendix}


\begin{thebibliography}{93}%
\makeatletter
\providecommand \@ifxundefined [1]{%
 \@ifx{#1\undefined}
}%
\providecommand \@ifnum [1]{%
 \ifnum #1\expandafter \@firstoftwo
 \else \expandafter \@secondoftwo
 \fi
}%
\providecommand \@ifx [1]{%
 \ifx #1\expandafter \@firstoftwo
 \else \expandafter \@secondoftwo
 \fi
}%
\providecommand \natexlab [1]{#1}%
\providecommand \enquote  [1]{``#1''}%
\providecommand \bibnamefont  [1]{#1}%
\providecommand \bibfnamefont [1]{#1}%
\providecommand \citenamefont [1]{#1}%
\providecommand \href@noop [0]{\@secondoftwo}%
\providecommand \href [0]{\begingroup \@sanitize@url \@href}%
\providecommand \@href[1]{\@@startlink{#1}\@@href}%
\providecommand \@@href[1]{\endgroup#1\@@endlink}%
\providecommand \@sanitize@url [0]{\catcode `\\12\catcode `\$12\catcode
  `\&12\catcode `\#12\catcode `\^12\catcode `\_12\catcode `\%12\relax}%
\providecommand \@@startlink[1]{}%
\providecommand \@@endlink[0]{}%
\providecommand \url  [0]{\begingroup\@sanitize@url \@url }%
\providecommand \@url [1]{\endgroup\@href {#1}{\urlprefix }}%
\providecommand \urlprefix  [0]{URL }%
\providecommand \Eprint [0]{\href }%
\providecommand \doibase [0]{http://dx.doi.org/}%
\providecommand \selectlanguage [0]{\@gobble}%
\providecommand \bibinfo  [0]{\@secondoftwo}%
\providecommand \bibfield  [0]{\@secondoftwo}%
\providecommand \translation [1]{[#1]}%
\providecommand \BibitemOpen [0]{}%
\providecommand \bibitemStop [0]{}%
\providecommand \bibitemNoStop [0]{.\EOS\space}%
\providecommand \EOS [0]{\spacefactor3000\relax}%
\providecommand \BibitemShut  [1]{\csname bibitem#1\endcsname}%
\let\auto@bib@innerbib\@empty
\bibitem [{\citenamefont {Toner}\ \emph {et~al.}(2005)\citenamefont {Toner},
  \citenamefont {Tu},\ and\ \citenamefont
  {Ramaswamy}}]{toner2005hydrodynamics}%
  \BibitemOpen
  \bibfield  {author} {\bibinfo {author} {\bibfnamefont {J.}~\bibnamefont
  {Toner}}, \bibinfo {author} {\bibfnamefont {Y.}~\bibnamefont {Tu}}, \ and\
  \bibinfo {author} {\bibfnamefont {S.}~\bibnamefont {Ramaswamy}},\ }\href@noop
  {} {\bibfield  {journal} {\bibinfo  {journal} {Annals of Physics}\ }\textbf
  {\bibinfo {volume} {318}},\ \bibinfo {pages} {170} (\bibinfo {year}
  {2005})}\BibitemShut {NoStop}%
\bibitem [{\citenamefont {Juelicher}\ \emph {et~al.}(2007)\citenamefont
  {Juelicher}, \citenamefont {Kruse}, \citenamefont {Prost},\ and\
  \citenamefont {Joanny}}]{juelicher2007active}%
  \BibitemOpen
  \bibfield  {author} {\bibinfo {author} {\bibfnamefont {F.}~\bibnamefont
  {Juelicher}}, \bibinfo {author} {\bibfnamefont {K.}~\bibnamefont {Kruse}},
  \bibinfo {author} {\bibfnamefont {J.}~\bibnamefont {Prost}}, \ and\ \bibinfo
  {author} {\bibfnamefont {J.-F.}\ \bibnamefont {Joanny}},\ }\href@noop {}
  {\bibfield  {journal} {\bibinfo  {journal} {Phys. Rep.}\ }\textbf {\bibinfo
  {volume} {449}},\ \bibinfo {pages} {3} (\bibinfo {year} {2007})}\BibitemShut
  {NoStop}%
\bibitem [{\citenamefont {Joanny}\ and\ \citenamefont
  {Prost}(2009)}]{joanny2009active}%
  \BibitemOpen
  \bibfield  {author} {\bibinfo {author} {\bibfnamefont {J.-F.}\ \bibnamefont
  {Joanny}}\ and\ \bibinfo {author} {\bibfnamefont {J.}~\bibnamefont {Prost}},\
  }\href@noop {} {\bibfield  {journal} {\bibinfo  {journal} {HFSP J}\ }\textbf
  {\bibinfo {volume} {3}},\ \bibinfo {pages} {94} (\bibinfo {year}
  {2009})}\BibitemShut {NoStop}%
\bibitem [{\citenamefont {Ramaswamy}(2010)}]{ramaswamy2010mechanics}%
  \BibitemOpen
  \bibfield  {author} {\bibinfo {author} {\bibfnamefont {S.}~\bibnamefont
  {Ramaswamy}},\ }\href@noop {} {\bibfield  {journal} {\bibinfo  {journal}
  {Annu. Rev. Condens. Matter Phys.}\ }\textbf {\bibinfo {volume} {1}},\
  \bibinfo {pages} {323} (\bibinfo {year} {2010})}\BibitemShut {NoStop}%
\bibitem [{\citenamefont {Schweitzer}\ and\ \citenamefont
  {Farmer}(2003)}]{schweitzer2003brownian}%
  \BibitemOpen
  \bibfield  {author} {\bibinfo {author} {\bibfnamefont {F.}~\bibnamefont
  {Schweitzer}}\ and\ \bibinfo {author} {\bibfnamefont {J.~D.}\ \bibnamefont
  {Farmer}},\ }\href@noop {} {\emph {\bibinfo {title} {Brownian agents and
  active particles: collective dynamics in the natural and social sciences}}},\
  Vol.~\bibinfo {volume} {1}\ (\bibinfo  {publisher} {Springer},\ \bibinfo
  {year} {2003})\BibitemShut {NoStop}%
\bibitem [{\citenamefont {Vicsek}\ and\ \citenamefont
  {Zafeiris}(2012)}]{vicsek2012collective}%
  \BibitemOpen
  \bibfield  {author} {\bibinfo {author} {\bibfnamefont {T.}~\bibnamefont
  {Vicsek}}\ and\ \bibinfo {author} {\bibfnamefont {A.}~\bibnamefont
  {Zafeiris}},\ }\href@noop {} {\bibfield  {journal} {\bibinfo  {journal}
  {Phys. Rep.}\ }\textbf {\bibinfo {volume} {517}},\ \bibinfo {pages} {71}
  (\bibinfo {year} {2012})}\BibitemShut {NoStop}%
\bibitem [{\citenamefont {Strogatz}(2004)}]{strogatz2004sync}%
  \BibitemOpen
  \bibfield  {author} {\bibinfo {author} {\bibfnamefont {S.}~\bibnamefont
  {Strogatz}},\ }\href@noop {} {\emph {\bibinfo {title} {Sync: The emerging
  science of spontaneous order}}}\ (\bibinfo  {publisher} {Penguin UK},\
  \bibinfo {year} {2004})\BibitemShut {NoStop}%
\bibitem [{\citenamefont {De~Magistris}\ and\ \citenamefont
  {Marenduzzo}(2015)}]{de2015introduction}%
  \BibitemOpen
  \bibfield  {author} {\bibinfo {author} {\bibfnamefont {G.}~\bibnamefont
  {De~Magistris}}\ and\ \bibinfo {author} {\bibfnamefont {D.}~\bibnamefont
  {Marenduzzo}},\ }\href@noop {} {\bibfield  {journal} {\bibinfo  {journal}
  {Physica A}\ }\textbf {\bibinfo {volume} {418}},\ \bibinfo {pages} {65}
  (\bibinfo {year} {2015})}\BibitemShut {NoStop}%
\bibitem [{\citenamefont {Krishnan}\ \emph {et~al.}(2010)\citenamefont
  {Krishnan}, \citenamefont {Deshpande},\ and\ \citenamefont
  {Kumar}}]{krishnan2010rheology}%
  \BibitemOpen
  \bibfield  {author} {\bibinfo {author} {\bibfnamefont {J.~M.}\ \bibnamefont
  {Krishnan}}, \bibinfo {author} {\bibfnamefont {A.~P.}\ \bibnamefont
  {Deshpande}}, \ and\ \bibinfo {author} {\bibfnamefont {P.~B.~S.}\
  \bibnamefont {Kumar}},\ }\href@noop {} {\emph {\bibinfo {title} {Rheology of
  complex fluids}}}\ (\bibinfo  {publisher} {Springer},\ \bibinfo {year}
  {2010})\BibitemShut {NoStop}%
\bibitem [{\citenamefont {Bottinelli}\ \emph {et~al.}(2016)\citenamefont
  {Bottinelli}, \citenamefont {Sumpter},\ and\ \citenamefont
  {Silverberg}}]{bottinelli2016emergent}%
  \BibitemOpen
  \bibfield  {author} {\bibinfo {author} {\bibfnamefont {A.}~\bibnamefont
  {Bottinelli}}, \bibinfo {author} {\bibfnamefont {D.~T.~J.}\ \bibnamefont
  {Sumpter}}, \ and\ \bibinfo {author} {\bibfnamefont {J.~L.}\ \bibnamefont
  {Silverberg}},\ }\href@noop {} {\bibfield  {journal} {\bibinfo  {journal}
  {Phys. Rev. Lett.}\ }\textbf {\bibinfo {volume} {117}},\ \bibinfo {pages}
  {228301} (\bibinfo {year} {2016})}\BibitemShut {NoStop}%
\bibitem [{\citenamefont {Helbing}\ and\ \citenamefont
  {Molnar}(1995)}]{helbing1995social}%
  \BibitemOpen
  \bibfield  {author} {\bibinfo {author} {\bibfnamefont {D.}~\bibnamefont
  {Helbing}}\ and\ \bibinfo {author} {\bibfnamefont {P.}~\bibnamefont
  {Molnar}},\ }\href@noop {} {\bibfield  {journal} {\bibinfo  {journal} {Phys.
  Rev. E}\ }\textbf {\bibinfo {volume} {51}},\ \bibinfo {pages} {4282}
  (\bibinfo {year} {1995})}\BibitemShut {NoStop}%
\bibitem [{\citenamefont {Garcimart{\'\i}n}\ \emph {et~al.}(2015)\citenamefont
  {Garcimart{\'\i}n}, \citenamefont {Pastor}, \citenamefont {Ferrer},
  \citenamefont {Ramos}, \citenamefont {Mart{\'\i}n-G{\'o}mez},\ and\
  \citenamefont {Zuriguel}}]{garcimartin2015flow}%
  \BibitemOpen
  \bibfield  {author} {\bibinfo {author} {\bibfnamefont {A.}~\bibnamefont
  {Garcimart{\'\i}n}}, \bibinfo {author} {\bibfnamefont {J.~M.}\ \bibnamefont
  {Pastor}}, \bibinfo {author} {\bibfnamefont {L.~M.}\ \bibnamefont {Ferrer}},
  \bibinfo {author} {\bibfnamefont {J.~J.}\ \bibnamefont {Ramos}}, \bibinfo
  {author} {\bibfnamefont {C.}~\bibnamefont {Mart{\'\i}n-G{\'o}mez}}, \ and\
  \bibinfo {author} {\bibfnamefont {I.}~\bibnamefont {Zuriguel}},\ }\href@noop
  {} {\bibfield  {journal} {\bibinfo  {journal} {Phys. Rev. E}\ }\textbf
  {\bibinfo {volume} {91}},\ \bibinfo {pages} {022808} (\bibinfo {year}
  {2015})}\BibitemShut {NoStop}%
\bibitem [{\citenamefont {Marchetti}\ \emph {et~al.}(2013)\citenamefont
  {Marchetti}, \citenamefont {Joanny}, \citenamefont {Ramaswamy}, \citenamefont
  {Liverpool}, \citenamefont {Prost}, \citenamefont {Rao},\ and\ \citenamefont
  {Simha}}]{marchetti2013hydrodynamics}%
  \BibitemOpen
  \bibfield  {author} {\bibinfo {author} {\bibfnamefont {M.~C.}\ \bibnamefont
  {Marchetti}}, \bibinfo {author} {\bibfnamefont {J.~F.}\ \bibnamefont
  {Joanny}}, \bibinfo {author} {\bibfnamefont {S.}~\bibnamefont {Ramaswamy}},
  \bibinfo {author} {\bibfnamefont {T.~B.}\ \bibnamefont {Liverpool}}, \bibinfo
  {author} {\bibfnamefont {J.}~\bibnamefont {Prost}}, \bibinfo {author}
  {\bibfnamefont {M.}~\bibnamefont {Rao}}, \ and\ \bibinfo {author}
  {\bibfnamefont {R.~A.}\ \bibnamefont {Simha}},\ }\href@noop {} {\bibfield
  {journal} {\bibinfo  {journal} {Rev. Mod. Phys.}\ }\textbf {\bibinfo {volume}
  {85}},\ \bibinfo {pages} {1143} (\bibinfo {year} {2013})}\BibitemShut
  {NoStop}%
\bibitem [{\citenamefont {Ballerini}\ \emph {et~al.}(2008)\citenamefont
  {Ballerini}, \citenamefont {Cabibbo}, \citenamefont {Candelier},
  \citenamefont {Cavagna}, \citenamefont {Cisbani}, \citenamefont {Giardina},
  \citenamefont {Lecomte}, \citenamefont {Orlandi}, \citenamefont {Parisi},
  \citenamefont {Procaccini} \emph {et~al.}}]{ballerini2008interaction}%
  \BibitemOpen
  \bibfield  {author} {\bibinfo {author} {\bibfnamefont {M.}~\bibnamefont
  {Ballerini}}, \bibinfo {author} {\bibfnamefont {N.}~\bibnamefont {Cabibbo}},
  \bibinfo {author} {\bibfnamefont {R.}~\bibnamefont {Candelier}}, \bibinfo
  {author} {\bibfnamefont {A.}~\bibnamefont {Cavagna}}, \bibinfo {author}
  {\bibfnamefont {E.}~\bibnamefont {Cisbani}}, \bibinfo {author} {\bibfnamefont
  {I.}~\bibnamefont {Giardina}}, \bibinfo {author} {\bibfnamefont
  {V.}~\bibnamefont {Lecomte}}, \bibinfo {author} {\bibfnamefont
  {A.}~\bibnamefont {Orlandi}}, \bibinfo {author} {\bibfnamefont
  {G.}~\bibnamefont {Parisi}}, \bibinfo {author} {\bibfnamefont
  {A.}~\bibnamefont {Procaccini}},  \emph {et~al.},\ }\href@noop {} {\bibfield
  {journal} {\bibinfo  {journal} {Proc. Nat. Acad. Sci.}\ }\textbf {\bibinfo
  {volume} {105}},\ \bibinfo {pages} {1232} (\bibinfo {year}
  {2008})}\BibitemShut {NoStop}%
\bibitem [{\citenamefont {Becco}\ \emph {et~al.}(2006)\citenamefont {Becco},
  \citenamefont {Vandewalle}, \citenamefont {Delcourt},\ and\ \citenamefont
  {Poncin}}]{becco2006experimental}%
  \BibitemOpen
  \bibfield  {author} {\bibinfo {author} {\bibfnamefont {C.}~\bibnamefont
  {Becco}}, \bibinfo {author} {\bibfnamefont {N.}~\bibnamefont {Vandewalle}},
  \bibinfo {author} {\bibfnamefont {J.}~\bibnamefont {Delcourt}}, \ and\
  \bibinfo {author} {\bibfnamefont {P.}~\bibnamefont {Poncin}},\ }\href@noop {}
  {\bibfield  {journal} {\bibinfo  {journal} {Physica A}\ }\textbf {\bibinfo
  {volume} {367}},\ \bibinfo {pages} {487} (\bibinfo {year}
  {2006})}\BibitemShut {NoStop}%
\bibitem [{\citenamefont {Calovi}\ \emph {et~al.}(2014)\citenamefont {Calovi},
  \citenamefont {Lopez}, \citenamefont {Ngo}, \citenamefont {Sire},
  \citenamefont {Chat{\'e}},\ and\ \citenamefont
  {Theraulaz}}]{calovi2014swarming}%
  \BibitemOpen
  \bibfield  {author} {\bibinfo {author} {\bibfnamefont {D.~S.}\ \bibnamefont
  {Calovi}}, \bibinfo {author} {\bibfnamefont {U.}~\bibnamefont {Lopez}},
  \bibinfo {author} {\bibfnamefont {S.}~\bibnamefont {Ngo}}, \bibinfo {author}
  {\bibfnamefont {C.}~\bibnamefont {Sire}}, \bibinfo {author} {\bibfnamefont
  {H.}~\bibnamefont {Chat{\'e}}}, \ and\ \bibinfo {author} {\bibfnamefont
  {G.}~\bibnamefont {Theraulaz}},\ }\href@noop {} {\bibfield  {journal}
  {\bibinfo  {journal} {New J. Phys.}\ }\textbf {\bibinfo {volume} {16}},\
  \bibinfo {pages} {015026} (\bibinfo {year} {2014})}\BibitemShut {NoStop}%
\bibitem [{\citenamefont {Peruani}\ \emph {et~al.}(2012)\citenamefont
  {Peruani}, \citenamefont {Starru{\ss}}, \citenamefont {Jakovljevic},
  \citenamefont {S{\o}gaard-Andersen}, \citenamefont {Deutsch},\ and\
  \citenamefont {B{\"a}r}}]{peruani2012collective}%
  \BibitemOpen
  \bibfield  {author} {\bibinfo {author} {\bibfnamefont {F.}~\bibnamefont
  {Peruani}}, \bibinfo {author} {\bibfnamefont {J.}~\bibnamefont
  {Starru{\ss}}}, \bibinfo {author} {\bibfnamefont {V.}~\bibnamefont
  {Jakovljevic}}, \bibinfo {author} {\bibfnamefont {L.}~\bibnamefont
  {S{\o}gaard-Andersen}}, \bibinfo {author} {\bibfnamefont {A.}~\bibnamefont
  {Deutsch}}, \ and\ \bibinfo {author} {\bibfnamefont {M.}~\bibnamefont
  {B{\"a}r}},\ }\href@noop {} {\bibfield  {journal} {\bibinfo  {journal} {Phys.
  Rev. Lett.}\ }\textbf {\bibinfo {volume} {108}},\ \bibinfo {pages} {098102}
  (\bibinfo {year} {2012})}\BibitemShut {NoStop}%
\bibitem [{\citenamefont {Schaller}\ \emph {et~al.}(2010)\citenamefont
  {Schaller}, \citenamefont {Weber}, \citenamefont {Semmrich}, \citenamefont
  {Frey},\ and\ \citenamefont {Bausch}}]{schaller2010polar}%
  \BibitemOpen
  \bibfield  {author} {\bibinfo {author} {\bibfnamefont {V.}~\bibnamefont
  {Schaller}}, \bibinfo {author} {\bibfnamefont {C.}~\bibnamefont {Weber}},
  \bibinfo {author} {\bibfnamefont {C.}~\bibnamefont {Semmrich}}, \bibinfo
  {author} {\bibfnamefont {E.}~\bibnamefont {Frey}}, \ and\ \bibinfo {author}
  {\bibfnamefont {A.~R.}\ \bibnamefont {Bausch}},\ }\href@noop {} {\bibfield
  {journal} {\bibinfo  {journal} {Nature}\ }\textbf {\bibinfo {volume} {467}},\
  \bibinfo {pages} {73} (\bibinfo {year} {2010})}\BibitemShut {NoStop}%
\bibitem [{\citenamefont {Sumino}\ \emph {et~al.}(2012)\citenamefont {Sumino},
  \citenamefont {Nagai}, \citenamefont {Shitaka}, \citenamefont {Tanaka},
  \citenamefont {Yoshikawa}, \citenamefont {Chat{\'e}},\ and\ \citenamefont
  {Oiwa}}]{sumino2012large}%
  \BibitemOpen
  \bibfield  {author} {\bibinfo {author} {\bibfnamefont {Y.}~\bibnamefont
  {Sumino}}, \bibinfo {author} {\bibfnamefont {K.~H.}\ \bibnamefont {Nagai}},
  \bibinfo {author} {\bibfnamefont {Y.}~\bibnamefont {Shitaka}}, \bibinfo
  {author} {\bibfnamefont {D.}~\bibnamefont {Tanaka}}, \bibinfo {author}
  {\bibfnamefont {K.}~\bibnamefont {Yoshikawa}}, \bibinfo {author}
  {\bibfnamefont {H.}~\bibnamefont {Chat{\'e}}}, \ and\ \bibinfo {author}
  {\bibfnamefont {K.}~\bibnamefont {Oiwa}},\ }\href@noop {} {\bibfield
  {journal} {\bibinfo  {journal} {Nature}\ }\textbf {\bibinfo {volume} {483}},\
  \bibinfo {pages} {448} (\bibinfo {year} {2012})}\BibitemShut {NoStop}%
\bibitem [{\citenamefont {Sanchez}\ \emph {et~al.}(2012)\citenamefont
  {Sanchez}, \citenamefont {Chen}, \citenamefont {DeCamp}, \citenamefont
  {Heymann},\ and\ \citenamefont {Dogic}}]{sanchez2012spontaneous}%
  \BibitemOpen
  \bibfield  {author} {\bibinfo {author} {\bibfnamefont {T.}~\bibnamefont
  {Sanchez}}, \bibinfo {author} {\bibfnamefont {D.~T.~N.}\ \bibnamefont
  {Chen}}, \bibinfo {author} {\bibfnamefont {S.~J.}\ \bibnamefont {DeCamp}},
  \bibinfo {author} {\bibfnamefont {M.}~\bibnamefont {Heymann}}, \ and\
  \bibinfo {author} {\bibfnamefont {Z.}~\bibnamefont {Dogic}},\ }\href@noop {}
  {\bibfield  {journal} {\bibinfo  {journal} {Nature}\ }\textbf {\bibinfo
  {volume} {491}},\ \bibinfo {pages} {431} (\bibinfo {year}
  {2012})}\BibitemShut {NoStop}%
\bibitem [{\citenamefont {Ramaswamy}\ and\ \citenamefont
  {Rao}(2001)}]{ramaswamy2001physics}%
  \BibitemOpen
  \bibfield  {author} {\bibinfo {author} {\bibfnamefont {S.}~\bibnamefont
  {Ramaswamy}}\ and\ \bibinfo {author} {\bibfnamefont {M.}~\bibnamefont
  {Rao}},\ }\href@noop {} {\bibfield  {journal} {\bibinfo  {journal} {C. R.
  Acad. Sci. Paris-Series IV}\ }\textbf {\bibinfo {volume} {2}},\ \bibinfo
  {pages} {817} (\bibinfo {year} {2001})}\BibitemShut {NoStop}%
\bibitem [{\citenamefont {Ramaswamy}\ \emph {et~al.}(2000)\citenamefont
  {Ramaswamy}, \citenamefont {Toner},\ and\ \citenamefont
  {Prost}}]{ramaswamy2000nonequilibrium}%
  \BibitemOpen
  \bibfield  {author} {\bibinfo {author} {\bibfnamefont {S.}~\bibnamefont
  {Ramaswamy}}, \bibinfo {author} {\bibfnamefont {J.}~\bibnamefont {Toner}}, \
  and\ \bibinfo {author} {\bibfnamefont {J.}~\bibnamefont {Prost}},\
  }\href@noop {} {\bibfield  {journal} {\bibinfo  {journal} {Phys. Rev. Lett.}\
  }\textbf {\bibinfo {volume} {84}},\ \bibinfo {pages} {3494} (\bibinfo {year}
  {2000})}\BibitemShut {NoStop}%
\bibitem [{\citenamefont {Menon}(2010)}]{menon2010active}%
  \BibitemOpen
  \bibfield  {author} {\bibinfo {author} {\bibfnamefont {G.~I.}\ \bibnamefont
  {Menon}},\ }\href@noop {} {\bibfield  {journal} {\bibinfo  {journal}
  {Rheology of Complex Fluids}\ ,\ \bibinfo {pages} {193}} (\bibinfo {year}
  {2010})}\BibitemShut {NoStop}%
\bibitem [{\citenamefont {Surrey}\ \emph {et~al.}(2001)\citenamefont {Surrey},
  \citenamefont {N{\'e}d{\'e}lec}, \citenamefont {Leibler},\ and\ \citenamefont
  {Karsenti}}]{surrey2001physical}%
  \BibitemOpen
  \bibfield  {author} {\bibinfo {author} {\bibfnamefont {T.}~\bibnamefont
  {Surrey}}, \bibinfo {author} {\bibfnamefont {F.}~\bibnamefont
  {N{\'e}d{\'e}lec}}, \bibinfo {author} {\bibfnamefont {S.}~\bibnamefont
  {Leibler}}, \ and\ \bibinfo {author} {\bibfnamefont {E.}~\bibnamefont
  {Karsenti}},\ }\href@noop {} {\bibfield  {journal} {\bibinfo  {journal}
  {Science}\ }\textbf {\bibinfo {volume} {292}},\ \bibinfo {pages} {1167}
  (\bibinfo {year} {2001})}\BibitemShut {NoStop}%
\bibitem [{\citenamefont {Deseigne}\ \emph {et~al.}(2010)\citenamefont
  {Deseigne}, \citenamefont {Dauchot},\ and\ \citenamefont
  {Chat{\'e}}}]{deseigne2010collective}%
  \BibitemOpen
  \bibfield  {author} {\bibinfo {author} {\bibfnamefont {J.}~\bibnamefont
  {Deseigne}}, \bibinfo {author} {\bibfnamefont {O.}~\bibnamefont {Dauchot}}, \
  and\ \bibinfo {author} {\bibfnamefont {H.}~\bibnamefont {Chat{\'e}}},\
  }\href@noop {} {\bibfield  {journal} {\bibinfo  {journal} {Phys. Rev. Lett.}\
  }\textbf {\bibinfo {volume} {105}},\ \bibinfo {pages} {098001} (\bibinfo
  {year} {2010})}\BibitemShut {NoStop}%
\bibitem [{\citenamefont {Bricard}\ \emph {et~al.}(2013)\citenamefont
  {Bricard}, \citenamefont {Caussin}, \citenamefont {Desreumaux}, \citenamefont
  {Dauchot},\ and\ \citenamefont {Bartolo}}]{bricard2013emergence}%
  \BibitemOpen
  \bibfield  {author} {\bibinfo {author} {\bibfnamefont {A.}~\bibnamefont
  {Bricard}}, \bibinfo {author} {\bibfnamefont {J.~B.}\ \bibnamefont
  {Caussin}}, \bibinfo {author} {\bibfnamefont {N.}~\bibnamefont {Desreumaux}},
  \bibinfo {author} {\bibfnamefont {O.}~\bibnamefont {Dauchot}}, \ and\
  \bibinfo {author} {\bibfnamefont {D.}~\bibnamefont {Bartolo}},\ }\href@noop
  {} {\bibfield  {journal} {\bibinfo  {journal} {Nature}\ }\textbf {\bibinfo
  {volume} {503}},\ \bibinfo {pages} {95} (\bibinfo {year} {2013})}\BibitemShut
  {NoStop}%
\bibitem [{\citenamefont {Thutupalli}\ \emph {et~al.}(2011)\citenamefont
  {Thutupalli}, \citenamefont {Seemann},\ and\ \citenamefont
  {Herminghaus}}]{thutupalli2011swarming}%
  \BibitemOpen
  \bibfield  {author} {\bibinfo {author} {\bibfnamefont {S.}~\bibnamefont
  {Thutupalli}}, \bibinfo {author} {\bibfnamefont {R.}~\bibnamefont {Seemann}},
  \ and\ \bibinfo {author} {\bibfnamefont {S.}~\bibnamefont {Herminghaus}},\
  }\href@noop {} {\bibfield  {journal} {\bibinfo  {journal} {New J. Phys.}\
  }\textbf {\bibinfo {volume} {13}},\ \bibinfo {pages} {073021} (\bibinfo
  {year} {2011})}\BibitemShut {NoStop}%
\bibitem [{\citenamefont {Popkin}(2016)}]{popkin2016physics}%
  \BibitemOpen
  \bibfield  {author} {\bibinfo {author} {\bibfnamefont {G.}~\bibnamefont
  {Popkin}},\ }\href@noop {} {\bibfield  {journal} {\bibinfo  {journal}
  {Nature}\ }\textbf {\bibinfo {volume} {529}},\ \bibinfo {pages} {16}
  (\bibinfo {year} {2016})}\BibitemShut {NoStop}%
\bibitem [{\citenamefont {Prost}\ \emph {et~al.}(2015)\citenamefont {Prost},
  \citenamefont {J{\"u}licher},\ and\ \citenamefont
  {Joanny}}]{prost2015active}%
  \BibitemOpen
  \bibfield  {author} {\bibinfo {author} {\bibfnamefont {J.}~\bibnamefont
  {Prost}}, \bibinfo {author} {\bibfnamefont {F.}~\bibnamefont {J{\"u}licher}},
  \ and\ \bibinfo {author} {\bibfnamefont {J.-F.}\ \bibnamefont {Joanny}},\
  }\href@noop {} {\bibfield  {journal} {\bibinfo  {journal} {Nature Physics}\
  }\textbf {\bibinfo {volume} {11}},\ \bibinfo {pages} {111} (\bibinfo {year}
  {2015})}\BibitemShut {NoStop}%
\bibitem [{\citenamefont {Needleman}\ and\ \citenamefont
  {Dogic}(2017)}]{needleman2017active}%
  \BibitemOpen
  \bibfield  {author} {\bibinfo {author} {\bibfnamefont {D.}~\bibnamefont
  {Needleman}}\ and\ \bibinfo {author} {\bibfnamefont {Z.}~\bibnamefont
  {Dogic}},\ }\href@noop {} {\bibfield  {journal} {\bibinfo  {journal} {Nature
  Reviews Materials}\ }\textbf {\bibinfo {volume} {2}},\ \bibinfo {pages} {1}
  (\bibinfo {year} {2017})}\BibitemShut {NoStop}%
\bibitem [{\citenamefont {Xi}\ \emph {et~al.}(2019)\citenamefont {Xi},
  \citenamefont {Saw}, \citenamefont {Delacour}, \citenamefont {Lim},\ and\
  \citenamefont {Ladoux}}]{xi2019material}%
  \BibitemOpen
  \bibfield  {author} {\bibinfo {author} {\bibfnamefont {W.}~\bibnamefont
  {Xi}}, \bibinfo {author} {\bibfnamefont {T.~B.}\ \bibnamefont {Saw}},
  \bibinfo {author} {\bibfnamefont {D.}~\bibnamefont {Delacour}}, \bibinfo
  {author} {\bibfnamefont {C.~T.}\ \bibnamefont {Lim}}, \ and\ \bibinfo
  {author} {\bibfnamefont {B.}~\bibnamefont {Ladoux}},\ }\href@noop {}
  {\bibfield  {journal} {\bibinfo  {journal} {Nature Reviews Materials}\
  }\textbf {\bibinfo {volume} {4}},\ \bibinfo {pages} {23} (\bibinfo {year}
  {2019})}\BibitemShut {NoStop}%
\bibitem [{\citenamefont {Trepat}\ and\ \citenamefont
  {Sahai}(2018)}]{trepat2018mesoscale}%
  \BibitemOpen
  \bibfield  {author} {\bibinfo {author} {\bibfnamefont {X.}~\bibnamefont
  {Trepat}}\ and\ \bibinfo {author} {\bibfnamefont {E.}~\bibnamefont {Sahai}},\
  }\href@noop {} {\bibfield  {journal} {\bibinfo  {journal} {Nature Physics}\
  }\textbf {\bibinfo {volume} {14}},\ \bibinfo {pages} {671} (\bibinfo {year}
  {2018})}\BibitemShut {NoStop}%
\bibitem [{\citenamefont {Dell’Arciprete}\ \emph {et~al.}(2018)\citenamefont
  {Dell’Arciprete}, \citenamefont {Blow}, \citenamefont {Brown},
  \citenamefont {Farrell}, \citenamefont {Lintuvuori}, \citenamefont {McVey},
  \citenamefont {Marenduzzo},\ and\ \citenamefont {Poon}}]{dell2018growing}%
  \BibitemOpen
  \bibfield  {author} {\bibinfo {author} {\bibfnamefont {D.}~\bibnamefont
  {Dell’Arciprete}}, \bibinfo {author} {\bibfnamefont {M.~L.}\ \bibnamefont
  {Blow}}, \bibinfo {author} {\bibfnamefont {A.~T.}\ \bibnamefont {Brown}},
  \bibinfo {author} {\bibfnamefont {F.~D.~C.}\ \bibnamefont {Farrell}},
  \bibinfo {author} {\bibfnamefont {J.~S.}\ \bibnamefont {Lintuvuori}},
  \bibinfo {author} {\bibfnamefont {A.~F.}\ \bibnamefont {McVey}}, \bibinfo
  {author} {\bibfnamefont {D.}~\bibnamefont {Marenduzzo}}, \ and\ \bibinfo
  {author} {\bibfnamefont {W.~C.~K.}\ \bibnamefont {Poon}},\ }\href@noop {}
  {\bibfield  {journal} {\bibinfo  {journal} {Nature Communications}\ }\textbf
  {\bibinfo {volume} {9}},\ \bibinfo {pages} {4190} (\bibinfo {year}
  {2018})}\BibitemShut {NoStop}%
\bibitem [{\citenamefont {P{\'e}rez-Gonz{\'a}lez}\ \emph
  {et~al.}(2019)\citenamefont {P{\'e}rez-Gonz{\'a}lez}, \citenamefont {Alert},
  \citenamefont {Blanch-Mercader}, \citenamefont {G{\'o}mez-Gonz{\'a}lez},
  \citenamefont {Kolodziej}, \citenamefont {Bazellieres}, \citenamefont
  {Casademunt},\ and\ \citenamefont {Trepat}}]{perez2019active}%
  \BibitemOpen
  \bibfield  {author} {\bibinfo {author} {\bibfnamefont {C.}~\bibnamefont
  {P{\'e}rez-Gonz{\'a}lez}}, \bibinfo {author} {\bibfnamefont {R.}~\bibnamefont
  {Alert}}, \bibinfo {author} {\bibfnamefont {C.}~\bibnamefont
  {Blanch-Mercader}}, \bibinfo {author} {\bibfnamefont {M.}~\bibnamefont
  {G{\'o}mez-Gonz{\'a}lez}}, \bibinfo {author} {\bibfnamefont {T.}~\bibnamefont
  {Kolodziej}}, \bibinfo {author} {\bibfnamefont {E.}~\bibnamefont
  {Bazellieres}}, \bibinfo {author} {\bibfnamefont {J.}~\bibnamefont
  {Casademunt}}, \ and\ \bibinfo {author} {\bibfnamefont {X.}~\bibnamefont
  {Trepat}},\ }\href@noop {} {\bibfield  {journal} {\bibinfo  {journal} {Nature
  Physics}\ }\textbf {\bibinfo {volume} {15}},\ \bibinfo {pages} {79} (\bibinfo
  {year} {2019})}\BibitemShut {NoStop}%
\bibitem [{\citenamefont {Cates}\ and\ \citenamefont
  {Tailleur}(2015)}]{cates2015motility}%
  \BibitemOpen
  \bibfield  {author} {\bibinfo {author} {\bibfnamefont {M.~E.}\ \bibnamefont
  {Cates}}\ and\ \bibinfo {author} {\bibfnamefont {J.}~\bibnamefont
  {Tailleur}},\ }\href@noop {} {\bibfield  {journal} {\bibinfo  {journal}
  {Annu. Rev. Condens. Matter Phys.}\ }\textbf {\bibinfo {volume} {6}},\
  \bibinfo {pages} {219} (\bibinfo {year} {2015})}\BibitemShut {NoStop}%
\bibitem [{\citenamefont {Brambilla}\ \emph {et~al.}(2013)\citenamefont
  {Brambilla}, \citenamefont {Ferrante}, \citenamefont {Birattari},\ and\
  \citenamefont {Dorigo}}]{brambilla2013swarm}%
  \BibitemOpen
  \bibfield  {author} {\bibinfo {author} {\bibfnamefont {M.}~\bibnamefont
  {Brambilla}}, \bibinfo {author} {\bibfnamefont {E.}~\bibnamefont {Ferrante}},
  \bibinfo {author} {\bibfnamefont {M.}~\bibnamefont {Birattari}}, \ and\
  \bibinfo {author} {\bibfnamefont {M.}~\bibnamefont {Dorigo}},\ }\href@noop {}
  {\bibfield  {journal} {\bibinfo  {journal} {Swarm Intelligence}\ }\textbf
  {\bibinfo {volume} {7}},\ \bibinfo {pages} {1} (\bibinfo {year}
  {2013})}\BibitemShut {NoStop}%
\bibitem [{\citenamefont {Dorigo}\ \emph {et~al.}(2014)\citenamefont {Dorigo},
  \citenamefont {Birattari},\ and\ \citenamefont
  {Brambilla}}]{dorigo2014swarm}%
  \BibitemOpen
  \bibfield  {author} {\bibinfo {author} {\bibfnamefont {M.}~\bibnamefont
  {Dorigo}}, \bibinfo {author} {\bibfnamefont {M.}~\bibnamefont {Birattari}}, \
  and\ \bibinfo {author} {\bibfnamefont {M.}~\bibnamefont {Brambilla}},\
  }\href@noop {} {\bibfield  {journal} {\bibinfo  {journal} {Scholarpedia}\
  }\textbf {\bibinfo {volume} {9}},\ \bibinfo {pages} {1463} (\bibinfo {year}
  {2014})}\BibitemShut {NoStop}%
\bibitem [{\citenamefont {Gompper}\ \emph {et~al.}(2020)\citenamefont
  {Gompper}, \citenamefont {Winkler}, \citenamefont {Speck}, \citenamefont
  {Solon}, \citenamefont {Nardini}, \citenamefont {Peruani}, \citenamefont
  {L{\"o}wen}, \citenamefont {Golestanian}, \citenamefont {Kaupp},
  \citenamefont {Alvarez} \emph {et~al.}}]{gompper20202020}%
  \BibitemOpen
  \bibfield  {author} {\bibinfo {author} {\bibfnamefont {G.}~\bibnamefont
  {Gompper}}, \bibinfo {author} {\bibfnamefont {R.~G.}\ \bibnamefont
  {Winkler}}, \bibinfo {author} {\bibfnamefont {T.}~\bibnamefont {Speck}},
  \bibinfo {author} {\bibfnamefont {A.}~\bibnamefont {Solon}}, \bibinfo
  {author} {\bibfnamefont {C.}~\bibnamefont {Nardini}}, \bibinfo {author}
  {\bibfnamefont {F.}~\bibnamefont {Peruani}}, \bibinfo {author} {\bibfnamefont
  {H.}~\bibnamefont {L{\"o}wen}}, \bibinfo {author} {\bibfnamefont
  {R.}~\bibnamefont {Golestanian}}, \bibinfo {author} {\bibfnamefont {U.~B.}\
  \bibnamefont {Kaupp}}, \bibinfo {author} {\bibfnamefont {L.}~\bibnamefont
  {Alvarez}},  \emph {et~al.},\ }\href@noop {} {\bibfield  {journal} {\bibinfo
  {journal} {Journal of Physics}\ }\textbf {\bibinfo {volume} {32}},\ \bibinfo
  {pages} {193001} (\bibinfo {year} {2020})}\BibitemShut {NoStop}%
\bibitem [{\citenamefont {Wensink}\ \emph {et~al.}(2012)\citenamefont
  {Wensink}, \citenamefont {Dunkel}, \citenamefont {Heidenreich}, \citenamefont
  {Drescher}, \citenamefont {Goldstein}, \citenamefont {L{\"o}wen},\ and\
  \citenamefont {Yeomans}}]{wensink2012meso}%
  \BibitemOpen
  \bibfield  {author} {\bibinfo {author} {\bibfnamefont {H.~H.}\ \bibnamefont
  {Wensink}}, \bibinfo {author} {\bibfnamefont {J.}~\bibnamefont {Dunkel}},
  \bibinfo {author} {\bibfnamefont {S.}~\bibnamefont {Heidenreich}}, \bibinfo
  {author} {\bibfnamefont {K.}~\bibnamefont {Drescher}}, \bibinfo {author}
  {\bibfnamefont {R.~E.}\ \bibnamefont {Goldstein}}, \bibinfo {author}
  {\bibfnamefont {H.}~\bibnamefont {L{\"o}wen}}, \ and\ \bibinfo {author}
  {\bibfnamefont {J.~M.}\ \bibnamefont {Yeomans}},\ }\href@noop {} {\bibfield
  {journal} {\bibinfo  {journal} {Proc. Nat. Acad. Sci.}\ }\textbf {\bibinfo
  {volume} {109}},\ \bibinfo {pages} {14308} (\bibinfo {year}
  {2012})}\BibitemShut {NoStop}%
\bibitem [{\citenamefont {Vicsek}\ \emph {et~al.}(1995)\citenamefont {Vicsek},
  \citenamefont {Czir{\'o}k}, \citenamefont {Ben-Jacob}, \citenamefont
  {Cohen},\ and\ \citenamefont {Shochet}}]{vicsek1995novel}%
  \BibitemOpen
  \bibfield  {author} {\bibinfo {author} {\bibfnamefont {T.}~\bibnamefont
  {Vicsek}}, \bibinfo {author} {\bibfnamefont {A.}~\bibnamefont {Czir{\'o}k}},
  \bibinfo {author} {\bibfnamefont {E.}~\bibnamefont {Ben-Jacob}}, \bibinfo
  {author} {\bibfnamefont {I.}~\bibnamefont {Cohen}}, \ and\ \bibinfo {author}
  {\bibfnamefont {O.}~\bibnamefont {Shochet}},\ }\href@noop {} {\bibfield
  {journal} {\bibinfo  {journal} {Phys. Rev. Lett.}\ }\textbf {\bibinfo
  {volume} {75}},\ \bibinfo {pages} {1226} (\bibinfo {year}
  {1995})}\BibitemShut {NoStop}%
\bibitem [{\citenamefont {Toner}\ and\ \citenamefont
  {Tu}(1995)}]{toner1995long}%
  \BibitemOpen
  \bibfield  {author} {\bibinfo {author} {\bibfnamefont {J.}~\bibnamefont
  {Toner}}\ and\ \bibinfo {author} {\bibfnamefont {Y.}~\bibnamefont {Tu}},\
  }\href@noop {} {\bibfield  {journal} {\bibinfo  {journal} {Phys. Rev. Lett.}\
  }\textbf {\bibinfo {volume} {75}},\ \bibinfo {pages} {4326} (\bibinfo {year}
  {1995})}\BibitemShut {NoStop}%
\bibitem [{\citenamefont {Toner}\ and\ \citenamefont
  {Tu}(1998)}]{toner1998flocks}%
  \BibitemOpen
  \bibfield  {author} {\bibinfo {author} {\bibfnamefont {J.}~\bibnamefont
  {Toner}}\ and\ \bibinfo {author} {\bibfnamefont {Y.}~\bibnamefont {Tu}},\
  }\href@noop {} {\bibfield  {journal} {\bibinfo  {journal} {Phys. Rev. E}\
  }\textbf {\bibinfo {volume} {58}},\ \bibinfo {pages} {4828} (\bibinfo {year}
  {1998})}\BibitemShut {NoStop}%
\bibitem [{\citenamefont {Toner}(2012)}]{toner2012reanalysis}%
  \BibitemOpen
  \bibfield  {author} {\bibinfo {author} {\bibfnamefont {J.}~\bibnamefont
  {Toner}},\ }\href@noop {} {\bibfield  {journal} {\bibinfo  {journal} {Phys.
  Rev. E}\ }\textbf {\bibinfo {volume} {86}},\ \bibinfo {pages} {031918}
  (\bibinfo {year} {2012})}\BibitemShut {NoStop}%
\bibitem [{\citenamefont {Ginelli}(2016)}]{ginelli2016physics}%
  \BibitemOpen
  \bibfield  {author} {\bibinfo {author} {\bibfnamefont {F.}~\bibnamefont
  {Ginelli}},\ }\href@noop {} {\bibfield  {journal} {\bibinfo  {journal} {Eur.
  Phys. J. Special Topics}\ }\textbf {\bibinfo {volume} {225}},\ \bibinfo
  {pages} {2099} (\bibinfo {year} {2016})}\BibitemShut {NoStop}%
\bibitem [{\citenamefont {Solon}\ \emph
  {et~al.}(2015{\natexlab{a}})\citenamefont {Solon}, \citenamefont
  {Chat{\'e}},\ and\ \citenamefont {Tailleur}}]{solon2015phase}%
  \BibitemOpen
  \bibfield  {author} {\bibinfo {author} {\bibfnamefont {A.~P.}\ \bibnamefont
  {Solon}}, \bibinfo {author} {\bibfnamefont {H.}~\bibnamefont {Chat{\'e}}}, \
  and\ \bibinfo {author} {\bibfnamefont {J.}~\bibnamefont {Tailleur}},\
  }\href@noop {} {\bibfield  {journal} {\bibinfo  {journal} {Phys. Rev. Lett.}\
  }\textbf {\bibinfo {volume} {114}},\ \bibinfo {pages} {068101} (\bibinfo
  {year} {2015}{\natexlab{a}})}\BibitemShut {NoStop}%
\bibitem [{\citenamefont {Fodor}\ and\ \citenamefont
  {Marchetti}(2018)}]{fodor2018statistical}%
  \BibitemOpen
  \bibfield  {author} {\bibinfo {author} {\bibfnamefont {E.}~\bibnamefont
  {Fodor}}\ and\ \bibinfo {author} {\bibfnamefont {M.~C.}\ \bibnamefont
  {Marchetti}},\ }\href@noop {} {\bibfield  {journal} {\bibinfo  {journal}
  {Physica A}\ }\textbf {\bibinfo {volume} {504}},\ \bibinfo {pages} {106}
  (\bibinfo {year} {2018})}\BibitemShut {NoStop}%
\bibitem [{\citenamefont {Gr{\'e}goire}\ and\ \citenamefont
  {Chat{\'e}}(2004)}]{gregoire2004onset}%
  \BibitemOpen
  \bibfield  {author} {\bibinfo {author} {\bibfnamefont {G.}~\bibnamefont
  {Gr{\'e}goire}}\ and\ \bibinfo {author} {\bibfnamefont {H.}~\bibnamefont
  {Chat{\'e}}},\ }\href@noop {} {\bibfield  {journal} {\bibinfo  {journal}
  {Phys. Rev. Lett.}\ }\textbf {\bibinfo {volume} {92}},\ \bibinfo {pages}
  {025702} (\bibinfo {year} {2004})}\BibitemShut {NoStop}%
\bibitem [{\citenamefont {Solon}\ \emph
  {et~al.}(2015{\natexlab{b}})\citenamefont {Solon}, \citenamefont {Caussin},
  \citenamefont {Bartolo}, \citenamefont {Chat{\'e}},\ and\ \citenamefont
  {Tailleur}}]{solon2015pattern}%
  \BibitemOpen
  \bibfield  {author} {\bibinfo {author} {\bibfnamefont {A.~P.}\ \bibnamefont
  {Solon}}, \bibinfo {author} {\bibfnamefont {J.~B.}\ \bibnamefont {Caussin}},
  \bibinfo {author} {\bibfnamefont {D.}~\bibnamefont {Bartolo}}, \bibinfo
  {author} {\bibfnamefont {H.}~\bibnamefont {Chat{\'e}}}, \ and\ \bibinfo
  {author} {\bibfnamefont {J.}~\bibnamefont {Tailleur}},\ }\href@noop {}
  {\bibfield  {journal} {\bibinfo  {journal} {Phys. Rev. E}\ }\textbf {\bibinfo
  {volume} {92}},\ \bibinfo {pages} {062111} (\bibinfo {year}
  {2015}{\natexlab{b}})}\BibitemShut {NoStop}%
\bibitem [{\citenamefont {K{\"u}rsten}\ and\ \citenamefont
  {Ihle}(2020)}]{kursten2020dry}%
  \BibitemOpen
  \bibfield  {author} {\bibinfo {author} {\bibfnamefont {R.}~\bibnamefont
  {K{\"u}rsten}}\ and\ \bibinfo {author} {\bibfnamefont {T.}~\bibnamefont
  {Ihle}},\ }\href@noop {} {\bibfield  {journal} {\bibinfo  {journal} {Phys.
  Rev. Lett.}\ }\textbf {\bibinfo {volume} {125}},\ \bibinfo {pages} {188003}
  (\bibinfo {year} {2020})}\BibitemShut {NoStop}%
\bibitem [{\citenamefont {Chat{\'e}}\ \emph {et~al.}(2008)\citenamefont
  {Chat{\'e}}, \citenamefont {Ginelli}, \citenamefont {Gr{\'e}goire},\ and\
  \citenamefont {Raynaud}}]{chate2008collective}%
  \BibitemOpen
  \bibfield  {author} {\bibinfo {author} {\bibfnamefont {H.}~\bibnamefont
  {Chat{\'e}}}, \bibinfo {author} {\bibfnamefont {F.}~\bibnamefont {Ginelli}},
  \bibinfo {author} {\bibfnamefont {G.}~\bibnamefont {Gr{\'e}goire}}, \ and\
  \bibinfo {author} {\bibfnamefont {F.}~\bibnamefont {Raynaud}},\ }\href@noop
  {} {\bibfield  {journal} {\bibinfo  {journal} {Phys. Rev. E}\ }\textbf
  {\bibinfo {volume} {77}},\ \bibinfo {pages} {046113} (\bibinfo {year}
  {2008})}\BibitemShut {NoStop}%
\bibitem [{\citenamefont {Solon}\ and\ \citenamefont
  {Tailleur}(2013)}]{solon2013revisiting}%
  \BibitemOpen
  \bibfield  {author} {\bibinfo {author} {\bibfnamefont {A.~P.}\ \bibnamefont
  {Solon}}\ and\ \bibinfo {author} {\bibfnamefont {J.}~\bibnamefont
  {Tailleur}},\ }\href@noop {} {\bibfield  {journal} {\bibinfo  {journal}
  {Phys. Rev. Lett.}\ }\textbf {\bibinfo {volume} {111}},\ \bibinfo {pages}
  {078101} (\bibinfo {year} {2013})}\BibitemShut {NoStop}%
\bibitem [{\citenamefont {Solon}\ and\ \citenamefont
  {Tailleur}(2015)}]{solon2015flocking}%
  \BibitemOpen
  \bibfield  {author} {\bibinfo {author} {\bibfnamefont {A.~P.}\ \bibnamefont
  {Solon}}\ and\ \bibinfo {author} {\bibfnamefont {J.}~\bibnamefont
  {Tailleur}},\ }\href@noop {} {\bibfield  {journal} {\bibinfo  {journal}
  {Phys. Rev. E}\ }\textbf {\bibinfo {volume} {92}},\ \bibinfo {pages} {042119}
  (\bibinfo {year} {2015})}\BibitemShut {NoStop}%
\bibitem [{\citenamefont {Chatterjee}\ \emph
  {et~al.}(2020{\natexlab{a}})\citenamefont {Chatterjee}, \citenamefont
  {Mangeat}, \citenamefont {Paul},\ and\ \citenamefont
  {Rieger}}]{chatterjee2020flocking}%
  \BibitemOpen
  \bibfield  {author} {\bibinfo {author} {\bibfnamefont {S.}~\bibnamefont
  {Chatterjee}}, \bibinfo {author} {\bibfnamefont {M.}~\bibnamefont {Mangeat}},
  \bibinfo {author} {\bibfnamefont {R.}~\bibnamefont {Paul}}, \ and\ \bibinfo
  {author} {\bibfnamefont {H.}~\bibnamefont {Rieger}},\ }\href@noop {}
  {\bibfield  {journal} {\bibinfo  {journal} {EPL}\ }\textbf {\bibinfo {volume}
  {130}},\ \bibinfo {pages} {66001} (\bibinfo {year}
  {2020}{\natexlab{a}})}\BibitemShut {NoStop}%
\bibitem [{\citenamefont {Mangeat}\ \emph {et~al.}(2020)\citenamefont
  {Mangeat}, \citenamefont {Chatterjee}, \citenamefont {Paul},\ and\
  \citenamefont {Rieger}}]{mangeat2020flocking}%
  \BibitemOpen
  \bibfield  {author} {\bibinfo {author} {\bibfnamefont {M.}~\bibnamefont
  {Mangeat}}, \bibinfo {author} {\bibfnamefont {S.}~\bibnamefont {Chatterjee}},
  \bibinfo {author} {\bibfnamefont {R.}~\bibnamefont {Paul}}, \ and\ \bibinfo
  {author} {\bibfnamefont {H.}~\bibnamefont {Rieger}},\ }\href@noop {}
  {\bibfield  {journal} {\bibinfo  {journal} {Phys. Rev. E}\ }\textbf {\bibinfo
  {volume} {102}},\ \bibinfo {pages} {042601} (\bibinfo {year}
  {2020})}\BibitemShut {NoStop}%
\bibitem [{\citenamefont {Chatterjee}\ \emph {et~al.}(2022)\citenamefont
  {Chatterjee}, \citenamefont {Mangeat},\ and\ \citenamefont
  {Rieger}}]{chatterjee2022polar}%
  \BibitemOpen
  \bibfield  {author} {\bibinfo {author} {\bibfnamefont {S.}~\bibnamefont
  {Chatterjee}}, \bibinfo {author} {\bibfnamefont {M.}~\bibnamefont {Mangeat}},
  \ and\ \bibinfo {author} {\bibfnamefont {H.}~\bibnamefont {Rieger}},\
  }\href@noop {} {\bibfield  {journal} {\bibinfo  {journal} {EPL}\ }\textbf
  {\bibinfo {volume} {138}},\ \bibinfo {pages} {41001} (\bibinfo {year}
  {2022})}\BibitemShut {NoStop}%
\bibitem [{\citenamefont {Solon}\ \emph {et~al.}(2022)\citenamefont {Solon},
  \citenamefont {Chat{\'e}}, \citenamefont {Toner},\ and\ \citenamefont
  {Tailleur}}]{solon2022susceptibility}%
  \BibitemOpen
  \bibfield  {author} {\bibinfo {author} {\bibfnamefont {A.}~\bibnamefont
  {Solon}}, \bibinfo {author} {\bibfnamefont {H.}~\bibnamefont {Chat{\'e}}},
  \bibinfo {author} {\bibfnamefont {J.}~\bibnamefont {Toner}}, \ and\ \bibinfo
  {author} {\bibfnamefont {J.}~\bibnamefont {Tailleur}},\ }\href@noop {}
  {\bibfield  {journal} {\bibinfo  {journal} {Phys. Rev. Lett.}\ }\textbf
  {\bibinfo {volume} {128}},\ \bibinfo {pages} {208004} (\bibinfo {year}
  {2022})}\BibitemShut {NoStop}%
\bibitem [{\citenamefont {Peled}\ \emph {et~al.}(2021)\citenamefont {Peled},
  \citenamefont {Ryan}, \citenamefont {Heidenreich}, \citenamefont {B{\"a}r},
  \citenamefont {Ariel},\ and\ \citenamefont {Be'Er}}]{peled2021heterogeneous}%
  \BibitemOpen
  \bibfield  {author} {\bibinfo {author} {\bibfnamefont {S.}~\bibnamefont
  {Peled}}, \bibinfo {author} {\bibfnamefont {S.~D.}\ \bibnamefont {Ryan}},
  \bibinfo {author} {\bibfnamefont {S.}~\bibnamefont {Heidenreich}}, \bibinfo
  {author} {\bibfnamefont {M.}~\bibnamefont {B{\"a}r}}, \bibinfo {author}
  {\bibfnamefont {G.}~\bibnamefont {Ariel}}, \ and\ \bibinfo {author}
  {\bibfnamefont {A.}~\bibnamefont {Be'Er}},\ }\href@noop {} {\bibfield
  {journal} {\bibinfo  {journal} {Phys. Rev. E}\ }\textbf {\bibinfo {volume}
  {103}},\ \bibinfo {pages} {032413} (\bibinfo {year} {2021})}\BibitemShut
  {NoStop}%
\bibitem [{\citenamefont {Chatterjee}\ \emph {et~al.}(2023)\citenamefont
  {Chatterjee}, \citenamefont {Mangeat}, \citenamefont {Woo}, \citenamefont
  {Rieger},\ and\ \citenamefont {Noh}}]{TSVM2023}%
  \BibitemOpen
  \bibfield  {author} {\bibinfo {author} {\bibfnamefont {S.}~\bibnamefont
  {Chatterjee}}, \bibinfo {author} {\bibfnamefont {M.}~\bibnamefont {Mangeat}},
  \bibinfo {author} {\bibfnamefont {C.~U.}\ \bibnamefont {Woo}}, \bibinfo
  {author} {\bibfnamefont {H.}~\bibnamefont {Rieger}}, \ and\ \bibinfo {author}
  {\bibfnamefont {J.~D.}\ \bibnamefont {Noh}},\ }\href@noop {} {\bibfield
  {journal} {\bibinfo  {journal} {Phys. Rev. E}\ }\textbf {\bibinfo {volume}
  {107}},\ \bibinfo {pages} {024607} (\bibinfo {year} {2023})}\BibitemShut
  {NoStop}%
\bibitem [{\citenamefont {Geyer}\ \emph {et~al.}(2019)\citenamefont {Geyer},
  \citenamefont {Martin}, \citenamefont {Tailleur},\ and\ \citenamefont
  {Bartolo}}]{geyer2019freezing}%
  \BibitemOpen
  \bibfield  {author} {\bibinfo {author} {\bibfnamefont {D.}~\bibnamefont
  {Geyer}}, \bibinfo {author} {\bibfnamefont {D.}~\bibnamefont {Martin}},
  \bibinfo {author} {\bibfnamefont {J.}~\bibnamefont {Tailleur}}, \ and\
  \bibinfo {author} {\bibfnamefont {D.}~\bibnamefont {Bartolo}},\ }\href@noop
  {} {\bibfield  {journal} {\bibinfo  {journal} {Phys. Rev. X}\ }\textbf
  {\bibinfo {volume} {9}},\ \bibinfo {pages} {031043} (\bibinfo {year}
  {2019})}\BibitemShut {NoStop}%
\bibitem [{\citenamefont {Fily}\ and\ \citenamefont
  {Marchetti}(2012)}]{fily2012athermal}%
  \BibitemOpen
  \bibfield  {author} {\bibinfo {author} {\bibfnamefont {Y.}~\bibnamefont
  {Fily}}\ and\ \bibinfo {author} {\bibfnamefont {M.~C.}\ \bibnamefont
  {Marchetti}},\ }\href@noop {} {\bibfield  {journal} {\bibinfo  {journal}
  {Phys. Rev. Lett.}\ }\textbf {\bibinfo {volume} {108}},\ \bibinfo {pages}
  {235702} (\bibinfo {year} {2012})}\BibitemShut {NoStop}%
\bibitem [{\citenamefont {Romanczuk}\ \emph {et~al.}(2012)\citenamefont
  {Romanczuk}, \citenamefont {B{\"a}r}, \citenamefont {Ebeling}, \citenamefont
  {Lindner},\ and\ \citenamefont {Schimansky-Geier}}]{romanczuk2012active}%
  \BibitemOpen
  \bibfield  {author} {\bibinfo {author} {\bibfnamefont {P.}~\bibnamefont
  {Romanczuk}}, \bibinfo {author} {\bibfnamefont {M.}~\bibnamefont {B{\"a}r}},
  \bibinfo {author} {\bibfnamefont {W.}~\bibnamefont {Ebeling}}, \bibinfo
  {author} {\bibfnamefont {B.}~\bibnamefont {Lindner}}, \ and\ \bibinfo
  {author} {\bibfnamefont {L.}~\bibnamefont {Schimansky-Geier}},\ }\href@noop
  {} {\bibfield  {journal} {\bibinfo  {journal} {Eur. Phys. J. Special Topics}\
  }\textbf {\bibinfo {volume} {202}},\ \bibinfo {pages} {1} (\bibinfo {year}
  {2012})}\BibitemShut {NoStop}%
\bibitem [{\citenamefont {Gr{\'e}goire}\ \emph {et~al.}(2003)\citenamefont
  {Gr{\'e}goire}, \citenamefont {Chat{\'e}},\ and\ \citenamefont
  {Tu}}]{gregoire2003moving}%
  \BibitemOpen
  \bibfield  {author} {\bibinfo {author} {\bibfnamefont {G.}~\bibnamefont
  {Gr{\'e}goire}}, \bibinfo {author} {\bibfnamefont {H.}~\bibnamefont
  {Chat{\'e}}}, \ and\ \bibinfo {author} {\bibfnamefont {Y.}~\bibnamefont
  {Tu}},\ }\href@noop {} {\bibfield  {journal} {\bibinfo  {journal} {Physica
  D}\ }\textbf {\bibinfo {volume} {181}},\ \bibinfo {pages} {157} (\bibinfo
  {year} {2003})}\BibitemShut {NoStop}%
\bibitem [{\citenamefont {Mart{\'\i}n-G{\'o}mez}\ \emph
  {et~al.}(2018)\citenamefont {Mart{\'\i}n-G{\'o}mez}, \citenamefont {Levis},
  \citenamefont {D{\'\i}az-Guilera},\ and\ \citenamefont
  {Pagonabarraga}}]{martin2018collective}%
  \BibitemOpen
  \bibfield  {author} {\bibinfo {author} {\bibfnamefont {A.}~\bibnamefont
  {Mart{\'\i}n-G{\'o}mez}}, \bibinfo {author} {\bibfnamefont {D.}~\bibnamefont
  {Levis}}, \bibinfo {author} {\bibfnamefont {A.}~\bibnamefont
  {D{\'\i}az-Guilera}}, \ and\ \bibinfo {author} {\bibfnamefont
  {I.}~\bibnamefont {Pagonabarraga}},\ }\href@noop {} {\bibfield  {journal}
  {\bibinfo  {journal} {Soft Matter}\ }\textbf {\bibinfo {volume} {14}},\
  \bibinfo {pages} {2610} (\bibinfo {year} {2018})}\BibitemShut {NoStop}%
\bibitem [{\citenamefont {Sese-Sansa}\ \emph {et~al.}(2018)\citenamefont
  {Sese-Sansa}, \citenamefont {Pagonabarraga},\ and\ \citenamefont
  {Levis}}]{sese2018velocity}%
  \BibitemOpen
  \bibfield  {author} {\bibinfo {author} {\bibfnamefont {E.}~\bibnamefont
  {Sese-Sansa}}, \bibinfo {author} {\bibfnamefont {I.}~\bibnamefont
  {Pagonabarraga}}, \ and\ \bibinfo {author} {\bibfnamefont {D.}~\bibnamefont
  {Levis}},\ }\href@noop {} {\bibfield  {journal} {\bibinfo  {journal} {EPL}\
  }\textbf {\bibinfo {volume} {124}},\ \bibinfo {pages} {30004} (\bibinfo
  {year} {2018})}\BibitemShut {NoStop}%
\bibitem [{\citenamefont {Peruani}\ \emph {et~al.}(2011)\citenamefont
  {Peruani}, \citenamefont {Klauss}, \citenamefont {Deutsch},\ and\
  \citenamefont {Voss-Boehme}}]{peruani2011traffic}%
  \BibitemOpen
  \bibfield  {author} {\bibinfo {author} {\bibfnamefont {F.}~\bibnamefont
  {Peruani}}, \bibinfo {author} {\bibfnamefont {T.}~\bibnamefont {Klauss}},
  \bibinfo {author} {\bibfnamefont {A.}~\bibnamefont {Deutsch}}, \ and\
  \bibinfo {author} {\bibfnamefont {A.}~\bibnamefont {Voss-Boehme}},\
  }\href@noop {} {\bibfield  {journal} {\bibinfo  {journal} {Phys. Rev. Lett.}\
  }\textbf {\bibinfo {volume} {106}},\ \bibinfo {pages} {128101} (\bibinfo
  {year} {2011})}\BibitemShut {NoStop}%
\bibitem [{\citenamefont {Kourbane-Houssene}\ \emph {et~al.}(2018)\citenamefont
  {Kourbane-Houssene}, \citenamefont {Erignoux}, \citenamefont {Bodineau},\
  and\ \citenamefont {Tailleur}}]{kourbane2018exact}%
  \BibitemOpen
  \bibfield  {author} {\bibinfo {author} {\bibfnamefont {M.}~\bibnamefont
  {Kourbane-Houssene}}, \bibinfo {author} {\bibfnamefont {C.}~\bibnamefont
  {Erignoux}}, \bibinfo {author} {\bibfnamefont {T.}~\bibnamefont {Bodineau}},
  \ and\ \bibinfo {author} {\bibfnamefont {J.}~\bibnamefont {Tailleur}},\
  }\href@noop {} {\bibfield  {journal} {\bibinfo  {journal} {Phys. Rev. Lett.}\
  }\textbf {\bibinfo {volume} {120}},\ \bibinfo {pages} {268003} (\bibinfo
  {year} {2018})}\BibitemShut {NoStop}%
\bibitem [{\citenamefont {Giomi}\ \emph {et~al.}(2013)\citenamefont {Giomi},
  \citenamefont {Bowick}, \citenamefont {Ma},\ and\ \citenamefont
  {Marchetti}}]{giomi2013defect}%
  \BibitemOpen
  \bibfield  {author} {\bibinfo {author} {\bibfnamefont {L.}~\bibnamefont
  {Giomi}}, \bibinfo {author} {\bibfnamefont {M.~J.}\ \bibnamefont {Bowick}},
  \bibinfo {author} {\bibfnamefont {X.}~\bibnamefont {Ma}}, \ and\ \bibinfo
  {author} {\bibfnamefont {M.~C.}\ \bibnamefont {Marchetti}},\ }\href@noop {}
  {\bibfield  {journal} {\bibinfo  {journal} {Phys. Rev. Lett.}\ }\textbf
  {\bibinfo {volume} {110}},\ \bibinfo {pages} {228101} (\bibinfo {year}
  {2013})}\BibitemShut {NoStop}%
\bibitem [{\citenamefont {Siebert}\ \emph {et~al.}(2018)\citenamefont
  {Siebert}, \citenamefont {Dittrich}, \citenamefont {Schmid}, \citenamefont
  {Binder}, \citenamefont {Speck},\ and\ \citenamefont
  {Virnau}}]{criticalABP2018}%
  \BibitemOpen
  \bibfield  {author} {\bibinfo {author} {\bibfnamefont {J.~T.}\ \bibnamefont
  {Siebert}}, \bibinfo {author} {\bibfnamefont {F.}~\bibnamefont {Dittrich}},
  \bibinfo {author} {\bibfnamefont {F.}~\bibnamefont {Schmid}}, \bibinfo
  {author} {\bibfnamefont {K.}~\bibnamefont {Binder}}, \bibinfo {author}
  {\bibfnamefont {T.}~\bibnamefont {Speck}}, \ and\ \bibinfo {author}
  {\bibfnamefont {P.}~\bibnamefont {Virnau}},\ }\href@noop {} {\bibfield
  {journal} {\bibinfo  {journal} {Phys. Rev. E}\ }\textbf {\bibinfo {volume}
  {98}},\ \bibinfo {pages} {030601} (\bibinfo {year} {2018})}\BibitemShut
  {NoStop}%
\bibitem [{\citenamefont {Wysocki}\ and\ \citenamefont
  {Rieger}(2020)}]{capillary2020adam}%
  \BibitemOpen
  \bibfield  {author} {\bibinfo {author} {\bibfnamefont {A.}~\bibnamefont
  {Wysocki}}\ and\ \bibinfo {author} {\bibfnamefont {H.}~\bibnamefont
  {Rieger}},\ }\href@noop {} {\bibfield  {journal} {\bibinfo  {journal} {Phys.
  Rev. Lett.}\ }\textbf {\bibinfo {volume} {124}},\ \bibinfo {pages} {048001}
  (\bibinfo {year} {2020})}\BibitemShut {NoStop}%
\bibitem [{\citenamefont {Fruchart}\ \emph {et~al.}(2021)\citenamefont
  {Fruchart}, \citenamefont {Hanai}, \citenamefont {Littlewood},\ and\
  \citenamefont {Vitelli}}]{fruchart2021NR}%
  \BibitemOpen
  \bibfield  {author} {\bibinfo {author} {\bibfnamefont {M.}~\bibnamefont
  {Fruchart}}, \bibinfo {author} {\bibfnamefont {R.}~\bibnamefont {Hanai}},
  \bibinfo {author} {\bibfnamefont {P.~B.}\ \bibnamefont {Littlewood}}, \ and\
  \bibinfo {author} {\bibfnamefont {V.}~\bibnamefont {Vitelli}},\ }\href@noop
  {} {\bibfield  {journal} {\bibinfo  {journal} {Nature}\ }\textbf {\bibinfo
  {volume} {592}},\ \bibinfo {pages} {363} (\bibinfo {year}
  {2021})}\BibitemShut {NoStop}%
\bibitem [{\citenamefont {Codina}\ \emph {et~al.}(2022)\citenamefont {Codina},
  \citenamefont {Mahault}, \citenamefont {Chat{\'e}}, \citenamefont {Dobnikar},
  \citenamefont {Pagonabarraga},\ and\ \citenamefont {Shi}}]{codina2022small}%
  \BibitemOpen
  \bibfield  {author} {\bibinfo {author} {\bibfnamefont {J.}~\bibnamefont
  {Codina}}, \bibinfo {author} {\bibfnamefont {B.}~\bibnamefont {Mahault}},
  \bibinfo {author} {\bibfnamefont {H.}~\bibnamefont {Chat{\'e}}}, \bibinfo
  {author} {\bibfnamefont {J.}~\bibnamefont {Dobnikar}}, \bibinfo {author}
  {\bibfnamefont {I.}~\bibnamefont {Pagonabarraga}}, \ and\ \bibinfo {author}
  {\bibfnamefont {X.}~\bibnamefont {Shi}},\ }\href@noop {} {\bibfield
  {journal} {\bibinfo  {journal} {Phys. Rev. Lett.}\ }\textbf {\bibinfo
  {volume} {128}},\ \bibinfo {pages} {218001} (\bibinfo {year}
  {2022})}\BibitemShut {NoStop}%
\bibitem [{\citenamefont {Karmakar}\ \emph {et~al.}(2023)\citenamefont
  {Karmakar}, \citenamefont {Chatterjee}, \citenamefont {Mangeat},
  \citenamefont {Rieger},\ and\ \citenamefont {Paul}}]{karmakar2023jamming}%
  \BibitemOpen
  \bibfield  {author} {\bibinfo {author} {\bibfnamefont {M.}~\bibnamefont
  {Karmakar}}, \bibinfo {author} {\bibfnamefont {S.}~\bibnamefont
  {Chatterjee}}, \bibinfo {author} {\bibfnamefont {M.}~\bibnamefont {Mangeat}},
  \bibinfo {author} {\bibfnamefont {H.}~\bibnamefont {Rieger}}, \ and\ \bibinfo
  {author} {\bibfnamefont {R.}~\bibnamefont {Paul}},\ }\href@noop {} {\bibfield
   {journal} {\bibinfo  {journal} {Phys. Rev. E}\ }\textbf {\bibinfo {volume}
  {108}},\ \bibinfo {pages} {014604} (\bibinfo {year} {2023})}\BibitemShut
  {NoStop}%
\bibitem [{\citenamefont {Bray}(2002)}]{alan_bray}%
  \BibitemOpen
  \bibfield  {author} {\bibinfo {author} {\bibfnamefont {A.~J.}\ \bibnamefont
  {Bray}},\ }\href@noop {} {\bibfield  {journal} {\bibinfo  {journal} {Advances
  in Physics}\ }\textbf {\bibinfo {volume} {51}},\ \bibinfo {pages} {481}
  (\bibinfo {year} {2002})}\BibitemShut {NoStop}%
\bibitem [{\citenamefont {Bray}(1994)}]{bray1993theory}%
  \BibitemOpen
  \bibfield  {author} {\bibinfo {author} {\bibfnamefont {A.}~\bibnamefont
  {Bray}},\ }\href@noop {} {\bibfield  {journal} {\bibinfo  {journal} {Advances
  in Physics}\ }\textbf {\bibinfo {volume} {43}},\ \bibinfo {pages} {357}
  (\bibinfo {year} {1994})}\BibitemShut {NoStop}%
\bibitem [{\citenamefont {Puri}(2009)}]{sanjay_puri}%
  \BibitemOpen
  \bibfield  {author} {\bibinfo {author} {\bibfnamefont {S.}~\bibnamefont
  {Puri}}\ }(\bibinfo  {publisher} {CRC press},\ \bibinfo {year} {2009})\
  p.~\bibinfo {pages} {13}\BibitemShut {NoStop}%
\bibitem [{\citenamefont {Ahmad}\ \emph {et~al.}(2012)\citenamefont {Ahmad},
  \citenamefont {Corberi}, \citenamefont {Das}, \citenamefont {Lippiello},
  \citenamefont {Puri},\ and\ \citenamefont {Zannetti}}]{aging2012}%
  \BibitemOpen
  \bibfield  {author} {\bibinfo {author} {\bibfnamefont {S.}~\bibnamefont
  {Ahmad}}, \bibinfo {author} {\bibfnamefont {F.}~\bibnamefont {Corberi}},
  \bibinfo {author} {\bibfnamefont {S.~K.}\ \bibnamefont {Das}}, \bibinfo
  {author} {\bibfnamefont {E.}~\bibnamefont {Lippiello}}, \bibinfo {author}
  {\bibfnamefont {S.}~\bibnamefont {Puri}}, \ and\ \bibinfo {author}
  {\bibfnamefont {M.}~\bibnamefont {Zannetti}},\ }\href@noop {} {\bibfield
  {journal} {\bibinfo  {journal} {Phys. Rev. E}\ }\textbf {\bibinfo {volume}
  {86}},\ \bibinfo {pages} {061129} (\bibinfo {year} {2012})}\BibitemShut
  {NoStop}%
\bibitem [{\citenamefont {Shrivastav}\ \emph {et~al.}(2014)\citenamefont
  {Shrivastav}, \citenamefont {Kumar}, \citenamefont {Banerjee},\ and\
  \citenamefont {Puri}}]{puri2014rfim}%
  \BibitemOpen
  \bibfield  {author} {\bibinfo {author} {\bibfnamefont {G.~P.}\ \bibnamefont
  {Shrivastav}}, \bibinfo {author} {\bibfnamefont {M.}~\bibnamefont {Kumar}},
  \bibinfo {author} {\bibfnamefont {V.}~\bibnamefont {Banerjee}}, \ and\
  \bibinfo {author} {\bibfnamefont {S.}~\bibnamefont {Puri}},\ }\href@noop {}
  {\bibfield  {journal} {\bibinfo  {journal} {Phys. Rev. E}\ }\textbf {\bibinfo
  {volume} {90}},\ \bibinfo {pages} {032140} (\bibinfo {year}
  {2014})}\BibitemShut {NoStop}%
\bibitem [{\citenamefont {Kumar}\ \emph {et~al.}(2017)\citenamefont {Kumar},
  \citenamefont {Chatterjee}, \citenamefont {Paul},\ and\ \citenamefont
  {Puri}}]{kumar2017ordering}%
  \BibitemOpen
  \bibfield  {author} {\bibinfo {author} {\bibfnamefont {M.}~\bibnamefont
  {Kumar}}, \bibinfo {author} {\bibfnamefont {S.}~\bibnamefont {Chatterjee}},
  \bibinfo {author} {\bibfnamefont {R.}~\bibnamefont {Paul}}, \ and\ \bibinfo
  {author} {\bibfnamefont {S.}~\bibnamefont {Puri}},\ }\href@noop {} {\bibfield
   {journal} {\bibinfo  {journal} {Phys. Rev. E}\ }\textbf {\bibinfo {volume}
  {96}},\ \bibinfo {pages} {042127} (\bibinfo {year} {2017})}\BibitemShut
  {NoStop}%
\bibitem [{\citenamefont {Chatterjee}\ \emph
  {et~al.}(2020{\natexlab{b}})\citenamefont {Chatterjee}, \citenamefont
  {Sutradhar}, \citenamefont {Puri},\ and\ \citenamefont {Paul}}]{rbcm}%
  \BibitemOpen
  \bibfield  {author} {\bibinfo {author} {\bibfnamefont {S.}~\bibnamefont
  {Chatterjee}}, \bibinfo {author} {\bibfnamefont {S.}~\bibnamefont
  {Sutradhar}}, \bibinfo {author} {\bibfnamefont {S.}~\bibnamefont {Puri}}, \
  and\ \bibinfo {author} {\bibfnamefont {R.}~\bibnamefont {Paul}},\ }\href@noop
  {} {\bibfield  {journal} {\bibinfo  {journal} {Phys. Rev. E}\ }\textbf
  {\bibinfo {volume} {101}},\ \bibinfo {pages} {032128} (\bibinfo {year}
  {2020}{\natexlab{b}})}\BibitemShut {NoStop}%
\bibitem [{\citenamefont {Wittkowski}\ \emph {et~al.}(2014)\citenamefont
  {Wittkowski}, \citenamefont {Tiribocchi}, \citenamefont {Stenhammar},
  \citenamefont {Allen}, \citenamefont {Marenduzzo},\ and\ \citenamefont
  {Cates}}]{wittkowski2014scalar}%
  \BibitemOpen
  \bibfield  {author} {\bibinfo {author} {\bibfnamefont {R.}~\bibnamefont
  {Wittkowski}}, \bibinfo {author} {\bibfnamefont {A.}~\bibnamefont
  {Tiribocchi}}, \bibinfo {author} {\bibfnamefont {J.}~\bibnamefont
  {Stenhammar}}, \bibinfo {author} {\bibfnamefont {R.~J.}\ \bibnamefont
  {Allen}}, \bibinfo {author} {\bibfnamefont {D.}~\bibnamefont {Marenduzzo}}, \
  and\ \bibinfo {author} {\bibfnamefont {M.~E.}\ \bibnamefont {Cates}},\
  }\href@noop {} {\bibfield  {journal} {\bibinfo  {journal} {Nature
  Communications}\ }\textbf {\bibinfo {volume} {5}},\ \bibinfo {pages} {4351}
  (\bibinfo {year} {2014})}\BibitemShut {NoStop}%
\bibitem [{\citenamefont {Pattanayak}\ \emph
  {et~al.}(2021{\natexlab{a}})\citenamefont {Pattanayak}, \citenamefont
  {Mishra},\ and\ \citenamefont {Puri}}]{pattanayak2021AMB}%
  \BibitemOpen
  \bibfield  {author} {\bibinfo {author} {\bibfnamefont {S.}~\bibnamefont
  {Pattanayak}}, \bibinfo {author} {\bibfnamefont {S.}~\bibnamefont {Mishra}},
  \ and\ \bibinfo {author} {\bibfnamefont {S.}~\bibnamefont {Puri}},\
  }\href@noop {} {\bibfield  {journal} {\bibinfo  {journal} {Phys. Rev. E}\
  }\textbf {\bibinfo {volume} {104}},\ \bibinfo {pages} {014606} (\bibinfo
  {year} {2021}{\natexlab{a}})}\BibitemShut {NoStop}%
\bibitem [{\citenamefont {Pattanayak}\ \emph
  {et~al.}(2021{\natexlab{b}})\citenamefont {Pattanayak}, \citenamefont
  {Mishra},\ and\ \citenamefont {Puri}}]{pattanayak2021domain}%
  \BibitemOpen
  \bibfield  {author} {\bibinfo {author} {\bibfnamefont {S.}~\bibnamefont
  {Pattanayak}}, \bibinfo {author} {\bibfnamefont {S.}~\bibnamefont {Mishra}},
  \ and\ \bibinfo {author} {\bibfnamefont {S.}~\bibnamefont {Puri}},\
  }\href@noop {} {\bibfield  {journal} {\bibinfo  {journal} {Soft Materials}\
  }\textbf {\bibinfo {volume} {19}},\ \bibinfo {pages} {286} (\bibinfo {year}
  {2021}{\natexlab{b}})}\BibitemShut {NoStop}%
\bibitem [{\citenamefont {Mishra}\ \emph {et~al.}(2014)\citenamefont {Mishra},
  \citenamefont {Puri},\ and\ \citenamefont {Ramaswamy}}]{mishra2014aspects}%
  \BibitemOpen
  \bibfield  {author} {\bibinfo {author} {\bibfnamefont {S.}~\bibnamefont
  {Mishra}}, \bibinfo {author} {\bibfnamefont {S.}~\bibnamefont {Puri}}, \ and\
  \bibinfo {author} {\bibfnamefont {S.}~\bibnamefont {Ramaswamy}},\ }\href@noop
  {} {\bibfield  {journal} {\bibinfo  {journal} {Philosophical Transactions of
  the Royal Society A: Mathematical, Physical and Engineering Sciences}\
  }\textbf {\bibinfo {volume} {372}},\ \bibinfo {pages} {20130364} (\bibinfo
  {year} {2014})}\BibitemShut {NoStop}%
\bibitem [{\citenamefont {Das}\ \emph {et~al.}(2018)\citenamefont {Das},
  \citenamefont {Mishra},\ and\ \citenamefont {Puri}}]{das2018ordering}%
  \BibitemOpen
  \bibfield  {author} {\bibinfo {author} {\bibfnamefont {R.}~\bibnamefont
  {Das}}, \bibinfo {author} {\bibfnamefont {S.}~\bibnamefont {Mishra}}, \ and\
  \bibinfo {author} {\bibfnamefont {S.}~\bibnamefont {Puri}},\ }\href@noop {}
  {\bibfield  {journal} {\bibinfo  {journal} {EPL}\ }\textbf {\bibinfo {volume}
  {121}},\ \bibinfo {pages} {37002} (\bibinfo {year} {2018})}\BibitemShut
  {NoStop}%
\bibitem [{\citenamefont {Saha}\ \emph {et~al.}(2020)\citenamefont {Saha},
  \citenamefont {Agudo-Canalejo},\ and\ \citenamefont
  {Golestanian}}]{saha2020scalar}%
  \BibitemOpen
  \bibfield  {author} {\bibinfo {author} {\bibfnamefont {S.}~\bibnamefont
  {Saha}}, \bibinfo {author} {\bibfnamefont {J.}~\bibnamefont
  {Agudo-Canalejo}}, \ and\ \bibinfo {author} {\bibfnamefont {R.}~\bibnamefont
  {Golestanian}},\ }\href@noop {} {\bibfield  {journal} {\bibinfo  {journal}
  {Phys. Rev. X}\ }\textbf {\bibinfo {volume} {10}},\ \bibinfo {pages} {041009}
  (\bibinfo {year} {2020})}\BibitemShut {NoStop}%
\bibitem [{\citenamefont {Rouzaire}\ and\ \citenamefont
  {Levis}(2022)}]{rouzaire2022dynamics}%
  \BibitemOpen
  \bibfield  {author} {\bibinfo {author} {\bibfnamefont {Y.}~\bibnamefont
  {Rouzaire}}\ and\ \bibinfo {author} {\bibfnamefont {D.}~\bibnamefont
  {Levis}},\ }\href@noop {} {\bibfield  {journal} {\bibinfo  {journal}
  {Frontiers in Physics}\ }\textbf {\bibinfo {volume} {10}},\ \bibinfo {pages}
  {976515} (\bibinfo {year} {2022})}\BibitemShut {NoStop}%
\bibitem [{\citenamefont {Dittrich}\ \emph {et~al.}(2023)\citenamefont
  {Dittrich}, \citenamefont {Midya}, \citenamefont {Virnau},\ and\
  \citenamefont {Das}}]{dittrich2023growth}%
  \BibitemOpen
  \bibfield  {author} {\bibinfo {author} {\bibfnamefont {F.}~\bibnamefont
  {Dittrich}}, \bibinfo {author} {\bibfnamefont {J.}~\bibnamefont {Midya}},
  \bibinfo {author} {\bibfnamefont {P.}~\bibnamefont {Virnau}}, \ and\ \bibinfo
  {author} {\bibfnamefont {S.~K.}\ \bibnamefont {Das}},\ }\href@noop {}
  {\bibfield  {journal} {\bibinfo  {journal} {Phys. Rev. E}\ }\textbf {\bibinfo
  {volume} {108}},\ \bibinfo {pages} {024609} (\bibinfo {year}
  {2023})}\BibitemShut {NoStop}%
\bibitem [{\citenamefont {Caporusso}\ \emph {et~al.}(2023)\citenamefont
  {Caporusso}, \citenamefont {Cugliandolo}, \citenamefont {Digregorio},
  \citenamefont {Gonnella}, \citenamefont {Levis},\ and\ \citenamefont
  {Suma}}]{caporusso2023dynamics}%
  \BibitemOpen
  \bibfield  {author} {\bibinfo {author} {\bibfnamefont {C.~B.}\ \bibnamefont
  {Caporusso}}, \bibinfo {author} {\bibfnamefont {L.~F.}\ \bibnamefont
  {Cugliandolo}}, \bibinfo {author} {\bibfnamefont {P.}~\bibnamefont
  {Digregorio}}, \bibinfo {author} {\bibfnamefont {G.}~\bibnamefont
  {Gonnella}}, \bibinfo {author} {\bibfnamefont {D.}~\bibnamefont {Levis}}, \
  and\ \bibinfo {author} {\bibfnamefont {A.}~\bibnamefont {Suma}},\ }\href@noop
  {} {\bibfield  {journal} {\bibinfo  {journal} {Phys. Rev. Lett.}\ }\textbf
  {\bibinfo {volume} {131}},\ \bibinfo {pages} {068201} (\bibinfo {year}
  {2023})}\BibitemShut {NoStop}%
\bibitem [{\citenamefont {Katyal}\ \emph {et~al.}(2020)\citenamefont {Katyal},
  \citenamefont {Dey}, \citenamefont {Das},\ and\ \citenamefont
  {Puri}}]{Coarsening2020VM}%
  \BibitemOpen
  \bibfield  {author} {\bibinfo {author} {\bibfnamefont {N.}~\bibnamefont
  {Katyal}}, \bibinfo {author} {\bibfnamefont {S.}~\bibnamefont {Dey}},
  \bibinfo {author} {\bibfnamefont {D.}~\bibnamefont {Das}}, \ and\ \bibinfo
  {author} {\bibfnamefont {S.}~\bibnamefont {Puri}},\ }\href@noop {} {\bibfield
   {journal} {\bibinfo  {journal} {The European Physical Journal E}\ }\textbf
  {\bibinfo {volume} {43}},\ \bibinfo {pages} {10} (\bibinfo {year}
  {2020})}\BibitemShut {NoStop}%
\bibitem [{\citenamefont {Dikshit}\ and\ \citenamefont
  {Mishra}(2023)}]{Dikshit_2023}%
  \BibitemOpen
  \bibfield  {author} {\bibinfo {author} {\bibfnamefont {S.}~\bibnamefont
  {Dikshit}}\ and\ \bibinfo {author} {\bibfnamefont {S.}~\bibnamefont
  {Mishra}},\ }\href@noop {} {\bibfield  {journal} {\bibinfo  {journal} {EPL}\
  }\textbf {\bibinfo {volume} {143}},\ \bibinfo {pages} {17001} (\bibinfo
  {year} {2023})}\BibitemShut {NoStop}%
\bibitem [{\citenamefont {Metropolis}\ \emph {et~al.}(1953)\citenamefont
  {Metropolis}, \citenamefont {Rosenbluth}, \citenamefont {Rosenbluth},
  \citenamefont {Teller},\ and\ \citenamefont
  {Teller}}]{metropolis1953equation}%
  \BibitemOpen
  \bibfield  {author} {\bibinfo {author} {\bibfnamefont {N.}~\bibnamefont
  {Metropolis}}, \bibinfo {author} {\bibfnamefont {A.~W.}\ \bibnamefont
  {Rosenbluth}}, \bibinfo {author} {\bibfnamefont {M.~N.}\ \bibnamefont
  {Rosenbluth}}, \bibinfo {author} {\bibfnamefont {A.~H.}\ \bibnamefont
  {Teller}}, \ and\ \bibinfo {author} {\bibfnamefont {E.}~\bibnamefont
  {Teller}},\ }\href@noop {} {\bibfield  {journal} {\bibinfo  {journal} {J.
  Chem. Phys.}\ }\textbf {\bibinfo {volume} {21}},\ \bibinfo {pages} {1087}
  (\bibinfo {year} {1953})}\BibitemShut {NoStop}%
\bibitem [{\citenamefont {Newman}\ and\ \citenamefont
  {Barkema}(1999)}]{newman1999monte}%
  \BibitemOpen
  \bibfield  {author} {\bibinfo {author} {\bibfnamefont {M.~E.~J.}\
  \bibnamefont {Newman}}\ and\ \bibinfo {author} {\bibfnamefont
  {G.}~\bibnamefont {Barkema}},\ }\href@noop {} {\emph {\bibinfo {title} {Monte
  Carlo methods in statistical physics}}}\ (\bibinfo  {publisher} {Clarendon
  Press},\ \bibinfo {year} {1999})\BibitemShut {NoStop}%
\bibitem [{\citenamefont {Landau}\ and\ \citenamefont
  {Binder}(2021)}]{landau2021guide}%
  \BibitemOpen
  \bibfield  {author} {\bibinfo {author} {\bibfnamefont {D.}~\bibnamefont
  {Landau}}\ and\ \bibinfo {author} {\bibfnamefont {K.}~\bibnamefont
  {Binder}},\ }\href@noop {} {\emph {\bibinfo {title} {A guide to Monte Carlo
  simulations in statistical physics}}}\ (\bibinfo  {publisher} {Cambridge
  university press},\ \bibinfo {year} {2021})\BibitemShut {NoStop}%
\end{thebibliography}

\begin{thebibliography}{59}%
\makeatletter
\providecommand \@ifxundefined [1]{%
 \@ifx{#1\undefined}
}%
\providecommand \@ifnum [1]{%
 \ifnum #1\expandafter \@firstoftwo
 \else \expandafter \@secondoftwo
 \fi
}%
\providecommand \@ifx [1]{%
 \ifx #1\expandafter \@firstoftwo
 \else \expandafter \@secondoftwo
 \fi
}%
\providecommand \natexlab [1]{#1}%
\providecommand \enquote  [1]{``#1''}%
\providecommand \bibnamefont  [1]{#1}%
\providecommand \bibfnamefont [1]{#1}%
\providecommand \citenamefont [1]{#1}%
\providecommand \href@noop [0]{\@secondoftwo}%
\providecommand \href [0]{\begingroup \@sanitize@url \@href}%
\providecommand \@href[1]{\@@startlink{#1}\@@href}%
\providecommand \@@href[1]{\endgroup#1\@@endlink}%
\providecommand \@sanitize@url [0]{\catcode `\\12\catcode `\$12\catcode
  `\&12\catcode `\#12\catcode `\^12\catcode `\_12\catcode `\%12\relax}%
\providecommand \@@startlink[1]{}%
\providecommand \@@endlink[0]{}%
\providecommand \url  [0]{\begingroup\@sanitize@url \@url }%
\providecommand \@url [1]{\endgroup\@href {#1}{\urlprefix }}%
\providecommand \urlprefix  [0]{URL }%
\providecommand \Eprint [0]{\href }%
\providecommand \doibase [0]{http://dx.doi.org/}%
\providecommand \selectlanguage [0]{\@gobble}%
\providecommand \bibinfo  [0]{\@secondoftwo}%
\providecommand \bibfield  [0]{\@secondoftwo}%
\providecommand \translation [1]{[#1]}%
\providecommand \BibitemOpen [0]{}%
\providecommand \bibitemStop [0]{}%
\providecommand \bibitemNoStop [0]{.\EOS\space}%
\providecommand \EOS [0]{\spacefactor3000\relax}%
\providecommand \BibitemShut  [1]{\csname bibitem#1\endcsname}%
\let\auto@bib@innerbib\@empty
\bibitem [{\citenamefont {Marchetti}\ \emph {et~al.}(2013)\citenamefont
  {Marchetti}, \citenamefont {Joanny}, \citenamefont {Ramaswamy}, \citenamefont
  {Liverpool}, \citenamefont {Prost}, \citenamefont {Rao},\ and\ \citenamefont
  {Simha}}]{marchetti2013hydrodynamics}%
  \BibitemOpen
  \bibfield  {author} {\bibinfo {author} {\bibfnamefont {M.~C.}\ \bibnamefont
  {Marchetti}}, \bibinfo {author} {\bibfnamefont {J.~F.}\ \bibnamefont
  {Joanny}}, \bibinfo {author} {\bibfnamefont {S.}~\bibnamefont {Ramaswamy}},
  \bibinfo {author} {\bibfnamefont {T.~B.}\ \bibnamefont {Liverpool}}, \bibinfo
  {author} {\bibfnamefont {J.}~\bibnamefont {Prost}}, \bibinfo {author}
  {\bibfnamefont {M.}~\bibnamefont {Rao}}, \ and\ \bibinfo {author}
  {\bibfnamefont {R.~A.}\ \bibnamefont {Simha}},\ }\href@noop {} {\bibfield
  {journal} {\bibinfo  {journal} {Rev. Mod. Phys.}\ }\textbf {\bibinfo {volume}
  {85}},\ \bibinfo {pages} {1143} (\bibinfo {year} {2013})}\BibitemShut
  {NoStop}%
\bibitem [{\citenamefont {Shaebani}\ \emph {et~al.}(2020)\citenamefont
  {Shaebani}, \citenamefont {Wysocki}, \citenamefont {Winkler}, \citenamefont
  {Gompper},\ and\ \citenamefont {Rieger}}]{shaebani2020computational}%
  \BibitemOpen
  \bibfield  {author} {\bibinfo {author} {\bibfnamefont {M.~R.}\ \bibnamefont
  {Shaebani}}, \bibinfo {author} {\bibfnamefont {A.}~\bibnamefont {Wysocki}},
  \bibinfo {author} {\bibfnamefont {R.~G.}\ \bibnamefont {Winkler}}, \bibinfo
  {author} {\bibfnamefont {G.}~\bibnamefont {Gompper}}, \ and\ \bibinfo
  {author} {\bibfnamefont {H.}~\bibnamefont {Rieger}},\ }\href@noop {}
  {\bibfield  {journal} {\bibinfo  {journal} {Nat. Rev. Phys.}\ }\textbf
  {\bibinfo {volume} {2}},\ \bibinfo {pages} {181} (\bibinfo {year}
  {2020})}\BibitemShut {NoStop}%
\bibitem [{\citenamefont {Bottinelli}\ \emph {et~al.}(2016)\citenamefont
  {Bottinelli}, \citenamefont {Sumpter},\ and\ \citenamefont
  {Silverberg}}]{bottinelli2016emergent}%
  \BibitemOpen
  \bibfield  {author} {\bibinfo {author} {\bibfnamefont {A.}~\bibnamefont
  {Bottinelli}}, \bibinfo {author} {\bibfnamefont {D.~T.~J.}\ \bibnamefont
  {Sumpter}}, \ and\ \bibinfo {author} {\bibfnamefont {J.~L.}\ \bibnamefont
  {Silverberg}},\ }\href@noop {} {\bibfield  {journal} {\bibinfo  {journal}
  {Phys. Rev. Lett.}\ }\textbf {\bibinfo {volume} {117}},\ \bibinfo {pages}
  {228301} (\bibinfo {year} {2016})}\BibitemShut {NoStop}%
\bibitem [{\citenamefont {Helbing}\ and\ \citenamefont
  {Molnar}(1995)}]{helbing1995social}%
  \BibitemOpen
  \bibfield  {author} {\bibinfo {author} {\bibfnamefont {D.}~\bibnamefont
  {Helbing}}\ and\ \bibinfo {author} {\bibfnamefont {P.}~\bibnamefont
  {Molnar}},\ }\href@noop {} {\bibfield  {journal} {\bibinfo  {journal} {Phys.
  Rev. E}\ }\textbf {\bibinfo {volume} {51}},\ \bibinfo {pages} {4282}
  (\bibinfo {year} {1995})}\BibitemShut {NoStop}%
\bibitem [{\citenamefont {Garcimart{\'\i}n}\ \emph {et~al.}(2015)\citenamefont
  {Garcimart{\'\i}n}, \citenamefont {Pastor}, \citenamefont {Ferrer},
  \citenamefont {Ramos}, \citenamefont {Mart{\'\i}n-G{\'o}mez},\ and\
  \citenamefont {Zuriguel}}]{garcimartin2015flow}%
  \BibitemOpen
  \bibfield  {author} {\bibinfo {author} {\bibfnamefont {A.}~\bibnamefont
  {Garcimart{\'\i}n}}, \bibinfo {author} {\bibfnamefont {J.~M.}\ \bibnamefont
  {Pastor}}, \bibinfo {author} {\bibfnamefont {L.~M.}\ \bibnamefont {Ferrer}},
  \bibinfo {author} {\bibfnamefont {J.~J.}\ \bibnamefont {Ramos}}, \bibinfo
  {author} {\bibfnamefont {C.}~\bibnamefont {Mart{\'\i}n-G{\'o}mez}}, \ and\
  \bibinfo {author} {\bibfnamefont {I.}~\bibnamefont {Zuriguel}},\ }\href@noop
  {} {\bibfield  {journal} {\bibinfo  {journal} {Phys. Rev. E}\ }\textbf
  {\bibinfo {volume} {91}},\ \bibinfo {pages} {022808} (\bibinfo {year}
  {2015})}\BibitemShut {NoStop}%
\bibitem [{\citenamefont {Ballerini}\ \emph {et~al.}(2008)\citenamefont
  {Ballerini}, \citenamefont {Cabibbo}, \citenamefont {Candelier},
  \citenamefont {Cavagna}, \citenamefont {Cisbani}, \citenamefont {Giardina},
  \citenamefont {Lecomte}, \citenamefont {Orlandi}, \citenamefont {Parisi},
  \citenamefont {Procaccini} \emph {et~al.}}]{ballerini2008interaction}%
  \BibitemOpen
  \bibfield  {author} {\bibinfo {author} {\bibfnamefont {M.}~\bibnamefont
  {Ballerini}}, \bibinfo {author} {\bibfnamefont {N.}~\bibnamefont {Cabibbo}},
  \bibinfo {author} {\bibfnamefont {R.}~\bibnamefont {Candelier}}, \bibinfo
  {author} {\bibfnamefont {A.}~\bibnamefont {Cavagna}}, \bibinfo {author}
  {\bibfnamefont {E.}~\bibnamefont {Cisbani}}, \bibinfo {author} {\bibfnamefont
  {I.}~\bibnamefont {Giardina}}, \bibinfo {author} {\bibfnamefont
  {V.}~\bibnamefont {Lecomte}}, \bibinfo {author} {\bibfnamefont
  {A.}~\bibnamefont {Orlandi}}, \bibinfo {author} {\bibfnamefont
  {G.}~\bibnamefont {Parisi}}, \bibinfo {author} {\bibfnamefont
  {A.}~\bibnamefont {Procaccini}},  \emph {et~al.},\ }\href@noop {} {\bibfield
  {journal} {\bibinfo  {journal} {Proc. Nat. Acad. Sci.}\ }\textbf {\bibinfo
  {volume} {105}},\ \bibinfo {pages} {1232} (\bibinfo {year}
  {2008})}\BibitemShut {NoStop}%
\bibitem [{\citenamefont {Becco}\ \emph {et~al.}(2006)\citenamefont {Becco},
  \citenamefont {Vandewalle}, \citenamefont {Delcourt},\ and\ \citenamefont
  {Poncin}}]{becco2006experimental}%
  \BibitemOpen
  \bibfield  {author} {\bibinfo {author} {\bibfnamefont {C.}~\bibnamefont
  {Becco}}, \bibinfo {author} {\bibfnamefont {N.}~\bibnamefont {Vandewalle}},
  \bibinfo {author} {\bibfnamefont {J.}~\bibnamefont {Delcourt}}, \ and\
  \bibinfo {author} {\bibfnamefont {P.}~\bibnamefont {Poncin}},\ }\href@noop {}
  {\bibfield  {journal} {\bibinfo  {journal} {Physica A}\ }\textbf {\bibinfo
  {volume} {367}},\ \bibinfo {pages} {487} (\bibinfo {year}
  {2006})}\BibitemShut {NoStop}%
\bibitem [{\citenamefont {Calovi}\ \emph {et~al.}(2014)\citenamefont {Calovi},
  \citenamefont {Lopez}, \citenamefont {Ngo}, \citenamefont {Sire},
  \citenamefont {Chat{\'e}},\ and\ \citenamefont
  {Theraulaz}}]{calovi2014swarming}%
  \BibitemOpen
  \bibfield  {author} {\bibinfo {author} {\bibfnamefont {D.~S.}\ \bibnamefont
  {Calovi}}, \bibinfo {author} {\bibfnamefont {U.}~\bibnamefont {Lopez}},
  \bibinfo {author} {\bibfnamefont {S.}~\bibnamefont {Ngo}}, \bibinfo {author}
  {\bibfnamefont {C.}~\bibnamefont {Sire}}, \bibinfo {author} {\bibfnamefont
  {H.}~\bibnamefont {Chat{\'e}}}, \ and\ \bibinfo {author} {\bibfnamefont
  {G.}~\bibnamefont {Theraulaz}},\ }\href@noop {} {\bibfield  {journal}
  {\bibinfo  {journal} {New J. Phys.}\ }\textbf {\bibinfo {volume} {16}},\
  \bibinfo {pages} {015026} (\bibinfo {year} {2014})}\BibitemShut {NoStop}%
\bibitem [{\citenamefont {Steager}\ \emph {et~al.}(2008)\citenamefont
  {Steager}, \citenamefont {Kim},\ and\ \citenamefont
  {Kim}}]{steager2008dynamics}%
  \BibitemOpen
  \bibfield  {author} {\bibinfo {author} {\bibfnamefont {E.~B.}\ \bibnamefont
  {Steager}}, \bibinfo {author} {\bibfnamefont {C.-B.}\ \bibnamefont {Kim}}, \
  and\ \bibinfo {author} {\bibfnamefont {M.~J.}\ \bibnamefont {Kim}},\
  }\href@noop {} {\bibfield  {journal} {\bibinfo  {journal} {Physics of
  Fluids}\ }\textbf {\bibinfo {volume} {20}} (\bibinfo {year}
  {2008})}\BibitemShut {NoStop}%
\bibitem [{\citenamefont {Peruani}\ \emph {et~al.}(2012)\citenamefont
  {Peruani}, \citenamefont {Starru{\ss}}, \citenamefont {Jakovljevic},
  \citenamefont {S{\o}gaard-Andersen}, \citenamefont {Deutsch},\ and\
  \citenamefont {B{\"a}r}}]{peruani2012collective}%
  \BibitemOpen
  \bibfield  {author} {\bibinfo {author} {\bibfnamefont {F.}~\bibnamefont
  {Peruani}}, \bibinfo {author} {\bibfnamefont {J.}~\bibnamefont
  {Starru{\ss}}}, \bibinfo {author} {\bibfnamefont {V.}~\bibnamefont
  {Jakovljevic}}, \bibinfo {author} {\bibfnamefont {L.}~\bibnamefont
  {S{\o}gaard-Andersen}}, \bibinfo {author} {\bibfnamefont {A.}~\bibnamefont
  {Deutsch}}, \ and\ \bibinfo {author} {\bibfnamefont {M.}~\bibnamefont
  {B{\"a}r}},\ }\href@noop {} {\bibfield  {journal} {\bibinfo  {journal} {Phys.
  Rev. Lett.}\ }\textbf {\bibinfo {volume} {108}},\ \bibinfo {pages} {098102}
  (\bibinfo {year} {2012})}\BibitemShut {NoStop}%
\bibitem [{\citenamefont {Giavazzi}\ \emph {et~al.}(2018)\citenamefont
  {Giavazzi}, \citenamefont {Paoluzzi}, \citenamefont {Macchi}, \citenamefont
  {Bi}, \citenamefont {Scita}, \citenamefont {Manning}, \citenamefont
  {Cerbino},\ and\ \citenamefont {Marchetti}}]{giavazzi2018flocking}%
  \BibitemOpen
  \bibfield  {author} {\bibinfo {author} {\bibfnamefont {F.}~\bibnamefont
  {Giavazzi}}, \bibinfo {author} {\bibfnamefont {M.}~\bibnamefont {Paoluzzi}},
  \bibinfo {author} {\bibfnamefont {M.}~\bibnamefont {Macchi}}, \bibinfo
  {author} {\bibfnamefont {D.}~\bibnamefont {Bi}}, \bibinfo {author}
  {\bibfnamefont {G.}~\bibnamefont {Scita}}, \bibinfo {author} {\bibfnamefont
  {M.~L.}\ \bibnamefont {Manning}}, \bibinfo {author} {\bibfnamefont
  {R.}~\bibnamefont {Cerbino}}, \ and\ \bibinfo {author} {\bibfnamefont
  {M.~C.}\ \bibnamefont {Marchetti}},\ }\href@noop {} {\bibfield  {journal}
  {\bibinfo  {journal} {Soft Matter}\ }\textbf {\bibinfo {volume} {14}},\
  \bibinfo {pages} {3471} (\bibinfo {year} {2018})}\BibitemShut {NoStop}%
\bibitem [{\citenamefont {Schaller}\ \emph {et~al.}(2010)\citenamefont
  {Schaller}, \citenamefont {Weber}, \citenamefont {Semmrich}, \citenamefont
  {Frey},\ and\ \citenamefont {Bausch}}]{schaller2010polar}%
  \BibitemOpen
  \bibfield  {author} {\bibinfo {author} {\bibfnamefont {V.}~\bibnamefont
  {Schaller}}, \bibinfo {author} {\bibfnamefont {C.}~\bibnamefont {Weber}},
  \bibinfo {author} {\bibfnamefont {C.}~\bibnamefont {Semmrich}}, \bibinfo
  {author} {\bibfnamefont {E.}~\bibnamefont {Frey}}, \ and\ \bibinfo {author}
  {\bibfnamefont {A.~R.}\ \bibnamefont {Bausch}},\ }\href@noop {} {\bibfield
  {journal} {\bibinfo  {journal} {Nature}\ }\textbf {\bibinfo {volume} {467}},\
  \bibinfo {pages} {73} (\bibinfo {year} {2010})}\BibitemShut {NoStop}%
\bibitem [{\citenamefont {Sumino}\ \emph {et~al.}(2012)\citenamefont {Sumino},
  \citenamefont {Nagai}, \citenamefont {Shitaka}, \citenamefont {Tanaka},
  \citenamefont {Yoshikawa}, \citenamefont {Chat{\'e}},\ and\ \citenamefont
  {Oiwa}}]{sumino2012large}%
  \BibitemOpen
  \bibfield  {author} {\bibinfo {author} {\bibfnamefont {Y.}~\bibnamefont
  {Sumino}}, \bibinfo {author} {\bibfnamefont {K.~H.}\ \bibnamefont {Nagai}},
  \bibinfo {author} {\bibfnamefont {Y.}~\bibnamefont {Shitaka}}, \bibinfo
  {author} {\bibfnamefont {D.}~\bibnamefont {Tanaka}}, \bibinfo {author}
  {\bibfnamefont {K.}~\bibnamefont {Yoshikawa}}, \bibinfo {author}
  {\bibfnamefont {H.}~\bibnamefont {Chat{\'e}}}, \ and\ \bibinfo {author}
  {\bibfnamefont {K.}~\bibnamefont {Oiwa}},\ }\href@noop {} {\bibfield
  {journal} {\bibinfo  {journal} {Nature}\ }\textbf {\bibinfo {volume} {483}},\
  \bibinfo {pages} {448} (\bibinfo {year} {2012})}\BibitemShut {NoStop}%
\bibitem [{\citenamefont {Sanchez}\ \emph {et~al.}(2012)\citenamefont
  {Sanchez}, \citenamefont {Chen}, \citenamefont {DeCamp}, \citenamefont
  {Heymann},\ and\ \citenamefont {Dogic}}]{sanchez2012spontaneous}%
  \BibitemOpen
  \bibfield  {author} {\bibinfo {author} {\bibfnamefont {T.}~\bibnamefont
  {Sanchez}}, \bibinfo {author} {\bibfnamefont {D.~T.~N.}\ \bibnamefont
  {Chen}}, \bibinfo {author} {\bibfnamefont {S.~J.}\ \bibnamefont {DeCamp}},
  \bibinfo {author} {\bibfnamefont {M.}~\bibnamefont {Heymann}}, \ and\
  \bibinfo {author} {\bibfnamefont {Z.}~\bibnamefont {Dogic}},\ }\href@noop {}
  {\bibfield  {journal} {\bibinfo  {journal} {Nature}\ }\textbf {\bibinfo
  {volume} {491}},\ \bibinfo {pages} {431} (\bibinfo {year}
  {2012})}\BibitemShut {NoStop}%
\bibitem [{\citenamefont {Vicsek}\ \emph {et~al.}(1995)\citenamefont {Vicsek},
  \citenamefont {Czir{\'o}k}, \citenamefont {Ben-Jacob}, \citenamefont
  {Cohen},\ and\ \citenamefont {Shochet}}]{vicsek1995novel}%
  \BibitemOpen
  \bibfield  {author} {\bibinfo {author} {\bibfnamefont {T.}~\bibnamefont
  {Vicsek}}, \bibinfo {author} {\bibfnamefont {A.}~\bibnamefont {Czir{\'o}k}},
  \bibinfo {author} {\bibfnamefont {E.}~\bibnamefont {Ben-Jacob}}, \bibinfo
  {author} {\bibfnamefont {I.}~\bibnamefont {Cohen}}, \ and\ \bibinfo {author}
  {\bibfnamefont {O.}~\bibnamefont {Shochet}},\ }\href@noop {} {\bibfield
  {journal} {\bibinfo  {journal} {Phys. Rev. Lett.}\ }\textbf {\bibinfo
  {volume} {75}},\ \bibinfo {pages} {1226} (\bibinfo {year}
  {1995})}\BibitemShut {NoStop}%
\bibitem [{\citenamefont {Toner}\ and\ \citenamefont
  {Tu}(1995)}]{toner1995long}%
  \BibitemOpen
  \bibfield  {author} {\bibinfo {author} {\bibfnamefont {J.}~\bibnamefont
  {Toner}}\ and\ \bibinfo {author} {\bibfnamefont {Y.}~\bibnamefont {Tu}},\
  }\href@noop {} {\bibfield  {journal} {\bibinfo  {journal} {Phys. Rev. Lett.}\
  }\textbf {\bibinfo {volume} {75}},\ \bibinfo {pages} {4326} (\bibinfo {year}
  {1995})}\BibitemShut {NoStop}%
\bibitem [{\citenamefont {Toner}\ and\ \citenamefont
  {Tu}(1998)}]{toner1998flocks}%
  \BibitemOpen
  \bibfield  {author} {\bibinfo {author} {\bibfnamefont {J.}~\bibnamefont
  {Toner}}\ and\ \bibinfo {author} {\bibfnamefont {Y.}~\bibnamefont {Tu}},\
  }\href@noop {} {\bibfield  {journal} {\bibinfo  {journal} {Phys. Rev. E}\
  }\textbf {\bibinfo {volume} {58}},\ \bibinfo {pages} {4828} (\bibinfo {year}
  {1998})}\BibitemShut {NoStop}%
\bibitem [{\citenamefont {Toner}(2012)}]{toner2012reanalysis}%
  \BibitemOpen
  \bibfield  {author} {\bibinfo {author} {\bibfnamefont {J.}~\bibnamefont
  {Toner}},\ }\href@noop {} {\bibfield  {journal} {\bibinfo  {journal} {Phys.
  Rev. E}\ }\textbf {\bibinfo {volume} {86}},\ \bibinfo {pages} {031918}
  (\bibinfo {year} {2012})}\BibitemShut {NoStop}%
\bibitem [{\citenamefont {Solon}\ \emph
  {et~al.}(2015{\natexlab{a}})\citenamefont {Solon}, \citenamefont
  {Chat{\'e}},\ and\ \citenamefont {Tailleur}}]{solon2015phase}%
  \BibitemOpen
  \bibfield  {author} {\bibinfo {author} {\bibfnamefont {A.~P.}\ \bibnamefont
  {Solon}}, \bibinfo {author} {\bibfnamefont {H.}~\bibnamefont {Chat{\'e}}}, \
  and\ \bibinfo {author} {\bibfnamefont {J.}~\bibnamefont {Tailleur}},\
  }\href@noop {} {\bibfield  {journal} {\bibinfo  {journal} {Phys. Rev. Lett.}\
  }\textbf {\bibinfo {volume} {114}},\ \bibinfo {pages} {068101} (\bibinfo
  {year} {2015}{\natexlab{a}})}\BibitemShut {NoStop}%
\bibitem [{\citenamefont {Gr{\'e}goire}\ and\ \citenamefont
  {Chat{\'e}}(2004)}]{gregoire2004onset}%
  \BibitemOpen
  \bibfield  {author} {\bibinfo {author} {\bibfnamefont {G.}~\bibnamefont
  {Gr{\'e}goire}}\ and\ \bibinfo {author} {\bibfnamefont {H.}~\bibnamefont
  {Chat{\'e}}},\ }\href@noop {} {\bibfield  {journal} {\bibinfo  {journal}
  {Phys. Rev. Lett.}\ }\textbf {\bibinfo {volume} {92}},\ \bibinfo {pages}
  {025702} (\bibinfo {year} {2004})}\BibitemShut {NoStop}%
\bibitem [{\citenamefont {Solon}\ \emph
  {et~al.}(2015{\natexlab{b}})\citenamefont {Solon}, \citenamefont {Caussin},
  \citenamefont {Bartolo}, \citenamefont {Chat{\'e}},\ and\ \citenamefont
  {Tailleur}}]{solon2015pattern}%
  \BibitemOpen
  \bibfield  {author} {\bibinfo {author} {\bibfnamefont {A.~P.}\ \bibnamefont
  {Solon}}, \bibinfo {author} {\bibfnamefont {J.~B.}\ \bibnamefont {Caussin}},
  \bibinfo {author} {\bibfnamefont {D.}~\bibnamefont {Bartolo}}, \bibinfo
  {author} {\bibfnamefont {H.}~\bibnamefont {Chat{\'e}}}, \ and\ \bibinfo
  {author} {\bibfnamefont {J.}~\bibnamefont {Tailleur}},\ }\href@noop {}
  {\bibfield  {journal} {\bibinfo  {journal} {Phys. Rev. E}\ }\textbf {\bibinfo
  {volume} {92}},\ \bibinfo {pages} {062111} (\bibinfo {year}
  {2015}{\natexlab{b}})}\BibitemShut {NoStop}%
\bibitem [{\citenamefont {K{\"u}rsten}\ and\ \citenamefont
  {Ihle}(2020)}]{kursten2020dry}%
  \BibitemOpen
  \bibfield  {author} {\bibinfo {author} {\bibfnamefont {R.}~\bibnamefont
  {K{\"u}rsten}}\ and\ \bibinfo {author} {\bibfnamefont {T.}~\bibnamefont
  {Ihle}},\ }\href@noop {} {\bibfield  {journal} {\bibinfo  {journal} {Phys.
  Rev. Lett.}\ }\textbf {\bibinfo {volume} {125}},\ \bibinfo {pages} {188003}
  (\bibinfo {year} {2020})}\BibitemShut {NoStop}%
\bibitem [{\citenamefont {Peruani}\ \emph {et~al.}(2011)\citenamefont
  {Peruani}, \citenamefont {Klauss}, \citenamefont {Deutsch},\ and\
  \citenamefont {Voss-Boehme}}]{peruani2011traffic}%
  \BibitemOpen
  \bibfield  {author} {\bibinfo {author} {\bibfnamefont {F.}~\bibnamefont
  {Peruani}}, \bibinfo {author} {\bibfnamefont {T.}~\bibnamefont {Klauss}},
  \bibinfo {author} {\bibfnamefont {A.}~\bibnamefont {Deutsch}}, \ and\
  \bibinfo {author} {\bibfnamefont {A.}~\bibnamefont {Voss-Boehme}},\
  }\href@noop {} {\bibfield  {journal} {\bibinfo  {journal} {Phys. Rev. Lett.}\
  }\textbf {\bibinfo {volume} {106}},\ \bibinfo {pages} {128101} (\bibinfo
  {year} {2011})}\BibitemShut {NoStop}%
\bibitem [{\citenamefont {Solon}\ and\ \citenamefont
  {Tailleur}(2013)}]{solon2013revisiting}%
  \BibitemOpen
  \bibfield  {author} {\bibinfo {author} {\bibfnamefont {A.~P.}\ \bibnamefont
  {Solon}}\ and\ \bibinfo {author} {\bibfnamefont {J.}~\bibnamefont
  {Tailleur}},\ }\href@noop {} {\bibfield  {journal} {\bibinfo  {journal}
  {Phys. Rev. Lett.}\ }\textbf {\bibinfo {volume} {111}},\ \bibinfo {pages}
  {078101} (\bibinfo {year} {2013})}\BibitemShut {NoStop}%
\bibitem [{\citenamefont {Solon}\ and\ \citenamefont
  {Tailleur}(2015)}]{solon2015flocking}%
  \BibitemOpen
  \bibfield  {author} {\bibinfo {author} {\bibfnamefont {A.~P.}\ \bibnamefont
  {Solon}}\ and\ \bibinfo {author} {\bibfnamefont {J.}~\bibnamefont
  {Tailleur}},\ }\href@noop {} {\bibfield  {journal} {\bibinfo  {journal}
  {Phys. Rev. E}\ }\textbf {\bibinfo {volume} {92}},\ \bibinfo {pages} {042119}
  (\bibinfo {year} {2015})}\BibitemShut {NoStop}%
\bibitem [{\citenamefont {Ishibashi}\ and\ \citenamefont
  {Sakaguchi}(2022)}]{ishibashi2022solitary}%
  \BibitemOpen
  \bibfield  {author} {\bibinfo {author} {\bibfnamefont {K.}~\bibnamefont
  {Ishibashi}}\ and\ \bibinfo {author} {\bibfnamefont {H.}~\bibnamefont
  {Sakaguchi}},\ }\href@noop {} {\bibfield  {journal} {\bibinfo  {journal} {J.
  Phys. Soc. Jpn.}\ }\textbf {\bibinfo {volume} {91}},\ \bibinfo {pages}
  {034003} (\bibinfo {year} {2022})}\BibitemShut {NoStop}%
\bibitem [{\citenamefont {Chatterjee}\ \emph {et~al.}(2020)\citenamefont
  {Chatterjee}, \citenamefont {Mangeat}, \citenamefont {Paul},\ and\
  \citenamefont {Rieger}}]{chatterjee2020flocking}%
  \BibitemOpen
  \bibfield  {author} {\bibinfo {author} {\bibfnamefont {S.}~\bibnamefont
  {Chatterjee}}, \bibinfo {author} {\bibfnamefont {M.}~\bibnamefont {Mangeat}},
  \bibinfo {author} {\bibfnamefont {R.}~\bibnamefont {Paul}}, \ and\ \bibinfo
  {author} {\bibfnamefont {H.}~\bibnamefont {Rieger}},\ }\href@noop {}
  {\bibfield  {journal} {\bibinfo  {journal} {EPL}\ }\textbf {\bibinfo {volume}
  {130}},\ \bibinfo {pages} {66001} (\bibinfo {year} {2020})}\BibitemShut
  {NoStop}%
\bibitem [{\citenamefont {Mangeat}\ \emph {et~al.}(2020)\citenamefont
  {Mangeat}, \citenamefont {Chatterjee}, \citenamefont {Paul},\ and\
  \citenamefont {Rieger}}]{mangeat2020flocking}%
  \BibitemOpen
  \bibfield  {author} {\bibinfo {author} {\bibfnamefont {M.}~\bibnamefont
  {Mangeat}}, \bibinfo {author} {\bibfnamefont {S.}~\bibnamefont {Chatterjee}},
  \bibinfo {author} {\bibfnamefont {R.}~\bibnamefont {Paul}}, \ and\ \bibinfo
  {author} {\bibfnamefont {H.}~\bibnamefont {Rieger}},\ }\href@noop {}
  {\bibfield  {journal} {\bibinfo  {journal} {Phys. Rev. E}\ }\textbf {\bibinfo
  {volume} {102}},\ \bibinfo {pages} {042601} (\bibinfo {year}
  {2020})}\BibitemShut {NoStop}%
\bibitem [{\citenamefont {Chatterjee}\ \emph {et~al.}(2022)\citenamefont
  {Chatterjee}, \citenamefont {Mangeat},\ and\ \citenamefont
  {Rieger}}]{chatterjee2022polar}%
  \BibitemOpen
  \bibfield  {author} {\bibinfo {author} {\bibfnamefont {S.}~\bibnamefont
  {Chatterjee}}, \bibinfo {author} {\bibfnamefont {M.}~\bibnamefont {Mangeat}},
  \ and\ \bibinfo {author} {\bibfnamefont {H.}~\bibnamefont {Rieger}},\
  }\href@noop {} {\bibfield  {journal} {\bibinfo  {journal} {EPL}\ }\textbf
  {\bibinfo {volume} {138}},\ \bibinfo {pages} {41001} (\bibinfo {year}
  {2022})}\BibitemShut {NoStop}%
\bibitem [{\citenamefont {Solon}\ \emph {et~al.}(2022)\citenamefont {Solon},
  \citenamefont {Chat{\'e}}, \citenamefont {Toner},\ and\ \citenamefont
  {Tailleur}}]{solon2022susceptibility}%
  \BibitemOpen
  \bibfield  {author} {\bibinfo {author} {\bibfnamefont {A.}~\bibnamefont
  {Solon}}, \bibinfo {author} {\bibfnamefont {H.}~\bibnamefont {Chat{\'e}}},
  \bibinfo {author} {\bibfnamefont {J.}~\bibnamefont {Toner}}, \ and\ \bibinfo
  {author} {\bibfnamefont {J.}~\bibnamefont {Tailleur}},\ }\href@noop {}
  {\bibfield  {journal} {\bibinfo  {journal} {Phys. Rev. Lett.}\ }\textbf
  {\bibinfo {volume} {128}},\ \bibinfo {pages} {208004} (\bibinfo {year}
  {2022})}\BibitemShut {NoStop}%
\bibitem [{\citenamefont {Romanczuk}\ \emph {et~al.}(2012)\citenamefont
  {Romanczuk}, \citenamefont {B{\"a}r}, \citenamefont {Ebeling}, \citenamefont
  {Lindner},\ and\ \citenamefont {Schimansky-Geier}}]{romanczuk2012active}%
  \BibitemOpen
  \bibfield  {author} {\bibinfo {author} {\bibfnamefont {P.}~\bibnamefont
  {Romanczuk}}, \bibinfo {author} {\bibfnamefont {M.}~\bibnamefont {B{\"a}r}},
  \bibinfo {author} {\bibfnamefont {W.}~\bibnamefont {Ebeling}}, \bibinfo
  {author} {\bibfnamefont {B.}~\bibnamefont {Lindner}}, \ and\ \bibinfo
  {author} {\bibfnamefont {L.}~\bibnamefont {Schimansky-Geier}},\ }\href@noop
  {} {\bibfield  {journal} {\bibinfo  {journal} {Eur. Phys. J. Special Topics}\
  }\textbf {\bibinfo {volume} {202}},\ \bibinfo {pages} {1} (\bibinfo {year}
  {2012})}\BibitemShut {NoStop}%
\bibitem [{\citenamefont {Cates}\ and\ \citenamefont
  {Tailleur}(2015)}]{cates2015motility}%
  \BibitemOpen
  \bibfield  {author} {\bibinfo {author} {\bibfnamefont {M.~E.}\ \bibnamefont
  {Cates}}\ and\ \bibinfo {author} {\bibfnamefont {J.}~\bibnamefont
  {Tailleur}},\ }\href@noop {} {\bibfield  {journal} {\bibinfo  {journal}
  {Annu. Rev. Condens. Matter Phys.}\ }\textbf {\bibinfo {volume} {6}},\
  \bibinfo {pages} {219} (\bibinfo {year} {2015})}\BibitemShut {NoStop}%
\bibitem [{\citenamefont {Gr{\'e}goire}\ \emph {et~al.}(2003)\citenamefont
  {Gr{\'e}goire}, \citenamefont {Chat{\'e}},\ and\ \citenamefont
  {Tu}}]{gregoire2003moving}%
  \BibitemOpen
  \bibfield  {author} {\bibinfo {author} {\bibfnamefont {G.}~\bibnamefont
  {Gr{\'e}goire}}, \bibinfo {author} {\bibfnamefont {H.}~\bibnamefont
  {Chat{\'e}}}, \ and\ \bibinfo {author} {\bibfnamefont {Y.}~\bibnamefont
  {Tu}},\ }\href@noop {} {\bibfield  {journal} {\bibinfo  {journal} {Physica
  D}\ }\textbf {\bibinfo {volume} {181}},\ \bibinfo {pages} {157} (\bibinfo
  {year} {2003})}\BibitemShut {NoStop}%
\bibitem [{\citenamefont {Mart{\'\i}n-G{\'o}mez}\ \emph
  {et~al.}(2018)\citenamefont {Mart{\'\i}n-G{\'o}mez}, \citenamefont {Levis},
  \citenamefont {D{\'\i}az-Guilera},\ and\ \citenamefont
  {Pagonabarraga}}]{martin2018collective}%
  \BibitemOpen
  \bibfield  {author} {\bibinfo {author} {\bibfnamefont {A.}~\bibnamefont
  {Mart{\'\i}n-G{\'o}mez}}, \bibinfo {author} {\bibfnamefont {D.}~\bibnamefont
  {Levis}}, \bibinfo {author} {\bibfnamefont {A.}~\bibnamefont
  {D{\'\i}az-Guilera}}, \ and\ \bibinfo {author} {\bibfnamefont
  {I.}~\bibnamefont {Pagonabarraga}},\ }\href@noop {} {\bibfield  {journal}
  {\bibinfo  {journal} {Soft Matter}\ }\textbf {\bibinfo {volume} {14}},\
  \bibinfo {pages} {2610} (\bibinfo {year} {2018})}\BibitemShut {NoStop}%
\bibitem [{\citenamefont {Sese-Sansa}\ \emph {et~al.}(2018)\citenamefont
  {Sese-Sansa}, \citenamefont {Pagonabarraga},\ and\ \citenamefont
  {Levis}}]{sese2018velocity}%
  \BibitemOpen
  \bibfield  {author} {\bibinfo {author} {\bibfnamefont {E.}~\bibnamefont
  {Sese-Sansa}}, \bibinfo {author} {\bibfnamefont {I.}~\bibnamefont
  {Pagonabarraga}}, \ and\ \bibinfo {author} {\bibfnamefont {D.}~\bibnamefont
  {Levis}},\ }\href@noop {} {\bibfield  {journal} {\bibinfo  {journal} {EPL}\
  }\textbf {\bibinfo {volume} {124}},\ \bibinfo {pages} {30004} (\bibinfo
  {year} {2018})}\BibitemShut {NoStop}%
\bibitem [{\citenamefont {Berthier}\ \emph {et~al.}(2019)\citenamefont
  {Berthier}, \citenamefont {Flenner},\ and\ \citenamefont
  {Szamel}}]{berthier2019glassy}%
  \BibitemOpen
  \bibfield  {author} {\bibinfo {author} {\bibfnamefont {L.}~\bibnamefont
  {Berthier}}, \bibinfo {author} {\bibfnamefont {E.}~\bibnamefont {Flenner}}, \
  and\ \bibinfo {author} {\bibfnamefont {G.}~\bibnamefont {Szamel}},\
  }\href@noop {} {\bibfield  {journal} {\bibinfo  {journal} {J. Chem. Phys.}\
  }\textbf {\bibinfo {volume} {150}} (\bibinfo {year} {2019})}\BibitemShut
  {NoStop}%
\bibitem [{\citenamefont {Kourbane-Houssene}\ \emph {et~al.}(2018)\citenamefont
  {Kourbane-Houssene}, \citenamefont {Erignoux}, \citenamefont {Bodineau},\
  and\ \citenamefont {Tailleur}}]{kourbane2018exact}%
  \BibitemOpen
  \bibfield  {author} {\bibinfo {author} {\bibfnamefont {M.}~\bibnamefont
  {Kourbane-Houssene}}, \bibinfo {author} {\bibfnamefont {C.}~\bibnamefont
  {Erignoux}}, \bibinfo {author} {\bibfnamefont {T.}~\bibnamefont {Bodineau}},
  \ and\ \bibinfo {author} {\bibfnamefont {J.}~\bibnamefont {Tailleur}},\
  }\href@noop {} {\bibfield  {journal} {\bibinfo  {journal} {Phys. Rev. Lett.}\
  }\textbf {\bibinfo {volume} {120}},\ \bibinfo {pages} {268003} (\bibinfo
  {year} {2018})}\BibitemShut {NoStop}%
\bibitem [{\citenamefont {Kolmogorov}(1936)}]{kolmogorov1936markov}%
  \BibitemOpen
  \bibfield  {author} {\bibinfo {author} {\bibfnamefont {A.~N.}\ \bibnamefont
  {Kolmogorov}},\ }\href@noop {} {\bibfield  {journal} {\bibinfo  {journal}
  {Math. Ann.}\ }\textbf {\bibinfo {volume} {112}},\ \bibinfo {pages} {155}
  (\bibinfo {year} {1936})}\BibitemShut {NoStop}%
\bibitem [{\citenamefont {Liu}\ and\ \citenamefont
  {Nagel}(2010)}]{liu2010jamming}%
  \BibitemOpen
  \bibfield  {author} {\bibinfo {author} {\bibfnamefont {A.~J.}\ \bibnamefont
  {Liu}}\ and\ \bibinfo {author} {\bibfnamefont {S.~R.}\ \bibnamefont
  {Nagel}},\ }\href@noop {} {\bibfield  {journal} {\bibinfo  {journal} {Annu.
  Rev. Condens. Matter Phys.}\ }\textbf {\bibinfo {volume} {1}},\ \bibinfo
  {pages} {347} (\bibinfo {year} {2010})}\BibitemShut {NoStop}%
\bibitem [{\citenamefont {Digregorio}\ \emph {et~al.}(2018)\citenamefont
  {Digregorio}, \citenamefont {Levis}, \citenamefont {Suma}, \citenamefont
  {Cugliandolo}, \citenamefont {Gonnella},\ and\ \citenamefont
  {Pagonabarraga}}]{digregorio2018full}%
  \BibitemOpen
  \bibfield  {author} {\bibinfo {author} {\bibfnamefont {P.}~\bibnamefont
  {Digregorio}}, \bibinfo {author} {\bibfnamefont {D.}~\bibnamefont {Levis}},
  \bibinfo {author} {\bibfnamefont {A.}~\bibnamefont {Suma}}, \bibinfo {author}
  {\bibfnamefont {L.~F.}\ \bibnamefont {Cugliandolo}}, \bibinfo {author}
  {\bibfnamefont {G.}~\bibnamefont {Gonnella}}, \ and\ \bibinfo {author}
  {\bibfnamefont {I.}~\bibnamefont {Pagonabarraga}},\ }\href@noop {} {\bibfield
   {journal} {\bibinfo  {journal} {Phys. Rev. Lett.}\ }\textbf {\bibinfo
  {volume} {121}},\ \bibinfo {pages} {098003} (\bibinfo {year}
  {2018})}\BibitemShut {NoStop}%
\bibitem [{\citenamefont {Van Der~Linden}\ \emph {et~al.}(2019)\citenamefont
  {Van Der~Linden}, \citenamefont {Alexander}, \citenamefont {Aarts},\ and\
  \citenamefont {Dauchot}}]{van2019interrupted}%
  \BibitemOpen
  \bibfield  {author} {\bibinfo {author} {\bibfnamefont {M.~N.}\ \bibnamefont
  {Van Der~Linden}}, \bibinfo {author} {\bibfnamefont {L.~C.}\ \bibnamefont
  {Alexander}}, \bibinfo {author} {\bibfnamefont {D.~G. A.~L.}\ \bibnamefont
  {Aarts}}, \ and\ \bibinfo {author} {\bibfnamefont {O.}~\bibnamefont
  {Dauchot}},\ }\href@noop {} {\bibfield  {journal} {\bibinfo  {journal} {Phys.
  Rev. Lett.}\ }\textbf {\bibinfo {volume} {123}},\ \bibinfo {pages} {098001}
  (\bibinfo {year} {2019})}\BibitemShut {NoStop}%
\bibitem [{\citenamefont {Caporusso}\ \emph {et~al.}(2020)\citenamefont
  {Caporusso}, \citenamefont {Digregorio}, \citenamefont {Levis}, \citenamefont
  {Cugliandolo},\ and\ \citenamefont {Gonnella}}]{caporusso2020motility}%
  \BibitemOpen
  \bibfield  {author} {\bibinfo {author} {\bibfnamefont {C.~B.}\ \bibnamefont
  {Caporusso}}, \bibinfo {author} {\bibfnamefont {P.}~\bibnamefont
  {Digregorio}}, \bibinfo {author} {\bibfnamefont {D.}~\bibnamefont {Levis}},
  \bibinfo {author} {\bibfnamefont {L.~F.}\ \bibnamefont {Cugliandolo}}, \ and\
  \bibinfo {author} {\bibfnamefont {G.}~\bibnamefont {Gonnella}},\ }\href@noop
  {} {\bibfield  {journal} {\bibinfo  {journal} {Phys. Rev. Lett.}\ }\textbf
  {\bibinfo {volume} {125}},\ \bibinfo {pages} {178004} (\bibinfo {year}
  {2020})}\BibitemShut {NoStop}%
\bibitem [{\citenamefont {Geyer}\ \emph {et~al.}(2019)\citenamefont {Geyer},
  \citenamefont {Martin}, \citenamefont {Tailleur},\ and\ \citenamefont
  {Bartolo}}]{geyer2019freezing}%
  \BibitemOpen
  \bibfield  {author} {\bibinfo {author} {\bibfnamefont {D.}~\bibnamefont
  {Geyer}}, \bibinfo {author} {\bibfnamefont {D.}~\bibnamefont {Martin}},
  \bibinfo {author} {\bibfnamefont {J.}~\bibnamefont {Tailleur}}, \ and\
  \bibinfo {author} {\bibfnamefont {D.}~\bibnamefont {Bartolo}},\ }\href@noop
  {} {\bibfield  {journal} {\bibinfo  {journal} {Phys. Rev. X}\ }\textbf
  {\bibinfo {volume} {9}},\ \bibinfo {pages} {031043} (\bibinfo {year}
  {2019})}\BibitemShut {NoStop}%
\bibitem [{\citenamefont {Ses{\'e}-Sansa}\ \emph {et~al.}(2021)\citenamefont
  {Ses{\'e}-Sansa}, \citenamefont {Levis},\ and\ \citenamefont
  {Pagonabarraga}}]{sese2021phase}%
  \BibitemOpen
  \bibfield  {author} {\bibinfo {author} {\bibfnamefont {E.}~\bibnamefont
  {Ses{\'e}-Sansa}}, \bibinfo {author} {\bibfnamefont {D.}~\bibnamefont
  {Levis}}, \ and\ \bibinfo {author} {\bibfnamefont {I.}~\bibnamefont
  {Pagonabarraga}},\ }\href@noop {} {\bibfield  {journal} {\bibinfo  {journal}
  {Physical Review E}\ }\textbf {\bibinfo {volume} {104}},\ \bibinfo {pages}
  {054611} (\bibinfo {year} {2021})}\BibitemShut {NoStop}%
\bibitem [{\citenamefont {Levis}\ \emph {et~al.}(2017)\citenamefont {Levis},
  \citenamefont {Codina},\ and\ \citenamefont
  {Pagonabarraga}}]{levis2017active}%
  \BibitemOpen
  \bibfield  {author} {\bibinfo {author} {\bibfnamefont {D.}~\bibnamefont
  {Levis}}, \bibinfo {author} {\bibfnamefont {J.}~\bibnamefont {Codina}}, \
  and\ \bibinfo {author} {\bibfnamefont {I.}~\bibnamefont {Pagonabarraga}},\
  }\href@noop {} {\bibfield  {journal} {\bibinfo  {journal} {Soft Matter}\
  }\textbf {\bibinfo {volume} {13}},\ \bibinfo {pages} {8113} (\bibinfo {year}
  {2017})}\BibitemShut {NoStop}%
\bibitem [{\citenamefont {Kerner}\ and\ \citenamefont
  {Rehborn}(1997)}]{kerner1997experimental}%
  \BibitemOpen
  \bibfield  {author} {\bibinfo {author} {\bibfnamefont {B.~S.}\ \bibnamefont
  {Kerner}}\ and\ \bibinfo {author} {\bibfnamefont {H.}~\bibnamefont
  {Rehborn}},\ }\href@noop {} {\bibfield  {journal} {\bibinfo  {journal} {Phys.
  Rev. Lett.}\ }\textbf {\bibinfo {volume} {79}},\ \bibinfo {pages} {4030}
  (\bibinfo {year} {1997})}\BibitemShut {NoStop}%
\bibitem [{\citenamefont {Henkes}\ \emph {et~al.}(2011)\citenamefont {Henkes},
  \citenamefont {Fily},\ and\ \citenamefont {Marchetti}}]{henkes2011active}%
  \BibitemOpen
  \bibfield  {author} {\bibinfo {author} {\bibfnamefont {S.}~\bibnamefont
  {Henkes}}, \bibinfo {author} {\bibfnamefont {Y.}~\bibnamefont {Fily}}, \ and\
  \bibinfo {author} {\bibfnamefont {M.~C.}\ \bibnamefont {Marchetti}},\
  }\href@noop {} {\bibfield  {journal} {\bibinfo  {journal} {Phys. Rev. E}\
  }\textbf {\bibinfo {volume} {84}},\ \bibinfo {pages} {040301} (\bibinfo
  {year} {2011})}\BibitemShut {NoStop}%
\bibitem [{\citenamefont {Hecht}(2012)}]{hecht2012new}%
  \BibitemOpen
  \bibfield  {author} {\bibinfo {author} {\bibfnamefont {F.}~\bibnamefont
  {Hecht}},\ }\href@noop {} {\bibfield  {journal} {\bibinfo  {journal} {J.
  Numer. Math.}\ }\textbf {\bibinfo {volume} {20}},\ \bibinfo {pages} {251}
  (\bibinfo {year} {2012})}\BibitemShut {NoStop}%
\bibitem [{\citenamefont {Zienkiewicz}\ \emph {et~al.}(1977)\citenamefont
  {Zienkiewicz}, \citenamefont {Taylor}, \citenamefont {Nithiarasu},\ and\
  \citenamefont {Zhu}}]{zienkiewicz1977finite}%
  \BibitemOpen
  \bibfield  {author} {\bibinfo {author} {\bibfnamefont {O.~C.}\ \bibnamefont
  {Zienkiewicz}}, \bibinfo {author} {\bibfnamefont {R.~L.}\ \bibnamefont
  {Taylor}}, \bibinfo {author} {\bibfnamefont {P.}~\bibnamefont {Nithiarasu}},
  \ and\ \bibinfo {author} {\bibfnamefont {J.~Z.}\ \bibnamefont {Zhu}},\
  }\href@noop {} {\bibfield  {journal} {\bibinfo  {journal} {Chap}\ }\textbf
  {\bibinfo {volume} {6}},\ \bibinfo {pages} {96} (\bibinfo {year}
  {1977})}\BibitemShut {NoStop}%
\bibitem [{\citenamefont {Karmakar}\ \emph {et~al.}(2023)\citenamefont
  {Karmakar}, \citenamefont {Chatterjee}, \citenamefont {Mangeat},
  \citenamefont {Rieger},\ and\ \citenamefont {Paul}}]{karmakar2023zenodo}%
  \BibitemOpen
  \bibfield  {author} {\bibinfo {author} {\bibfnamefont {M.}~\bibnamefont
  {Karmakar}}, \bibinfo {author} {\bibfnamefont {S.}~\bibnamefont
  {Chatterjee}}, \bibinfo {author} {\bibfnamefont {M.}~\bibnamefont {Mangeat}},
  \bibinfo {author} {\bibfnamefont {H.}~\bibnamefont {Rieger}}, \ and\ \bibinfo
  {author} {\bibfnamefont {R.}~\bibnamefont {Paul}},\ }\href {\doibase
  https://doi.org/10.5281/zenodo.7942534} {\bibfield  {journal} {\bibinfo
  {journal} {Zendo}\ } (\bibinfo {year} {2023}),\
  https://doi.org/10.5281/zenodo.7942534}\BibitemShut {NoStop}%
\bibitem [{\citenamefont {Weinrib}\ and\ \citenamefont
  {Halperin}(1983)}]{weinrib1983critical}%
  \BibitemOpen
  \bibfield  {author} {\bibinfo {author} {\bibfnamefont {A.}~\bibnamefont
  {Weinrib}}\ and\ \bibinfo {author} {\bibfnamefont {B.~I.}\ \bibnamefont
  {Halperin}},\ }\href@noop {} {\bibfield  {journal} {\bibinfo  {journal}
  {Phys. Rev. B}\ }\textbf {\bibinfo {volume} {27}},\ \bibinfo {pages} {413}
  (\bibinfo {year} {1983})}\BibitemShut {NoStop}%
\bibitem [{\citenamefont {Dotsenko}(1995)}]{dotsenko1995critical}%
  \BibitemOpen
  \bibfield  {author} {\bibinfo {author} {\bibfnamefont {V.~S.}\ \bibnamefont
  {Dotsenko}},\ }\href@noop {} {\bibfield  {journal} {\bibinfo  {journal}
  {Physics-Uspekhi}\ }\textbf {\bibinfo {volume} {38}},\ \bibinfo {pages} {457}
  (\bibinfo {year} {1995})}\BibitemShut {NoStop}%
\bibitem [{\citenamefont {Kumar}\ \emph {et~al.}(2017)\citenamefont {Kumar},
  \citenamefont {Chatterjee}, \citenamefont {Paul},\ and\ \citenamefont
  {Puri}}]{kumar2017ordering}%
  \BibitemOpen
  \bibfield  {author} {\bibinfo {author} {\bibfnamefont {M.}~\bibnamefont
  {Kumar}}, \bibinfo {author} {\bibfnamefont {S.}~\bibnamefont {Chatterjee}},
  \bibinfo {author} {\bibfnamefont {R.}~\bibnamefont {Paul}}, \ and\ \bibinfo
  {author} {\bibfnamefont {S.}~\bibnamefont {Puri}},\ }\href@noop {} {\bibfield
   {journal} {\bibinfo  {journal} {Phys. Rev. E}\ }\textbf {\bibinfo {volume}
  {96}},\ \bibinfo {pages} {042127} (\bibinfo {year} {2017})}\BibitemShut
  {NoStop}%
\bibitem [{\citenamefont {Forg{\'a}cs}\ \emph {et~al.}(2021)\citenamefont
  {Forg{\'a}cs}, \citenamefont {Lib{\'a}l}, \citenamefont {Reichhardt},\ and\
  \citenamefont {Reichhardt}}]{forgacs2021active}%
  \BibitemOpen
  \bibfield  {author} {\bibinfo {author} {\bibfnamefont {P.}~\bibnamefont
  {Forg{\'a}cs}}, \bibinfo {author} {\bibfnamefont {A.}~\bibnamefont
  {Lib{\'a}l}}, \bibinfo {author} {\bibfnamefont {C.}~\bibnamefont
  {Reichhardt}}, \ and\ \bibinfo {author} {\bibfnamefont {C.~J.~O.}\
  \bibnamefont {Reichhardt}},\ }\href@noop {} {\bibfield  {journal} {\bibinfo
  {journal} {Phys. Rev. E}\ }\textbf {\bibinfo {volume} {104}},\ \bibinfo
  {pages} {044613} (\bibinfo {year} {2021})}\BibitemShut {NoStop}%
\bibitem [{\citenamefont {Reichhardt}\ and\ \citenamefont
  {Reichhardt}(2014)}]{reichhardt2014active}%
  \BibitemOpen
  \bibfield  {author} {\bibinfo {author} {\bibfnamefont {C.}~\bibnamefont
  {Reichhardt}}\ and\ \bibinfo {author} {\bibfnamefont {C.~J.~O.}\ \bibnamefont
  {Reichhardt}},\ }\href@noop {} {\bibfield  {journal} {\bibinfo  {journal}
  {Phys. Rev. E}\ }\textbf {\bibinfo {volume} {90}},\ \bibinfo {pages} {012701}
  (\bibinfo {year} {2014})}\BibitemShut {NoStop}%
\bibitem [{\citenamefont {Karmakar}\ \emph {et~al.}(shed)\citenamefont
  {Karmakar}, \citenamefont {Chatterjee}, \citenamefont {Rieger},\ and\
  \citenamefont {Paul}}]{karmakarunpublished}%
  \BibitemOpen
  \bibfield  {author} {\bibinfo {author} {\bibfnamefont {M.}~\bibnamefont
  {Karmakar}}, \bibinfo {author} {\bibfnamefont {S.}~\bibnamefont
  {Chatterjee}}, \bibinfo {author} {\bibfnamefont {H.}~\bibnamefont {Rieger}},
  \ and\ \bibinfo {author} {\bibfnamefont {R.}~\bibnamefont {Paul}},\
  }\href@noop {} {\enquote {\bibinfo {title} {Disorder active potts model},}\ }
  (\bibinfo {year} {unpublished})\BibitemShut {NoStop}%
\bibitem [{\citenamefont {Merrigan}\ \emph {et~al.}(2020)\citenamefont
  {Merrigan}, \citenamefont {Ramola}, \citenamefont {Chatterjee}, \citenamefont
  {Segall}, \citenamefont {Shokef},\ and\ \citenamefont
  {Chakraborty}}]{merrigan2020arrested}%
  \BibitemOpen
  \bibfield  {author} {\bibinfo {author} {\bibfnamefont {C.}~\bibnamefont
  {Merrigan}}, \bibinfo {author} {\bibfnamefont {K.}~\bibnamefont {Ramola}},
  \bibinfo {author} {\bibfnamefont {R.}~\bibnamefont {Chatterjee}}, \bibinfo
  {author} {\bibfnamefont {N.}~\bibnamefont {Segall}}, \bibinfo {author}
  {\bibfnamefont {Y.}~\bibnamefont {Shokef}}, \ and\ \bibinfo {author}
  {\bibfnamefont {B.}~\bibnamefont {Chakraborty}},\ }\href@noop {} {\bibfield
  {journal} {\bibinfo  {journal} {Phys. Rev. R.}\ }\textbf {\bibinfo {volume}
  {2}},\ \bibinfo {pages} {013260} (\bibinfo {year} {2020})}\BibitemShut
  {NoStop}%
\bibitem [{\citenamefont {Caprini}\ \emph {et~al.}(2020)\citenamefont
  {Caprini}, \citenamefont {Marconi},\ and\ \citenamefont
  {Puglisi}}]{caprini2020spontaneous}%
  \BibitemOpen
  \bibfield  {author} {\bibinfo {author} {\bibfnamefont {L.}~\bibnamefont
  {Caprini}}, \bibinfo {author} {\bibfnamefont {U.~M.~B.}\ \bibnamefont
  {Marconi}}, \ and\ \bibinfo {author} {\bibfnamefont {A.}~\bibnamefont
  {Puglisi}},\ }\href@noop {} {\bibfield  {journal} {\bibinfo  {journal} {Phys.
  Rev. Lett.}\ }\textbf {\bibinfo {volume} {124}},\ \bibinfo {pages} {078001}
  (\bibinfo {year} {2020})}\BibitemShut {NoStop}%
\bibitem [{\citenamefont {Farrell}\ \emph {et~al.}(2012)\citenamefont
  {Farrell}, \citenamefont {Marchetti}, \citenamefont {Marenduzzo},\ and\
  \citenamefont {Tailleur}}]{farrell2012pattern}%
  \BibitemOpen
  \bibfield  {author} {\bibinfo {author} {\bibfnamefont {F.~D.~C.}\
  \bibnamefont {Farrell}}, \bibinfo {author} {\bibfnamefont {M.~C.}\
  \bibnamefont {Marchetti}}, \bibinfo {author} {\bibfnamefont {D.}~\bibnamefont
  {Marenduzzo}}, \ and\ \bibinfo {author} {\bibfnamefont {J.}~\bibnamefont
  {Tailleur}},\ }\href@noop {} {\bibfield  {journal} {\bibinfo  {journal}
  {Phys. Rev. Lett.}\ }\textbf {\bibinfo {volume} {108}},\ \bibinfo {pages}
  {248101} (\bibinfo {year} {2012})}\BibitemShut {NoStop}%
\end{thebibliography}

\begin{thebibliography}{60}%
\makeatletter
\providecommand \@ifxundefined [1]{%
 \@ifx{#1\undefined}
}%
\providecommand \@ifnum [1]{%
 \ifnum #1\expandafter \@firstoftwo
 \else \expandafter \@secondoftwo
 \fi
}%
\providecommand \@ifx [1]{%
 \ifx #1\expandafter \@firstoftwo
 \else \expandafter \@secondoftwo
 \fi
}%
\providecommand \natexlab [1]{#1}%
\providecommand \enquote  [1]{``#1''}%
\providecommand \bibnamefont  [1]{#1}%
\providecommand \bibfnamefont [1]{#1}%
\providecommand \citenamefont [1]{#1}%
\providecommand \href@noop [0]{\@secondoftwo}%
\providecommand \href [0]{\begingroup \@sanitize@url \@href}%
\providecommand \@href[1]{\@@startlink{#1}\@@href}%
\providecommand \@@href[1]{\endgroup#1\@@endlink}%
\providecommand \@sanitize@url [0]{\catcode `\\12\catcode `\$12\catcode
  `\&12\catcode `\#12\catcode `\^12\catcode `\_12\catcode `\%12\relax}%
\providecommand \@@startlink[1]{}%
\providecommand \@@endlink[0]{}%
\providecommand \url  [0]{\begingroup\@sanitize@url \@url }%
\providecommand \@url [1]{\endgroup\@href {#1}{\urlprefix }}%
\providecommand \urlprefix  [0]{URL }%
\providecommand \Eprint [0]{\href }%
\providecommand \doibase [0]{http://dx.doi.org/}%
\providecommand \selectlanguage [0]{\@gobble}%
\providecommand \bibinfo  [0]{\@secondoftwo}%
\providecommand \bibfield  [0]{\@secondoftwo}%
\providecommand \translation [1]{[#1]}%
\providecommand \BibitemOpen [0]{}%
\providecommand \bibitemStop [0]{}%
\providecommand \bibitemNoStop [0]{.\EOS\space}%
\providecommand \EOS [0]{\spacefactor3000\relax}%
\providecommand \BibitemShut  [1]{\csname bibitem#1\endcsname}%
\let\auto@bib@innerbib\@empty
\bibitem [{\citenamefont {Ramaswamy}(2010)}]{ramaswamy}%
  \BibitemOpen
  \bibfield  {author} {\bibinfo {author} {\bibfnamefont {S.}~\bibnamefont
  {Ramaswamy}},\ }\href@noop {} {\bibfield  {journal} {\bibinfo  {journal}
  {Annual Review of Condensed Matter Physics}\ }\textbf {\bibinfo {volume}
  {1}},\ \bibinfo {pages} {323} (\bibinfo {year} {2010})}\BibitemShut {NoStop}%
\bibitem [{\citenamefont {Marchetti}\ \emph {et~al.}(2013)\citenamefont
  {Marchetti}, \citenamefont {Joanny}, \citenamefont {Ramaswamy}, \citenamefont
  {Liverpool}, \citenamefont {Prost}, \citenamefont {Rao},\ and\ \citenamefont
  {Simha}}]{marchetti2013hydrodynamics}%
  \BibitemOpen
  \bibfield  {author} {\bibinfo {author} {\bibfnamefont {M.~C.}\ \bibnamefont
  {Marchetti}}, \bibinfo {author} {\bibfnamefont {J.~F.}\ \bibnamefont
  {Joanny}}, \bibinfo {author} {\bibfnamefont {S.}~\bibnamefont {Ramaswamy}},
  \bibinfo {author} {\bibfnamefont {T.~B.}\ \bibnamefont {Liverpool}}, \bibinfo
  {author} {\bibfnamefont {J.}~\bibnamefont {Prost}}, \bibinfo {author}
  {\bibfnamefont {M.}~\bibnamefont {Rao}}, \ and\ \bibinfo {author}
  {\bibfnamefont {R.~A.}\ \bibnamefont {Simha}},\ }\href@noop {} {\bibfield
  {journal} {\bibinfo  {journal} {Reviews of Modern Physics}\ }\textbf
  {\bibinfo {volume} {85}},\ \bibinfo {pages} {1143} (\bibinfo {year}
  {2013})}\BibitemShut {NoStop}%
\bibitem [{\citenamefont {Needleman}\ and\ \citenamefont
  {Dogic}(2017)}]{needleman2017active}%
  \BibitemOpen
  \bibfield  {author} {\bibinfo {author} {\bibfnamefont {D.}~\bibnamefont
  {Needleman}}\ and\ \bibinfo {author} {\bibfnamefont {Z.}~\bibnamefont
  {Dogic}},\ }\href@noop {} {\bibfield  {journal} {\bibinfo  {journal} {Nature
  Reviews Materials}\ }\textbf {\bibinfo {volume} {2}},\ \bibinfo {pages} {1}
  (\bibinfo {year} {2017})}\BibitemShut {NoStop}%
\bibitem [{\citenamefont {Gompper}\ \emph {et~al.}(2020)\citenamefont
  {Gompper}, \citenamefont {Winkler}, \citenamefont {Speck}, \citenamefont
  {Solon}, \citenamefont {Nardini}, \citenamefont {Peruani}, \citenamefont
  {L{\"o}wen}, \citenamefont {Golestanian}, \citenamefont {Kaupp},
  \citenamefont {Alvarez} \emph {et~al.}}]{gompper20202020}%
  \BibitemOpen
  \bibfield  {author} {\bibinfo {author} {\bibfnamefont {G.}~\bibnamefont
  {Gompper}}, \bibinfo {author} {\bibfnamefont {R.~G.}\ \bibnamefont
  {Winkler}}, \bibinfo {author} {\bibfnamefont {T.}~\bibnamefont {Speck}},
  \bibinfo {author} {\bibfnamefont {A.}~\bibnamefont {Solon}}, \bibinfo
  {author} {\bibfnamefont {C.}~\bibnamefont {Nardini}}, \bibinfo {author}
  {\bibfnamefont {F.}~\bibnamefont {Peruani}}, \bibinfo {author} {\bibfnamefont
  {H.}~\bibnamefont {L{\"o}wen}}, \bibinfo {author} {\bibfnamefont
  {R.}~\bibnamefont {Golestanian}}, \bibinfo {author} {\bibfnamefont {U.~B.}\
  \bibnamefont {Kaupp}}, \bibinfo {author} {\bibfnamefont {L.}~\bibnamefont
  {Alvarez}},  \emph {et~al.},\ }\href@noop {} {\bibfield  {journal} {\bibinfo
  {journal} {Journal of Physics: Condensed Matter}\ }\textbf {\bibinfo {volume}
  {32}},\ \bibinfo {pages} {193001} (\bibinfo {year} {2020})}\BibitemShut
  {NoStop}%
\bibitem [{\citenamefont {Ballerini}\ \emph {et~al.}(2008)\citenamefont
  {Ballerini}, \citenamefont {Cabibbo}, \citenamefont {Candelier},
  \citenamefont {Cavagna}, \citenamefont {Cisbani}, \citenamefont {Giardina},
  \citenamefont {Lecomte}, \citenamefont {Orlandi}, \citenamefont {Parisi},
  \citenamefont {Procaccini}, \citenamefont {Viale},\ and\ \citenamefont
  {Zdravkovic}}]{Ballerini}%
  \BibitemOpen
  \bibfield  {author} {\bibinfo {author} {\bibfnamefont {M.}~\bibnamefont
  {Ballerini}}, \bibinfo {author} {\bibfnamefont {N.}~\bibnamefont {Cabibbo}},
  \bibinfo {author} {\bibfnamefont {R.}~\bibnamefont {Candelier}}, \bibinfo
  {author} {\bibfnamefont {A.}~\bibnamefont {Cavagna}}, \bibinfo {author}
  {\bibfnamefont {E.}~\bibnamefont {Cisbani}}, \bibinfo {author} {\bibfnamefont
  {I.}~\bibnamefont {Giardina}}, \bibinfo {author} {\bibfnamefont
  {V.}~\bibnamefont {Lecomte}}, \bibinfo {author} {\bibfnamefont
  {A.}~\bibnamefont {Orlandi}}, \bibinfo {author} {\bibfnamefont
  {G.}~\bibnamefont {Parisi}}, \bibinfo {author} {\bibfnamefont
  {A.}~\bibnamefont {Procaccini}}, \bibinfo {author} {\bibfnamefont
  {M.}~\bibnamefont {Viale}}, \ and\ \bibinfo {author} {\bibfnamefont
  {V.}~\bibnamefont {Zdravkovic}},\ }\href@noop {} {\bibfield  {journal}
  {\bibinfo  {journal} {Proceedings of the National Academy of Sciences}\
  }\textbf {\bibinfo {volume} {105}},\ \bibinfo {pages} {1232} (\bibinfo {year}
  {2008})}\BibitemShut {NoStop}%
\bibitem [{\citenamefont {Becco}\ \emph {et~al.}(2006)\citenamefont {Becco},
  \citenamefont {Vandewalle}, \citenamefont {Delcourt},\ and\ \citenamefont
  {Poncin}}]{Becco}%
  \BibitemOpen
  \bibfield  {author} {\bibinfo {author} {\bibfnamefont {C.}~\bibnamefont
  {Becco}}, \bibinfo {author} {\bibfnamefont {N.}~\bibnamefont {Vandewalle}},
  \bibinfo {author} {\bibfnamefont {J.}~\bibnamefont {Delcourt}}, \ and\
  \bibinfo {author} {\bibfnamefont {P.}~\bibnamefont {Poncin}},\ }\href@noop {}
  {\bibfield  {journal} {\bibinfo  {journal} {Physica A: Statistical Mechanics
  and its Applications}\ }\textbf {\bibinfo {volume} {367}},\ \bibinfo {pages}
  {487} (\bibinfo {year} {2006})}\BibitemShut {NoStop}%
\bibitem [{\citenamefont {Calovi}\ \emph {et~al.}(2014)\citenamefont {Calovi},
  \citenamefont {Lopez}, \citenamefont {Ngo}, \citenamefont {Sire},
  \citenamefont {Chat{\'{e}}},\ and\ \citenamefont {Theraulaz}}]{Calovi}%
  \BibitemOpen
  \bibfield  {author} {\bibinfo {author} {\bibfnamefont {D.~S.}\ \bibnamefont
  {Calovi}}, \bibinfo {author} {\bibfnamefont {U.}~\bibnamefont {Lopez}},
  \bibinfo {author} {\bibfnamefont {S.}~\bibnamefont {Ngo}}, \bibinfo {author}
  {\bibfnamefont {C.}~\bibnamefont {Sire}}, \bibinfo {author} {\bibfnamefont
  {H.}~\bibnamefont {Chat{\'{e}}}}, \ and\ \bibinfo {author} {\bibfnamefont
  {G.}~\bibnamefont {Theraulaz}},\ }\href@noop {} {\bibfield  {journal}
  {\bibinfo  {journal} {New Journal of Physics}\ }\textbf {\bibinfo {volume}
  {16}},\ \bibinfo {pages} {015026} (\bibinfo {year} {2014})}\BibitemShut
  {NoStop}%
\bibitem [{\citenamefont {Steager}\ \emph {et~al.}(2008)\citenamefont
  {Steager}, \citenamefont {Kim},\ and\ \citenamefont {Kim}}]{Steager}%
  \BibitemOpen
  \bibfield  {author} {\bibinfo {author} {\bibfnamefont {E.~B.}\ \bibnamefont
  {Steager}}, \bibinfo {author} {\bibfnamefont {C.~B.}\ \bibnamefont {Kim}}, \
  and\ \bibinfo {author} {\bibfnamefont {M.~J.}\ \bibnamefont {Kim}},\
  }\href@noop {} {\bibfield  {journal} {\bibinfo  {journal} {Physics of
  Fluids}\ }\textbf {\bibinfo {volume} {20}},\ \bibinfo {pages} {073601}
  (\bibinfo {year} {2008})}\BibitemShut {NoStop}%
\bibitem [{\citenamefont {Peruani}\ \emph {et~al.}(2012)\citenamefont
  {Peruani}, \citenamefont {Starru{\ss}}, \citenamefont {Jakovljevic},
  \citenamefont {S{\o}gaard-Andersen}, \citenamefont {Deutsch},\ and\
  \citenamefont {Bär}}]{Peruani}%
  \BibitemOpen
  \bibfield  {author} {\bibinfo {author} {\bibfnamefont {F.}~\bibnamefont
  {Peruani}}, \bibinfo {author} {\bibfnamefont {J.}~\bibnamefont
  {Starru{\ss}}}, \bibinfo {author} {\bibfnamefont {V.}~\bibnamefont
  {Jakovljevic}}, \bibinfo {author} {\bibfnamefont {L.}~\bibnamefont
  {S{\o}gaard-Andersen}}, \bibinfo {author} {\bibfnamefont {A.}~\bibnamefont
  {Deutsch}}, \ and\ \bibinfo {author} {\bibfnamefont {M.}~\bibnamefont
  {Bär}},\ }\href@noop {} {\bibfield  {journal} {\bibinfo  {journal} {Physical
  Review Letters}\ }\textbf {\bibinfo {volume} {108}} (\bibinfo {year}
  {2012})}\BibitemShut {NoStop}%
\bibitem [{\citenamefont {Schaller}\ \emph {et~al.}(2010)\citenamefont
  {Schaller}, \citenamefont {Weber}, \citenamefont {Semmrich}, \citenamefont
  {Frey},\ and\ \citenamefont {Bausch}}]{schaller2010polar}%
  \BibitemOpen
  \bibfield  {author} {\bibinfo {author} {\bibfnamefont {V.}~\bibnamefont
  {Schaller}}, \bibinfo {author} {\bibfnamefont {C.}~\bibnamefont {Weber}},
  \bibinfo {author} {\bibfnamefont {C.}~\bibnamefont {Semmrich}}, \bibinfo
  {author} {\bibfnamefont {E.}~\bibnamefont {Frey}}, \ and\ \bibinfo {author}
  {\bibfnamefont {A.~R.}\ \bibnamefont {Bausch}},\ }\href@noop {} {\bibfield
  {journal} {\bibinfo  {journal} {Nature}\ }\textbf {\bibinfo {volume} {467}},\
  \bibinfo {pages} {73} (\bibinfo {year} {2010})}\BibitemShut {NoStop}%
\bibitem [{\citenamefont {Sumino}\ \emph {et~al.}(2012)\citenamefont {Sumino},
  \citenamefont {Nagai}, \citenamefont {Shitaka}, \citenamefont {Tanaka},
  \citenamefont {Yoshikawa}, \citenamefont {Chat{\'e}},\ and\ \citenamefont
  {Oiwa}}]{sumino2012large}%
  \BibitemOpen
  \bibfield  {author} {\bibinfo {author} {\bibfnamefont {Y.}~\bibnamefont
  {Sumino}}, \bibinfo {author} {\bibfnamefont {K.~H.}\ \bibnamefont {Nagai}},
  \bibinfo {author} {\bibfnamefont {Y.}~\bibnamefont {Shitaka}}, \bibinfo
  {author} {\bibfnamefont {D.}~\bibnamefont {Tanaka}}, \bibinfo {author}
  {\bibfnamefont {K.}~\bibnamefont {Yoshikawa}}, \bibinfo {author}
  {\bibfnamefont {H.}~\bibnamefont {Chat{\'e}}}, \ and\ \bibinfo {author}
  {\bibfnamefont {K.}~\bibnamefont {Oiwa}},\ }\href@noop {} {\bibfield
  {journal} {\bibinfo  {journal} {Nature}\ }\textbf {\bibinfo {volume} {483}},\
  \bibinfo {pages} {448} (\bibinfo {year} {2012})}\BibitemShut {NoStop}%
\bibitem [{\citenamefont {Sanchez}\ \emph {et~al.}(2012)\citenamefont
  {Sanchez}, \citenamefont {Chen}, \citenamefont {DeCamp}, \citenamefont
  {Heymann},\ and\ \citenamefont {Dogic}}]{sanchez2012spontaneous}%
  \BibitemOpen
  \bibfield  {author} {\bibinfo {author} {\bibfnamefont {T.}~\bibnamefont
  {Sanchez}}, \bibinfo {author} {\bibfnamefont {D.~T.~N.}\ \bibnamefont
  {Chen}}, \bibinfo {author} {\bibfnamefont {S.~J.}\ \bibnamefont {DeCamp}},
  \bibinfo {author} {\bibfnamefont {M.}~\bibnamefont {Heymann}}, \ and\
  \bibinfo {author} {\bibfnamefont {Z.}~\bibnamefont {Dogic}},\ }\href@noop {}
  {\bibfield  {journal} {\bibinfo  {journal} {Nature}\ }\textbf {\bibinfo
  {volume} {491}},\ \bibinfo {pages} {431} (\bibinfo {year}
  {2012})}\BibitemShut {NoStop}%
\bibitem [{\citenamefont {Vicsek}\ and\ \citenamefont
  {Zafeiris}(2012)}]{collectivemotion}%
  \BibitemOpen
  \bibfield  {author} {\bibinfo {author} {\bibfnamefont {T.}~\bibnamefont
  {Vicsek}}\ and\ \bibinfo {author} {\bibfnamefont {A.}~\bibnamefont
  {Zafeiris}},\ }\href@noop {} {\bibfield  {journal} {\bibinfo  {journal}
  {Physics Reports}\ }\textbf {\bibinfo {volume} {517}},\ \bibinfo {pages} {71}
  (\bibinfo {year} {2012})}\BibitemShut {NoStop}%
\bibitem [{\citenamefont {Shaebani}\ \emph {et~al.}(2020)\citenamefont
  {Shaebani}, \citenamefont {Wysocki}, \citenamefont {Winkler}, \citenamefont
  {Gompper},\ and\ \citenamefont {Rieger}}]{shaebani2020computational}%
  \BibitemOpen
  \bibfield  {author} {\bibinfo {author} {\bibfnamefont {M.~R.}\ \bibnamefont
  {Shaebani}}, \bibinfo {author} {\bibfnamefont {A.}~\bibnamefont {Wysocki}},
  \bibinfo {author} {\bibfnamefont {R.~G.}\ \bibnamefont {Winkler}}, \bibinfo
  {author} {\bibfnamefont {G.}~\bibnamefont {Gompper}}, \ and\ \bibinfo
  {author} {\bibfnamefont {H.}~\bibnamefont {Rieger}},\ }\href@noop {}
  {\bibfield  {journal} {\bibinfo  {journal} {Nature Reviews Physics}\ }\textbf
  {\bibinfo {volume} {2}},\ \bibinfo {pages} {181} (\bibinfo {year}
  {2020})}\BibitemShut {NoStop}%
\bibitem [{\citenamefont {De~Magistris}\ and\ \citenamefont
  {Marenduzzo}(2015)}]{de2015introduction}%
  \BibitemOpen
  \bibfield  {author} {\bibinfo {author} {\bibfnamefont {G.}~\bibnamefont
  {De~Magistris}}\ and\ \bibinfo {author} {\bibfnamefont {D.}~\bibnamefont
  {Marenduzzo}},\ }\href@noop {} {\bibfield  {journal} {\bibinfo  {journal}
  {Physica A: Statistical Mechanics and its Applications}\ }\textbf {\bibinfo
  {volume} {418}},\ \bibinfo {pages} {65} (\bibinfo {year} {2015})}\BibitemShut
  {NoStop}%
\bibitem [{\citenamefont {Krishnan}\ \emph {et~al.}(2010)\citenamefont
  {Krishnan}, \citenamefont {Deshpande},\ and\ \citenamefont
  {Kumar}}]{krishnan2010rheology}%
  \BibitemOpen
  \bibfield  {author} {\bibinfo {author} {\bibfnamefont {J.~M.}\ \bibnamefont
  {Krishnan}}, \bibinfo {author} {\bibfnamefont {A.~P.}\ \bibnamefont
  {Deshpande}}, \ and\ \bibinfo {author} {\bibfnamefont {P.~B.~S.}\
  \bibnamefont {Kumar}},\ }\href@noop {} {\emph {\bibinfo {title} {Rheology of
  complex fluids}}}\ (\bibinfo  {publisher} {Springer},\ \bibinfo {year}
  {2010})\BibitemShut {NoStop}%
\bibitem [{\citenamefont {Vicsek}\ \emph {et~al.}(1995)\citenamefont {Vicsek},
  \citenamefont {Czir{\'o}k}, \citenamefont {Ben-Jacob}, \citenamefont
  {Cohen},\ and\ \citenamefont {Shochet}}]{Vicsek}%
  \BibitemOpen
  \bibfield  {author} {\bibinfo {author} {\bibfnamefont {T.}~\bibnamefont
  {Vicsek}}, \bibinfo {author} {\bibfnamefont {A.}~\bibnamefont {Czir{\'o}k}},
  \bibinfo {author} {\bibfnamefont {E.}~\bibnamefont {Ben-Jacob}}, \bibinfo
  {author} {\bibfnamefont {I.}~\bibnamefont {Cohen}}, \ and\ \bibinfo {author}
  {\bibfnamefont {O.}~\bibnamefont {Shochet}},\ }\href@noop {} {\bibfield
  {journal} {\bibinfo  {journal} {Physical Review Letters}\ }\textbf {\bibinfo
  {volume} {75}},\ \bibinfo {pages} {1226} (\bibinfo {year}
  {1995})}\BibitemShut {NoStop}%
\bibitem [{\citenamefont {Toner}\ and\ \citenamefont
  {Tu}(1995)}]{toner1995long}%
  \BibitemOpen
  \bibfield  {author} {\bibinfo {author} {\bibfnamefont {J.}~\bibnamefont
  {Toner}}\ and\ \bibinfo {author} {\bibfnamefont {Y.}~\bibnamefont {Tu}},\
  }\href@noop {} {\bibfield  {journal} {\bibinfo  {journal} {Physical Review
  Letters}\ }\textbf {\bibinfo {volume} {75}},\ \bibinfo {pages} {4326}
  (\bibinfo {year} {1995})}\BibitemShut {NoStop}%
\bibitem [{\citenamefont {Toner}\ and\ \citenamefont
  {Tu}(1998)}]{toner1998flocks}%
  \BibitemOpen
  \bibfield  {author} {\bibinfo {author} {\bibfnamefont {J.}~\bibnamefont
  {Toner}}\ and\ \bibinfo {author} {\bibfnamefont {Y.}~\bibnamefont {Tu}},\
  }\href@noop {} {\bibfield  {journal} {\bibinfo  {journal} {Physical Review
  E}\ }\textbf {\bibinfo {volume} {58}},\ \bibinfo {pages} {4828} (\bibinfo
  {year} {1998})}\BibitemShut {NoStop}%
\bibitem [{\citenamefont {Solon}\ and\ \citenamefont
  {Tailleur}(2013)}]{AIM_solon_intro}%
  \BibitemOpen
  \bibfield  {author} {\bibinfo {author} {\bibfnamefont {A.~P.}\ \bibnamefont
  {Solon}}\ and\ \bibinfo {author} {\bibfnamefont {J.}~\bibnamefont
  {Tailleur}},\ }\href@noop {} {\bibfield  {journal} {\bibinfo  {journal}
  {Physical Review Letters}\ }\textbf {\bibinfo {volume} {111}},\ \bibinfo
  {pages} {078101} (\bibinfo {year} {2013})}\BibitemShut {NoStop}%
\bibitem [{\citenamefont {Solon}\ and\ \citenamefont
  {Tailleur}(2015)}]{solon2015flocking}%
  \BibitemOpen
  \bibfield  {author} {\bibinfo {author} {\bibfnamefont {A.~P.}\ \bibnamefont
  {Solon}}\ and\ \bibinfo {author} {\bibfnamefont {J.}~\bibnamefont
  {Tailleur}},\ }\href@noop {} {\bibfield  {journal} {\bibinfo  {journal}
  {Physical Review E}\ }\textbf {\bibinfo {volume} {92}},\ \bibinfo {pages}
  {042119} (\bibinfo {year} {2015})}\BibitemShut {NoStop}%
\bibitem [{\citenamefont {Toner}\ \emph {et~al.}(2005)\citenamefont {Toner},
  \citenamefont {Tu},\ and\ \citenamefont {Ramaswamy}}]{toner2005170}%
  \BibitemOpen
  \bibfield  {author} {\bibinfo {author} {\bibfnamefont {J.}~\bibnamefont
  {Toner}}, \bibinfo {author} {\bibfnamefont {Y.}~\bibnamefont {Tu}}, \ and\
  \bibinfo {author} {\bibfnamefont {S.}~\bibnamefont {Ramaswamy}},\ }\href@noop
  {} {\bibfield  {journal} {\bibinfo  {journal} {Annals of Physics}\ }\textbf
  {\bibinfo {volume} {318}},\ \bibinfo {pages} {170} (\bibinfo {year}
  {2005})},\ \bibinfo {note} {special Issue}\BibitemShut {NoStop}%
\bibitem [{\citenamefont {Giomi}\ \emph {et~al.}(2013)\citenamefont {Giomi},
  \citenamefont {Bowick}, \citenamefont {Ma},\ and\ \citenamefont
  {Marchetti}}]{giomi2013defect}%
  \BibitemOpen
  \bibfield  {author} {\bibinfo {author} {\bibfnamefont {L.}~\bibnamefont
  {Giomi}}, \bibinfo {author} {\bibfnamefont {M.~J.}\ \bibnamefont {Bowick}},
  \bibinfo {author} {\bibfnamefont {X.}~\bibnamefont {Ma}}, \ and\ \bibinfo
  {author} {\bibfnamefont {M.~C.}\ \bibnamefont {Marchetti}},\ }\href@noop {}
  {\bibfield  {journal} {\bibinfo  {journal} {Physical Review Letters}\
  }\textbf {\bibinfo {volume} {110}},\ \bibinfo {pages} {228101} (\bibinfo
  {year} {2013})}\BibitemShut {NoStop}%
\bibitem [{\citenamefont {Cates}\ and\ \citenamefont {Tailleur}(2015)}]{MIPS}%
  \BibitemOpen
  \bibfield  {author} {\bibinfo {author} {\bibfnamefont {M.~E.}\ \bibnamefont
  {Cates}}\ and\ \bibinfo {author} {\bibfnamefont {J.}~\bibnamefont
  {Tailleur}},\ }\href@noop {} {\bibfield  {journal} {\bibinfo  {journal}
  {Annual Review of Condensed Matter Physics}\ }\textbf {\bibinfo {volume}
  {6}},\ \bibinfo {pages} {219} (\bibinfo {year} {2015})}\BibitemShut {NoStop}%
\bibitem [{\citenamefont {Solon}\ \emph {et~al.}(2015)\citenamefont {Solon},
  \citenamefont {Chat\'e},\ and\ \citenamefont {Tailleur}}]{solonVM}%
  \BibitemOpen
  \bibfield  {author} {\bibinfo {author} {\bibfnamefont {A.~P.}\ \bibnamefont
  {Solon}}, \bibinfo {author} {\bibfnamefont {H.}~\bibnamefont {Chat\'e}}, \
  and\ \bibinfo {author} {\bibfnamefont {J.}~\bibnamefont {Tailleur}},\
  }\href@noop {} {\bibfield  {journal} {\bibinfo  {journal} {Physical Review
  Letters}\ }\textbf {\bibinfo {volume} {114}},\ \bibinfo {pages} {068101}
  (\bibinfo {year} {2015})}\BibitemShut {NoStop}%
\bibitem [{\citenamefont {Siebert}\ \emph {et~al.}(2018)\citenamefont
  {Siebert}, \citenamefont {Dittrich}, \citenamefont {Schmid}, \citenamefont
  {Binder}, \citenamefont {Speck},\ and\ \citenamefont
  {Virnau}}]{criticalABP2018}%
  \BibitemOpen
  \bibfield  {author} {\bibinfo {author} {\bibfnamefont {J.~T.}\ \bibnamefont
  {Siebert}}, \bibinfo {author} {\bibfnamefont {F.}~\bibnamefont {Dittrich}},
  \bibinfo {author} {\bibfnamefont {F.}~\bibnamefont {Schmid}}, \bibinfo
  {author} {\bibfnamefont {K.}~\bibnamefont {Binder}}, \bibinfo {author}
  {\bibfnamefont {T.}~\bibnamefont {Speck}}, \ and\ \bibinfo {author}
  {\bibfnamefont {P.}~\bibnamefont {Virnau}},\ }\href@noop {} {\bibfield
  {journal} {\bibinfo  {journal} {Physical Review E}\ }\textbf {\bibinfo
  {volume} {98}},\ \bibinfo {pages} {030601} (\bibinfo {year}
  {2018})}\BibitemShut {NoStop}%
\bibitem [{\citenamefont {Wysocki}\ and\ \citenamefont
  {Rieger}(2020)}]{capillary2020adam}%
  \BibitemOpen
  \bibfield  {author} {\bibinfo {author} {\bibfnamefont {A.}~\bibnamefont
  {Wysocki}}\ and\ \bibinfo {author} {\bibfnamefont {H.}~\bibnamefont
  {Rieger}},\ }\href@noop {} {\bibfield  {journal} {\bibinfo  {journal}
  {Physical Review Letters}\ }\textbf {\bibinfo {volume} {124}},\ \bibinfo
  {pages} {048001} (\bibinfo {year} {2020})}\BibitemShut {NoStop}%
\bibitem [{\citenamefont {Mangeat}\ \emph {et~al.}(2020)\citenamefont
  {Mangeat}, \citenamefont {Chatterjee}, \citenamefont {Paul},\ and\
  \citenamefont {Rieger}}]{Raja_APM}%
  \BibitemOpen
  \bibfield  {author} {\bibinfo {author} {\bibfnamefont {M.}~\bibnamefont
  {Mangeat}}, \bibinfo {author} {\bibfnamefont {S.}~\bibnamefont {Chatterjee}},
  \bibinfo {author} {\bibfnamefont {R.}~\bibnamefont {Paul}}, \ and\ \bibinfo
  {author} {\bibfnamefont {H.}~\bibnamefont {Rieger}},\ }\href@noop {}
  {\bibfield  {journal} {\bibinfo  {journal} {Physical Review E}\ }\textbf
  {\bibinfo {volume} {102}},\ \bibinfo {pages} {042601} (\bibinfo {year}
  {2020})}\BibitemShut {NoStop}%
\bibitem [{\citenamefont {Chatterjee}\ \emph
  {et~al.}(2020{\natexlab{a}})\citenamefont {Chatterjee}, \citenamefont
  {Mangeat}, \citenamefont {Paul},\ and\ \citenamefont
  {Rieger}}]{chatterjee2020flocking}%
  \BibitemOpen
  \bibfield  {author} {\bibinfo {author} {\bibfnamefont {S.}~\bibnamefont
  {Chatterjee}}, \bibinfo {author} {\bibfnamefont {M.}~\bibnamefont {Mangeat}},
  \bibinfo {author} {\bibfnamefont {R.}~\bibnamefont {Paul}}, \ and\ \bibinfo
  {author} {\bibfnamefont {H.}~\bibnamefont {Rieger}},\ }\href@noop {}
  {\bibfield  {journal} {\bibinfo  {journal} {Europhysics Letters}\ }\textbf
  {\bibinfo {volume} {130}},\ \bibinfo {pages} {66001} (\bibinfo {year}
  {2020}{\natexlab{a}})}\BibitemShut {NoStop}%
\bibitem [{\citenamefont {K{\"u}rsten}\ and\ \citenamefont
  {Ihle}(2020)}]{kursten2020dry}%
  \BibitemOpen
  \bibfield  {author} {\bibinfo {author} {\bibfnamefont {R.}~\bibnamefont
  {K{\"u}rsten}}\ and\ \bibinfo {author} {\bibfnamefont {T.}~\bibnamefont
  {Ihle}},\ }\href@noop {} {\bibfield  {journal} {\bibinfo  {journal} {Physical
  Review Letters}\ }\textbf {\bibinfo {volume} {125}},\ \bibinfo {pages}
  {188003} (\bibinfo {year} {2020})}\BibitemShut {NoStop}%
\bibitem [{\citenamefont {Fruchart}\ \emph {et~al.}(2021)\citenamefont
  {Fruchart}, \citenamefont {Hanai}, \citenamefont {Littlewood},\ and\
  \citenamefont {Vitelli}}]{fruchart2021NR}%
  \BibitemOpen
  \bibfield  {author} {\bibinfo {author} {\bibfnamefont {M.}~\bibnamefont
  {Fruchart}}, \bibinfo {author} {\bibfnamefont {R.}~\bibnamefont {Hanai}},
  \bibinfo {author} {\bibfnamefont {P.~B.}\ \bibnamefont {Littlewood}}, \ and\
  \bibinfo {author} {\bibfnamefont {V.}~\bibnamefont {Vitelli}},\ }\href@noop
  {} {\bibfield  {journal} {\bibinfo  {journal} {Nature}\ }\textbf {\bibinfo
  {volume} {592}},\ \bibinfo {pages} {363} (\bibinfo {year}
  {2021})}\BibitemShut {NoStop}%
\bibitem [{\citenamefont {Solon}\ \emph {et~al.}(2022)\citenamefont {Solon},
  \citenamefont {Chat{\'e}}, \citenamefont {Toner},\ and\ \citenamefont
  {Tailleur}}]{solon2022susceptibility}%
  \BibitemOpen
  \bibfield  {author} {\bibinfo {author} {\bibfnamefont {A.}~\bibnamefont
  {Solon}}, \bibinfo {author} {\bibfnamefont {H.}~\bibnamefont {Chat{\'e}}},
  \bibinfo {author} {\bibfnamefont {J.}~\bibnamefont {Toner}}, \ and\ \bibinfo
  {author} {\bibfnamefont {J.}~\bibnamefont {Tailleur}},\ }\href@noop {}
  {\bibfield  {journal} {\bibinfo  {journal} {Physical Review Letters}\
  }\textbf {\bibinfo {volume} {128}},\ \bibinfo {pages} {208004} (\bibinfo
  {year} {2022})}\BibitemShut {NoStop}%
\bibitem [{\citenamefont {Chatterjee}\ \emph {et~al.}(2022)\citenamefont
  {Chatterjee}, \citenamefont {Mangeat},\ and\ \citenamefont
  {Rieger}}]{chatterjee2022polar}%
  \BibitemOpen
  \bibfield  {author} {\bibinfo {author} {\bibfnamefont {S.}~\bibnamefont
  {Chatterjee}}, \bibinfo {author} {\bibfnamefont {M.}~\bibnamefont {Mangeat}},
  \ and\ \bibinfo {author} {\bibfnamefont {H.}~\bibnamefont {Rieger}},\
  }\href@noop {} {\bibfield  {journal} {\bibinfo  {journal} {Europhysics
  Letters}\ }\textbf {\bibinfo {volume} {138}},\ \bibinfo {pages} {41001}
  (\bibinfo {year} {2022})}\BibitemShut {NoStop}%
\bibitem [{\citenamefont {Codina}\ \emph {et~al.}(2022)\citenamefont {Codina},
  \citenamefont {Mahault}, \citenamefont {Chat{\'e}}, \citenamefont {Dobnikar},
  \citenamefont {Pagonabarraga},\ and\ \citenamefont {Shi}}]{codina2022small}%
  \BibitemOpen
  \bibfield  {author} {\bibinfo {author} {\bibfnamefont {J.}~\bibnamefont
  {Codina}}, \bibinfo {author} {\bibfnamefont {B.}~\bibnamefont {Mahault}},
  \bibinfo {author} {\bibfnamefont {H.}~\bibnamefont {Chat{\'e}}}, \bibinfo
  {author} {\bibfnamefont {J.}~\bibnamefont {Dobnikar}}, \bibinfo {author}
  {\bibfnamefont {I.}~\bibnamefont {Pagonabarraga}}, \ and\ \bibinfo {author}
  {\bibfnamefont {X.}~\bibnamefont {Shi}},\ }\href@noop {} {\bibfield
  {journal} {\bibinfo  {journal} {Physical Review Letters}\ }\textbf {\bibinfo
  {volume} {128}},\ \bibinfo {pages} {218001} (\bibinfo {year}
  {2022})}\BibitemShut {NoStop}%
\bibitem [{\citenamefont {Chatterjee}\ \emph {et~al.}(2023)\citenamefont
  {Chatterjee}, \citenamefont {Mangeat}, \citenamefont {Woo}, \citenamefont
  {Rieger},\ and\ \citenamefont {Noh}}]{TSVM2023}%
  \BibitemOpen
  \bibfield  {author} {\bibinfo {author} {\bibfnamefont {S.}~\bibnamefont
  {Chatterjee}}, \bibinfo {author} {\bibfnamefont {M.}~\bibnamefont {Mangeat}},
  \bibinfo {author} {\bibfnamefont {C.~U.}\ \bibnamefont {Woo}}, \bibinfo
  {author} {\bibfnamefont {H.}~\bibnamefont {Rieger}}, \ and\ \bibinfo {author}
  {\bibfnamefont {J.~D.}\ \bibnamefont {Noh}},\ }\href@noop {} {\bibfield
  {journal} {\bibinfo  {journal} {Physical Review E}\ }\textbf {\bibinfo
  {volume} {107}},\ \bibinfo {pages} {024607} (\bibinfo {year}
  {2023})}\BibitemShut {NoStop}%
\bibitem [{\citenamefont {Karmakar}\ \emph {et~al.}(2023)\citenamefont
  {Karmakar}, \citenamefont {Chatterjee}, \citenamefont {Mangeat},
  \citenamefont {Rieger},\ and\ \citenamefont {Paul}}]{karmakar2023jamming}%
  \BibitemOpen
  \bibfield  {author} {\bibinfo {author} {\bibfnamefont {M.}~\bibnamefont
  {Karmakar}}, \bibinfo {author} {\bibfnamefont {S.}~\bibnamefont
  {Chatterjee}}, \bibinfo {author} {\bibfnamefont {M.}~\bibnamefont {Mangeat}},
  \bibinfo {author} {\bibfnamefont {H.}~\bibnamefont {Rieger}}, \ and\ \bibinfo
  {author} {\bibfnamefont {R.}~\bibnamefont {Paul}},\ }\href@noop {} {\bibfield
   {journal} {\bibinfo  {journal} {Physical Review E}\ }\textbf {\bibinfo
  {volume} {108}},\ \bibinfo {pages} {014604} (\bibinfo {year}
  {2023})}\BibitemShut {NoStop}%
\bibitem [{\citenamefont {Bray}(2002)}]{alan_bray}%
  \BibitemOpen
  \bibfield  {author} {\bibinfo {author} {\bibfnamefont {A.~J.}\ \bibnamefont
  {Bray}},\ }\href@noop {} {\bibfield  {journal} {\bibinfo  {journal} {Advances
  in Physics}\ }\textbf {\bibinfo {volume} {51}},\ \bibinfo {pages} {481}
  (\bibinfo {year} {2002})}\BibitemShut {NoStop}%
\bibitem [{\citenamefont {Bray}(1994)}]{bray1993theory}%
  \BibitemOpen
  \bibfield  {author} {\bibinfo {author} {\bibfnamefont {A.~J.}\ \bibnamefont
  {Bray}},\ }\href@noop {} {\bibfield  {journal} {\bibinfo  {journal} {Advances
  in Physics}\ }\textbf {\bibinfo {volume} {43}},\ \bibinfo {pages} {357}
  (\bibinfo {year} {1994})}\BibitemShut {NoStop}%
\bibitem [{\citenamefont {Puri}(2009)}]{sanjay_puri}%
  \BibitemOpen
  \bibfield  {author} {\bibinfo {author} {\bibfnamefont {S.}~\bibnamefont
  {Puri}}\ }(\bibinfo  {publisher} {CRC press},\ \bibinfo {year} {2009})\
  p.~\bibinfo {pages} {13}\BibitemShut {NoStop}%
\bibitem [{\citenamefont {Ahmad}\ \emph {et~al.}(2012)\citenamefont {Ahmad},
  \citenamefont {Corberi}, \citenamefont {Das}, \citenamefont {Lippiello},
  \citenamefont {Puri},\ and\ \citenamefont {Zannetti}}]{aging2012}%
  \BibitemOpen
  \bibfield  {author} {\bibinfo {author} {\bibfnamefont {S.}~\bibnamefont
  {Ahmad}}, \bibinfo {author} {\bibfnamefont {F.}~\bibnamefont {Corberi}},
  \bibinfo {author} {\bibfnamefont {S.~K.}\ \bibnamefont {Das}}, \bibinfo
  {author} {\bibfnamefont {E.}~\bibnamefont {Lippiello}}, \bibinfo {author}
  {\bibfnamefont {S.}~\bibnamefont {Puri}}, \ and\ \bibinfo {author}
  {\bibfnamefont {M.}~\bibnamefont {Zannetti}},\ }\href@noop {} {\bibfield
  {journal} {\bibinfo  {journal} {Physical Review E}\ }\textbf {\bibinfo
  {volume} {86}},\ \bibinfo {pages} {061129} (\bibinfo {year}
  {2012})}\BibitemShut {NoStop}%
\bibitem [{\citenamefont {Shrivastav}\ \emph {et~al.}(2014)\citenamefont
  {Shrivastav}, \citenamefont {Kumar}, \citenamefont {Banerjee},\ and\
  \citenamefont {Puri}}]{puri2014rfim}%
  \BibitemOpen
  \bibfield  {author} {\bibinfo {author} {\bibfnamefont {G.~P.}\ \bibnamefont
  {Shrivastav}}, \bibinfo {author} {\bibfnamefont {M.}~\bibnamefont {Kumar}},
  \bibinfo {author} {\bibfnamefont {V.}~\bibnamefont {Banerjee}}, \ and\
  \bibinfo {author} {\bibfnamefont {S.}~\bibnamefont {Puri}},\ }\href@noop {}
  {\bibfield  {journal} {\bibinfo  {journal} {Physical Review E}\ }\textbf
  {\bibinfo {volume} {90}},\ \bibinfo {pages} {032140} (\bibinfo {year}
  {2014})}\BibitemShut {NoStop}%
\bibitem [{\citenamefont {Kumar}\ \emph {et~al.}(2017)\citenamefont {Kumar},
  \citenamefont {Chatterjee}, \citenamefont {Paul},\ and\ \citenamefont
  {Puri}}]{kumar2017ordering}%
  \BibitemOpen
  \bibfield  {author} {\bibinfo {author} {\bibfnamefont {M.}~\bibnamefont
  {Kumar}}, \bibinfo {author} {\bibfnamefont {S.}~\bibnamefont {Chatterjee}},
  \bibinfo {author} {\bibfnamefont {R.}~\bibnamefont {Paul}}, \ and\ \bibinfo
  {author} {\bibfnamefont {S.}~\bibnamefont {Puri}},\ }\href@noop {} {\bibfield
   {journal} {\bibinfo  {journal} {Physical Review E}\ }\textbf {\bibinfo
  {volume} {96}},\ \bibinfo {pages} {042127} (\bibinfo {year}
  {2017})}\BibitemShut {NoStop}%
\bibitem [{\citenamefont {Chatterjee}\ \emph
  {et~al.}(2020{\natexlab{b}})\citenamefont {Chatterjee}, \citenamefont
  {Sutradhar}, \citenamefont {Puri},\ and\ \citenamefont {Paul}}]{rbcm}%
  \BibitemOpen
  \bibfield  {author} {\bibinfo {author} {\bibfnamefont {S.}~\bibnamefont
  {Chatterjee}}, \bibinfo {author} {\bibfnamefont {S.}~\bibnamefont
  {Sutradhar}}, \bibinfo {author} {\bibfnamefont {S.}~\bibnamefont {Puri}}, \
  and\ \bibinfo {author} {\bibfnamefont {R.}~\bibnamefont {Paul}},\ }\href@noop
  {} {\bibfield  {journal} {\bibinfo  {journal} {Physical Review E}\ }\textbf
  {\bibinfo {volume} {101}},\ \bibinfo {pages} {032128} (\bibinfo {year}
  {2020}{\natexlab{b}})}\BibitemShut {NoStop}%
\bibitem [{\citenamefont {Wittkowski}\ \emph {et~al.}(2014)\citenamefont
  {Wittkowski}, \citenamefont {Tiribocchi}, \citenamefont {Stenhammar},
  \citenamefont {Allen}, \citenamefont {Marenduzzo},\ and\ \citenamefont
  {Cates}}]{wittkowski2014scalar}%
  \BibitemOpen
  \bibfield  {author} {\bibinfo {author} {\bibfnamefont {R.}~\bibnamefont
  {Wittkowski}}, \bibinfo {author} {\bibfnamefont {A.}~\bibnamefont
  {Tiribocchi}}, \bibinfo {author} {\bibfnamefont {J.}~\bibnamefont
  {Stenhammar}}, \bibinfo {author} {\bibfnamefont {R.~J.}\ \bibnamefont
  {Allen}}, \bibinfo {author} {\bibfnamefont {D.}~\bibnamefont {Marenduzzo}}, \
  and\ \bibinfo {author} {\bibfnamefont {M.~E.}\ \bibnamefont {Cates}},\
  }\href@noop {} {\bibfield  {journal} {\bibinfo  {journal} {Nature
  Communications}\ }\textbf {\bibinfo {volume} {5}},\ \bibinfo {pages} {4351}
  (\bibinfo {year} {2014})}\BibitemShut {NoStop}%
\bibitem [{\citenamefont {Pattanayak}\ \emph
  {et~al.}(2021{\natexlab{a}})\citenamefont {Pattanayak}, \citenamefont
  {Mishra},\ and\ \citenamefont {Puri}}]{pattanayak2021AMB}%
  \BibitemOpen
  \bibfield  {author} {\bibinfo {author} {\bibfnamefont {S.}~\bibnamefont
  {Pattanayak}}, \bibinfo {author} {\bibfnamefont {S.}~\bibnamefont {Mishra}},
  \ and\ \bibinfo {author} {\bibfnamefont {S.}~\bibnamefont {Puri}},\
  }\href@noop {} {\bibfield  {journal} {\bibinfo  {journal} {Physical Review
  E}\ }\textbf {\bibinfo {volume} {104}},\ \bibinfo {pages} {014606} (\bibinfo
  {year} {2021}{\natexlab{a}})}\BibitemShut {NoStop}%
\bibitem [{\citenamefont {Pattanayak}\ \emph
  {et~al.}(2021{\natexlab{b}})\citenamefont {Pattanayak}, \citenamefont
  {Mishra},\ and\ \citenamefont {Puri}}]{pattanayak2021domain}%
  \BibitemOpen
  \bibfield  {author} {\bibinfo {author} {\bibfnamefont {S.}~\bibnamefont
  {Pattanayak}}, \bibinfo {author} {\bibfnamefont {S.}~\bibnamefont {Mishra}},
  \ and\ \bibinfo {author} {\bibfnamefont {S.}~\bibnamefont {Puri}},\
  }\href@noop {} {\bibfield  {journal} {\bibinfo  {journal} {Soft Materials}\
  }\textbf {\bibinfo {volume} {19}},\ \bibinfo {pages} {286} (\bibinfo {year}
  {2021}{\natexlab{b}})}\BibitemShut {NoStop}%
\bibitem [{\citenamefont {Mishra}\ \emph {et~al.}(2014)\citenamefont {Mishra},
  \citenamefont {Puri},\ and\ \citenamefont {Ramaswamy}}]{mishra2014aspects}%
  \BibitemOpen
  \bibfield  {author} {\bibinfo {author} {\bibfnamefont {S.}~\bibnamefont
  {Mishra}}, \bibinfo {author} {\bibfnamefont {S.}~\bibnamefont {Puri}}, \ and\
  \bibinfo {author} {\bibfnamefont {S.}~\bibnamefont {Ramaswamy}},\ }\href@noop
  {} {\bibfield  {journal} {\bibinfo  {journal} {Philosophical Transactions of
  the Royal Society A: Mathematical, Physical and Engineering Sciences}\
  }\textbf {\bibinfo {volume} {372}},\ \bibinfo {pages} {20130364} (\bibinfo
  {year} {2014})}\BibitemShut {NoStop}%
\bibitem [{\citenamefont {Das}\ \emph {et~al.}(2018)\citenamefont {Das},
  \citenamefont {Mishra},\ and\ \citenamefont {Puri}}]{das2018ordering}%
  \BibitemOpen
  \bibfield  {author} {\bibinfo {author} {\bibfnamefont {R.}~\bibnamefont
  {Das}}, \bibinfo {author} {\bibfnamefont {S.}~\bibnamefont {Mishra}}, \ and\
  \bibinfo {author} {\bibfnamefont {S.}~\bibnamefont {Puri}},\ }\href@noop {}
  {\bibfield  {journal} {\bibinfo  {journal} {Europhysics Letters}\ }\textbf
  {\bibinfo {volume} {121}},\ \bibinfo {pages} {37002} (\bibinfo {year}
  {2018})}\BibitemShut {NoStop}%
\bibitem [{\citenamefont {Saha}\ \emph {et~al.}(2020)\citenamefont {Saha},
  \citenamefont {Agudo-Canalejo},\ and\ \citenamefont
  {Golestanian}}]{saha2020scalar}%
  \BibitemOpen
  \bibfield  {author} {\bibinfo {author} {\bibfnamefont {S.}~\bibnamefont
  {Saha}}, \bibinfo {author} {\bibfnamefont {J.}~\bibnamefont
  {Agudo-Canalejo}}, \ and\ \bibinfo {author} {\bibfnamefont {R.}~\bibnamefont
  {Golestanian}},\ }\href@noop {} {\bibfield  {journal} {\bibinfo  {journal}
  {Physical Review X}\ }\textbf {\bibinfo {volume} {10}},\ \bibinfo {pages}
  {041009} (\bibinfo {year} {2020})}\BibitemShut {NoStop}%
\bibitem [{\citenamefont {Rouzaire}\ and\ \citenamefont
  {Levis}(2022)}]{rouzaire2022dynamics}%
  \BibitemOpen
  \bibfield  {author} {\bibinfo {author} {\bibfnamefont {Y.}~\bibnamefont
  {Rouzaire}}\ and\ \bibinfo {author} {\bibfnamefont {D.}~\bibnamefont
  {Levis}},\ }\href@noop {} {\bibfield  {journal} {\bibinfo  {journal}
  {Frontiers in Physics}\ }\textbf {\bibinfo {volume} {10}},\ \bibinfo {pages}
  {976515} (\bibinfo {year} {2022})}\BibitemShut {NoStop}%
\bibitem [{\citenamefont {Dittrich}\ \emph {et~al.}(2023)\citenamefont
  {Dittrich}, \citenamefont {Midya}, \citenamefont {Virnau},\ and\
  \citenamefont {Das}}]{dittrich2023growth}%
  \BibitemOpen
  \bibfield  {author} {\bibinfo {author} {\bibfnamefont {F.}~\bibnamefont
  {Dittrich}}, \bibinfo {author} {\bibfnamefont {J.}~\bibnamefont {Midya}},
  \bibinfo {author} {\bibfnamefont {P.}~\bibnamefont {Virnau}}, \ and\ \bibinfo
  {author} {\bibfnamefont {S.~K.}\ \bibnamefont {Das}},\ }\href@noop {}
  {\bibfield  {journal} {\bibinfo  {journal} {Physical Review E}\ }\textbf
  {\bibinfo {volume} {108}},\ \bibinfo {pages} {024609} (\bibinfo {year}
  {2023})}\BibitemShut {NoStop}%
\bibitem [{\citenamefont {Caporusso}\ \emph {et~al.}(2023)\citenamefont
  {Caporusso}, \citenamefont {Cugliandolo}, \citenamefont {Digregorio},
  \citenamefont {Gonnella}, \citenamefont {Levis},\ and\ \citenamefont
  {Suma}}]{caporusso2023dynamics}%
  \BibitemOpen
  \bibfield  {author} {\bibinfo {author} {\bibfnamefont {C.~B.}\ \bibnamefont
  {Caporusso}}, \bibinfo {author} {\bibfnamefont {L.~F.}\ \bibnamefont
  {Cugliandolo}}, \bibinfo {author} {\bibfnamefont {P.}~\bibnamefont
  {Digregorio}}, \bibinfo {author} {\bibfnamefont {G.}~\bibnamefont
  {Gonnella}}, \bibinfo {author} {\bibfnamefont {D.}~\bibnamefont {Levis}}, \
  and\ \bibinfo {author} {\bibfnamefont {A.}~\bibnamefont {Suma}},\ }\href@noop
  {} {\bibfield  {journal} {\bibinfo  {journal} {Physical Review Letters}\
  }\textbf {\bibinfo {volume} {131}},\ \bibinfo {pages} {068201} (\bibinfo
  {year} {2023})}\BibitemShut {NoStop}%
\bibitem [{\citenamefont {Katyal}\ \emph {et~al.}(2020)\citenamefont {Katyal},
  \citenamefont {Dey}, \citenamefont {Das},\ and\ \citenamefont
  {Puri}}]{Coarsening2020VM}%
  \BibitemOpen
  \bibfield  {author} {\bibinfo {author} {\bibfnamefont {N.}~\bibnamefont
  {Katyal}}, \bibinfo {author} {\bibfnamefont {S.}~\bibnamefont {Dey}},
  \bibinfo {author} {\bibfnamefont {D.}~\bibnamefont {Das}}, \ and\ \bibinfo
  {author} {\bibfnamefont {S.}~\bibnamefont {Puri}},\ }\href@noop {} {\bibfield
   {journal} {\bibinfo  {journal} {The European Physical Journal E}\ }\textbf
  {\bibinfo {volume} {43}},\ \bibinfo {pages} {10} (\bibinfo {year}
  {2020})}\BibitemShut {NoStop}%
\bibitem [{\citenamefont {Dikshit}\ and\ \citenamefont
  {Mishra}(2023)}]{Dikshit_2023}%
  \BibitemOpen
  \bibfield  {author} {\bibinfo {author} {\bibfnamefont {S.}~\bibnamefont
  {Dikshit}}\ and\ \bibinfo {author} {\bibfnamefont {S.}~\bibnamefont
  {Mishra}},\ }\href@noop {} {\bibfield  {journal} {\bibinfo  {journal}
  {Europhysics Letters}\ }\textbf {\bibinfo {volume} {143}},\ \bibinfo {pages}
  {17001} (\bibinfo {year} {2023})}\BibitemShut {NoStop}%
\bibitem [{\citenamefont {Majumder}\ and\ \citenamefont
  {Das}(2011)}]{2dIsing_growth}%
  \BibitemOpen
  \bibfield  {author} {\bibinfo {author} {\bibfnamefont {S.}~\bibnamefont
  {Majumder}}\ and\ \bibinfo {author} {\bibfnamefont {S.~K.}\ \bibnamefont
  {Das}},\ }\href@noop {} {\bibfield  {journal} {\bibinfo  {journal} {Physical
  Review E}\ }\textbf {\bibinfo {volume} {84}},\ \bibinfo {pages} {021110}
  (\bibinfo {year} {2011})}\BibitemShut {NoStop}%
\bibitem [{\citenamefont {Chatterjee}\ \emph {et~al.}(2018)\citenamefont
  {Chatterjee}, \citenamefont {Puri},\ and\ \citenamefont
  {Paul}}]{chatterjee2018clock}%
  \BibitemOpen
  \bibfield  {author} {\bibinfo {author} {\bibfnamefont {S.}~\bibnamefont
  {Chatterjee}}, \bibinfo {author} {\bibfnamefont {S.}~\bibnamefont {Puri}}, \
  and\ \bibinfo {author} {\bibfnamefont {R.}~\bibnamefont {Paul}},\ }\href@noop
  {} {\bibfield  {journal} {\bibinfo  {journal} {Physical Review E}\ }\textbf
  {\bibinfo {volume} {98}},\ \bibinfo {pages} {032109} (\bibinfo {year}
  {2018})}\BibitemShut {NoStop}%
\bibitem [{\citenamefont {Dey}\ \emph {et~al.}(2012)\citenamefont {Dey},
  \citenamefont {Das},\ and\ \citenamefont
  {Rajesh}}]{Supravat2012spatstructGNF}%
  \BibitemOpen
  \bibfield  {author} {\bibinfo {author} {\bibfnamefont {S.}~\bibnamefont
  {Dey}}, \bibinfo {author} {\bibfnamefont {D.}~\bibnamefont {Das}}, \ and\
  \bibinfo {author} {\bibfnamefont {R.}~\bibnamefont {Rajesh}},\ }\href@noop {}
  {\bibfield  {journal} {\bibinfo  {journal} {Physical Review Letters}\
  }\textbf {\bibinfo {volume} {108}},\ \bibinfo {pages} {238001} (\bibinfo
  {year} {2012})}\BibitemShut {NoStop}%
\bibitem [{\citenamefont {Yurke}\ \emph {et~al.}(1993)\citenamefont {Yurke},
  \citenamefont {Pargellis}, \citenamefont {Kovacs},\ and\ \citenamefont
  {Huse}}]{yurke1993coarsening}%
  \BibitemOpen
  \bibfield  {author} {\bibinfo {author} {\bibfnamefont {B.}~\bibnamefont
  {Yurke}}, \bibinfo {author} {\bibfnamefont {A.~N.}\ \bibnamefont
  {Pargellis}}, \bibinfo {author} {\bibfnamefont {T.}~\bibnamefont {Kovacs}}, \
  and\ \bibinfo {author} {\bibfnamefont {D.~A.}\ \bibnamefont {Huse}},\
  }\href@noop {} {\bibfield  {journal} {\bibinfo  {journal} {Physical Review
  E}\ }\textbf {\bibinfo {volume} {47}},\ \bibinfo {pages} {1525} (\bibinfo
  {year} {1993})}\BibitemShut {NoStop}%
\bibitem [{\citenamefont {Benvegnen}\ \emph {et~al.}(2022)\citenamefont
  {Benvegnen}, \citenamefont {Chat{\'e}}, \citenamefont {Krapivsky},
  \citenamefont {Tailleur},\ and\ \citenamefont
  {Solon}}]{benvegnen2022flocking}%
  \BibitemOpen
  \bibfield  {author} {\bibinfo {author} {\bibfnamefont {B.}~\bibnamefont
  {Benvegnen}}, \bibinfo {author} {\bibfnamefont {H.}~\bibnamefont
  {Chat{\'e}}}, \bibinfo {author} {\bibfnamefont {P.~L.}\ \bibnamefont
  {Krapivsky}}, \bibinfo {author} {\bibfnamefont {J.}~\bibnamefont {Tailleur}},
  \ and\ \bibinfo {author} {\bibfnamefont {A.}~\bibnamefont {Solon}},\
  }\href@noop {} {\bibfield  {journal} {\bibinfo  {journal} {Physical Review
  E}\ }\textbf {\bibinfo {volume} {106}},\ \bibinfo {pages} {054608} (\bibinfo
  {year} {2022})}\BibitemShut {NoStop}%
\bibitem [{\citenamefont {Press}(2007)}]{press2007numerical}%
  \BibitemOpen
  \bibfield  {author} {\bibinfo {author} {\bibfnamefont {W.~H.}\ \bibnamefont
  {Press}},\ }\href@noop {} {\emph {\bibinfo {title} {Numerical recipes: The
  art of scientific computing}}}\ (\bibinfo  {publisher} {Cambridge university
  press},\ \bibinfo {year} {2007})\BibitemShut {NoStop}%
\end{thebibliography}

\begin{thebibliography}{38}%
\makeatletter
\providecommand \@ifxundefined [1]{%
 \@ifx{#1\undefined}
}%
\providecommand \@ifnum [1]{%
 \ifnum #1\expandafter \@firstoftwo
 \else \expandafter \@secondoftwo
 \fi
}%
\providecommand \@ifx [1]{%
 \ifx #1\expandafter \@firstoftwo
 \else \expandafter \@secondoftwo
 \fi
}%
\providecommand \natexlab [1]{#1}%
\providecommand \enquote  [1]{``#1''}%
\providecommand \bibnamefont  [1]{#1}%
\providecommand \bibfnamefont [1]{#1}%
\providecommand \citenamefont [1]{#1}%
\providecommand \href@noop [0]{\@secondoftwo}%
\providecommand \href [0]{\begingroup \@sanitize@url \@href}%
\providecommand \@href[1]{\@@startlink{#1}\@@href}%
\providecommand \@@href[1]{\endgroup#1\@@endlink}%
\providecommand \@sanitize@url [0]{\catcode `\\12\catcode `\$12\catcode
  `\&12\catcode `\#12\catcode `\^12\catcode `\_12\catcode `\%12\relax}%
\providecommand \@@startlink[1]{}%
\providecommand \@@endlink[0]{}%
\providecommand \url  [0]{\begingroup\@sanitize@url \@url }%
\providecommand \@url [1]{\endgroup\@href {#1}{\urlprefix }}%
\providecommand \urlprefix  [0]{URL }%
\providecommand \Eprint [0]{\href }%
\providecommand \doibase [0]{http://dx.doi.org/}%
\providecommand \selectlanguage [0]{\@gobble}%
\providecommand \bibinfo  [0]{\@secondoftwo}%
\providecommand \bibfield  [0]{\@secondoftwo}%
\providecommand \translation [1]{[#1]}%
\providecommand \BibitemOpen [0]{}%
\providecommand \bibitemStop [0]{}%
\providecommand \bibitemNoStop [0]{.\EOS\space}%
\providecommand \EOS [0]{\spacefactor3000\relax}%
\providecommand \BibitemShut  [1]{\csname bibitem#1\endcsname}%
\let\auto@bib@innerbib\@empty
\bibitem [{\citenamefont {Ramaswamy}(2010)}]{ramaswamy2010mechanics}%
  \BibitemOpen
  \bibfield  {author} {\bibinfo {author} {\bibfnamefont {S.}~\bibnamefont
  {Ramaswamy}},\ }\href@noop {} {\bibfield  {journal} {\bibinfo  {journal}
  {Annual Review of Condensed Matter Physics}\ }\textbf {\bibinfo {volume}
  {1}},\ \bibinfo {pages} {323} (\bibinfo {year} {2010})}\BibitemShut {NoStop}%
\bibitem [{\citenamefont {Marchetti}\ \emph {et~al.}(2013)\citenamefont
  {Marchetti}, \citenamefont {Joanny}, \citenamefont {Ramaswamy}, \citenamefont
  {Liverpool}, \citenamefont {Prost}, \citenamefont {Rao},\ and\ \citenamefont
  {Simha}}]{marchetti2013hydrodynamics}%
  \BibitemOpen
  \bibfield  {author} {\bibinfo {author} {\bibfnamefont {M.~C.}\ \bibnamefont
  {Marchetti}}, \bibinfo {author} {\bibfnamefont {J.~F.}\ \bibnamefont
  {Joanny}}, \bibinfo {author} {\bibfnamefont {S.}~\bibnamefont {Ramaswamy}},
  \bibinfo {author} {\bibfnamefont {T.~B.}\ \bibnamefont {Liverpool}}, \bibinfo
  {author} {\bibfnamefont {J.}~\bibnamefont {Prost}}, \bibinfo {author}
  {\bibfnamefont {M.}~\bibnamefont {Rao}}, \ and\ \bibinfo {author}
  {\bibfnamefont {R.~A.}\ \bibnamefont {Simha}},\ }\href@noop {} {\bibfield
  {journal} {\bibinfo  {journal} {Reviews of Modern Physics}\ }\textbf
  {\bibinfo {volume} {85}},\ \bibinfo {pages} {1143} (\bibinfo {year}
  {2013})}\BibitemShut {NoStop}%
\bibitem [{\citenamefont {Gompper}\ \emph {et~al.}(2020)\citenamefont
  {Gompper}, \citenamefont {Winkler}, \citenamefont {Speck}, \citenamefont
  {Solon}, \citenamefont {Nardini}, \citenamefont {Peruani}, \citenamefont
  {L{\"o}wen}, \citenamefont {Golestanian}, \citenamefont {Kaupp},
  \citenamefont {Alvarez} \emph {et~al.}}]{gompper20202020}%
  \BibitemOpen
  \bibfield  {author} {\bibinfo {author} {\bibfnamefont {G.}~\bibnamefont
  {Gompper}}, \bibinfo {author} {\bibfnamefont {R.~G.}\ \bibnamefont
  {Winkler}}, \bibinfo {author} {\bibfnamefont {T.}~\bibnamefont {Speck}},
  \bibinfo {author} {\bibfnamefont {A.}~\bibnamefont {Solon}}, \bibinfo
  {author} {\bibfnamefont {C.}~\bibnamefont {Nardini}}, \bibinfo {author}
  {\bibfnamefont {F.}~\bibnamefont {Peruani}}, \bibinfo {author} {\bibfnamefont
  {H.}~\bibnamefont {L{\"o}wen}}, \bibinfo {author} {\bibfnamefont
  {R.}~\bibnamefont {Golestanian}}, \bibinfo {author} {\bibfnamefont {U.~B.}\
  \bibnamefont {Kaupp}}, \bibinfo {author} {\bibfnamefont {L.}~\bibnamefont
  {Alvarez}},  \emph {et~al.},\ }\href@noop {} {\bibfield  {journal} {\bibinfo
  {journal} {Journal of Physics: Condensed Matter}\ }\textbf {\bibinfo {volume}
  {32}},\ \bibinfo {pages} {193001} (\bibinfo {year} {2020})}\BibitemShut
  {NoStop}%
\bibitem [{\citenamefont {Needleman}\ and\ \citenamefont
  {Dogic}(2017)}]{needleman2017active}%
  \BibitemOpen
  \bibfield  {author} {\bibinfo {author} {\bibfnamefont {D.}~\bibnamefont
  {Needleman}}\ and\ \bibinfo {author} {\bibfnamefont {Z.}~\bibnamefont
  {Dogic}},\ }\href@noop {} {\bibfield  {journal} {\bibinfo  {journal} {Nature
  Reviews Materials}\ }\textbf {\bibinfo {volume} {2}},\ \bibinfo {pages} {1}
  (\bibinfo {year} {2017})}\BibitemShut {NoStop}%
\bibitem [{\citenamefont {Vernerey}\ \emph {et~al.}(2019)\citenamefont
  {Vernerey}, \citenamefont {Benet}, \citenamefont {Blue}, \citenamefont
  {Fajrial}, \citenamefont {Sridhar}, \citenamefont {Lum}, \citenamefont
  {Shakya}, \citenamefont {Song}, \citenamefont {Thomas},\ and\ \citenamefont
  {Borden}}]{vernerey2019biological}%
  \BibitemOpen
  \bibfield  {author} {\bibinfo {author} {\bibfnamefont {F.~J.}\ \bibnamefont
  {Vernerey}}, \bibinfo {author} {\bibfnamefont {E.}~\bibnamefont {Benet}},
  \bibinfo {author} {\bibfnamefont {L.}~\bibnamefont {Blue}}, \bibinfo {author}
  {\bibfnamefont {A.~K.}\ \bibnamefont {Fajrial}}, \bibinfo {author}
  {\bibfnamefont {S.~L.}\ \bibnamefont {Sridhar}}, \bibinfo {author}
  {\bibfnamefont {J.~S.}\ \bibnamefont {Lum}}, \bibinfo {author} {\bibfnamefont
  {G.}~\bibnamefont {Shakya}}, \bibinfo {author} {\bibfnamefont {K.~H.}\
  \bibnamefont {Song}}, \bibinfo {author} {\bibfnamefont {A.~N.}\ \bibnamefont
  {Thomas}}, \ and\ \bibinfo {author} {\bibfnamefont {M.~A.}\ \bibnamefont
  {Borden}},\ }\href@noop {} {\bibfield  {journal} {\bibinfo  {journal}
  {Advances in Colloid and Interface Science}\ }\textbf {\bibinfo {volume}
  {263}},\ \bibinfo {pages} {38} (\bibinfo {year} {2019})}\BibitemShut
  {NoStop}%
\bibitem [{\citenamefont {Ghosh}\ \emph {et~al.}(2021)\citenamefont {Ghosh},
  \citenamefont {Somasundar},\ and\ \citenamefont {Sen}}]{ghosh2021enzymes}%
  \BibitemOpen
  \bibfield  {author} {\bibinfo {author} {\bibfnamefont {S.}~\bibnamefont
  {Ghosh}}, \bibinfo {author} {\bibfnamefont {A.}~\bibnamefont {Somasundar}}, \
  and\ \bibinfo {author} {\bibfnamefont {A.}~\bibnamefont {Sen}},\ }\href@noop
  {} {\bibfield  {journal} {\bibinfo  {journal} {Annual Review of Condensed
  Matter Physics}\ }\textbf {\bibinfo {volume} {12}},\ \bibinfo {pages} {177}
  (\bibinfo {year} {2021})}\BibitemShut {NoStop}%
\bibitem [{\citenamefont {B{\"a}r}\ \emph {et~al.}(2020)\citenamefont
  {B{\"a}r}, \citenamefont {Gro{\ss}mann}, \citenamefont {Heidenreich},\ and\
  \citenamefont {Peruani}}]{bar2020self}%
  \BibitemOpen
  \bibfield  {author} {\bibinfo {author} {\bibfnamefont {M.}~\bibnamefont
  {B{\"a}r}}, \bibinfo {author} {\bibfnamefont {R.}~\bibnamefont
  {Gro{\ss}mann}}, \bibinfo {author} {\bibfnamefont {S.}~\bibnamefont
  {Heidenreich}}, \ and\ \bibinfo {author} {\bibfnamefont {F.}~\bibnamefont
  {Peruani}},\ }\href@noop {} {\bibfield  {journal} {\bibinfo  {journal}
  {Annual Review of Condensed Matter Physics}\ }\textbf {\bibinfo {volume}
  {11}},\ \bibinfo {pages} {441} (\bibinfo {year} {2020})}\BibitemShut
  {NoStop}%
\bibitem [{\citenamefont {Shaebani}\ \emph {et~al.}(2020)\citenamefont
  {Shaebani}, \citenamefont {Wysocki}, \citenamefont {Winkler}, \citenamefont
  {Gompper},\ and\ \citenamefont {Rieger}}]{shaebani2020computational}%
  \BibitemOpen
  \bibfield  {author} {\bibinfo {author} {\bibfnamefont {M.~R.}\ \bibnamefont
  {Shaebani}}, \bibinfo {author} {\bibfnamefont {A.}~\bibnamefont {Wysocki}},
  \bibinfo {author} {\bibfnamefont {R.~G.}\ \bibnamefont {Winkler}}, \bibinfo
  {author} {\bibfnamefont {G.}~\bibnamefont {Gompper}}, \ and\ \bibinfo
  {author} {\bibfnamefont {H.}~\bibnamefont {Rieger}},\ }\href@noop {}
  {\bibfield  {journal} {\bibinfo  {journal} {Nature Reviews Physics}\ }\textbf
  {\bibinfo {volume} {2}},\ \bibinfo {pages} {181} (\bibinfo {year}
  {2020})}\BibitemShut {NoStop}%
\bibitem [{\citenamefont {Sanchez}\ \emph {et~al.}(2012)\citenamefont
  {Sanchez}, \citenamefont {Chen}, \citenamefont {DeCamp}, \citenamefont
  {Heymann},\ and\ \citenamefont {Dogic}}]{sanchez2012spontaneous}%
  \BibitemOpen
  \bibfield  {author} {\bibinfo {author} {\bibfnamefont {T.}~\bibnamefont
  {Sanchez}}, \bibinfo {author} {\bibfnamefont {D.~T.~N.}\ \bibnamefont
  {Chen}}, \bibinfo {author} {\bibfnamefont {S.~J.}\ \bibnamefont {DeCamp}},
  \bibinfo {author} {\bibfnamefont {M.}~\bibnamefont {Heymann}}, \ and\
  \bibinfo {author} {\bibfnamefont {Z.}~\bibnamefont {Dogic}},\ }\href@noop {}
  {\bibfield  {journal} {\bibinfo  {journal} {Nature}\ }\textbf {\bibinfo
  {volume} {491}},\ \bibinfo {pages} {431} (\bibinfo {year}
  {2012})}\BibitemShut {NoStop}%
\bibitem [{\citenamefont {Peruani}\ \emph {et~al.}(2012)\citenamefont
  {Peruani}, \citenamefont {Starru{\ss}}, \citenamefont {Jakovljevic},
  \citenamefont {S{\o}gaard-Andersen}, \citenamefont {Deutsch},\ and\
  \citenamefont {B{\"a}r}}]{peruani2012collective}%
  \BibitemOpen
  \bibfield  {author} {\bibinfo {author} {\bibfnamefont {F.}~\bibnamefont
  {Peruani}}, \bibinfo {author} {\bibfnamefont {J.}~\bibnamefont
  {Starru{\ss}}}, \bibinfo {author} {\bibfnamefont {V.}~\bibnamefont
  {Jakovljevic}}, \bibinfo {author} {\bibfnamefont {L.}~\bibnamefont
  {S{\o}gaard-Andersen}}, \bibinfo {author} {\bibfnamefont {A.}~\bibnamefont
  {Deutsch}}, \ and\ \bibinfo {author} {\bibfnamefont {M.}~\bibnamefont
  {B{\"a}r}},\ }\href@noop {} {\bibfield  {journal} {\bibinfo  {journal}
  {Physical Review Letters}\ }\textbf {\bibinfo {volume} {108}},\ \bibinfo
  {pages} {098102} (\bibinfo {year} {2012})}\BibitemShut {NoStop}%
\bibitem [{\citenamefont {Garcimart{\'\i}n}\ \emph {et~al.}(2015)\citenamefont
  {Garcimart{\'\i}n}, \citenamefont {Pastor}, \citenamefont {Ferrer},
  \citenamefont {Ramos}, \citenamefont {Mart{\'\i}n-G{\'o}mez},\ and\
  \citenamefont {Zuriguel}}]{garcimartin2015flow}%
  \BibitemOpen
  \bibfield  {author} {\bibinfo {author} {\bibfnamefont {A.}~\bibnamefont
  {Garcimart{\'\i}n}}, \bibinfo {author} {\bibfnamefont {J.~M.}\ \bibnamefont
  {Pastor}}, \bibinfo {author} {\bibfnamefont {L.~M.}\ \bibnamefont {Ferrer}},
  \bibinfo {author} {\bibfnamefont {J.~J.}\ \bibnamefont {Ramos}}, \bibinfo
  {author} {\bibfnamefont {C.}~\bibnamefont {Mart{\'\i}n-G{\'o}mez}}, \ and\
  \bibinfo {author} {\bibfnamefont {I.}~\bibnamefont {Zuriguel}},\ }\href@noop
  {} {\bibfield  {journal} {\bibinfo  {journal} {Physical Review E}\ }\textbf
  {\bibinfo {volume} {91}},\ \bibinfo {pages} {022808} (\bibinfo {year}
  {2015})}\BibitemShut {NoStop}%
\bibitem [{\citenamefont {Becco}\ \emph {et~al.}(2006)\citenamefont {Becco},
  \citenamefont {Vandewalle}, \citenamefont {Delcourt},\ and\ \citenamefont
  {Poncin}}]{becco2006experimental}%
  \BibitemOpen
  \bibfield  {author} {\bibinfo {author} {\bibfnamefont {C.}~\bibnamefont
  {Becco}}, \bibinfo {author} {\bibfnamefont {N.}~\bibnamefont {Vandewalle}},
  \bibinfo {author} {\bibfnamefont {J.}~\bibnamefont {Delcourt}}, \ and\
  \bibinfo {author} {\bibfnamefont {P.}~\bibnamefont {Poncin}},\ }\href@noop {}
  {\bibfield  {journal} {\bibinfo  {journal} {Physica A: Statistical Mechanics
  and its Applications}\ }\textbf {\bibinfo {volume} {367}},\ \bibinfo {pages}
  {487} (\bibinfo {year} {2006})}\BibitemShut {NoStop}%
\bibitem [{\citenamefont {Calovi}\ \emph {et~al.}(2014)\citenamefont {Calovi},
  \citenamefont {Lopez}, \citenamefont {Ngo}, \citenamefont {Sire},
  \citenamefont {Chat{\'e}},\ and\ \citenamefont
  {Theraulaz}}]{calovi2014swarming}%
  \BibitemOpen
  \bibfield  {author} {\bibinfo {author} {\bibfnamefont {D.~S.}\ \bibnamefont
  {Calovi}}, \bibinfo {author} {\bibfnamefont {U.}~\bibnamefont {Lopez}},
  \bibinfo {author} {\bibfnamefont {S.}~\bibnamefont {Ngo}}, \bibinfo {author}
  {\bibfnamefont {C.}~\bibnamefont {Sire}}, \bibinfo {author} {\bibfnamefont
  {H.}~\bibnamefont {Chat{\'e}}}, \ and\ \bibinfo {author} {\bibfnamefont
  {G.}~\bibnamefont {Theraulaz}},\ }\href@noop {} {\bibfield  {journal}
  {\bibinfo  {journal} {New Journal of Physics}\ }\textbf {\bibinfo {volume}
  {16}},\ \bibinfo {pages} {015026} (\bibinfo {year} {2014})}\BibitemShut
  {NoStop}%
\bibitem [{\citenamefont {Steager}\ \emph {et~al.}(2008)\citenamefont
  {Steager}, \citenamefont {Kim},\ and\ \citenamefont
  {Kim}}]{steager2008dynamics}%
  \BibitemOpen
  \bibfield  {author} {\bibinfo {author} {\bibfnamefont {E.~B.}\ \bibnamefont
  {Steager}}, \bibinfo {author} {\bibfnamefont {C.~B.}\ \bibnamefont {Kim}}, \
  and\ \bibinfo {author} {\bibfnamefont {M.~J.}\ \bibnamefont {Kim}},\
  }\href@noop {} {\bibfield  {journal} {\bibinfo  {journal} {Physics of
  Fluids}\ }\textbf {\bibinfo {volume} {20}} (\bibinfo {year}
  {2008})}\BibitemShut {NoStop}%
\bibitem [{\citenamefont {Schaller}\ \emph {et~al.}(2010)\citenamefont
  {Schaller}, \citenamefont {Weber}, \citenamefont {Semmrich}, \citenamefont
  {Frey},\ and\ \citenamefont {Bausch}}]{schaller2010polar}%
  \BibitemOpen
  \bibfield  {author} {\bibinfo {author} {\bibfnamefont {V.}~\bibnamefont
  {Schaller}}, \bibinfo {author} {\bibfnamefont {C.}~\bibnamefont {Weber}},
  \bibinfo {author} {\bibfnamefont {C.}~\bibnamefont {Semmrich}}, \bibinfo
  {author} {\bibfnamefont {E.}~\bibnamefont {Frey}}, \ and\ \bibinfo {author}
  {\bibfnamefont {A.~R.}\ \bibnamefont {Bausch}},\ }\href@noop {} {\bibfield
  {journal} {\bibinfo  {journal} {Nature}\ }\textbf {\bibinfo {volume} {467}},\
  \bibinfo {pages} {73} (\bibinfo {year} {2010})}\BibitemShut {NoStop}%
\bibitem [{\citenamefont {Sumino}\ \emph {et~al.}(2012)\citenamefont {Sumino},
  \citenamefont {Nagai}, \citenamefont {Shitaka}, \citenamefont {Tanaka},
  \citenamefont {Yoshikawa}, \citenamefont {Chat{\'e}},\ and\ \citenamefont
  {Oiwa}}]{sumino2012large}%
  \BibitemOpen
  \bibfield  {author} {\bibinfo {author} {\bibfnamefont {Y.}~\bibnamefont
  {Sumino}}, \bibinfo {author} {\bibfnamefont {K.~H.}\ \bibnamefont {Nagai}},
  \bibinfo {author} {\bibfnamefont {Y.}~\bibnamefont {Shitaka}}, \bibinfo
  {author} {\bibfnamefont {D.}~\bibnamefont {Tanaka}}, \bibinfo {author}
  {\bibfnamefont {K.}~\bibnamefont {Yoshikawa}}, \bibinfo {author}
  {\bibfnamefont {H.}~\bibnamefont {Chat{\'e}}}, \ and\ \bibinfo {author}
  {\bibfnamefont {K.}~\bibnamefont {Oiwa}},\ }\href@noop {} {\bibfield
  {journal} {\bibinfo  {journal} {Nature}\ }\textbf {\bibinfo {volume} {483}},\
  \bibinfo {pages} {448} (\bibinfo {year} {2012})}\BibitemShut {NoStop}%
\bibitem [{\citenamefont {Veigel}\ and\ \citenamefont
  {Schmidt}(2011)}]{veigel2011moving}%
  \BibitemOpen
  \bibfield  {author} {\bibinfo {author} {\bibfnamefont {C.}~\bibnamefont
  {Veigel}}\ and\ \bibinfo {author} {\bibfnamefont {C.~F.}\ \bibnamefont
  {Schmidt}},\ }\href@noop {} {\bibfield  {journal} {\bibinfo  {journal}
  {Nature Reviews Molecular Cell Biology}\ }\textbf {\bibinfo {volume} {12}},\
  \bibinfo {pages} {163} (\bibinfo {year} {2011})}\BibitemShut {NoStop}%
\bibitem [{\citenamefont {Wong}\ \emph {et~al.}(2016)\citenamefont {Wong},
  \citenamefont {Dey},\ and\ \citenamefont {Sen}}]{wong2016synthetic}%
  \BibitemOpen
  \bibfield  {author} {\bibinfo {author} {\bibfnamefont {F.}~\bibnamefont
  {Wong}}, \bibinfo {author} {\bibfnamefont {K.~K.}\ \bibnamefont {Dey}}, \
  and\ \bibinfo {author} {\bibfnamefont {A.}~\bibnamefont {Sen}},\ }\href@noop
  {} {\bibfield  {journal} {\bibinfo  {journal} {Annual Review of Materials
  Research}\ }\textbf {\bibinfo {volume} {46}},\ \bibinfo {pages} {407}
  (\bibinfo {year} {2016})}\BibitemShut {NoStop}%
\bibitem [{\citenamefont {Vicsek}\ \emph {et~al.}(1995)\citenamefont {Vicsek},
  \citenamefont {Czir{\'o}k}, \citenamefont {Ben-Jacob}, \citenamefont
  {Cohen},\ and\ \citenamefont {Shochet}}]{vicsek1995novel}%
  \BibitemOpen
  \bibfield  {author} {\bibinfo {author} {\bibfnamefont {T.}~\bibnamefont
  {Vicsek}}, \bibinfo {author} {\bibfnamefont {A.}~\bibnamefont {Czir{\'o}k}},
  \bibinfo {author} {\bibfnamefont {E.}~\bibnamefont {Ben-Jacob}}, \bibinfo
  {author} {\bibfnamefont {I.}~\bibnamefont {Cohen}}, \ and\ \bibinfo {author}
  {\bibfnamefont {O.}~\bibnamefont {Shochet}},\ }\href@noop {} {\bibfield
  {journal} {\bibinfo  {journal} {Physical Review Letters}\ }\textbf {\bibinfo
  {volume} {75}},\ \bibinfo {pages} {1226} (\bibinfo {year}
  {1995})}\BibitemShut {NoStop}%
\bibitem [{\citenamefont {Toner}\ and\ \citenamefont
  {Tu}(1995)}]{toner1995long}%
  \BibitemOpen
  \bibfield  {author} {\bibinfo {author} {\bibfnamefont {J.}~\bibnamefont
  {Toner}}\ and\ \bibinfo {author} {\bibfnamefont {Y.}~\bibnamefont {Tu}},\
  }\href@noop {} {\bibfield  {journal} {\bibinfo  {journal} {Physical Review
  Letters}\ }\textbf {\bibinfo {volume} {75}},\ \bibinfo {pages} {4326}
  (\bibinfo {year} {1995})}\BibitemShut {NoStop}%
\bibitem [{\citenamefont {Toner}\ and\ \citenamefont
  {Tu}(1998)}]{toner1998flocks}%
  \BibitemOpen
  \bibfield  {author} {\bibinfo {author} {\bibfnamefont {J.}~\bibnamefont
  {Toner}}\ and\ \bibinfo {author} {\bibfnamefont {Y.}~\bibnamefont {Tu}},\
  }\href@noop {} {\bibfield  {journal} {\bibinfo  {journal} {Physical Review
  E}\ }\textbf {\bibinfo {volume} {58}},\ \bibinfo {pages} {4828} (\bibinfo
  {year} {1998})}\BibitemShut {NoStop}%
\bibitem [{\citenamefont {Toner}(2012)}]{toner2012reanalysis}%
  \BibitemOpen
  \bibfield  {author} {\bibinfo {author} {\bibfnamefont {J.}~\bibnamefont
  {Toner}},\ }\href@noop {} {\bibfield  {journal} {\bibinfo  {journal}
  {Physical Review E}\ }\textbf {\bibinfo {volume} {86}},\ \bibinfo {pages}
  {031918} (\bibinfo {year} {2012})}\BibitemShut {NoStop}%
\bibitem [{\citenamefont {Solon}\ \emph
  {et~al.}(2015{\natexlab{a}})\citenamefont {Solon}, \citenamefont
  {Chat{\'e}},\ and\ \citenamefont {Tailleur}}]{solon2015phase}%
  \BibitemOpen
  \bibfield  {author} {\bibinfo {author} {\bibfnamefont {A.~P.}\ \bibnamefont
  {Solon}}, \bibinfo {author} {\bibfnamefont {H.}~\bibnamefont {Chat{\'e}}}, \
  and\ \bibinfo {author} {\bibfnamefont {J.}~\bibnamefont {Tailleur}},\
  }\href@noop {} {\bibfield  {journal} {\bibinfo  {journal} {Physical Review
  Letters}\ }\textbf {\bibinfo {volume} {114}},\ \bibinfo {pages} {068101}
  (\bibinfo {year} {2015}{\natexlab{a}})}\BibitemShut {NoStop}%
\bibitem [{\citenamefont {Chat{\'e}}\ \emph {et~al.}(2008)\citenamefont
  {Chat{\'e}}, \citenamefont {Ginelli}, \citenamefont {Gr{\'e}goire},\ and\
  \citenamefont {Raynaud}}]{chate2008collective}%
  \BibitemOpen
  \bibfield  {author} {\bibinfo {author} {\bibfnamefont {H.}~\bibnamefont
  {Chat{\'e}}}, \bibinfo {author} {\bibfnamefont {F.}~\bibnamefont {Ginelli}},
  \bibinfo {author} {\bibfnamefont {G.}~\bibnamefont {Gr{\'e}goire}}, \ and\
  \bibinfo {author} {\bibfnamefont {F.}~\bibnamefont {Raynaud}},\ }\href@noop
  {} {\bibfield  {journal} {\bibinfo  {journal} {Physical Review E}\ }\textbf
  {\bibinfo {volume} {77}},\ \bibinfo {pages} {046113} (\bibinfo {year}
  {2008})}\BibitemShut {NoStop}%
\bibitem [{\citenamefont {Solon}\ \emph
  {et~al.}(2015{\natexlab{b}})\citenamefont {Solon}, \citenamefont {Caussin},
  \citenamefont {Bartolo}, \citenamefont {Chat{\'e}},\ and\ \citenamefont
  {Tailleur}}]{solon2015pattern}%
  \BibitemOpen
  \bibfield  {author} {\bibinfo {author} {\bibfnamefont {A.~P.}\ \bibnamefont
  {Solon}}, \bibinfo {author} {\bibfnamefont {J.~B.}\ \bibnamefont {Caussin}},
  \bibinfo {author} {\bibfnamefont {D.}~\bibnamefont {Bartolo}}, \bibinfo
  {author} {\bibfnamefont {H.}~\bibnamefont {Chat{\'e}}}, \ and\ \bibinfo
  {author} {\bibfnamefont {J.}~\bibnamefont {Tailleur}},\ }\href@noop {}
  {\bibfield  {journal} {\bibinfo  {journal} {Physical Review E}\ }\textbf
  {\bibinfo {volume} {92}},\ \bibinfo {pages} {062111} (\bibinfo {year}
  {2015}{\natexlab{b}})}\BibitemShut {NoStop}%
\bibitem [{\citenamefont {K{\"u}rsten}\ and\ \citenamefont
  {Ihle}(2020)}]{kursten2020dry}%
  \BibitemOpen
  \bibfield  {author} {\bibinfo {author} {\bibfnamefont {R.}~\bibnamefont
  {K{\"u}rsten}}\ and\ \bibinfo {author} {\bibfnamefont {T.}~\bibnamefont
  {Ihle}},\ }\href@noop {} {\bibfield  {journal} {\bibinfo  {journal} {Physical
  Review Letters}\ }\textbf {\bibinfo {volume} {125}},\ \bibinfo {pages}
  {188003} (\bibinfo {year} {2020})}\BibitemShut {NoStop}%
\bibitem [{\citenamefont {Solon}\ \emph {et~al.}(2022)\citenamefont {Solon},
  \citenamefont {Chat{\'e}}, \citenamefont {Toner},\ and\ \citenamefont
  {Tailleur}}]{solon2022susceptibility}%
  \BibitemOpen
  \bibfield  {author} {\bibinfo {author} {\bibfnamefont {A.}~\bibnamefont
  {Solon}}, \bibinfo {author} {\bibfnamefont {H.}~\bibnamefont {Chat{\'e}}},
  \bibinfo {author} {\bibfnamefont {J.}~\bibnamefont {Toner}}, \ and\ \bibinfo
  {author} {\bibfnamefont {J.}~\bibnamefont {Tailleur}},\ }\href@noop {}
  {\bibfield  {journal} {\bibinfo  {journal} {Physical Review Letters}\
  }\textbf {\bibinfo {volume} {128}},\ \bibinfo {pages} {208004} (\bibinfo
  {year} {2022})}\BibitemShut {NoStop}%
\bibitem [{\citenamefont {Chatterjee}\ \emph {et~al.}(2022)\citenamefont
  {Chatterjee}, \citenamefont {Mangeat},\ and\ \citenamefont
  {Rieger}}]{chatterjee2022polar}%
  \BibitemOpen
  \bibfield  {author} {\bibinfo {author} {\bibfnamefont {S.}~\bibnamefont
  {Chatterjee}}, \bibinfo {author} {\bibfnamefont {M.}~\bibnamefont {Mangeat}},
  \ and\ \bibinfo {author} {\bibfnamefont {H.}~\bibnamefont {Rieger}},\
  }\href@noop {} {\bibfield  {journal} {\bibinfo  {journal} {Europhysics
  Letters}\ }\textbf {\bibinfo {volume} {138}},\ \bibinfo {pages} {41001}
  (\bibinfo {year} {2022})}\BibitemShut {NoStop}%
\bibitem [{\citenamefont {Chatterjee}\ \emph {et~al.}(2020)\citenamefont
  {Chatterjee}, \citenamefont {Mangeat}, \citenamefont {Paul},\ and\
  \citenamefont {Rieger}}]{chatterjee2020flocking}%
  \BibitemOpen
  \bibfield  {author} {\bibinfo {author} {\bibfnamefont {S.}~\bibnamefont
  {Chatterjee}}, \bibinfo {author} {\bibfnamefont {M.}~\bibnamefont {Mangeat}},
  \bibinfo {author} {\bibfnamefont {R.}~\bibnamefont {Paul}}, \ and\ \bibinfo
  {author} {\bibfnamefont {H.}~\bibnamefont {Rieger}},\ }\href@noop {}
  {\bibfield  {journal} {\bibinfo  {journal} {Europhysics Letters}\ }\textbf
  {\bibinfo {volume} {130}},\ \bibinfo {pages} {66001} (\bibinfo {year}
  {2020})}\BibitemShut {NoStop}%
\bibitem [{\citenamefont {Mangeat}\ \emph {et~al.}(2020)\citenamefont
  {Mangeat}, \citenamefont {Chatterjee}, \citenamefont {Paul},\ and\
  \citenamefont {Rieger}}]{mangeat2020flocking}%
  \BibitemOpen
  \bibfield  {author} {\bibinfo {author} {\bibfnamefont {M.}~\bibnamefont
  {Mangeat}}, \bibinfo {author} {\bibfnamefont {S.}~\bibnamefont {Chatterjee}},
  \bibinfo {author} {\bibfnamefont {R.}~\bibnamefont {Paul}}, \ and\ \bibinfo
  {author} {\bibfnamefont {H.}~\bibnamefont {Rieger}},\ }\href@noop {}
  {\bibfield  {journal} {\bibinfo  {journal} {Physical Review E}\ }\textbf
  {\bibinfo {volume} {102}},\ \bibinfo {pages} {042601} (\bibinfo {year}
  {2020})}\BibitemShut {NoStop}%
\bibitem [{\citenamefont {Codina}\ \emph {et~al.}(2022)\citenamefont {Codina},
  \citenamefont {Mahault}, \citenamefont {Chat{\'e}}, \citenamefont {Dobnikar},
  \citenamefont {Pagonabarraga},\ and\ \citenamefont {Shi}}]{codina2022small}%
  \BibitemOpen
  \bibfield  {author} {\bibinfo {author} {\bibfnamefont {J.}~\bibnamefont
  {Codina}}, \bibinfo {author} {\bibfnamefont {B.}~\bibnamefont {Mahault}},
  \bibinfo {author} {\bibfnamefont {H.}~\bibnamefont {Chat{\'e}}}, \bibinfo
  {author} {\bibfnamefont {J.}~\bibnamefont {Dobnikar}}, \bibinfo {author}
  {\bibfnamefont {I.}~\bibnamefont {Pagonabarraga}}, \ and\ \bibinfo {author}
  {\bibfnamefont {X.}~\bibnamefont {Shi}},\ }\href@noop {} {\bibfield
  {journal} {\bibinfo  {journal} {Physical Review Letters}\ }\textbf {\bibinfo
  {volume} {128}},\ \bibinfo {pages} {218001} (\bibinfo {year}
  {2022})}\BibitemShut {NoStop}%
\bibitem [{\citenamefont {Benvegnen}\ \emph {et~al.}(2023)\citenamefont
  {Benvegnen}, \citenamefont {Granek}, \citenamefont {Ro}, \citenamefont
  {Yaacoby}, \citenamefont {Chat{\'e}}, \citenamefont {Kafri}, \citenamefont
  {Mukamel}, \citenamefont {Solon},\ and\ \citenamefont
  {Tailleur}}]{benvegnen2023metastability}%
  \BibitemOpen
  \bibfield  {author} {\bibinfo {author} {\bibfnamefont {B.}~\bibnamefont
  {Benvegnen}}, \bibinfo {author} {\bibfnamefont {O.}~\bibnamefont {Granek}},
  \bibinfo {author} {\bibfnamefont {S.}~\bibnamefont {Ro}}, \bibinfo {author}
  {\bibfnamefont {R.}~\bibnamefont {Yaacoby}}, \bibinfo {author} {\bibfnamefont
  {H.}~\bibnamefont {Chat{\'e}}}, \bibinfo {author} {\bibfnamefont
  {Y.}~\bibnamefont {Kafri}}, \bibinfo {author} {\bibfnamefont
  {D.}~\bibnamefont {Mukamel}}, \bibinfo {author} {\bibfnamefont
  {A.}~\bibnamefont {Solon}}, \ and\ \bibinfo {author} {\bibfnamefont
  {J.}~\bibnamefont {Tailleur}},\ }\href@noop {} {\bibfield  {journal}
  {\bibinfo  {journal} {Physical Review Letters}\ }\textbf {\bibinfo {volume}
  {131}},\ \bibinfo {pages} {218301} (\bibinfo {year} {2023})}\BibitemShut
  {NoStop}%
\bibitem [{\citenamefont {Solon}\ and\ \citenamefont
  {Tailleur}(2015)}]{solon2015flocking}%
  \BibitemOpen
  \bibfield  {author} {\bibinfo {author} {\bibfnamefont {A.~P.}\ \bibnamefont
  {Solon}}\ and\ \bibinfo {author} {\bibfnamefont {J.}~\bibnamefont
  {Tailleur}},\ }\href@noop {} {\bibfield  {journal} {\bibinfo  {journal}
  {Physical Review E}\ }\textbf {\bibinfo {volume} {92}},\ \bibinfo {pages}
  {042119} (\bibinfo {year} {2015})}\BibitemShut {NoStop}%
\bibitem [{\citenamefont {Benvegnen}\ \emph {et~al.}(2022)\citenamefont
  {Benvegnen}, \citenamefont {Chat\'e}, \citenamefont {Krapivsky},
  \citenamefont {Tailleur},\ and\ \citenamefont {Solon}}]{solon1dAIM}%
  \BibitemOpen
  \bibfield  {author} {\bibinfo {author} {\bibfnamefont {B.}~\bibnamefont
  {Benvegnen}}, \bibinfo {author} {\bibfnamefont {H.}~\bibnamefont {Chat\'e}},
  \bibinfo {author} {\bibfnamefont {P.~L.}\ \bibnamefont {Krapivsky}}, \bibinfo
  {author} {\bibfnamefont {J.}~\bibnamefont {Tailleur}}, \ and\ \bibinfo
  {author} {\bibfnamefont {A.}~\bibnamefont {Solon}},\ }\href@noop {}
  {\bibfield  {journal} {\bibinfo  {journal} {Physical Review E}\ }\textbf
  {\bibinfo {volume} {106}},\ \bibinfo {pages} {054608} (\bibinfo {year}
  {2022})}\BibitemShut {NoStop}%
\bibitem [{\citenamefont {Xue}\ \emph {et~al.}(2023)\citenamefont {Xue},
  \citenamefont {Li}, \citenamefont {Chen}, \citenamefont {Chen},\ and\
  \citenamefont {Han}}]{xue2023machine}%
  \BibitemOpen
  \bibfield  {author} {\bibinfo {author} {\bibfnamefont {T.}~\bibnamefont
  {Xue}}, \bibinfo {author} {\bibfnamefont {X.}~\bibnamefont {Li}}, \bibinfo
  {author} {\bibfnamefont {X.}~\bibnamefont {Chen}}, \bibinfo {author}
  {\bibfnamefont {L.}~\bibnamefont {Chen}}, \ and\ \bibinfo {author}
  {\bibfnamefont {Z.}~\bibnamefont {Han}},\ }\href@noop {} {\bibfield
  {journal} {\bibinfo  {journal} {Machine Learning: Science and Technology}\
  }\textbf {\bibinfo {volume} {4}},\ \bibinfo {pages} {015028} (\bibinfo {year}
  {2023})}\BibitemShut {NoStop}%
\bibitem [{\citenamefont {K{\"u}rsten}\ \emph {et~al.}(2020)\citenamefont
  {K{\"u}rsten}, \citenamefont {Stroteich}, \citenamefont {H{\'e}rnandez},\
  and\ \citenamefont {Ihle}}]{kursten2020multiple}%
  \BibitemOpen
  \bibfield  {author} {\bibinfo {author} {\bibfnamefont {R.}~\bibnamefont
  {K{\"u}rsten}}, \bibinfo {author} {\bibfnamefont {S.}~\bibnamefont
  {Stroteich}}, \bibinfo {author} {\bibfnamefont {M.~Z.}\ \bibnamefont
  {H{\'e}rnandez}}, \ and\ \bibinfo {author} {\bibfnamefont {T.}~\bibnamefont
  {Ihle}},\ }\href@noop {} {\bibfield  {journal} {\bibinfo  {journal} {Physical
  Review Letters}\ }\textbf {\bibinfo {volume} {124}},\ \bibinfo {pages}
  {088002} (\bibinfo {year} {2020})}\BibitemShut {NoStop}%
\bibitem [{\citenamefont {Chatterjee}\ \emph {et~al.}(2018)\citenamefont
  {Chatterjee}, \citenamefont {Puri},\ and\ \citenamefont
  {Paul}}]{clockmodel2018}%
  \BibitemOpen
  \bibfield  {author} {\bibinfo {author} {\bibfnamefont {S.}~\bibnamefont
  {Chatterjee}}, \bibinfo {author} {\bibfnamefont {S.}~\bibnamefont {Puri}}, \
  and\ \bibinfo {author} {\bibfnamefont {R.}~\bibnamefont {Paul}},\ }\href@noop
  {} {\bibfield  {journal} {\bibinfo  {journal} {Physical Review E}\ }\textbf
  {\bibinfo {volume} {98}},\ \bibinfo {pages} {032109} (\bibinfo {year}
  {2018})}\BibitemShut {NoStop}%
\bibitem [{\citenamefont {Chatterjee}\ \emph {et~al.}(shed)\citenamefont
  {Chatterjee}, \citenamefont {Karmakar}, \citenamefont {Mangeat},
  \citenamefont {Paul},\ and\ \citenamefont
  {Rieger}}]{swarnajit2023metastability}%
  \BibitemOpen
  \bibfield  {author} {\bibinfo {author} {\bibfnamefont {S.}~\bibnamefont
  {Chatterjee}}, \bibinfo {author} {\bibfnamefont {M.}~\bibnamefont
  {Karmakar}}, \bibinfo {author} {\bibfnamefont {M.}~\bibnamefont {Mangeat}},
  \bibinfo {author} {\bibfnamefont {R.}~\bibnamefont {Paul}}, \ and\ \bibinfo
  {author} {\bibfnamefont {H.}~\bibnamefont {Rieger}},\ }\href@noop {}
  {\enquote {\bibinfo {title} {Metastability of ordered phase in discretized
  flocking},}\ } (\bibinfo {year} {unpublished})\BibitemShut {NoStop}%
\end{thebibliography}

\begin{thebibliography}{4}%
\makeatletter
\providecommand \@ifxundefined [1]{%
 \@ifx{#1\undefined}
}%
\providecommand \@ifnum [1]{%
 \ifnum #1\expandafter \@firstoftwo
 \else \expandafter \@secondoftwo
 \fi
}%
\providecommand \@ifx [1]{%
 \ifx #1\expandafter \@firstoftwo
 \else \expandafter \@secondoftwo
 \fi
}%
\providecommand \natexlab [1]{#1}%
\providecommand \enquote  [1]{``#1''}%
\providecommand \bibnamefont  [1]{#1}%
\providecommand \bibfnamefont [1]{#1}%
\providecommand \citenamefont [1]{#1}%
\providecommand \href@noop [0]{\@secondoftwo}%
\providecommand \href [0]{\begingroup \@sanitize@url \@href}%
\providecommand \@href[1]{\@@startlink{#1}\@@href}%
\providecommand \@@href[1]{\endgroup#1\@@endlink}%
\providecommand \@sanitize@url [0]{\catcode `\\12\catcode `\$12\catcode
  `\&12\catcode `\#12\catcode `\^12\catcode `\_12\catcode `\%12\relax}%
\providecommand \@@startlink[1]{}%
\providecommand \@@endlink[0]{}%
\providecommand \url  [0]{\begingroup\@sanitize@url \@url }%
\providecommand \@url [1]{\endgroup\@href {#1}{\urlprefix }}%
\providecommand \urlprefix  [0]{URL }%
\providecommand \Eprint [0]{\href }%
\providecommand \doibase [0]{http://dx.doi.org/}%
\providecommand \selectlanguage [0]{\@gobble}%
\providecommand \bibinfo  [0]{\@secondoftwo}%
\providecommand \bibfield  [0]{\@secondoftwo}%
\providecommand \translation [1]{[#1]}%
\providecommand \BibitemOpen [0]{}%
\providecommand \bibitemStop [0]{}%
\providecommand \bibitemNoStop [0]{.\EOS\space}%
\providecommand \EOS [0]{\spacefactor3000\relax}%
\providecommand \BibitemShut  [1]{\csname bibitem#1\endcsname}%
\let\auto@bib@innerbib\@empty
\bibitem [{\citenamefont {Mangeat}\ \emph {et~al.}(2020)\citenamefont
  {Mangeat}, \citenamefont {Chatterjee}, \citenamefont {Paul},\ and\
  \citenamefont {Rieger}}]{mangeat2020flocking}%
  \BibitemOpen
  \bibfield  {author} {\bibinfo {author} {\bibfnamefont {M.}~\bibnamefont
  {Mangeat}}, \bibinfo {author} {\bibfnamefont {S.}~\bibnamefont {Chatterjee}},
  \bibinfo {author} {\bibfnamefont {R.}~\bibnamefont {Paul}}, \ and\ \bibinfo
  {author} {\bibfnamefont {H.}~\bibnamefont {Rieger}},\ }\href@noop {}
  {\bibfield  {journal} {\bibinfo  {journal} {Phys. Rev. E}\ }\textbf {\bibinfo
  {volume} {102}},\ \bibinfo {pages} {042601} (\bibinfo {year}
  {2020})}\BibitemShut {NoStop}%
\bibitem [{\citenamefont {Inc}(2017)}]{inc2017mathematica}%
  \BibitemOpen
  \bibfield  {author} {\bibinfo {author} {\bibfnamefont {W.~R.}\ \bibnamefont
  {Inc}},\ }\href@noop {} {\bibfield  {journal} {\bibinfo  {journal}
  {Champaign, I. L.}\ }\textbf {\bibinfo {volume} {473}} (\bibinfo {year}
  {2017})}\BibitemShut {NoStop}%
\bibitem [{\citenamefont {Kourbane-Houssene}\ \emph {et~al.}(2018)\citenamefont
  {Kourbane-Houssene}, \citenamefont {Erignoux}, \citenamefont {Bodineau},\
  and\ \citenamefont {Tailleur}}]{kourbane2018exact}%
  \BibitemOpen
  \bibfield  {author} {\bibinfo {author} {\bibfnamefont {M.}~\bibnamefont
  {Kourbane-Houssene}}, \bibinfo {author} {\bibfnamefont {C.}~\bibnamefont
  {Erignoux}}, \bibinfo {author} {\bibfnamefont {T.}~\bibnamefont {Bodineau}},
  \ and\ \bibinfo {author} {\bibfnamefont {J.}~\bibnamefont {Tailleur}},\
  }\href@noop {} {\bibfield  {journal} {\bibinfo  {journal} {Phys. Rev. Lett.}\
  }\textbf {\bibinfo {volume} {120}},\ \bibinfo {pages} {268003} (\bibinfo
  {year} {2018})}\BibitemShut {NoStop}%
\bibitem [{\citenamefont {Chatterjee}\ \emph {et~al.}(2020)\citenamefont
  {Chatterjee}, \citenamefont {Mangeat}, \citenamefont {Paul},\ and\
  \citenamefont {Rieger}}]{chatterjee2020flocking}%
  \BibitemOpen
  \bibfield  {author} {\bibinfo {author} {\bibfnamefont {S.}~\bibnamefont
  {Chatterjee}}, \bibinfo {author} {\bibfnamefont {M.}~\bibnamefont {Mangeat}},
  \bibinfo {author} {\bibfnamefont {R.}~\bibnamefont {Paul}}, \ and\ \bibinfo
  {author} {\bibfnamefont {H.}~\bibnamefont {Rieger}},\ }\href@noop {}
  {\bibfield  {journal} {\bibinfo  {journal} {EPL}\ }\textbf {\bibinfo {volume}
  {130}},\ \bibinfo {pages} {66001} (\bibinfo {year} {2020})}\BibitemShut
  {NoStop}%
\end{thebibliography}
\end{document}